\documentclass[preprint,aps,a4paper,amssymb,superscriptaddress,nofootinbib]{revtex4}
\usepackage{amsmath}
\usepackage{array}
\usepackage{cancel}
\usepackage[utf8]{inputenc}
\usepackage[export]{adjustbox} 
\usepackage{accents}
\usepackage{color}
\usepackage{graphicx}
\usepackage{bbold}
\usepackage{centernot} 		
\usepackage{tikz}		
\usepackage[compat=1.1.0]{tikz-feynman} 
\usepackage{hyperref} 
\hypersetup{
        colorlinks=true,
        linkcolor=blue,
        citecolor=red,
        filecolor=magenta,
        urlcolor=cyan
}
\DeclareFontFamily{OT1}{pzc}{}
\DeclareFontShape{OT1}{pzc}{m}{it}{<-> s * [1.20] pzcmi7t}{}
\DeclareMathAlphabet{\mathpzc}{OT1}{pzc}{m}{it}

\renewcommand{\cal}{\mathpzc}
\newcommand{\Case}[2]{{\textstyle \frac{#1}{#2}}}

\newcommand{\dsl}[1]{{\centernot{#1}}}	
\newcommand{\cross}[1]{{\mathchar'26\mkern-9mu {#1}}} 
\newcommand*\circled[1]{\tikz[baseline=(char.base)]{
            \node[shape=circle,draw,inner sep=2pt] (char) {#1};}}
\numberwithin{equation}{section}
 
\begin{document}
%
\preprint{~\hfill\today}

\title{Lectures on Introduction to Quantum Field Theory}
\author{Ghanashyam Date} \email{gdate@cmi.ac.in, ghanashyamdate@gmail.com}
\affiliation{Chennai Mathematical Institute\\ H1, SIPCOT IT Park,
Siruseri, Kelambakkam 603103, INDIA.}
%
\begin{abstract} 
These are lecture notes of the QFT-I course I gave in an online mode at
Chennai Mathematical Institute. The course focussed on the free
relativistic quantum fields, their interactions in the perturbative
scattering framework, standard computations of QED processes, radiative
corrections at 1-loop with renormalization and an introduction to the
toolbox of path integrals.
\end{abstract}


\maketitle

\tableofcontents

\newpage

\section{What is Quantum Field Theory?}\label{WhatIsQFT}

There is of course no short answer to this question. There are different
facets of quantum fields. An aspect that is used extensively in
condensed matter physics is the `QFT as a framework for computing
processes in interacting many body systems'. Here, each of the many
bodies typically has its internal energy states eg atoms/molecules at
various lattice sites, localized states in periodic potentials etc. The
interactions take place by making transitions among these internal
states. Since the internal states are discrete, the interactions also
involve discrete exchanges of energy/momentum/spin etc which is
described by a {\em quantum field} eg., $[\Psi(\vec{r}),
\Psi^{\dagger}(\vec{r}')] \sim \delta^3(\vec{r} - \vec{r}')$. With large
number of bodies involved, a quantum field capable of indefinite number
of transitions is a well suited framework.

There is another aspect which makes quantized fields {\em essential}
when we want to describe interactions of even few bodies in a
relativistically covariant manner. Special relativity imposes two
crucial modifications: (a) the notion of causality is modified by the
finite upper speed limit and (b) equivalence of mass and energy demands
that the rest masses be also included in the energy conservation. The
latter one allows annihilation and creation of `particles' at the
expense of energy. The framework of non-relativistic quantum mechanics
is not capable of accommodating these possibilities. Why is it so?

For this, we need to be sharper about what is meant by ``relativistic
covariance''.

\subsection{Relativistic Covariance}\label{RelCov}
Special relativity posits that the space-time that is appropriate arena
for describing physical processes is the Minkowski space-time -
$\mathbb{R}^4$ with the flat metric. This allows the space-time to be
described in terms of coordinates $x^{\mu} \leftrightarrow (t, x^i)$ and
metric $\eta^{\mu\nu} = diag( -1, 1, 1, 1) = \eta_{\mu\nu}$. All
inertial observers, use such a space-time and relate their descriptions
of the experiments by using the coordinate transformations: $x^{'\mu} =
\Lambda^{\mu}_{~\nu}x^{\nu} + a^{\mu}$. The $\Lambda^{\mu}_{~\nu}$
denote the Lorentz transformations. These transformations, termed
Poincare transformations, preserve the space-time metric. Under
composition of transformations, these form the Poincare group. 

In general, {\em covariance of a `structure' with respect to some group}
means that there is a homomorphism of that group, onto the invariance
group of the structure. Here, invariance group of the structure means
the group of transformations of the structure which preserves the
structure. For instance, if the `structure' is a vector space, then its
invariance group must be the group of linear transformations. For the
quantum framework, the structure is a projective Hilbert space (Hilbert
space modulo non-zero scaling) and the basic observable quantities are
the transition probabilities: $|\langle\psi|\phi\rangle|^2/(\|\psi\|
\|\phi\|)$. It is a theorem due to Wigner that the transformations of
physical states (elements of the projective Hilbert space) preserving
the transition probabilities can be represented by linear or anti-linear
transformation of the Hilbert space, $\psi' = A\psi$, satisfying
$\langle\psi'|\phi'\rangle = \langle\psi|\phi\rangle ~ \mbox{or} ~
\langle\psi'|\phi'\rangle = \langle\phi|\psi\rangle.$ The former defines
$A$ to be a {\em unitary} operator while the latter defines an {\em
anti-unitary} operator. The anti-unitary operator is necessarily
anti-linear as well. The only anti-unitary operator one encounters is
the {\em time reversal} operator (more on this later). {\em Thus
covariance requires the Hilbert space of the quantum system to carry a
unitary representation of the Poincare group.} 

If $g \in G \to R(g)$ denotes a group action which satisfies the
composition rule: $R(g_1\cdot g_2) = R(g_1)\cdot R(g_2)$, $R(g)$
constitutes a representation of the group $G$. On a Hilbert space, we
denote a group action as $|\psi'\rangle := g\cdot|\psi\rangle :=
U(g)|\psi\rangle$. It follows that $g_1\cdot(g_2\cdot|\psi\rangle) =
U(g_1)( U(g_2)|\psi\rangle) = U(g_1\cdot g_2)|\psi\rangle = (g_1\cdot
g_2)\cdot|\psi\rangle$. This action also induces a natural action on the
operators via $\langle\psi|g\cdot A|\psi\rangle := \langle
g^{-1}\cdot\psi|A|g^{-1}\cdot\psi\rangle \leftrightarrow A' := g\cdot A
:= U(g)AU(g)^{\dagger}$. Check that (a) this is indeed a homomorphism:
$(g_1\cdot g_2)\cdot A = g_1\cdot(g_2\cdot A)$ and (b)
$\langle\psi'|A'|\phi'\rangle = \langle\psi|A|\phi\rangle$. All
algebraic relations among the operators are preserved under this group
action.  Incidentally, this is exactly how one defines the action of
diffeomorphisms on functions on a manifold: $[g\cdot f](p) :=
f(g^{-1}\cdot p)$ (this is not the pull back definition). All this
essentially says that covariance is implemented through a representation
of the group. {\em Which representation?}

\subsection{Compatibility with quantum conditions} \label{QCondns}
Any {\em particular} quantum system is however distinguished by what
Dirac called {\em quantum conditions} eg. $[q^i, p_j] = i\hbar
\delta^i_{~j}$. It is on the Hilbert space carrying a representation of
the quantum conditions that the covariance group must be represented.
To appreciate what this means, let us take the specific examples of the
Galilean group and the Poincare group and try to seek unitary
representations on the Hilbert space of a single particle with the above
quantum conditions with, $i, j = 1,2,3$. Taking the generators of the
infinitesimal transformations to be represented by self-adjoint
operators, the Lie algebras of the two groups have the following
commutation relations: 
\begin{eqnarray*}
\mbox{Galilean transformations} & : & x'^i = R^{i}_{~j}x^j + v^i t +
a^i~~,~~ t' = t + b ~~~,~~~  R^i_{~m}R^j_{~n}\delta^{mn} = \delta^{ij}
;\\
\mbox{Generators} & : & P_0, P_i, C_i, M_{ij}~ ; ~~~ \mbox{with
non-vanishing commutators:}\\
\big[M_{ij}, M_{kl}\big] & = & i\bigg(\delta_{ik}M_{jl} -
(i\leftrightarrow j) - (k\leftrightarrow l) + (i,k \leftrightarrow
j,l)\bigg) ; \\
\big[M_{ij}, P_k\big] & = & i\bigg(\delta_{ik}P_j - \delta_{jk}P_i\bigg)
~~ ; ~~
\big[M_{ij}, C_k\big] ~ = ~ i\bigg(\delta_{ik}C_j - \delta_{jk}C_i\bigg)
; \\
\big[C_i, P_j\big] & = & i M\delta_{ij} ~~,~~ \big[C_i, P_0\big] ~ = ~ i
P_i \ . \\
\mbox{Poincare transformations} & : &  x'^{\mu} :=
\Lambda^{\mu}_{~\nu}x^{\nu} + a^{\mu} ~~~ ,  ~~~
\Lambda^{\mu}_{~\alpha}\Lambda^{\nu}_{~\beta}\eta^{\alpha\beta} =
\eta^{\mu\nu} , \eta = diag(-1, 1, 1, 1) ;  \\
\mbox{Generators} & : & P_0, P_i, K_i, M_{ij}~ ; ~~~ \mbox{with
non-vanishing commutators:}\\
\big[M_{ij}, M_{kl}\big] & = & i\bigg(\delta_{ik}M_{jl} -
(i\leftrightarrow j) - (k\leftrightarrow l) + (i,k \leftrightarrow
j,l)\bigg) ; \\
\big[M_{ij}, P_k\big] & = & i\bigg(\delta_{ik}P_j - \delta_{jk}P_i\bigg)
~~ ; ~~
\big[M_{ij}, K_k\big] ~ = ~ i\bigg(\delta_{ik}K_j - \delta_{jk}K_i\bigg)
; \\
\big[K_i, P_j\big] & = & i \delta_{ij}P_0 ~~,~~ \big[K_i, P_0\big] ~ = ~
i P_i  ~~~ \mbox{and,} \\
\big[K_i, K_j\big] & = & i M_{ij} \ .
\end{eqnarray*}
Both have the same number of generators, 10, and almost the same Lie
algebra except that the Galilean algebra has a `central extension'
denoted by the (mass) parameter, $M$ and the `Galilean boost' generators
$C_i$ commute among themselves unlike the `Lorentz boost' generators
$K_i$ . 

These can be expressed in terms of the basic operators $q^i, p_j$ as
follows\footnote{Thomas Jordan, 0810.4637 and 1st reference from it. The
	reference 1 shows that if $q^i$ is taken to represent position
	and transform under Galilean group accordingly, then the
	Galilean brackets suffice to fix the remaining generators,
	including $P$ up to a constant. This also shows that there are
	no functions on the phase space which are invariant under the
Galilean group apart form constant functions.}. 
\begin{eqnarray*}
\mbox{Galilean} & ~~,~~ & \mbox{Poincare} \\
M_{ij} := q_ip_j - q_j p_i ~,~ P_i := p_i & , & M_{ij} := q_ip_j - q_j
p_i ~,~ P_i := p_i ~ ; \\
C_i := Mq_i ~~,~~ P_0 := \frac{\vec{p}\cdot\vec{p}}{2M} & , & K_i :=
q_iP_0 ~~,~~ P_0 := \sqrt{\vec{p}\cdot\vec{p} + M^2} ~ .
\end{eqnarray*}
The parameter $M$ is naturally identified with the mass of our particle
system. Notice that only the boost generators and the $P_0$ generators
are different in the two groups.

{\em Ex:} Verify that the above definitions of the generators, satisfy
the respective algebras, up to factors of $\hbar$.

{\em Ex:} Under infinitesimal group action, the infinitesimal change in
any operator is given by $\delta_{\vec{\epsilon}}A := i[A,
\vec{\epsilon}\cdot\vec{T}]$, where $\vec{\epsilon}$ denotes the
infinitesimal parameters and $\vec{T}$ denotes the generators. Show that
the infinitesimal changes, $\delta_{\vec{\epsilon}}q^i,
\delta_{\vec{\epsilon}}p_i$ are at the most {\em linear} in $q^i, p_i$
for the Galilean case but not so for the Poincare case for the $K_i,
P_0$.

In particular, the Poincare group action is {\em non-linear} on the
basic operators. This is problematic for the following reason. At the
classical level, the observables are sufficiently smooth functions on
the phase space and the set of all observables can be given the
structure of an {\em algebra}: i.e. a vector space with a multiplication
(product of two functions) defined. A similar feature holds in the
quantum case. Observables are self-adjoint operators. The set of
observables too can be given the structure of an algebra (product of two
operators). In both cases, the algebras can be thought of as being
generated by the basic observables. The action of a group on the basic
variables induces a corresponding action on all operators. There are
domain issues in the quantum case, but much more importantly, the
multiplication is non-commutative which leads to the ordering issues.
\subsection{Manifest Covariance} \label{ManCov}
When the basic observables transform non-linearly, it is much harder to
deduce the transformation properties of the other operators. This is
compounded further by the non-commutativity in the quantum case. When
the transformations of basic variables are linear, the non-commutativity
issue exists but is usually easier to manage. With this in mind we now
put the additional requirement that the basic observables transform {\em
linearly} i.e. in effect tensorially (modulo inhomogeneous action of
translations). This is sometimes called the requirement of {\em manifest
covariance}.

Invoking manifest covariance discards Poincare group action on the phase
space of a particle. The action of the Galilean group is however
permissible.

This could of course be generalized to $N-$particles, i,j = 1,\ldots 3N.
We see that non-relativistic quantum theory of $N-$particles is not
capable of implementing manifest Poincare covariance. and we can expect
potential hazards in incorporating the relativistic causality and the
creation/annihilation processes.

Could we not extend the quantum conditions to a relativistically
covariant form? 

For instance, {\em define} a quantum system by postulating quantum
conditions as $[q^{\mu}, p_{\nu}] = i\hbar\delta^{\mu}_{~\nu}$. Then the
Poincare generators can be taken to be $M_{\mu\nu} := q_{\mu}p_{\nu} -
q_{\nu}p_{\mu} $ and $P_{\mu} := p_{\mu}$. It is trivial to check the
Poincare Lie algebra. These also show that the basic observables are
tensors of ranks as indicated by their index positions. This could be a
model for a relativistic particle. However, now there is another problem
in {\em quantum} theory.

From the basic commutators, we can define the creation/annihilation
operators: $a^{\mu} := \Case{1}{\sqrt{2\hbar}} (q^{\mu} + i
\eta^{\mu\nu}p_{\nu})$ and its adjoint $(a^{\mu})^{\dagger}$, satisfying
the commutation relations $[a^{\mu}, (a^{\nu})^{\dagger}] =
\eta^{\mu\nu}$. As usual, the states are labeled by the eigenvalues of
the number operators and the vacuum state is defined by
$a^{\mu}|0\rangle = 0\ \forall \ \mu$. This is a Lorentz covariant form
of defining the vacuum state. It follows immediately that $\langle
0|[a^0, (a^0)^{\dagger}]|0\rangle = \|(a^0)^{\dagger}|0\rangle\|^2 =
\eta^{00} = -1$. 

Additionally, if we were to interpret $x^0$ as a ``time'' coordinate so
that $p_0$ is interpreted as energy, then {\em the energy eigenvalues
must necessarily be unbounded below}. The Stone-von Neumann theorem
implies that if the canonical commutation relations arise from the
infinitesimal form of the Weyl relations, then the spectra of both the
conjugate operators consists of all real numbers.

Thus, {\em the manifestly Lorentz covariant quantum conditions {\em
cannot} be realized on a Hilbert space!}

Let us turn attention to systems with infinitely many degrees of
freedom. The best known example is classical electrodynamics. This is
also relativistically covariant (in fact, led to special relativity!).
Its consequence, the electromagnetic waves, surprisingly show
particulate behavior.  While classically, an accelerated charge radiates
waves at the frequency of its mechanical motion, in atomic systems the
electrons only radiate at frequencies related by energy differences
between levels.  When photographic films are seen at low intensity
light, the `light marks' are localized. In interference experiments, the
interference pattern builds up grain by grain. The photo-electric effect
also shows a threshold behavior. All these are very particulate
manifestations. The observation of intensity correlations, the $(g-2)$
measurement and the Lamb shift all point to going beyond classical
electrodynamics. The validation by these high precision measurements
support the quantum electrodynamical theory - a quantum field theory.
The most recent example is of course the prediction of the Higgs
particle and its discovery.  These are positive arguments in favor of
quantum field theoretic framework being well suited for a quantum theory
in relativistic regimes.

What {\em is} a QFT framework? Let us keep in mind the familiar
classical electromagnetic field. As an example, consider a reflective
cavity in which some classical electromagnetic field exist. This field
satisfies the source free, first order in time,  Maxwell equations which
imply that the electric and magnetic fields satisfy a wave equation. Let
us put some boundary conditions (the cavity is closed, say). We can
write the general solution of the wave equation as a linear combination
of an infinite set of {\em mode functions}, which are by themselves
convenient solutions satisfying the boundary condition. Typically, this
selects a set of frequencies/wavelength. The expansion coefficients
encode the relative weightage of the mode functions. One can express the
energy and the momentum contained in the cavity field, in terms of the
expansion coefficients with the disappearance of the mode functions -
their memory remains only in the frequencies/wavelengths. Classically,
the expansion coefficients are complex numbers without any particular
restriction (except that the total energy/momentum should be finite!).
Schematically, we quantize the electromagnetic field by putting hats on
the fields or equivalently on the {\em expansion coefficients}. The
quantization procedure requires imposition of quantum conditions which
make the expansion coefficients similar to creation/annihilation
operators - {\em and we have the quanta of the cavity field.} Any
exchange of energy/momentum within the cavity or with any outside
environment is done in terms of these quanta, typically referred to as
{\em photons}. The many body facet mentioned at the beginning does
essentially this process in reverse.

It is useful to recall two equivalent ways of thinking about quantum
harmonic oscillator. We may think of it as a single particle performing
motion which is somehow ``discretized'' {\em or} as a collection of
``quanta'', each carrying an energy $\hbar\omega$ and the `particle
motion' being understood as consequences of emission/absorption of these
quanta. It is the latter view which is more convenient in the context of
a field.

In nutshell then, {\em A quantum field is a means to hold or supply an
arbitrary collection of quanta and all interactions are transactions of
energy/momentum/charges etc in terms of these quanta.} 

This view is certainly borne out in perturbative analysis but by no
means the most general one. 

\newpage

\section{The Poincare group and its representation: mass and spin/helicity}
\label{ParticleReprens}

\subsection{Poincare group, Lie algebra and Casimir invariants}
{\bf The Poincare group:} The Poincare group or inhomogeneous Lorentz
group is defined by the transformations of space-time coordinates,
\begin{equation}
	x'^{\mu} = \Lambda^{\mu}_{~\nu}x^\nu + a^{\mu} ~,~
	\Lambda^{\mu}_{~\alpha}\Lambda^{\nu}_{~\beta}\eta^{\alpha\beta}
	= \eta^{\mu\nu} ~,~ \eta^{\mu\nu} = diag(-1, 1, 1, 1) =
	\eta_{\mu\nu} ~,~ a^{\mu} \in \mathbb{R}^4.
\end{equation}
The defining conditions on $\Lambda^{\mu}_{~\nu}$ imply that,
\begin{eqnarray}
	\eta_{\mu\nu}\Lambda^{\mu}_{~\alpha}\Lambda^{\nu}_{~\beta} =
	\eta_{\alpha\beta} = \Lambda_{\nu\alpha}\Lambda^{\nu}_{~\beta} &
	\Rightarrow & \delta^{\alpha}_{~\beta} =
	\Lambda_{\nu}^{~\alpha}\Lambda^{\nu}_{~\beta} \\
	\therefore \Lambda_{\nu}^{~\alpha} =
	(\Lambda^{-1})^{\alpha}_{~\nu} ~ ~ \leftrightarrow ~ ~
	(\Lambda^{-1})_{\nu}^{~\alpha} = \Lambda^{\alpha}_{~\nu} & &
	\mbox{Note the index positions}
\end{eqnarray}
It is convenient to view $\Lambda^{\mu}_{~\nu}$ as matrices. It is then
important to adopt a consistent convention regarding the index
positions.  Viewed as matrices, in the last equation,
$\Lambda_{\nu}^{~\alpha}$ is the {\em left inverse} of $\Lambda$ while
$\Lambda^{\alpha}_{~\nu}$ is the {\em right inverse} of $\Lambda$.

The {\em infinitesimal Transformations} are defined by taking
$\Lambda^{\mu}_{~\nu} = \delta^{\mu}_{~\nu} + \omega^{\mu}_{~\nu}$ and
$a^{\mu} = \epsilon^{\mu}$. The defining equation for Lorentz
transformations then imply $\omega^{\mu\nu} = - \omega^{\nu\mu}\ ,
\omega^{\mu\nu} := \omega^{\mu}_{~\alpha}\eta^{\alpha\nu}$. This also
explains the parameter counting of 6 for the Lorentz and 4 for the
translations.

Thus, the Poincare group is a 10 parameter continuous group (actually a
Lie group) with the 6 parameters in $\Lambda$ constituting the {\em
Lorentz} subgroup and the 4 parameters $a^{\mu}$ constituting the {\em
translations} subgroup.  We denote a generic element of the group as
$(\Lambda, a)$.

Clearly $det \Lambda = \pm 1$. The transformations with positive
determinant are called {\em proper Lorentz} transformations while those
with negative one are called {\em improper}. Furthermore,
$-(\Lambda^{0}_{~0})^2 + \sum^{3}_{i=1}(\Lambda^0_{~i})^2 = -1
\Rightarrow \Lambda^0_{~0} = \pm \sqrt{1 + \sum_i(\Lambda^0_{~i})^2}$.
Those with positive $\Lambda^0_{~0}$ are called {\em orthochronous}
Lorentz transformations.  The identity transformation, being both proper
and orthochronous, the subgroup of proper, orthochronous transformations
is the continuous subgroup connected to the identity. The two improper
transformations with $\Lambda = diag(1, -1, -1, -1)$ and $\Lambda =
diag(-1, 1, 1, 1)$ are called {\em space inversion} and {\em time
reversal} transformations. It can be shown that any Lorentz
transformation can be obtained from a proper, orthochronous
transformation followed by space inversion and/or time reversal. 

{\bf Group composition:} The composition of two Poincare transformations
is defined as: $(\Lambda_2, a_2)\cdot(\Lambda_1, a_1) =
(\Lambda_2\Lambda_1, \Lambda_2 a_1 + a_2)$. It follows that
$(\mathbb{1}, 0)$ is the identity and $(\Lambda, a)^{-1} =
(\Lambda^{-1}, -\Lambda^{-1}a)$.

{\em Check:} $~(\Lambda, a)\cdot(\mathbb{1}, 0)\cdot(\Lambda,
a)^{-1} = (\mathbb{1}, \Lambda b)$ ~ i.e.~  $g H g^{-1} \in H ~ \forall~
g$. Such a subgroup $H$ is called an {\em invariant or normal subgroup}.
Thus the translations subgroup is a normal subgroup of the Poincare
group and the {\em Poincare group is said to be a semi-direct product of
the Lorentz and the translations groups}. 

{\em Check:} {\em Any} group element can be {\em uniquely} written as:
\[(\Lambda, a) = (\Lambda, 0)\cdot(\mathbb{1}, \Lambda^{-1}a) =
(\mathbb{1}, a)\cdot(\Lambda, 0)\].
While it is possible to proceed with the abstract identification of the
Lie algebra via the vector fields generating the infinitesimal
transformation, we can simplify by directly considering a {\em unitary
representation} of the Poincare group. That is, a set of unitary
operators $U(g): \cal{H} \to \cal{H}$ such that (i) $U^{\dagger}(g)U(g)
= \mathbb{1} = U(g)U(g)^{\dagger}$ and (ii) $U(g_2 g_1) = U(g_2)
U(g_1)~, ~ \forall g's \in G$. To allow for the possibility of infinite
dimensional representations, we use operators on Hilbert space. We also
restrict to {\em irreducible representations} i.e. the representation
space has no proper subspace which is invariant under the $U(g)$
operators. It is a result (from Schur's lemma) that if $U(g)$ is a
unitary, irreducible representation, then the only bounded operator that
commutes with all $U(g)$ operators is a multiple of the identity
operator. Its converse also holds, namely, if the only operator that
commutes with all the $U(g)$'s is a multiple of the identity operator,
then the unitary representation is irreducible.  A simple application of
this result is that unitary, irreducible representations of abelian
groups are one dimensional. Hence the translation subgroup has only
1-dimensional irreducible representations.

We define the infinitesimal generators as,
\begin{equation}
	U(\Lambda, a) = U(\delta^{\mu}_{~\alpha} +
	\omega^{\mu}_{~\alpha}, \epsilon^{\mu}) := \mathbb{1} +
	\frac{i}{2}\omega_{\mu\nu}M^{\mu\nu} - i \epsilon_{\mu}P^{\mu} +
	o(\omega^2, \epsilon^2) ~ ~,~ ~ M, P \mbox{~are self-adjoint .}
\end{equation}
The commutation relations among the generators are obtained from using
the homomorphism property, $U(\Lambda, a)^{-1} U(\lambda, b) U(\Lambda,
a) = U[ (\Lambda, a)^{-1}\cdot(\lambda, b)\cdot(\Lambda, a) ]$ and
taking $\lambda = \mathbb{1} + \omega, b = \epsilon$. Using the
definition of the generators, it follows,
\begin{eqnarray} \omega_{\mu\nu}U(\Lambda, a)^{-1} M^{\mu\nu} U(\Lambda,
	a) & = &
	(\Lambda^{-1}\omega\Lambda)_{\alpha\beta}M^{\alpha\beta} - 2
	(\Lambda^{-1}\omega a)_{\alpha}P^{\alpha} \\
	\epsilon_{\mu}U(\Lambda, a)^{-1}P^{\mu}(\Lambda, a) & = &
	(\Lambda^{-1}\epsilon)_{\alpha}P^{\alpha}\ .  ~ ~ \mbox{Using}
	\nonumber \\
	(\Lambda^{-1}\epsilon)_{\alpha} & = &
	\eta_{\alpha\sigma}(\Lambda^{-1})^{\sigma}_{~\mu}\epsilon^{\mu}
	= (\Lambda^{-1})_{\alpha}^{~\mu}\epsilon_{\mu} =
	\Lambda^{\mu}_{~\alpha}\epsilon^{\alpha} , \\
	(\Lambda^{-1}\omega\Lambda)_{\alpha\beta} =
	\omega_{\mu\nu}\Lambda^{\mu}_{~\alpha}\Lambda^{\nu}_{~\beta} &
	\mbox{and} & (\Lambda^{-1}\omega a)_{\alpha} =
	\omega_{\mu\nu}\Lambda^{\mu}_{~\alpha}a^{\nu} \ ; \nonumber \\
	U(\Lambda, a)^{-1}M^{\mu\nu}U(\Lambda, a) & = &
	\Lambda^{\mu}_{~\alpha}\Lambda^{\nu}_{~\beta}M^{\alpha\beta} -
	(\Lambda^{\mu}_{~\alpha}a^{\nu} -
	\Lambda^{\nu}_{~\alpha}a^{\mu})P^{\alpha} ,
	\label{LorentzTensors} \\
	U(\Lambda, a)^{-1} P^{\mu} U(\Lambda, a) & = &
	\Lambda^{\mu}_{~\nu}P^{\nu} \nonumber 
\end{eqnarray}
The last two equations show that under the `adjoint' action of the
group, the generators transform as Lorentz tensors.  To deduce the
commutation relations, take $\Lambda, a$ to be infinitesimal i.e.
$U(\Lambda, a)^{\pm 1} = \mathbb{1} \pm
\frac{i}{2}\omega_{\alpha\beta}M^{\alpha\beta} \mp i
\epsilon_{\alpha}P^{\alpha}. $ Substitution and reading off coefficients
gives,
\begin{eqnarray} 
	\left[M^{\mu\nu}, M^{\alpha\beta}\right] & = & i
	\left(\eta^{\mu\alpha}M^{\nu\beta} - (\mu \leftrightarrow \nu)
	- (\alpha \leftrightarrow \beta) + (\mu,\alpha \leftrightarrow
\nu,\beta) \right) \\
	\left[M^{\mu\nu}, P^{\alpha}\right] & = & i
	\big(\eta^{\mu\alpha}P^{\nu} - \eta^{\nu\alpha}P^{\mu}\big) \\
	\left[P^{\mu}, P^{\nu}\right] & = & 0\ .
\end{eqnarray}
It is convenient to introduce the notation, $K^i := M^{i0}$ and $J_i :=
\frac{1}{2}\epsilon_{ijk}M^{jk} \leftrightarrow M^{ij} =
\epsilon^{ijk}J_k$ and also $H := P^0.$ Then the Poincare commutators
take the form (non-zero commutators only),
\begin{eqnarray}\label{PoincareLieAlgebra}
	\left[J_i, J_j\right] ~ = ~ i\epsilon_{ij}^{~ ~k}J_k & , &
	\left[J_i, K_j\right] ~ = ~ i\epsilon_{ij}^{~ ~ l}K_l \\
	\left[K_i, K_j\right] ~ = ~ - i\epsilon_{ij}^{~ ~l}J_l & , &
	\left[J_i, P_j\right] ~ = ~ i\epsilon_{ij}^{~ ~ k}P_k \\
	\left[K_i, P_j\right] ~ = ~ - iH \delta_{ij} & , & \left[K_i,
	H\right] ~ = ~ iP_i 
\end{eqnarray}
Define the {\em Pauli-Lubanski vector}, $W_{\mu} :=
\frac{1}{2}\epsilon_{\mu\nu\alpha\beta}M^{\nu\alpha}P^{\beta}~ , ~
\epsilon_{0123} := 1$. It follows that $[P^{\mu}, W^{\nu}] = 0\ ,\
[M^{\mu\nu}, W^{\lambda}] = i\big(\eta^{\mu\alpha}W^{\nu} -
\eta^{\nu\alpha}W^{\mu}\big)$ and $[W_{\mu}, W_{\nu}] = -i
\epsilon_{\mu\nu\alpha\beta}W^{\alpha}P^{\beta}$. These relations imply
that $P^2 := \eta_{\mu\nu}P^{\mu}P^{\nu} ~,~ W^2 :=
\eta^{\mu\nu}W_{\mu}W_{\nu}$ commute with all the generators of the
Poincare group and hence also with all the group elements of the proper,
orthochronous Poincare group. These are the two independent {\em Casimir
invariants} of the Poincare group and must be multiples of identity
operator in irreducible representations. These multiples serve to label
the unitary, irreducible representations.
\subsection{Representations of the proper, orthochronous Poincare group:} 
Each of its representations induces a representation of its Lie algebra
\ref{PoincareLieAlgebra}. Since the $P^{\mu}$'s commute, their
simultaneous eigenvectors can be taken as a group of labels for a basis. 
\begin{equation} 
	P^{\mu}|k^{\mu}, \xi\rangle ~ = ~ k^{\mu}|k^{\mu}, \xi\rangle ~
	~ \Rightarrow  ~  P^2 |k^{\mu}, \xi\rangle ~ = ~ k^2|k^{\mu},
	\xi\rangle ~ ~ , ~ ~ k^2 = \eta_{\mu\nu}k^{\mu}k^{\nu} \ .
\end{equation}
Here, $\xi$ collectively labels the degenerate eigenvectors and $k^2$ is
the value of the Casimir $P^2$. Thus, first level of classification is
done by the value of the Lorentz invariant $k^2$. There are four classes
that arise naturally:

(i) $m^2 := - k^2 > 0$ (massive), (ii) $k^2 = 0, k^{\mu} \neq 0$
(massless), (iii) $k^2 > 0$ (tachyonic), and $~ ~ ~ ~ ~ $(iv) $k^{\mu} =
0$ (vacuum). The vacuum representation is just a single vector. The
tachyonic representations have not showed up in experiments yet.

$\bullet$ Since $P^{\mu}$ transforms as a Lorentz vector, so does
$k^{\mu}$ and for $k^2 \le 0$, the $sgn(k^0)$ distinguishes two further
subclasses of the massive and the massless representations. 

{\em We restrict to the massive and the massless representations with
$k^0 > 0$ as being physically relevant.}

$ \bullet$ Under a Lorentz transformation, $(\Lambda, 0)$, $k^{\mu}$
goes to another point on the hyperboloid $k^2 = $ constant. How does the
collective label $\xi$ transform? We would like to deduce the action of
$U(\Lambda, a)$ on the vectors $|k^{\mu}, \xi\rangle$. Since
$U(\mathbb{1}, a) = exp{ -i a^{\mu}P_{\mu}} $, we already know that
$U(\mathbb{1}, a)|k^{\mu}, \xi\rangle = exp{ -i a^{\mu}k_{\mu}}|k^{\mu},
\xi\rangle$ and  we focus on the Lorentz transformations $U(\Lambda, 0)
=: U(\Lambda)$.

The Lorentz transformation relation (\ref{LorentzTensors}) imply,
\begin{eqnarray}
	P^{\mu}\big( U(\Lambda)|k, \xi\rangle\big) & = &
	\Lambda^{\mu}_{~\nu}U(\Lambda)P^{\nu}|k, \xi\rangle =
	\big(\Lambda^{\mu}_{~\nu}k^{\nu}\big) \big(U(\Lambda)|k,
	\xi\rangle\big) ~ ~ ~ ~ \Rightarrow \nonumber \\
	U(\Lambda)|k, \xi\rangle & = & \sum_{\xi'}C_{\xi, \xi'}|\Lambda
	k, \xi'\rangle \ .
\end{eqnarray}
We want to find the coefficients $C_{\xi, \xi'}$ corresponding to
irreducible representations of the {\em Lorentz group}. In principle,
these coefficients could (and do) depend on $k$ label, however, their
number (the degree of degeneracy) must be the same over a given
hyperboloid. This is essentially a continuity argument - the label $k$
changes continuously while the degree of degeneracy is integer valued. 

Let $\hat{k}$ be some convenient, fixed vector on the $k^2 \leq 0$
hyperboloid and let $L(k)$ be some arbitrarily chosen Lorentz
transformation such that $k^{\mu} = L(k)^{\mu}_{~\nu}\hat{k}^{\nu}$. The
Lorentz transformation $L$ depends explicitly on $k$ (and also
implicitly on $\hat{k}$) and also is not unique. For instance, for
$\hat{k} = (M, \vec{0})$, any $k$ can be obtained by a combination of a
rotation and a boost transformation. Having made a choice of $L(k)$,
define,
\begin{equation} \label{NkDefn}
	|k, \xi\rangle := N(k)U(L(k))|\hat{k}, \xi\rangle
\end{equation}
Since this is a definition, the same label $\xi$ appears on both sides.
It follows,
\begin{equation*}
	U(\Lambda)|k, \xi\rangle = N(k) U(\Lambda
	L(k))|\hat{k},\xi\rangle ~ = ~ N(k)\left[ U(L(\Lambda k))\cdot
	U( (L(\Lambda k))^{-1} )\right] U(\Lambda)\cdot U(L(k))|\hat{k},
	\xi\rangle \ .
\end{equation*}
The last three factors of $U$ combine to give $U(L^{-1}(\Lambda
k))\cdot\Lambda\cdot L(k)$, which acting on $\hat{k}$ takes $\hat{k} \to
k \to \Lambda k \to \hat{k}$ since $(\Lambda k)^{\mu} =[L(\Lambda
k)]^{\mu}_{~\nu}\hat{k}^{\nu}$.  Denote: $W(\Lambda, k) :=
L^{-1}(\Lambda k)\cdot\Lambda\cdot L(k)$. It follows that $W(\Lambda,
k)\hat{k} = \hat{k}\ \forall \Lambda$.

The Lorentz transformations, $W(\Lambda, k)$ leaving the $\hat{k}$
invariant, form a subgroup of the Lorentz transformations. It is known
as the {\em little group of $\hat{k}$} or {\em stability subgroup of
$\hat{k}$}. Defining a representation $D(W)$ by, $U(W(\Lambda,
k))|\hat{k}, \xi\rangle  := \sum_{\xi'}D_{\xi'\xi}(W)|\hat{k},
\xi'\rangle$ , we get,
\begin{eqnarray}
	U(\Lambda)|k,\xi\rangle & = & N(k)U(L(\Lambda
	k))\sum_{\xi'}D_{\xi' \xi}(W)|\hat{k}, \xi'\rangle \nonumber \\
	& = & N(k)\sum_{\xi'}D_{\xi' \xi}(W)\left[U(L(\Lambda
	k))|\hat{k}, \xi'\rangle.\right] \nonumber \\
	\therefore U(\Lambda)|k,\xi\rangle & = & \frac{N(k)}{N(\Lambda
	k)}\sum_{\xi'}D_{\xi' \xi}(W)|\Lambda k, \xi'\rangle
\end{eqnarray}
In the last equation, we have used the definition (\ref{NkDefn}).  The
irreducible representations of the little group can thus be used to
characterize the irreducible representations of the Lorentz group. We
are not done yet, the normalization factors $N(k)$ need to be
determined.

{\bf Determination of $N(k)$:}

For different $\hat{k}, \hat{k}'$, the corresponding little groups are
different. The unitarity of the Poincare (and hence of the Lorentz
group) representation implies the representations of the little groups
be also unitary. We can {\em choose} $\langle\hat{k}', \xi'|\hat{k},
\xi\rangle = \delta_{\xi, \xi'}\delta^3(\hat{k} - \hat{k}')$ and we
would like this to be preserved under Lorentz transformations i.e. with
the hatted $k$'s being replaced by their Lorentz transforms. This is
possible only for a specific choice of $N(k)$ which we determine now.

Note that this requirement associates the ortho-normalization with the
hyperboloids determined by $\hat{k}^2, (\hat{k}')^2$. Since a delta
function is defined in the context of an integration, we are implicitly
envisaging an integration over the hyperboloids and {\em not} over
$\mathbb{R}^4$. The integration is to be Lorentz invariant. This can be
inferred as follows.
\begin{eqnarray}
	\int_{\mathrm{mass shell}}\theta(k^0)f(k) & := &
	\int_{\mathbb{R}^4}d^4k \theta(k^0)\delta(k^2 + m^2)f(k) ~ = ~
	\int d^3k dk^0 \theta(k^0)\delta(- (k^0)^2 + \vec{k}\cdot\vec{k}
	+ m^2)f(k) \nonumber \\
	& = & \int d^3k \frac{f(\sqrt{\vec{k}\ ^2 + m^2}, \vec{k})}{2
	\sqrt{\vec{k}\ ^2 + m^2}}  ~ ~ ~ \because \delta(f(x)) = \sum_i
	\frac{\delta(x - x_i)}{|f'(x_i)|} \ . 
\end{eqnarray}
Although there are two roots from the delta function of the mass shell,
the $\theta(k^0)$ picks out only one term with $\omega_k := k^0 := +
\sqrt{\vec{k}\cdot\vec{k} + m^2}$. The $\frac{d^3k}{2\omega_k}$ is the
{\em Lorentz invariant volume element (measure)} on the mass shell.
Since we have,
\begin{equation}
	f(\vec{k}) = \int_{\mathbb{R}^3}d^3k'
	f(\vec{k}')\delta^3(\vec{k} - \vec{k}') = \int_{\mathbb{R}^3}
	\frac{d^3k'}{2\omega_k} f(\vec{k})\left[2\omega_k
	\delta^3(\vec{k} - \vec{k}')\right] ,
\end{equation} 
we identify the {\em invariant delta function} on the mass shell as:
\[
\delta_{inv}^3(\vec{k} - \vec{k}') := 2\omega_k\delta^3(\vec{k} -
\vec{k}') = 2\sqrt{\vec{k}^2 + m^2}\delta^3(\vec{k} - \vec{k}')  .
\]

Let $k = L(k)\hat{k}$ and {\em define} $k' := L(k)\hat{k}'$ {\em with
the same} $L(k)$.
\[
	\langle k', \xi'|k, \xi\rangle = N^*(k')N(k)\langle \hat{k}',
	\xi'|U^{\dagger}(L(k))\ U(L(k))|\hat{k}, \xi\rangle =
	|N(k)|^2\delta_{\xi',\xi}\delta^3(\hat{k}' - \hat{k}) . 
\]
We have used unitarity of $U(L(k))$.  But we also know that
$\delta^3(\vec{k}' - \vec{k})$ and $\delta(\hat{k}' - \hat{k})$ are
related by the same Lorentz transformation. Therefore using invariant
delta function we have,
\begin{eqnarray}
	\omega_k\delta^3(\vec{k}' - \vec{k}) & = &
	\omega_{\hat{k}}\delta^3(\hat{k}' - \hat{k})  \\
	\therefore \langle k', \xi'|k, \xi\rangle & = &
	|N(k)|^2\delta_{\xi',\xi}\left[\frac{k^0}{\hat{k}^0}\right]
	\delta^3(\vec{k}' - \vec{k}) ~ ~ \mbox{and the choice,} \\
	|N(k)| & := & \sqrt{\frac{\hat{k}^0}{k^0}} ~ ~ \Rightarrow ~ ~
	\\
	\langle k', \xi'|k, \xi\rangle & = &
	\delta_{\xi',\xi}\delta^3(\vec{k} - \vec{k}') ~ ~ \mbox{and}~
	,\\
	U(\Lambda)|k, \xi\rangle & = & \sqrt{\frac{(\Lambda
	k)^0}{k^0}}\sum_{\xi'}D_{\xi'\xi}(W(\Lambda, k))|\Lambda k,
	\xi'\rangle 
\end{eqnarray}
\subsection{Little groups}
It remains to determine the little groups for the various cases.

\underline{The vacuum representation ($\mathbf{k^{\mu} = 0}$) :} 

The little group is all of the Lorentz group.  However the
representation is the trivial one: $U(\Lambda) = \mathbb{1}, \ \forall ~
\Lambda$.

\underline{Massive representation ($\mathbf{k^2 = - m^2}$) :}
Choose $\hat{k}^{\mu} = (M, \vec{0})$. The little group is defined by

$\Lambda^{\mu}_{~\nu}\hat{k}^{\nu} = \hat{k}^{\mu} \Rightarrow
\Lambda^0_{~0} = 1, \Lambda^i_{~0} = 0$. The defining condition on
$\Lambda, \eta^{00} = \Lambda^0_{~0}\Lambda^{0}_{~0}\eta^{00} +
(\Lambda^0_{~i})^2\delta^{ii},$ gives $\Lambda^i_{~0} = 0$. Hence
$\Lambda$ is block diagonal, with $\delta^{ij} = \Lambda^{i}_{~k}
\Lambda^j_{~l}\delta^{kl}$ and the little group is $SO(3)$, the group of
rotations. Its unitary representations are all finite dimensional and
labeled by its Casimir, $J^2 = j(j+1), j \in \mathbb{N}/2$. The half
integer ones come from the double cover $SU(2)$. The $D_{\xi'\xi}$ are
the usual $(2s+1)$ dimensional representations. 

The representations of this class are thus labeled by `mass', m and
`spin' s.

\underline{Massless representation ($\mathbf{k^2 = 0}$) :}

Choose $\hat{k}^{\mu} = (k, 0, 0, k), k > 0$. It is more convenient to
determine the generators of the little group. These are some linear
combinations of the Lorentz generators $M^{\alpha\beta}$. Consider the
commutator,
\begin{eqnarray*}
\left[\epsilon_{\alpha\beta}M^{\alpha\beta}, P^{\lambda}\right] & = &
i\big(\epsilon^{\lambda}_{~\beta}P^{\beta} -
\epsilon_{\alpha}^{~\lambda}P^{\alpha}\big) ~ ~ ~ \Rightarrow \\
\left[\epsilon_{\alpha\beta}M^{\alpha\beta}, P^{\lambda}\right]|\hat{k},
\xi\rangle & = & i\big(\epsilon^{\lambda}_{~\beta}\hat{k}^{\beta} -
\epsilon_{\alpha}^{~\lambda}\hat{k}^{\alpha}\big)|\hat{k}, \xi\rangle ~
= ~ 2i\epsilon^{\lambda}_{~\alpha}\hat{k}^{\alpha}.
\end{eqnarray*}
The l.h.s. of the above equation vanishes because, the generators of the
little group, $\epsilon\cdot M$ leave the eigenstate of $\hat{k}$
invariant. The r.h.s. Then gives the conditions on $\epsilon$'s.
Explicitly, $0 = \epsilon^{\lambda}_{~\alpha}\hat{k}^{\alpha} =
(\epsilon^{\lambda}_{~0} + \epsilon^{\lambda}_{~3})k) \Rightarrow
\epsilon_{\alpha 0} + \epsilon_{\alpha 3} = 0$. This in turn gives
$\epsilon_{03} = 0, \epsilon_{01} = \epsilon_{13}, \epsilon_{02} =
\epsilon_{23}$. Hence,
\begin{eqnarray}
	\frac{1}{2}\epsilon_{\alpha\beta}M^{\alpha\beta}|_{little} & = &
	\epsilon_{12}M^{12} + \epsilon{13}(M^{13} - M^{01}) +
	\epsilon_{23}(M^{23} - M^{02}) \nonumber \\
	& := & \epsilon_1(J_1 - K_2) + \epsilon_2(-J_2 - K_2) +
	\epsilon_3J_3 
\end{eqnarray}
It is conventional to denote: $A := J_2 + K_1, B := -J_1 + K_2$ and
write the little algebra of $\hat{k} = k(1, 0, 0, 1)$ as,
\begin{equation}
	[A, B] = 0 ~,~ [J_3, A] = iB ~,~ [J_3, B] = -iA \ .
\end{equation}
This is the Euclidean group in 2 dimensions: two translations and one
rotation. Its representations are not as familiar. To study these,
define $R(\theta) := exp{\ i\theta J_3}$. It is easy to check that
$R(\theta)\ A\ R^{-1}(\theta) = A cos(\theta) - B sin(\theta)$ and
$R(\theta)\ B\ R^{-1}(\theta) = A sin(\theta) + B cos(\theta)$.

Consider a representation $D$ in which $A|\hat{k}, a, b\rangle =
a|\hat{k}, a, b\rangle$ and $B|\hat{k}, a, b\rangle = b|\hat{k}, a,
b\rangle$. Define $|\hat{k}, a, b, \theta\rangle := R(\theta)|\hat{k},
a, b\rangle$. It follows that,
\[
	A|\hat{k}, a, b, \theta\rangle = (a cos\theta - b
	sin\theta)|\hat{k}, a, b, \theta\rangle  ~ , ~ ~ B|\hat{k}, a, b,
	\theta\rangle = (a sin\theta + b cos\theta)|\hat{k}, a, b,
	\theta\rangle  ~ . 
\]
Hence, if $a, b$ are non-zero, then for each $\theta$, the states
$|\hat{k}, a, b, \theta\rangle$ are simultaneous eigenstates of $A$ and
$B$. Such a representation has the
continuous parameter $\theta$ and no such additional parameter is seen
in observations. To avoid such a parameter, we {\em restrict the
representation of the Little group to those for which $a = b = 0$} or $A,
B$ vanish in the representation. The little group then effectively
reduces to just $U(1)$ generated by $J_3$. Its irreducible
representations are labeled by {\em half integers}. We denote these
representations as, $J_3|\hat{k}, \sigma\rangle = \sigma|\hat{k},
\sigma\rangle, \sigma \in \frac{1}{2} \mathbb{Z}$. The label $\sigma$ is
called {\em helicity}. The one dimensional representation is denoted as
$D_{\sigma'\sigma}$. The representations with $a, b$ nonzero, are called
``continuous spin representations''. 

{\bf In summary: }

\begin{center}
\fbox{
\begin{minipage}{0.9\textwidth}
\begin{eqnarray} \label{ReprnSummary}
	|k, \sigma\rangle & := &
	\sqrt{\frac{\hat{k}^0}{k^0}}U(L(k))|\hat{k}, \sigma\rangle \\ 
	\mbox{\underline{For $m > 0$:}} & & \nonumber \\
	U(\Lambda)|k, \sigma\rangle & = & \sqrt{\frac{(\Lambda
	k)^0}{k^0}} \sum_{\sigma'}
	D^j_{\sigma\sigma'}\left[W(\Lambda,k)\right] |\Lambda k,
	\sigma'\rangle  ~ ~ ~ where, \nonumber \\
	W(\Lambda, k) & := & L^{-1}(\Lambda k)\cdot\Lambda\cdot L(k) ~ ~
	~ , ~ ~ ~ L(k)\hat{k} := k ~ ~ ~ \sigma ~\mbox{is the $j_3$
	eigenvalue} \\
	\mbox{\underline{For $m = 0$:}} & & \nonumber \\
	U(\Lambda)|k, \sigma\rangle & = & \sqrt{\frac{(\Lambda
	k)^0}{k^0}} exp\left\{i \sigma\theta(\Lambda, k)\right\}|\Lambda
	k, \sigma\rangle  ~ ~ ~ where, \mbox{$\sigma$ is the helicity
	label,}\nonumber \\
	W(\Lambda, k) & := & L^{-1}(\Lambda k)\cdot\Lambda\cdot L(k) ~
	:= ~ S(\alpha(\Lambda, k), \beta(\Lambda, k))\ R(\theta(\Lambda,
	k)) 
\end{eqnarray}
\end{minipage}
}
\end{center}

Let us evaluate the second Casimir invariant of the Poincare group,
$W^2$. For this we evaluate $W_{\mu}|\hat{k}, \sigma\rangle$.

(i) $\mathbf{-\hat{k}^2 = m^2 \neq 0}$: For $\hat{k} = (m, 0, 0, 0)$,
	
$W_{\mu}|\hat{k}, \sigma\rangle = \frac{1}{2}\epsilon_{\mu\nu\alpha
0}M^{\nu\alpha} m |\hat{k}, \sigma\rangle$. This vanishes for $\mu = 0$
and for $\mu = i$, we get $W_{i}|\hat{k}, \sigma\rangle =
-\frac{1}{2}\epsilon_{0ijk}M^{jk} m |\hat{k}, \sigma\rangle = - m
J_i|\hat{k}, \sigma\rangle$. Therefore,
\[
	W^2|\hat{k}, \sigma\rangle = m^2\ j(j+1)|\hat{k}, \sigma\rangle.
\]

(ii) $\mathbf{-\hat{k}^2 = 0}$: For $\hat{k} = k(1,0,0,1)$, 

$W_{\mu}|\hat{k}, \sigma\rangle = \frac{1}{2}(\epsilon_{\mu\nu\alpha 0}
+ \epsilon_{\mu\nu\alpha 3}M^{\nu\alpha})\ k |\hat{k}, \sigma\rangle$.
This implies,
\begin{eqnarray}
W_0|\hat{k}, \sigma\rangle ~ = ~ k\sigma |\hat{k}, \sigma\rangle & , &
W_1|\hat{k}, \sigma\rangle ~ = ~ k(-J_1 + K_2) |\hat{k}, \sigma\rangle =
0 \\
W_3|\hat{k}, \sigma\rangle ~ = ~ - k\sigma |\hat{k}, \sigma\rangle  & ,
& W_2|\hat{k}, \sigma\rangle ~ = ~ k(-J_2 - K_1) |\hat{k}, \sigma\rangle
= 0 .  \end{eqnarray}
Hence $W^2|\hat{k}, \sigma\rangle = [- (k\sigma)^2 + (-k\sigma)^2]
|\hat{k}, \sigma\rangle = 0$, for our choice of representations.

Note that the action of the translations subgroup on the Poincare
representation is given by, $U(\mathbb{1},a)|k, \sigma\rangle =
exp{-ia^{\mu} k_{\mu}}|k, \sigma\rangle$ while that of the Lorentz group
is given in the box. Since any element of the Poincare group can be
uniquely written as a product of a Lorentz transformation times a
translation, we have specified the full action. Since this action is
determined by the choice of the unitary, irreducible representation of
the little group, the Poincare representation is said to be {\em
induced} by a representation of the little group.

{\em These representations are taken to identify {\em elementary
quanta}}.
\subsection{The Discrete Subgroups and their actions} 
Recall that we have two improper Lorentz transformations, the space
inversion $\cal{P}: \Lambda^{\mu}_{~\nu} = diag(1, -1, -1, -1)$ and the
time reversal $\cal{T}: \Lambda^{\mu}_{~\nu} = diag(-1, 1, 1, 1)$. All
Lorentz transformations can be generated from the proper, orthochronous
transformations combined with either or both of these.  Clearly these
transformations are order 2 i.e. $\cal{P}^{-1} = \cal{P}$ and
$\cal{T}^{-1} = \cal{T}$. Let us assume that these have an action on the
physical state space preserving probabilities. Then by Wigner's theorem,
these are represented either as linear and unitary or anti-linear and
anti-unitary operators on the Hilbert space. {\em Let us use the same
symbols to denote the corresponding operators which should be clear from
the context.} 

The homomorphism property, 
\begin{equation}\label{HomomorphismProp}
	U(\Lambda, a)^{-1}U(\lambda, b)U(\Lambda, a) = U\left(
	(\Lambda,a)^{-1}\cdot(\lambda,b)\cdot (\lambda,a)\right)
\end{equation}
gives us the action of space inversion and time reversal on the proper,
orthochronous transformations. For $a = 0$ and $\Lambda = \cal{P}
~\mathrm{or}~ \cal{T}$ leads to:
\begin{equation} \label{DiscreteOnPoincare}
	\cal{P}^{-1}U(\lambda, b)\cal{P} ~ = ~
	U(\cal{P}^{-1}\lambda\cal{P}, \cal{P}^{-1}b) ~ ~ , ~ ~ 
	\cal{T}^{-1}U(\lambda, b)\cal{T} ~ = ~
	U(\cal{T}^{-1}\lambda\cal{T}, \cal{T}^{-1}b) .
\end{equation}

The same homomorphism (\ref{HomomorphismProp}), together with the
infinitesimal form, $U(\mathbb{1} + \omega, \epsilon) = \mathbb{1} +
\Case{i}{2}\omega_{\alpha\beta}M^{\alpha\beta} -i
\epsilon_{\alpha}P^{\alpha}$ for $(U(\lambda,b)$, gives us the
	relations:
\[
	U^{-1}(\Lambda) M^{\mu\nu}U(\Lambda) =
	\Lambda^{\mu}_{~\alpha}\Lambda^{\nu}_{~\beta}M^{\alpha\beta} ~
	~,~ ~ U^{-1}(\Lambda) P^{\mu} U(\Lambda) =
	\Lambda^{\mu}_{~\nu}P^{\nu} .
\]
Following the same steps as before but {\em not} canceling the factors
of $i$ to allow for anti-linearity, we get,
\begin{eqnarray}
	\cal{P}^{-1}(iP^0)\cal{P} = iP^0  & , &
	\cal{P}^{-1}(iP^i)\cal{P} = -iP^i  \label{InversionDefn}\\
	\cal{T}^{-1}(iP^0)\cal{T} = -iP^0  & , &
	\cal{T}^{-1}(iP^i)\cal{T} = iP^i \label{TimeReversalDefn}
\end{eqnarray}

Let if possible $\cal{P}$ be anti-linear. Then it {\em anti-commutes}
with $P^0$ which represents the energy. This in turn means that for
every positive energy state, there is a negative energy state. This
conflicts with the energy being bounded below. Identical implication
follows if $\cal{T}$ is {\em linear}!  Hence, {\em the space inversion
operator must be linear and unitary while the time reversal operator
must be anti-linear and anti-unitary.}

With this understood, the action of these discrete operators on the
Poincare generators is given by,
\begin{eqnarray}\label{DiscreteOnGenerators}
	\cal{P}^{-1}P^0\cal{P} = P^0 ~,~ \cal{P}^{-1}P^i\cal{P} = - P^i & , & 
	\cal{P}^{-1}J_i\cal{P} = J_i ~,~ \cal{P}^{-1}K_i\cal{P} = - K_i
	\label{SpaceInversionAction}\\
	\cal{T}^{-1}P^0\cal{T} = P^0 ~,~ \cal{T}^{-1}P^i\cal{T} = - P^i & , & 
	\cal{T}^{-1}J_i\cal{T} = - J_i ~,~ \cal{T}^{-1}K_i\cal{T} =  K_i
	\label{TimeReversalAction}
\end{eqnarray}

From these defining actions, we can determine how these operators act on
the unitary, irreducible representations of the Poincare group.

It is easy to see that under space inversion and time reversal, the two
Casimir invariants are invariant: $\cal{P}^{-1}P^2\cal{P} = P^2$ and
$\cal{P}^{-1}W^2\cal{P} = W^2$. This is obvious from the action of these
operators on the generators which transform as their tensor indices
indicate. The Casimir invariants are Lorentz {\em scalars} (not
pseudo-scalars) and hence invariant under inversions and time reversal.
Thus both parity and time reversal will not mix different irreducible
representations. To evaluate their actions on individual vectors, we
note that the irreducible representations are characterized in terms of
$\hat{k}$ vector and an eigenvalue(s) of $J_3$ generator.

{\bf Action of $\cal{P}:$}

\underline{$m > 0$}: 

Here, $\hat{k} = (m, 0, 0, 0)$ and $\sigma = -s, -s+1, \ldots s-1, s$,
$s$ being the spin. Since $\cal{P}$ commutes with $P^0$, anti-commutes
with $P^i$ and commutes with $J_3$, we see that $\cal{P}|\hat{k},
\sigma\rangle$ also has the same eigenvalues of the $P^{\mu}, J_3$.
Hence, $\boxed{\cal{P}|\hat{k}, \sigma\rangle = \eta_{\sigma}|\hat{k},
\sigma\rangle}$, with $\eta_{\sigma}$ being a phase (since the vectors
are non-degenerate). Next, from the usual angular momentum algebra, we
know
\begin{eqnarray*}
	(J_1 \pm iJ_2)|\hat{k}, \sigma\rangle & = & \sqrt{(s\mp\sigma)(s
	\pm \sigma + 1)}|\hat{k},\sigma\pm 1\rangle ~ \Rightarrow \\ 
	(J_1 \pm iJ_2)(\cal{P}|\hat{k}, \sigma\rangle) & = &
	\sqrt{(s\mp\sigma)(s \pm \sigma + 1)}(\cal{P}|\hat{k},\sigma\pm
	1\rangle) ~ \Rightarrow ~ \eta_{\sigma} = \eta_{\sigma\pm 1} \ .
\end{eqnarray*}
Thus $\eta_{\sigma}$ is independent of $\sigma$ and equal $\pm 1$ since
$\cal{P}^2 = \mathbb{1}$. The phase $\eta$ is called the {\em intrinsic
parity} of the representation. What about action on the states $|k,
\sigma\rangle$ ? Recall that these states are obtained as
(\ref{ReprnSummary}),
\[
	|k, \sigma\rangle =
	\sqrt{\frac{\hat{k}^0}{k^0}}U(L(k))|\hat{k},\sigma\rangle ~ ~,~
	~ k := L(k)\hat{k} ~ ~ ~ \mbox{for a chosen $L(k)$}.
\]
Acting by the matrix $\cal{P}$ on the defining equation for the Lorentz
transformation $L(k)$, gives $(\cal{P}k) =
(\cal{P}L(k)\cal{P}^{-1})(\cal{P}\hat{k})$ which implies that
$(\cal{P}L(k)\cal{P}^{-1}) = L(\cal{P}k)$. The homomorphism property,
(\ref{HomomorphismProp}), then gives $\cal{P}^{-1}U(L(k))\cal{P} =
U(L(\cal{P}k))$. Acting on $|k, \sigma\rangle$ gives,
\begin{equation}
	\cal{P}|k, \sigma\rangle ~ = ~
	\sqrt{\frac{\hat{k}^0}{k^0}}U(L(\cal{P}k))\eta|\hat{k},\sigma\rangle
	~ ~ = ~ ~ \eta|\cal{P}k, \sigma\rangle . 
\end{equation}
Thus $\cal{P}$ acts on all vectors of the representation with the same
$\eta$ and thus, {\em $\eta$ is a property of the irreducible
representation}.

\underline{$m = 0$}: 

Now $\hat{k} = (k, 0, 0, k)$ and the helicity $\sigma$ is the
eigenvalues of the $J_3$, the angular momentum direction along the
momentum direction. Since ${\cal P}$ changes the sign of the momentum
but not of the angular momentum, a state $|\hat{k},\sigma\rangle \propto
|-\hat{k}, -\sigma\rangle$. Unlike the massive case, the direction of
the reference momentum $\hat{k}$ is changed and this makes it
inconvenient to infer the action of parity on a general state. For this
reason, it is useful to choose a rotation $R_2$ about, say, the 2-axis
to bring back $-\hat{k} \to \hat{k}$.  Hence consider a new operator
${\cal P}' := U_2{\cal P}$, with $U_2 := e^{-i\pi J_2}$. But $U_2$ also
rotates $J_3$ (just as $ \hat{k}$) and hence the helicity is unchanged
by the $U_2$ operation.  Hence, $\boxed{ {\cal P}'|\hat{k},
\sigma\rangle = \eta_{\sigma}|\hat{k}, -\sigma\rangle }$. Here
$\eta_{\sigma}$ is some other phase factor. What about the action of
$\cal{P}$ on a general state $|p, \sigma\rangle$ defined in the equation
(\ref{ReprnSummary})? Note that we are using $p := (|p|, \vec{p})$
instead of $k$ to denote the general vector.	

Let $R(p, \hat{k})$ be the rotation that rotates the reference $\hat{k}$
in the direction of the general vector $p = (|\vec{p}|, \vec{p})$. Let
$B(p,\hat{k})$ denote the boost along the 3-direction which changes the
reference magnitude $k$ to $|\vec{p}|$. We can thus go from $\hat{k}$ to
$p$ by first using the boost $B$ followed by the rotation $R(p,
\hat{k}):  |p,\sigma\rangle \sim U\big( R(p, \hat{k})B(p,
\hat{k})\big)|\hat{k}, \sigma\rangle$. Noting that $\cal{P}'$ commutes
with boosts along the 3-axis, we have,
\begin{eqnarray*}
U(B(p, \hat{k})) & = & \cal{P}'U(B(p, \hat{k}))(\cal{P}')^{-1} =
U_2\cal{P}B\cal{P}^{-1}U_2^{-1} ~ ~ \Rightarrow ~ ~ 
\cal{P}U(B(p, \hat{k})) ~ = ~ U_2^{-1}U(B)U_2\cal{P} \\
\therefore \cal{P}|p, \sigma\rangle & \sim & U\big(R(p,\hat{k})\big)
\cal{P} U\big(B(p,\hat{k})\big) |\hat{k},\sigma\rangle ~ = ~ U\big(R(p,
\hat{k})R_2^{-1}B\big) \ \big(\eta_{\sigma}|\hat{k}, -\sigma\rangle\big)
\end{eqnarray*}
Now, although $R(p, \hat{k})R_2^{-1}$ is a rotation that takes the
3-axis along  $-\vec{p}$, its unitary representative is not quite
$U(R(-\vec{p}, \hat{k}))$. It introduces additional phases. I refer you
to the equation (2.6.21) in \cite{Weinberg}. The net result is that,
$\boxed{\cal{P}|p,\sigma\rangle = \eta_{\sigma}exp\{\mp
i\pi\sigma\}|\cal{P}(p),-\sigma\rangle.} $ The additional phase $\mp
\pi\sigma$ correlates with the sign of the $2-$component of the momentum
$\vec{p}$.

{\bf Action of $\cal{T}$}:

\underline{$m > 0$}:

Now $\cal{T}$ commutes with $P^0$, anti commutes with $P^i$ and $J_i$.
\[
	\therefore P^i\cal{T}|\hat{k}, \sigma \rangle = 0 ~ , ~
	P^0\cal{T}|\hat{k}, \sigma \rangle = m |\hat{k}, \sigma \rangle
	~, ~ J_3|\hat{k}, \sigma \rangle = - \sigma |\hat{k}, \sigma
	\rangle .
\]
The last one implies that $\cal{T}|\hat{k}, \sigma \rangle =
\zeta_{\sigma}|\hat{k}, -\sigma \rangle$, where $\zeta$ is a phase
factor. From the angular momentum algebra we get,
\begin{eqnarray*}
	(J_1 \pm iJ_2)|\hat{k}, \sigma\rangle & = & \sqrt{(s\mp\sigma)(s
	\pm \sigma + 1)}|\hat{k},\sigma\pm 1\rangle ~ \Rightarrow \\ 
	(- J_1 \pm iJ_2)(\cal{T}|\hat{k}, \sigma\rangle) & = &
	\sqrt{(s\mp\sigma)(s \pm \sigma + 1)}(\cal{T}|\hat{k},\sigma\pm
	1\rangle) \nonumber \\
	& = & \sqrt{(s\mp\sigma)(s \pm \sigma + 1)}\zeta_{\sigma\pm
1}|\hat{k}, - \sigma\mp 1\rangle) 
	~ \Rightarrow -\zeta_{\sigma} = \zeta_{\sigma\pm 1} \ .
\end{eqnarray*}
$\zeta_{\sigma + 1} = \zeta_{\sigma -1}$. Choosing the phase for say
$\zeta := \zeta_{\sigma = j}$, the phase for $j-1, j-2, \dots$ alternate
between $\pm\zeta$. Thus we can write the phase as $\zeta_{\sigma} =
\zeta (-1)^{j - \sigma}$, $\zeta$ is an arbitrarily chosen phase.  The
phase $\zeta$ can actually be absorbed away by putting $|\hat{k}, \sigma
\rangle' := \sqrt{\zeta}|\hat{k}, \sigma \rangle$. The action of
$\cal{T}$ on the new vector gives,
\[
	\cal{T}|\hat{k},\sigma \rangle' = \sqrt{\zeta^*}\cal{T}|\hat{k},
	\sigma \rangle = |\zeta|(-1)^{j- \sigma}|\hat{k},-\sigma \rangle
	\ .
\]
\underline{Note:} A similar manipulation with the $\eta$ phase will
retain the phase in the redefined vectors and thus the intrinsic parity
cannot be absorbed away. The anti-linearity of $\cal{T}$ is responsible
for this feature. We may retain the irrelevant phase $\zeta$.

The action of time reversal operator on the general vectors $|k, \sigma
\rangle$ proceeds in exactly the same manner as above leading to,
$\cal{T}|k,\sigma \rangle = \zeta(-1)^{j-\sigma}|k,-\sigma \rangle$. It
follows immediately that $\cal{T}^2|k,\sigma \rangle =
(-1)^{2j}|k,\sigma \rangle$ which distinguishes different spins.
	
\underline{$m = 0$}: 

The analysis here proceeds in an exactly same manner as that for the
parity and we just note the final result as:
$\boxed{\cal{T}|p,\sigma\rangle = \xi_{\sigma}{\sigma}exp\{\mp
i\pi\sigma\}|\cal{T}(p),\sigma\rangle.} $ The additional phase $\mp
\pi\sigma$ correlates with the sign of the $2-$component of the momentum
$\vec{p}$.

This completes the basic definitions related to the Poincare group and
its unitary, irreducible representations. 

These representations specify the attributes that we may assign to
elementary quanta. However, these are {\em not} suitable for manifest
covariance. Firstly, the label $\sigma$ transforms covariantly only
under the little group and not the full Lorentz group. Secondly, the
label $k$ is restricted to the positive hyperboloid and not the full
$\mathbb{R}^4$. This makes it inconvenient to take Fourier transform to
link them with space-time coordinates. For this we need to construct
representations on vector valued functions on the space-time. This is
done next.

\newpage
\section{Representations suitable for manifest covariance: ``field
representations''} \label{FieldReprens}

As noted earlier, the abstractly classified unitary, irreducible
representations of the Poincare group are not suitable for manifest
Lorentz covariance. For this, we begin by defining vector space of
suitably chosen function and define an action (homomorphism) of the
Poincare group. These will be reducible in general and irreducibility
will be imposed by partial differential equation (``field equations'').
To make the action unitary, requires an inner product to be defined.
This will further show a `doubling of representations' leading to the
particle/anti-particle interpretation.

Let $\Psi^A(x)$ be a finite dimensional column vector of a suitable
class of complex/real valued functions of the space-time coordinates
$x^{\mu}$. The defining action of the Poincare group is: $x \to gx :=
x_{(\Lambda,a)} = \Lambda x + a$. It induces a corresponding action on
the functions which we denote as: $\Psi \to (g\Psi) := \Psi_g$,
\[
	\Psi_g(x) = D(h(g))\Psi(g^{-1}x) ~ ~ \leftrightarrow ~ ~
	\Psi'_{(\Lambda,a)}(x) = D^A_{\ B}(h(\lambda,a))\Psi^B(
	(\Lambda,a)^{-1}x)
\]
Here, $D$ is a finite dimensional representation of the {\em Lorentz
group}, $h(\Lambda,a)$ is a map from the Poincare group to the Lorentz
group. The above action is required to be a homomorphism, i.e.
$(g\Psi)(x) = D(h(g))\Psi(g^{-1}x)$ such that
\begin{eqnarray}
	\left(g'\left(g\Psi\right)\right)(x) & = & (g'g\Psi)(x) ~
	\forall~ g,g' \in \mathrm{Poincare \ and}~ \forall ~ x .  \\
	l.h.s. & = & D(h(g'))\ (g\Psi)(g'^{-1}x) = D(h(g')) D(h(g))
	\Psi(g^{-1}g'^{-1}x) \nonumber \\
	& = & D(h(g')h(g)) \Psi( (g'g)^{-1}x ) . \\
	r.h.s. & = & D(h(g'g))\Psi(g'g)^{-1}x) ~ ~ \therefore ~ h(g)'s ~
	\mbox{must satisfy,} \\
	D(h(g'))\ D(h(g)) & = & D(h(g'g)) ~ \forall ~ g, g' \in
	~\mbox{Poincare. Equivalently,} \\
	D( h(g')h(g) ) & = & D (h(g'g) )  ~ ~ \mbox{which implies,}
	\nonumber \\
	h(g')h(g) & = & h(g'g) ~ , ~ \mbox{or $h(g)$ must be a
	homomorphism.} 
\end{eqnarray}

Since our primary focus is on the Lorentz group, and $h(g) =
h( (\mathbb{1},a)\cdot(\Lambda,0)) = h( (\mathbb{1},a)) h( (\Lambda,0)
)$, we may {\em choose} $h( (\mathbb{1}, a) ) = \mathbb{1}$. This
reduces the homomorphism from Poincare into Lorentz, to a homomorphism
from Lorentz to Lorentz. We can and do take, this homomorphism to the
identity homomorphism. With this we write,
\begin{equation}
	\Psi_{(\Lambda,a)}(x) ~ := ~ D(\Lambda) \Psi( \Lambda^{-1}(x -
	a) ).
\end{equation}
\subsection{Induced representation of Poincare from inducing
representation of Lorentz}
We have thus got an {\em induced representation} of the Poincare group
from an {\em inducing representation} $D(\Lambda)$ of the Lorentz group
and we take $D$ to be an irreducible representation. There is no
requirement of unitarity as $\Psi^A(x)$ is {\em not} a quantum
mechanical wave function. Additionally, it is a fact that the Lorentz
group has {\em no non-trivial finite dimensional, unitary
representations}. Our task now reduces to finding out all possible,
irreducible, finite dimensional representations of the Lorentz group.

To appreciate irreducibility, consider the infinitesimal action:
$(\Lambda, a) = (\mathbb{1} + \omega, \epsilon), \ D(\Lambda) =
D(\mathbb{1} + \frac{i}{2}\omega_{\alpha\beta}T^{\alpha\beta})$ where,
$T$ represent the Lorentz generators in the $D$ representation.  Then,
\begin{equation}
	\Psi_{\mathbb{1} + \omega, \epsilon}(x) = (\mathbb{1} +
	\frac{i}{2}\omega\cdot T)(\Psi(x - \omega x - \epsilon)) ~ = ~
	\Psi(x) + \frac{i}{2}\omega\cdot T\Psi(x) -
	(\omega^{\alpha}_{~\beta}x^{\beta} +
	\epsilon^{\alpha})\partial_{\alpha}\Psi|_x  
\end{equation}
This defines the Poincare generators acting on the $\Psi(x)$.
Explicitly,
\begin{eqnarray}
	M^{\alpha\beta}\Psi & := & (T^{\alpha\beta})^A_{~B}\Psi^B(x) +i
	(\eta^{\gamma\alpha}x^{\beta} -
	\eta^{\gamma\beta}x^{\alpha})\partial_{\gamma}\Psi^A(x) \\
	P_{\mu}\Psi & := & -i \partial_{\mu}\Psi^A(x)
\end{eqnarray}

The two Casimir invariants, $P^2, W^2$ acting on $\Psi$ should evaluate
to some constants to satisfy the necessary condition for irreducibility
of the Poincare representation. The $P^2$ Casimir is easy to evaluate
and gives,
\begin{equation} \label{Casimir1onPsi}
(P^2\Psi)^A = \eta^{\mu\nu}(-i)^2\partial_{\mu}\partial_{\nu}\Psi^A ~ =
~ - \Box\Psi^A ~ ~ ~ ~ ~ ~ \Box := -\partial_0^2 + \nabla^2 \ .
\end{equation}
For the second Casimir we have,
\begin{eqnarray}
(W_{\mu}\Psi)^A & = & \frac{1}{2} \epsilon_{\mu\nu\alpha\beta}
(M^{\nu\alpha}P^{\beta}\Psi)^A \nonumber \\
& = & \frac{1}{2} \epsilon_{\mu\nu\alpha\beta}
\left[(T^{\nu\alpha})^A_{~B} + i\delta^A_{~B} \left(\eta^{\gamma\nu}
x^{\alpha} - \eta^{\gamma\alpha} x^{\nu}\right) \partial_{\gamma}\right]
\left[-i\delta^B_{~C} \partial^{\beta}\right]\Psi^C \nonumber \\
& = & -\frac{i}{2} \epsilon_{\mu\nu\alpha\beta} (T^{\nu\alpha})^A_{~B}
\partial^{\beta}\Psi^B + \epsilon_{\mu\nu\alpha\beta} x^{\alpha}
\partial^{\nu}\partial^{\beta}\Psi^A
 = -\frac{i}{2} \epsilon_{\mu\nu\alpha\beta} (T^{\nu\alpha})^A_{~B}
 \partial^{\beta}\Psi^B + 0
\end{eqnarray}
Note that the 2-derivative term has canceled. Then,
\begin{eqnarray}
(W^2\Psi)^A & = & \frac{1}{2}\epsilon^{\mu\nu'\alpha'\beta'}
(M_{\nu'\alpha'}P_{\beta'})^A_{~B}(W_{\mu}\Psi)^B  \nonumber \\
& = & \frac{1}{2} \epsilon^{\mu}_{~\nu'\alpha'\beta'}
\left[(T^{\nu'\alpha'})^A_{~B} + i\delta^A_{~B}
(\eta^{\gamma\nu'}x^{\alpha'} - \eta^{\gamma\alpha'}x^{\nu'})
\partial_{\gamma}\right] \left[-i\delta^{B}_{~C}\partial^{\beta'}
\right] \left(W_{\mu}\Psi\right)^C \nonumber \\
& = & -\frac{i}{2} \epsilon^{\mu}_{~\nu'\alpha'\beta'}
(T^{\nu'\alpha'})^A_{~B} \partial^{\beta'}(W_{\mu}\Psi)^B +
\epsilon^{\mu}_{~\nu'\alpha'\beta'} x^{\alpha'} \partial^{\nu'}
\partial^{\beta'}(W_{\mu}\Psi)^A \nonumber \\
& = & - \frac{1}{4}\epsilon^{\mu}_{~\nu'\alpha'\beta'}
\epsilon_{\mu\nu\alpha\beta} (T^{\nu'\alpha'} T^{\nu\alpha})^A_{~B}
\partial^{\beta'}\partial^{\beta}\Psi^B 
\end{eqnarray}
Notice that $W^2\Psi = 0$ for non-one dimensional $D$ representation.

As illustrations consider two examples: (i) $D$ is the trivial representation
of the Lorentz group and (ii) $D$ is the defining representation of the
Lorentz group.

\underline{$D(\Lambda) = 1$}:  Now $W^2\Psi = 0$ while $P^2\Psi =
-\Box\Psi = -m^2\Psi$. Thus, irreducibility requires that $\Psi$ must
satisfy $(\Box - m^2)\Psi(x) = 0$ and we recognize this as the
Klein-Gordon equation. The $\Psi$ transforms as: $\Psi_{(\Lambda, a)}(x)
= \Psi(\Lambda^{-1}(x - a))$.

\underline{$D(\Lambda)$ is the defining representation}: This means
$\Lambda^{\mu}_{~\nu} = \delta^{\mu}_{~\nu} + \omega^{\mu}_{~\nu} :=
\delta^{\mu}_{~\nu} + \frac{i}{2} (\omega_{\alpha\beta}
T^{\alpha\beta})^{\mu}_{~\nu}$ and hence 
\[
	(T^{\alpha\beta})^{\mu}_{~\nu} := -i (\eta^{\alpha\mu}
	\delta^{\beta}_{~\nu} - \eta^{\beta\mu} \delta^{\alpha}_{~\nu}) 
\]
This gives, \[
	(T^{\nu'\alpha'}T^{\nu\alpha})^{\rho}_{~\sigma} = -\left[
		\eta^{\nu'\rho}\eta^{\nu\alpha'}\delta^{\alpha}_{~\sigma}
		-
		\eta^{\alpha'\rho}\eta^{\nu\nu'}\delta^{\alpha}_{~\sigma}
		-
		\eta^{\nu'\rho}\eta^{\alpha\alpha'}\delta^{\nu}_{~\sigma}
	+
\eta^{\alpha'\rho}\eta^{\alpha\nu'}\delta^{\nu}_{~\sigma}\right] 
\]
This is manifestly antisymmetric in $(\nu'\alpha')$ and $(\nu\alpha)$.
Contraction with the epsilons give equal contributions and cancel the
factor of 4, leading to $(W^2\Psi)^{\rho} =
\epsilon^{\mu\rho\nu}_{~~~\beta'} \epsilon_{\mu\nu\sigma\beta}
\partial^{\beta\beta'}\Psi^{\sigma}$.

Recall the general identity: 
\begin{equation} \label{EpsilonIdentity}
	\epsilon^{a_1\cdots a_j\ a_{j+1}\cdots a_n} \epsilon_{a_1\cdots
	a_j\ b_{j+1}\cdots b_n} ~ = ~ (-1)^s\ (n-j)!\ j! \
	\delta^{[a_{j+1}}_{~ ~b_{j+1}}\ \cdots\ \delta^{a_n]}_{~ ~b_n} \
	,
\end{equation}
where $s$ is the index of the metric i.e. number of negative eigenvalues
of the metric. In our case the metric is the Minkowski metric and $s =
1$. This gives,
\[
	\epsilon^{\mu\rho\nu}_{~ ~ ~ \beta'}\epsilon_{\mu\nu\sigma\beta}
	~ = ~ - \epsilon^{\mu\nu\rho}_{~ ~ ~
	\beta'}\epsilon_{\mu\nu\sigma\beta} ~ = ~ -(-1)\ 2!\ 2!\
	\frac{1}{2}\ (\delta^{\rho}_{\sigma}\eta_{\beta'\beta} -
	\eta_{\beta'\sigma}\delta^{\rho}_{\beta}) ,
\]
leading to,
\begin{eqnarray}\label{CasimirsOnVector}
(W^2\Psi)^{\rho} & = & 2(\delta^{\rho}_{\sigma}\eta^{\beta'\beta} -
\eta^{\beta\rho}\delta^{\beta'}_{\sigma}) \partial^2_{~
\beta'\beta}\Psi^{\sigma} ~ = ~ 2 \left[ \Box \Psi^{\rho} -
	\partial^{\rho}(\partial_{\sigma}\Psi^{\sigma}\right] \ , ~
	\mbox{and} \\
(P^2\Psi)^{\rho} & = & - \Box\Psi^{\rho}
\end{eqnarray}
\underline{Note:} We have specified the action of the Poincare group on
$\Psi^{A}$. How does $\partial_{\mu}\Psi^A$ transform? We can think of
the derivative as a new quantity with two indices. The index $A$ will
transform by $(T^{\alpha\beta})^A_{~B}$ as before while the index $\mu$
will transform by the {\em defining} representation of the Lorentz
group, $(T^{\alpha\beta})_{\mu}^{~\nu} = +i(\eta^{\alpha\nu}
\delta^{\beta}_{\mu} - \eta^{\beta\nu}\delta^{\alpha}_{\mu})$.  In the
above calculation of the second $W^{\mu}$, this was missed. However it
may be checked that this extra contribution vanishes in the $W^2$
calculation.
\subsection{Irreducibility and field equations}
The Poincare representation will be irreducible provided the Casimir invariants
have fixed values. Let $P^2\Psi^{\rho} = -m^2\Psi{\rho}$ and
$W^2\Psi^{\rho} = C\Psi^{\rho}$. For non-zero mass, with spin $j$, we know that $W^2$
equals $m^2\ j(j+1)$ and hence $C = m^2j(j+1)$. Thus $\Psi^{\rho}$
satisfies,
\[
	\Box \psi^{\rho} = m^2\Psi^{\rho} ~ ~,~ ~ 2(m^2\Psi^{\rho} -
	\partial^{\rho}\partial\cdot\Psi) = m^2 j(j+1)\Psi^{\rho}
\]

We have two cases to consider. The defining vector representation of the
Lorentz group, splits as $1 \oplus 0$ under the rotation subgroup. Hence
$j = 1 \ or\ 0$. For $j = 1$, taking the divergence of the second
equation implies $\partial\cdot\Psi = 0$. For $j = 0$, this argument
gives an identity. However, vanishing of $W^2$ gives
$\partial_{\rho}\partial\cdot\Psi = m^2\psi_{\rho}$. Differentiating by
$\partial_{\sigma}$ and anti-symmetrising in the indices gives,
$\partial_{\rho}\Psi_{\sigma} = \partial_`{\sigma}\Psi_{\rho}
\Rightarrow \Psi_{\rho} = \partial_{\rho}\Phi$ for some scalar $\Phi$
which is defined up to a constant. This scalar satisfies $(\Box -
m^2)\Phi = $constant which can be taken to be zero. Thus, 

{\em if the defining representation is to give the massive, spin one
	irreducible representation, $\Psi^{\rho}$ must satisfy: $(\Box -
	m^2)\Psi^{\rho} = 0 = \partial\cdot\Psi$. If it is to give the
	massive, spin zero irreducible representation, then $\Psi_{\rho}
	= \partial_{\rho}\Phi$ with $(\Box - m^2)\Phi = 0$.}

For $m = 0$ and our restriction on the representations of the little
group, the Casimir conditions give, $\Box \Psi_{\mu} = 0$ and
$\partial_{\mu}\partial_{\nu}\Psi^{\nu} = 0$. Introduce a new field,
$A_{\mu} := \Psi_{\mu} + \partial_{\mu}\Lambda$. It follows that,
\[
	\Box A_{\mu} = 0 + \partial_{\mu}\Box\Lambda ~ ~,~
	\partial_{\mu}\partial^{\nu}A_{\nu} = 0 +
	\partial_{\mu}\Box\Lambda  ~ ~ ~ \Rightarrow \Box A_{\mu} -
		\partial_{\mu}\partial^{\nu}A_{\nu} = 0 ~ =
		\partial^{\nu}\big(\partial_{\nu}A_{\mu} -
		\partial_{\mu}A_{\nu}\big) 
\]
We recognize the last equality as the source free Maxwell equation, {\em
including its gauge invariance}: $A_{\mu} \rightarrow A_{\mu} +
\partial_{\mu}\Lambda$!

These examples show that the induced representation $\Psi^{A}$ of the
{\em Poincare group} defined through the inducing representation of the
{\em Lorentz} subgroup, is in general {\em reducible}. Requiring that
the Poincare Casimir invariants, evaluated on $\Psi$ gives the values
corresponding to the unitary, irreducible representation of the Poincare
group, imposes {\em differential conditions} on $\Psi$ eg the
Klein-Gordon equation and the subsidiary conditions. Usually, these
equations are proposed as the {\em field equations}. Here we see them
arising as irreducibility conditions.

Let us return to the classification of the finite dimensional,
representations of the Lorentz group.

We already know that the defining representation acts as: $v'^{\mu} =
\Lambda^{\mu}_{~\nu}v^{\nu} \leftrightarrow v'_{\mu} =
\Lambda_{\mu}^{~\nu} v_{\nu} = (\Lambda^{-1})^{\nu}_{~\mu} v_{\nu}$.
From these we can trivially construct other irreducible representations
by taking tensor products and subtracting suitable `traces'. For
instance, a rank-2 tensor $V^{\mu\nu}$ transforms as $V'^{\mu\nu} =
\Lambda^{\mu}_{~\alpha}\Lambda^{\nu}_{~\beta}V^{\alpha\beta}$. This is a
reducible representation though. Why? Because its symmetrised and
anti-symmetrised parts transform among themselves and thus form
invariant subspaces. Furthermore, tensors of the form $\eta^{\mu\nu}C$
also transforms into the same form. Hence the symmetric combination
splits further into traceless and trace part. Thus we get the space
$\cal{V}$ of the second rank tensors decomposes as:
\[
\cal{V} = \cal{V}_{anti-sym} \oplus \cal{V}_{sym, traceless} \oplus
\cal{V}_{trace} , ~ ~ \mbox{each being an irreducible representation.}
\]
Analogous construction can be carried out for higher rank tensors. These
however do {\em not} exhaust all the finite dimensional irreducible
representations. There are the spinor representations which are missed.
It is a result that {\em all finite dimensional representations of
pseudo-orthogonal groups, $SO(p,q), p+q \ge 2$ can be constructed from
the irreducible representations of the corresponding Clifford algebra}.
\subsection{Clifford Algebras and Spinor representations}
\label{Clifford}
Clifford algebras are algebras generated by elements, $\{\mathbb{1},
\gamma^{\mu}, \mu = 0, \ldots, (p+q-1)\}$ satisfying $\gamma^{\mu}
\gamma^{\nu} + \gamma^{\nu} \gamma^{\mu} = 2
\bar{\eta}^{\mu\nu}\mathbb{1}, \bar{\eta}^{\mu\nu} = diag(-1, \dots ,
-1, +1, \dots +1)$. There are $p$ negative and $q$ positive signs.  We
have put a bar on the metric since we will relate it to the $\eta$
defining the Lorentz group for which $p = 1$ and $q = 3$. Note that for
$\mu \neq \nu$ the gamma's anti-commute, $(\gamma^0)^2 =
\bar{\eta}^{00}$ and $(\gamma^i)^2 = \bar{\eta}^{ii}$. Since the gamma's
anti-commute, their bilinear satisfy commutation relations.

Let $\Sigma^{\mu\nu} := a(\gamma^{\mu}\gamma^{\nu} -
\gamma^{\nu}\gamma^{\mu})$. We will choose the proportionality constant
$a$ suitably. Consider,
\begin{eqnarray}
\left[\Sigma^{\mu\nu}, \Sigma^{\alpha\beta}\right] & = & a^2\left\{
	\left[\gamma^{\mu}\gamma^{\nu},
	\gamma^{\alpha}\gamma^{\beta}\right] - (\mu \leftrightarrow \nu)
	- (\alpha \leftrightarrow \beta) + (\mu,\alpha \leftrightarrow
\nu,\beta)\right\} \\
\gamma^{\mu}\gamma^{\nu}\gamma^{\alpha}\gamma^{\beta} & = &
2\bar{\eta}^{\nu\alpha}\gamma^{\mu}\gamma^{\beta} -
2\bar{\eta}^{\mu\alpha}\gamma^{\nu}\gamma^{\beta} +
2\bar{\eta}^{\nu\beta}\gamma^{\alpha}\gamma^{\mu} -
2\bar{\eta}^{\mu\beta}\gamma^{\alpha}\gamma^{\nu}  \nonumber \\
\therefore \frac{1}{2}\left[\gamma^{\mu}\gamma^{\nu},
\gamma^{\alpha}\gamma^{\beta}\right] & = & \left[ -
	\bar{\eta}^{\mu\alpha}\gamma^{\nu}\gamma^{\beta} +
	\bar{\eta}^{\nu\alpha}\gamma^{\mu}\gamma^{\beta} -
	\bar{\eta}^{\mu\beta}\gamma^{\alpha}\gamma^{\nu} +
\bar{\eta}^{\nu\beta}\gamma^{\alpha}\gamma^{\mu} \right]\nonumber \\
\therefore \left[\Sigma^{\mu\nu}, \Sigma^{\alpha\beta}\right] & = &
(4a)\left\{ - \bar{\eta}^{\mu\alpha}\Sigma^{\nu\beta} + (\mu
\leftrightarrow \nu) + (\alpha \leftrightarrow \beta) - (\mu,\alpha
\leftrightarrow \nu,\beta)\right\}
\end{eqnarray}
This has the same form as the algebra satisfied by the Lorentz
generators: $\left[M^{\mu\nu}, M^{\alpha\beta}\right] =
i(\eta^{\mu\alpha}M^{\nu\beta} - - +)$, the $M$'s were defined through
$U(\Lambda = 1 + \omega) = \mathbb{1} + \Case{i}{2} \omega_{\mu\nu}
M^{\mu\nu}$. Thus we need $-4a\bar{\eta}^{\mu\nu} = i\eta^{\mu\nu}$.
There are two ways to satisfy this and we {\em choose our conventions
as: $\bar{\eta} = -\eta$ and $a = \Case{i}{4}$.} Thus,
\begin{equation}\label{CliffordDefns}
\boxed{ \gamma^{\mu}\gamma^{\nu} + \gamma^{\nu}\gamma^{\mu} = -
	2\eta^{\mu\nu} ~ ~ ~ , ~ ~ ~ \Sigma^{\mu\nu} :=
	\frac{i}{4}(\gamma^{\mu}\gamma^{\nu} - \gamma^{\nu}\gamma^{\mu})
	~ ~=~  ~ \frac{i}{4}\left[\gamma^{\mu}, \gamma^{\nu}\right] .  }
\end{equation}

The $\Sigma$'s satisfy the Lorenz Lie algebra.  Furthermore, the
definitions imply that $\boxed{[\Sigma^{\mu\nu}, \gamma^{\lambda}] =
i(\eta^{\mu\lambda}\gamma^{\nu} - \eta^{\nu\lambda}\gamma^{\mu})}$ i.e.
{\em the $\gamma$'s transform as Lorentz vectors}.

It is useful to define $\boxed{\gamma_5 :=
+i\gamma^0\gamma^1\gamma^2\gamma^3}$ Just from the Clifford algebra and
definitions it follows that $\gamma^2_5 = + \mathbb{1}$,
$\gamma_5\gamma^{\mu} = - \gamma^{\mu}\gamma_5$, and $[\gamma_5,
\Sigma^{\mu\nu}] = 0$.

\underline{Claim:} {\em It is always possible to choose the $\gamma$'s to be
finite dimensional unitary matrices.}

This is based on the following facts: (a) The set of elements
$ G := \{\pm\mathbb{1}, \pm \gamma^{\mu}, \pm\gamma^{\mu}\gamma{\nu},
	\pm\gamma^{\mu}\gamma^{\nu}\gamma^{\lambda}, \dots, \pm
\gamma_5\}$ with all indices distinct in the products, forms a finite
group of 32 elements (in our 4 dimensional case); (b) All representations of
finite groups can be made unitary; (c) representation theory of finite
groups applied to the group $G$ gives that there is exactly one,
non-trivial, unitary, irreducible representation of G and hence of the
Clifford algebra and that its dimension is $2^{[4/2]} = 4$. The results
also extends to other dimensions. 

Unitarity of the $\gamma$'s and their squares being $\pm\mathbb{1}$,
imply further that $(\gamma^0)^{\dagger} = \gamma^0$ and
$(\gamma^i)^{\dagger} = - \gamma^i$. This in turn gives
$(\Sigma^{0i})^{\dagger} = - \Sigma^{0i}\ , \ (\Sigma^{ij})^{\dagger} =
\Sigma^{ij}$ and $\gamma_5^{\dagger} = \gamma_5$. 

Hence, the representation $D$ of the Lorentz group provided by
$\Sigma$'s is (i) {\em non-unitary} and (ii) {\em reducible}, since
$\gamma_5$ is Hermitian and commutes with the generators.

Since $\gamma$'s are finite dimensional, their traces are defined and
all of them including $\gamma_5$ are traceless. In particular, this
implies that $\gamma_5$ has two eigenvalues equal to +1 and other two
equal to -1. Hence $\Case{1\pm\gamma_5}{2}$ are projection matrices and
reduce the 4 dimensional representation of the Lorentz group into two
irreducible representations of dimensions 2.

By convention, $\Psi_L := \frac{1 - \gamma_5}{2}\Psi , \leftrightarrow
\gamma_5\Psi_L = -\Psi_L$ is  called a {\em left handed Weyl spinor}
while $\Psi_R := \frac{1 + \gamma_5}{2}\Psi , \leftrightarrow
\gamma_5\Psi_R = + \Psi_R$ is  called a {\em right handed Weyl spinor}. 

The 4-component $\Psi^A$'s are called {\em Dirac spinors} while the
2-component projections, $\Psi_{\pm} := \Case{1\pm\gamma_5}{2}\Psi$ are
called {\em Weyl spinors}. Choosing $D$ to be generated by the
$\Sigma_{\pm} := \Case{1\pm\gamma_5}{2}\Sigma$ give irreducible
representations of the {\em Lorentz group} and in turn an irreducible
representation of the Poincare group . 

Returning to Poincare representation, consider the combination
$\gamma^{\mu}P_{\mu}$, apparently a Lorentz scaler. Indeed, recalling
that $[M^{\mu\nu}, P_{\lambda}] = +i(\delta^{\mu}_{\lambda}P^{\nu} -
\delta^{\nu}_{\lambda}P^{\mu})$, it follows that,
\[
	\left[T^{\mu\nu}, \gamma^{\lambda}P_{\lambda}\right] =
	\left[\Sigma^{\mu\nu}, \gamma^{\lambda}\right]P_{\lambda} +
	\gamma^{\lambda}\left[M^{\mu\nu}, P_{\lambda}\right] = 0\ .
\]
We have used $T$ to denote a general representation, $M$ to denote the
defining representation and $\Sigma$ to denote the reducible spinor
representation of the Lorentz group. We use the conventional
abbreviation $a\!\!\!/ := \gamma^{\mu}a_{\mu}$ for all 4-vectors
$a_{\mu}$.

Consider the Poincare Casimir $P^2$. Irreducibility forces $P^2\Psi =
-m^2\Psi$. Let $m \neq 0$. The combinations, $m\mathbb{1} \pm p\!\!\!/ $
commutes with the Lorentz generators and also satisfy, $(m \pm
p\!\!\!/)^2 = m^2 + (p\!\!\!/)^2 \pm 2m p\!\!\!/ = 2(m \pm p\!\!\!/)$.
Therefore, $\pi_{\pm} := \Case{m \pm p\!\!\!/}{2m}$ is a projection
matrix operator (since $P$ is an operator) that also commutes with the
Lorentz generators. It trivially commutes with the Poincare generators
as well. Hence the Poincare representation induced by the Dirac spinor
is {\em reducible} in yet another way, the irreducible subspaces being
provided by $(\pm m + p\!\!\!/)\Psi = 0 = (- i\gamma^{\mu}\partial_{\mu}
\pm m)\Psi = 0$. This is just the Dirac equation! Note that the
projection property holds {\em only for non-zero mass}.

\underline{Note:} The projectors $\frac{1 \pm \gamma_5}{2}$ reduce the
$D$ representation of the Lorentz group and each of these induces an
irreducible representation of the Poincare group. By contrast, $
\frac{m \pm \dsl{p}}{2m}$, do {\em not} reduce the $D$ representation, but
nevertheless reduces the Poincare representations. Are either of these
Poincare representations further reduced by the other projector? The
answer is `No' because $\gamma_5$ anti-commutes with $\dsl{p}$ and thus
exchanges the two projectors. 

\underline{Note:} Consider the action of $\dsl{p}$ on the left(right)
handed Weyl spinors. It follows immediately that $\gamma_5(\dsl{p})
\Psi_{L/R} = \pm \dsl{p} \Psi_{R/L}$. Therefore, if we restrict to any
one irreducible Weyl representation, then $\dsl{p}$ must annihilate it.
That is {\em Weyl spinors satisfy the massless Dirac equation - also
known as the Weyl equation.} Conversely, if we have a Dirac spinor
$\Psi$ that satisfies the Weyl equation, then its decomposition into its
Weyl spinors also satisfy the Weyl equation individually: $\dsl{p}(\Psi_L
+ \Psi_R) = 0 \Rightarrow \dsl{p}\Psi_{L/R} = 0$.

{\bf Exercises:} 
\begin{enumerate}
	\item Form the anti-symmetrized products of the
		$\gamma$-matrices, satisfying the Clifford algebra
		$\{\gamma^{\mu}, \gamma^{\nu}\} = -2\eta^{\mu\nu}$,

		$\Gamma^{\mu} := \gamma^{\mu}\\
		\Gamma^{\mu\nu} := \gamma^{\mu}\gamma^{\nu} -
		\gamma^{\nu}\gamma^{\mu} \\
		\Gamma^{\mu\nu\lambda} :=
		\gamma^{\mu}\gamma^{\nu}\gamma^{\lambda} + 
		\gamma^{\nu}\gamma^{\lambda}\gamma^{\mu} + 
		\gamma^{\lambda}\gamma^{\mu}\gamma^{\nu} - 
		\gamma^{\nu}\gamma^{\mu}\gamma^{\lambda} - 
		\gamma^{\mu}\gamma^{\lambda}\gamma^{\nu} - 
		\gamma^{\lambda}\gamma^{\nu}\gamma^{\mu} \\
		\Gamma^{\mu\nu\alpha\beta} :=
		\gamma^{[\mu}\gamma^{\nu}\gamma^{\alpha}\gamma^{\beta]}$
		with no numerical overall factors. Since the indices
		take 4 values only, we cannot have any more antisymmetric
		products. Together with $\mathbb{1}$, this set comprises
		of 16 matrices.

		Show that products of any number of $\gamma$'s can be
		expressed  in terms of the  these 16 matrices together
		with products of $\eta$'s. Conclude that these 16
		matrices constitute a basis for arbitrary products of
		$\gamma$'s. Since they have different Lorentz
		transformation properties, they are all independent. The
		vector space of $k\times k$ matrices has dimension
		$k^2$, hence the minimum matrix order for the $\gamma$'s
		is 4 and they are necessarily irreducible representation
		of the Clifford algebra. Hence our $\gamma$'s are
		$4\times 4$. Analogous arguments hold for Clifford
		algebra of $n$ dimensions (indices taking $n$ values).
	\item 	For spinors satisfying the Dirac equation with $M\neq
		0$, evaluate the Pauli-Lubanski scalar,
		$W_{\mu}W^{\mu}\Psi$.
	\item 	The Lorentz $J_i, K_i$ satisfy the algebras $[J_i, J_j]
		= i\epsilon_{ij}^{~l} J_l; ~ [K_i, K_j] = -
		i\epsilon_{ij}^{~l} K_l; ~ [J_i, K_j] =
		i\epsilon_{ij}^{~l} J_l.$ Define: $A_i :=
		\Case{1}{2}(J_i + i K_i), B_i := \Case{1}{2}(J_i -
		iK_i)$. Check that the $[A_i, B_j] =
		i\epsilon_{ij}^{~l}A_l, \ , [B_i, B_j] =
		i\epsilon_{ij}^{~l}B_l, ~ , [A_i, B_j] = 0.$ This the
		$M^{\mu\nu}$ Lie algebra is equivalent to the direct sum
		of two, mutually commuting $SU(2)$ algebras. They have
		the Casimir invariants $\vec{A}^2, \vec{B}^2$. Evaluate
		these for the spinor representation provided by
		$\Sigma^{\mu\nu}$'s.
\end{enumerate}

In summary:

\begin{center}
\fbox{
	\begin{minipage}{\textwidth}
\begin{eqnarray}
\Psi_{\Lambda,a}(x) & := & D(\Lambda)\Psi(\Lambda^{-1}x - \Lambda^{-1}a)
\ \mbox{(Reducible) Poincare action} \\
& & \mbox{Require Poincare Casimir invariants to take constant values} \nonumber \\
(\Box - m^2)\Phi = 0 & : & \mbox{ Klein-Gordon equation (scalar)
} \\
(\Box - m^2)v^{\mu} = 0, \partial\cdot v = 0 & : & \mbox{ Proca
equation (massive vector)} \\
\Box A^{\mu} = 0, \partial\cdot A = 0 & \leftrightarrow &
\partial_{\mu}F^{\mu\nu} = 0\ , \ F_{\mu\nu} := \partial_{\mu}A_{n} -
\partial_{\nu}A_{\mu} ~ \mbox{Maxwell equation)} \\
(-i\dsl{\partial} + m)\Psi = 0 & : & \mbox{Dirac equation, $m = 0$ is
Weyl equation} \\ \nonumber
\end{eqnarray}
\end{minipage}
}
\end{center}

\newpage
\section{Unitary representations on the space of solutions: anti-particles}
\label{AntiParticles}

At this stage, we have two sets of irreducible representations of the
Poincare group - the ``Particle'' representations $\{|k, \sigma
\rangle\}$, and the ``field'' representations $\Psi^{A}(x)$ satisfying
the appropriate field equations (or irreducibility conditions). The
former are unitary while the latter have no such notion defined for
them. We fill this gap now.

The space of vector valued functions, $\Psi^A$ can be made a complex
vector space easily enough, but we need to define an inner product to
define the notion of unitary. Since the irreducible representations are
solutions of the field equations and the equations are linear, we
consider the complex vector space of the {\em solutions of the field
equations} and look for an inner product. 
%

Consider the {\bf massive scalar field} first:
$(\partial^{\mu}\partial_{\mu} - m^2)\Phi(x) = 0, \Phi_{\Lambda,a}(x) =
\Phi(\Lambda^{-1}(x - a))$. Let $u, v$ be two solutions of the
Klein-Gordon equation. Then,
\[
	u^*(\Box -m^2)v - v(\Box - m^2)u^* =
	\partial_{\mu}\left(u^*\partial^{\mu}v -
	v\partial^{\mu}u^*\right) = 0 .
\]
Hence, $\boxed{J^{\mu} := \lambda\left(u^*\partial^{\mu}v -
v\partial^{\mu}u^* \right)}$ is a conserved current. This has an
immediate implication. Consider a 4 dimensional region bounded by two
hypersurfaces of constant value of the time coordinate. More generally,
these are two Cauchy surfaces. Restricting to solutions which vanish
sufficiently rapidly as one approaches asymptotic infinity along the
space-like directions (eg $|\vec{x}| \to \infty$), 
\[
	\int_{region}d^4x\partial_{\mu}J^{\mu} = 0 = \int_{\Sigma_2 \cup
	\Sigma_1}d^3x J^0 ~ ~ \Rightarrow ~ ~ \int_{\Sigma_2}d^3J^0 =
	\int_{\Sigma_1}d^3x J^0 \ .
\]
Here we have used that both the Cauchy surfaces have their normals
directed `outward' (future pointing for the later one and past pointing
for the earlier one). This suggests that we define a candidate inner
product between two solutions as,
\begin{equation}
	(v, u) ~ := ~ \lambda\int_{\Sigma}d^3x\ J^0(v,u)~ , ~ J^0(v,u)
	:= v^*\partial^0u - u\partial^0v^* \ , ~ \Sigma ~ \mbox{a Cauchy
	surface.}
\end{equation}
The conserved current implies that the inner product is independent of
the Cauchy surface.

\underline{Note:} On a Cauchy surface, the solution and its time
derivative form an initial data and the inner product is really defined
on these data. However the $\Sigma-$independence of the inner product
allows us to think of this as an inner product on the space of
solutions. 

Is $(v, u)$ really an inner product? It is (i) linear in $u$ and
anti-linear in $v$; (ii) $(v, u)^* = (u, v)$ {\em provided $\lambda^* =
- \lambda$} and (iii) $(u, u) \ge 0$ with equality for $u = 0$?

To check the third property, let us write a solution of the field
equation in the form $u(t,\vec{x}) = e^{\pm i\omega
t}u_{\omega}(\vec{x})\ , \omega > 0$. Let us {\em conventionally  call
$e^{-i\omega t}$ as a positive frequency solution}. Then, $(u, u) =
\lambda (-2 i\omega) \int_{\Sigma}d^3x |u_{\omega}(\vec{x})|^2$. The
inner product then satisfies the crucial third property provided we
choose $\lambda = +i|\lambda| := i$. The absolute value of $\lambda$ has
been taken to be one as it only affects the normalization of the
solutions.

Thus, with the convention adopted, the $(v, u)$ with $\lambda = i$ is
indeed an inner product on the {\em subspace of positive frequency}
solution. {\em Equally well}, the choice $\lambda = -i$ defines an inner
product on the {\em subspace of negative frequency} solutions.

Consider a family of solution, the plane wave solution, $u_{\vec{k}}(x)
:= A_{\vec{k}}e^{ik\cdot x}, \ k\cdot x := -k^0t + \vec{k}\cdot\vec{x},
\ k^0 := \sqrt{\vec{k}^2 + m^2} =: \omega_{\vec{k}}$. Substitution
gives,
\begin{eqnarray}
	(u_{\vec{k}'}, u_{\vec{k}}) & = & i A^*_{\vec{k}'}A_{\vec{k}}
	\int_{\Sigma_t}d^3x (-ik^0 - ik^{'0})e^{i(k - k')x} \\
	& = & +A^*_{\vec{k}'}A_{\vec{k}}(k^0 + k^{'0})e^{-i(k^0 -
	k^{'0})t} \left( (2\pi)^3\delta^3(\vec{k} - \vec{k}')\right) \\
	& = &
	\left[(2\pi)^3|A_{\vec{k}}|^2\right]\left[(2k^0)\delta^3(\vec{k}
	- \vec{k}')\right]
\end{eqnarray}
Since $k^0$ depends only on the magnitude of $\vec{k}$, the delta
function forces the frequencies to be equal. We recognize the second
square bracket as the Lorentz invariant delta function and {\em choose}
the normalization constant to $A_{\vec{k}} := (2\pi)^{-3/2}$ so that,
\begin{eqnarray} \label{Orthonormality0}
	u_{\vec{k}}(x) & := &
	\frac{1}{(2\pi)^{3/2}}e^{ik\cdot x} ~ ~,~ ~ k\cdot
	x := - \sqrt{\vec{k}^2 + m^2}\ t + \vec{k}\cdot\vec{x} ~ ~,~ ~
	\vec{k} \in \mathbb{R}^3 \\
	(u_{\vec{k}'}, u_{\vec{k}}) & = & \delta^3_{inv}(\vec{k} -
	\vec{k}') ~ ~ = ~ ~ (2k^0)\delta^3(\vec{k} - \vec{k}') \ .
\end{eqnarray}
The above family of solutions formally form an orthonormal set in the
space of {\em positive frequency solutions}. Technically, these do {\em
not} belong to the space of solutions which have to die off suitably at
spatial infinity. This is understood as usual by either using `box
normalization' or forming wave-packets. Manipulations done using the
above do not lead to any inconsistency.

Now we are ready to check if the Poincare action on the above inner
product space is unitary. We will check this by showing that under the
Poincare action, the orthonormality is preserved.

Consider, 
\begin{eqnarray}
[u_{\vec{k}}(x)]_{\Lambda,a} & = & u_{\vec{k}}( \Lambda^{-1}(x - a) ) ~
= ~ \frac{1}{(2\pi)^{3/2}}e^{i k\cdot \Lambda^{-1}(x - a)} \\
& = & \frac{1}{(2\pi)^{3/2}}e^{i (\Lambda k)\cdot(x - a)} ~ ~ \because
k_{\mu}(\Lambda^{-1})^{\mu}_{~\nu}x^{\nu} =
(\Lambda_{\nu}^{~\mu}k_{\mu})x^{\nu}  = (\Lambda k)\cdot x \nonumber \\
\therefore \left[u_{\vec{k}}(x)\right]_{\Lambda,a} & = & e^{-i(\Lambda
k)\cdot a}\ u_{\Lambda k}(x) ~ ~ \mbox{and,} \\
\left( [u_{\vec{k}'}]_{\Lambda,a} \ , \ [u_{\vec{k}}]_{\Lambda,a}\right)
& = & e^{i(\vec{\Lambda(k - k')}) \cdot a} \left( [u_{\vec{\Lambda k}'}]
\ , \ [u_{\vec{\Lambda k}}] \right) ~ = ~ e^{i(\vec{\Lambda(k - k')})
\cdot a}\ \delta_{inv}^3(\vec{\Lambda(k - k')}) \nonumber \\
& = & \left( [u_{\vec{k}'}]\ , \ [u_{\vec{k}}]\right) ~ ~ ~ \mbox{Hence,
unitarity!}
\end{eqnarray}
\underline{Note:}  The orthochronous Lorentz group preserves the sign of
the frequency, $k^0 > 0 => (\Lambda k)^0 > 0$, and hence maps positive
(negative) frequency solutions into positive (negative) frequency
solutions. Thus we see that {\em the space of solutions of the
irreducibility condition (field equations) itself decomposes into {\em
two} Lorentz invariant subspaces of positive/negative frequency
solutions.} There is no contradiction with the Casimir being constant -
we just happen to have {\em two} unitary representations which have the
{\em same} values of the Poincare (orthochronous, proper) Casimir invariants.
These in fact represent `particle' and `anti-particle' representations
respectively.

The plane wave orthonormal basis, $\{ u_{\vec{k}}(x)\}$ is in one-to-one
and onto correspondence with the `particle basis' $\{|k \rangle\}$.

The same features are exhibited by the other solution spaces, it remains
to identify a conserved current on the space of solutions, define an
inner product, obtain an orthonormal set and show unitarity. Since all
field equations imply that the fields always satisfy the Klein-Gordon
equation, we will always have the positive/negative frequency subspaces
and the particle/anti-particle identification.
 
Consider the {\bf Proca equation with divergence condition}: $(\Box -
m^2)v^{\mu} = 0 = \partial\cdot v$. Let $V$ denote the space of
solutions of these equations. As before, let $v^{\mu}$ and $u^{\mu}$ be
two solutions. Then it follows that $\boxed{J^{\mu} :=
\lambda(v^{*\nu}\partial^{\mu}u_{\nu} -
u^{\nu}\partial^{\mu}v^*_{\nu})}$ is conserved exactly as before. The
divergence condition just selects a subspace of the space of solution of
the Klein-Gordon equation. The inner product is defined as (with the
same convention of positive frequency solutions),
\[
	(v, u) ~ := ~ i\int_{\Sigma_t}d^3x \left[v^{*\nu}
	\partial^{\mu}u_{\nu} - u^{\nu}\partial^{\mu}v^*_{\nu}\right]
\]
Consider the family of solutions, 
\begin{eqnarray}
	u^{\mu}_{\vec{k}}(x) & := &
	\frac{1}{(2\pi)^{3/2}}\varepsilon^{\mu}(k) e^{ik\cdot x} ~ , ~
	k_{\mu}\varepsilon^{\mu}(k) = 0 ~,~ k\cdot x := -k^0\ t +
	\vec{k}\cdot\vec{x} \\ 
\left(u_{\vec{k}'}, u_{\vec{k}}\right) & = & \frac{1}{(2\pi)^3}(k^0 +
k^{'0})\left[ \varepsilon^{*\mu} (k') (\varepsilon_{\mu} (k)
	\right]e^{-i(k^0 - k^{'0})t} \left[(2\pi)^3\delta^3(\vec{k} -
	\vec{k}')\right] \nonumber \\
	& = & \left[ \varepsilon^{*\mu} (k') (\varepsilon_{\mu} (k)
		\right] \delta^3_{inv}(\vec{k}' - \vec{k}) ~ ~ , ~ ~
		\mbox{here} ~ k^0 := \sqrt{\vec{k}^2 + m^2} \ .
\end{eqnarray}
The polarizations $\varepsilon(k)$ satisfying the transversality
condition, $k\cdot\varepsilon(k) = 0$ selects 3 independent vectors for
$m \neq 0$ and 2 independent vectors for $m = 0$ thanks to the
equivalence of $A_{\mu}$ and $A_{\mu} + \partial_{\mu}\Lambda$. To
distinguish the different polarization vectors, we introduce an
additional label, $a$ taking 3 and 2 values respectively for the massive
and massless cases. We choose them to satisfy the orthonormality
relations, $\boxed{\varepsilon^{*}_{\mu}(k,a)\varepsilon^{\mu}(k,b) =
\delta_{ab}}$ and of course $\varepsilon(k, a)\cdot k = 0\ \forall\ a =
1,2,3$.  For massless case, $a$ is usually denoted by $\lambda$ and
takes only two values. The plane wave family of solutions so defined,
constitute an orthonormal set.

Under the Poincare action,
\begin{eqnarray}
\left[u^{\mu}_{\vec{k},a}(x)\right]_{\Lambda,b} & = &
\Lambda^{\mu}_{~\nu}u^{\nu}_{\vec{k},a}(\Lambda^{-1}(x - b)) ~ = ~
\frac{1}{(2\pi)^{3/2}}\Lambda^{\mu}_{~\nu}\varepsilon^{\nu}
(\vec{k},a)e^{ik\Lambda^{-1}(x -b)} \nonumber \\
& = & e^{-i(\Lambda k)\cdot
b}(\Lambda^{\mu}_{~\nu}\varepsilon^{\nu}(k,a))e^{i(\Lambda k)\cdot x}  ~
~ \mbox{But,}\nonumber \\
(\Lambda k)_{\mu}(\Lambda^{\mu}_{~\nu}\varepsilon^{\nu}(k,a)) & = &
k_{\mu}\varepsilon^{\mu}(k,a) ~ = ~ 0 ~ \therefore ~ \mbox{r.h.s. = }~
e^{i(\Lambda k)\cdot b)}\varepsilon^{\mu}(\Lambda k), a)e^{i(\Lambda
k)\cdot x } \nonumber \\
\left[u^{\mu}_{\vec{k},a}(x)\right]_{\Lambda,b} & = & e^{i(\Lambda
k)\cdot b)}u^{\mu}_{\vec{\Lambda k},a}(x) \ .
\end{eqnarray}
Noting further that $(\Lambda\varepsilon(k',a))^* \cdot
(\Lambda\varepsilon(k,b)) = \varepsilon^*(\Lambda k', a) \cdot
\varepsilon(\Lambda k, b)$, unitarity of the Poincare action now follows
in exactly the same manner as for the scalar. As in the case of the
scalar, here too we have two unitary, irreducible representations of the
(orthochronous, proper) Poincare group corresponding to the `particle'
and `anti-particle' tags. For the zero mass case, the only difference is
that the polarizations are restricted to the two spatially transverse
directions.

The case of the spinors is more interesting. We follow the same strategy
of looking for a conserved current and a corresponding inner product, but
we will do this for the Dirac equation which is a first order
differential equation. We have:
\[
	\Psi^A_{\Lambda,a}(x) = D^A_{~B}(\Lambda)\Psi^B(\Lambda^{-1}(x -
	a)) ~ ~,~ ~ (-i\gamma^{\mu}\partial_{\mu} \pm m)\Psi = 0\ .
\]
Note that there are {\em two} equations.

To look for a conserved current, we need to consider the (matrix)
adjoint of the Dirac equation. The adjoint will involve
$\gamma^{\dagger}$ which are inconvenient for manipulations. However, it
follows readily that $\boxed{(\gamma^{\mu})^{\dagger} =
\gamma^0\gamma^{\mu}\gamma^0}$. This suggests that we use the {\em Dirac
conjugate} defined as: $\bar{\Psi} := \Psi^{\dagger}\gamma^0$. It
follows that
\[
(-i\gamma^{\mu}\partial_{\mu} \pm m)\Psi = 0 ~ ~ \leftrightarrow ~ ~
i\partial_{\mu}\bar{\Psi}\gamma^{\mu} \pm m \bar{\Psi} = 0\ .
\]
Using these equations for two solutions, $\Psi_1, \Psi_2$ and their
Dirac conjugates, it follows that $\boxed{J^{\mu}(\Psi_2, \Psi_1) :=
\lambda(\bar{\Psi}_2\gamma^{\mu}\Psi_1)}$ is the conserved current for
both $\pm\ m$ equations. With the choice $\lambda = +1$, the inner
product on the solutions of Dirac equations is defined as: 
\[
(\Psi_2, \Psi_1) := \int_{\Sigma_t}d^3x \bar{\Psi}_2\gamma^0\Psi_1 ~ ~
\leftrightarrow ~ ~ (\Psi, \Psi) = \int_{\Sigma_t}d^3x\Psi^{\dagger}\Psi
~ \ge ~ 0 \ .
\]
\underline{Caution:} For $\gamma_0$ in the above equation, $(\Psi, \Psi)
\le 0$.

To construct a family of orthonormal solutions, consider the `plain wave
ansatz', $\Phi^A(\vec{k},\sigma,x) := \frac{1}{(2\pi)^{3/2}}
\varphi^A(k,\sigma)e^{ik\cdot x}$. The Dirac equations then require
$(\dsl{k} \pm m)\varphi(k,\sigma) = 0$ or $\dsl{k}\varphi = \mp
m\varphi$. Multiplying by $\dsl{k}$ and using $\dsl{k}\dsl{k} = - k^2$
leads to $k^2 + m^2 = 0 \leftrightarrow k^0 := \pm \sqrt{\vec{k}^2 +
m^2}$. We fix $k^0$ to be positive and call a solution with $e^{ik\cdot
x}$ as having positive frequency and a solution with $e^{-ik\cdot x}$ as
having negative frequency. The {\em two} Dirac equations can now be
taken as a single Dirac equation with $+m$, admitting positive and
negative frequency solutions. Explicitly, we refine the ansatz as
$\Phi^A_+(\vec{k},\sigma,x) := \frac{1}{(2\pi)^{3/2}}
u^A(k,\sigma)e^{ik\cdot x}$ and $\Phi^A_-(\vec{k},\sigma,x) :=
\frac{1}{(2\pi)^{3/2}} v^A(k,\sigma)e^{-ik\cdot x}$. Both satisfying
$(-i \dsl{\partial} + m)\Phi_{\pm} = 0$. This implies that $(\dsl{k} +
m)u(k,\sigma) = 0 = (\dsl{k} - m)v(k,\sigma)$. The $u, v$ spinors get
their $k-$dependence through these defining equations. These spinors are
eigen-spinors of $\dsl{k}$ with eigenvalues $\pm\ m$. Since the trace of
the $\gamma$'s vanishes, $\dsl{k}$ is traceless too and hence each
eigenvalue is doubly degenerate i.e. the `$\sigma$' label in the
eigen-spinors takes {\em two} values. These will be linked to the
helicities later on.

Substitution of these solutions in the inner product leads to,
\begin{eqnarray}
	(\Phi(k',\sigma',), \Phi(k,\sigma)) & = & \int_{\Sigma_t}d^3x
	\frac{1}{(2\pi)^3} u^{\dagger}(k',\sigma')u(k,\sigma)e^{-i(k'
	-k)\cdot x} \nonumber \\
	& = & u^{\dagger}(k',\sigma')u(k,\sigma)e^{-i(k^0 -
	k^{'0})t}\delta^3(\vec{k} - \vec{k}') \\
	\therefore (\Phi(k',\sigma',), \Phi(k,\sigma)) & = &
	\left[\frac{u^{\dagger}(k',\sigma')u(k,\sigma)}{2k^0}\right]
	\delta^3_{inv}(\vec{k} - \vec{k}') ~ ~ := ~ ~
	\delta_{\sigma',\sigma}\delta^3_{inv}(\vec{k} - \vec{k}') \ .
\end{eqnarray}
We have chosen a normalization of the $u-$spinors. Identical
orthonormality relations follow for the $v-$spinors.

Under Lorentz transformations, the solutions transform as,
\[
\left[\Phi(k,\sigma)\right]_{\Lambda}(x) ~ = ~
\left[D(\Lambda)u(k,\sigma)\right] \frac{e^{i(\Lambda k)\cdot
x)}}{(2\pi)^{3/2}}  ~ ~ ~ \stackrel{?}{=} ~ \Phi(\Lambda k, \sigma)(x) 
\]
The last equality will hold if $D(\Lambda)u(k,\sigma) = u(\Lambda k,
\sigma)$ i.e. if $(\dsl{(\Lambda k)} + m) \left[D(\Lambda) u(k,\sigma)
\right] = 0$.

We already have $[\Sigma^{\mu\nu}, \gamma^{\lambda}] =
i(\eta^{\mu\lambda}\gamma^{n} - \eta^{\nu\lambda}\gamma^{\mu})$. This
implies (may be checked by taking the infinitesimal form of
$D(\Lambda)$), 
\[
D^{-1}(\Lambda)\gamma^{\mu}D(\Lambda) = \Lambda^{\mu}_{~\nu}\gamma^{\nu}
~ \Rightarrow ~ \gamma^{\mu}D(\Lambda) =
\Lambda^{\mu}_{~\nu}D(\Lambda)\gamma^{n} \ .
\]

It follows, 
\begin{eqnarray}
\left[\dsl{(\Lambda k)} + m\right] \left(D(\Lambda) u(k,\sigma) \right)
& = & \left[(\Lambda k)_{\mu}\gamma^{\mu}D(\Lambda) + D(\Lambda)
m\right]u(k,\sigma) \\
& = & \left[(\Lambda k)_{\mu}\ \Lambda^{\mu}_{~\nu}
D(\Lambda)\gamma^{\nu} + D(\Lambda) m\right] u(k,\sigma) \nonumber \\
& = & D(\Lambda) \left[\Lambda_{\mu}^{~\alpha}
k_{\alpha}\Lambda^{\mu}_{~\nu} \gamma^{\nu} + m\right]u(k,\sigma) ~ ~
\mbox{but, $\Lambda_{\mu}^{~\alpha} \Lambda^{\mu}_{~\nu} =
\delta^{\alpha}_{~\nu}$ } \nonumber \\
\therefore \left[\dsl{(\Lambda k)} + m\right] \left(D(\Lambda)
u(k,\sigma) \right) & = & D(\Lambda)\left[k_{\nu}\gamma^{\nu} +
m\right]u(k,\sigma) ~ = ~ 0. ~ \mbox{(Covariance Result)} ~ ~ 
\end{eqnarray}
Thus, indeed we have $\boxed{D(\Lambda)u(k,\sigma) = u(\Lambda k,
\sigma)}$ and hence, 
\[
\left[\Phi(k,\sigma)\right]_{\Lambda,a}(x) ~ = ~ e^{-i(\Lambda k)\cdot
a}\Phi(\Lambda k, \sigma)(x) \ .
\]

Under Lorentz transformations then the orthonormality relation
transforms as,
\begin{eqnarray}
(\Phi(k',\sigma')_{\Lambda,a}, \Phi(k,\sigma)_{\Lambda,a}) & = &
e^{i\Lambda(k' - k)\cdot a} \left(\Phi(\Lambda k',\sigma'), \Phi(\Lambda
k, \sigma)\right) \nonumber \\ 
& = & \left[ \frac{u^{\dagger} (\Lambda k', \sigma') u(\Lambda k,
\sigma)}{2(\Lambda k)^0} \right] \delta^3_{inv} (\vec{k}' - \vec{k}) ~ =
~ \delta_{\sigma', \sigma}\delta^3_{inv} (\vec{k}' - \vec{k})\ .
\end{eqnarray}
In the last equality we have used the normalization of the $u$ spinors.
This proves the unitarity of the Poincare representation on the positive
frequency solutions. As before, Lorentz transformations do not mix the
subspaces of the positive/negative frequency solutions.

The normalization definition does not look Lorentz invariant, but it is.
Both $u^{\dagger}u = \bar{u}\gamma^0u$ and $k^0$ transform the same way.
A more convenient form will be displayed later on.

\underline{Note:} The covariance result allows us to define the $u, v$
spinors from their definition in the rest frame (for massive case)i,
$\hat{k} = (m, \vec{0})$. In this frame, the spinors satisfy the
equations: $-m\gamma^0 u( \hat{k}, \sigma) = -m u( \hat{k}, \sigma)\ ,$
and $-m\gamma^0 v( \hat{k}, \sigma) = +m v( \hat{k}, \sigma)$. Thus
these spinors at $\hat{k}$ are eigen-spinors of $\gamma^0$ with
eigenvalues $\pm 1$ respectively. $\gamma^0$ being Hermitian, these
spinors are orthogonal: $u^{\dagger}(\hat{k},\sigma)v(\hat{k}, \sigma) =
0$. As noted before, these eigenvalues are doubly degenerate and
$\sigma$ labels these. For the massless case, $\hat{k} = (k, 0, 0, k)$
and $u, v$ are eigen-spinors of $J_3$ generator of the Little group.

\underline{Note:} Incidentally, covariance result and the identification
of $u(\hat{k},\sigma), v(\hat{k},\sigma)$ as eigen-spinors of $\gamma^0$
also leads to a {\em completeness relation} as follows. Given that
$\gamma^0 u(\hat{k},\sigma) = u(\hat{k}, \sigma)\ , \gamma^0 v(\hat{k},
\sigma) = - v(\hat{k},\sigma)$ and each eigen-space being two
dimensional, $\Rightarrow$ completeness relations (`spectral
representation' - $A = \sum_n \lambda_n |n \rangle \langle n|$\ ) takes
the form 
\begin{eqnarray*}
\sum_{\sigma}u(\hat{k},\sigma)u^{\dagger}(\hat{k},\sigma) =
\frac{\mathbb{1} + \gamma^0}{2} & ~,~ &
\sum_{\sigma}v(\hat{k},\sigma)v^{\dagger}(\hat{k},\sigma) =
\frac{\mathbb{1} - \gamma^0}{2} \\
\sum_{\sigma}u(\hat{k},\sigma)\bar{u}(\hat{k},\sigma) = \frac{\gamma^0 +
\mathbb{1}}{2} ~ = ~ \frac{-\dsl{\hat{k}} + m}{2m} & ~,~ &
\sum_{\sigma}v(\hat{k},\sigma)\bar{v}(\hat{k},\sigma) = \frac{\gamma^0 -
\mathbb{1}}{2} ~ = ~ - \frac{\dsl{\hat{k}} + m}{2m} \\
\sum_{\sigma}u(\vec{k},\sigma)\bar{u}(\vec{k},\sigma) = \frac{-\dsl{{k}}
+ m}{2m} & ~,~ & \sum_{\sigma}v(\vec{k},\sigma)\bar{v}(\vec{k},\sigma) =
- \frac{\dsl{{k}} + m}{2m} \\
\end{eqnarray*}
In the second equation, we used $\gamma^0 = -\Case{\dsl{\hat{k}}}{m}$
and in the last equation we multiplied by $D(L)$ on the left and
$D^{-1}(L)$ on the right and used the covariance result. Here $L$ is a
boost which takes $\hat{k} \rightarrow \vec{k}$. Notice that multiplying
by $u(\vec{k},\sigma')$ and $v(\vec{k},\sigma')$ respectively, and using
the equations satisfied by the spinors, gives
$\boxed{\bar{u}(\vec{k},\sigma) u(\vec{k},\sigma') =
\delta_{\sigma,\sigma'} = - \bar{v}(\vec{k},\sigma)
v(\vec{k},\sigma')}$. These {\em are consistent} with the previously
chosen normalizations: $u^{\dagger}(\vec{k},\sigma)u(\vec{k},\sigma') =
2k^0 \delta_{\sigma,\sigma'} = v^{\dagger}(\vec{k},\sigma)
v(\vec{k},\sigma')$. 

Thus, for all the field equations we have found unitary, irreducible
representations of the Poincare group. We have also discovered that the
{\em field representations automatically also include `anti-particles'}.
This is really a consequence of the requirement of {\em manifest
Poincare covariance} which necessitates the field representations.

\underline{Note:} From the expressions above, it should be clear that
the ket vectors $|k,\sigma \rangle$ are in one-to-one correspondence
with the ``Plane wave solutions'' with positive {\em and} negative
frequencies, displayed above.

We need to study the behavior of these representations under the
space inversion, time reversal and the new possibility of `charge
conjugation' (particle-particle exchange). 

\newpage
\section{Parity, time reversal and charge
conjugation}\label{DiscreteSyms}

For the field representations above, we focussed on the subgroup of the
Poincare group, connected to the identity. The remaining elements of the
Poincare group - improper and non-orthochronous are generated by the two
transformations of space inversion  and time reversal. Their actions on
the generators is given in eq.(\ref{DiscreteOnGenerators}). These
relations continue to hold in {\em any} representation, in particular
also on the field representations. Looking at the general form of the
Poincare action on the $\Psi^A(x)$'s, $\Psi^A_{\Lambda,a}(x) =
D^A_{~B}(\Lambda)\Psi^B(\Lambda^{-1}(x-a))$, we see that the action on
the space-time point is common to {\em all} $D(\Lambda)$ representations
and we may focus on the $D(\Lambda)$ part of it.  Recalling that the
time reversal operation is anti-linear and anti-unitary, we have to be
careful about taking complex conjugates.  Thus we define the actions as,
\begin{eqnarray}
	\Psi^A_{\cal{P}}(x) & := & D^A_{~B}(\cal{P})\Psi^B(t, -\vec{x})
	~ ~ , ~ ~ D(\cal{P}) =: \Pi \\ 
	\Psi^A_{\cal{T}}(x) & := & D^A_{~B}(\cal{T})(\Psi^B(-t,
	\vec{x}))^* ~ ~ , ~ ~ D(\cal{T}) =: \tau , 
\end{eqnarray}
It is customary to denote the anti-unitary, anti-linear operator
$\cal{T}$ as $\tau K$ where $\tau$ is a unitary operator and  $K$ takes
complex conjugate of numbers on its right. $\cal{T}^2 = \mathbb{1}$
implies $\tau\tau^* = \mathbb{1}$.

Consider the translation generators. These act as differential operators
on the $\Psi$'s: $P_{\mu}\Psi^A(x) = -i\partial_{\mu}\Psi^A$. Consider
the $\cal{P}$ and $\cal{T}$ actions on $P_{\mu}$ as given in eqn.
(\ref{DiscreteOnGenerators}). For instance,
\begin{eqnarray}
(\cal{T}^{-1}P_0\cal{T})\Psi^A(t,\vec{x}) & = &
\cal{T}^{-1}\{-i\partial_0 (\tau^A_{\ B}\Psi^B(-t, \vec{x})^*)\} ~ = ~
\cal{T}^{-1}\{+i\partial_{-t} (\tau^A_{\ B}\Psi^B(-t, \vec{x})^*)\}
\nonumber \\
& = & -i \partial_{t}(\tau\tau^*)^A_{\ B} \Psi^B(t,\vec{x}) ~ = ~
P_0\Psi^A(t,\vec{x})\ .
\end{eqnarray}
The time has been reversed twice and the complex conjugation has been
effected twice. For the spatial components, there is no `t' reversal and
the $\cal{T}$ introduces a sign. Space inversion is likewise straight
forward. We now focus on the Lorentz generators only.

The scalar case has no non-trivial matrices and the vector case is
simpler and is left as an exercise. The spinner case is non-trivial.  We
have the Lorentz generators:
\[
	\Sigma^{\mu\nu} = \frac{i}{4}\left[\gamma^{\mu},
	\gamma^{\nu}\right] ~ \leftrightarrow ~ J_i =
	\frac{\epsilon_{ijk}}{2}\Sigma^{jk} ~ , ~ K^i = \Sigma^{i0} \ .
\]
The $\Pi$ and $\tau$ matrices are also $4\times 4$ matrices. We note a
result.

\underline{Result:} {\em Any $4\times 4$ matrix can be written as a
	linear combination of the 16 $\Gamma$ matrices, $\{\Gamma\} =
	\{\mathbb{1}, \gamma^{\mu},\gamma^{[\mu}\gamma^{\nu]},
	\gamma^{[\mu}\gamma^{\nu}\gamma^{\lambda]}, \gamma_5\}$, the
square brackets denoting anti-symmetrization (16 = 1 + 4 + 6 + 4 + 1).} 

The result follows by showing that the 16 matrices are linearly
independent.

We now deduce the $\Pi$ and $\tau$ matrices from the commutation
relations (see eq.(\ref{DiscreteOnGenerators})):
\[
\Sigma_{jk}\Pi = \Pi\Sigma_{jk} ~,~ \Sigma_{0i}\Pi = - \Pi\Sigma_{0i} ~
~ , ~ ~ \Sigma_{jk}\tau = - \tau(\Sigma_{jk})^* ~,~ \Sigma_{0i}\tau =
\tau(\Sigma_{0i})^* \ . 
\]

The determination of $\Pi$ is straight forward. Commutation with
$\Sigma_{ij}$ implies that $\Pi$ is a linear combination of $\mathbb{1},
\gamma_5, \gamma_0$. Anti-commutation with $\Sigma_{0i} =
\Case{i}{2}\gamma_0\gamma_i$ implies that only $\gamma_0$ survives and
$\Pi = c\gamma^0$. Now $\Pi^2 = \mathbb{1}$ implies $c = \pm 1$. Hence,
$\boxed{(\Psi_{\cal{P}})^A(x) = \pm (\gamma^0)^A_{~B}\Psi^B(t,
-\vec{x})}$. It follows immediately that {\em left(right) handed Weyl
spinor becomes right(left) handed Weyl spinor under space
inversion.}

The determination of $\tau$ involves complex conjugation. From the
unitarity of the $\gamma$ matrices and the defining Clifford relations,
the hermiticity properties are determined: $\gamma_0^{\dagger} =
\gamma_0, \ \gamma_i^{\dagger} = - \gamma_i$. The transpose/complex
conjugation properties however depend on the explicit choice of the
$\gamma$'s. The $\tau$ matrix thus depends on the explicit
representation of the $\gamma$'s. There are \underline{three} commonly
employed representations:

\begin{eqnarray} \label{GammaRepresentations}
	\mbox{Dirac-Pauli} & : & \gamma^0 := \left(\begin{array}{cc}
	\mathbb{1} & \mathbb{0} \\ \mathbb{0} & -\mathbb{1}
	\end{array}\right)\ ,\
	\gamma^i := \left(\begin{array}{cc} \mathbb{0} & \sigma^i \\
	-\sigma^i & \mathbb{0} \end{array}\right)\ ,\
	\gamma_5 := \left(\begin{array}{cc} \mathbb{0} & \mathbb{1} \\
	\mathbb{1} & \mathbb{0} \end{array}\right) \ ; \\
	\mbox{Weyl} & : & \gamma^0 := \left(\begin{array}{cc} \mathbb{0}
	& \mathbb{1} \\ \mathbb{1} & \mathbb{0} \end{array}\right)\ ,\
	\gamma^i := \left(\begin{array}{cc} \mathbb{0} & \sigma^i \\
	-\sigma^i & \mathbb{0} \end{array}\right)\ ,\
	\gamma_5 := \left(\begin{array}{cc} -\mathbb{1} & \mathbb{0} \\
	\mathbb{0} & \mathbb{1} \end{array}\right) \ ; \\
	\mbox{Majorana} & : & \gamma^0 := \left(\begin{array}{cc}
	\mathbb{0} & \sigma_2 \\ \sigma_2 & \mathbb{0}
	\end{array}\right)\ ,\
	\gamma^1 := \left(\begin{array}{cc} i\sigma_3 & \mathbb{0} \\
	\mathbb{0} & i\sigma_3 \end{array}\right)\ ,\
	\gamma^2 := \left(\begin{array}{cc} \mathbb{0} & -\sigma_2 \\
	\sigma_2 & \mathbb{0} \end{array}\right) \ ; \\
	& : & \gamma^3 := \left(\begin{array}{cc} -i\sigma_1 &
	\mathbb{0} \\ \mathbb{0} & -i\sigma_1 \end{array}\right)\ ,\
	\gamma_5 := \left(\begin{array}{cc} \sigma_2 & \mathbb{0} \\
	\mathbb{0} & -\sigma_2 \end{array}\right)\ ,\
	\mbox{All purely imaginary.} ~ ~ 
\end{eqnarray}
In the Dirac-Pauli and Weyl representations, only $\gamma_2$ is
imaginary. The $\tau$ matrix thus satisfies:
\begin{eqnarray*}
	& & \Sigma_{12}\tau = - \tau\Sigma_{12} ~,~ \Sigma_{23}\tau = -
	\tau\Sigma_{23} ~,~ \Sigma_{31}\tau =  \tau\Sigma_{31} ~, \\
	& & \Sigma_{01}\tau = - \tau\Sigma_{01} ~,~ \Sigma_{02}\tau =
	\tau\Sigma_{02} ~,~ \Sigma_{03}\tau = - \tau\Sigma_{03} ~,
\end{eqnarray*}
The last relation in the first line suggests $\tau = \lambda \gamma^0
\gamma^2$. It also checks with all other relations. The $\lambda$ is
determined as follows. 

The operator $D(\cal{T})$ is anti-unitary. Hence,
\[
(D(\cal{T})D(\cal{T})\Psi_1, D(\cal{T})D(\cal{T})\Psi_2) ~ = ~
(D(\cal{T})\Psi_2, D(\cal{T})\Psi_1) ~ ~ = ~ ~ (\Psi_1, \Psi_2) ~ ~
\Rightarrow ~ ~ D^2(\cal{T}) ~ \mbox{is unitary}.
\] Next, $D^2(\cal{T}) = (\tau K)(\tau K) = \tau\tau^* =
|\lambda|^2(\gamma^0\gamma^2\gamma^0(-\gamma^2)) = -
|\lambda|^2\mathbb{1}$. The unitarity then implies that $|\lambda| = 1~$
i.e. $\lambda$ is a phase and $D^2(\cal{T}) = -\mathbb{1}$ in the spinor
representation.

In the Majorana representation, all $\gamma$'s are imaginary. Hence,
$\Sigma_{\mu\nu}^* = - \Sigma_{\mu\nu}.$ Hence $\Sigma_{jk}$ commute
with $\tau$ and anti-commute with $\Sigma_{0i}$. This is exactly as for
the space inversion and we deduce that $D(\cal{T}) = \lambda\gamma^0 K$.
Once again $D^2(\cal{T})$ is unitary and equal to $-\mathbb{1}$ and
$\lambda$ is a phase.

For vector representation, the `$A$' index will be a tensorial index and
it is left as an exercise to work out the $D$ matrices for space
inversion and time reversal.

The existence of the anti-particle representations suggest one more
discrete transformation of order 2, called {\em Charge Conjugation}.

Recall that the anti-particle representation is the subspace of negative
frequency solutions. These subspaces are spanned by plane wave solutions
and involve the operation of complex conjugation which takes $e^{ik\cdot
x} \to e^{-ik\cdot x}$. For the non-trivial $D(\Lambda)$
representations, complex conjugation also takes the complex conjugate of
the $D(\Lambda)$ matrices. However, a complex conjugation of a solution
need not be a solution again (especially for spinor as we will see) and
hence the charge conjugation must also involve additional
transformations over and above complex conjugation. For the
Klein-Gordon, Proca and Maxwell equations, complex conjugate of a
solution is also a solution as the differential operators are real. For
the Dirac equation we need to do further work.

For instance, let $\Psi$ be a positive frequency solution of the Dirac
equation, $(-i\gamma^{\mu}\partial_{\mu} + m)\Psi = 0$. Taking complex
conjugate of the equation, we get $(+i(\gamma^{\mu})^*\partial_{\mu} +
m)\Psi^* = 0$. {\em If} we could find an invertible matrix $B$ such that
$(\gamma^{\mu})^* = -B\gamma^{\mu}B^{-1}$, {\em then} $\Psi^c :=
(B^{-1}\Psi^*)$ satisfies the same Dirac equation and of course is a
negative frequency solution. $\Psi^c$ is called the {\em charge
conjugate}\footnote{The terminology comes when coupling to external
electromagnetic field is considered by the minimal substitution
$\partial_{\mu} \to \partial_{\mu} -ieA_{\mu}$. Under complex
conjugation, $e \to -e$. So if $\Psi$ is thought of as charge `e'
solution then $\Psi^c$ is a charge `-e' solution.} of the $\Psi$.
Apparently, $B$ depends on the choice of explicit $\gamma$ matrices.

As a preparation for subsequent development, we have a subsection
on properties of $\gamma$ matrices.
\subsection{Representations of the Clifford algebra and relations among
them}
Quite generally, for any group, given a (matrix) representation $R(G)$,
we have three other representations, namely, $R^*(G), (R^T)^{-1}(G)$ and
$(R^{\dagger})^{-1}(G)$. From the basic relation $R(g_1)R(g_2) =
R(g_1.g_2)$, taking complex conjugate, transpose inverse and adjoint
inverse immediately verifies the assertion. If in addition, the
representation $R(G)$ is {\em unitary} (always true for finite groups
and compact Lie groups), then $(R^T)^{-1}(G) =  R^*(G)$ and
$(R^{\dagger})^{-1}(G) = R(G)$. $R(G)$ and $R^*(G)$ are then the only
independent representations. These are either equivalent, $R^*(g) =
SR(g)S^{-1}$ or inequivalent. It turns out that for unitary, irreducible
representations, if $R$ and $R^*$ are equivalent, then $S$ is either
symmetric or anti-symmetric. The following terminology ensues.

For unitary, irreducible representations:
\begin{eqnarray*}
	R(G) \ \mbox{is complex \hspace{0.55cm}} & \mbox{if} & R^*(G) 
	\nsim R(G) \\
	R(G) \ \mbox{is pseudo-real \hspace{0.0cm}} & \mbox{if} & R^*(G) 
	= S R^*(G)
	S^{-1} ~ ~,~ ~ S^T = - S \\
	R(G) \ \mbox{is real \hspace{1.4cm}} & \mbox{if} & R^*(G) = S
	R^*(G) S^{-1} ~ ~,~ ~ S^T =  S 
\end{eqnarray*}

This is relevant for representations of internal symmetry groups as well
as the Clifford group of the 32 elements. The $D(\Lambda)$
representations are finite dimensional but not unitary since Lorentz
group does not have unitary finite dimensional representations.

For the Clifford group in 4 dimensions (and in even number of
dimensions) there is only one (up-to unitary equivalence) non-trivial
representation which is 4 dimensional (or $2^{N/2}$ dimensional for
$N$-even number of dimensions). Since $\pm\gamma^*, \pm\gamma^T$ and
$\pm\gamma^{\dagger}$ all satisfy the same Clifford algebra and the
representation is unique, $\exists$ matrices $B, C, D$ such that,
\[
	(\gamma^{\mu})^* = - B\gamma^{\mu}B^{-1} ~,~ (\gamma^{\mu})^T =
	- C\gamma^{\mu}C^{-1} ~,~ (\gamma^{\mu})^{\dagger} = +
	D\gamma^{\mu}D^{-1} ~.
\]
The choice of signs above is conventional. Notice that replacing any of
the $B, C, D$ matrices by multiplying by $\gamma_5$ on the right,
reverses the signs. 

It follows immediately that,
\[
	-(\Sigma^{\mu\nu})^* ~ = ~ B\Sigma^{\mu\nu}B^{-1} ~ ~,~ ~
	-(\Sigma^{\mu\nu})^T ~ = ~ C\Sigma^{\mu\nu}C^{-1} ~ ~,~ ~
	(\Sigma^{\mu\nu})^* ~ = ~ D\Sigma^{\mu\nu}D^{-1}
\]
Given a representation $D(\Lambda)$ we have the
$(D^{\dagger})^{-1}(\Lambda)\ , (D^T)^{-1}(\Lambda)$ and $D^*(\Lambda)$
representations. The infinitesimal forms are $(\mathbb{1} +
\frac{i}{2}\omega\cdot\Sigma), (\mathbb{1} -
\frac{i}{2}\omega\cdot\Sigma)^{\dagger}, (\mathbb{1} -
\frac{i}{2}\omega\cdot\Sigma^T), (\mathbb{1} -
\frac{i}{2}\omega\cdot\Sigma^*)$ which imply that the corresponding
generators are $\Sigma^{\mu\nu}, (\Sigma^{\mu\nu})^{\dagger},
-(\Sigma^{\mu\nu})^T, -(\Sigma^{\mu\nu})^*$. The $B, C, D$ matrices
precisely relate these generators. 

These relations can be exponentiated to get corresponding relations
among the $D(\Lambda), D^{\dagger}(\Lambda), D^T(\Lambda),
D^*(\Lambda)$, namely, 
\[
	D^{\dagger}(\Lambda) = DD^{-1}(\Lambda)D^{-1} ~ ~,~ ~
	D^{T}(\Lambda) = CD^{-1}(\Lambda)C^{-1} ~ ~,~ ~ D^{*}(\Lambda) =
	BD(\Lambda)B^{-1} ~ .
\]
These can be checked easily by using the series form of the exponentials
derived from $D(\Lambda) = \sum_{k=0}^{\infty} \frac{(i \omega\cdot
\Sigma)^k}{k!}$. 
\subsection{Dirac-Majorana-Charge Conjugates}
This allows us to construct Lorentz invariant bi-linears  from the
spinors. For instance, define the {\em Majorana conjugate} $\tilde{\Psi}
:= \Psi^T C$. Then,
\[
	(\tilde{\Psi})_{\Lambda} := (\Psi_{\Lambda}^T) C ~ = ~
	\Psi^T(D^T(\Lambda)) C ~ = ~ \Psi^T C D^{-1}(\Lambda) ~ = ~
	\tilde{\Psi}D^{-1}(\Lambda) \ .
\]
Thus the Majorana conjugate, like the Dirac conjugate $\bar{\psi} :=
\Psi^{\dagger}D$, transforms by $D^{-1}(\Lambda)$. Recall that the
charge conjugate, $\Psi^{c}$ transforms by $D(\Lambda)$. {\em
	Consequently, $\bar{\Psi}\Psi, \bar{\Psi}\Psi^c, \tilde{\Psi}
\Psi, \tilde{\Psi} \Psi^c$ are all Lorentz scalars.}

Recalling that $D^{-1}(\Lambda)\gamma^{\mu}D(\Lambda) =
\Lambda^{\mu}_{~\nu}\gamma^{\nu}$, it is easy to see that,
\begin{eqnarray*}
	\bar{\Psi}\Psi & \mbox{is a} & \mbox{Scalar} \\
	\bar{\Psi}\gamma^{\mu}\Psi & \mbox{is a} & \mbox{Vector} \\
	\bar{\Psi}\gamma^{\mu}\gamma^{\nu}\Psi & \mbox{is a} &
	\mbox{Tensor of rank 2} \\
	\bar{\Psi}\gamma^{\mu}\gamma_5\Psi & \mbox{is a} & \mbox{Axial
	(or pseudo) vector} \\
	\bar{\Psi}\gamma_5\Psi & \mbox{is a} & \mbox{Pseudo-scalar} 
\end{eqnarray*}
Combine these with the definitions,
\begin{eqnarray*}
	\mbox{Dirac Spinor} & : & \Psi = \Psi ~ ~,~ ~ (\Psi)_{\Lambda} =
	D(\Lambda)\Psi ; \\
	\mbox{Dirac Conjugate} & : & \bar{\Psi} = \Psi^{\dagger}D ~ ~,~
	~ (\bar{\Psi})_{\Lambda} = \bar{\Psi}D^{-1}(\Lambda) ; \\
	\mbox{Charge Conjugate} & : & \Psi^c = B^{-1}\Psi^* ~ ~,~ ~
	(\Psi^c)_{\Lambda} = D(\Lambda)\Psi^c ; \\
	\mbox{Majorana Conjugate} & : & \tilde{\Psi} = \Psi^TC ~ ~,~ ~
	(\tilde{\Psi})_{\Lambda} = \tilde{\Psi}D^{-1}(\Lambda) . 
\end{eqnarray*}
Thus, the same Lorentz transformation properties hold with $\Psi \to
\Psi^c$ and/or $\bar{\Psi} \to \tilde{\Psi}$. The last two need a little
explanation.

We defined $\gamma_5 := i \gamma^0\gamma^1\gamma^2\gamma^3 =
\Case{i}{4!}\varepsilon_{\mu\nu\alpha\beta} \gamma^{\mu} \gamma^{\nu}
\gamma^{\alpha} \gamma^{\beta}$, where $\varepsilon$ is the completely
antisymmetric symbol, with $\varepsilon_{0123} := 1$. This symbol is
also used to define the determinant of a matrix, eg,
$\boxed{\varepsilon_{\mu'\nu'\alpha'\beta'} \Lambda^{\mu'}_{~\mu}
	\Lambda^{\nu'}_{~\nu} \Lambda^{\alpha'}_{~\alpha}
\Lambda^{\beta'}_{~\beta} = det(\Lambda)
\varepsilon_{\mu\nu\alpha\beta}}$. It follows immediately that
$\boxed{D^{-`}(\Lambda)\gamma_5D(\Lambda) = det(\Lambda)\gamma_5}$.

Consider now an axial vector combination,
\[
	[\bar{\Psi}\gamma^{\mu}\gamma_5\Psi]_{\Lambda} ~ = ~
	\bar{\Psi}D^{-1}(\Lambda)\gamma^{\mu}\gamma_5D(\Lambda)\Psi ~ =
	~ \Lambda^{\mu}_{~\nu} \bar{\Psi} \gamma^{\nu} D^{-1} \gamma_5
	D(\Lambda)\Psi ~ = ~ det(\Lambda) \Lambda^{\mu}_{~\nu}
	\bar{\Psi} \gamma^{\nu} \gamma_5 \Psi \ .
\]
Note that for proper Lorentz transformations, the axial vector and the
pseudo-scalar transform as vector and scalar respectively.

Quantities that transform as tensors but with an extra factor of
determinant of the transformation are  called {\em pseudo-tensors}.
There are also tensor looking quantities that transform with additional
factor of $|det|^w$ and these are  called {\em tensor densities of
weight $w$.}

Let us return to determine the matrices $B, C, D$.

Since the $\gamma$'s are unitary, we can always choose any of the $B, C,
D$ matrices to be unitary. Quite generally, if $R' = S R S^{-1}$ for two
irreducible, equivalent unitary representations, then $R'^{\dagger}R' =
\mathbb{1} \Rightarrow R^{\dagger}S^{\dagger}S = S^{\dagger}S R
\Rightarrow S^{\dagger} S = \alpha\mathbb{1}$. Positivity of
$S^{\dagger}S$ gives $\alpha > 0$. Hence $S' := S/\sqrt{\alpha}$ is a
unitary matrix. This proves the claim. 

Next, $\gamma^{\mu} = (\gamma^{\mu*})^* = -B^*\gamma^{\mu*}(B^{-1})^* =
+ B^*B \gamma^{m} (B^*B)^{-1} \Rightarrow B^*B = \lambda\mathbb{1}$ with
$\lambda$ being real since $B^*B$ is.  The determinant of $B$ being a
phase (unitarity of $B$) fixes $\lambda := \epsilon_B := \pm 1$ and
$\boxed{B^*B = \epsilon_B\mathbb{1}}$. Similar manipulation with
$\gamma^{\mu} = (\gamma^{\mu T})^T$ leads to $C^{-1} C^T =
\lambda\mathbb{1}$. Unitarity gives $C^{-1}C^T = C^{\dagger}C^T =
(C^T)^*C^T$, hence $\lambda$ is real and determinant gives $\lambda^4 =
1$. Hence, $\boxed{C^T = \epsilon_C C,  \epsilon_C = \pm 1}$. For
$\gamma = (\gamma^{\dagger})^{\dagger}$ we get $D^{\dagger} = \lambda D$
for some phase $\lambda$. Since $D$ is determined to within a phase, we
can define $D' := e^{i\alpha}D$ and choose $\alpha$ so that $\lambda =
e^{-2i\alpha}$ to get $(D')^{\dagger} = D'$. Thus, without loss of
generality, we can always choose $D^{\dagger} = D$. Now, knowing the
hermiticity properties of the $\gamma$'s, we get $\boxed{D = \gamma^0}$,
independent of any choice of explicit $\gamma$ matrices.

The $\epsilon_B, \epsilon_C$ are correlated since $\gamma^T =
(\gamma^*)^{\dagger} = (\gamma^{\dagger})^*$.
$\gamma^T = (\gamma^*)^{\dagger}$ leads to,
\[
	-C\gamma C^{-1} ~ = ~ -(B^{\dagger})^{-1}D\gamma
	D^{-1}B^{\dagger} ~ ~ \Rightarrow ~ ~ C = \lambda B D .
\]
$\gamma^T = (\gamma^{\dagger})^*$ leads to,
\[
	-C\gamma C^{-1} ~ = ~ -(D^{*})B\gamma
	B^{-1}(D^{-1})^* ~ ~ \Rightarrow ~ ~ C = \lambda' D^* B .
\]
Eliminating $C$ gives $BD = \Case{\lambda'}{\lambda}D^*B$ or $D^* =
\Case{\lambda}{\lambda'}B D B^{-1}$. However, $D = \gamma^0$ can always
be achieved as shown above. Therefore, $\lambda' = - \lambda$ and
$C = \lambda B D = - \lambda D^* B$. Next,
\[
	C^T = \lambda D^T B^T = \lambda(D^{\dagger})^*(B^{\dagger})^* =
	\lambda D^* (B^*)^{-1} = \lambda D^* (\epsilon_B B) = -
	\epsilon_B C ~ \Rightarrow ~ \boxed{\epsilon_C = -\epsilon_B .}
\]
Note that this is independent of explicit $\gamma$ matrices and also
independent of the phase $\lambda$! We still need to determine
$\epsilon_B$, say.

{\em Claim:} The $\epsilon_B$ does not depend on the choice of $\gamma$
matrices.

This is easily proved. Let $\gamma'$ and $\gamma$ be two distinct
choices of explicit $\gamma$-matrices. Both being unitary are related by
some unitary matrix, $S$ as $\gamma = S\gamma' S^{-1}$. Then
substitution in $\gamma^* = -B\gamma B^{-1}$ gives, $(\gamma')^* = -
B'\gamma' (B')^{-1}$ with $B' := (S^*)^{-1} B S$. It follows that
$(B')^* B' = (S^{-1}B^*S^*)( (S^*)^{-1}B S ) = \epsilon_B\mathbb{1}$,
proving the claim. It therefore suffices to choose an explicit
representation and evaluate $\epsilon_B$.  

Referring to say the Dirac-Pauli representation, see
(\ref{GammaRepresentations}), only $\gamma^2$ is pure imaginary. Hence,
$\gamma^{0,1,3}B = -B\gamma^{0,1,3}$ and $\gamma^2B = +B\gamma^2$.
Furthermore, $\gamma_5^* = \gamma_5 \Rightarrow \gamma_5B = - B\gamma_5$
as well. Therefore $B$ must be made up of odd number of $\gamma$'s. By
inspection, $B = \alpha\gamma^2$ satisfies the conditions. Unitarity of
$B$ restricts $\alpha$ to be a phase and $B^*B =
|\alpha|^2(-\gamma^2)(\gamma^2) = \mathbb{1} \Rightarrow \epsilon_B =
+1$ and $\epsilon_C = -1$. Also $C = \lambda BD =
\lambda\alpha\gamma^2\gamma^0 = (-\lambda\alpha)\gamma^0\gamma^2$. {\em
The phases $\lambda,\ \alpha$ are arbitrary and convention dependent}.
We {\em choose} $\lambda = 1$ and $\alpha = i$ so that $\boxed{B =
i\gamma^2 ~,~ C = i\gamma^2\gamma^0\ .}$

This also verifies for the Weyl representation. For Majorana
representation, all $\gamma^{\mu}$'s are imaginary and thus {\em
commute} with $B$. Hence $B$ must be a phase multiple of $\mathbb{1}$.
This is also consistent with commutation with $\gamma_5$. Clearly
$\epsilon_B = +1$ and $C = (\lambda\alpha)\gamma^0$.

This completes the discussion of representations and relations among
them.

It is customary to introduce a {\em charge conjugation operator},
$\cal{C} := B^{-1} K$, where $K$ instructs to take the complex conjugate
of the numbers on its right. This operator is anti-linear and since $B$
is unitary, it is also anti-unitary operator. It follows,
\[
	\cal{C}\Sigma^{\mu\nu} = B^{-1}(\Sigma^{\mu\nu})^*K =
	B^{-1}(-B\Sigma^{\mu\nu}B^{-1})K = - \Sigma^{\mu\nu}\cal{C} ~
	\Rightarrow ~ \cal{C}D(\Lambda) = D(\Lambda)\cal{C} \ .
\]
Thus charge conjugation acts invariantly on Lorentz representations.
{\em Provided} $\cal{C}^2 = \mathbb{1}$, we have $(\Case{\mathbb{1} \pm
\cal{C}}{2})^2 = \Case{\mathbb{1} \pm \cal{C}}{2}$ and the Lorentz
representation can be reduced further. The corresponding subspaces
satisfy $\cal{C} \psi = \pm \Psi$ and these spinors are  called
{\em Majorana spinors}. Thus, Majorana spinors are Dirac spinors which
satisfy $\cal{C} \Psi = \pm \Psi \leftrightarrow B^{-1}\psi^* = \pm
\Psi$. So is $\cal{C}^2 = \mathbb{1}$? We have $\cal{C}^2 =
B^{-1}KB^{-1}K = (B^*B)^{-1}K^2 = \epsilon_B\mathbb{1} = \mathbb{1}$.
Therefore, we {\em do} have Majorana spinors (in 4 dimensions).

Do we have Weyl-Majorana spinors? Well, $\cal{C}\gamma_5 =
B^{-1}\gamma_5^*K = - \gamma_5B^{-1}K = -\gamma_5 \cal{C}$ and we {\em
cannot have} Weyl-Majorana spinors in 4 dimensions. 

A similar analysis can be carried out for spinors in any dimensions and
with any metric signature. This may be seen for instance in
\cite{Wetterich}.

\newpage
%
\section{Relativistic Actions: classical fields}
\label{ActionFormulation}

So far we focused on the representation theory of the Poincare group.
The abstract, algebraic approach revealed the attributes of these
representations, namely mass and spin/helicity. In order to have a
framework which is manifestly Poincare covariant, we constructed and
analyzed representations on vector valued (complex in general) function,
$\Psi^A$ - henceforth these will be generically referred to as {\em
fields}. The irreducibility condition emerged as ``field equations''.
These equations are all homogeneous, linear and with at the most two
derivatives. They all admit plane waves and their linear combinations as
solutions, but no other `phenomenon' involving (say) different types of
waves modifying their propagation etc. Intuitively, there are {\em no
interactions}.

We would like to have a framework which is not only Poincare covariant
but also involves ``interaction'' or non-trivial ``dynamics''. The well
tested and successful strategy is to have an {\em action formulation}.
Let us quickly note several advantages of having an action formulation:
\begin{enumerate}
	\item The equations of motion are retrieved invoking a
		variational principle as the Euler-Lagrange equations of
		motion;
	\item ``Interactions'' can be understood naturally as leading to
		non-linear and/or coupled equations which can be easily
		introduced as more than quadratic order terms in the
		action;
	\item Covariance of equations of motion (or dynamics) can be
		easily incorporated by requiring appropriate invariance
		of the action;
	\item The Noether's theorem gives a recipe for obtaining
		quantities conserved by equations of motion (and hence
		during interactions as well);
	\item It leads to a canonical framework of symplectic structure
		(``Poisson brackets'') and a Hamiltonian evolution. This
		provides a systematic method of identifying ``degrees of
		freedom'' (eg Dirac's theory of constrained systems). 

		A canonical structure is already inherent in the quantum
		framework: the imaginary part of the inner product
		provides the symplectic structure while the Schrodinger
		equation gives a Hamiltonian evolution;
	\item Path integral quantization - very well suited for gauge
		theories especially the non-abelian ones - has action as
		\underline{the} central quantity.
\end{enumerate}

Without further ado, let us proceed with an action formulation. Our
first aim is to obtain the Klein-Gordon, the Dirac and the Proca/Maxwell
equations. Because the equations are local, partial differential
equations, the action must be an integral over the Minkowski space-time,
of a Lagrangian density built out of the Poincare covariant fields and
their derivatives i.e. it must of the form $S = \int
d^4x\sqrt{|det(\eta)|} \cal{L} $ with $\cal{L}$ is a Lorentz scalar
built out of the fields. It is  called the {\em Lagrangian density}. Here,
$\eta$ denotes the Minkowski metric and is necessary to absorb the
Jacobian of Lorentz transforms of the space-time coordinates. The
absolute value of the determinant happens to be 1 and is suppressed
throughout. A variational principle has the form,
\[
	\delta S[\Phi] := S[\Phi + \delta\Phi] - S[\Phi] ~ = ~ \int d^4x
	\delta \cal{L} ~ := ~ \int d^4x ~ \delta\Phi ~ `` ~\frac{\delta
	\cal{L}}{\delta\Phi} ~ " \ . 
\]
Extremization of the action under arbitrary variation $\delta\Phi$ leads
to the Euler-Lagrange equations: $\frac{\delta \cal{L}}{\delta\Phi} =
0$. Since the equation we want to derive are linear, it suffices to have
$\cal{L}$ to be quadratic in the fields. Furthermore, the equations have
no more than two derivatives which can be obtained by restricting to
first derivatives in the Lagrangian.

Let us begin with the scalar field, 
\begin{eqnarray*}
	\phi_{\Lambda,a}(x) & = & \phi(\Lambda^{-1}(x - a)) ~ ~ , ~ ~
	\Lambda_{\mu}^{~\alpha}\partial_{\alpha}\phi(\Lambda^{-1}(x -
	a)) ~ ~ \Rightarrow ~ ~ \\
	\left[\partial^{\mu}\phi\partial_{\mu}\phi\right]_{\Lambda,a} &
	= & \eta^{\mu\nu} \left[\partial_{\mu} \phi\partial_{\nu}
	\phi\right]_{\Lambda,a} ~ = ~ \eta^{\mu\nu}
	\Lambda_{\mu}^{~\alpha} \Lambda_{\nu}^{~\beta}
	\left[\partial_{\alpha} \phi\partial_{\beta}\phi\right]
	(\Lambda^{-1}(x-a)) \\
	& = &
	\eta^{\alpha\beta}\partial_{\alpha}\phi\partial_{\beta}\phi
	(\Lambda^{-1}(x-a)) ~ ~ = ~ ~ \left[\partial^{\alpha}\phi
	\partial_{\alpha}\phi\right] (\Lambda^{-1}(x-a)) \ .
\end{eqnarray*}
When integrated over the space-time, we can change the dummy variable $x
\to \Lambda x + a$ which restores the argument of the $\phi$'s without
changing the integration measure. Hence $\int d^4x \partial^{\mu}\phi
\partial_{\mu}\phi$ is clearly Poincare invariant. Since all fields have
the same shift of the space-time argument, we will suppress the
translation part and focus on the Lorentz. This works as long as there
are no externally prescribed fields/functions, which break translation
invariance. Thus, let
\begin{eqnarray}
S[\phi] & := & \int_{M}d^4x\left[ -\frac{1}{2}\partial^{\mu}
\phi\partial_{\mu}\phi - \frac{1}{2}m^2\phi^2\right] ~ ~ =: ~ ~ \int_M
d^4x \cal{L}(\phi, \partial\phi) \\
\delta S\left[\phi\right] & = & \int_M d^4x \left[ - \partial_{\mu}\phi
\partial^{\mu}\delta\phi - m^2\phi \delta\phi\right] \nonumber \\
& = & \int_M d^4x \left[ - \partial_{\mu}(\partial^{\mu}\phi \delta\phi)
+ (\Box\phi - m^2\phi)\delta\phi \right] \nonumber \\
& = & \int_M d^4x \ \delta\phi\left[(\Box - m^2)\phi\right] -
\int_{\partial M}d^3x \ n_{\mu}(\partial^{\mu}\phi \delta\phi)
\end{eqnarray}
The integration is over all of Minkowski space-time which has no
boundary and hence the second term should be zero. In practice, such
space-times extending to infinite coordinates  are handled/defined putting
the system in a large box and taking a limit. The fields must satisfy
suitable boundary conditions so that the boundary contribution again
vanishes. Alternatively requiring the field to vanish fast enough in the
asymptotic regions, makes the boundary term vanish. We can always choose
the variational principle to set $\delta\phi = 0$ at the
boundary/asymptotic regions. While the boundary contribution are
important in some context, we will restrict ourselves to boundary
contribution being zero.

Demanding that the action be stationary for arbitrary variations of the
field, we get the equation of motion $(\Box - m^2)\phi = 0$.
Incidentally, we will get the same equation of motion from another
Lagrangian $\cal{L}' = \cal{L} + \partial_{\mu}\Lambda^{\mu}$. Thus,
several Lagrangians can give the same equations of motion (although the
Hamiltonian formulation will vary with the Lagrangians.)

Consider now a vector field. For the massive case, the equations we want
are $(\Box - m^2)v^{\mu} = 0 = \partial_{\mu}v^{\mu}$.

Consider $\cal{L} = aF^{\mu\nu}F_{\mu\nu} + b m^2v^\mu v_\mu,$ where
$F_{\mu\nu} := \partial_{\mu}v_{\nu} - \partial_{\nu}v_{\mu}$ and $a, b$
are non-zero constants to be determined.
\begin{eqnarray}
	\delta\cal{L} & = & 2a F^{\mu\nu}(\partial_{\mu}\delta v_{\nu} -
	\partial_{\nu}v_{\mu}) + 2b m^2v^{\mu}\delta v_{\mu} ~ ~ = ~ ~
	4aF^{\mu\nu}\partial_{\mu}\delta v_{\nu} + 2 b m^2 v^{\nu}\delta
	v_{\nu} \nonumber \\
	& = & \partial_{\mu}\left(4a F^{\mu\nu}\delta v_{\nu}\right) -
	4a(\partial_{\mu}F^{\mu\nu})\delta v_{\nu} + 2b m^2
	v^{\nu}\delta v_{\nu} ~ ~ \Rightarrow ~ ~
	-4a\partial_{\mu}F^{\mu\nu} + 2b m^2 v^{\nu} ~ = ~ 0 .\nonumber
	\\
	\therefore 0 & = & \left(\Box v^{\nu} - \frac{b}{2a}m^2
	v^{\nu}\right) - \partial^{\nu}\partial\cdot v ~ ~ ~ ~ \mbox{
	taking divergence and choosing $b = 2a$ gives} \nonumber \\
	0 & = & (\Box - m^2)v^{\nu} ~ \mbox{and} ~ \partial\cdot v = 0 .
\end{eqnarray}
We got $b = 2a$ and we {\em choose} $b = -\Case{1}{2}$ similar to the
scalar field case. The reason for this choice will be clear little
later.

{\em Note:} For $m = 0$, we denote the vector field by $A^{\mu}$. Now we
cannot conclude $\partial\cdot A = 0$, and the equation we get is:
$\partial_{\mu}F^{\mu\nu} = 0$ which is just the Maxwell equation. It
has the well known {\em gauge invariance}: $A_{\mu} \to A_{\mu} +
\partial_{\mu}\Lambda$ which allows us to take one of the components of
$A_{\mu}$ to be zero.

{\em Exercise:} Another natural choice of the Lagrangian is: $\cal{L}' =
a \partial^{\mu}v^{\nu}\partial_{\mu}v_{\nu}  + b m^2 v_{\mu}v^{\mu} + c
(\partial_{\mu}v^{\mu})^2$. Repeat the steps and deduce the choices
for the constants so as to get the Proca equations. Check that $\cal{L},
\cal{L}'$ differ by a divergence. 

Lastly, let us consider the action for getting the Dirac equation. We
have already noted the Lorentz transformations of $\Psi, \bar{\Psi},
\gamma^{\mu}, \partial_{\mu}$. From these it follows that we can have
the following Lorentz invariant terms, quadratic  in the fields and with
a single derivative: $\bar{\Psi}\gamma^{\alpha}\partial_{\alpha}\Psi,\
\partial^{\alpha}\bar{\Psi}\partial_{\alpha}\Psi,\ \bar{\Psi}\Psi$ and
the same terms with a $\gamma_5$ inserted between the spinors. The terms
with the $\gamma_5$ are all pseudo-scalars and may be dropped if parity
(space-inversion plus a rotation) is required to be a symmetry. We
assume so for the present. The candidate Lagrangian is then,
\begin{equation}
	\cal{L} := a \bar{\Psi}\gamma^{\alpha}\partial_{\alpha}\Psi + b
	\partial^{\alpha}\bar{\Psi}\partial_{\alpha}\Psi + c m
	\bar{\Psi}\Psi ~\mbox{and}~ \delta_{\bar{\Psi}}\cal{L} =
	\delta\bar{\Psi}[ a
	\dsl{\partial}\Psi - b\Box\Psi  + c m\Psi] .
\end{equation}
For the choice $a = -i, b = 0$ and $c = 1$, we get the Dirac equation.
If we varied $\psi$, then for the same choice, we will get the conjugate
Dirac equation: $i\partial_{\alpha}\bar{\Psi}\gamma^{\alpha} + m
\bar{\Psi} = 0$.

In summary, we have $S = \int_M d^4x\sqrt{|det\eta|}\cal{L}$ with,

\begin{center}
\fbox{ 
\begin{minipage}{0.8\textwidth}
	\begin{eqnarray} \label{LagrangiansEqn}
	\cal{L}_{scalar} & = & -
	\frac{1}{2}\partial^{\mu}\phi\partial_{\mu}\phi  -
	\frac{1}{2}m^2 \phi^2 \ ; \\
	\cal{L}_{vector} & = & - \frac{1}{4}F^{\mu\nu}F_{\mu\nu} -
	\frac{1}{2}m^2 v^{\mu}v_{\mu}\ ; \\
	\cal{L}_{spinor} & = & -i\bar{\Psi}\gamma^{\mu}\partial_{\mu} +
	m \bar{\psi}\Psi \ . \\ \nonumber
\end{eqnarray}
\end{minipage}
}
\end{center}

$\delta S = 0$ leads to the field equations implementing  the
irreducibility condition. For massless vector, the action is invariant
under the gauge transformation: $A_{\mu} \to A_{\mu} +
\partial_{\mu}\Lambda$. The actions are also invariant under parity.

{\em Note:} We have taken the scalar and vector fields to be real. We
could have complex scalar field (say) with two field equations: $(\Box -
m^2)\phi = 0 = (\Box - m^2)\phi^*$. The complex field may be considered
as two real fields, $\phi = \phi_1 + i\phi_2$ and two terms may be
included in the action. Alternatively, the complex scalar field equations
may be derived from $\cal{L} = - \partial^{\mu}\phi^*\partial_{\mu}\phi
- m^2\phi^*\phi$. 

Notice that all actions are {\em real} (for spinors it is convenient to
take matrix hermitian conjugate). We will always take the actions to be
real so that $e^{\Case{i}{\hbar}S}$ will be a phase. Only when
dissipation of energy/momentum/angular momentum is to be incorporated we
need to take the action to be complex. In this course, we will not do
so.
\subsection{Variational Principle, Symmetries of the Action and
Noether's theorem} \label{Nother}
Let us denote a generic field by $X$ with all indices suppressed. Let an
action be expressed as, $S[X] := \int_Md^4x \cal{L}(X, \partial X)$.
Let $\delta X := \epsilon \chi(X(x))$ be an arbitrary, infinitesimal
variation of the field $X(x)$. Then,
\begin{eqnarray*}
\delta S & = & \int_M d^4x\left(\frac{\delta \cal{L}}{\delta X}\delta X
+ \frac{\delta \cal{L}}{\delta X_{\mu}} \delta X_{\mu} \right) ~ = ~
\int_M d^4x\left( \frac{\delta \cal{L}}{\delta X}\delta X + \frac{\delta
\cal{L}}{\delta X_{\mu}} \partial_{\mu}\delta X \right) \\ 
& = & \int_M d^4x\left( \frac{\delta\cal{L}}{\delta X} -
\partial_{\mu}\frac{\delta\cal{L}}{\delta X_{,\mu}}\right)\delta X + 
\int_M d^4x \partial_{\mu}\left( \frac{\delta \cal{L}}{\delta
X_{,\mu}}\delta X\right) 
\end{eqnarray*} 
Here, $X_{,\mu} = \partial_{\mu}X$. The $\Case{\delta \cal{L}}{\delta
X}$ is really like partial derivative, but since both $\cal{L}$ and $X$
depend on the coordinate, these should be mentioned too. We have kept
these implicit and used the $\delta$ as a reminder. We follow this
customary practice.  The last term is a divergence and can be expressed
as a boundary integral: $\int_{\partial M}d^3y n_{\mu}\Case{\delta
\cal{L}}{\delta X_{,\mu}}\delta X$. This is typically dropped/vanishes
for various reasons. 

{\em Note:} For an action principle to be well defined (this
includes the specification of the class of variations), it is necessary
that the boundary contribution must vanish. If in some cases it does not
vanish, additional `surface action' needs to be added to cancel the total
boundary contribution. This is typically encountered in gravitational
actions on manifolds with boundaries.

To summarize: {\em a variational principle asserts that $\delta S[X]$
vanishes for arbitrary variation $\delta X$ around $X$ iff $\Case{\delta
\cal{L}}{\delta X} - \partial_{\mu}\Case{\delta \cal{L}}{\delta
X_{,\mu}} = 0$ i.e. iff $X(x)$ is a solution of the equation of motion.}

We now consider a different situation. We consider special, restricted
variations such that $\delta S[X] =  0$ at {\em all fields $X(x)$}. Such
variations are  called {\em infinitesimal symmetries of the action}.
This is translated in terms of the Lagrangian density as,
\[
	\delta S[X] ~ = ~ \int_M d^4x \delta \cal{L}  ~ = ~ 0 ~ ~
	\mbox{if either $\delta \cal{L} = 0$ ~ ~ or ~ $\delta \cal{L} =
	\partial_{\mu}\delta \Lambda^{\mu}$ , }
\] where $\Lambda^{\mu}$ is some 4-vector which vanishes/falls off on
the boundary. Now we have the Noether's theorem:

\underline{Noether's Theorem:} For every infinitesimal symmetry of the
action, $\exists$ a conserved current, conserved on every solution of
the equation of motion.

\underline{Proof:} We already have the infinitesimal variation of the
$\cal{L}$ which must equal a divergence i.e. for a symmetry variation,
\[
\left( \frac{\delta\cal{L}}{\delta X} -
\partial_{\mu}\frac{\delta\cal{L}}{\delta X_{,\mu}}\right)\delta X + 
\partial_{\mu}\left( \frac{\delta \cal{L}}{\delta X_{,\mu}}\delta
X\right) ~ = ~ \delta \cal{L} ~ = ~ \partial_{\mu}\Lambda^{\mu} \  ~ ~
\Rightarrow ,
\]
\[
\left( \frac{\delta\cal{L}}{\delta X} - \partial_{\mu}
\frac{\delta\cal{L}} {\delta X_{,\mu}}\right) \delta X +
\partial_{\mu}\left( \frac{\delta \cal{L}}{\delta X_{,\mu}}\delta X  -
\partial_{\mu}\Lambda^{\mu} \right) ~ = ~ 0 \ .
\]
Thus, if $X$ satisfies the equation of motion, the first term vanish and
we have a conserved current, $\boxed{\delta J^{\mu} := \frac{\delta
\cal{L}} {\delta X_{,\mu}} \delta X  - \partial_{\mu}\Lambda^{\mu} ~ , ~
\partial_{\mu}\delta J^{\mu} = 0 }$

This is neat prescription to discover conserved currents and their
corresponding conserved charges: $\delta Q := \int_{\partial M}d^3
\sigma\ n^{\mu}\delta J_{\mu}$. Discovering conserved currents by
inspection of the equation of motion is easy only in the simplest of
cases.

Let us see an example, particularly a symmetry variation induced by
infinitesimal Poincare transformations for which $\delta S = 0$ by
construction. Consider a real scalar field action for simplicity.

Under an infinitesimal Poincare transformation, 
\begin{eqnarray*}
\delta \phi_{\Lambda,a}(x) & := & \phi(\Lambda^{-1}(x - a)) - \phi(x) ~
= ~ \phi( (\delta^{\mu}_{~\nu} - \omega^{\mu}_{~\nu})(x -
\epsilon)^{\nu}) - \phi(x) \\
& = & (-\omega^{\mu}_{~\nu}x^{\nu} - \epsilon^{\mu})\partial_{\mu}\phi ~
~ := ~ ~ - \xi^{\mu}\partial_{\mu}\phi(x) ~ ~,~ ~ \xi^{\mu} :=
\omega^{\mu}_{~\nu}x^{\nu} + \epsilon^{\mu} \ .
\end{eqnarray*}
This leads to
\begin{eqnarray*}
\delta \cal{L} & = & - \partial^{\mu} \phi \partial_{\mu}
(-\xi\cdot\partial\phi) - m^2 \phi (-\xi\cdot\partial\phi) ~ = ~
\partial^{\mu}\phi\partial_{\mu}(\xi^{\nu}\partial_{\nu}\phi) + m^2\phi
\xi^{\nu}\partial_{\nu}\phi \\
& = & (\partial_{\mu}\xi^{\nu})\partial^{\mu}\phi\partial_{\nu}\phi +
\xi^{\nu}\left(\partial^{\mu}\phi\partial^2_{\nu\mu}\phi +
m^2\phi\partial_{\nu}\phi \right) \\
& = & \frac{1}{2} (\partial_{\mu}\xi_{\nu} + \partial_{\nu} \xi_{\mu})
\partial^{\mu}\phi \partial^{\nu}\phi + \xi^{\nu} \left(\frac{1}{2}
	\frac{\delta \cal{L}}{\delta\phi_{,\mu}} \partial_{\nu}
	\phi_{,\mu} + \frac{1}{2} m^2 \frac{\delta \cal{L}}{\delta\phi}
\partial_{\nu}\phi\right)    
\end{eqnarray*}
Explicitly, $\partial_{\mu}\xi_{\nu} =
\partial_{\mu}(\omega_{\nu\alpha}x^{\alpha} + \epsilon^{\alpha}) =
\omega_{\nu\mu} + 0$ which is antisymmetric in $\mu \leftrightarrow \nu$
and hence the first term vanishes. The second term is just
$\xi^{\nu}\partial_{\nu}\cal{L}$.
\[
\therefore \delta \cal{L} = \xi^{\mu}\partial_{\mu}\cal{L} =
\partial_{\mu}(\xi^{\mu}\cal{L}) - (\partial\cdot \xi)\cal{L} =
\partial_{\mu}\delta\Lambda^{\mu} + 0 \ , ~ \mbox{or} ~ ~ \boxed{\delta
	\cal{L} = \partial_{\mu}\delta\Lambda^{\mu}~ ~,~ ~
\delta\Lambda^{\mu} = \xi^{\mu}\cal{L}} \ .
\]

\underline{Note:} We ensured that $\cal{L}$ is a Lorentz scalar, but it
is {\em not} a translation scalar. Action gets it Poincare invariance
thanks to the $\int d^4x$. Since all fields under translations shift
their coordinates by $a$, in all cases we will have $\delta \cal{L} =
\xi\cdot\partial \cal{L}$ provided $\xi_{\mu}$ satisfies
$\partial_{\mu}\xi_{\nu} + \partial_{\nu}\xi_{\mu} = 0$ i.e. $\xi^{\mu}$
is a ``Killing vector'' of the Minkowski metric.

By Noether's theorem, we get 
\[
	\delta J^{\mu} = \frac{\delta
	\cal{L}}{\delta\phi_{,\mu}}\delta\phi - \delta\Lambda^{\mu} =
	\partial^{\mu}\phi( - \xi^{\nu}\partial_{\nu}\phi) -
	\xi^{\mu}\cal{L} ~ = ~
	-\xi_{\nu}\left(\partial^{\mu}\phi\partial^{\nu}\phi +
	\eta^{\mu\nu}\cal{L}\right) ~ =: ~ - \xi_{\nu}T^{\mu\nu} \ .
\]
Conservation of the Noether current gives $(\partial_{\mu} \xi_{\nu})
T^{\mu\nu} + \xi_{\nu}\partial_{\mu}T^{\mu\nu} = 0$. Provided
$T^{\mu\nu}$ is symmetric (as it is here), the first term vanishes and
independence of $\xi^{\mu}$, we get $\partial_{\mu}T^{\mu\nu} = 0$. The
tensor, $\boxed{T^{\mu\nu} := \partial^{\mu}\phi\partial^{\nu}\phi +
\eta^{\mu\nu} \cal{L}}$ is called the {\em canonical stress tensor} for
the scalar field.

Thus, for each Killing vector $\xi^{\mu}$, we have $J^{\mu} :=
T^{\mu\nu}\xi_{\nu}$ which is conserved on the solutions of the field
equation. We define the corresponding charges as, $Q^{\mu}(\xi) ~ := ~
\int_{\Sigma_t}d^3 x T^{0\nu}\xi_{\nu}$ which are independent of the
hypersurfaces $\Sigma_t$.

Here are practice exercises:

(1) For the Proca Lagrangian, show that $\delta \cal{L} =
\xi^{\mu}\partial_{\mu}\cal{L}$ for $\xi^{\mu} = \omega^{\mu}_{~\nu}
x^{\nu} + \epsilon^{\mu}, \ \omega, \epsilon$ being infinitesimal, and
obtain the canonical stress tensor.

(2) For $m = 0$, show that this stress tensor is not gauge invariant and
one needs to ``improve'' it to get a symmetric, gauge invariant and
conserved stress tensor. Guess it.

(3) For the scalar and the Maxwell field, obtain the conserved charges
corresponding to the 10 Killing vectors. The charges corresponding to
the translations are the energy and momentum whiles those corresponding
to the $\omega_{ij}$ are the angular momentum components.

\subsection{Conserved Poincare charges for scalar field solutions}
Let us see an explicit example of Noether charges for a general solution
of the Klein-Gordon equation. We noted above that the canonical stress
tensor for the scalar field is given by,
\begin{eqnarray*}
	T_{\mu\nu} ~ = ~ \partial_{\mu}{\phi}\partial_{\nu}\phi +
	\eta_{\mu\nu}\cal{L} & , & \cal{L} ~ := ~ - \frac{1}{2}
	\partial^{\mu}{\phi} \partial_{\mu}{\phi} - \frac{1}{2}m^2
	\phi^2 \\ 
	\partial_{\mu}{T^{\mu\nu}} ~ = ~ 0 & , & ~ \forall ~ ~ ~ (\Box -
	m^2)\phi = 0 \ . \\
	\mbox{Define:} ~ ~ M^{\mu\nu\lambda} ~ := ~
	T^{\mu\nu}x^{\lambda} - T^{\mu\lambda}x^{\nu} & ~ \Rightarrow &
	\partial_{\mu}{M^{\mu\nu\lambda}} ~ = ~ T^{\lambda\nu} -
	T^{\nu\lambda} \\
	\therefore \partial_{\mu}{M^{\mu\nu\lambda}} = 0 ~ \mbox{on
	solutions} ~ & ~\mbox{iff}~ & \partial_{\mu}{T^{\mu\nu}} = 0 ~
	\mbox{{\em and}} ~ T^{\mu\nu} = T^{\nu\mu}
\end{eqnarray*}
These 6 conserved currents ($M^{\mu\nu\lambda}$ are antisymmetric in the
last two indices) together with the conserved stress tensor give the 10
Poincare charges:
\[
	P^{\mu} ~ := ~ \int_{\Sigma_t}d^3 x T^{0\mu} ~ ~ ~ , ~ ~ ~
	M^{\mu\nu} ~ := ~ \int_{\Sigma_t}d^3 x M^{0\mu\nu} \ . 
\]

We had noted the plane wave solutions of the Klein-Gordon equation,
labeled by $\vec{k} \in \mathbb{R}^3$ namely, 
\[
	u_{\vec{k}}(x) = \frac{1}{\sqrt{2\omega_{\vec{k}}(2\pi)^3}}e^{i
	k\cdot x} ~ ~ , ~ ~ k\cdot x := -\omega_{\vec{k}}t +
	\vec{k}\cdot \vec{x} ~ ~,~ ~ \omega_{\vec{k}} := \sqrt{\vec{k}^2
	+ m^2} 
\]
and their complex conjugates. The general solution is thus expressed as
$\boxed{\phi(x) = \int d^3k[ a(\vec{k})u_{\vec{k}}(x) +
a^*(\vec{k})u^*_{\vec{k}}(x)] }$ with $a(\vec{k})$ being complex numbers
with a suitable dependence on $\vec{k}$ so that the field has the
appropriate boundary behavior. The general solution is manifestly real.

To compute the energy-momentum and angular momentum charges, we have
integration of expressions quadratic in the solutions. So we note: $
\partial_{\mu}u_{\vec{k}}(x) = ik_{\mu}u_{\vec{k}}(x) ~,~
\partial_{\mu}{u^*_{\vec{k}}}(x) = - ik_{\mu}u^*_{\vec{k}}(x) ~,~ k^0 =
\omega_{\vec{k}} \ $ and the orthogonality relations: 
\begin{eqnarray*}
	\int d^3x\ u^*_{\vec{k}}(x) u_{\vec{k}'}(x) ~ = ~
	\frac{e^{i(\omega_{\vec{k}} -
	\omega_{\vec{k}'})t}}{\sqrt{2\omega_{\vec{k}}}\sqrt{2\omega_{\vec{k}'}}}
	& = & \frac{\delta^3(\vec{k} - \vec{k}')}{2\omega_{\vec{k}}} ~ =
	~ \delta_{inv}^3(\vec{k} - \vec{k}') \\
	\int d^3x\ u_{\vec{k}}(x) u_{\vec{k}'}(x) ~ = ~
	e^{-2i\omega_{\vec{k}}t}\delta^3_{inv}(\vec{k} + \vec{k}') & , & 
	\int d^3x\ u^*_{\vec{k}}(x) u^*_{\vec{k}'}(x) ~ = ~
	e^{2i\omega_{\vec{k}}t}\delta^3_{inv}(\vec{k} + \vec{k}')\ .
\end{eqnarray*}

We need,
\begin{eqnarray}
	P^{\mu} & = & \int d^3x \partial^{0}{\phi} \partial^{\mu}{\phi}
	+ \eta^{0\mu}\left( -\frac{1}{2} \partial^{\alpha}{\phi}
	\partial_{\alpha}{\phi} - \frac{1}{2}m^2\phi^2\right) \\
	M^{\mu\nu} & = & \int d^3x \left\{ \left( \partial^{0}{\phi}
	\partial^{\mu}{\phi} + \eta^{0\mu} \cal{L}\right)x^{\nu} -
\left( \partial^{0}{\phi} \partial^{\nu}{\phi} + \eta^{0\nu}
\cal{L}\right)x^{\mu} \right\}
\end{eqnarray}
And we have,
\begin{eqnarray}
\cal{L} & = & -\frac{1}{2}\int d^3k\int d^3k'\left\{
\left(a(\vec{k})a(\vec{k}') u_{\vec{k}}u_{\vec{k}'} +
a^*(\vec{k})a^*(\vec{k}') u^*_{\vec{k}}u^*_{\vec{k}'} \right)
\left(-\vec{k}\cdot \vec{k}' + m^2\right) \right. \nonumber \\
& & \mbox{\hspace{3.0cm}} \left. \left(a(\vec{k})a^*(\vec{k}')
u_{\vec{k}}u^*_{\vec{k}'} + a^*(\vec{k})a(\vec{k}')
u^*_{\vec{k}}u_{\vec{k}'} \right) \left(+\vec{k}\cdot \vec{k}' +
m^2\right) \right\} \\
\partial^{0}{\phi} \partial^{\mu}{\phi} & = & \int d^3k\int d^3k'\left\{
a(\vec{k})a(\vec{k}') u_{\vec{k}}u_{\vec{k}'} +
a^*(\vec{k})a^*(\vec{k}') u^*_{\vec{k}}u^*_{\vec{k}'} \right. \nonumber \\
& & \mbox{\hspace{3.0cm}} \left. - a(\vec{k})a^*(\vec{k}')
u_{\vec{k}}u^*_{\vec{k}'} - a^*(\vec{k})a(\vec{k}')
u^*_{\vec{k}}u_{\vec{k}'} \right\} \left(-\vec{k}^0
\vec{k}^{'\mu}\right)
\end{eqnarray}

Substituting and carrying out the $\int d^3x$ using the orthogonality
relations gives,
\begin{eqnarray}
P^i & = & \int d^3k \int d^3k'\left[ \frac{k^0k^i}{2\omega_{\vec{k}}}
\left\{ a(\vec{k})a(- \vec{k})e^{-2i\omega_{\vec{k}} t} +
a^*(\vec{k})a^*(-\vec{k})e^{2i\omega_{\vec{k}}}\right\} \delta^3(\vec{k}
+ \vec{k}') \right.  \nonumber \\
	& & \left. \mbox{\hspace{3.0cm}}
	\frac{k^0k^i}{2\omega_{\vec{k}}} \left\{ a(\vec{k})a^*(\vec{k})
+ a^*(\vec{k})a(\vec{k})\right\} \delta^3(\vec{k} - \vec{k}') \right]
\nonumber \\
	& = & \frac{1}{2}\int d^3k k^i\left\{ a(\vec{k})a(-
	\vec{k})e^{-2i\omega_{\vec{k}} t} +
a^*(\vec{k})a^*(-\vec{k})e^{2i\omega_{\vec{k}} t} +
a(\vec{k})a^*(\vec{k}) + a^*(\vec{k})a(\vec{k}) \right\}  ~ ~ 
\end{eqnarray}
The first two, time decedent terms in the braces are symmetric under
$\vec{k} \leftrightarrow -\vec{k}$ while the last two time independent
term go just change their argument. Hence, under symmetric integration,
the time dependent terms vanish and we are left with,
\begin{equation} \label{FieldMomentum}
P^i = \int d^3k \ k^i \ \frac{a(\vec{k})a^*(\vec{k}) -
a(-\vec{k})a^*(-\vec{k})}{2} 
\end{equation} 
Since the $a$'s are ordinary complex numbers the factor of 2 cancels and
a new factor arises from explicit anti-symmetrization w.r.t. $\vec{k}$.
The claimed conserved quantity is manifestly independent of $t$.

The calculation of $P^0$ proceeds much the same way. We have the
$t-$dependent term proportional to $\delta^3(\vec{k} + \vec{k}')$ while
the $t-$independent term is proportional to $\delta^3(\vec{k} -
\vec{k}')$. We get,
\begin{eqnarray}
	P^0 & = & \int \frac{d^3k}{2\omega_{\vec{k}}}\left[
	-\omega^2_{\vec{k}} \left\{
a(\vec{k})a(-\vec{k})e^{-2i\omega_{\vec{k}} t} +
a^*(\vec{k})a^*(-\vec{k})e^{2i\omega_{\vec{k}} t}\right\} +
\omega^2_{\vec{k}} \left\{2 a(\vec{k})a^*(\vec{k})\right\} \right.
\nonumber \\
& & \mbox{\hspace{1.5cm}} + \frac{1}{2}\left\{
a(\vec{k})a(-\vec{k})e^{-2i\omega_{\vec{k}} t} +
a^*(\vec{k})a^*(-\vec{k})e^{2i\omega_{\vec{k}} t} \right\}\left(
\omega_{\vec{k}}^2 + \vec{k}^2 + m^2\right) \nonumber \\ 
& & \mbox{\hspace{1.5cm}} + \left.  \frac{1}{2}\left\{ 2
a(\vec{k})a^*(\vec{k})\right\}\left(- \omega^2_{\vec{k}} + \vec{k}^2 +
m^2\right) \right]
\end{eqnarray}
The $t-$dependent terms cancel while the $t-$independent terms give,
\begin{equation}\label{FieldEnergy}
	P^0 ~ = ~ \int d^3k \ \omega_{\vec{k}}\  a(\vec{k}) a^*(\vec{k})
\end{equation}

The calculation of $M^{\mu\nu}$ is similar. The new feature is the
explicit $x^{\lambda}$ in the integral. In the $M^{ij}$ we have $\int
T^{0i}x^j - T^{0j}x^i$. We trade $x^i$ for $ \partial_{k_i}$ by noting
that
\[
	\partial_{k_i}u_{\vec{k}}(x) = u_{\vec{k}}(x)\left\{ -i t
	\frac{k^j}{\omega_{\vec{k}}} + i x^j\right\}  ~ ~ or ~ ~
	x^ju_{\vec{k}} = -i \partial_{k_j}{u} + \frac{t
	k^j}{\omega_{\vec{k}}} u_{\vec{k}} \ . 
\]
As before the spatial integration will produce $\delta$ functions and
also derivatives of $\delta$ functions from the $ \partial_{k_i}$. While
doing the $k'-$integration, we need to flip the derivative as per the
rules of $\delta$ function. As before, the $e^{\pm 2i
\omega_{\vec{k}}t}$ will turn out to be symmetric under $ \vec{k}
\leftrightarrow -\vec{k}$ and will not contribute. Filling in the
algebra leads to the final results,
\begin{eqnarray}
	M^{ij} & = & i\int d^3k\left[a(\vec{k})\left(k^i \partial_{k_j}
	- k^j \partial_{k_i}\right)a^*_{\vec{k}} - a^*(\vec{k})\left(k^i
\partial_{k_j} - k^j \partial_{k_i}\right)a_{\vec{k}}\right] \\
	M^{0i} & = & - \frac{i}{2}\int d^3k\
	\omega_{\vec{k}}\left\{a_{\vec{k}} \partial_{k_i}{a^*_{\vec{k}}}
	- a_{\vec{k}} \partial_{k_i}{a^*_{\vec{k}}} \right\} \ .
\end{eqnarray}
All are manifestly independent of time.

Apart from verifying that the conserved quantities are indeed time
independent, the expressions for energy and momentum show that these
quantities are a sum (integral) of energy-momentum of individual
solutions. Thus each of these plane waves, at least heuristically be
thought of as carrying an energy $\omega_{\vec{k}}$ and momentum
$\vec{k}$. Note that there is no $\hbar$ yet so these are mathematically
defined conserved quantities much like the energy-momentum fluxes of
electromagnetic fields identified via the Poynting theorem. This is
helpful for interpreting the solutions of the field equations as
physical entities carrying energy-momentum. To strengthen this further,
we would like to see if the fields and the action can be cast in the
form of a ``dynamical system''.

\newpage
\section{Fourier decompositions of fields: collection of harmonic
oscillators}\label{FieldDecompositions}

In the previous section, we wrote the general solution of the
Klein-Gordon equation. Now we want to see if the action takes the `form
of a dynamical system'. The meaning will be clear shortly. Consider
again the scalar field. Let $\Sigma_t$ be a constant $t$ surface. On
$\Sigma_t$, the set of functions, $\{\varphi_{\vec{k}} = \Case{1}{(2\pi)^{3/2}}
e^{i\vec{k}\cdot \vec{x}},\ \vec{k}  \in \mathbb{R}^3\}$ forms a
complete, orthonormal set of functions, $\int_{\Sigma_t}d^3x\
\varphi^*_{\vec{k}}(\vec{x}) \varphi_{\vec{k}'}(\vec{x}) = \delta^3(\vec{k} -
\vec{k}')$. Let us {\em Fourier decompose} the field as,
\begin{equation} \label{FourierDecomposition}
\phi(t, \vec{x}) := \int_{\mathbb{R}^3}d^3k\
f_{\vec{k}}(t)\varphi_{\vec{k}}(\vec{x}) . 
\end{equation}

The reality of the field, $\phi^*(x) = \phi(x) \Rightarrow
f_{\vec{k}}^*(t) = f_{-\vec{k}}(t)$. The expansion coefficient functions
are determined by the equations of motion. It is trivial to check that
$(\Box - m^2)\phi(x) = 0 \Rightarrow \int d^3k (-\ddot{f}_{\vec{k}} -
\vec{k}^2 f_{\vec{k}} - m^2 f_{\vec{k}}) \varphi_{\vec{k}}(x) = 0$. By
independence of the basis functions, we get $\ddot{f}_{\vec{k}} +
\omega^2_{\vec{k}}f_{\vec{k}} = 0 \ \forall \ \vec{k} \in \mathbb{R}^3 \
, \omega^2_{\vec{k}} := \vec{k}^2 + m^2$. Its general solution is
$f_{\vec{k}}(t) = a_{\vec{k}}e^{-i\omega_{\vec{k}} t} + b_{\vec{k}}
e^{i\omega_{\vec{k}} t}$. The reality condition $f^*_{\vec{k}} =
f_{-\vec{k}}$ gives $a^*_{\vec{k}} = b_{-\vec{k}}$ or $b_{\vec{k}} =
a^*_{-\vec{k}}$. Substitution gives,
\begin{eqnarray}
\phi_{soln}(x) & = & \int_{\mathbb{R}^3}d^3k
\left(a_{\vec{k}}e^{-i\omega_{\vec{k}}t} +
b_{\vec{k}}e^{i\omega_{\vec{k}}t}\right) \frac{e^{i\vec{k}\cdot
\vec{x}}}{\sqrt{2\omega_{\vec{k}}(2\pi)^3}} \nonumber \\
& = & \int
\frac{d^3k}{\sqrt{2\omega_{\vec{k}}(2\pi)^3}}\left(a_{\vec{k}}
e^{-i\omega_{\vec{k}}t + i\vec{k}\cdot \vec{x}} +
b_{-\vec{k}}e^{i\omega_{\vec{k}}t - i\vec{k}\cdot \vec{x}}\right)
\nonumber \\
\therefore \phi_{soln}(x) & = & \int d^3k\left(a_{\vec{k}}
u_{\vec{k}}(x) + a^*_{\vec{k}} u^*_{\vec{k}}(x) \right) ~ ~ , ~ ~
u_{\vec{k}} := \frac{e^{ik\cdot x}}{\sqrt{2\omega_{\vec{k}}(2\pi)^3}}
\label{ModeDecomposition}
\end{eqnarray}
This is exactly the same form we had for the general solution. We have
adjusted the normalization factor for future convenience. The last
equation will be referred to as a {\em mode decomposition}, with
$u_{\vec{k}}(x)$'s denoting the {\em mode functions}.

Let us rewrite the Fourier decomposition in a manifestly real form:
\[
\phi(t, \vec{x}) = \int_{\mathbb{R}^3/2}d^3k\ \left[
f_{\vec{k}}(t)\varphi_{\vec{k}}(\vec{x}) +
f^*_{\vec{k}}(t)\varphi^*_{\vec{k}}(\vec{x})\right]
\] 
We have separated the $-\vec{k}$ tagged terms and used the reality
condition. The integration is over {\em positive half} of
$\mathbb{R}^3$.  Substituting the Fourier decomposition into the
Lagrangian, we get,
\begin{eqnarray}
L = \int_{\Sigma_t}d^3x \cal{L} & = &
\int_{\Sigma_t}d^3x\left[-\frac{1}{2}\left( -\dot{\phi}^2 +
(\nabla\phi)^2 \right) - \frac{m^2}{2}\phi^2\right] \nonumber \\
\dot{\phi}(t, \vec{x}) & = & \int_{\mathbb{R}^3/2}d^3k\ \left[
\dot{f}_{\vec{k}}(t)\varphi_{\vec{k}}(\vec{x}) +
\dot{f}^*_{\vec{k}}(t)\varphi^*_{\vec{k}}(\vec{x})\right] \nonumber \\
\partial_{j}{\phi}(t, \vec{x}) & = & \int_{\mathbb{R}^3/2}d^3k\
(ik_j)\left[ f_{\vec{k}}(t)\varphi_{\vec{k}}(\vec{x}) -
f^*_{\vec{k}}(t)\varphi^*_{\vec{k}}(\vec{x})\right] ~ ~ \mbox{using
ortho-normality, we get}\nonumber \\
L & = & \int_{\mathbb{R}^3/2} d^3k
\left[\dot{f}_{\vec{k}}\dot{f}^*_{\vec{k}} -
\omega^2_{\vec{k}}f_{\vec{k}}f^*_{\vec{k}}\right] ~ ~ ,  ~ ~
\omega^2_{\vec{k}} = \vec{k}^2 + m^2 \ . 
\end{eqnarray}

Let $f_{\vec{k}}(t) : = \frac{1}{\sqrt{2}} (q_{\vec{k}}(t) + i
q'_{\vec{k}}(t))$, where $q, q'$ are real and $\vec{k}$ are in the
positive half of $\mathbb{R}^3$.  Substitution in the Lagrangian
gives,
\[
L ~ = ~ \int_{\mathbb{R}^3/2} d^3k \ \frac{1}{2}\left[\left(\dot{q}^2_k -
\omega_{\vec{k}}^2 q^2_{\vec{k}}\right) + \left(\dot{q}^{'2}_k -
\omega_{\vec{k}}^2 q^{'2}_{\vec{k}}\right)\right] \ .
\]
{\em Denoting} $q'_{\vec{k}} =: q_{-\vec{k}}$, we can express the
Lagrangian as,
\begin{equation}
L ~ = ~ \int_{\mathbb{R}^3} d^3k \ \frac{1}{2}\left[\left(\dot{q}^2_k -
\omega_{\vec{k}}^2 q^2_{\vec{k}}\right) \right] \ .
\end{equation}
$\vec{k}$ now range over full $\mathbb{R}^3$.

The Lagrangian is manifestly expressed as a sum (integral) of
Lagrangians for harmonic oscillators $q_{\vec{k}}$ each with frequency
$\omega_{\vec{k}} \ \mbox{for}\  \vec{k} \in \mathbb{R}^3$. One can
easily pass to the Hamiltonian form by defining $p_{\vec{k}} :=
\frac{\partial L}{\partial \dot{q}_{\vec{k}}} = \dot{q}_{\vec{k}}$. We
get,
\[
	H ~ = ~ \int_{\mathbb{R}^3} d^3k \left[ \frac{1}{2}\left(
	p^2_{\vec{k}} + \omega^2_{\vec{k}}q^2_{\vec{k}}\right) \right] \
	.
\]
To complete the canonical form, introduce the usual basic Poisson
brackets: $\{q_{\vec{k}}, q_{\vec{k}'}\} = 0 = \{p_{\vec{k}},
p_{\vec{k}'}\}$ and $\{q_{\vec{k}}, p_{\vec{k}'}\} = \delta^3(\vec{k} -
\vec{k}'), \ \vec{k} \in \mathbb{R}^3$.

Introduce the field, $\pi(t, \vec{x}) := \dot{\phi}(t, \vec{x})$ and
{\em define} a Poisson brackets among the fields $\phi, \pi$ fields from
the Poisson brackets among the $q_{\vec{k}}$ and $p_{\vec{k}}$ using the
Fourier decompositions. Noting that $f_{\vec{k}} =
\Case{1}{\sqrt{2}}(q_{\vec{k}} + i q_{-\vec{k}})$ and $\dot{f}_{\vec{k}}
= \Case{1}{\sqrt{2}}(p_{\vec{k}} + i p_{-\vec{k}}) \ , \vec{k} \in
\mathbb{R}^3/2$. It follows,
\begin{eqnarray}
\{\phi(t, \vec{x}), \pi(t, \vec{y}) & = & \int_{\mathbb{R}^3/2}d^3k
	\int_{\mathbb{R}^3/2}d^3k'
	\{f_{\vec{k}}\varphi_{\vec{k}}(\vec{x}) +
		f^*_{\vec{k}}\varphi^*_{\vec{k}}(\vec{x}) \ , \
		\dot{f}_{\vec{k}'}\varphi_{\vec{k}'}(\vec{y}) +
\dot{f}^*_{\vec{k}'}\varphi^*_{\vec{k}'}) \} \nonumber \\ 
& = & \int_{\mathbb{R}^3/2}d^3k \int_{\mathbb{R}^3/2}d^3k' \nonumber \\
& & \mbox{\hspace{1.0cm}} \left[ \varphi_{\vec{k}}(\vec{x})
\varphi_{\vec{k}'}(\vec{y}) \{ f_{\vec{k}}, f_{\vec{k}'} \} +
\varphi_{\vec{k}}(\vec{x}) \varphi^*_{\vec{k}'}(\vec{y}) \{ f_{\vec{k}},
f^*_{\vec{k}'} \} \right.  \nonumber \\
& & \mbox{\hspace{1.0cm}} \left.  + \varphi^*_{\vec{k}}(\vec{x})
\varphi_{\vec{k}'}(\vec{y}) \{ f^*_{\vec{k}}, f_{\vec{k}'} \} +
\varphi^*_{\vec{k}}(\vec{x}) \varphi^*_{\vec{k}'}(\vec{y}) \{
f^*_{\vec{k}}, f^*_{\vec{k}'} \} \right] \nonumber \\
& = & \int_{\mathbb{R}^3/2}d^3k \left[ \varphi_{\vec{k}}(\vec{x})
\varphi^*_{\vec{k}}(\vec{y}) + \varphi^*_{\vec{k}}(\vec{x})
\varphi_{\vec{k}}(\vec{y}) \right] = \int_{\mathbb{R}^3}d^3k \
\varphi_{\vec{k}}(\vec{x}) \varphi^*_{\vec{k}}(\vec{y}) \\
\therefore \{\phi(t, \vec{x}), \pi(t, \vec{y}) & = & \delta^3(\vec{x} -
	\vec{y})
\end{eqnarray}
The remaining Poisson brackets $\{\phi, \phi\}, \{\pi, \pi\}$, defined
similarly, are zero.

{\em Note:} The same $t$ is taken for the fields since the $q$'s and
$p$'s are also at the same time.

Starting from the Lagrangian, using the Fourier decomposition, we saw
explicitly that the Lagrangian can be expressed as a sum of Lagrangians
of harmonic oscillators. Furthermore, we could define Poisson brackets
among the fields {\em from} the same oscillator system. It is apparent
now that {\em a field satisfying the Klein-Gordon equation as
irreducibility condition, can be given a canonical formulation wherein
it appears as a system of infinitely many, uncoupled harmonic
oscillators.}

\underline{Note:} The integration domains appearing above can be
comfusing. For real valued fields, using a {\em manifestly real form} of
a Fourier decomposition, the $\vec{k} \in \mathbb{R}^3/2$. Once the {\em
solutions for the time dependent coefficients} of the Fourier
decomposition are subtituted back in the Fourier decomposition, we get
the {\em mode decomposition} of the field, with $\vec{k} \in
\mathbb{R}$. For complex valued fields, in both Fourier and mode
decompositions, $\vec{k} \in \mathbb{R}^3$.

We now repeat the exercise for the Maxwell field and the Dirac field. We
will use the same orthonormal set of $\varphi_{\vec{k}}(\vec{x})$ in
developing the Fourier decomposition. The steps being very similar, we
will be brief and suppress the integration range of the $\vec{k}$.
\subsection{Maxwell Field}
Knowing the plane wave solutions having two polarizations, we write the
Fourier decomposition of the vector field $A_{\mu}(t, \vec{x})$ as,
\[
	A_{\mu}(t,\vec{x}) ~ = ~ \int d^3k\ \varepsilon_{\mu}(\vec{k},a)
	f_{\vec{k}, a}(t) \varphi_{\vec{k}}(\vec{x}) ~ ~ , ~ ~
	\varphi_{\vec{k}}(\vec{x}) = \frac{e^{i\vec{k} \cdot \vec{x}}}
	{(2\pi)^{3/2}} \ .
\]
The $\varepsilon_{\mu}$ is a polarization vector which depends on
$\vec{k}$ while the label $a$ enumerates the numbers of polarization
vectors. A priori, there are {\em four} independent polarization vectors
(a tetrad) and $a$ takes 4 values. One of the Maxwell equations fixes
the $0^{th}$ component of all polarization vectors in terms of the
spatial components. One of the second set of equations is trivially true
for the {\em longitudinal} polarization, $\vec{\varepsilon}(\vec{k})
\propto \vec{k}$ and we are left with only two independent equations for
the $f_{\vec{k},a}$ corresponding to the transverse polarizations. We
could have kept the $f(t, a)$ inside the polarization vector, but this
is more convenient. We will suppress the sum-over-$a$ till the final
expression.

This leads to, 
\begin{eqnarray*}
F_{\mu\nu} ~ = ~ \partial_{\mu}{A_{\nu}} - \partial_{\nu}{A_{\mu}} & = &
\int d^3k \left[\varepsilon_{\nu} \partial_{\mu}{(f\varphi)} -
\varepsilon_{\mu} \partial_{\nu}({f\varphi})\right]. \\
\therefore F_{0i} ~ = ~ \int d^3k\ \varphi_{\vec{k}}(\vec{x})\left\{
\varepsilon_i\dot{f}_{\vec{k}} - ik_i \varepsilon_0\ f_{\vec{k}}\right\}
& , & F_{ij} ~ = ~ \int d^3k\ f_{\vec{k}}\ \varphi_{\vec{k}} (\vec{x})
\left\{ i (k_i\varepsilon_j - k_j\varepsilon_i) \right\}
\end{eqnarray*}
The equations of motion, using independence of $\varphi_{\vec{k}}
(\vec{x})$ give,
\begin{eqnarray*}
	\partial^{i}{F_{i0}} = 0 & : &
	\varepsilon_0(\vec{k},a)f_{\vec{k}}(t) = -
	\frac{i\vec{k}\cdot\vec{\varepsilon}}{\vec{k}^2}\dot{f}_{\vec{k}}(t)
	\ \Rightarrow ~ \varepsilon_0\dot{f}_{\vec{k}} = -
	\frac{i\vec{k} \cdot \vec{\varepsilon}}
	{\vec{k}^2}\ddot{f}_{\vec{k}}(t) \\
	\partial^0F_{0i} + \partial^jF_{ji} = 0 & : &
	\varepsilon_i\left(-\ddot{f}_{\vec{k}} -
	\vec{k}^2f_{\vec{k}}\right) + ik_i(\varepsilon_0
	\dot{f}_{\vec{k}}) +
	(\vec{k}\cdot\vec{\varepsilon})k_if_{\vec{k}} = 0 \\
	\therefore 0 & = & \ddot{f}_{\vec{k}}\left\{-\varepsilon_i +
	\frac{k_ik^j\varepsilon_j}{\vec{k}^2}\right\} +
	\vec{k}^2f_{\vec{k}}\left\{-\varepsilon_i +
	\frac{k_ik^j\varepsilon_j}{\vec{k}^2}\right\} \\
	\mbox{or,\hspace{1.0cm}} 0 & = & \left[\left(-\delta_i^{~j} +
	\frac{k_ik^j}{\vec{k}^2}\right)\varepsilon_j\right]
	\left[\ddot{f}_{\vec{k}} + \vec{k}^2f_{\vec{k}}\right] \ .
\end{eqnarray*}
The prefactor is projector, projecting a 3-vector onto the plane
perpendicular to $\vec{k}$. Thus,
$\boxed{\underline{\varepsilon}_i(\vec{k},a) := \left(\delta_i^{~j} -
\frac{k_i k^j}{\vec{k}^2}\right)\varepsilon_j(\vec{k},a)}$ define {\em
the transverse polarizations} and there are two independent ones. Since
the prefactor is non-zero, the Maxwell equations imply that
$f_{\vec{k}}$ satisfy the same harmonic oscillator equation as before,
with $\omega^2_{\vec{k}} = \vec{k}^2$ and the mass is zero. It is
important to remember that transverse polarization,
$\underline{\varepsilon}_i(\vec{k}, a)$, is defined by $\vec{k}$ and
hence acquire the $\vec{k}$ dependence as well as the label $a$. Hence
the $f_{\vec{k},a}$ satisfy the oscillator equations only for the
transverse polarization. We may not display it always to avoid clutter. 

The next task is to compute the Lagrangian, $\cal{L} =
-\Case{1}{4}F_{\mu\nu}F^{\mu\nu} = \Case{1}{2}(F_{0i})^2 -
\Case{1}{4}(F_{ij})^2$. 
\begin{eqnarray*}
	F_{0i}^2 & = & \int d^3k\int d^3k'\left\{\varepsilon_i
	\dot{f}_{\vec{k}} - ik_i\varepsilon_0 \right\}
	\left\{\varepsilon'_i\dot{f}_{\vec{k}'} -ik'_i\varepsilon'_0
	f_{\vec{k}}\right\} \varphi_{\vec{k}}(\vec{x})
	\varphi_{\vec{k}'}(\vec{x})
\end{eqnarray*}
Eliminating $\varepsilon_0f_{\vec{k}} = -i
\Case{\vec{k}\cdot\vec{\varepsilon}}{\vec{k}^2}\dot{f}_{\vec{k}}$ and
likewise $\varepsilon'_0 f_{\vec{k}'}$, simplifies the product of the
$\{\dots\}$ to\footnote{This is equivalent to using one of the Maxwell
equations, the Gauss law equation which is a constraint.},
\[
\{\dots\}\{\dots\} ~ = ~ \dot{f}_{\vec{k}} \dot{f}_{\vec{k}'} \left[
	\vec{\varepsilon} \cdot \vec{\varepsilon}' +
	\vec{k}\cdot\vec{k}'
	\frac{\vec{k}\cdot\vec{\varepsilon}}{\vec{k}^2}
	\frac{\vec{k}'\cdot\vec{\varepsilon}'}{\vec{k}^{'2}} -
(\vec{\varepsilon} \cdot\vec{k}')
\frac{\vec{\varepsilon}'\cdot\vec{k}'}{(\vec{k}')^2} -
(\vec{\varepsilon}' \cdot\vec{k})
\frac{\vec{\varepsilon}\cdot\vec{k}}{(\vec{k})^2} \right]
\]
Integration over $\vec{x}$ gives $\delta^3(\vec{k} + \vec{k}')$ which
cancels two term and we are left with,
\begin{eqnarray*}
\frac{1}{2}\int_{\Sigma_t}d^3x\ (F_{0i})^2 & = & \frac{1}{2} \int
d^3k\left[ \vec{\varepsilon}(\vec{k}) \cdot \vec{\varepsilon}'(-\vec{k})
- \frac{(\vec{\varepsilon}(\vec{k})\cdot\vec{k})
(\vec{\varepsilon}'(-\vec{k})\cdot\vec{k})} {\vec{k}^2}\right]
\dot{f}_{\vec{k}} \dot{f}_{-\vec{k}}  \\
\mbox{But} ~ ~ \left[\dots\right] & = & \varepsilon_i(\vec{k})
\left(\delta^{ij} - \frac{k^i k^j}{\vec{k}^2}\right)
\varepsilon_j(-\vec{k}) ~ ~ = ~ ~ \vec{\underline{\varepsilon}}(\vec{k})
\cdot\vec{\underline{\varepsilon}}(-\vec{k}) \\
\therefore \frac{1}{2}\int_{\Sigma_t}d^3x\ (F_{0i})^2 & = & \int d^3k\
\frac{1}{2} \vec{\underline{\varepsilon}}(\vec{k})
\cdot\vec{\underline{\varepsilon}}(-\vec{k}) \
\dot{f}_{\vec{k}}\dot{f}_{-\vec{k}} ~ ~ ~ ~ \mbox{and similarly,} \\
-\frac{1}{4}\int d^3x (F_{ij})^2 & = & \int d^3k\ -\frac{\vec{k}^2}{2}\
\vec{\underline{\varepsilon}}(\vec{k})
\cdot\vec{\underline{\varepsilon}}(-\vec{k}) \
{f}_{\vec{k}}{f}_{-\vec{k}}  
\end{eqnarray*}
The dot product of the transverse polarization \underline{includes} the
sum over the polarizations. The longitudinal polarization has dropped
out explicitly. Making it explicit, we have,
\begin{equation}
	L_{Maxwell} ~ = ~ \int_{\mathbb{R}^3/2} d^3k\ \sum_{a,b = 1}^{2}
	\left[ \vec{\underline{\varepsilon}}(\vec{k},a)
	\cdot\vec{\underline{\varepsilon}}(-\vec{k},b)\right]\left[
	\dot{f}_{\vec{k},a}\dot{f}_{-\vec{k},b} -
\omega^2_{\vec{k}}f_{\vec{k},a}f_{-\vec{k},b} \right]
\end{equation}
We may now choose the transverse polarizations such that
$\boxed{\vec{\underline{\varepsilon}}(\vec{k},a)
\cdot\vec{\underline{\varepsilon}}(-\vec{k},b) = \delta_{a,b}}$ (note
the $\pm \vec{k}$ for both polarizations). Using this, the Maxwell
action becomes,
\begin{eqnarray}
S[A] & = & \int dt\int_{\mathbb{R}^3/2}d^3k \sum_{a = 1,2}\left[
\dot{f}_{\vec{k},a}\dot{f}_{-\vec{k},a} -
\omega^2_{\vec{k}}f_{\vec{k},a}f_{-\vec{k},a} \right] \\
& = & \int dt\int_{\mathbb{R}^3}d^3k \frac{1}{2}\sum_{a = 1,2}\left[
\dot{q}^2_{\vec{k},a} - \omega^2_{\vec{k}}q^2_{\vec{k},a}\right] ~ ~ ~ ~
\mbox{for,} \\
A_{\mu}(t,\vec{x}) & = & \int_{\mathbb{R}^3/2} d^3k\ \sum_a
\varepsilon_{\mu}(\vec{k},a) f_{\vec{k}, a}(t)
\varphi_{\vec{k}}(\vec{x}) +
\varepsilon^*_{\mu}(\vec{k},a)f^*_{\vec{k},a}(t)
\varphi^*_{\vec{k}}(\vec{x})  ~ ~ ~ ~\mbox{with,}~ ~ \\
\varepsilon_0(\vec{k},a)f_{\vec{k},a}(t) & = & -i
\frac{\vec{k}\cdot\vec{\varepsilon}(\vec{k},a)}{\vec{k}^2}\dot{f}_{\vec{k},a}
~ ~,~ ~ \underline{\varepsilon}_i(\vec{k},a)) := \left(\delta_i^{~j} -
\frac{k_ik^j}{\vec{k}^2}\right)\varepsilon_j(\vec{k},a) \\
\delta_{a,b} & = & \vec{\underline{\varepsilon}}(\vec{k},a)
\cdot\vec{\underline{\varepsilon}}^*(\vec{k},b) ~ ~ ~ ~
\mbox{(Normalization)}.
\end{eqnarray}
As in the case of the scalar, we have gone through the introducing the
`real $q_{\vec{k}a}$' degrees of freedom and expressed the Maxwell
action too is a sum of {\em twice as many} harmonic oscillators. The
canonical form goes through as well. 

{\em Note:} If we were to begin with the Maxwell action and attempt to
get its Hamiltonian formulation, we would obtain the Gauss Law equation
of motion as a constraint equation. By using the Fourier decomposition,
we have explicitly solved this constraint and eliminated the
$\varepsilon_0$ polarization. Once this is done, we get the transverse
projector which eliminates the longitudinal polarization. This gives the
action involving only the physical degrees of freedom.
\subsection{Dirac Field}
Lastly, let us consider the Dirac action. Here the Lagrangian is first
order in the derivatives and we do not expect a harmonic oscillator
form. We will also find it clearer to write the Fourier decomposition
using the positive half in $\vec{k}$ space. We take the Fourier
decomposition of the spinor as,
\[
\psi(t, \vec{x}) = \int_{\mathbb{R}^3/2}d^3k\ \left[\sum_{\sigma}
u(\vec{k}, \sigma)f_{\vec{k},\sigma}(t)\right]
\varphi_{\vec{k}}(\vec{x}) + \left[ \sum_{\sigma} v(\vec{k},
\sigma)g_{\vec{k},\sigma}(t)\right] \varphi^*_{\vec{k}}(\vec{x})
\]
Substitution in the Dirac equation, keeping summation over $\sigma$
implicit, gives,
\begin{eqnarray*}
0 = -i\gamma^{\mu} \partial_{\mu}{\psi} + m \psi & = & \int d^3k \left[
\left\{-i\gamma^0 u_{\vec{k}}\dot{f}_{\vec{k}} +
(\vec{k}\cdot\vec{\gamma} +
m)u_{\vec{k}}f_{\vec{k}}\right\}\varphi_{\vec{k}} \right. \\
& & \mbox{\hspace{1.5cm}} \left. + \left\{-i\gamma^0
v_{\vec{k}}\dot{g}_{\vec{k}} + (- \vec{k}\cdot\vec{\gamma} +
m)v_{\vec{k}}g_{\vec{k}}\right\}\varphi^*_{\vec{k}}\right] ~ ~
\end{eqnarray*}
Linear independence of the basis functions gives two equations:
\[
-i\gamma^0 u_{\vec{k}}\dot{f}_{\vec{k}} + (\vec{k}\cdot\vec{\gamma} +
m)u_{\vec{k}}f_{\vec{k}} ~ = 0 = ~ -i\gamma^0
v_{\vec{k}}\dot{g}_{\vec{k}} + (- \vec{k}\cdot\vec{\gamma} +
m)v_{\vec{k}}g_{\vec{k}} \ .
\]
Differentiating w.r.t. time once more, multiplying by $-i\gamma^0$on the
left and using the above equations again leads to,
\[
	-\sum_{\sigma}u_{\vec{k},\sigma}\left\{\ddot{f}_{\vec{k},\sigma}
	+ \omega^2_{\vec{k}}f_{\vec{k},\sigma}\right\} ~ = 0 = ~
	-\sum_{\sigma}v_{\vec{k},\sigma}\left\{\ddot{g}_{\vec{k},\sigma}
	+ \omega^2_{\vec{k}}g_{\vec{k},\sigma}\right\} ~ ~,~ ~
	\omega^2_{\vec{k}} := \vec{k}^2 + m^2\ .
\]
Since $u_{\vec{k},\sigma}, v_{\vec{k},\sigma}$ are presumed linearly
independent, each of the $f$ and the $g$ functions satisfy the same,
familiar harmonic oscillator equation, with solutions $e^{\pm i
\omega_{\vec{k}} t}$. Of course one of the solutions for each of $f, g$
is spurious since the original equations are first order. Substituting
$f_{\vec{k},\sigma} \sim e^{-i\omega_{\vec{k}}t}$ and
$g_{\vec{k},\sigma} \sim e^{+i\omega_{\vec{k}}t}$ in the first order
equations {\em requires} the $u, v$ spinors to satisfy: $\boxed{(\dsl{k}
+ m)u_{\vec{k},\sigma} = 0 = (\dsl{k} - m)v_{\vec{k},\sigma}}$. Choice
of the other signs does not give a Lorentz covariant equation for the
spinors. Hence we restrict to: $\boxed{\dot{f}_{\vec{k},\sigma} =
-i\omega_{\vec{k}}f_{\vec{k},\sigma} \ , \ \dot{g}_{\vec{k},\sigma} =
+i\omega_{\vec{k}}g_{\vec{k},\sigma}}$. These equations satisfied by the
spinors are analogous to the transversality conditions we got on the
polarization tensors. These are consequences of the Dirac equation i.e.
hold for {\em solutions}.

To compute the Lagrangian, we need the decomposition of the Dirac
conjugate, 
\[
	\bar{\psi}(t, \vec{x}) = \int_{\mathbb{R}^3/2}d^3k\
	\left[\left\{\sum_{\sigma} \bar{u}(\vec{k},
		\sigma)f^*_{\vec{k},\sigma}(t)\right\}
		\varphi^*_{\vec{k}}(\vec{x}) + \left\{ \sum_{\sigma}
		\bar{v}(\vec{k}, \sigma)g^*_{\vec{k},\sigma}(t)\right\}
	\varphi_{\vec{k}}(\vec{x})\right]
\]

We will now choose the $u, v$ spinors to be solutions of the equations:
$(\dsl{k} + m)u(\vec{k},\sigma) = 0 = (\dsl{k} - m)v(\vec{k},\sigma)$
and express the Dirac action in terms of the $f, g$ functions alone.
\begin{eqnarray*}
%
-i\gamma^{\mu}\partial_{\mu}\psi  & = & \int_{\mathbb{R}^3/2} d^3k\left[
	\left\{-i\gamma^0 u(\vec{k},\sigma) \dot{f}_{\vec{k},\sigma}
		\varphi_{\vec{k}} (\vec{x}) + -i\gamma^0
		v(\vec{k},\sigma) \dot{g}_{\vec{k},\sigma}
\varphi^*_{\vec{k}} (\vec{x})\right\} \right.  \\
& & \mbox{\hspace{1.0cm}} \left. + \left\{ k_i\gamma^i u(\vec{k},\sigma)
f_{\vec{k},\sigma} \varphi_{\vec{k}} (\vec{x}) - k_i\gamma^i
v(\vec{k},\sigma) g_{\vec{k},\sigma} \varphi^*_{\vec{k}} (\vec{x})
\right\} \right] \\
\therefore (-i\gamma^{\mu} \partial_{\mu} + m)\psi & = & \int d^3k
\left[\left\{-i\gamma^0u(\vec{k},\sigma)\dot{f}_{\vec{k},\sigma} +
k_i\gamma^iu(\vec{k},\sigma)f_{\vec{k},\sigma} \right.\right. \\
& & \mbox{\hspace{5.0cm}} \left. + m u(\vec{k},\sigma)
f_{\vec{k},\sigma}\right\} \varphi_{\vec{k}}(\vec{x}) \\
& & \mbox{\hspace{1.0cm}} +
\left\{-i\gamma^0v(\vec{k},\sigma)\dot{g}_{\vec{k},\sigma} -
k_i\gamma^iv(\vec{k},\sigma)g_{\vec{k},\sigma} \right. \\
& & \mbox{\hspace{5.0cm}}\left.\left. + m v(\vec{k},\sigma)
g_{\vec{k},\sigma}\right\} \varphi^*_{\vec{k}}(\vec{x})\right] \\
\therefore L & = &\int_{\Sigma_t}d^3x \bar{\psi}(-i\gamma^{\mu}
\partial_{\mu} + m)\psi ~ = ~ \int_{\mathbb{R}^3/2}
d^3k'\int_{\mathbb{R}^3/2} d^3k\int_{\Sigma_t} d^3x \\
& & \left[\left\{\bar{u}(\vec{k}',
\sigma)f^*_{\vec{k}',\sigma'}(t)\right\} \varphi^*_{\vec{k}'}(\vec{x}) +
\left\{ \bar{v}(\vec{k}', \sigma')g^*_{\vec{k}',\sigma'}(t)\right\}
\varphi_{\vec{k}'}(\vec{x}) \right]\times\\
& & \left[\left\{-i\gamma^0u(\vec{k},\sigma)\dot{f}_{\vec{k},\sigma} +
k_i\gamma^iu(\vec{k},\sigma)f_{\vec{k},\sigma} + m u(\vec{k},\sigma)
f_{\vec{k},\sigma}\right\} \varphi_{\vec{k}}(\vec{x}) \right. \\
& & \left. + \left\{-i\gamma^0v(\vec{k},\sigma)\dot{g}_{\vec{k},\sigma}
- k_i\gamma^iv(\vec{k},\sigma)g_{\vec{k},\sigma} + m v(\vec{k},\sigma)
g_{\vec{k},\sigma}\right\} \varphi^*_{\vec{k}}(\vec{x})\right]
\end{eqnarray*}
Thanks to the integration domain of the $\vec{k}, \vec{k}'$
integrations, only $\varphi^*_{\vec{k}'}\varphi_{\vec{k}}$ terms
contribute $\delta^{3}(\vec{k} - \vec{k}')$. Hence only $\bar{u}$ pairs
with $u$ and $\bar{v}$ pairs with $v$. Using the equations satisfied by
the spinors, we eliminate $(k_i\gamma^i + m)u =
+\omega_{\vec{k}}\gamma^0 u$ and $(- k_i\gamma^i + m)v =
-\omega_{\vec{k}}\gamma^0 v$. This leads to,
\begin{eqnarray}
L_{Dirac} & = & \int_{\mathbb{R}^3/2} d^3k \left\{ \sum_{\sigma,\sigma'}
\left[\bar{u}(\vec{k},\sigma') \gamma^0 u(\vec{k},\sigma)\right]\left[
f^*_{\vec{k},\sigma'}(-i\dot{f}_{\vec{k},\sigma} +
\omega_{\vec{k}}f_{\vec{k},\sigma})\right]\right. \nonumber \\
& & \left. \mbox{\hspace{1.5cm}} + \sum_{\sigma,\sigma'}
\left[\bar{v}(\vec{k},\sigma') \gamma^0 v(\vec{k},\sigma) \right]\left[
g^*_{\vec{k},\sigma'} (-i\dot{g}_{\vec{k},\sigma} -
\omega_{\vec{k}}g_{\vec{k},\sigma})\right]\right\}
\end{eqnarray}
Choosing the normalization of the spinors so that $\bar{u}(\vec{k},
\sigma')\gamma^0 u(\vec{k}, \sigma) = \delta_{\sigma,\sigma'} = \pm
\bar{v}(\vec{k}, \sigma')\gamma^0 v(\vec{k}, \sigma)$ gives the Dirac
action as,
\begin{equation}
	S[\bar{\psi}, \psi] = \int dt \int_{\mathbb{R}^3/2}d^3k\  (-i)
	\sum_{\sigma} \left[
		f^*_{\vec{k},\sigma}(\dot{f}_{\vec{k},\sigma} +
		i\omega_{\vec{k}} f_{\vec{k},\sigma}) \pm
		g^*_{\vec{k},\sigma}(\dot{g}_{\vec{k},\sigma} -
	i\omega_{\vec{k}}g_{\vec{k},\sigma})\right] 
\end{equation}
We can group the $f, g$'s together to get a uniform Lagrangian with
$\vec{k} \in \mathbb{R}^3$. Denote $f_{-\vec{k},\sigma} :=
g^*_{\vec{k},\sigma}\ \forall\ \vec{k} \in \mathbb{R}^3/2$. Do a partial
integration of the $g^*_{\vec{k},\sigma}\dot{g}_{\vec{k},\sigma} =
f_{-\vec{k},\sigma}\dot{f}^*_{-\vec{k},\sigma}$. This makes the relative
signs the same in both the terms. Choosing the `-' sign for the
normalization condition for the $v$-spinors, allows us to combine the
two terms and extend the integration domain to full $\mathbb{R}^3$.
Thus, we get,
\begin{eqnarray}
S_{Dirac}[\bar{\psi}, \psi] & = & \int dt\int_{\mathbb{R}^3}d^3k
\sum_{\sigma}\left[ f^*_{\vec{k},\sigma}( -i\dot{f}_{\vec{k},\sigma} +
\omega_{\vec{k}}f_{\vec{k},\sigma})\right] ~ ~ \mbox{for} \\
\psi(t, \vec{x}) & = & \int_{\mathbb{R}^3/2}d^3k\ \sum_{\sigma} \left[
u(\vec{k}, \sigma)f_{\vec{k},\sigma}(t) \varphi_{\vec{k}}(\vec{x}) +
v(\vec{k},
\sigma)f^*_{-\vec{k},\sigma}(t)\varphi^*_{\vec{k}}(\vec{x})\right] \\
0 & = & (\dsl{k} + m)u(\vec{k},\sigma) ~ = ~ (-\dsl{k} + m)
v(\vec{k},\sigma)  ~ ~ ~ ~ \mbox{normalized as,} \\
\delta_{\sigma;,\sigma} & = & \bar{u}(\vec{k},\sigma') \gamma^0
u(\vec{k},\sigma)  ~ = ~ - \bar{v}(\vec{k},\sigma') \gamma^0
v(\vec{k},\sigma) \ .
\end{eqnarray}
Note that the action is real again.

\underline{An aside:} It is an interesting exercise to express the action
in terms of the real $q$ type variables as we did for the scalar.  Since
the action is manifestly a sum over variables with uncoupled labels
$\sigma, \vec{k}$, we focus on just one of the terms. Thus consider the
manifestly real Lagrangian, $L(f, f^*) := \Case{1}{2}\{ -i f^*\dot{f} +
i\dot{f}^* f\} + \omega f^*f $. Introduce the real variables $x, y$ as:
$f := x + iy , \dot{f} = \dot{x} + i \dot{y}$. The Lagrangian then takes
the form,
\[
	L(x,y, \dot{x}, \dot{y}) = x\dot{y} - y\dot{x} + \omega(x^2 +
	y^2) ~ ~ \Rightarrow ~ ~ p_x = -y ~ , ~ p_y = +x
\]
Notice that the equations defining the conjugate momenta {\em cannot} be
inverted to solve for the velocities - {\em we have a constrained
system} \cite{ConstrainedSys}. This is not the place to discuss it in
detail, the reference gives the necessary background. I will just list
the steps.
\begin{eqnarray*}
\phi & := & p_x + y  ~ ~ , ~ ~ \chi ~ := ~ p_y - x ~
\mbox{\hspace{2.0cm} (primary constraints);} \\
H_{canonical} & := & p_x\dot{x} + p_y\dot{y} - L ~ = ~ -\omega(x^2 +
y^2); \\
H_{total} & = & -\omega(x^2 + y^2) + \lambda\phi + \mu\chi ~
\mbox{Preservation of primary constraints $\Rightarrow $} \\
H_{total} & = & -\omega(xp_y - yp_x) ~ ~ \because \lambda = \omega y~ ,
~ \mu = -\omega x\ ; \\
\mbox{Dirac brackets} & : & \{x, p_x\}_* = \{y, p_y\}_* = \frac{1}{2} ~
, ~ \{x, y\}_* = \{p_x, p_x\}_* = - \frac{1}{2} \ . ~ ~ ~ ~\mbox{Use:}
\\
H_{total} & = & -\omega(x^2 + y^2) ~ ~ ~ ~ \mbox{and Dirac brackets for
equations of motion:} ~ \\
\dot{x} & = & \omega\ y ~ ~ , ~ ~ \dot{y} = - \omega\ x ~ ; ~
\mbox{finally drop $p_x, p_y$ and set $x := \frac{p}{\sqrt{2}}\ ,\ y :=
\frac{q}{\sqrt{2}}$,} \\
H_{total} & = & -\frac{1}{2} \omega(p^2 + q^2) ~ , ~ \{q, p \} = 1 ~ , ~
\dot{p} = \omega\ q ~,~ \dot{q} = - \omega\ p \ .
\end{eqnarray*}
Re-introducing the $(\vec{k}, \sigma)$ labels shows that the spinor
field Hamiltonian too is a sum of harmonic oscillators for each of the
$\vec{k} \in \mathbb{R}^3$ and $\sigma = \pm$ labels. 

{\em Puzzle: Why is the Hamiltonian of a wrong sign?} Is it merely a
choice of the overall sign in the Lagrangian density (will not change
the equation of motion)? What is the rationale for a choice of sign? We
will not pursue this classical formulation further here.

{\em Remarks:} The scalar and the vector fields that we discussed have
been real. This was reflected in the Fourier decomposition with complex
conjugate coefficients. What if we have a complex field? Well, we can
always write the Fourier decomposition as (see the spinorial case),
\[
	\phi(t,\vec{x}) ~ = ~ \int_{\mathbb{R}^3/2}d^3k\left[
	f_{\vec{k}}(t)\varphi_{\vec{k}}(\vec{x}) +
g^*_{\vec{k}}(t)\varphi_{\vec{k}}(\vec{x})\right]
\]
For real scalar field, the reality condition simply identifies
$g_{\vec{k}}(t) = f_{-\vec{k}}(t)$. For a complex field, there is no
such identification. We will then have {\em two} harmonic oscillators
for each $\vec{k} \in \mathbb{R}^3$. For the spinorial case, we had a
constrained system and hence half the number of oscillators.

{\em Remark:} Consider the time-Fourier transform of a scalar field:
$\phi(\omega,\vec{x}) := \Case{1}{\sqrt{2\pi}}\int_{-\infty}^{\infty}dt
\ e^{-i\omega t}\phi(t,\vec{x})$. The Fourier transform of its complex
conjugate is given by: $\phi_c(\omega) := \Case{1}{\sqrt{2\pi}}
\int_{-\infty}^{\infty}dt \ e^{+i\omega t}\phi^*(t,\vec{x})$. Clearly,
$\phi_c(\omega) = \phi^*(-\omega)$. Therefore, a positive frequency
$\phi(t,\vec{x})$ has $\phi(\omega, \vec{x}) = 0,$ for $\omega < 0$.
This immediately also give that $\phi^*(t,\vec{x})$ {\em is} a negative
frequency field, i.e. $\phi_c(\omega, \vec{x}) = 0$ for $\omega > 0$.
This gets reflected in the {\em mode decomposition}.

While the Fourier decomposition displays the ``degrees of freedom'' of
the field, the mode decomposition will turn out to be convenient for
passage to quantum fields. To appreciate this, consider a single
oscillator (a single $\vec{k}$). The degrees of freedom view gives:
$\ddot{q} + \omega^2q^2 = 0, p = \dot{q}, H = \Case{1}{2}(p^2 + q^2),
\{q, p\} = 1$. The mode decomposition gives ($a, a^*$ are constants), 
\begin{eqnarray*}
q(t) ~ = ~ \underline{a}e^{-i\omega t} + \underline{a}^*e^{i\omega t} ~
& \Rightarrow & ~ p(t) = -i\omega\left(\underline{a}e^{-i\omega t} -
\underline{a}^*e^{i\omega t}\right)~ ~ \mbox{inverting gives,} \\
\underline{a} ~ = ~ \frac{e^{i\omega t}}{2}\left(q + i
\frac{p}{\omega}\right) ~ & , & ~ \underline{a}^* ~ = ~
\frac{e^{-i\omega t}}{2}\left(q - i \frac{p}{\omega}\right) \\
\therefore \{\underline{a}, \underline{a}^*\} ~ = ~ -\frac{i}{2\omega} ~
~ & \mbox{Define:} & ~ a := \sqrt{2\omega}\underline{a}~ , ~ a^* :=
\sqrt{2\omega}\underline{a}^* ~ ~ \Rightarrow \\
\{a, a^*\} ~ = ~ -i ~ ~ & , & ~ ~ q(t) = \left(\frac{e^{-i\omega
t}}{\sqrt{2\omega}}\right) a + \left(\frac{e^{i\omega
t}}{\sqrt{2\omega}}\right) a^* 
\end{eqnarray*}
Notice the normalization factor of $\sqrt{2\omega}$. This is correlated
with the normalization of the Poisson bracket of $a, a^*$ (and of course
has nothing to do with the Lorentz covariance!). As per the usual
canonical quantization procedure, $\{ , \} \to -i/(\hbar = 1)\times[ ,
]$. This would lead to $[a, a^{\dagger}] = 1$.

Keeping this in mind, we summarize the mode decompositions for the
fields:

\begin{center}
\fbox{
\begin{minipage}{0.9\textwidth}
\begin{eqnarray}
\phi(x) & = & \int_{\mathbb{R}^3} \frac{d^3k} {\sqrt{2\omega(2\pi)^3}}
\left[ a_{\vec{k}}e^{ik\cdot x} + b^*_{\vec{k}}e^{-ik\cdot x}\right] ~ ~
k\cdot x := -\omega_{\vec{k}}t + \vec{k}\cdot \vec{x} \\
& & \mbox{ \hspace{6.0cm} If $\phi^* = \phi$ then $b_{\vec{k}} =
a_{\vec{k}}$.  } \nonumber \\
& & \nonumber \\
A_{\mu}(x) & = & \int_{\mathbb{R}^3}\frac{d^3k}
{\sqrt{2\omega(2\pi)^3}} \sum_{\lambda=1,2} \left[ a_{\vec{k},\lambda}
	\underline{\varepsilon}_{\mu} (\vec{k},\lambda) e^{ik\cdot x} +
	a^*_{\vec{k},\lambda} \underline{\varepsilon}^*_{\mu}
(\vec{k},\lambda)e^{-ik\cdot x}\right] ~ ~ \\
& & \underline{\varepsilon}_0(\vec{k},\lambda) ~ := ~ -\frac{\vec{k}
\cdot \vec{\varepsilon} (\vec{k},\lambda)}{\omega_{\vec{k}}} ~ ~ ~ ~ , ~
~ ~ ~ \omega_{\vec{k}} = |\vec{k}| \nonumber \\
& & \underline{\varepsilon}_i(\vec{k},\lambda) ~ := ~
\left(\delta_i^{~j} - \frac{k_i\ k^j} {\vec{k}^2}\right)
\varepsilon_j(\vec{k},\lambda) ~ ~ , ~ ~ \underline{\vec{\varepsilon}}
(\vec{k},\lambda) \cdot \underline{\vec{\varepsilon}}^*
(\vec{k},\lambda') ~ = ~ \delta_{\lambda,\lambda'} \nonumber \\
& &  \nonumber \\
\psi(x) & = & \int_{\mathbb{R}^3}\frac{d^3k} {\sqrt{2\omega(2\pi)^3}}
\sum_{\sigma =\pm} \left[ b_{\vec{k},\sigma} u(\vec{k},\sigma)e^{ik\cdot
x} + d^*_{\vec{k},\sigma} v(\vec{k},\sigma) e^{-ik\cdot x} \right] \\
& & \mbox{\hspace{2.0cm}} (\dsl{k} + m) u(\vec{k},\sigma) ~ = 0 = ~
(-\dsl{k} + m)v(\vec{k},\sigma) \nonumber \\
& & \mbox{\hspace{2.0cm}} \bar{u}(\vec{k},\sigma)u(\vec{k},\sigma') ~ =
~ \delta_{\sigma,\sigma'} ~ = ~ -
\bar{v}(\vec{k},\sigma)v(\vec{k},\sigma') \nonumber \\
& & \sum_{\sigma}u(\vec{k}, \sigma)\bar{u}(\vec{k},\sigma) = \frac{-
\dsl{k} + m}{2m} ~ ~,~ ~ \sum_{\sigma}v(\vec{k},
\sigma)\bar{v}(\vec{k},\sigma) = - \frac{ \dsl{k} + m}{2m}  \nonumber
\end{eqnarray}
\end{minipage} }
\end{center}
For a real scalar field, we note,
\begin{equation}\label{ModesToFieldPBs}
\{a_{\vec{k}}, a^*_{\vec{k}'} \} = -i \delta^3(\vec{k} -
\vec{k}')  ~ ~ ~ \leftrightarrow ~ ~ ~ \{\phi(t, \vec{x}),
\pi(t, \vec{x}')\} = \delta^3(\vec{x} - \vec{x}') \ .
\end{equation}

\underline{NOTE:} ~ {\em So far whatever we have done has {\em no quantum}
in it. We have obtained a classical theory of non-interacting fields}.

\subsection{Interaction with source: Green's functions}
A simplest kind of `interaction' we are familiar with, eg. from
classical electrodynamics, is interaction with a ``source''. So let us
consider a real scalar field interacting with a source $J(x)$ which is a
real function. The interaction is described by the
equation\footnote{Follows from the Lagrangian ${\cal L} = {\cal
L}_{scalar} + J(x)\phi(x)\ $ with ${\cal L}_{scalar}$ given in
(\ref{LagrangiansEqn}).}:
\begin{eqnarray*}
(\Box - m^2)\phi(x) & = & - J(x) ~ ~ \leftrightarrow ~ ~ (\Box_x -
m^2)G(x,x') = - \delta^4(x - x')  ~ ~ \mbox{Green's Function} \\
\Rightarrow \phi(x) & = & + \int_{Source}d^4x'\ G(x, x')J(x') 
\end{eqnarray*}
Poincare invariance of the defining equation for Green's function implies
that it is a Lorentz invariant function of $(x - x')$. It can be
represented as,
\begin{equation}\label{GreenFnDefn}
	\boxed{
G(x-x') ~ = ~ \int_{\mathbb{R}^4}\frac{d^4k}{(2\pi)^4} \frac{e^{i
k\cdot(x - x')}}{k^2 + m^2} ~ ~,~ ~ k^2 = -(k^0)^2 + \vec{k}^2 ~ . 
}
\end{equation}
The denominator vanishes for $k^2 = -m^2$ and thus we need a {\em
definition} of the integral and hence of the Green's function. 

The usual method is to define $\int d^4k = \int d^3k \int dk^0$ and
interpret the $k^0$ integral as a contour integral along a suitably
chosen contour. Each choice is supposed to reflect a ``boundary
condition'' for the wave operator. In the complex $k^0$ plane, there are
poles at $k^0 = \pm\ \omega_{\vec{k}} := \sqrt{\vec{k} ^2 + m^2}$. The
integral is sought to be defined by shifting the poles off the real
axis. This can be done in 4 ways: (i) shift both poles in the Lower Half
Plane (LHP), $k^0 \to k^0 + i\epsilon$, (ii) shift both poles in the
UHP, $k^0 \to k^0 - i\epsilon$, (iii) shift the pole at
+$\omega_{\vec{k}}$ in the LHP and shift the pole at $-\omega_{\vec{k}}$
in the UHP, $\omega^2_{\vec{k}} \to \omega^2_{\vec{k}} - i\epsilon$ and
(iv) reverse of the (iii). To see the consequences of any of these
choices, let us consider the special case of a source localized at the
space-time origin: $J(t, x) = \delta^4(x)$. The solution takes the form,
\begin{eqnarray}\label{InhomogeneousSoln}
	\phi(x) & = & \int dt'\int d^3x'\int \frac{d^4k}{(2\pi)^4}
	\frac{e^{ik\cdot x}}{k^2 + m^2}
	e^{-i(-k^0t'+\vec{k}\cdot\vec{x}')}\delta(t')\delta^3(\vec{x}')
	\nonumber \\
	& = & \int
	\frac{d^3k}{(2\pi)^3}e^{i\vec{k}\cdot\vec{x}}\left[\int
	\frac{dk^0}{2\pi} ~ \frac{e^{-ik^0 t}}{(k^0)^2 -
\omega_k^2}\right] \mbox{\hspace{1.0cm} and, }\\
	\left[\dots\right] & = & \frac{1}{2\omega_{\vec{k}}}\int
	\frac{dk^0}{2\pi} \left(\frac{1}{k_0 - \omega_{\vec{k}}} -
	\frac{1}{k^0 + \omega_{\vec{k}}}\right) e^{-ik^0 t} \nonumber 
\end{eqnarray}
Without the $i\epsilon$'s, both sides of the above equations are real.
Now we introduce the $i\epsilon$'s. The contour integrals will have the
segment of a semicircle at infinity whose contribution should vanish.
On the semicircle at large $R$, $k^0 = R e^{i\theta}$, the integrand
behaves as $R^{-2}e^{R sin(\theta) t}$ and the measure provides factor
of $R$.  This suffices for convergence for $t=0$. However, for a
non-zero $t$, we can get an exponential divergence unless $\theta$ is
restricted appropriately. This dictates how the contour should be closed
in UHP or LHP.  Since $\phi(x)$ is a function of the Lorentz invariant
$x^2 = -t^2 + \vec{x}^2$, we can consider the two cases as: (a) $x^2 \le
0$ (sign of $t$ is invariant) and we may evaluate the integrals for
$\vec{x} = 0$; (b) $x^2 > 0$ and we can evaluate the integrals for $t =
0$.

(i) {\em Both poles to LHP:} To pickup the residues, we should close the
contour in the LHP. For space-like interval, taking $t = 0$, we see that
the residues at the two poles cancel each other and the solution
vanishes. For time-like intervals, $t \ne 0$ but we can take $\vec{x} =
0$.  To have a vanishing contribution from the semicircle at infinity,
we have to restrict $t > 0$. The evaluation of the  integral is trivial
and gives,
\[
\left[\dots\right]_{Retarded} ~ = ~ -\frac{1}{\omega_{\vec{k}}}\
\theta(t) sin(\omega_{\vec{k}} t) \ .
\]

(ii) {\em Both poles to UHP:} To pick-up the residues, the contour
should be closed in the UHP and for time-like interval, we need to
restrict to $t < 0$. The evaluation gives,
\[
\left[\dots\right]_{Advanced} ~ = ~ \frac{1}{\omega_{\vec{k}}}\
\theta(-t) sin( \omega_{\vec{k}} t) \ .
\]
As before, for space-like interval, the solution vanishes.

(iii) {\em Positive pole to LHP and negative to UHP:} Now closing the
contour in either of UHP/LHP will always pick up a contribution. For
time-like interval, for LHP closure, we have to take $t> 0$ and we will
pick up the contribution from the first term. Likewise, for the UHP
closure, we have to take $t < 0$ and we will pick up the contribution
from the second term. Evaluation gives, for time-like or light-like
separation
\[
\left[\dots\right]_{Feynman} ~ = ~ -i\frac{1}{2\omega_{\vec{k}}}\
\left[ \theta(t)e^{-i\omega_{\vec{k}} t}  + \theta(-t)
e^{+i\omega_{\vec{k}} t} \right]  ~ = ~ -i\frac{1}{2\omega_{\vec{k}}}\
e^{-i\omega_{\vec{k}} |t|} \left[ \theta(t)+ \theta(-t)\right].
\]
For space-like interval, taking $t = 0$, we get the {\em non-zero}
answer, $\left[\dots\right]_{Feynman} = -i\Case{1}{2\omega_{\vec{k}}}$
for closure in either LHP or UHP.

The case (iv) is similar to (iii) and is obtained in an obvious manner.
The solution $\phi(x)$ is obtained by the spatial Fourier transform as
given above (\ref{InhomogeneousSoln}).

{\em Remarks:} There are two properties of the Retarded and the Advanced
choices that stand out. In both cases, the solution $\phi(x)$ (a) is real
for a real source function and (b) it reflects the expected {\em causal}
behavior - vanishing outside the future/past light cones. By contrast,
the Feynman choice gives a non-real $\phi(x)$ even for a real source
function and the solution is non-vanishing outside the light cones! The
same holds for the choice (iv). {\em What do we make of this?}

The retarded and advanced choices support the interpretation that the
field can be regarded as an {\em observable} responding to a source in a
causally consistent manner. This is what we would have in a classical
field theory. The Feynman choice however disallows such an
interpretation - {\em the particular solution $\phi(x)$ {\em cannot} be
interpreted as causally consistent response to a source.  }

Both the causal and the Feynman Green's functions appear naturally when
$\phi(x)$ is promoted to be an operator field i.e. in {\em Quantum Field
Theory}. In the scattering theory, section \ref{ScatteringTheory}, the
causal Green's function are used in articulating the {\em asymptotic
conditions} while the Feynman Green's function (or Feynman propagator)
appears naturally in the Lehmann-Symanzik-Zimmermann (LSZ) reduction of
the scattering matrix.

For completeness, we note the full Green's functions\footnote{The
	overall signs are convention dependent. It is easiest to match it
for the massless case. For detail see: \cite{ZhangEtAl2008,
GreinerReinhardt}}: 

\begin{center}
\fbox{ 
\begin{minipage}{0.9\textwidth}
\begin{eqnarray}
G_{ret}(x) & = & +\frac{\theta(t)}{2\pi}\delta(x^2) -
\theta(t)\theta(-x^2)\frac{mJ_1(m\sqrt{-x^2})}{4\pi\sqrt{-x^2}} \\
G_{adv}(x) & = &  +\frac{\theta(-t)}{2\pi}\delta(x^2) -
\theta(-t)\theta(-x^2)\frac{mJ_1(m\sqrt{-x^2})}{4\pi\sqrt{-x^2}}\\
G_{Feynman}(x^2 < 0) & = & + \frac{\delta(x^2)}{4\pi} -
\frac{m}{8\pi\sqrt{-x^2}}H^{(2)}_1(m\sqrt{-x^2}) \nonumber \\
G_{Feynman}(x^2 > 0) & = & +\frac{i}{4\pi^2}\frac{m}{\sqrt{x^2}}K_1(m
\sqrt{x^2}) \\ \nonumber 
\end{eqnarray}
\end{minipage}
}
\end{center}

\newpage
\section{Covariance of quantum fields and relativistic causality}
\label{QuantumFields}

So far we have studied representations of the Poincare group and seen
emergence of a ``field dynamical system'' - an infinite collection of
harmonic oscillators. In our attempt to incorporate interactions with
classical sources, we encountered the Green's functions. We met the
expected and familiar retarded and advanced Green's functions, but also
encountered the mathematical possibility of the Feynman alternative
which does not gel with a classical field theory interpretation. Taking
it as a hint of an opportunity, we attempt a `quantum' interpretation.
Here, the field as a collection of harmonic oscillators gives a strong
clue: {\em use the mode decomposition and promote the expansion
coefficients to operators}. We will then have a collection of creation
and annihilation operators and can tag the states of the system by the
number of quanta. We already saw (\ref{ModesToFieldPBs}) in the simpler
case of a real scalar field that postulating Poisson brackets for the
coefficients and using the completeness of mode functions , we can
deduce the Poisson brackets for the fields themselves. The same method
works for operators as well. That is, not only do we promote the fields
to operators but we also postulate the necessary commutation relations.
\subsection{Poincare Covariance of Quantum fields}
The first feature, fields as operators, immediately modifies the
implementation of covariance: a covariant quantum field requires
specific transformations for the coefficient functions and conversely.
Transformation laws of the mode functions do not suffice.

To see this, consider a linear combination of classical fields,
$\psi^A(x) := \sum_n c_n \psi^A_n(x)$. A group action on $\psi^A$ is
deduced from that on the individual $\psi^A_n$: $g\cdot \psi^A_n(x) =
D^A_{~B}(g)\psi^B_n(g^{-1}x)$, 
\[
g\cdot\psi^A(x) := \sum_n c_n g\cdot \psi^A_n(x) = \sum_n
D^A_{~B}(g)\psi^B_n(g^{-1}x) = D^A_{~B}(g) \psi^B(g^{-1}x). 
\]
This of course presumes quite naturally that the coefficients are {\em
invariant} under the group action. When a similar linear combination is
promoted to an operator by making the $c_n$ as operators, the above
procedure breaks down. 
\[
	U(g)\hat{\psi}^A(x)U^{\dagger}(g) := \sum_n
	\left[U(g)\hat{c}_nU(g)^{\dagger}\right]\psi^A_n(x)
	\stackrel{?}{=}D^A_{~B}(g^{-1})\hat{\psi}^B(gx) \ .
\]
Evidently, if we assume the $\hat{c}_n$ to be invariant, then the field
cannot be covariant. So the $\hat{c}_n$ should transform in a manner
that shifts the group action on to the $\psi_n^A(x)$ and then the
$\sum_n$ to work out correctly so as to produce the expected right hand
side\footnote{The homomorphism dictates the expected form on the
r.h.s.}.  Taking now $c_n \to \hat{a}, \hat{a}^{\dagger} , \sum_n \to
\int \Case{d^3k}{\dots}$ and using the mode expansion form, we observe
that {\em if} we postulate the transformation laws for the $\hat{a},
\hat{a}^{\dagger}$ as,
\[
	U(g)\hat{a}(\vec{k}, \sigma)U^{\dagger}(g) :=
	\hat{a}(\vec{\Lambda k}, ``D(\Lambda)" \sigma) ~ ~ , ~ ~
	U(g)\hat{a}^{\dagger}(\vec{k}, \sigma)U^{\dagger}(g) :=
	\hat{a}^{\dagger}(\vec{\Lambda k}, ``D(\Lambda)" \sigma) ~ ,
\]
then we can shift the changed labels from the operators to the mode
function labels using the change of integration (and summation)
variables. Provided the integration measure is invariant, the right hand
side takes the expected form and we have the {\em covariance of the
quantum field}. Note that this essentially fixes the form of the mode
decomposition! In equations (for a Dirac field say),
\begin{eqnarray*}
\left[U(g)\hat{\psi}U^{\dagger}(g)\right](x) & := & \int \frac{d^3k}
{\sqrt{2\omega_{\vec{k}} (2\pi)^3}} \sum_{\sigma} \left[
	(U\hat{b}_{\vec{k},\sigma} U^{\dagger}) u(\vec{k},\sigma)
	e^{ik\cdot x} + (U\hat{d}^{\dagger}_{\vec{k},\sigma}
U^{\dagger}) v(\vec{k},\sigma) e^{-ik\cdot x} \right] \\
& = & \int \frac{d^3k} {\sqrt{2\omega_{\vec{k}} (2\pi)^3}} \sum_{\sigma}
\left[ \hat{b}_{\vec{\Lambda k},\sigma_{\Lambda}}  u(\vec{k},\sigma)
e^{ik\cdot x} + \hat{d}^{\dagger}_{\vec{\Lambda k},\sigma_{\Lambda}}
v(\vec{k},\sigma) e^{-ik\cdot x} \right] \\
& = & \int \frac{d^3(\Lambda^{-1}k)}
{\sqrt{2\omega_{\vec{(\Lambda^{-1}k)}} (2\pi)^3}}
\sum_{\sigma_{\Lambda^{-1}}} \left[ \hat{b}_{\vec{k},\sigma}
u(\vec{(\Lambda^{-1}k)},\sigma_{\Lambda^{-1}}) e^{i(\Lambda^{-1}k)\cdot
x} \right. \\ 
& & \mbox{\hspace{4.0cm}} \left. ~ + \
\hat{d}^{\dagger}_{\vec{k},\sigma}
v(\vec{(\Lambda^{-1}k)},\sigma_{\Lambda^{-1}}) e^{-i(\Lambda^{-1}k)\cdot
x} \right] \\
& = & \int \frac{d^3k} {\sqrt{2\omega_{\vec{k}} (2\pi)^3}} \sum_{\sigma}
\left[ \hat{b}_{\vec{k},\sigma} \left\{D(\Lambda^{-1})
u(\vec{k},\sigma)\right\} e^{ik\cdot (\Lambda x)} \right. \\
& & \mbox{\hspace{4.0cm}}\left.  + \ \hat{d}^{\dagger}_{\vec{k},\sigma}
	\left\{ D(\Lambda^{-1}) v(\vec{k},\sigma)\right\}
	e^{-ik\cdot(\Lambda x}  \right] \\
& = & D(\Lambda^{-1})\hat{\psi}(\Lambda x)
\end{eqnarray*}

\underline{Note:} If we postulate a Poincare invariant state $|0
\rangle$ and define $|\vec{k}, \sigma \rangle :=
b^{\dagger}(\vec{k},\sigma)|0 \rangle$, then these states transform
precisely as per the particle representation. Thus the postulated
group action on the $b, d$ operators are precisely as needed for
building up the particle representations and we will use it shortly.
\subsection{Space Inversion, Time Reversal, Charge Conjugation of Dirac
Field} 
Let us quickly see how the discrete symmetries act on quantum fields.
Again, we will consider a Dirac field for illustration and now suppress
the `hats' on the operators.

{\bf Space Inversion:} 

We want $\Psi_{\cal{P}}(t, \vec{x}) := \cal{P}\Psi\cal{P}^{\dagger} =
D(\cal{P})\Psi(t, -\vec{x})$. From the defining relations for $\cal{P}$,
namely, $\cal{P} P_i \cal{P}^{\dagger} = - P_i$ and $\cal{P} J_i
\cal{P}^{\dagger} = J_i$, we postulate,
\[
	\cal{P}b^{\dagger}(\vec{k},\sigma)\cal{P}^{\dagger} := \eta
	b^{\dagger}(-\vec{k},\sigma) ~ ~ , ~ ~ 
	\cal{P}d^{\dagger}(\vec{k},\sigma)\cal{P}^{\dagger} := \eta
	d^{\dagger}(-\vec{k},\sigma) ~ ~ , ~ ~ \eta^2 = \pm 1 \ .
\]
As noted while discussing the discrete symmetry actions on the particle
representations, the phase is $\sigma$ independent.
	
Substituting the mode expansion and using these postulated actions
gives,
\begin{eqnarray*}
\cal{P}\Psi^{\dagger}(\vec{k},\sigma)\cal{P}^{\dagger} & = & \int
[d^3k]_{inv} \sum_{\sigma} \left\{ \eta^*b(\vec{k},\sigma)
	u_{-\vec{k},\sigma}e^{ik\cdot (\cal{P}x)} + \eta
	d^{\dagger}(\vec{k},\sigma) v_{-\vec{k},\sigma}e^{-ik\cdot
(\cal{P}x)} \right\}
\end{eqnarray*}

Claim: The  $u, v$ spinors satisfy, $u(-\vec{k},\sigma) =
\gamma^0u(\vec{k},\sigma)$ and $v(-\vec{k},\sigma) = - \gamma^0
v(\vec{k},\sigma)$.

This follows by noting that
\begin{eqnarray*}
0 & = & ( \dsl{k} + m)u(\vec{k},\sigma) = (-\omega \gamma^0 +
\vec{k}\cdot\vec{\gamma} + m)u(\vec{k},\sigma) ~ , ~ u(\vec{k},\sigma) =
\gamma^2 u(\vec{k},\sigma) \\
\therefore 0 & = & (-\omega\gamma^0\gamma^0 +
\vec{k}\cdot\vec{\gamma}\gamma^0 + m\gamma^0)(\gamma^0
u(\vec{k},\sigma)) \\
& = & (-\omega\gamma^0 + (-\vec{k}\cdot\vec{\gamma} + m)(\gamma^0
	u(\vec{k},\sigma)) ~ ~ \Rightarrow  \gamma^0u(\vec{k},\sigma)
	\propto u(-\vec{k},\sigma)
\end{eqnarray*}
Similarly, $\gamma^0 v(\vec{k},\sigma) \propto v(-\vec{k},\sigma)$. The
proportionality factors are fixed by noting that for $\vec{k} =
\hat{k}$, the $u, v$ spinors are eigen-spinors of $\gamma^0$ with
eigenvalues $\pm 1$. This implies,
\begin{eqnarray*}
\cal{P}b^{\dagger}(\vec{k},\sigma)\cal{P}^{\dagger} & = & \gamma^0\ \int
[d^3]_{inv} \sum_{\sigma} \left\{ \eta^*b(\vec{k},\sigma)
	u_{\vec{k},\sigma}e^{ik\cdot (\cal{P}x)} - \eta
	d^{\dagger}(\vec{k},\sigma) v_{\vec{k},\sigma}e^{-ik\cdot
(\cal{P}x)} \right\}
\end{eqnarray*}
Clearly $\eta$ should be imaginary for the right hand side to have the
field $\psi(\cal{P}x)$. We choose $\eta = -i$ and deduce that,
$\boxed{\cal{P}\Psi(x)\cal{P}^{\dagger} = i\gamma^0 \Psi(\cal{P} x)\}\ ,
i.e., D(\cal{P}) = i\gamma^0} ~ \mbox{and}~  \eta^2 = -1$.

{\bf Time reversal:}

We want $\Psi_{\cal{T}}(t,\vec{x}) := \cal{T}\Psi \cal{T}^{\dagger} =
D(\cal{T})\Psi(-t,\vec{x})$. The defining relations where both $P_i$ and
$J_i$ change sign, we postulate,
\[
	\cal{T}b^{\dagger}(\vec{k},\sigma)\cal{T}^{\dagger} ~ = ~
	\xi_{\sigma}b^{\dagger}(-\vec{k},-\sigma) ~ ~ , ~ ~ 
	\cal{T}d^{\dagger}(\vec{k},\sigma)\cal{T}^{\dagger} ~ = ~
	\xi_{\sigma}d^{\dagger}(-\vec{k},-\sigma) ~  . 
\]
As noted while discussing the discrete symmetry actions on the particle
representations, the phase is $\sigma$ dependent.

Substituting the mode decomposition and noting the $\cal{T}$ operator is
anti-unitary, we get,
\begin{eqnarray*}
[\cal{T}\Psi \cal{T}^{\dagger}](x) & = & \int[d^3k]_{inv} \sum_{\sigma}
\left\{\xi^*_{\sigma}b(-\vec{k},-\sigma)u^*(\vec{k},\sigma)e^{-ik\cdot
x} + \xi_{\sigma}d^{\dagger}(-\vec{k},-\sigma)v^*(\vec{k},\sigma)
e^{ik\cdot x}\right\} \\
& = & \int[d^3k]_{inv} \sum_{\sigma}
\left\{\xi^*_{-\sigma}b(\vec{k},\sigma)u^*(-\vec{k},-\sigma)e^{-ik\cdot(
\cal{T}x)} + \xi_{-\sigma} d^{\dagger} (\vec{k},\sigma)
v^*(-\vec{k},-\sigma) e^{ik\cdot(\cal{T}x)}\right\} \\
\end{eqnarray*} 
Now we need to use the relations,
\[
	u^*(-\vec{k}, -\sigma) = -\sigma C\gamma_5u(\vec{k},\sigma) ~ ~
	, ~ ~ 
	v^*(-\vec{k}, -\sigma) = -\sigma C\gamma_5v(\vec{k},\sigma) ~ ~
	. 
\]
Once again these are proved from the defining equations for the spinors.

Clearly, if we choose $\xi_{\sigma} = \sigma (=\pm 1)$, the we get
$\Psi(\cal{T}x)$ on the right hand side and we get, $\boxed{
	\cal{T}\Psi(x)\cal{T}^{\dagger} = C\gamma_5 \Psi(\cal{T}x), i.e.
	D(\cal{T}) = C\gamma_5\ , \ C\gamma^{\mu}C^{\dagger} = -
(\gamma^{\mu})^T }~$.

{\bf Charge conjugation:}

Lastly, consider the charge conjugation action. This is defined through
the particle-anti-particle exchange and we postulate,
\begin{equation} \label{ChargeCongugationDefn}
	\cal{C} b^{\dagger}_{\vec{k},\sigma}\cal{C}^{\dagger} = \xi
	d^{\dagger}_{\vec{k},\sigma} ~ ~ , ~ ~ 
	\cal{C} d^{\dagger}_{\vec{k},\sigma}\cal{C}^{\dagger} = \xi
	b^{\dagger}_{\vec{k},\sigma} ~ . 
\end{equation}

Substitution of mode expansion gives,
\begin{eqnarray*}
[\cal{C}\Psi\cal{C}^{\dagger}](x) & = & \int [d^3k]_{inv}\sum_{\sigma}
\left\{\xi^*d_{\vec{k},\sigma}u(\vec{k},\sigma)e^{ik\cdot x} + \xi
	b^{\dagger}_{\vec{k},\sigma}v(\vec{k},\sigma)e^{-ik\cdot
x}\right\}
\end{eqnarray*}

Now we need to use: $v(\vec{k},\sigma) =
C^{\dagger}\bar{u}^T(\vec{k},\sigma)$ and recall that the $C$ matrix is
given by $C = i\gamma^2\gamma^0\ , \ C^T = -C\ , \ C^{\dagger}C =
\mathbb{1}$ so that $v = - C\bar{u}^T $ and $ u = -C \bar{v}^T$.  This
gives,
\begin{eqnarray*}
[\cal{C}\Psi\cal{C}^{\dagger}](x) & = & \int [d^3k]_{inv}\sum_{\sigma}
\left\{\xi^*d_{\vec{k},\sigma}(-C\bar{v}^T(\vec{k},\sigma))e^{ik\cdot x}
	+ \xi
	b^{\dagger}_{\vec{k},\sigma}(-C\bar{u}^T(\vec{k},\sigma))e^{-ik\cdot
x}\right\} ~ ~ \mbox{and,} \\
\bar{\Psi}^T(x) & = & \int [d^3k]_{inv}\sum_{\sigma} \left\{
b^{\dagger}_{\vec{k},\sigma}(\bar{u}^T(\vec{k},\sigma))e^{-ik\cdot x} +
d_{\vec{k},\sigma}(\bar{v}^T(\vec{k},\sigma))e^{ik\cdot x} \right\}
\end{eqnarray*} 
Hence, if we choose $\xi^* = \xi = -1$, then we get, $\boxed{
\cal{C}\Psi(x)\cal{C}^{\dagger} = + C (\bar{\Psi})^T(x) } $~ .
Note that {\em$\cal{C}$ operator is unitary!}

From these it is easy to get the transformations of Dirac conjugates.
Here is a summary:
\begin{center}
\fbox{
\begin{minipage}{0.9\textwidth}
	\begin{eqnarray} \label{CPTtransforms}
	\cal{P}\Psi\cal{P}^{\dagger}(x) = i\gamma^0 \Psi(\cal{P}x) &
	\mbox{\hspace{1.0cm} , \hspace{1.0cm}} &
	\cal{P}\bar{\Psi}\cal{P}^{\dagger}(x) = -i
	\bar{\Psi}(\cal{P}x)\gamma^0 \ ; \\ 
	\cal{T}\Psi\cal{T}^{\dagger}(x) = C \gamma_5\Psi(\cal{T}x) &
	\mbox{\hspace{1.0cm} , \hspace{1.0cm}} &
	\cal{T}\bar{\Psi}\cal{T}^{\dagger}(x) =
	\bar{\Psi}(\cal{T}x)\gamma_5C^{\dagger} \ ; \\ 
	\cal{C}\Psi\cal{C}^{\dagger}(x) = C \bar{\Psi}^T(x) &
	\mbox{\hspace{1.0cm} , \hspace{1.0cm}} &
	\cal{C}\bar{\Psi}\cal{C}^{\dagger}(x) = \Psi^T(x)C
	\ ; \\ 
	C\gamma^{\mu}C^{\dagger} = - (\gamma^{\mu})^T ~ , ~ C^{\dagger}
	C = \mathbb{1} &  \mbox{\hspace{1.0cm} , \hspace{1.0cm}} & C^T =
	-C ~ , ~ C := i\gamma^2\gamma^0 \\ \nonumber
\end{eqnarray}
\end{minipage}
}
\end{center}

We are now ready to verify one of the important general theorem of
relativistic quantum field theory, {\em the CPT theorem}. We verify it
for the spinorial quantum field.
\subsection{CPT theorem for Dirac Field}\label{CPTThm} 
We want to consider the combined action of the discrete transformations.
Let us choose one ordering, say, $\cal{C} \cal{P} \cal{T}$. Then,
\begin{eqnarray*}
(\cal{C}\cal{P}\cal{T})\Psi(x)(\cal{C}\cal{P}\cal{T})^{\dagger} & = &
\cal{C} \cal{P} (C\gamma_5\Psi(\cal{T}x))
\cal{P}^{\dagger}\cal{T}^{\dagger} ~ = ~ C\gamma_5\cal{C}
(i\gamma^0\Psi(\cal{P}\cal{T}x)) \cal{C}^{\dagger} \\
& = & i C\gamma_5\gamma^0C\ \bar{\Psi}^T(-x) ~ ~ ~ \because
\cal{P}\cal{T}x = -x \\
(\cal{C}\cal{P}\cal{T}) \bar{\Psi}(x) (\cal{C}\cal{P}\cal{T})^{\dagger}
& = & \cal{C} \cal{P} (\bar{\Psi}(\cal{T}x)\gamma_5C^{\dagger})
\cal{P}^{\dagger} \cal{T}^{\dagger} ~ = ~ \cal{C} (\bar{\Psi}
(\cal{P}\cal{T}x)) (-i\gamma^0) \gamma_5C^{\dagger}) \cal{C}^{\dagger}
\\ 
& = & i \Psi^T(-x) C\gamma^0\gamma_5 C ~ ~ ~ \because C^{\dagger}
= -C 
\end{eqnarray*}

Next, consider the CPT transform of a general bilinear of the form
$\bar{\Psi}(x) A \Psi(x)$.  Using
$(\cal{C}\cal{P}\cal{T})^{\dagger}(\cal{C}\cal{P}\cal{T}) = \cal{1}$,
the CPT transform will be the CPT transform of the Dirac conjugate, that
of the matrix $A$ and that of $\Psi$. The CPT transform of $A$ is just
$A^*$ since only the complex conjugate acts on $A$.
\begin{eqnarray}\label{CPT-Bilinears}
\left[(\cal{C}\cal{P}\cal{T}) \bar{\Psi}(x)
(\cal{C}\cal{P}\cal{T})^{\dagger}\right] \left[(\cal{C}\cal{P}\cal{T}) A
(\cal{C}\cal{P}\cal{T})^{\dagger}\right] \left[ (\cal{C}\cal{P}\cal{T})
\Psi(x) (\cal{C}\cal{P}\cal{T})^{\dagger}\right] \nonumber \\ 
=  \left[i \Psi^T(-x) C\gamma^0\gamma_5C\right] \left[ A^* \right]
\left[ iC\gamma_5\gamma^0C\bar{\Psi}^T(-x)\right] \mbox{\hspace{1.8cm}}
\nonumber \\
= i^2\left[\bar{\Psi}(-x)(C\gamma_5\gamma^0C)^T\ A^{\dagger}\
(C\gamma^0\gamma_5C)^T \Psi(-x)\right]^T \times (-1) \nonumber \\
= + \bar{\Psi}(-x)\left\{(C\gamma_5\gamma^0C)^T A^{\dagger}
(C\gamma^0\gamma_5C)^T\right\} \Psi(-x) \mbox{\hspace{1.65cm}} 
\end{eqnarray}
The explicit $(-1)$ in the third line is due to the {\em fermionic}
nature of the spinors which we will see shortly in the discussion of the
spin-statistics theorem.

The factors of the gamma matrices simplify as,
\[
	C\gamma_5\gamma^0C = +\gamma_5C\gamma^0C = -
	\gamma_5C\gamma^0C^{\dagger} = +\gamma_5(\gamma^0)^T =
	\gamma_5\gamma^0 ~ ~ \mbox{and} ~ ~ C\gamma^0\gamma_5C =
	\gamma^0\gamma_5 \ .
\]
\begin{equation} \label{CPTOfBilinear}
(\cal{C}\cal{P}\cal{T}) \left[\bar{\Psi}(x) A \Psi(x)\right]
(\cal{C}\cal{P}\cal{T})^{\dagger} ~ = ~ \bar{\Psi}(-x)\left\{
(\gamma^0\gamma_5) A^{\dagger} (\gamma_5\gamma^0)\right\}\Psi(-x)
\end{equation}
It remains to evaluate the middle braces for $A = \mathbb{1},
\gamma^{\mu}, \gamma^{\mu} \gamma^{\nu}, \gamma^{\mu} \gamma_5$ and
$\gamma_5$. These are obtained easily as,
\begin{eqnarray*}
	\gamma^0\gamma_5\mathbb{1}\gamma_5\gamma^0 & = & \mathbb{1} \\
	\gamma^0\gamma_5\gamma^{\mu}\gamma_5\gamma^0 & = & -\gamma^{\mu}
	\\
	\gamma^0\gamma_5\Sigma^{\mu,\nu}\gamma_5\gamma^0 & = &
	+\Sigma^{\mu\nu} \\
	\gamma^0\gamma_5\gamma^{\mu}\gamma_5\gamma_5\gamma^0 & = &
	-\gamma^{\mu} \\
	\gamma^0\gamma_5(i\gamma_5)\gamma_5\gamma^0 & = &  \gamma_5 
\end{eqnarray*}

The Lagrangian density is made up of the bi-linears and also involves
derivatives. For a $\partial_{\mu}\Psi(x)$, there is an extra minus sign
due to $\cal{P}\cal{T} x = -x$. Thus we see that {\em every tensor
	index, on $\gamma$ or on $ \partial$, gives a minus sign under
CPT transform.} For a Hermitian Lagrangian (coefficients are
appropriately real or pure imaginary) which is a Lorentz scalar (so even
rank tensor), CPT contributes no extra sign and only changes the
space-time argument i.e.  $\boxed{\cal{C}\cal{P}\cal{T}) \cal{L}(x)
(\cal{C}\cal{P} \cal{T})^{\dagger} = \cal{L}(-x) }$. Thus, 

{\em a Hermitian action, invariant under proper and orthochronous
Lorentz transformations, is invariant under the discrete CPT
transformations.}

We have explicitly verified it for the quantum, Dirac field.
Verification for scalar and Maxwell field is left as an exercise. 

\underline{Note:} Had we not used the extra minus sign attributed to the
fermionic nature of the fields, we would have got the CPT transform of
the Lagrangian to be minus of $\cal{L}(-x)$ and got CPT non-invariance
of the action! This minus sign is tightly connected with the
spin-statistics theorem which we discuss next.
\subsection{Relativistic Causality and the Spin-Statistics Theorem}
\label{SpinStatistics}
Now we come to the second feature of quantum fields - the commutation
relations. Here we get another surprise. The Pauli exclusion principle
follows from the requirement of relativistic causality!

The Fourier decomposition suggested field as a collection of harmonic
oscillators. A natural quantization procedure is to postulate
$[a_{\vec{k}}, a_{\vec{k}'}^{\dagger}] = \delta^{3}(\vec{k} -
\vec{k}')$. This has a potential problem for $\vec{k} = \vec{k}'$. The
usual way this is interpreted is to imagine the field being confined to
a large box satisfying periodic or Dirichlet boundary conditions. This
discretizes the momentum labels, $\vec{k} \to \vec{k}_n \sim 2\pi
\vec{n}/L$ and also replaces the delta function by Kronecker delta.
Another way is to choose a suitable, countable orthonormal set of
functions $\{\varphi_n(\vec{x})\}$ such that $\nabla^2 \varphi_n = -
\omega_n^2\varphi_n$. Completeness of the $\varphi_n$'s allows
translating $[a_m, a_n^{\dagger}]$ to $[\Phi(t,\vec{x}), \Pi(t,
\vec{y})] = \delta^3(x-y)$. We will implicitly assume such a procedure
and proceed formally using the delta function.

From the postulated commutation relations we have, for each $\vec{k}$,
\begin{eqnarray*}
N_{\vec{k}} := a_{\vec{k}}^{\dagger}a_{\vec{k}} & , & [N_{\vec{k}},
a_{\vec{k}}] = - a_{\vec{k}}\ , \ [N_{\vec{k}}, a_{\vec{k}}^{\dagger}] =
a_{\vec{k}}^{\dagger}\ , \\
N_{\vec{k}}|n_{\vec{k}} \rangle = n_{\vec{k}}|n_{\vec{k}} \rangle & , &
n_{\vec{k}} = 0, 1, \dots ; ~ ~ ~  \langle m_{\vec{k}}|n_{\vec{k}}
\rangle = \delta_{m_{\vec{k}}, n_{\vec{k}}} \\
|n_{\vec{k}} \rangle := \frac{(a_{\vec{k}}^{\dagger})^{n_{\vec{k}}}}
{\sqrt{(n_{\vec{k}})!}}| 0_{\vec{k}} \rangle & , &
a_{\vec{k}}|n_{\vec{k}} \rangle = \sqrt{n_{\vec{k}}}|n_{\vec{k}} - 1
\rangle ~ , ~ a^{\dagger}_{\vec{k}} |n_{\vec{k}} \rangle =
\sqrt{n_{\vec{k}} +1 }|n_{\vec{k}} + 1 \rangle 
\end{eqnarray*}
The state label $n_{\vec{k}}$ is interpreted as denoting the number of
quanta, created by $a^{\dagger}_{\vec{k}}$ and destroyed by
$a_{\vec{k}}$.

\underline{Note:} In this case of a scalar field, all quanta have the
same mass parameter, but different $k_{\vec{k}}$ label. For a vector
field, we have the polarization label $\lambda$ and for the spinor field
we have the spin projection label $\sigma$. All quanta of a given mass
$m$ and spin/helicity are identical.

Now we add a postulate of {\em relativistic causality} \cite{Weinberg}
also referred to as {\em micro-causality}, namely, {\em all pairs of fields
and their adjoints, commute or anti-commute for space-like separation.} 

This requirement comes about as follow. In a Lorentz invariant theory,
all observables would be Lorentz tensors of various ranks. Spinors being
double valued quantities, they themselves are {\em not observables}.
Spinorial fields must come in even powers (single valued) in an
observable. As per general principles of quantum theory, namely,
simultaneously measurable observables must commute and the notion of
relativistic causality that simultaneously measurable observables must
be space-like separated, suggest that for a space-time dependent
operators to be observables, they must commute for space-like
separation. Operators built out of spinorial fields being at least
quadratic in the spinorial fields, allows possible anti-commutation
rules for spinor fields\footnote{Weinberg bases his argument in favor
	of relativistic causality by invoking Lorentz invariance of the
scattering matrix.}. 

All our mode decompositions have annihilation operators as coefficients
of the positive frequency modes and the creation operators as
coefficients of the negative frequency modes. Let us denote these parts
as $\Psi_+(x)$ and $\Psi_-(x)$ respectively. Consider a general linear
combination of these fields: $\Psi(x) := \alpha\Psi_+(x) + \beta
\Psi_-{x}$ and let $\Psi^{\dagger}(x)$ denote its adjoint. Let all
observables be built out of $\Psi(x)$ and $\Psi^{\dagger}(x)$.  Let $[A,
B]_{\pm} := AB \pm BA$, the anti-commutator and commutator respectively.
The requirement of micro-causality is then stated as: $\boxed{[\Psi(x),
	\Psi(y)]_{\pm} = 0 = [\Psi(x), \Psi^{\dagger}(y)]_{\pm} ~
\mathrm{for} ~ (x-y)^2 > 0}$. This requirement is quite restrictive, as
seen below.

Let us begin with a real scalar field. We have, 
\[
\phi_+(x) := \int \frac{d^3k}{\sqrt{2\omega_{\vec{k}}(2\pi)^3}}
a_{\vec{k}} e^{ik\cdot x} ~ , ~ \phi_-(x) := \int
\frac{d^3k}{\sqrt{2\omega_{\vec{k}}(2\pi)^3}} a^{\dagger}_{\vec{k}}
e^{-ik\cdot x} ~ ~ = ~ ~ \phi_+^{\dagger}(x)\ .
\]
\[
	\therefore [\phi_+(x), \phi_-(y)]_{\pm} ~ = ~ \int
	\frac{d^3k}{\sqrt{2\omega_{\vec{k}}(2\pi)^3}}\int
	\frac{d^3k'}{\sqrt{2\omega_{\vec{k}'}(2\pi)^3}} e^{ik\cdot x -
	ik'\cdot y}[a_{\vec{k}}, a^{\dagger}_{\vec{k}'}]_{\pm}
\]

For the scalar field, we use the commutator given above. Let us assume
the same form for the anti-commutator i.e. we assume, $[a_{\vec{k}},
a^{\dagger}_{\vec{k}'}]_{\pm} = \delta^3(\vec{k} - \vec{k}')$.  This
gives,
\begin{eqnarray*}
[\phi_+(x), \phi_-(y)]_{\pm} & = & \int
\frac{d^3k}{2\omega_{\vec{k}}(2\pi)^3} e^{ik\cdot(x - y)} ~  =: ~
\Delta_+(x-y) ~ \mbox{is Poincare invariant. For} \\
\mbox{space-like separation} & , &  \mbox{choose $(x-y)^2 =
(\vec{x}-\vec{y})^2 =: r^2$ }, \\
\mbox{Thus~ ~} \Delta_+(x^2) & = & \frac{1}{(2\pi)^3} \int_0^{\infty}
dk\frac{k^2}{2\sqrt{k^2 + m^2}}\int_{-1}^{1}
dcos(\theta)\int_0^{2\pi}d\varphi e^{ikrcos\theta} \\ 
& = & \frac{1}{4\pi^2}\int_0^{\infty}dk\frac{k^2}{\sqrt{k^2 +
m^2}}\frac{sin(kr)}{kr}  ~ ~ ~ \mbox{put $k = m\alpha$,} \\
\Delta_+(x^2) & = & \frac{1}{4\pi^2}\frac{m}{r}\int_0^{\infty}d\alpha
\frac{\alpha}{\sqrt{\alpha^2 + 1}} sin(mr\alpha) \\
& = &  - \frac{1}{4\pi^2} \frac{1}{r}\frac{\partial~}{\partial r}
\int_0^{\infty}d\alpha \frac{\alpha}{\sqrt{\alpha^2 + 1}} cos(mr\alpha)
\\
\end{eqnarray*}

Thus, $\boxed{\Delta_+(x^2 > 0) ~ = ~ K_1(m\sqrt{x^2}) ~~
\mbox{(Modified Bessel function of the second kind)}}$, and it is {\em
non-zero for space-like separation}.

Define $\phi(x) := \alpha\phi_+(x) + \beta\phi_-(x) \ , \
\phi^{\dagger}(y) = \alpha^*\phi_-(y) + \beta^*\phi_+(y)$. It follows,
\begin{eqnarray*}
	\left[\phi(x), \phi(y)\right]_{\pm} & = & \alpha\beta
	\left[\phi_+(x),\phi_-(y)\right]_{\pm} +
	\beta\alpha\left[\phi_-(x), \phi_+(y)\right]_{\pm} \\
	& = & \alpha\beta\left( \Delta_+(x-y) \pm \Delta_+(y-x)\right)
	\\
	& = & \alpha\beta(1\pm 1)\Delta_+(x-y)  ~ ~ ~ ~ \mbox{since
	$\Delta_+$ is symmetric in $x,y$.} \\
	\left[\phi(x), \phi^{\dagger}(y)\right]_{\pm} & = & |\alpha|^2
	\left[\phi_+(x),\phi_-(y)\right]_{\pm} +
	|\beta|^2\left[\phi_-(x), \phi_+(y)\right]_{\pm} \\
	& = & (|\alpha|^2 \pm |\beta|^2)\Delta_+(x-y)
\end{eqnarray*}

To satisfy the requirement of causality, both the brackets must vanish
i.e. in both equations we {\em must} choose the {\em minus} sign i.e. a
{\em commutator} as well as the condition $|\alpha| = |\beta|$. This is
quite a strong restriction on both the sign as well as the coefficients
$\alpha, \beta$.

Let us consider the spinor field now. Let,
\begin{eqnarray*} 
\psi_+(x)  :=  \int\frac{d^3k} {\sqrt{2\omega_{\vec{k}}(2\pi)^3}}
\sum_{\sigma} (u(\vec{k},\sigma) e^{ik\cdot x}) b_{\vec{k},\sigma} ~ , ~
\mbox{\hspace{2.0cm} $ \longleftrightarrow $ \hspace{2.0cm} } \\
(\psi_+)^{\dagger}(x)  :=  \int\frac{d^3k}
{\sqrt{2\omega_{\vec{k}}(2\pi)^3}} \sum_{\sigma} (u^*(\vec{k},\sigma)
e^{-ik\cdot x}) b^{\dagger}_{\vec{k},\sigma} ~ ; \\
\psi_-(x)  :=  \int \frac{d^3k} {\sqrt{2\omega_{\vec{k}}(2\pi)^3}}
\sum_{\sigma} (v(\vec{k},\sigma) e^{-ik\cdot x})
d^{\dagger}_{\vec{k},\sigma} ~ , ~
\mbox{\hspace{2.0cm} $ \longleftrightarrow $ \hspace{2.0cm} } \\
(\psi_-)^{\dagger}(x)
:=  \int \frac{d^3k} {\sqrt{2\omega_{\vec{k}}(2\pi)^3}} \sum_{\sigma}
(v^*(\vec{k},\sigma) e^{+ik\cdot x}) d_{\vec{k},\sigma} ~ .
\end{eqnarray*}
The Dirac index is suppressed and the dagger refers to the adjoint of
the operators and not of the spinors.

Let $\psi(x) := \mu \psi_+(x) + \nu \psi_-(x)$ and $\psi^{\dagger}(y) :=
\mu^* \psi^{\dagger}_+(y) + \nu^*\psi^{\dagger}_-(y)$. Then,
\begin{eqnarray*}
\left[\psi_{\alpha}(x), \psi^{\dagger}_{\beta}(y)\right]_{\pm} & = &
\int \frac{d^3k}{2\omega_{\vec{k}}(2\pi)^3} \sum_{\sigma}\left\{
|\mu|^2\
u_{\alpha}(\vec{k},\sigma)u^*_{\beta}(\vec{k},\sigma)e^{ik\cdot(x-y)}
\right.  \\
& & \left. \mbox{\hspace{2.7cm}} \pm |\nu|^2
v_{\alpha}(\vec{k},\sigma)v^*_{\beta}(\vec{k},\sigma)e^{-ik\cdot(x-y)}
\right\}
\end{eqnarray*} 
The sums over $\sigma$ are given by,
\begin{eqnarray*}
\sum_{\sigma}u_{\alpha}(\vec{k},\sigma)u^*_{\beta}(\vec{k},\sigma) & = &
\sum_{\sigma}\left[u(\vec{k},\sigma)\bar{u}(\vec{k},\sigma) \gamma^0
\right]_{\alpha\beta} ~ = ~ \frac{-\dsl{k} + m}{2m}\gamma^0 ~ ~
\Rightarrow \\
\sum_{\sigma}u_{\alpha}(\vec{k},\sigma)u^*_{\beta}(\vec{k},\sigma)
e^{ik\cdot(x-y)} & = & (2m)^{-1}(i\dsl{\partial} +
m)\gamma^0e^{ik\cdot(x-y)} ~ ~ ~ ~ \mbox{and likewise,} \\
\sum_{\sigma}v_{\alpha}(\vec{k},\sigma)v^*_{\beta}(\vec{k},\sigma) & = &
\sum_{\sigma}\left[v(\vec{k},\sigma)\bar{v}(\vec{k},\sigma) \gamma^0
\right]_{\alpha\beta} ~ = ~ \frac{-\dsl{k} - m}{2m}\gamma^0 ~ ~
\Rightarrow \\
\sum_{\sigma}v_{\alpha}(\vec{k},\sigma)v^*_{\beta}(\vec{k},\sigma)
e^{-ik\cdot(x-y)} & = & -(2m)^{-1}(i\dsl{\partial} +
m)\gamma^0e^{-ik\cdot(x-y)}  ~ ~ ;
\end{eqnarray*}
and the integration over $\vec{k}$ just gives the $\Delta_+( (x-y)^2 )$
function computed above. Hence,
\[
\left[\psi_{\alpha}(x), \psi^{\dagger}_{\beta}(y)\right]_{\pm} ~ = ~
(|\mu|^2 \mp |\nu|^2)\frac{i\dsl{\partial} + m}{2m}\gamma^0 \Delta_+(
(x-y)^2 ) 
\]
Once again, the right hand side vanishes for space-like separation
provided the upper signs are chosen {\em and} $|\mu| = |\nu|$. Thus, the
for spinors we have to choose anti-commutation relations and the weights
of the positive and negative frequency solutions must be the same up to a
phase factor. We may just take $\mu = \nu = 1$ which gives back our
previous mode decomposition.

\underline{Note:} While writing the mode decomposition, we did not
bother about the normalizations of the coefficients, $b, d^*$ etc. Now,
in choosing the brackets, $[b_{\vec{k},\sigma}, b^{\dagger}_{\vec{k}',
\sigma'}]_+ = \delta_{\sigma,\sigma'} \delta^3(\vec{k} - \vec{k}')$ and
ditto for the $d$'s, we have fixed these normalizations. We could still
have arbitrary multiple in each term. The requirement of causality
restricts this freedom to the above.

The two examples considered above, generalize to other representations
as well and what we have is the celebrated {\em spin-statistics
theorem:}

{\em A relativistic quantum field theory satisfies the requirement of
causality provided integer spin/helicity field quantum conditions use
commutators (bosons) while the half integer spin/helicity field quantum
conditions use anti-commutators (fermions).}

\underline{Note:} While discussing quantum statistics in statistical
mechanics, we invoke the additional attribute of {\em
indistinguishability} among identical particles and require the
Bose-Einstein/Fermi-Dirac statistics. Identical particles are defined by
having identical intrinsic attributes such as mass, spin, color, flavor
etc. The (in)distinguishability arises from a spatial localization and
subsequent ability to tag them through, say, a scattering process. This
is lost when their wave functions overlap, the `quantum' becomes
operative and quantum statistics becomes essential. In the above
discussion, no such additional property emerged.  We already have the
`quantum' operative so indistinguishability is also operative. It is the
demand of relativistic causality that forced the spin-statistics
correlation.

With the anti-commutators around, $[b_{\vec{k},\sigma} ,
b^{\dagger}_{\vec{k}',\sigma'} ]_+ = \delta_{\sigma,\sigma'}
\delta^3(\vec{k} - \vec{k}'), \ [b_{\vec{k},\sigma},
b_{\vec{k}',\sigma'}]_+ = 0$, we define the number operators for each
$(\vec{k},\sigma)$ as before: $N := b^{\dagger}b$ It follows that $[N,
b]_- = -b\ , \ [N, b^{\dagger}]_- = +b^{\dagger}$. Eigenvalues of $N$
continue to be integers but thanks to $(b^{\dagger})^2 = 0$, they take
only two values, $0, 1$. The corresponding eigenstates are: $|0 \rangle$
and $|1 \rangle$. This is the incorporation of the Pauli exclusion
principle. 

Consider states of two modes (scalar for simplicity), say $\vec{k}_1,
\vec{k}_2$. A general basis state would be $|\vec{k}_1,\dots\vec{k}_1,
\vec{k}_2,\dots\vec{k}_2 \rangle \sim (a^{\dagger}_{\vec{k}_1})^m
(a^{\dagger}_{\vec{k}_2})^n|0,0 \rangle$. 

If the $a$'s anti-commute, then $m,n \le 1$ and we have 4 possible
states: $|0,0 \rangle, |\vec{k}_1, 0 \rangle, |0, \vec{k}_2 \rangle,
|\vec{k}_1, \vec{k}_2 \rangle$. Thanks to anti-commutation, $|\vec{k}_2,
\vec{k}_1 \rangle = - |\vec{k}_1, \vec{k}_2 \rangle$. For bosonic
operators, there is no minus sign under exchange. Generalizing from
this, it is clear that (anti-)commutation relations among the creation
operators automatically ensure complete (anti-)symmetry under
permutations of the labels.

To understand the extra minus sign in the third line of eq.
(\ref{CPT-Bilinears}), momentarily write the $\bar{\Psi}$ as
$\bar{\Psi}_1$ and $\Psi$ as $\Psi_2$. Then in the third line of the
equation, we see an explicit exchange of $\Psi_1$ and $\Psi_2$. The
fermionic commutation relation then gives that extra minus sign. As
noted before, this is also important for the CPT theorem.

%
{\em To summarize:} Quantum fields are operators {\em tagged by
space-time points} and obeying certain commutation relations. We may
think of them as a collection of creation-annihilation operators for
each mode (solution of the equations of motion). A general state of a
quantum field is a linear combination of basis states which are
multi-quanta states. Each quantum with momentum label $\vec{k}$ and a
spin/helicity label $\sigma$ carries energy-momentum-angular momentum as
given by the Poincare charges. As per quantum field theory, all
processes involve exchanges of quanta among different quantum fields.
\newpage
\section{States of Free Quantum Fields: Particles, 
coherence and coherent states} \label{QuantumFieldStates}

While we argued that particle dynamics with position-momentum is not
compatible with relativity, we observe particles in a variety of
experiments eg as tracks in bubble chambers, photographic emulsions,
stacks of particle counters etc and also use the particle view in
designing accelerator beams. So the more sophisticated relativistic
framework should also have a suitable description of ``particles''.
Since quantum fields are operators, the description must be in terms of
{\em states of a quantum field}.  In the section
(\ref{FieldDecompositions}), we saw that from a dynamical view point, a
(free) field is an infinite collection of non-interacting harmonic
oscillators.  Furthermore, its quantization follows from the
quantization of oscillators and A quantum field emerges as a linear
combination of creation-annihilation operators of its modes. Its states
are built by the action of the creation operators on vacuum state of
each mode. We have to identify ``particles'' in this huge space of
states of a quantum field. 
\subsection{Particle/anti-particle wave packets} \label{WavePackets}
Let $|0\rangle$ denote the unique vacuum state annihilated by {\em all}
annihilation operators $a_{\vec{k}}$. We can generate ``1-particle
states'' by taking a linear combination of the form $\sum_{\vec{k}}
\alpha(\vec{k}) a^{\dagger}_{\vec{k}} |0\rangle$, ``2-particle states''
by taking a linear combination of the form $\sum_{\vec{k}, \vec{l}}
\alpha(\vec{k}, \vec{l})a^{\dagger}_{\vec{k}} a^{\dagger}_{\vec{l}}
|0\rangle$ and so on.  The totality of all such states forms the state
space of a free quantum field. For the $n-particle$ states, the
(anti-)commutators of the creation operators automatically take care of
(anti-)symmetrization of the states. Notice that 1-particle state does
{\em not} mean a {\em single} $a^{\dagger}_{\vec{k}}|0\rangle$, but a
linear combination of several creation operators. These allow us to form
{\em wave packets}. 

As an explicit example, let us construct a wave packet of an
anti-fermion with some {\em momentum and spin distribution}. We have the
Dirac field,
\begin{eqnarray*}
	\Psi(x) & = & \int [d^3k]\ \left( b_{\vec{k},\sigma}
	u(\vec{k},\sigma) e^{ik\cdot x} + d^{\dagger}_{\vec{k},\sigma}
v(\vec{k},\sigma) e^{-ik\cdot x}\right)  ~ ~ , \mbox{where,} \\
	\big[d^3k\big] & := &
	\frac{d^3k}{\sqrt{2\omega_{\vec{k}}(2\pi)^3}} ~ ~ , ~ ~
	\mbox{and} ~ ~ (\dsl{k} + m)u = 0 = (-\dsl{k} + m)v .
\end{eqnarray*}
Acting on the vacuum, it will create a {\em 1-anti-fermion} state, the
linear combination from the $d^{\dagger}_{\vec{k},\sigma}$ creation
operators.  Let $f(\vec{k'},\sigma')$ be a suitable complex valued
function. Define
\begin{eqnarray}
	\langle f| & := & \int d^3k' \sum_{\sigma'} f(\vec{k}',\sigma')\
	\langle 0| d_{\vec{k}',\sigma'} ~ ~ , ~ ~ \int d^3k\sum_{\sigma}
	|f(\vec{k}, \sigma)|^2 = 1 \ . \\
	(\chi)_c(x) & := & \langle f|\Psi(x)|0\rangle \nonumber \\
	& = & \int [d^3k]\sum_{\sigma}f(\vec{k},
	\sigma)v(\vec{k},\sigma) e^{-ik\cdot x}  ~ ~ \because ~ ~
	\langle 0|d_{\vec{k}',\sigma'}
	d^{\dagger}_{\vec{k},\sigma}|0\rangle =
	\delta_{\sigma,\sigma'}\delta^3(\vec{k}'-\vec{k})
	\label{AntiWaveFn}
\end{eqnarray}
The so constructed $(\chi)_c(x)$ is a spinor-valued function of $x$
which satisfies the Dirac equation. The $v(\vec{k},\sigma)$ spinor
signifies that it is an anti-fermion wave packet and the suffix $c$ on
$\chi$ is a reminder. How do we get a fermion wave packet?  Well, in
place of $\Psi(x)$ operator, we use its charge conjugate,
$C\bar{\Psi}^T(x)$ (see \ref{CPTtransforms}), and in place of the
$d^{\dagger}_{\vec{k},\sigma}$ in $\langle f|$ we use
$b_{\vec{k},\sigma}$. The corresponding fermion wave packet takes the
form,
\begin{eqnarray}
	\langle g| & := & \int d^3k' \sum_{\sigma'} g(\vec{k}',\sigma')\
	\langle 0| b_{\vec{k}',\sigma'} ~ ~ , ~ ~ \int d^3k\sum_{\sigma}
	|g(\vec{k}, \sigma)|^2 = 1 \ . \\
	\chi(x) & := & \langle g|C\bar{\Psi}^T(x)|0\rangle ~ ~
	\mbox{and} ~ ~ \langle 0|b_{\vec{k}',\sigma'}
	b^{\dagger}_{\vec{k},\sigma}|0\rangle =
	\delta_{\sigma,\sigma'}\delta^3(\vec{k}'-\vec{k}) ~ ~
	\Rightarrow \nonumber \\
	& = & \int [d^3k]\sum_{\sigma}g(\vec{k},
	\sigma)u^c(\vec{k},\sigma) e^{-ik\cdot x}  \label{WaveFn}
\end{eqnarray}
where, $u^c(\vec{k},\sigma) = (i\gamma^2 u^*(\vec{k},\sigma))$ is the
charge conjugate of the $u(\vec{k},\sigma)$ spinor. 

The generalization for wave packets of other fields should be obvious.
All these wave packets are positive or negative frequency solutions of
the corresponding field equations. We already have discussed the inner
products on these spaces and we may view these as the usual probability
amplitudes. Of course we do not have the analogue of position operator
unless we restrict to a non-relativistic approximation/limit. Suffice it
to say that one can use Dirac equation to analyze beams of relativistic
fermions, as is done for instance in \cite{JagannathanKhan}.
\subsection{Correlation Functions and Coherence of
states}\label{CorrCoh}
There are of course more general states which are not labeled by the
number of quanta. Even for a single oscillator, states with a given
occupation number is only a class of states. We can have finite linear
combination or infinite linear combinations of these numbers states, can
have squeezed states, coherent states etc. They are distinguished by
their properties and utility. A some what more general characterization
of quantum states is in terms of {\em correlation functions}. Let us see
an example of this.
 
Consider a real scalar quantum field. We have its mode decomposition and
had also defined the $\phi_{\pm}(x)$ fields consisting of the
positive/negative frequency parts. Consider the measurement of the
operators $\phi_-(x)\phi_+(y)$.  Its average value is {\em some state}
which is kept implicit, is denoted as 
\begin{eqnarray*}
	G(x,x') & := & \langle\phi_-(x)\phi_+(y) \rangle ~ ~
	\mbox{\hspace{2.0cm}(a correlation function)} \\
	& = & \int \frac{d^3k}{\sqrt{2\omega_{\vec{k}}(2\pi)^3}} \int
	\frac{d^3k'}{\sqrt{2\omega_{\vec{k}'}(2\pi)^3}}
	e^{-i\vec{k}\cdot x + i\vec{k}'\cdot x'} \langle
	a^{\dagger}_{\vec{k}}a_{\vec{k}'} \rangle
\end{eqnarray*}
where, $\langle a^{\dagger}_{\vec{k}}a_{\vec{k}'} \rangle := Tr(\rho
a^{\dagger}_{\vec{k}}a_{\vec{k}'})$ and thus the measured averaged value
depends on the density matrix or ``state'', $\rho$. The usual
Schrodinger equation for the ket vectors becomes the equation $id_t\rho
= [H, \rho]$ whose solution is: $\rho(t) = e^{-itH}\rho_0e^{itH}$ for a
time independent Hamiltonian. These are the only Hamiltonians considered
below. For example, the Hamiltonian of our scalar field is given by the
quantum version of (\ref{FieldEnergy}). 

A time-independent density matrix $\rho$ is said to represent a {\em
stationary} state. Thus for a stationary state, $[\rho, H] = 0$ which
implies, $\rho = e^{-itH}\rho e^{itH} \ \forall\ t$.
\begin{eqnarray*}
Tr\left[\rho a^{\dagger}_{\vec{k}}a_{\vec{k}'}\right] & = &
Tr\left[e^{-itH}\rho e^{itH}a^{\dagger}_{\vec{k}}a_{\vec{k}'}\right] ~ =
~ Tr\left[\rho e^{itH}a^{\dagger}_{\vec{k}}a_{\vec{k}'}e^{-itH}\right]
\\
& = & Tr\left[\rho (e^{itH}a^{\dagger}_{\vec{k}}e^{-itH})\
(e^{itH}a_{\vec{k}'}e^{-itH})\right] \\
& = & Tr\left[\rho a^{\dagger}_{\vec{k}} a_{\vec{k}'} \right]
e^{it(\omega_{\vec{k}} - \omega_{\vec{k}'})} ~ ~ ~ ~ ~ \because ~ H \sim
\int[d^3k''] \omega_{\vec{k}''}N_{\vec{k}''}
\end{eqnarray*}

The left hand side is time independent, so $Tr\left[\rho
a^{\dagger}_{\vec{k}} a_{\vec{k}'} \right] = 0$ unless $\omega_{\vec{k}}
= \omega_{\vec{k}'}$ i.e. unless $|\vec{k}| = |\vec{k}'|$. For the same
frequency, the trace need not vanish and we have a matrix in the labels
$\vec{k}, \vec{k}'$. This matrix is hermitian ($\rho$ is hermitian) and
can be diagonalised. Hence, without loss of generality,
$\boxed{\mbox{for a stationary state,} ~ Tr\left[\rho
a^{\dagger}_{\vec{k}} a_{\vec{k}'} \right] =: \overline{n_{\vec{k}}} \
\delta_{\vec{k}, \vec{k}'}}\ $ , where $\overline{n_{\vec{k}}}$ {\em
denotes the average number of quanta in the $\vec{k}$ mode}. This leads
to,
\[
G(x, x') ~ = ~ \int \frac{d^3k}{2\omega_{\vec{k}} (2\pi)^3} \
\overline{n_{\vec{k}}}\  e^{i\omega_{\vec{k}}(t - t') -
i\vec{k}\cdot(\vec{x} - \vec{x}')}\ \mbox{\hspace{1.0cm} (In a
stationary state)}.
\]
Clearly, {\em a stationary state leads to {\em translationl invariance}
of the above correlation function and this correlation function contains
the information of the average number of quanta in each mode.}

Assume now that the average number of quanta is vanishingly small
outside a small neighbourhood of some direction $\hat{k}_0$
(collimation) and also outside a small neighbourhood of
$\omega_{\vec{k}_0}$ (monochromaticity).  We may then approximate the
integral as ,
\[
	G(x, x') ~ \approx ~ \frac{N_0}{2\omega_{\vec{k}_0}(2\pi)^3} \
	e^{i\omega_{\vec{k}_0}(t - t') - i\vec{k}_0\cdot(\vec{x} -
	\vec{x}')}\   ~ ~ , ~ ~ N_0 \approx \int_{ngbd(\vec{k}_0)}
	\frac{d^3k}{2\omega_{\vec{k}} (2\pi)^3} \ \overline{n_{\vec{k}}}
\]
It is apparent that the correlation function has a factorised form:
$G(x, x') \approx \varphi^*(x)\varphi(x')~ , ~ \varphi(x) :=
\sqrt{\Case{N_0} {2\omega_{\vec{k}_0} (2\pi)^3} }e^{ik_0\cdot x}$ which
is a plane wave. A state $\rho$ with such a form of the correlation
function is said to have a {\em first order coherence}\footnote{$k^{th}$
	order coherence is defined in terms of the factorization
	property of $G^{n}(x_1,\ldots, x_n, y_1,\ldots, y_n) := \langle
	\phi_-(x_1),\ldots, \phi_+(y_n)\rangle = E^*(x_1)\ldots
	E^*(x_n)\ E(y_1)\ldots E(y_n)$, where $E(x)$ is independent of
	$n$ for all $1 \le n \le k$ \cite{Glauber}. The coherent states,
eigenstates of the annihilation operators, familiar from the harmonic
oscillator have infinite order coherence.}.  

Notice that the factorised form results when the integral can be {\em
approximated}. This in turn is valid for $|\vec{x} - \vec{x}'| \approx
(2\pi)/|\vec{k}_0|$ and $|t - t'| \approx (2\pi)/\omega_{\vec{k}_0}$.
Furthermore, the imperfections in the monochromaticity and collimation
affect the approximation and also limit the validity of the first order
coherence. This introduces the ideas of {\em coherence time} and {\em
coherence length} as time and space intervals over which the
factorization property or coherence property holds. Thus a stationary
state with first order coherence, with finite coherence length and time,
may be interpreted as a ``beam'' along the direction $\hat{k}_0$ with
intensity proportional to $N_0\ \omega_{\vec{k}_0}$. 

Let us continue further and assume perfect collimation along say the
x-axis. The beam may have a distribution of frequencies though. The
phase\footnote{The formula is given for a massless field. More
	generally, we will have $\tau(\omega) =
t-t'-\sqrt{1-m^2/\omega^2}(x-x')$. For very high frequency/energy, the
mass can be neglected.} then becomes $i\omega_{\vec{k}}(t - t' -x + x')$
and $G(x,x') =: G(\tau),\ \tau := (t - t' -x + x')$. Since we have a
single direction, we can denote the dependence on $\vec{k}$ by the
frequency $\omega$.  Suppose further that $\overline{n_{\vec{k}}} =:
\overline{n_{\omega}}$ has a Gaussian dependence on $\omega$:
$\overline{n_{\omega}} = A\ exp{[ - \Case{(\omega -
\omega_0)^2}{2\sigma^2} ]}$. Now the integral can be done exactly and
get, 
\[
G(\tau) ~ = ~ A \int_0^{\infty}\frac{d\omega}{\omega(2\pi)}e^{[ -
\frac{(\omega - \omega_0)^2} {2\sigma^2} ]} e^{i\omega\tau} ~ = ~
G(0)e^{i\omega_0\tau}e^{-\frac{\sigma^2\tau^2}{2}} \ .
\]
Thus, a highly collimated beam with a frequency range of $\sigma$ around
$\omega_0$, has the correlation function $G(x,x')$ decaying with $\tau =
(t - t' - x + x') \sim \sigma^{-1}$. For example, if we have say a
laser beam with $\omega_0 \sim 10^{14}$ Hz and $\sigma \sim 10^6$ Hz,
then $\tau \sim 10^{-6}$ seconds or about $10^2$ meters. Since the beam
is presumed to be a stationary state, the $\tau$ does {\em not} vary
with the time, $t$ (or $t'$). However, beyond the coherence length, the
beam will show frequency dispersion.  Note that first order coherence
suffices for the notions of coherence length and time.

{\em Remark:} In this example, we have stipulated the properties of
stationarity and 1$^{st}$ order coherence to restrict the implicit state
of the quantum field. These stipulations do not single out a particular
state, but permits a subset of states $\rho$. All such states suffice to
describe a beam with stable properties over the coherence interval.
\subsection{Coherent States}\label{CohStates}
There is a very important class of states of quantum fields, especially
for electromagnetic field, namely the class of coherent states. These
are familiar from the quantum harmonic oscillator. Let us recall their
construction briefly.
\subsubsection{An aside: Harmonic Oscillator States}
With the usual creation-annihilation operators, $[a, a^{\dagger}] = 1$
we take the oscillator variables as: 
\begin{eqnarray*}
	q(t) & := & a e^{-i\omega t} + a^{\dagger} e^{i\omega t} := q_+(t) +
	q_-(t) ~ , \\
	p(t) & := & -\frac{i}{2}a e^{-i\omega t} +
	\frac{i}{2} a^{\dagger} e^{-\omega t}  ~ \leftrightarrow ~ [q(t),
	p(t)] - i\ .
\end{eqnarray*}
Define:

(i) {\em Displacement operators:} For each $\alpha \in \mathbb{C},$
\begin{eqnarray*}
D(\alpha) & := & e^{\alpha a^{\dagger} - \alpha^* a} ~ = ~ e^{\alpha
a^{\dagger}}\ e^{-\alpha^* a}\ e^{- \Case{|\alpha|^2}{2}} ~ ~ ~ ~
\mbox{Using the BCH formula,} \\
\therefore D^{\dagger}(\alpha) & = & e^{- \Case{|\alpha|^2}{2}}
e^{-\alpha a^{\dagger}}\ e^{\alpha^* a} ~ ~ = ~ ~ D(-\alpha) ~ ~ \mbox{~
~ and~ ~ } ~ ~ D^{\dagger}(\alpha)D(\alpha) = \mathbb{1} \\
D^{\dagger}(\alpha)\ a\ D(\alpha) & = & a + \alpha\mathbb{1}
\mbox{\hspace{0.5cm} ~ , ~ \hspace{0.5cm}} D^{\dagger}(\alpha)\
a^{\dagger}\ D(\alpha) ~ = ~ a^{\dagger} + \alpha^*\mathbb{1} ~ ,
\mbox{~ and ~}\\
D(\alpha + \beta) & = & D(\alpha)\ D(\beta)\ e^{ -
\Case{i}{2}(\alpha\beta^* - \alpha^*\beta)} \ .
\end{eqnarray*}

(ii) {\em Squeezing operators:} for each complex number $\epsilon := r
e^{i\phi}$, 
\begin{eqnarray*}
S(\epsilon) ~ := ~ e^{\Case{1}{2}(\epsilon^*a^2 -
\epsilon(a^{\dagger})^2)} ~  ,  ~ S^{\dagger}(\epsilon) ~ = ~
S(-\epsilon) &  ~ , ~  & S^{\dagger}(\epsilon) S(\epsilon) = \mathbb{1}
\\
S^{\dagger}(\epsilon)\ a\ S(\epsilon) ~ = ~ a\ \mathrm{ch}(r) -
a^{\dagger} e^{-2i\phi}\mathrm{sh}(r) &  ~ , ~ &  S^{\dagger}(\epsilon)\
a^{\dagger}\ S(\epsilon) ~ = ~ a^{\dagger} \mathrm{ch}(r) -  a\
e^{2i\phi}\mathrm{sh}(r) 
\end{eqnarray*}

These operators define a two parameter family of states, {\em squeezed
coherent states}, as
\[
|\alpha, \epsilon\rangle ~ := ~ D(\alpha) S(\epsilon) | 0\rangle ~ ~ ; ~
~ \epsilon = 0 \leftrightarrow\ \mbox{(coherent states)},~ \alpha = 0
\leftrightarrow\ \mbox{(squeezed states)}.
\]
For this family of states, it is straightforward to see,
\begin{eqnarray*}
\langle\alpha,\epsilon| a^2 |\alpha,\epsilon\rangle & = & -
e^{-2i\phi}\mathrm{ch}(r)\mathrm{sh}(r) + \alpha^2 ~ ~ , ~
~\boxed{\langle\alpha,\epsilon|a|\alpha,\epsilon\rangle ~ = ~ \alpha ~ ,
~ \forall ~ \epsilon} \\
\langle\alpha,\epsilon| (a^{\dagger})^2 |\alpha,\epsilon\rangle & = & -
e^{2i\phi}\mathrm{ch}(r)\mathrm{sh}(r) + (\alpha^*)^2 ~ ~ , ~
~\langle\alpha,\epsilon|a^{\dagger}|\alpha,\epsilon\rangle ~ = ~
\alpha^* \\
\langle\alpha,\epsilon|N|\alpha,\epsilon\rangle & = & |\alpha|^2 +
\mathrm{sh}^2(r) \\
a|\alpha,\epsilon\rangle & = & -
\mathrm{sh}(r)e^{-2i\phi}D(\alpha)S(\epsilon)a^{\dagger}|0\rangle +
\alpha|\alpha,\epsilon\rangle \ .
\end{eqnarray*}
The last equation shows that only for the squeezing parameter, $r = 0$,
$|\alpha,0\rangle$ is an eigenstate of the annihilation operator which
is the usual definition of coherent states.

The Heisenberg uncertainty relation takes the form, 
\begin{eqnarray*}
	A := 1 + 2\langle N\rangle - 2|\langle a\rangle|^2 & = & 1 + 2
	\mathrm{sh}^2(r) \\
	B := (\Delta a)^2 e^{-2i\omega t} + (\Delta
	a^{\dagger})^2e^{2i\omega t} & = & -2
	\mathrm{ch}(r)\mathrm{sh}(r)\mathrm{cos}(\omega t + \phi) \\
	(\Delta q)^2(t) = A + B & ~ , ~ & 4 (\Delta p)^2(t) = A - B 
\end{eqnarray*}
Clearly, for the coherent states $r = 0 \Rightarrow A = 1, B = 0$ and
$(\Delta q)^2(\Delta p)^2 = \Case{1}{4}$ which saturates the uncertainty
bound. The uncertainties are also time independent.  Some of the basic
properties of the coherent states are: 
\begin{eqnarray}\label{CoherentStatesProps}
|\alpha\rangle & = & |\alpha, 0\rangle = D(\alpha)|0\rangle ~ ~
\Rightarrow ~ ~ \langle\alpha|\alpha\rangle = 1 \\
|\alpha\rangle & = & e^{-\Case{|\alpha|^2}{2}}\sum_{n = 0}^{\infty}
\frac{\alpha^n}{\sqrt{n!}} |n\rangle \\
\langle\beta|\alpha\rangle & = & e^{- \Case{|\alpha|^2 + |\beta|^2}{2} +
\alpha\beta^* } \\
\mathbb{1} & = & \frac{1}{\pi}\int d^2\alpha\
|\alpha\rangle\langle\alpha| ~ = ~ \sum_{n\ge 0}| n\rangle\langle n | \
. 
\end{eqnarray}
\subsubsection{Correlation functions}
This subsection is based on \cite{WallsMilburn,Glauber}.

We begin with the definition:
\begin{eqnarray}\label{GnDefn}
G^n(t_1, \dots t_n; t_{n+1} \dots t_{2n}) & := & \langle q_-(t_1)\dots
q_-(t_n)\ q_+(t_{n+1}) \dots q_+(t_{2n})\rangle \\
& = & \left[e^{i\omega(t_1 + \dots t_n - t_{n+1} \dots - t_{2n})}\right]
Tr\left[\rho(a^{\dagger})^n a^n\right]
\end{eqnarray}
Since the $\hat{q}_-$ and $\hat{q}_+$ commute among themselves, the
ordering of $(t_1, \dots t_n)$ and $(t_{n+1} \dots t_{2n})$ is
unimportant. Note that for (and only for) $\rho = |0\rangle\langle0|$,
the correlation functions vanish identically. In the following, this
density matrix is {\em excluded}.

We invoke the general result: $\boxed{ \forall ~ \hat{A}\ ,\ \mbox{and}\
\forall\ \rho\ ,\ Tr[\rho A^{\dagger} A] \ \ge\  0 \ .} $

Taking $A := \sum_i \alpha_i \hat{q}_+(t_i)$, 
\[
	\langle A^{\dagger} A\rangle = \sum_{i,j}\alpha^*_i G^1(t_i,
	t_j)\alpha_j \ \ge 0 \ \forall \ \alpha_i's \ .
\]
Hence, the matrix $G^1(t_i, t_j)$ is non-negative. So its determinant is
non-negative. Thus,
\[
	G^1(t_i, t_i)G^1(t_j, t_j) - G^1(t_i, t_j)G^1(t_j, t_i) \ge 0 \
	.
\]
From the definition of $G^1$, it follows that $G^1(t_j, t_i) = G^1(t_i,
t_j)^*$. Therefore $G^1(t, t)$ is real and in fact non-negative.  The
non-negative determinant condition becomes $\boxed{ |G^1(t_i, t_j)|^2
\le G^1(t_i, t_i) G^1(t_j, t_j) . } $ This leads to the definition of
the {\em normalized correlation functions}, 
\begin{equation}
g^n(t_1, \dots, t_{2n}) ~ := ~ \frac{G^n(t_1, \dots, t_{2n})}
{\sqrt{G^1(t_1, t_1) \dots G^1(t_{2n}, t_{2n})}} ~  , 
\end{equation}
The determinant condition may be expressed as, $ |g^1(t_i, t_j)| \le 1 \
$ . Using the explicit definition, 
\[
	g^1(t_i, t_j)  = \frac{e^{i \omega(t_i - t_j)}Tr(\rho
	a^{\dagger} a)}{\sqrt{(Tr(\rho a^{\dagger} a)\ Tr(\rho
		a^{\dagger} a)}} ~ = ~ e^{i \omega(t_i - t_j)} ~ ~
		\therefore ~ ~ |g^1(t_i, t_j)| = 1. 
\]

${\bf k^{th} ~ order ~ Coherence:}$

A state $\rho$ is said to have {\em $k^{th}$-order coherence}, if
$|g^n(t_1,\dots,t_{2n})| = 1\ \forall\ 1 \le n \le k$ .

\underline{Remark:} This is a property of a state, defined with respect
to a specific kind of correlation function. Thus we could have different
notions of coherence, if we adopt different correlation functions. The
choice of this specific class of correlation functions has its roots in
the measurements in optics which involves electromagnetic field. The
correlation functions are also defined for classical fields (with the
averages defined as ensemble averages over repeated observations) and
have very similar properties as the quantum counterparts. The
distinctions begin to show up when the intensities of the light beams
is decreased to almost ``single photon'' level. 

\underline{Remark:} In light of the result in the preceding line, {\em
every} state has 1st order coherence.  It follows immediately that for
$k^{th}$ order coherence, $|G^n(t_1, \dots, t_{2n})|^2 = \prod_i
G^1(t_i, t_i)$.

\underline{Result:} {\em A state has a} $k^{th}$ {\em order coherence iff
there exist a function $E(t)$, independent of $n$, such that}

~\hfill $G^n(t_1,\dots,t_{2n}) = E^*(t_1)\dots E^*(t_n)E(t_{n+1})\dots
E(t_{2n}) \ \forall \ 1 \le n \le k $. \hfill~ 

The proof is easy. First prove it for $G^1(t_1, t_2)$. From this the
result for $G^n$ follows. This factorization property is another
characterization of $k^{th}$ order coherence.

Among the pure states, we have states which are finite linear
combinations of number states and those with infinite linear
combinations. Let $|\psi\rangle$ be a state with finite linear
combinations of the number eigenstates. Let $K$ be the maximum
eigenvalue of $N$ appearing in the linear combination. Then
$a^n|\psi\rangle = 0\ \forall\ n > K$. Since, for pure states, $|g^n| =
\Case{\langle\psi|(a^{\dagger})^n
a^n|\psi\rangle}{(\langle\psi|a^{\dagger} a|\psi\rangle)^n}$, it follows
that $|g^n| = 0\ \forall \ n > K$. {\em Hence, finite linear
	combinations are necessarily partially coherent i.e. have
	maximum order of coherence to be $K$. A fully coherent state
must have infinite linear combinations.}

But not all infinite linear combinations have full coherence! Here are
two example.

\underline{Coherent states $|\alpha,0\rangle$:} We have $|g^n| =
\Case{(\alpha^*)^n \alpha^n}{(|\alpha|^2)^n} = 1 \ \forall\ n \ge 1$ and
coherent states are fully coherent. These are the {\em only} non-vacuum
states which have infinite order coherence. This may be checked by
considering an arbitrary, infinitesimal perturbation of a coherent
state, $|\Psi\rangle := (1 - \epsilon/2)|\alpha\rangle +
\epsilon/2|\phi\rangle, |\phi\rangle \neq |0\rangle$ and computing
$\langle\Psi|(a^{\dagger})^n a^n |\Psi\rangle - (\langle\Psi|
a^{\dagger}a|\Psi\rangle)^n$ to first order in $\epsilon$.
  
\underline{Squeezed vacuum $|0, \epsilon\rangle$:} Now,
\[
|g^2| = \frac{\langle 0|S^{\dagger}(\epsilon)(a^{\dagger})^2 a^2
S(\epsilon)|0\rangle}{\sqrt{(\langle 0|S^{\dagger}(\epsilon)a^{\dagger}
a S(\epsilon) |0\rangle)^n}} 
\]
This evaluates to,
\begin{eqnarray*}
	\mbox{Numerator:} ~ ~ & = &\langle 0|S^{\dagger}a^{\dagger}S\
	S^{\dagger}a^{\dagger}S\ S^{\dagger}aS\ S^{\dagger}aS|0\rangle
	\\
& = &\langle 0|(\mathrm{ch}(r)a^{\dagger} - \mathrm{sh}(r)e^{2i\phi}a)
(\mathrm{ch}(r)a^{\dagger} - \mathrm{sh}(r)e^{2i\phi}a) \\
& & \mbox{\hspace{1.0cm}}(\mathrm{ch}(r)a -
\mathrm{sh}(r)e^{-2i\phi}a^{\dagger}) (\mathrm{ch}(r)a -
\mathrm{sh}(r)e^{-2i\phi}a^{\dagger})|0\rangle \\
& = & (\mathrm{sh}(r))^2\langle 0| a(\mathrm{ch}(r)a^{\dagger} -
\mathrm{sh}(r)e^{2i\phi}a)(\mathrm{ch}(r)a -
\mathrm{sh}(r)e^{-2i\phi}a^{\dagger})a^{\dagger}|0\rangle \\
& = & (\mathrm{sh}(r))^2\left[ (\mathrm{ch}(r))^2 + 2
(\mathrm{sh}(r))^2\right] \\
\mbox{Denominator:} ~ ~ & = & \langle 0|S^{\dagger} N S|0\rangle ~ = ~
(\mathrm{sh}(r))^2 ~ ~ \Longrightarrow ~ ~
\boxed{|g^2| ~ = ~ 1 + 3 (\mathrm{sh}(r))^2 ~  > 1 \ \forall\
|0,\epsilon\rangle .}
\end{eqnarray*}

Thus while full coherence requires infinite linear combinations, they do
not guarantee even a second order coherence!

\underline{Points to note as a summary:} 

\begin{itemize}
\item Quantum framework has vastly more number of states compared to the
	classical framework. The observed measurements are always
	averages with uncertainties. There are many states which have
	the same average of a given set of observable. While
	uncertainties serve to distinguish among these, alternative
	attribute are provided by the correlation functions and notion
	of coherence. There is at least one class of states, the
	coherent states, which have full coherence, have time
	independent uncertainties (for oscillator dynamics) and have the
	smallest uncertainties compatible with the Heisenberg relation.
	This class of states is large enough to provide (over-complete)
	basis for the Hilbert space.
\item We used a specific kind of correlation function and defined
	coherence of states w.r.t. these. These correlators have equal
	number of creation and annihilation operators which are ordered
	with annihilation operators to the right (the so-called {\em
	normal order}). This is prejudiced on typical detectors working
	by {\em absorbing} quanta. For opposite types of detectors,
	anti-normal order would be appropriate. The correlators with
	unequal number of $a, a^{\dagger}$ operators, {\em are used} in
	measurement of phase information.  

	{\em All these correlation functions are premised on processes
	being understood as emission and absorption of quanta.}
\item Generalizing, the coherent states of fields are essentially tensor
	product of single mode coherent states.  We may thus denote a
	coherent state of a scalar field as $|\{\alpha_{\vec{k}}\}
	\rangle$ with $\alpha_{\vec{k}} :=\langle\alpha_{\vec{k}}|
	a_{\vec{k}} |\alpha_{\vec{k}}\rangle$.  Then, 
	\[
	\varphi(x) :=\langle\{\alpha_{\vec{k}}\} |\hat{\phi(x)}|
	\{\alpha_{\vec{k}}\}\rangle ~ = ~ \int\left[d^3k\right]\left(
	\alpha_{\vec{k}}e^{ik\cdot x} + \alpha^*_{\vec{k}}e^{-ik\cdot x}
\right) .
	\]
	Thus every coherent state of the quantum field gives a classical
	solution determined by the complex parameters
	$\alpha_{\vec{k}}$'s.  Conversely, we may interpret a classical
	(free) field as the expectation value of the quantum field in
	the corresponding {\em coherent} state. It will faithfully mimic
	the usual classical description eg descriptions of wave
	solutions etc. This is how one understands the usual classical
	electromagnetic waves. 
\item There is also a generalization of coherent states for fermionic
	fields. This uses Grassmann variables, their algebra and
	calculus. The constructions are very analogous. I refer you to
	\cite{Glauber} 
\end{itemize}
\subsection{Evolution into a coherent state}
Since we see a large set of classical field solutions, a question arises
as to why does a quantum system, go into a coherent state? Under what
conditions are these states manifested? 

Under a free evolution (linear equation), a state which is say a finite
linear combination of the number states will continue to maintain the
finite linear combination and not evolve into a coherent state,
interactions are essential. It suffices to have an interaction with a
{\em classical} source i.e. a c-number source. This is demonstrated
most easily by using  using interaction picture for a single harmonic
oscillator.

\underline{Recall:}

Let $|\psi\rangle_S$ denote a state vector in the Schrodinger picture
and $A_S$ denote a generic operator without explicit time dependence and
let the Hamiltonian be also time independent. The {\em Schrodinger
picture} is defined by: $id_t|\psi(t)\rangle_S = H|\psi(t)\rangle ~ , ~
d_t A_S = 0$. Define a new `picture I' with an arbitrary unitary
operator $V(t)\ , V(0) = \mathbb{1}$ as, 
\begin{eqnarray*}
|\psi(t)\rangle_I  & := & V(t)|\psi(t)\rangle_S ~ ~ , ~ ~ A_I(t) :=
V^{\dagger}(t)A_SV(t) ~ ~ \Rightarrow \\
id_t|\psi(t)\rangle_I & = & id_tV^{\dagger}(t)|\psi(t)\rangle_S +
V^{\dagger}(t)H_S|\psi(t)\rangle_S ~ = ~ V^{\dagger}(t)\left[(-id_t +
	H_S\right]V(t)|\psi(t)\rangle_I \\
\therefore id_t|\psi(t)\rangle_I & = & \left[H_I - V^{\dagger}(t)
id_tV(t)\right]|\psi(t)\rangle_I  ~ ~ , ~ ~ H_I(t) :=
V^{\dagger}(t)H_SV(t) ~ . ~ ~ \mbox{Similarly,}\\
id_tA_I(t) & = & \left[A_I(t), V^{\dagger}(t)id_t V(t)\right] \ .
\end{eqnarray*} 
For the choice $d_t V(t)  = 0$, we get back the Schrodinger picture. For
the choice $id_tV(t) = V(t)H_I(t) \leftrightarrow id_t V(t) = H_SV(t)$,
we get $d_t|\psi(t)\rangle_I = 0 ~ , ~ id_tA_I(t) = \left[A_I(t),
H_I(t)\right]$ and we have the {\em Heisenberg picture}.

Let us split the Schrodinger Hamiltonian as $H_S := H_0 + H'$ where
$H_0$ is the ``free'' Hamiltonian and $H'$ is the ``interaction''
Hamiltonian. Now we choose, $id_t V_I(t) = H_0 V_I(t)$. This results in
the equations,
\begin{equation}
id_t|\psi(t)\rangle_I ~ = ~ (H_I - (H_0)_I)|\psi(t)\rangle_I ~ := ~
H'_I|\psi(t)\rangle ~ ~ , ~ ~ id_t A_I(t) = \left[A_I, (H_0)_I\right]
\end{equation}
This is the ``Dirac'' or the ``interaction'' picture wherein the states
evolve by interaction Hamiltonian while the operators evolve by the
free Hamiltonian. Notice that $(H_0)_I(t) = H_0$.

Consider now a quantum harmonic oscillator interacting with a ``classical
source'' i.e. $H = (\Case{1}{2}\hat{p}^2 + \Case{1}{2}\omega^2
\hat{q}^2) + J(t)\hat{q}(t)$, where $J(t)$ is an externally prescribed
real function of time. We express the Hamiltonian as,
\[
	H = \omega(a^{\dagger} a + \frac{1}{2}) + J(t)(a + a^{\dagger})
	~ ~ := ~ ~ H_0 + H' \ .
\]

In the interaction picture, 
\begin{eqnarray*} 
	id_t a_I(t) & = & \left[a_I(t), H_0\right] ~ = ~
	\omega\left[a_I, a^{\dagger}_I a_I\right] ~ = ~ \omega a_I \\
	\therefore a_I(t) & = & e^{-i\omega t}a_I(0) ~ ~ , ~ ~
	a^{\dagger}_I(t) ~ = ~ e^{+i\omega t}a^{\dagger}_I(0) \\
	\therefore H'_I & = & J(t)e^{i\omega t} a_I(0) + J(t)e^{-i
	\omega t} a^{\dagger}_I(0) \ .
\end{eqnarray*}
The state vector evolves by $H'_I(t)$ which gives, $|\psi(t)\rangle =
exp\left( -i \int_0^t dt'H'_I(t')\right)|\psi(0)\rangle_I$. The
exponential simplifies as,
\begin{eqnarray*}
e^{-i\int_0^t dt' H'_I(t')} & = & e^{-i\int_0^tdt'\left[J(t')e^{i\omega
t'}a + e^{-i\omega t'}a^{\dagger}\right]} ~ ~ a := a_I(0)\ , \
a^{\dagger} := a^{\dagger}_I(0) \\ & = & e^{\alpha(t) a^{\dagger} -
\alpha^*(t) a} ~ ~ , ~ ~ \alpha(t) := -i\int_0^tdt' J(t')e^{-i\omega t'}
~ .  ~ \mbox{Evidently,}\\
|\psi(t)\rangle_I & = & D(\alpha(t))|\psi(0)\rangle_I \ .
\end{eqnarray*}
We see immediately that if $|\psi(0)\rangle_I = |0\rangle$, then at each
$t$, $|\psi(t)\rangle_I$ is the coherent state $|\alpha(t)\rangle$.

{\em Thus a quantum oscillator interacting with a classical source
(linearly coupled) will evolve into a coherent state from its ground
state.}

The generalization to a quantum field is immediate. A quantum field in
its ground state (vacuum state), upon interacting (linear coupling) with
a classical source, will evolve into a coherent state.

\underline{Remark:} You may try out different coupling to a classical
source eg $J(t)q^2$, to see what happens. This is indeed what happens
with a quantum field in a FRW expanding universe. The result is {\em
squeezed states} of the field modes, in exactly the same manner. If the
source is also an operator, then in general we do not expect the state
to evolve into a coherent state.

\newpage

\section{Quantum fields in Scattering Phenomena}\label{ScatteringTheory}

So far we have introduced and studied quantum fields in the
``non-interacting'' context in the sense, the equations of motion that
they satisfied were linear. 
We alluded to interactions of the fields with sources/detectors,
involving emission/absorption of quanta.  A more detailed version of
this is typically treated in interactions of quantum fields with
atoms/molecules which are non-relativistic system making transitions
among their bound states. There is another class of interactions which
are manifested in {\em scattering} phenomena. These can be studied in
high energy collisions and are a {\em direct tool to probe the basic
microscopic interactions.}

We first discuss typical scattering arrangement, identify the relevant
measurable physical quantities, set up a theoretical framework and see
how quantum fields gel with the framework.

A typical scattering experiment involves shooting particles at some
target to be studied and observing the deflected or emerging set of
particles. Instead of fixed target we could also have colliding beams.
It is usually the case that,

(i) we understand the projectile particles reasonably well and able to
control their properties such as energy, momentum, spin, charge,
shooting angles etc fairly accurately. It is presumed (a necessary
condition) that the presence of target and/or other particles do not
affect these properties significantly at least as long as they are well
separated (spatially and/or temporally);

(ii) after a large but finite time, we can observe the emergent
particles whose properties can be similarly ascertained and they too are
presumed well understood;

(iii) the correlations among the incoming and outgoing particles, though
determined by the intervening target region, can be {\em measured}
independent of the target.

If we wish to view the combined system of projectile, target and
emergent particles as a {\em single} dynamical system, then the
dynamical system must have three separate evolutions at least
approximately. These are: (a) evolution in the distant past ($t \to
-\infty)$; (b) evolution in the distant future ($t \to +\infty)$ and (c)
evolution during the intermediate time of interaction. For instance, in
a classical scattering, we will have a trajectory beginning and ending
in a region of the phase space, marked as ``asymptotic region''. In
these regions, the exact trajectory may be well approximated by a
different, usually ``free'' Hamiltonian. In a quantum setting, we may
replace a trajectory in phase space by one in the state space. It should
be noted right away that {\em not all trajectories are of the scattering
type - classically or quantum mechanically.}

What are the typically observed quantities?

In practice, it is not a single particle that is sent in but rather a
well collimated beam directed at the target region. After individual
particles emerge, there is a spread in their direction as the initial
conditions are not identical. Typically, these are detected at different
locations by detectors which have finite aperture and therefore the
state of the detected particles is also known approximately. Given the
aperture of a detector and its distance to the target, we know the solid
angle subtended at the scattering region. Thus, the basic measurements
are: (i) $I(\hat{n})d\Omega := $ number of particles or events detected
per second along the direction $\hat{n}$. This may be measured for each
species of particle separately or just the total; (ii) the prepared beam
has a flux $J := $ number of incident particles per second per unit
transverse area of the beam. Both these numbers are known directly. They
are obviously proportional and define: $\boxed{I(\hat{n})d\Omega :=
d\sigma(\hat{n},\dots) J . }$. The $d\sigma(\hat{n}, \dots)$ is called
the {\em differential cross-section} and $\dots$ refer to its dependence
on other properties such as energy etc. These reflect the averaged
attributes of the beam - average energy, momentum, polarization etc. The
detection can be made finer with particle detection in coincidence,
select specific polarization, charge etc. Depending upon the attributes
tagged {\em and included} in $I(\hat{n})d\Omega$ we get {\em
exclusive/inclusive cross-sections.} If one does not care even about the
angle, we have the {\em total cross-section}. These are typically used
in determining decay rates and in applications in statistical mechanics
as input from the microscopic dynamics.

In summary, the only information that is used about the beam is the flux
while finer details of the final state are used in defining various
differential cross-sections. Thus any cross-section is defined as the
ratio above and has dimensions of {\em area}. Experimentalist give these
measured numbers while a theorist has to choose a favorite model and
compute these  numbers. For this, we turn to the theoretical formulation
of scattering processes in the quantum framework. Here are some basic
observations.
\subsection{General scattering framework}
Let $\cal{H}$ be the Hilbert space of the combined system
(projectile-target-scattered/emerged). Let $H_0, H_0'$ and $H$ be the
three Hamiltonian operators governing the time evolution  in the remote
past, distant future and all through. For simplicity, we take $H_0' =
H_0$, though they could be different. The evolution generated by $H_0$ is
is said to be ``free''. All are assumed to have no explicit time
dependence. All are assumed to be self-adjoint.

Let their corresponding eigenvalue equations be, $H_0\phi_n = \epsilon_n
\phi_n$ and $H\psi_{\alpha} = \varepsilon_{\alpha} \psi_{\alpha}$. We
have the most general solutions expressed as,
\begin{eqnarray*}
	\psi(t) & = & \sum_{\alpha}c_{\alpha}e^{-i\varepsilon_{\alpha}
	t}\psi_{\alpha}(0) ~ ~ , ~ ~ \phi(t) = \sum_na_ne^{-i\epsilon_n
	t}\phi_n(0) \\
	\psi_{\alpha}(0) & = & \sum_n b_{\alpha,n}\phi_n(0) ~ ~ ~ ~
	\because \{\psi_{\alpha}(0)\} ~ , ~ \{\phi_n(0)\} ~, ~\mbox{are
	complete.} \\
	\therefore \psi(t) & = & \sum_{\alpha,
	n}c_{\alpha}b_{\alpha,n}e^{-i\varepsilon_{\alpha}t}\ \phi_n(0) ~
	~ , ~ ~ \mbox{all coefficients are time independent.}
\end{eqnarray*}

We are interested in those solutions $\psi(t)$ while resembles some free
solution as $t \to \pm \infty$. Thus, as $t \to -\infty$, 
\begin{eqnarray*}
\psi(t) - \phi(t) \to 0 & \Rightarrow & \sum_n \left\{ \sum_{\alpha}
c_{\alpha} b_{\alpha,n} e^{-i\varepsilon_{\alpha}t + i\epsilon_n t} -
a_n\right\} \phi_n(0) \to 0\ . \\
\therefore a_n & \to & \sum_{\alpha} c_{\alpha} b_{\alpha,n}
e^{i(\epsilon_n - \varepsilon_{\alpha})t} ~ ~ ~ ~ \forall ~ n \ .
\mbox{\ $t-$independence of coefficients implies,} \\
c_{\alpha}b_{\alpha,n} ~ (\mbox{no sum}) ~ & = 0 & ~ \mbox{whenever} ~
\varepsilon_{\alpha} \neq \epsilon_n \ .
\end{eqnarray*}

So, if $H_0, H$ do not have any eigenvalue in common, then for every
$c_{\alpha_*} \neq 0 ,  b_{\alpha_*, n} = 0 \ \forall \ n$. Hence, {\em
no} solution $\psi_{\alpha}(0)$ is approximable by a free evolution. For
a sufficiently general solution $\phi(t)$ to approximate {\em some}
exact solution $\psi(t)$, spectrum of $H_0$ should be a subset of the
spectrum of $H$. Conversely, every solution with $c_{\alpha} = 0$ for
every $\varepsilon_{\alpha} \notin Spec(H_0)$, can be asymptotically
approximated, by some free evolution. Such solutions are called {\em
scattering solutions}. 

\underline{Note:} Usually, the spectrum of $H_0$ is continuous and
bounded below. Its eigenstates span the Hilbert space. The scattering
states of $H$ however span only a subspace of $\cal{H}$. To span the
full Hilbert space we need to include the bound states of $H$. We have
been somewhat informal in the above argument, here are the sharper
definitions.

{\em Definition:} A solution $|\psi(t)\rangle$ is said to be {\em
incoming (outgoing)} if $\exists$ a solution $|\phi_{in}(t)\rangle \
(|\phi_{out}(t)\rangle)$ such that 
\[
	\lim_{t \to -\infty} \| \psi(t) - \phi_{in}(t) \| = 0 ~ ~
	(\lim_{t \to +\infty} \| \psi(t) - \phi_{out}(t) \|) .
\]
\underline{Note:} If the prepared beam or the detected scattered
particles are to be associated with a state of the full system, then it
is necessary that {\em at the most only one $|\psi(t)\rangle$ can be
	asymptote to a given $|\phi_{in}(t)\rangle$ or a given
$|\phi_{out}(t)\rangle$}. Establishing such a property for a model of
scattering is called the basic existence and uniqueness problem.

The solutions can be tagged by states at any particular instance for
example, $``|\psi(t=0)\rangle"$. We identify $\cal{H}_{in} := \{\psi \in
\cal{H}/\psi(t)\ \mbox{is incoming}\ \}$ and $\cal{H}_{out} := \{\psi
\in \cal{H}/\psi(t)\ \mbox{is outgoing}\ \}$.

Potentially it is possible to have (i) $\psi \notin \cal{H}_{in}, \psi
\notin \cal{H}_{out}$ (no scattering); (ii) $\psi \in \cal{H}_{in},$ but
$\psi \notin \cal{H}_{out}$ ('capture' process); (iii) $\psi \notin
\cal{H}_{in},$ but $\psi\in \cal{H}_{out}$ ('decay' process) and (iv)
$\psi \in \cal{H}_{in} \cap \cal{H}_{out}$ (scattering).

{\em Definition:} A scattering system is {\em weakly asymptotically
complete} if $\cal{H}_{in} = \cal{H}_{out}$. It may so happen that the
subspace of bound states, $\cal{H}_{bound}$ is such that $\cal{H} =
\cal{H}_{bound} \oplus \cal{H}_{in} \oplus \cal{H}_{out}$. Then the
system is said to be {\em asymptotically complete}.

These distinctions  have a bearing on the definition of $S-$matrix and
its unitarity.

Let $U(t), U_0(t)$ denote the unitary evolution operators corresponding
to $H_0, H$ respectively. Let $|\psi(t)\rangle =
U(t-t')|\psi(t')\rangle$ be an incoming solution.  Then $\exists \
\phi_{in}(t) = U_0(t-t')\phi_{in}(t')\rangle$ ($t'$ arbitrarily chosen
instant) such that,
\begin{eqnarray*}
	\mbox{As $t \to -\infty$ ~} & , &\|U(t-t')\psi(t') -
	U_0(t-t')\phi_{in}(t')\| ~ \to ~ 0 \\
	\therefore & & \|\psi(t') -
	U^{\dagger}(t-t')U_0(t-t')\phi_{in}(t')\| ~ \to ~ 0 \\
\therefore \lim_{t\to -\infty}\psi(t') & = & \lim_{t \to -\infty}
U^{\dagger}(t - t')U_0(t-t')\phi_{in}(t') \ .\\
\mbox{Define:~} \Omega_+ & := & \lim_{t \to -\infty} U^{\dagger}(t-t')
U_0(t-t') ~ = ~ \lim_{t \to -\infty} U^{\dagger}(t) U_0(t) ~ ~
\mbox{Using $U^{\dagger}(t) = U(-t)$,} \\
\mbox{we have,} & & \boxed{\Omega_+ ~ = ~ \lim_{t \to -\infty}
U(-t)U_0(t) ~ \mbox{and} ~ |\psi(t')\rangle =
\Omega_+|\phi_{in}(t')\rangle ~ \forall ~ t'} \ .
\end{eqnarray*}
Likewise, for an asymptotically outgoing solution, we have,
\[
\boxed{\Omega_- ~ = ~ \lim_{t \to +\infty} U(-t)U_0(t) ~ \mbox{and} ~
|\psi(t')\rangle = \Omega_-|\phi_{out}(t')\rangle ~ \forall ~ t'} \ .
\]
The $\Omega_{\pm}$ are called the {\em Moller operators}. They map free
solutions to asymptotic solutions. The assumption of existence and
uniqueness of scattering solutions, guarantees the existence of these
operators. Their adjoints are defined as, $\Omega_+^{\dagger} :=
\lim_{t\to -\infty} U_0(-t)U(t)\ , \ \Omega_-^{\dagger} := \lim_{t\to
+\infty} U_0(-t)U(t)$. The adjoints give $\phi_{in}(t') =
\Omega_+^{\dagger}\psi(t') ~ , ~ \phi_{out}(i') =
\Omega_-^{\dagger}\psi(t') ~ \ \forall \ t' .$

It follows that $\psi(t') = \Omega_{\pm}\Omega_{\pm}^{\dagger}\psi(t'),
\ \Omega_+^{\dagger}\Omega_+ \phi_{in}(t) = \phi_{in}(t')$ and
$\Omega_+^{\dagger}\Omega_+ \phi_{out}(t) = \phi_{out}(t')$. Since the
$\phi_{in/out}$ span the full Hilbert space, the last two equalities
imply that $\Omega_{\pm}^{\dagger}\Omega_{\pm} = \mathbb{1}$. The same
is not true for the $\psi(t)$'s and hence $\Omega_{\pm}
\Omega_{\pm}^{\dagger} \neq \mathbb{1}$ . Thus the Moller operators are
isometry operators but not unitary operators. For the special case of
asymptotic completeness, we can write $\Omega_{\pm}
\Omega_{\pm}^{\dagger} = \mathbb{1} - P_{bound}$ where $P_{bound}$ is
the projection operator on $\cal{H}_{bound}$. From the Schrodinger
equations satisfied by the $\psi(t)$ and $\phi(t)$, it is easy to see
that $\boxed{H\Omega_{\pm} = \Omega_{\pm}H_0 ~ \mbox{and} ~
H_0\Omega_{\pm}^{\dagger} = \Omega_{\pm}^{\dagger}H\ .}$ 

Consider now a solution which is both asymptotically incoming and
outgoing. Then we have, $\psi(t) = \Omega_+\phi_{in}(t)$ and
$\phi_{out}(t) = \Omega^{\dagger}_-\psi(t)$. Therefore,
\begin{eqnarray}
	\phi_{out}(t) & = & \Omega_-^{\dagger}\Omega_+ \phi_{in}(t) ~ ~
	, ~ ~ \phi_{in}(t) ~ = ~ \Omega_+^{\dagger}\Omega_-
	\phi_{out}(t) ~ ~ \mbox{and we define,} \\
	S & := & \Omega_-^{\dagger} \Omega_+ ~ ~ ~ \mbox{on
	$\left[Range(\Omega_+)\right]\ \cap \
\left[Domain(\Omega_-)\right] \ \neq \ \emptyset$ \ .}
\end{eqnarray}
Note that the range of $\Omega_+$ are states in $\cal{H}_{in}$ while the
domain of $\Omega_-^{\dagger}$ are states in $\cal{H}_{out}$. The {\em
scattering operator}, $S$, is thus well defined when weak asymptotic
completeness holds. Similarly, we define $S^{\dagger} :=
\Omega_+^{\dagger}\Omega_-$ which is also well defined when weak
asymptotic completeness holds. If in addition asymptotic completeness
holds, then
\begin{eqnarray}
	S^{\dagger}S & = & \Omega_+^{\dagger} \Omega_-
	\Omega_-^{\dagger} \Omega_+ ~ = ~ \Omega_+^{\dagger}(\mathbb{1}
	- P_{bound})\Omega_+ ~ = ~ \mathbb{1} - \mathbb{0} ~ ~
	\mbox{likewise,} \\
	S S^{\dagger} & = & \Omega_-^{\dagger} \Omega_+
	\Omega_+^{\dagger} \Omega_- ~ = ~ \Omega_-^{\dagger}(\mathbb{1}
	- P_{bound})\Omega_- ~ = ~ \mathbb{1} - \mathbb{0} ~ .
\end{eqnarray}
Thus the scattering operator is unitary when asymptotic completeness
holds. This is regardless of the Moller operators being non-unitary.
{\em The scattering operator $S$ maps from free solutions $\phi_{in}(t)$
to free solutions $\phi_{out}$}. Its matrix elements between the basis
of free solutions define the $S-matrix$. Explicitly, let
$\phi^a_{in}(t)$ and $\phi^b_{out}(t)$ denote incoming and out going
free solutions. Then,
\begin{equation} \label{SMatrixDefn1}
\boxed{	S_{ba} :=\langle\phi^b_{out}(t)|S|\phi^a_{in}(t)\rangle
=:\langle\phi^b_{out}(t)|\tilde{\phi}^a_{out}(t)\rangle \ . }
\end{equation}
The $\tilde{\phi}^a_{out}(t)$ state is a state that evolved from a
specific incoming state while $\phi^b_{out}(t)$ is an arbitrary free
state and $S_{ba}$ is the inner product between the two.

Hence, {\em $S_{ba}$ gives the probability amplitude for a free state
$|\phi_{in}(t)\rangle$ to evolve into an outgoing state
$|\phi_{out}^b(t)\rangle$ - in short a transition amplitude.}

The $S$ operator defined in terms of the Moller operators is manifestly
time independent. It matrix elements however seem to have a time
dependence through those of the $\phi_{in/out}(t)$. This apparent time
dependence is actually absent and can be seen as follows.

Recall the boxed relations among $\Omega_{\pm}, \Omega_{\pm}^{\dagger},
H$ and $H_0$. From these we get, 
\begin{eqnarray*}
SH_0 & = & \Omega_-^{\dagger}\Omega_+H_0 = \Omega_-^{\dagger}H\Omega_+ =
H_0\Omega_-^{\dagger}\Omega_+ = H_0 S. \\
id_tS_{ba} & = & id_t(\langle\phi^b_{out}(t)|S|\phi^a_{in}(t)\rangle) \\
& = & - \langle\phi^b_{out}(t)H_0 S |\phi^a_{in}(t)\rangle + 0 +
\langle\phi^b_{out}(t)iS H_0 |\phi^a_{in}(t)\rangle \\
& = & \langle\phi^b_{out}(t)|[S, H_0]|\phi^a_{in}(t)\rangle ~ = ~ 0.
\end{eqnarray*}

\underline{In summary:}
\begin{itemize}
\item A generic scattering system has $\cal{H}$ and the three
	Hamiltonians $H, H_0, H_0'$ specified;
\item There exist asymptotic subspaces of $\cal{H}, \cal{H}_{in},
	\cal{H}_{out}$ elements of which give full solutions approaching
	some free solutions as $t \to \pm\infty$;
\item Provided the system admits scattering states, there exist the
	Moller operators, $\Omega_{\pm}$ on $\cal{H}$ with ranges in
	$\cal{H}_{in/out}$ respectively;
\item If weak asymptotic completeness holds, $\exists$ the scattering
	operators $S: \cal{H}_{in} \to \cal{H}_{out}$ and its adjoint
	$S^{\dagger}$. The scattering operator $S$ is unitary provided
	asymptotic completeness holds.
\item Once $S_{ba}$ are known, the cross-section can be computed, since
	the number of scattered particles is proportional to
	$|S_{ba}|^2$, while the flux of the incoming particles is given
	by the probability current corresponding to $\phi_{in}(t)$.
\item Much of the rigorous analysis goes in addressing the existence and
	uniqueness properties of the proposed scattering system. Further
	details may be seen in \cite{ReedSimon, Newton}.

	In practice, rarely a scattering system is specified in
	sufficient details to address these issues, but they are at the
	very basic definitional level.
\end{itemize}

Returning to interacting quantum fields, we have the candidate free
states - the states of free quantum fields with known free Hamiltonian
$H_0$. We may specify the interacting system by additional Lorentz
invariant terms in the Lagrangian or specifying the Hamiltonian $H$.
However, we do not know what the Hilbert space of interacting quantum
fields is let alone if it admits scattering states. But we do know what
properties we would like to have for a scattering interpretation. The
task now is to make suitable assumptions and build a computational
recipe to define and compute the $S-$matrix elements.
\subsection{Scattering with quantum fields: Heisenberg Picture}
The above discussion has been in terms of the states evolving in time
i.e. in Schrodinger picture. For interacting quantum fields we do not
know the state space (except for the free fields). We do have Poincare
covariance  as a guide and this puts the field operators and their
Poincare transforms at the center stage. We are thus led to use the
Heisenberg picture and formulate the recipe in terms of the evolving
operators. As states do not evolve in Heisenberg picture, we postulate
certain properties for the states, postulate Poincare covariance on the
operator evolution and make specific additional assumption of
``asymptotic condition'' to incorporate the scattering feature. A
reference for this material is \cite{BjorkenDrellII}.

\begin{enumerate}
\item We choose the basis states of the interacting quantum fields to be
	labeled by mutually commuting, conserved Noether charges whose
	existence is guaranteed by the presumed symmetries. Among these
	symmetries is included the Poincare group (proper, orthochronous
	with/without discrete symmetries). The Poincare symmetry
	immediately gives eigenvalues of $P^{\mu}$ as a set of labels.
	Our first assumption is regarding the spectrum of $P^{\mu}$.
\item \begin{itemize}
	\item There exists a unique vacuum, eigenstate $|0\rangle$ with
		$P^{\mu}|0\rangle = 0$; 
	\item Eigenvalues, $p^{\mu}$, of $P^{\mu}$ lie in the forward
		light cone i.e.  $p^2 \le 0, p^0 > 0$;
	\item $\exists$ stable, {\em single particle states} with masses
		$m_i$: $P^{\mu}|p^{i}\rangle = p^{\mu}_i|p^i\rangle ~ ,
		~ p^2_i = - m_i^2 \ .$
	\item The eigenvalues $0$ and $m_i^2$ of $P^2$ are
		discrete in the sense of being isolated values. 
		The ``multi-particle'' states have their
		total momentum time-like but the corresponding $-p^2$
		can take any value above their total rest mass. 
		\begin{figure}[htbp]
		\begin{center}
			\scalebox{0.5}{\input{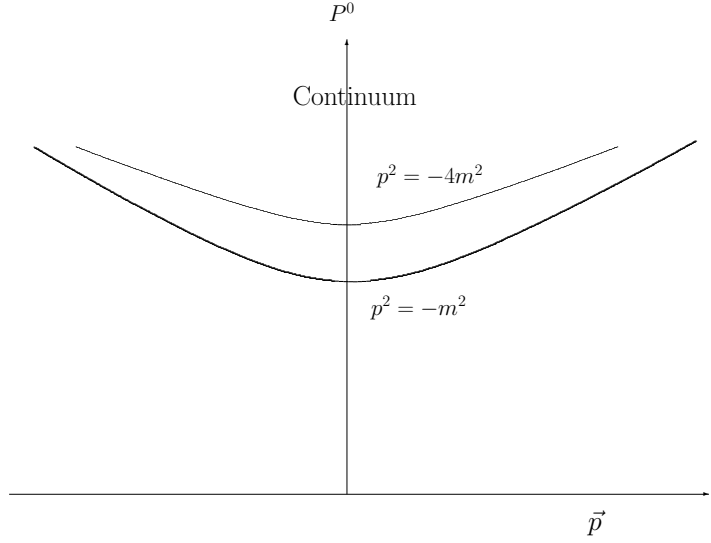}}
			\caption{Spectrum for a single, massive scalar
			field. The two particle states form a continuum
		above and including $-(p_1 + p_2)^2 = 4m^2$ hyperboloid.}
		\end{center}
		\end{figure}
	\end{itemize}

	\underline{Note:} If there are ``internal symmetries'' (i.e.
	non-space-time symmetries), then it is conceivable that the 
	zero eigenvalue of $P^{\mu}$ is degenerate, and carries some
	representation of the internal symmetry group (usually compact).
	Every one of these states must still be singlets of the Poincare
	group. But now, these vacua themselves carry some quantum
	numbers and quantum fields acting on these will create single
	particle states labeled by tensor product of representations of
	the internal symmetry group. This is not done and one stipulates
	the vacuum to be a singlet under {\em all symmetries}. 

	\underline{Note:} We have taken the stable particles to have a
	non-zero mass. This is a statement about the spectrum of $P^2$
	in the full, exact interacting quantum field theory and not at
	some approximation of it. This would seem to exclude the
	massless representations of the Poincare group for interacting
	quantum fields. In practice i.e. in perturbation theory we will
	use massless photons and see the accompanying infrared problems.

%
%

	%
\item \begin{itemize}
	\item An interacting field operator, denoted schematically by
		$\Phi(x)$ satisfies an operator equation of the form,
		$(\Box - m_0^2)\Phi(x) = J(x)$ where $J(x)$ is an
		operator built out of $\Phi$ and possibly other field
		operators. We re-write the field equation as $(\Box -
		m^2)\Phi(x) = J(x) - (m^2 - m_0^2)\Phi(x) =:
		\tilde{J}(x)$
	\item The field obeys the equal time (anti-)commutation
		relations:
		\[
			\left[\Phi(t, \vec{x}), \Phi(t, \vec{y})\right]
			= 0 = \left[\Pi(t, \vec{x}), \Pi(t,
			\vec{y})\right]  ~ , ~ \left[\Phi(t, \vec{x}),
			\Pi(t, \vec{y})\right] = i \delta^3(\vec{x} -
			\vec{y}) \ .
		\]
		If $J(x)$ has no derivatives of fields, the momentum
		field is $\Pi(t,\vec{x}) = \dot{\Phi}(t,\vec{x})$. This
		stipulates the quantized nature of the field while the
		equation of motion stipulates the nature of interaction.
	\item $\exists$ another set of fields, quantized $\Phi_{in}(x),
		\Phi_{out}(x)$, build out of the interacting field
		$\Phi(x)$, and satisfying the same set of equal time
		(anti-)commutation relations but obeying free field
		equations: $(\Box - m^2)\Phi_{in/out} = 0$ where $m$ is
		the ``physical'' mass. 

		The in and out fields thus have the same mode
		decomposition introducing the creation/annihilation
		operators which create/destroy stable particles with
		physical mass $m$. These fields also transform the same
		way as the field $\Phi(x)$ under the symmetries, in
		particular, $\Phi_{in}(x+a) = e^{ia\cdot
		P}\Phi_{in}(x)e^{-ia\cdot P}$.
	\end{itemize}
\item \underline{Asymptotic conditions:} The exact interacting quantum
	field $\Phi(x)$ is linked to the in and out fields by the
	asymptotic conditions,
	\[
\boxed{\left. \Phi(x)\right|_{t\to -\infty} ~ \longrightarrow
		\sqrt{Z}\Phi_{in}(x) ~ ~ , ~ ~ \left. \Phi(x)\right|_{t
		\to \infty} ~ \longrightarrow	\sqrt{Z} \Phi_{out}(x)}
	\]
	Here $Z$ is a possible normalization constant which we will show
	to be necessarily different from 1.

\underline{Note:} The asymptotic conditions are {\em not} operator
equations but are {\em weak} equations i.e. hold for arbitrary matrix
elements in the basis of normalized states.
\end{enumerate}
Below we elaborate the conditions and note a number of properties.

{\em Asymptotic conditions are consistent with Poincare covariance:} A
formal solution of the equation satisfied by $\Phi(x)$ is given by,
\begin{equation}\label{FormalSoln}
	\Phi(x) ~ = ~ \int d^4y\ G_{ret}(x-y,m)\tilde{J}(y) +
	\sqrt{Z}\Phi_{in}(x) .
\end{equation}
As $t\to -\infty$, the first term vanishes (property of retarded Green
function) recovering the asymptotic condition. This is a heuristic
argument since $J(y)$ it built out of $\Phi(x)$ and some iterative
procedure is implicit. The consistency with covariance is seen as,
\begin{eqnarray*}
	\sqrt{Z}\Phi_{in}(x-a) & = & \Phi(x-a) - \int d^4y\
	G_{ret}(x-a-y,m)\tilde{J}(y) \\
	& = & e^{-ia\cdot P}\Phi(x)e^{ia\cdot P} - \int d^4y\
	G_{ret}(x-y, m)\tilde{J}(y-a) \\
	& = & e^{-ia\cdot P}\left[\Phi(x) - \int d^4y\ G_{ret}(x-y,
	m)\tilde{J}(y)\right]e^{ia\cdot P} ~ = ~ \sqrt{Z}e^{-ia\cdot
	P}\Phi_{in}(x)e^{ia\cdot P} \ .
\end{eqnarray*}
\underline{Note:} Only the $G_{ret}$ is used in the formal solution and
not $G_{Feynman}$.

{\em $\Phi_{in}(x)$ creates states with physical mass $m$:} The
infinitesimal form of the transformation law for $\Phi_{in}(x)$ implies,
$[P^{\mu}, \Phi_{in}(x)] = -i\partial^{\mu}\Phi_{in}(x)$ and we also
have the equation of motion $(\Box - m^2)\Phi_{in} = 0$.

For any physical state $|k,\alpha\rangle, P^{\mu}|k,\alpha\rangle =
k^{\mu}|k, \alpha\rangle$. Therefore,
\begin{eqnarray*}
	-i\partial^{\mu}\langle k, \alpha|\Phi_{in}(x)|0\rangle & =
	&\langle k,\alpha|[P^{\mu}, \Phi_{in}(x)]|k,\alpha\rangle =
	k^{\mu}\langle k, \alpha|\Phi_{in}(x)|0\rangle \\
	\therefore -\Box \langle k, \alpha|\Phi_{in}(x)|0\rangle & = &
	k^2 \langle k, \alpha|\Phi_{in}(x)|0\rangle \\
	\therefore (-\Box + m^2)\langle k, \alpha|\Phi_{in}(x)|0\rangle
	& = 0 = & (k^2 + m^2)\langle k, \alpha|\Phi_{in}(x)|0\rangle ~
	\Rightarrow ~ k^2 = -m^2\ . 
\end{eqnarray*}
Thus, $\Phi_{in}(x)$ behaves exactly like a free field producing
physical states from the vacuum and we can be taken to have the usual
mode decomposition,
\begin{eqnarray}\label{InversionFormula0}
	\Phi_{in}(x) & = & \int \frac{d^3k} {\sqrt{2\omega_{\vec{k}}
	(2\pi)^3}} \left[a_{in}(\vec{k}) e^{ik\cdot x} +
	a_{in}^{\dagger}(\vec{k})e^{-ik\cdot x}\right] \nonumber \\
	\partial_0\Phi_{in}(x) & = &
	\int\frac{d^3k}{\sqrt{2\omega_{\vec{k}}(2\pi)^3}}
	\left[-i\omega_{\vec{k}} a_{in}(\vec{k})e^{ik\cdot x} +
	i\omega_{\vec{k}}a_{in}^{\dagger}e^{-ik\cdot x}\right] ~ ~
	\Rightarrow \nonumber \\
	a_{in}^{\dagger}(\vec{k}) & = & \frac{-i}
	{\sqrt{2\omega_{\vec{k}} (2\pi)^3}} \int d^3x \ e^{ik\cdot x}\
	\overleftrightarrow{\partial_0}\ \Phi_{in}(x) ~ ~ , ~ ~ f
	\overleftrightarrow{\partial}g := f\partial g - \partial(f) g \
	. \\
	\mbox{And,~} a_{in}(\vec{k}) & = & \frac{i}
	{\sqrt{2\omega_{\vec{k}} (2\pi)^3}} \int d^3x \ e^{-ik\cdot x}\
	\overleftrightarrow{\partial_0}\ \Phi_{in}(x) \nonumber 
\end{eqnarray} 
\underline{Note:} In writing the inversion formula, we have done the
spatial integration over a $t = $constant hypersurface $\Sigma_{t}$. The
choice of this hypersurface does not matter ($a_{in}^{\dagger}(\vec{k})$
is independent of $t$). {\em This is kept implicit in all the inversion
formulae below.}

It follows immediately, $\langle 0|\Phi_{in}(x)|0\rangle = 0$ and,
\begin{eqnarray*}
	\langle k,\alpha|\Phi_{in}(x)|0\rangle & = & \langle k,\alpha|
	e^{-ix\cdot P}\Phi_{in}(0)e^{ix\cdot P} |0\rangle ~ = ~
	e^{-ik\cdot x} \langle k,\alpha|\Phi_{in}(0)|0\rangle \\
	\langle k,\alpha|\Phi_{in}(0)|0\rangle & = & \frac{1}
	{\sqrt{2\omega_{\vec{k}}(2\pi)^2}} ~ \Rightarrow ~ \boxed{
	\langle k,\alpha|\Phi_{in}(x)|0\rangle = \frac{e^{-ik\cdot x}}
{\sqrt{2\omega_{\vec{k}} (2\pi)^3}} .  } 
\end{eqnarray*}
%

Suffice it to say that corresponding expressions exist for $\phi_{out}$.
In particular the heuristic expression take the form,
\[
	\Phi(x) = \sqrt{Z} \Phi_{out}(x) + \int d^4y\ G_{adv}(x-y,
	m^2)\tilde{J}(y).
\] 

{\em How far can we get with Lorentz covariance and the assumptions on
the spectrum?}
\subsection{Kallen-Lehmann Representation} \label{KallenLehmann}
Consider the vacuum expectation values of the commutator of interacting
field, $i\Delta'(x,x') := \langle 0|[\Phi(x), \Phi(x')]0\rangle$. We had
evaluated this for the free fields while considering the
spin-statistics, $\sim \Delta_+(x-x')$. We evaluate it as,
\begin{eqnarray}
	i\Delta(x,x') & = & \sum_n\langle 0|\Phi(x)|n\rangle\langle
	n|\Phi(x')|0\rangle - (x \leftrightarrow x') , \\
	& = & \sum_n\langle 0|\Phi(0)e^{ix\cdot P}|n\rangle\langle
	n|e^{-ix'\cdot P}\Phi(0)|0\rangle - (x \leftrightarrow x') , \\
	& = & \sum_n |\langle 0|\Phi(0)|n\rangle|^2\left(e^{i(x-x')\cdot
	k_n} - e^{-i(x-x')\cdot k_n}\right)  ~ ~  ~ \mbox{a function of
	$(x-x')$.} ~ ~ ~ ~
\end{eqnarray}
The states $|n\rangle$ contain the vacuum, the single particle states on
the isolated hyperboloid $k_n^2 = -m^2$ as well as the multi-particles
continuum. To collect terms with the same total $k^{\mu}_n$, insert $1 =
\int d^4q \delta^4(q - k_n)$ and write,
\begin{eqnarray*}
\Delta'(x-x') & = & - \frac{i}{(2\pi)^3}\int d^4q\left[(2\pi)^3 \sum_n
\delta^4(q - k_n) |\langle 0|\Phi(0)|n\rangle|^2 \left(e^{i(x-x')\cdot
k_n} - e^{-i(x-x') \cdot k_n}\right)\right] \\ 
& := &  - \frac{i}{(2\pi)^3}\int d^4q \left(e^{i(x-x')\cdot q} -
e^{-i(x-x')\cdot q}\right)\rho(q)  ~ ~ ~ ~ \mbox{where, } \\
\rho(q) & := & \left[(2\pi)^3 \sum_n \delta^4(q - k_n)|\langle
0|\Phi(0)|n\rangle|^2\right] ~ ~ ~ \mbox{(the spectral density).}
\end{eqnarray*}

\underline{Claim:} The spectral density is a scalar function of $q^2$.

\underline{Proof:} Under a Lorentz transformation, $U(\Lambda)$, we have
$U(\Lambda)|0\rangle = |0\rangle$ and $U(\Lambda)\Phi(0)U^{-1}(\Lambda)
= \Phi(\Lambda 0) = \Phi(0)$. Hence,
\[
	\rho(q) = (2\pi)^3\sum_n \delta^4(q - k_n)|\langle
	0|\Phi(0)U(\Lambda)|n\rangle|^2 \ .
\]
Next,
\begin{eqnarray*}
	\delta^4(k) & \sim & \int d^4x e^{ik\cdot x} = \int d^4(\Lambda
	x)e^{ik\cdot(\Lambda x)} = \int d^4x e^{i(\Lambda^{-1}k)\cdot x}
	\sim \delta^4(\Lambda^{-1}k) \ . \\
	\therefore \rho(q) & = & (2\pi)^3\sum_n\delta^4(\Lambda^{-1}(q -
	k_n)) |\langle 0|\Phi(0)U^{-1}(\Lambda)|n\rangle|^2 \ .
\end{eqnarray*}

Let $U(\Lambda)|n\rangle = |m\rangle$. Then $U^{-1}(\Lambda) P^{\mu}
U(\Lambda) = \Lambda^{\mu}_{~\nu} P^{\nu}$.  Acting on $|n\rangle$
implies, $P^{\mu}|m\rangle = k^{\mu}_{m}|m\rangle$ and we already had
$P^{\mu}|n\rangle = k^{\mu}_{n}|n\rangle$. This in turn implies,
$k^{\mu}_m U^{-1}(\Lambda)|m\rangle = \Lambda^{\mu}_{~\nu} k^{\nu}_n
|n\rangle$ or, $k^{\mu}_m = \Lambda^{\mu}_{~\nu} k^{\nu}_n$. Lowering
the Lorentz index gives, $(k_m)_{\mu} = (k_n)_{\nu}
(\Lambda^{-1})^{\nu}_{~\mu}$. Replacing the sum over $n$ by that over
$m$ gives,
\[
	\rho(q) = (2\pi)^3\sum_m \delta^4(k_m - \Lambda^{-1}q)|\langle
	0|\Phi(0)|m\rangle|^2 = \rho(\Lambda^{-1}q) \ .
\]
Thus, $\rho(q)$ can only depend on $q$ through $q^2$. 

Since the physical spectrum has $q^2 <  0$ with $q^0 > 0$, we write
$\rho(q) := \theta(q^0) \rho(q^2), \ \rho(q^2) = 0 ~ \mathrm{for} ~ q^2
> 0$.  Substituting in $\Delta'$ gives,
\begin{eqnarray}
\Delta'(x-x') & = & \frac{-i}{(2\pi)^3}\int d^4q \rho(q^2) \theta(q^0)
\left(e^{iq\cdot(x-x')} - e^{-iq\cdot(x-x')}\right) \nonumber\\
& = & \int_0^{\infty} d\sigma^2 \rho(\sigma^2) \left[
\frac{-i}{(2\pi)^3} \int d^4q \delta(q^2 - \sigma^2) \epsilon(q^0)
e^{iq\cdot(x-x')}\right] \ . \nonumber\\
\therefore \Delta'(x-x') & = & \int_0^{\infty}d\sigma^2\ \rho(\sigma^2)
\Delta(x-x', \sigma^2) \ . \label{KallenLehmannScalar}
\end{eqnarray}
Here we have introduced the function $\epsilon(x) = \theta(x) -
\theta(-x)$ to combine the two exponentials and also recognized that the
the square bracket is just $-i[\Phi_{in}(x), \Phi_{in}(x')]$ for $m^2 =
\sigma^2$. The last equation is known as the {\em Kallen-Lehmann
representation} or the {\em spectral representation} for the commutator
function $\Delta'(x-x')$ of the exact interacting quantum fields. This
has followed from Poincare covariance (separation of the
$x-$dependence), the assumptions regarding the spectrum of the
interacting system ($\rho(q^2>0)=0$) {\em and} normalizations used for
$\Phi_{in}$ in identifying the $\Delta(x-x',\sigma^2)$. 

We can separate the contribution of the 1-particle states at $\sigma^2 =
m^2$. We obtain the $\langle 0|\Phi(x)|k\rangle$ using the asymptotic
condition,
\begin{eqnarray}
\langle 0|\Phi(x)|k\rangle & = & \sqrt{Z}\langle 0|\Phi_{in}(x)|k\rangle
+ \int d^4y\ G_{ret}(x-y, m)\langle 0|\tilde{J}(y)|k\rangle ~ , ~
\mbox{but} \nonumber \\
\langle 0|\tilde{J}(y)|k\rangle & = &\langle 0|(\Box -
m^2)\Phi(x)|k\rangle = (\Box - m^2)\left(\langle 0|\Phi(0)|k\rangle
e^{ik\cdot x}\right) \nonumber \\
& = & (-k^2 - m^2)\left(\langle 0|\Phi(0)|k\rangle e^{ik\cdot x}\right)
~ = ~ 0 . ~ ~ \mbox{The first term then gives,}\nonumber \\
\langle 0|\Phi(x)|k\rangle & = & \sqrt{Z}\langle 0|\Phi_{in}(x)|k\rangle
= \sqrt{Z}\frac{e^{ik\cdot x}}{\sqrt{2\omega_{\vec{k}}(2\pi)^3}}
\Rightarrow |\langle 0|\Phi(x)|k\rangle|^2 =
\frac{Z}{2\omega_{\vec{k}}(2\pi)^3} \ . \nonumber \\
\therefore \rho(q^2)|_{1-particle} & = & (2\pi)^3\int d^3k \delta^4(k -
q) \frac{Z}{2\omega_{\vec{k}}(2\pi)^3} = Z \delta(q^2 - m^2)\theta(q^0)
\\
\therefore & & \boxed{\Delta'(x-x')  =  Z \Delta(x-x',m^2) +
\int_{m^2_1}^{\infty}d\sigma^2 \rho(\sigma^2) \Delta(x-x', \sigma^2)} \
. \label{KallenLehmannSeparated}
\end{eqnarray}
The lower limit on the integral is the smallest invariant mass of the
multi-particle continuum and $m^2_1 > m^2$ is assumed. 

We can now derive a bound on $Z$. Observe that 
\[
\lim_{t'\to t} (i\partial_t \Delta'(x-x')) = \langle 0|[\dot{\Phi}(x),
\Phi(x')]|0\rangle = -i\delta^3(x-x') = (i\partial_t \Delta(x-x',m^2)) \
.
\]
We are using the assumption that there are no derivative terms in $J(x)$
or equivalently, $\Pi(x) = \dot{\Phi}(x)$ and noting that the in/out
fields satisfy the same commutation relations.

Taking $-i\partial_t$ on eq.(\ref{KallenLehmannSeparated}) and taking
the limit $t' \to t$, we deduce
\begin{eqnarray}
-i \delta^3(\vec{x} - \vec{x}') & = & Z(-i \delta^3(\vec{x} - \vec{x}'))
+ \int_{m_1^2}^{\infty}d\sigma^2\ \rho(\sigma^2)\left[-i\delta^3(\vec{x}
- \vec{x}',\sigma)\right] \nonumber \\
Or, ~ 1 & = & Z + \int_{m_1^2}^{\infty}d\sigma^2 \ \rho(\sigma^2) ~ ~
\Longrightarrow \boxed{ 0 \le Z < 1 . }
\end{eqnarray}

The inequalities on $Z$ can be understood as follows. Since asymptotic
condition is needed to link $\Phi(x)$ with $\Phi_{in/out}(x)$, $Z \neq
0$. To understand the upper limit, notice that the free fields can only
produce 1-particles states since it is linear in creation/annihilation
operators. The interacting field has no such restriction and can produce
multi-particle states as well. Hence the probability for it to produce
1-particle states is {\em less} than that for the free fields i.e.
$|\langle 0|\Phi(x)|k\rangle|^2 < |\langle 0|\Phi_{in/out}(x)
|k\rangle|^2$. This results in $Z < 1$.

In principle the $Z$ factors for the in and out fields could have been
different in the asymptotic conditions. However, following identical
steps as above would lead us to the same relation
(\ref{KallenLehmannSeparated}) implying equality of the $Z$'s.

An important point to note is that $Z < 1$ has a physical reason, quite
independent of any infinities and their renormalization that we will
encounter later on.

\underline{Remark:} In the Kallen-Lehmann spectral representation we
considered the vacuum expectation value of the commutator of the fully
interacting quantum field presumed to satisfy the usual canonical
commutation relations. Because of this, the interacting quantum field
gets identified with the so called ``bare'' quantum field in the
renormalized perturbation theory. We could instead use a
``renormalized'' quantum field, $\Phi_R(x) := \Phi(x)/\sqrt{Z}$ and
correspondingly define, $\Delta'_R(x-x') := Z^{-1}\Delta'(x-x')$. It
follows that  $\rho_R(q^2) = Z^{-1}\rho(q^2)$ and
$[\Phi_R(x),\dot{\Phi}_R(x')] = iZ^{-1}\delta^3(x-x')$. In terms of the
renormalized quantities, there is no factor of $Z$ in equation
(\ref{KallenLehmannSeparated}). The manipulations leading to the bound
on $Z$ will now give,
\[
\lim_{t'\to t} (i\partial_t \Delta'_R(x-x')) = \langle
0|[\dot{\Phi}_R(x), \Phi_R(x')]|0\rangle = -iZ^{-1}\delta^3(x-x') =
Z^{-1}(i\partial_t \Delta(x-x',m^2)) \ ,
\]
and the eq.(\ref{KallenLehmannSeparated}) will lead to $1 =
Z\left[1+\int_0^{\infty}d\sigma^2\rho_R(\sigma^2)\right]$ implying again
that $Z < 1$. These bounds are true provided the spectral integral is
{\em finite}. In perturbation theory, the two point functions
(commutator or the Feynman propagator) of the unrenormalized fields are
Ultra-Violet divergent and the bound cannot be inferred (See section
10.7 of \cite{Weinberg}). The Kallen-Lehmann representation
eq.(\ref{KallenLehmannScalar}) itself does {\em not} depend on the
canonical commutation relations for the interacting field and can indeed
be derived for composite fields as well.
\subsection{The $S$-matrix and its properties} \label{SMatrix}
Having the $\Phi_{in/out}(x)$ and their corresponding mode expansions at
our disposal, we can now define the in/out states as, $|k\rangle_{in} :=
a^{\dagger}_{in}(\vec{k})|0\rangle$ and $|k\rangle_{out} :=
a^{\dagger}_{out}(\vec{k})|0\rangle$. Similarly, multi-particle states
can be defined. We can form wave packets for normalizable states and
take limit of infinite sharpness at the end. We bypass these steps and
work with the momentum eigenstates. Note that these states are {\em time
independent}. While we can certainly define the general in/out states
and take their spans, we need to (and do) make a further assumption that
these spans generate the full Hilbert space of the interacting quantum
fields\footnote{This automatically precludes bound states in the
interacting system and amounts to postulating asymptotic completeness
property mentioned earlier.}. The sets of in and out states just
constitute two orthonormal bases, conveniently transforming by
representations of the Poincare group. This guarantees existence of a
unitary operator connecting the two orthonormal bases. This is our
scattering operator in this Heisenberg picture formulation. Taking the
basis elements to be generated by {\em monomials of $a^{\dagger}_{in}
	(\vec{k}, \sigma, \dots)$ and $a^{\dagger}_{out} (\vec{k},
		\sigma, \dots)$}, the $S-$matrix is defined as,
\begin{equation}\label{SMatrixDefn2}
	S_{\beta\alpha} :=\langle\beta\ out|\alpha\ in\rangle ~ ~ =: ~
	~\langle\beta\ in|S|\alpha\ in\rangle\ 
\end{equation}
Note that $\alpha, \beta$ label {\em basis states} which are created by
{\em monomials} in $ a^{\dagger}_{in}, a^{\dagger}_{out}$. The
definition of $S$ operator shows that it preserves the labels:
$\langle\beta\ out| =:\langle\beta\ in|S$. We stipulate some conditions
on the $S-$matrix.

{\em Stability of the vacuum and 1-particle states:} We expect vacuum
state to suffer no scattering spontaneously except may be acquiring a
phase (uniqueness of vacuum allows a phase since rays are defined unto
phases). We choose the phase to be 1, and stipulate $\boxed{S_{00} =
1}$. Likewise presumed stability of single particle states also
disallows any spontaneous change (nothing to scatter against). Thus we
stipulate that $S_{k', k} = \delta^3(\vec{k}' - \vec{k})$. Scattering
takes place with multi-particle states.

\underline{Claim:} $~ \Phi_{out}(x) = S^{-1} \Phi_{in}(x) S$ .

\underline{Proof:} ~ Consider, $\langle \beta\ out | \Phi_{out}(x)
|\alpha\ in\rangle$ and evaluate in two ways.
\begin{eqnarray*}
\langle \beta\ out | \Phi_{out}(x) |\alpha\ in\rangle & = & (\langle
\beta\ in|S)\Phi_{out}(x)| \alpha \ in\rangle .
\end{eqnarray*}
Next, $\langle\beta'\ out| := \langle\beta\ out|\Phi_{out}(x)$ is a
linear combination of the out-basis elements.  Each of this can be
expressed as corresponding in-basis element times $S$. And the in-basis
element is also obtained by action of $\Phi_{in}(x)$. In equation,
\begin{eqnarray*}
\langle\beta'\ out| = \sum_{\gamma}C_{\gamma}\langle\gamma\ out| =
\sum_{\gamma}C_{\gamma}\left(\langle\gamma\ in|S\right) = \hspace{6.0cm}
& & \\
\left(\sum_{\gamma}C_{\gamma}\langle\gamma\ in|\right)S =:
\langle\beta'\ in|S = (\langle\beta\ in|\Phi_{in}(x)) S. 
\end{eqnarray*}
Taking inner product of the two expressions with $|\alpha\ in\rangle$
gives,
\[
\langle\beta\ out|\Phi_{out}(x)|\alpha\ in\rangle =\langle\beta\
in|S\Phi_{out}(x)|\alpha\ in\rangle = \langle\beta\
in|\Phi_{in}(x))S|\alpha\ in\rangle \ \forall \ \alpha, \beta.
\]
Completeness of the in-basis implies, $\boxed{\Phi_{out}(x) =
S^{-1}\Phi_{in}(x) S\ . }$ proving the claim.

\underline{Claim:} ~ {\em Covariance of the in/out fields gives
invariance of the scattering operator.}

\underline{Proof:} ~ We have,
\begin{eqnarray}
U(\Lambda,a)\Phi_{in}(x)U^{-1}(\Lambda,a) & = & \Phi_{in}(\Lambda x + a)
\nonumber \\
U(\Lambda,a)\Phi_{out}(x)U^{-1}(\Lambda,a) & = & \Phi_{out}(\Lambda x +
a)
~ ~ ~ ~\mbox{and} ~ ~ ~ ~\Phi_{out}(x) = S^{-1}\Phi_{in}(x) S \nonumber
\\
\therefore U (S^{-1}\Phi_{in}(x)S)U^{-1} & = & S^{-1}\Phi_{in}(\Lambda x
+ a)S ~ = ~ S^{-1}U\Phi_{in}(x)U^{-1} S \\ 
\therefore U^{-1}SUS^{-1}\Phi_{in}(x) & = & \Phi_{in}(x)U^{-1}SUS^{-1} ~
=> ~ \nonumber \\ U^{-1}S U S^{-1} & = & \mathbb{1} ~ or ~ \boxed{S
U(\Lambda,a) = U(\Lambda,a) S} \ .
\end{eqnarray}

\underline{Question:} Are the Schrodinger picture definitions of the
scattering matrix element (\ref{SMatrixDefn1}) and the Heisenberg
picture scattering matrix element defined using $\Phi_{out} =
S^{-1}\Phi_{in}S$, related? If so, how?

We now relate the S-matrix elements to vacuum expectation values of time
ordered products of the interacting fields - the
Lehmann-Symanzik-Zimmermann (LSZ) reduction formulae. This is followed
by covariant perturbation series leading to Feynman rules. We will first
detail these steps for the notationally simpler case of a scalar field
and then summarize the corresponding steps for the Dirac and the Maxwell
fields. 
\subsection{Lehmann-Symanzik-Zimmermann Reduction of $S-$matrix}
We have defined the $S-$matrix elements using the in and out states
which form the basis elements created by monomials in
$a^{\dagger}_{in}(\vec{k},\dots)$ and $a^{\dagger}_{out}(\vec{k},\dots)$
acting on the unique vacuum. The in and out fields are not independent
or uncorrelated, they are linked through the interacting field $\Phi(x)$
via the asymptotic condition. The LSZ reduction of the $S-$matrix
elements expresses them in terms of the interacting field explicitly.
For simplicity, let us first consider the case of an interacting
Hermitian scalar field.
\subsubsection{LSZ reduction for Klein-Gordon field}
Consider a matrix element of the form $\langle\beta\ out|\alpha,k\
in\rangle$ where we have separated the $k$ label in the in-state. The
definition gives,
\begin{eqnarray*}
	\langle\beta\ out|\alpha,k\ in\rangle & = &\langle\beta\
	out|a^{\dagger}_{in}(\vec{k})|\alpha\ in\rangle = \langle\beta\
	out|\left(a^{\dagger}_{in}(\vec{k}) - a^{\dagger}_{out}(\vec{k})
	+ a^{\dagger}_{out}(\vec{k}) \right)|\alpha\ in\rangle \ .
\end{eqnarray*}
The $a^{\dagger}_{out}(\vec{k})$ acting on the out-state will remove a
particle with label $\vec{k}$ if it is present in the label set $\beta$,
or will annihilate the out-state. For simplicity, let us assume that
there is no such state in the $\beta$ label. Then we have just added and
subtracted a 0. For the first two term in the brackets , use the
inversion formula (\ref{InversionFormula0}). 
\begin{eqnarray*}
\langle\beta\ out|\alpha,k\ in\rangle & = &\langle\beta\
out|\left[\frac{-i}{\sqrt{2\omega_{\vec{k}}(2\pi)^3}}\int d^3x
e^{ik\cdot x} \overleftrightarrow{\partial_0}\Phi_{in}(x) \right. \\
& & \mbox{\hspace{1.5cm}} \left. +
\frac{+i}{\sqrt{2\omega_{\vec{k}}(2\pi)^3}}\int d^3x e^{ik\cdot x}
\overleftrightarrow{\partial_0}\Phi_{out}(x)\right] |\alpha\ in\rangle .
\\
\left[\dots\right] & = & \left(\lim_{t \to +\infty} - \lim_{t\to
-\infty}\right)\left\{\frac{+i}{\sqrt{2\omega_{\vec{k}}(2\pi)^3}}\int
d^3x \frac{e^{ik\cdot x} \overleftrightarrow{\partial_0} \Phi(x)}
{\sqrt{Z}}\right\} \\
\therefore \sqrt{Z}\left[\dots\right] & = & \int d^4x\ \partial_0\left\{
e^{ik\cdot x} \overleftrightarrow{\partial_0} \Phi(x)\right\} \\
& = & \int d^4x\ \left\{ e^{ik\cdot x} {\partial^2_0} \Phi(x) -
\partial^2_0(e^{ik\cdot x})\Phi(x)\right\} ~ , ~ (-\partial_0^2
e^{ik\cdot x} = (-\nabla^2 + m^2)e^{ik\cdot x})\\
& = & \int d^4x\ e^{ik\cdot x}\left\{ {\partial^2_0} \Phi(x) +
(-\nabla^2 + m^2)\Phi(x)\right\} ~ , ~ (\nabla^2\ \mbox{flipped onto\ }
\Phi(x)). \\
\therefore \langle\beta\ out|\alpha,k\ in\rangle & = &
\frac{-i}{\sqrt{Z}}\int \frac{d^4x e^{ik\cdot x}}
{\sqrt{2\omega_{\vec{k}} (2\pi)^3}} (\Box - m^2)\langle\beta\
out|\Phi(x)|\alpha\ in\rangle \ . 
\end{eqnarray*}

Now let us separate a particle with label $k'$ from the out-state and
write $\langle\beta\ out| =\langle\gamma,k'\ out|$ the inner product on
the right hand side of the above equation. Consider,
\begin{eqnarray*}
\langle\gamma,k'\ out|\Phi(x)|\alpha\ in\rangle & = & \langle\gamma\
out|a_{out}(\vec{k}')\Phi(x)|\alpha\ in\rangle \\
& = & \langle\gamma\ out|(a_{out}(\vec{k}')\Phi(x) -
\Phi(x)a_{in}(\vec{k}') + \Phi(x)a_{in}(\vec{k}'))|\alpha\ in\rangle 
\end{eqnarray*}
As before, for simplicity, assume $\alpha$-label does not contain $k'$
and thus drop the third term in the brackets. Notice that we have added
and subtracted the zero term with $a_{in}(\vec{k}')$ to the right of
$\Phi(x)$ while $a_{out}(\vec{k}')$ is naturally to the left of
$\Phi(x)$. We need to maintain this order as we do not have the
commutation relations between the in/out field and the interacting
field. As before, using the inversion formula (\ref{InversionFormula0})
and using the asymptotic condition, we get
\begin{eqnarray*}
\langle\gamma,k'\ out|\Phi(x)|\alpha\ in\rangle & = & \langle\gamma\
out|\left[ \frac{i}{\sqrt{2\omega_{\vec{k}'}(2\pi)^3}}\int
	d^3x'e^{-ik'\cdot
	x'}\left\{\overleftrightarrow{\partial'_0}\Phi(x')_{out}\Phi(x)
	\right.  \right. \\ 
& & \mbox{\hspace{4.5cm}} \left.\left. -
\Phi(x)\overleftrightarrow{\partial'_0}\Phi(x')_{in}\right\}\right]
|\alpha\ in\rangle \\
\therefore \sqrt{Z}\left[\dots\right] & = & \frac{i}{\sqrt {2\omega_
{\vec{k}'} (2\pi)^3}}\int d^3x'\left\{ \lim_{t'\to \infty}(e^{-ik'\cdot
x'} \overleftrightarrow{\partial'_0}\Phi(x'))\Phi(x) \right. \\ 
& & \left. \mbox{\hspace{3.0cm}} - \lim_{t'\to -\infty} \Phi(x)
(e^{-ik'\cdot x'} \overleftrightarrow{\partial'_0} \Phi(x')) \right\} 
\end{eqnarray*}
It is convenient to define a time ordering instruction. Define a {\em
time ordered product} as: $~\hfill\boxed{T\left\{A(t_1,\vec{x})
B(t_2,\vec{y})\right\} := \theta(t_1 - t_2) A(t_1,\vec{x})
B(t_2,\vec{y}) + \theta(t_2-t_1) B(t_2,\vec{y}) A(t_1, \vec{x})
.}\hfill~$ 

Notice that the two limits already have the time ordering incorporated.
Hence we can combine the terms and write, 
\begin{eqnarray*}
\sqrt{Z}\left[\dots\right] & = & \frac{i}{\sqrt {2\omega_ {\vec{k}'}
(2\pi)^3}}\int d^3x'\left( \lim_{t'\to \infty} - \lim_{t'\to
-\infty}\right) e^{-ik'\cdot x'} \overleftrightarrow{\partial'_0}
T\left\{\Phi(x')\Phi(x)\right\}  \\ 
& = & \frac{i}{\sqrt {2\omega_ {\vec{k}'} (2\pi)^3}}\int d4x'
\partial'_0 \left( e^{-ik'\cdot x'} \overleftrightarrow{\partial'_0}
T\left\{\Phi(x')\Phi(x) \right\}\right)
\end{eqnarray*}
Proceeding exactly as before we get,
\begin{eqnarray}
\langle\gamma,k'\ out|\alpha,k\ in\rangle & = &
\left(\frac{-i}{\sqrt{Z}}\right)^2\int d^4x'\int d^4x \frac{e^{-ik'\cdot
x'}}{\sqrt{2\omega_{\vec{k}'}(2\pi)^3}}\frac{e^{ik\cdot
x}}{\sqrt{2\omega_{\vec{k}}(2\pi)^3}} \nonumber \\
& & \mbox{\hspace{0.0cm}} (\Box_{x'} - m^2)(\Box_x -
m^2)\langle\gamma,k'\ out|T\left\{\Phi(x')\Phi(x)\right\}|\alpha,k\
in\rangle 
\end{eqnarray}
%
%
The generalization is immediate and we state the master formula for the
$S-$matrix element as,
\begin{eqnarray}
\langle k_1,'\dots,k'_n\ out|k_1,\dots,k_m\ in\rangle & = &
\left(\frac{-i}{\sqrt{Z}} \right)^{m+n}\int d^4x'_1..d^4x'_n\int
d^4x_1..d^4x_m \nonumber \\
& & ~ ~ ~ ~ \prod_{j}^n \frac{e^{-ik'_j\cdot x'_j}}
{\sqrt{2\omega_{\vec{k}'} (2\pi)^3}} \prod_{j}^m \frac{e^{+ik_j\cdot
x_j}} {\sqrt{2\omega_{\vec{k}} (2\pi)^3}} (\Box_{x'_1} - m^2) \nonumber
\\
& & ~ ~ ~ ~ \dots(\Box_{x'_m} - m^2)(\Box_{x_1} - m^2)\dots(\Box_{x_n} -
m^2) \nonumber \\
& & ~ ~ ~ ~ \langle 0|T\left\{ \Phi(x'_1) \dots \Phi(x'_n) \Phi(x_1)
\dots \Phi(x_n) \right\}|0\rangle
\end{eqnarray}
The vacuum expectation value of the time ordered product of $n$ quantum
fields (interacting or free) is called an $n-${\em point function/Green
function/n-point correlation function}. 

To get the $S-$matrix elements, the n-point function is operated by the
free field equation expression, multiplied by the in/out mode functions
and integrated over all the space-time points.  Note that the time
ordering arises because the definition of $S-$matrix requires all the
creation operators, $a^{\dagger}_{in}(\vec{k})$, have to be on the right
while the annihilation operators, $a_{out}(\vec{k})$ have to be on the
left. This will also naturally lead to the Feynman Green's function
(free, 2-point function).

Let us note the steps that were followed to get the $S-$matrix elements
in terms of the n-point functions.
\begin{itemize}
\item Mode decomposition of the free field and the inversion formulae;
\item Equation of motion for the interacting field with a `source' term
	on the right hand side;
\item Equal time(anti-)commutation relations for both the interacting
	and the in/out fields;
\item Asymptotic conditions;
\item Reduction process;
\end{itemize}
\subsubsection{LSZ reduction for Dirac field}
We begin by recalling the mode decomposition.
\begin{eqnarray*}
\Psi(x) & = & \int \frac{d^3k}{\sqrt{2\omega_{\vec{k}}(\pi)^3}}
\left[b(\vec{k},\sigma)u(\vec{k},\sigma)e^{ik\cdot x} +
d^{\dagger}(\vec{k},\sigma)v(\vec{k},\sigma)e^{-ik\cdot x}\right] \\
\Psi^{\dagger}(x) & = & \int \frac{d^3k}
{\sqrt{2\omega_{\vec{k}}(\pi)^3}}
\left[b^{\dagger}(\vec{k},\sigma)u^{\dagger}(\vec{k},\sigma)e^{-ik\cdot
x} + d(\vec{k},\sigma)v^{\dagger}(\vec{k},\sigma)e^{ik\cdot x}\right] 
\end{eqnarray*} 
We have used the $\Psi^{\dagger}$ field instead of $\bar{\Psi}$ because
the inversion formulae take a more convenient form.

We had defined the $u, v$ spinors as,
\begin{eqnarray*}
	(\dsl{k}+m)u(\vec{k},\sigma) = 0 ~ , ~
	\bar{u}(\vec{k},\sigma)u(\vec{k},\sigma') ~ = ~ &
	\delta_{\sigma,\sigma'} & ~ = ~ - \bar{v}(\vec{k},\sigma)
	v(\vec{k},\sigma') ~ , ~ (-\dsl{k}+m)v(\vec{k},\sigma) = 0\\
	\sum_{\sigma}u(\vec{k},\sigma)\bar{u}(\vec{k},\sigma) =
	\frac{-\dsl{k}+m}{2m} & , &
	\sum_{\sigma}v(\vec{k},\sigma)\bar{v}(\vec{k},\sigma) =
	-\frac{\dsl{k}+m}{2m}
\end{eqnarray*}
We need to express these in terms of the adjoints rather than in terms
of the Dirac adjoints. This alternative for can be derived as follows.
We have,
\begin{eqnarray*}
	\bar{u}(\vec{k},\sigma)\gamma^{\mu}u(\vec{k},\sigma') & = &
	\Lambda^{\mu}_{~\nu} \bar{u}(\hat{k},\sigma) \gamma^{\mu}
	u(\hat{k},\sigma') ~  ,  ~ k^{\mu} = \Lambda^{\mu}_{~\nu}
	\hat{k}^{\nu} ~ , ~ \hat{k}^{\mu} = (m, \vec{0}) ~. ~ \mbox{For
	$\mu = 0 , $}\\
	u^{\dagger}(\vec{k},\sigma) u(\vec{k},\sigma') & = &
	\Lambda^0_{~\nu}u^{\dagger}(\hat{k},\sigma) \gamma^0
	\gamma^{\nu} u(\hat{k},\sigma') ~ = ~
	\Lambda^0_{~\nu}u^{\dagger}(\hat{k},\sigma) \gamma^{\nu}
	u(\hat{k},\sigma') ~ \because ~ \gamma^0u(\hat{k},\sigma') =
	u(\hat{k}, \sigma') \\
	& = & \Lambda^0_{~0}u^{\dagger}(\hat{k},
	\sigma) \gamma^0 u(\hat{k},\sigma') + \Lambda^0_{~i}u^{\dagger}(
	 \hat{k},\sigma)\gamma^iu(\hat{k},\sigma') \\
	 \therefore u^{\dagger}(\vec{k},\sigma)u(\vec{k},\sigma')
	 & = & \Lambda^0_{~0}\bar{u}(\hat{k},\sigma)u(\hat{k}, \sigma')
	 + 0 ~ = ~ \Lambda^0_{~0}\delta_{\sigma,\sigma'} + 0
\end{eqnarray*}
The second term in the last equation is zero because, for $\hat{k}$
spinors $\gamma^0 u(\hat{k},\sigma) = u(\hat{k},\sigma)$. This implies, 
\[
	u^{\dagger}\gamma^iu = u^{\dagger}\gamma^0\gamma^iu = -
	u^{\dagger}\gamma^i\gamma^0u = - u^{\dagger}\gamma^i u \ .
\]
Similarly, $v^{\dagger}(\vec{k},\sigma)v(\vec{k},\sigma') = -
\Lambda^0_{~0}\bar{v}(\hat{k}, \sigma)v(\hat{k}, \sigma') = +
\Lambda^0_{~0}\delta_{\sigma,\sigma'}$. But $\Lambda^0_{~0}$ is defined
through, $k^{\mu} = \Lambda^{\mu}_{~\nu}\hat{k}^{\nu}$. For $\mu = 0$
this gives, $\Lambda^0_{~0} = \Case{\omega_{\vec{k}}}{m}$. Our
orthonormality relations then take the form,
\[
	\boxed{u^{\dagger}(\vec{k},\sigma)u(\vec{k},\sigma') =
		\frac{\omega_{\vec{k}}}{m}\delta_{\sigma,\sigma'} =
		v^{\dagger}(\vec{k},\sigma)v(\vec{k},\sigma') ~ ~ , ~ ~
	u^{\dagger}v = 0 = v^{\dagger} u \ . }
\]

We can use these relations to derive the inversion formulae. For
instance, multiplying the mode decomposition for $\Psi(x)$ by
$u^{\dagger}(\vec{k},\sigma)e^{-ik\cdot x}$ and integrating over $d^3x$
yields,
\[
	b(\vec{k},\sigma) = \frac{2m}{\sqrt{2\omega_{\vec{k}}(2\pi)^3}}
	\int d^3x u^{\dagger}(\vec{k},\sigma)e^{-ik\cdot x}\Psi(x) 
\]
It is convenient to define 
\[
U(\vec{k}, \sigma) := 2m\frac{u(\vec{k}, \sigma)}
{\sqrt{2\omega_{\vec{k}} (2\pi)^3}} ~ ~ , ~ ~ V(\vec{k}, \sigma) :=
2m\frac{v(\vec{k}, \sigma)} {\sqrt{2\omega_{\vec{k}} (2\pi)^3}} 
\]
The inversion formulae are,
\begin{eqnarray}
b(\vec{k},\sigma) & = & \int d^3x e^{-ik\cdot x} U^{\dagger}
(\vec{k},\sigma) \Psi(x)  \\
d^{\dagger}(\vec{k},\sigma) & = & \int d^3x e^{+ik\cdot x} V^{\dagger}
(\vec{k},\sigma) \Psi(x)  \\
b^{\dagger}(\vec{k},\sigma) & = & \int d^3x e^{+ik\cdot x}
\Psi^{\dagger}(x) U(\vec{k},\sigma)  \\
d(\vec{k},\sigma) & = & \int d^3x e^{-ik\cdot x} \Psi^{\dagger}(x)
V(\vec{k},\sigma)
\end{eqnarray}

In the reduction process, we will have for instance,
\begin{eqnarray*}
b^{\dagger}_{in}(\vec{k},\sigma) - b^{\dagger}_{out} (\vec{k},\sigma) &
= & \int d^3x\ e^{ik\cdot x}\left(\Psi^{\dagger}(x)_{in} -
\Psi^{\dagger}_{out}(x) \right) U(\vec{k},\sigma) \\
\mbox{and using asymptotic condition} & \longrightarrow &
\frac{1}{\sqrt{Z}}\int d^4x\ \partial_0\left(e^{ik\cdot
x}\Psi^{\dagger}(x)\right) U(\vec{k},\sigma) 
\end{eqnarray*}
Next,
\begin{eqnarray*}
\partial_0(e^{ik\cdot x}\Psi^{\dagger}(x))U(\vec{k},\sigma) & = &
e^{ik\cdot x}(\partial_0\Psi^{\dagger} + ik_0\Psi^{\dagger})U ; \\
ik_0\Psi^{\dagger}U & = & i\bar{\Psi}(k_0\gamma^0)U = i\bar{\Psi}
(\dsl{k} - k_i\gamma^i)U = i\bar{\Psi}(-mU) + \bar{\Psi}\gamma^i(-ik_iU)
\\
\therefore e^{ik\cdot x}ik_0\Psi^{\dagger}U & = & e^{ik\cdot x}
(-i\bar{\Psi}mU) + \bar{\Psi} \gamma^i(-\partial_i e^{ik\cdot x}) U ~ =
~ e^{ik\cdot x}\left(-im\bar{\Psi} + \partial_i \bar{\Psi} \gamma^i
\right)U \\
\therefore \partial_0\left(e^{ik\cdot x} \Psi^{\dagger}\right)
U(\vec{k},\sigma) & = & e^{ik\cdot x} \left\{\partial_0\bar{\Psi}
\gamma^0 + \partial_i\bar{\Psi}\gamma^i - im \bar{\Psi}\right\} U ~ = ~
(-ie^{ik\cdot x}) \left\{ i\bar{\Psi}\overleftarrow{\dsl{\partial}} +
m\bar{\Psi}\right\}U
\end{eqnarray*}
Hence, instead of $(-\Box_x + m^2)$ acting on $\Phi(x)$, we will have
$(-2mi)(i\overleftarrow{\dsl{\partial}} + m)$ acting on $\bar{\Psi}(x)$
and $(-2mi)(-i\overrightarrow{\dsl{\partial}} + m)$ acting on $\Psi(x)$.
The wavefunction factors will be: 
\begin{eqnarray}
\mbox{Particle in in-state:~} U(\vec{k},\sigma)e^{ik\cdot x} & , &
\mbox{Particle in out-state:~} \bar{U}(\vec{k},\sigma)e^{-ik\cdot x} \\
\mbox{anti-Particle in in-state:~} \bar{V}(\vec{k},\sigma)e^{ik\cdot x}
& , & \mbox{anti-Particle in out-state:~} V(\vec{k},\sigma)e^{-ik\cdot
x} 
\end{eqnarray}

For multi-particle case we will again have time ordering instruction. For
fermions, it is defined with an relative minus sign, \\
$~\hfill\boxed{T\left\{\Psi_{\alpha}(x)\Psi_{\beta}(y)\right\} ~ := ~
\theta(x^0 - y^0)\Psi_{\alpha}(x)\Psi_{\beta}(y) - \theta(y^0 -
x^0)\Psi_{\beta}(y)\Psi_{\alpha}(x) }\hfill~$.

\noindent and $T\left\{\Psi_{\alpha}(x)\Psi_{\beta}(y)\right\} =
-T\left\{\Psi_{\beta}(y) \Psi_{\alpha}(x)\right\}$.  Thus for fermions,
changing the ordering inside the time ordering generates a minus sign
for each exchange. We summarize the general matrix element for $m$
fermions going into $n$ fermions as,
\begin{eqnarray}
	_{out}\langle .. \vec{k}'_{j'}\sigma'_{j'} .. |..
	\vec{k}_j\sigma_j ..  \rangle_{in} & = &
	\left(\frac{-i}{\sqrt{Z_{\Psi}}}\right)^{m+n} \prod_{j,j'}\int
	d^4x'_{j'} d^4x_j
	U(\vec{k}_j,\sigma_j)\bar{U}(\vec{k}'_{j'},\sigma'_{j'})
	\nonumber \\
	& & \mbox{\hspace{0.5cm}}e^{-i\sum_{j'}k'_{j'}\cdot
	x'_{j'}}(-i\dsl{\partial}'_{1} + m)\dots(-i\dsl{\partial}'_{n} +
	m)\nonumber \\
	& & \mbox{\hspace{1.5cm}}\langle 0|T\left\{
	\Psi(x_1')..\Psi(x'_n)\bar{\Psi}(x_1)..\bar{\Psi}(x_m)\right\}
	|0\rangle \nonumber \\
	& & \mbox{\hspace{2.5cm}}(-i\overleftarrow{\dsl{\partial}_{1}} +
	m)\dots(-i\overleftarrow{\dsl{\partial}_{m}} + m)
	e^{+i\sum_{j}k_{j}\cdot x_{j}}
\end{eqnarray}
For anti-fermions in the in/out states, the spinors $U, \bar{U}$ are
changed to $V, \bar{V}$ as appropriate.

\subsubsection{LSZ reduction for Maxwell field} 
This is a bosonic field so there is no additional minus sign. Like the
Dirac field, it has two polarizations and thanks to the zero mass, these
are transverse polarizations.  We begin by recalling the mode
decomposition.
\begin{eqnarray*}
A_{\mu}(x) & = & \int\frac{d^3k} {\sqrt{2\omega(2\pi)^3}}
\sum_{\lambda=1,2} \left[ a_{\vec{k},\lambda}
	\underline{\varepsilon}_{\mu} (\vec{k},\lambda) e^{ik\cdot x} +
	a^*_{\vec{k},\lambda} \underline{\varepsilon}^*_{\mu}
(\vec{k},\lambda)e^{-ik\cdot x}\right] ~ ~ \\
& & \underline{\varepsilon}_0(\vec{k},\lambda) ~ := ~ -\frac{\vec{k}
\cdot \vec{\varepsilon} (\vec{k},\lambda)}{\omega_{\vec{k}}} ~ ~ , ~ ~
\underline{\varepsilon}_i(\vec{k},\lambda) ~ := ~ \left(\delta_i^{~j} -
\frac{k_i\ k^j} {\vec{k}^2}\right) \varepsilon_j(\vec{k},\lambda) ~ ~ ,
~ ~ \omega_{\vec{k}} = |\vec{k}| \nonumber \\
& & \underline{\vec{\varepsilon}} (\vec{k},\lambda) \cdot
\underline{\vec{\varepsilon}}^* (\vec{k},\lambda') ~ = ~
\delta_{\lambda,\lambda'} ~ ~ , ~ ~ \sum_{\lambda}
\underline{\varepsilon}_i(\vec{k},\lambda)
\underline{\varepsilon}^*_j(\vec{k},\lambda)  = ~ \left(\delta_i^{~j} -
\frac{k_i\ k^j} {\vec{k}^2}\right)
\end{eqnarray*}

This gives the inversion formulae as,
\begin{eqnarray*}
	a(\vec{k},\lambda) ~ = ~ i\varepsilon_i(\vec{k},\lambda)\int
	d^3x e^{-ik\cdot x} \overleftrightarrow{\partial_0}A^i(x) & , &
	a^{\dagger}(\vec{k},\lambda) ~ = ~
	-i\varepsilon^*_i(\vec{k},\lambda)\int d^3x e^{+ik\cdot x}
	\overleftrightarrow{\partial_0}A^i(x)
\end{eqnarray*}

In the transverse/radiation/Coulomb gauge,
$\underline{\varepsilon}_{\mu}(\vec{k}, \lambda) =
\varepsilon_{\mu}(\vec{k}, \lambda) ~ , ~\vec{k}\cdot
\vec{\varepsilon}(\vec{k}, \lambda) = 0$ and the equations of motion are
$\Box A_i(x) = 0$. The commutation relations are:
\[
	[A_i(t,\vec{x}), \pi_j(t, \vec{y})] ~ = ~ i\left(\delta_{ij} -
	\frac{\partial_i\partial_j}{\nabla^2}\right) ~ := ~ i\int
	\frac{d^3k}{(2\pi)^3} e^{ik\cdot(x-y)}\left(\delta_{ij}
	-\frac{k_ik_j}{\vec{k}^2}\right) 
\]

The reduction formula goes similar to the scalar. We will have $(-\Box_x
+ m^2) \to -\Box_x$ and the wave function factors would be $\boxed{
\mathcal{E}_{\mu}(\vec{k}, \lambda) := \Case{\varepsilon(\vec{k},
\lambda)}{\sqrt{2\omega_{\vec{k}}(2\pi)^3}} . }$ Then the general
$S-$matrix element for $m \to n$ particles takes the form,
\begin{eqnarray}
	_{out}\langle .. \vec{k}'_{j'}\lambda'_{j'} .. |..
	\vec{k}_j\lambda ..  \rangle_{in} & = &
	\left(\frac{-i}{\sqrt{Z_{A}}}\right)^{m+n} \prod_{j,j'}\int
	d^4x'_{j'} d^4x_j \mathcal{E}(\vec{k}_j,\lambda)
	\mathcal{E}^*(\vec{k}'_{j'},\lambda'_{j'}) \nonumber \\
	& & \mbox{\hspace{0.5cm}}e^{-i\sum_{j'}k'_{j'}\cdot
	x'_{j'}}(-\Box'_{1})\dots(-\Box'_{n})\nonumber \\
	& & \mbox{\hspace{1.5cm}}\langle 0|T\left\{ A_{\mu'_1}(x_1')..
	A_{\mu'_n}(x'_n) A_{\mu_1}(x_1).. A_{\mu_m}(x_m)\right\}
	|0\rangle \nonumber \\
	& & \mbox{\hspace{2.5cm}}(-\overleftarrow{\Box_1}) \dots
	(-\overleftarrow{\Box_{m}}) e^{+i\sum_{j}k_{j}\cdot x_{j}}
\end{eqnarray}

Having related the  $S-$matrix elements to the $n-$point functions of
the interacting fields, out next task is to try see if these can be
expressed in terms of the free fields in some systematic, well defined
way. This is achieved in the so-called covariant perturbation series.
\newpage
\section{Covariant Perturbation Theory}\label{PerturbationTheory}

In formulating the scattering theory for interacting fields, we
postulated the in/out fields satisfying the free field equations. More
importantly, we also postulated them to satisfy the same equal time
(anti-)commutation relations. With some assumptions of existence, these
suffice to develop a perturbation scheme to compute the $n-$point
functions.

Since the interacting and the free (in/out) fields satisfy the same
basic quantum conditions, it is plausible that they are related by some
{\em unitary transformation}. For systems with finitely many degrees of
freedom, it is theorem that guarantees existence of canonical
transformations at the classical level and unitary transformations at
the quantum level (thanks to the Stone-von-Neumann theorem). For field
theories, there is no such guarantee and we need to postulate the
required existence. Their utility gives a post-facto justification for
the assumptions. Let,

$~\hfill\boxed{\Phi(t,\vec{x}) = U^{-1}(t)\Phi_{in}(t,\vec{x})U(t) ~ ~ ,
~ ~ \Pi(t,\vec{x}) = U^{-1}(t)\Pi_{in}(t,\vec{x})U(t) \ .}\hfill ~$.

The fields satisfy the equations of motion in the form,
\begin{eqnarray}\label{HamEqns}
\partial_t \Phi_{in}(x) = i[H_{in}(\Phi_{in},\Pi_{in}), \Phi_{in}(x)] ~
& , & ~ \partial_t \Pi_{in}(x) = i[H_{in}(\Phi_{in},\Pi_{in}),
\Pi_{in}(x)] \\
\partial_t \Phi(x) = i[H(\Phi,\Pi), \Phi(x)] ~ & , & ~ \partial_t \Pi(x)
= i[H(\Phi,\Pi), \Pi(x)] 
\end{eqnarray}

These enable us to derive an equation for $U(t)$.
\begin{eqnarray*}
\partial_t\Phi_{in} & = & \partial_t[U(t)\Phi U^{-1}(t)] =
\partial_tU\cdot U^{-1}\Phi_{in} + U(\partial_t\Phi)U^{-1} +
\Phi_{in}\cdot U\partial_tU^{-1} \\
& = & [\partial_tU\cdot U^{-1}, \Phi_{in}] + iU[H(\Phi, \Pi), \Phi]
U^{-1} ~ ~ \mbox{But,} \\
U [H(\Phi, \Pi), \Phi] U^{-1} & = & [UH(\Phi,\Pi]U^{-1}, U\Phi U^{-1}] ~
= ~ [H(\Phi_{in}, \Pi_{in}), \Phi_{in}] \\
\therefore \partial_t\Phi_{in} & = & [ \partial_tU\cdot U^{-1} +
i\left\{ H(\Phi_{in},\Pi_{in}) - H_{in}(\Phi_{in}, \Pi_{in}) +
H_{in}(\Phi_{in}, \Pi_{in})\right\}, \Phi_{in} ] \\
\partial_t\Phi_{in} & = & \partial_t\Phi_{in} + [\partial_tU\cdot U^{-1}
+ iH_I(\Phi_{in}, \Pi_{in}), \Phi_{in}] ~ ~ ~ ~ \mbox{Similarly,} \\
\partial_t\Pi_{in} & = & \partial_t\Pi_{in} + [\partial_tU\cdot U^{-1} +
iH_I(\Phi_{in}, \Pi_{in}), \Pi_{in}] ~ . 
\end{eqnarray*} \newpage
Since $\partial_tU\cdot U^{-1} + iH_I(\Phi_{in},\Pi_{in})$ commutes with
both $\Phi_{in}, \Pi_{in}$, it must be a multiple of identity, say
$E_0\mathbb{1}$. This gives the equation determining the $U(t)$ as,
\begin{equation}\label{UEqn}
	\boxed{i\partial_tU(t) ~ = ~ H'_I(\Phi_{in},\Pi_{in})U(t)} ~ , ~
	\boxed{H'_I(\Phi_{in}, \Pi_{in}) := H(\Phi_{in},\Pi_{in}) -
	H_{in}(\Phi_{in}, \Pi_{in}) + E_0(t)} 
\end{equation}
To solve the equation, it is convenient to define the combination
$U(t,t') := U(t)U^{-1}(t')$. It follows that this combination too
satisfies the same first order equation with the initial condition,
$U(t,t) = \mathbb{1}$.  The integral form of this equation is:
\[
	U(t,t') ~ = ~ \mathbb{1} - i\int_{t'}^{t}dt'' H'_I(t'')U(t'',t')
	\ .
\]
This is solved by iteration. 

Define: $U_0(t,t') = \mathbb{1} , $ and for $n \ge 1,\ U_n(t,t') :=
\mathbb{1} - i\int_{t'}^{t}dt_n H'(t_n)U_{n-1}(t_n, t')$. Then $U(t,t')
= \sum_n U_n(t, t')$ is the formal solution of the integral equation.

It is trivial to verify the formal solution. It is formal because no
conditions are imposed for convergence of the series. The first few
terms are,
\begin{eqnarray}
U_0(t,t') & = & \mathbb{1} \nonumber \\
U_1(t,t') & = & \mathbb{1} - i\int_{t'}^tdt_1H'(t_1)\mathbb{1} \nonumber
\\
U_2(t,t') & = & \mathbb{1} - \int_{t'}^tdt_2 H'(t_2)U_1(t_2,t')
\nonumber \\
& = & \mathbb{1} +(-i)\int_{t'}^tdt_2H'(t_2) +
(-i)^2\int_{t'}^tdt_2\int_{t'}^{t_2}dt_1H'(t_2)H'(t_1) \nonumber \\
U_3(t,t') & = & \mathbb{1} + (-i)\int_{t'}^tdt_3
H'(t_3)\left\{\mathbb{1} +(-i)\int_{t'}^tdt_2H'(t_2) \right. \nonumber
\\
& & \mbox{\hspace{4.5cm}} \left. +
(-i)^2\int_{t'}^tdt_2\int_{t'}^{t_2}dt_1H'(t_2)H'(t_1) \right\}
\nonumber \\
& = & \mathbb{1} + (-i)\int_{t'}^tdt_3 H'(t_3) + (-i)^2 \int_{t'}^t dt_3
\int_{t'}^{t_3}dt_2H'(t_3)H'(t_2)\nonumber \\
& & \mbox{\hspace{3.8cm}} + (-i)^3\int_{t'}^tdt_3 \int_{t'}^{t_3} dt_2
\int_{t'}^{t_2} dt_1 H'(t_3)H'(t_2)H'(t_1) ~ ~ ~ ~  \nonumber \\
\therefore U_n(t,t') & = & \mathbb{1} + \sum_{k=1}^n
(-i)^k\int_{t'}^tdt_k \int_{t'}^{t_k}dt_{k-1}\dots\int_{t'}^{t_2}dt_1
H'(t_k)H'(t_{k-1})\dots H'(t_1) 
\end{eqnarray}

Consider the term $\int_{t'}^tdt_n\int_{t'}^{t_n}dt_{n-1}$. By
interchanging the order f integration we can write it as ($H'_n :=
H'(t_n)$),
\begin{eqnarray*}
\int_{t'}^tdt_n\int_{t'}^{t_n}dt_{n-1}H'_{n}H'_{n-1} & = &
\int_{t'}^tdt_{n-1}\int_{t_{n-1}}^{t}dt_{n} H'_{n}H'_{n-1} ~ = ~
\int_{t'}^tdt_{n}\int_{t_{n}}^{t}dt_{n-1} H'_{n-1}H'_{n} \\ 
\therefore LHS = \frac{1}{2}(LHS + RHS) & = & \frac{1}{2}\left[
\int_{t'}^tdt_n\int_{t'}^{t_n}dt_{n-1}H'_{n}H'_{n-1} +
\int_{t'}^tdt_{n}\int_{t_{n}}^{t}dt_{n-1} H'_{n-1}H'_{n} \right] \\
& = & \frac{1}{2}\left[
\int_{t'}^tdt_n\int_{t'}^{t_n}dt_{n-1}T\left\{H'_{n}H'_{n-1}\right\}
\right. \\
& & \mbox{\hspace{3.0cm}} \left. +
\int_{t'}^tdt_{n}\int_{t_{n}}^{t}dt_{n-1} T\left\{H'_{n}H'_{n-1}\right\}
\right] \\ 
& = & \frac{1}{2}\int_{t'}^tdt_n \int_{t'}^{t} dt_{n-1} T
\left\{H'_{n}H'_{n-1}\right\} 
\end{eqnarray*}
Similarly, we can symmetrize the higher order terms to replace each
product by their time ordered products and then combine the integrals to
have the same limits of integrations. The $n^{th}$ order term then takes
the form: $\Case{1}{n!} \int_{t'}^tdt_n \int_{t'}^tdt_{n-1} \dots
\int_{t'}^tdt_1 T\{H'(t_n)H'(t_{n-1})\dots H'(t_1)\}$. We write the
iterated formal solution of the (\ref{UEqn}) as,
\begin{eqnarray} \label{USoln}
U(t,t') & := & T\left(exp\left\{-i\int_{t'}^tdt' H'_I(\Phi_{in}(t'),
\Pi_{in}(t')) \right\}\right) \\
& = & \mathbb{1} + \sum_{n=1}^{\infty}\frac{(-i)^n}{n!}
\int_{t'}^tdt_1\dots\int_{t'}^tdt_n\ T\left\{H'_I(t_1)\dots
H'_{I}(t_n)\right\} 
\end{eqnarray}
The solution is thus obtained entirely in terms of the `in' fields. Note
that the solution gives $U(t,t')$ and {\em not} $U(t)$. While the
equations satisfied by $U(t)$ and $U(t,t')$ are the same, there is no
``initial'' condition provided for $U(t)$ and in a sense it is
ill-defined. There is no canonical way to deduce/define  $U(t)$ from
$U(t,t')$. Fortunately, $U(t,t')$ suffice.

\underline{Note:} From the definition it follows the useful composition
property that $U(t,t') = U(t,t'')U(t'',t')$, regardless of any
inequalities for $t''$. This may also be verified directly.

Let us see an implication of the existence of $U(t)$.  Consider an
$n-$point function for a scalar, $G(x_1,\dots,x_n) =\langle 0| T\left\{
\Phi(x_1)\dots \Phi(x_n)\right\}|0\rangle $. Insert in this $\Phi(x) =
U^{-1}(t)\Phi_{in}(x)U(t)$ which gives,
\begin{eqnarray*}
\Phi(x_1)\Phi(x_2)\dots \Phi(x_n) & = & U^{-1}(t_1) \Phi_{in}(x_1)
U(t_1)\cdot U^{-1}(t_2) \Phi_{in}(x_2) U(t_2)..U^{-1}(t_n)
\Phi(x_n)U(t_n) \\
& = & U^{-1}(t) \left[U(t,t_1)\Phi_{in}(x_1) U(t_1, t_2)
\Phi_{in}(x_2)\dots \right. \\
& & \left. \mbox{\hspace{1.0cm}} U(t_{n-1}, t_n) \Phi_{in}(x_n)
U(t_n,-t) \right]U(-t) ~ ~ , ~ ~ U(t,t') := U(t)U^{-1}(t')
\end{eqnarray*}

Observe that in the limit $t \to \infty$, the factor $U^{-1}(t)$ goes to
the extreme left and $U(t)$ to the extreme right. All other space-time
arguments being finite, we can take these factors outside the time
ordering symbol. The product of the fields $\Phi(x)$'s is already under
time ordering which then allows us to arrange all the factors within
$T\{\dots\}$ to be conveniently grouped. In particular, all the $U(t_i,
t_j)$ can be combined using their composition property noted above.

With the limit of $t \to \infty$ implicit, we can write the $n-$point
function as,
\begin{equation}\label{GwithU}
G(x_1,\dots,x_n) ~ = ~\langle 0|U^{-1}(t)T\left\{\Phi_{in}(x_1) \dots
\Phi_{in}(x_n)\ exp\left(-i\int_{-t}^{t}dt' H'_{I}(t')\right)\right\}
U(t)|0\rangle \ .
\end{equation}

Next, we evaluate the $U(-t)|0\rangle$ in the limit $t\to \infty$ and
its adjoint. 

\underline{Claim:} $U(-t)|0\rangle \to \lambda_-|0\rangle$ as $t\to
\infty$.

\underline{Proof:} Consider an in-state containing at least one particle
of momentum $k, |\alpha,k\rangle_{in}$. Then,
\begin{eqnarray*}
	_{in}\langle\alpha,k|U(-t)|0\rangle & = &
	_{in}\langle\alpha|a_{in}(\vec{k})U(-t)|0\rangle \\
	& = & i\int_{\Sigma_{t'}}
	\frac{d^3x}{\sqrt{2\omega_{\vec{k}}(2\pi)^3}}e^{-ik\cdot
	x}\overleftrightarrow{\partial'_{0}}_{in}\langle \alpha|
	\Phi_{in}(-t',\vec{x})U(-t)|0\rangle
\end{eqnarray*}
We have used the inversion formula for $a_{in}$ by choosing an arbitrary
$\Sigma_{t'}$ hyper-surface and $a_{in}$ is of course independent of
this. 

In the integrand above, substitute $\Phi_{in}(t',\vec{x}) = U(t')
\Phi(t',\vec{x}) U^{-1}(t')$ and evaluate the time derivative.
\begin{eqnarray*}
\mbox{Integrand} & = & e^{-ik\cdot x}\langle\alpha| \left[\dot{U}(-t')
U^{-1}(-t') \Phi_{in}(-t',\vec{x}) U(-t) + \Phi_{in}(-t',\vec{x}) U(-t')
\partial_{t'}U^{-1}(t') U(-t) \right. \\
& & \left. + U(-t')\partial'_0\Phi(-t')U^{-1}(-t')U(-t) -
U(-t')\Phi(-t')U^{-1}(-t')U(-t)(\partial'_0(-ik\cdot x))\right]|0\rangle
\end{eqnarray*}
Now take choose $t' = t$. Then, the first two terms on the r.h.s.
combine as,
\[
[\dot{U}(-t)U^{-1}(-t), \Phi_{in}(-t, \vec{x})]U(-t) ~ = ~
-i[H'_I(\Phi_{in}, \Pi_{in}), \Phi_{in}(-t, \vec{x})]U(-t) ~ = ~ 0 \ .
\]
In the usual theories, the interaction Hamiltonian has no $\Pi$
dependence and hence at equal time, commutes with the field giving 0.

In the last two terms, the $U^{-1}(-t')U(-t)$ factor becomes
$\mathbb{1}$ for $t' = t$. These terms combine to give, $_{in}
\langle\alpha| U(-t) \left(e^{-ik\cdot x} \overleftrightarrow
{\partial_0} \Phi(-t) \right)|0\rangle$.  Taking the  limit $t \to
\infty$ and invoking the asymptotic condition, converts $\Phi(x)$ into
$\sqrt{Z}\Phi_{in}(x)$.  The spatial integration converts the expression
into $\sqrt{Z}_{in} \langle\alpha| U(-t) a_{in}(\vec{k})|0\rangle = 0$.

Thus we conclude that $_{in}\langle\alpha,k|U(-t)|0\rangle \to 0$ as $t
\to \infty$ for {\em all} states containing at least one particle.
$U(-t)|0\rangle$ must therefore be proportional to $|0\rangle$ itself
thereby proving the claim.

By similar reasoning, we obtain $U(t)|0\rangle \to \lambda_+|0\rangle$
as $t \to \infty$.

\underline{Corollary:}
\begin{equation} \label{GNorm}
\lambda_+^*\lambda_- = \left[\langle 0|T exp \left(-i
\int_{-\infty}^{\infty}dt' H'_I(t') \right) |0\rangle \right]^{-1}
\end{equation}
The proof is simple,
\begin{eqnarray*}
	\lambda_+^*\lambda_- & = & \lim_{t\to \infty}\langle
	0|U^{-1}(t)|0\rangle\langle 0|U(-t)|0\rangle \\
	& \approx & \langle 0|U^{-1}(t)|U(-t)|0\rangle ~ ~ \mbox{since
	intermediate states do not contribute by above result}\\
	& \approx & \langle 0|U(-t)U^{-1}(t)|0\rangle ~ = ~ \langle
	0|U(-t,t)|0\rangle \\
	& \approx & \langle 0|(U(t,-t))^{-1} |0\rangle ~ = ~ (\langle
	0|U(t,-t)|0\rangle)^{-1} \ .
\end{eqnarray*}
The final expression follows by substituting the formal solution in the
limit $t \to \infty$.

Using the equations (\ref{GwithU}, \ref{GNorm}), we express the $n-$
point function as,
\begin{equation}\label{GEqn}
	\boxed{ G(x_1,\dots,x_n) ~ = ~ \frac{\langle
	0|T\left\{\Phi_{in}(x_i)\dots\Phi_{in}(x_n)exp
	-i\int_{-\infty}^{\infty}dt' H_I(t')\right\}|0\rangle}{\langle
0|T\left\{exp -i\int_{-\infty}^{\infty}dt' H_I(t')\right\}|0\rangle} }
\end{equation}
Notice that the multiple of identity $E_0(t)\mathbb{1}$ has been
canceled between the numerator and the denominator before taking the
limit $t\to \infty$.

Suffice it to say that analogous expressions exist for other fields as
well. However we will not write them explicitly.

The reduction formulae give $S-$matrix elements in terms of the
$n-$point functions and the $n-$point functions have a perturbative
expansion in terms of the free fields alone. These two together are the
master formulae for computations. 

A further simplification is provided by the {\em Wick's Theorem}.
\subsection{Normal ordering and Wick's theorem}
Recall that the mode expansion of the {\em free} quantum fields can be
grouped as $\Phi(x) = \Phi_+(x) + \Phi_-(x) $ with  $\Phi_{\pm}(x)$
denoting the sum over positive/negative  frequency modes. The positive
frequency part is made up of annihilation operators while the negative
frequency part is made up of creation operators alone. For a Hermitian
quantum field, the negative frequency part is the Hermitian adjoint of
the positive frequency part. A product of fields at different points
will mix these parts and the idea is to bring all annihilation operators
to the right and the creation operators to the left. The vacuum
expectation value of so ordered groups will of course be zero. In this
process, several commutators of fields at different space-time points
are generated, but {\em they all are c-numbers}. This leads to a
simplification. 

We define {\em normal ordered products} of two {\em free} fields as:
\begin{eqnarray}
\mbox{For bosons~ ~}	:\Phi_-(x)\Phi_+(y): ~ = ~ \Phi_-(x)\Phi_+(y) ~
~ &,&~ ~ :a^{\dagger} a: ~ := ~ a^{\dagger} a \\
:\Phi_+(x)\Phi_-(y): ~ = ~ \Phi_-(y)\Phi_+(x) ~ ~ &,&~ ~ :a a^{\dagger}:
~ := ~ a^{\dagger} a \\
\mbox{For fermions~ ~}	:\Psi_-(x)\Psi_+(y): ~ = ~ \Psi_-(x)\Psi_+(y) ~
~ &,&~ ~ :b^{\dagger} b: ~ := ~ b^{\dagger} b \\
:\Psi_+(x)\Psi_-(y): ~ = ~ - \Psi_-(y)\Psi_+(x) ~ ~ &,&~ ~ :b
b^{\dagger}: ~ := ~ - b^{\dagger} b 
\end{eqnarray}
For product of two or more positive (or negative) frequency fields, the
normal ordering does not change order. It is immediately obvious that
$\langle 0| : Operator : |0\rangle = 0$. 

{\em Note that the normal ordering are meaningful {\em only} for
products of free fields (in/out). Hence the $in/out$ suffixes are
suppressed for these.} 

To get a general form for several bosonic as well as fermionic free
fields, notice that for a quantum field $Q(x) = Q_+(x)+ Q_-(x)$ a
product of $n$ fields generates terms with the $k$ number of $Q_+$'s
interspersed with $(n-k)$ number of $Q_-$'s, with the order of the
space-time points maintained, with $k = 0,\dots,n$. Under normal ordering
each of these terms will shift the $Q_+$'s to the right and $Q_-$'s to
the left, generating a permutation of the space-time points, say $p$. If
$Q$ is a fermionic field, the term with permutation $p$ will get a
factor of $\sigma_p = sgn(p)$ while a bosonic field will have $\sigma_p
= 1$. Note that for any given $k$, there will be several terms eg
$Q_1Q_2, Q_3Q_7,\dots$ etc. Each will generate its own permutation under
normal ordering. With this understood, we can write the general
expression for normal ordered product of $n-$fields as,

$~\hfill \boxed{
		:Q(x_1)\dots Q(x_n): ~ := ~ \sum_{A,B} \sigma_p
		\prod_{i\in A} Q_-(x_{p(i)}) \prod_{j\in B}
		Q_+(x_{p(j)})\ .
}\hfill ~$ \\

The $A,B$ are two groups of space-time points corresponding to the $k,
(n-k)$ mentioned above and the sum refers to various possible groupings
of the space-time points within each class. 

{\em Exercise:} Verify for $n = 4$.

Consider now the relation between time ordered products and normal
ordered products. For a single field, both orderings are trivial and
their vacuum expectation value vanishes: $\langle 0|T\{\phi_{in}(x)\}|0
\rangle = 0 =\langle 0|:\phi(x):|0\rangle$. 

For time ordered product of two fields, for any particular instances of
time they appear in a particular order which can be explicitly put in
the normal ordered form, eg. $a(\vec{k})a^{\dagger}(\vec{k}') =
``\mathbb{1}" + a^{\dagger}(\vec{k}')a(\vec{k}) =
:a(\vec{k})a^{\dagger}(\vec{k}'): + ``\mathbb{1}"$. The last term is a
c-number (multiple of the identity operator). Similarly, $\Phi(x)\Phi(y)
= :\Phi(x)\Phi(y): +\ c-$number.  And the c-number is trivial to
evaluate by just taking the vev (vacuum expectation value): $c-$number =
$\langle 0|\Phi(x)\Phi(y)|0\rangle$. Noting that $T\{\Phi(x)\Phi(y)\} =
\theta(x^0 - y^0)\Phi(x)\Phi(y) + \theta(y^0-x^0)\Phi(y)\Phi(x)$, we
get, $\boxed{T\{\Phi(x)\Phi(y)\} = :T\{\Phi(x)\Phi(y)\}: + \langle
0|T\{\Phi(x)\Phi(y)\}|0\rangle}$. Notice that normal ordering and time
ordering commute: $:T\{A(x)B(y)\}: = T\{:A(x)B(y):\}$, for all
non-trivial $A(x), B(y)$ operators i.e. for operators which are
polynomials in the basic field operators of degree greater that zero.
The generalization of this to product of arbitrary number of field is
the Wick's theorem. We state it explicitly for a scalar field.

\underline{Wick's Theorem:}
\begin{eqnarray*}
T\{\Phi(x_1)\dots\Phi(x_n)\} & = & :T\{\Phi(x_1)\dots\Phi(x_n)\}: \\
& & + \left[\langle 0|T\{\Phi(x_1)\Phi(x_2)\}|0\rangle
:T\{\Phi(x_3)\dots\Phi(x_n)\}: + \mbox{~permutations}\right] \\
& & + \left[\langle 0|T\{\Phi(x_1)\Phi(x_2)\}|0\rangle \langle
0|T\{\Phi(x_3)\Phi(x_4)\}|0\rangle \times \right. \\
& & \left. \mbox{\hspace{4.0cm}} :T\{\Phi(x_5)\dots\Phi(x_n)\}: +
\mbox{~permutations}\right] \\
& &  \mbox{\hspace{1.0cm}} \vdots~\vdots \\
& & + \left\{ 	\begin{array}{l}
		\left[\langle 0|T\{\Phi_1\Phi_2\}|0\rangle \dots \langle
		0|T\{\Phi_{n-1}\Phi_n\}|0\rangle + \mbox{~permutations}
	\right] \\
		\mbox{\hspace{8.0cm} ($n$ even)} \\
		\left[\langle 0|T\{\Phi_1\Phi_2\}|0\rangle \dots \langle
		0|T\{\Phi_{n-2}\Phi_{n-1}\}|0\rangle \Phi_n \right. \\
		\left. \mbox{\hspace{4.9cm}} + \mbox{~permutations}
		\right] ~ \mbox{($n$ odd)}
		\end{array}
	\right.
\end{eqnarray*}
In the last line we have used the abbreviation $\Phi_i := \Phi(x_i)$ and
we will use the same when convenient.

\underline{Proof:}

The proof is by induction and we have already seen the validity for $n =
1, 2.$ Assume the theorem is true for $n = n$. Include an extra field
$\Phi(x_{n+1})$ and choose $t_{n+1}$ to be the earliest instant so that
it stays on the extreme right. Then,
\begin{eqnarray*}
T\{\Phi_1\dots \Phi_{n+1}\} & = & T\{\Phi_1\dots \Phi_n\}\Phi_{n+1} \\
& = & \left[:T\{\Phi_1\dots \Phi_n\}: + \sum_{perm}\langle
0|T\{\Phi_1\Phi_2\}|0\rangle:T\{\dots\}: + \dots \right]\Phi_{n+1}~ , \\
:\Phi_1\dots \Phi_n:\Phi_{n+1} & = &
\left\{\sum_{A,B}\sigma_p\prod_{i\in A}\Phi^-_i\ \prod_{j\in
B}\Phi^+_j\right\}\left(\Phi^+_{n+1} + \Phi^-_{n+1}\right) \\
& = & \sum_{A,B}\sigma_p\prod_{i\in A}\Phi^-_i\ \prod_{j\in
B}\Phi^+_j\Phi^+_{n+1} \ + \\
& & \mbox{\hspace{1.0cm}} \sum_{A,B}\sigma_p\prod_{i\in A}\Phi^-_i\
\left\{\Phi_{n+1}^- \prod_{j\in B}\Phi^+_j + \left[\prod_{j\in
B}\Phi^+_j\ , \ \Phi_{n+1}^-\right]\right\}
\end{eqnarray*}
The first two terms in the last equation have the requisite normal
ordered form for $n+1$ fields. The third term has several commutators,
$[\Phi^+_j, \Phi^-_{n+1}]$ which again are c-numbers and can be
evaluated by taking vev. That is,
\begin{eqnarray*}
[\Phi^+_j, \Phi^-_{n+1}] & = & \langle 0|(\Phi^+_j\Phi^-_{n+1} -
\Phi^-_{n+1}\Phi^+_j)|0\rangle ~ = ~ \langle
0|\Phi^+_j\Phi^-_{n+1}|0\rangle + 0 \\
& = &\langle 0|\Phi_j\Phi_{n+1}|0\rangle ~ = ~\langle
0|T\left\{\Phi_j\Phi_{n+1}\right\}|0\rangle  ~ ~ , ~ ~ \mbox{since
$t_{n+1}$ is the earliest.} \\
\therefore\ :\Phi_1\dots \Phi_n:\Phi_{n+1} & = & :\Phi_1\dots
\Phi_n\Phi_{n+1}: + \sum_j\sigma_p :\Phi_1\dots\Phi_{j-1}
\Phi_{j+1}\dots\Phi_n:\langle 0|T\left\{ \Phi_j \Phi_{n+1}
\right\}|0\rangle
\end{eqnarray*}
Thus, inserting the time ordering, the first term, $:T\{\Phi_1\dots
\Phi_n\}:\Phi_{n+1}$ gets expressed as $:T\{\Phi_1\dots
\Phi_n\Phi_{n+1}\}: + \ \mbox{\{terms which have a structure similar to
the second term\}}$ and so on. Hence the induction hypothesis allows the
expression to hold for $n+1$ and the theorem is proved by induction.

{\em Exercise:} Verify the theorem for $n = 3, 4$.

Observe that $G(x_1,\dots,x_n) =\langle 0|T\{\Phi(x_1) \dots
\Phi(x_n)\}|0\rangle$ vanishes for odd $n$ and equals
$\sum_{perm}\sigma_p \langle 0|T\{\Phi(x_1) \Phi(x_2)\}|0\rangle\dots
\langle 0|T\{\Phi(x_{n-1}) \Phi(x_n)\}|0\rangle$ for even $n$. 

Thus, {\em $2n-$point function of free fields is the sum of products of
$n$, $2-$point functions of free fields. We denote:} $\boxed{\langle
0|T\{\Phi(x)\Phi(y)\}|0\rangle =: i\Delta_F(x-y) =: \mbox{(Feynman
propagator)}}$. 

We have the $S-$matrix elements in terms of the $n-$point functions of
interacting quantum fields, expressed in terms of vev of time ordered
products of free fields and thanks to Wick's theorem these are expressed
in terms of $2-$point functions of free fields. In short, the $S-$matrix
elements are obtained from a bunch of $2-$point functions of free
fields, together with some integrations and operations of $(\Box -
m^2)$.  It remains to have a convenient algorithm for computations.

A few remarks are in order. The normal ordering is introduced here as a
simplification step. It works because the postulated {\em unique}
Poincare invariant vacuum state is annihilated by all (free field)
annihilation operators.  This also {\em requires} the Poincare
generators of the free fields to be normal ordered\footnote{Note that
	while abelian symmetries could allow a non-zero multiple of
	identity for invariance (i.e.  $P^{\mu}|0\rangle =
	\alpha^{\mu}|0\rangle$), non-abelian symmetry generators have to
	annihilate the vacuum. Hence the Lorentz generators must
	annihilate the vacuum and then the commutator of $M^{\mu\nu}$
with $P^{\lambda}$ shows that the $\alpha^{\mu} = 0$ must also hold.
Poincare group being non-abelian must have its generators being normal
ordered for the vacuum to be invariant.}. The Poincare generators of the
interacting fields also need to annihilate the vacuum. Thanks to the
assumed unitary operator linking the free and the interacting fields,
these generators too are expressed in terms of the free field
creation-annihilation operator. These expressions, in particular the
Hamiltonian, must also be normal ordered for Poincare invariance. We
will make this explicit for the interaction Hamiltonian $H_I(\Phi_{in},
\Pi_{in})$.

In the quantum optics context, the normal ordering was used since the
detectors worked by absorbing a quantum from the field and in principle,
for a detector working by emission of a quantum would require
anti-normal ordering. Such an option is not available with manifest
Poincare invariance {\em with a unique invariant vacuum state}.

An explicit example of evaluation of an $n-$point function will be
useful. Consider $G_2(x_1, x_2)$ computed to second order in $H_I$.
Consider the numerator first. {\em All fields are `in' fields and the
`in' subscript is suppressed below.}
\begin{eqnarray*}
G_2(x_1,x_2)|_{Nr} & = &\langle 0|T\{\Phi(x_1)\Phi(x_2) \left(\mathbb{1}
	-i\int_{-\infty}^{\infty}dt'H_I(t') \right. \\
& & \left. \mbox{\hspace{3.0cm}} + \frac{(-i)^2}{2!}
\int_{-\infty}^{\infty} dt_1 dt_2 T\{H_I(t_1) H_I(t_2)\} \right)\}
|0\rangle \\
& = & \langle 0|T\{\phi(x_1)\Phi(x_2)\}|0\rangle - i
\int_{-\infty}^{\infty} dt_1\langle 0|T\{\Phi(x_1) \Phi(x_2) H_I(t_1)\}
|0\rangle \\
& & \mbox{\hspace{2.0cm}} + \frac{(-i)^2}{2} \int_{-\infty}^{\infty}
dt_1 dt_2 \langle 0|T\{ \Phi(x_1) \Phi(x_2) H_I(t_1) H_I(t_2) \}
|0\rangle
\end{eqnarray*}
Now let us take a specific interaction Hamiltonian, simplest non-trivial
being $H_I(t) := g\int d^3y:\Phi(t,\vec{y})^3:$ . As noted above, for
the case of Poincare invariant vacuum, the Poincare generators and hence
in particular the Hamiltonian must be normal ordered. The time
integrations combine with these spatial integration to give a space time
integral. Thus, we get,
\begin{eqnarray*}
G_2(x_1,x_2)|_{Nr} & = & \langle 0|T\{\phi(x_1)\Phi(x_2)\}|0\rangle - ig
\int\ d^4y\langle 0|T\{\Phi(x_1) \Phi(x_2) :\Phi^3(y):\} |0\rangle \\
& & \mbox{\hspace{0.5cm}} + \frac{(-i)^2}{2}g^2 \int\ d^4y_1\int d^4y_2
\langle 0|T\{ \Phi(x_1) \Phi(x_2) :\Phi^3(y_1)::\Phi^3(y_2): \}
|0\rangle \\
& = & i\Delta_F(x_1 - x_2) - \frac{g^2}{2} \int\ d^4y_1\int d^4y_2
\langle 0|T\{ \Phi(x_1) \Phi(x_2) :\Phi^3(y_1)::\Phi^3(y_2): \}
|0\rangle
\end{eqnarray*}
The first term is just the Feynman propagator and the $o(g)$ term
vanishes since it is a 5-point function. The non-trivial term is the
$o(g^2)$ term. This is not in the form of the Wick's theorem, in that
there are normal ordered factors. We can replace the normal ordered
product in terms of unordered product minus the product of two point
functions i.e. using the Wick's theorem in reverse form. For instance,
consider (for three distinct points)
\begin{eqnarray*}
:\Phi(y_1)\Phi(y'_1)\Phi(y''_1): & = & \Phi(y_1)\Phi(y'_1)\Phi(y''_1)
-\langle 0|\Phi(y_1)\Phi(y'_1)|0\rangle\Phi(y''_1) \\
& & \mbox{\hspace{2.0cm}} -\langle 0|\Phi(y'_1) \Phi(y''_1)|0\rangle
\Phi(y_1) -\langle 0|\Phi(y_1) \Phi(y''_1)|0\rangle \Phi(y'_1) 
\end{eqnarray*}
And likewise for the second factor of $:\Phi(y_2)^3:$. If we now
substitute these in the $o(g^2)$ term and eventually take the
coincidence limit of the primed arguments, we see that the two point
functions of coincident points cancel out. The net result result is that
in writing the permutation terms, we omit the Feynman propagators of
coincident points and this holds because we use the normal ordered
interaction Hamiltonian. If we did not normal order the interaction
Hamiltonian then these terms, which are formally divergent, will remain
and will have to be handled differently. This is what will happen in the
functional integral method which will be discussed later. Keeping this
in mind, the $o(g^2)$ term under vev becomes,
\begin{eqnarray*}
\langle 0|T\{\Phi(x_1)\Phi(x_2):\Phi^3(y_1): :\Phi^3(y_2):\}|0\rangle ~
\mbox{\hspace{8.0cm}$~$} & & \\
=
i\Delta_F(x_1-x_2).i\Delta_F(y_1-y_2).i\Delta_F(y_1-y_2).i\Delta_F(y_1-y_2)
+ \mbox{~permutations} & & \\
= (i^4)\left[\Delta_F(x_1-x_2)\Delta_F(y_1-y_2)^3 +
\Delta_F(x_1-y_1)\Delta_F(x_2-y_2)\Delta_F(y_1-y_2)^2 \right. & & \\ 
\left. + \Delta_F(x_1-y_2)\Delta_F(x_2-y_1)\Delta_F(y_1-y_2)^2 \right]
\end{eqnarray*}
There are no other `pairings' of space-time points as they would involve
Feynman propagators with coincident arguments.

The denominator of the $n-$point function has the same structure as the
numerator except for the $\Phi(x_1)\Phi(x_2)$ fields i.e.
\begin{eqnarray*}
G_2(x_1,x_2)|_{Dr} & = &\langle 0|0\rangle - ig \int\ d^4y\langle
0|T\{:\Phi^3(y):\} |0\rangle \\
& & \mbox{\hspace{0.5cm}} + \frac{(-i)^2}{2}g^2 \int\ d^4y_1\int d^4y_2
\langle 0|T\{ :\Phi^3(y_1)::\Phi^3(y_2): \} |0\rangle \\
& = & 1  - \frac{g^2}{2} \int\ d^4y_1\int d^4y_2 \langle 0|T\{
:\Phi^3(y_1): :\Phi^3(y_2): \} |0\rangle \\
& = & 1 + \frac{(-ig)^2}{2} (CF) \int\ d^4y_1\int d^4y_2\
(i\Delta_F(y_1-y_2))^3
\end{eqnarray*}
Here, $CF$ is the {\em combinatorial factor} - number of distinct ways
of effecting the same pairings - and equals $3!$ in the example above.

The generalization is fairly obvious. For definiteness consider the
interaction Hamiltonian {\em density} to be $H_I(\Phi(y)) :=
\Case{g}{k!} :\Phi^k(y):$.  Then, it is apparent that the numerator of a
general $n-$point function is a sum of the $m^{th}$ order contributions,
$m = 0,1,\dots$. The $m^{th}$ order contribution has the form:

$\bullet ~ \Case{(-ig)/k!)^m)}{m!}\int d^4y_1\dots d^4y_m \langle 0|
T\{\Phi(x_1)\dots\Phi(x_n) :\Phi^k(y_1): \dots: \Phi^k(y_m): \}|0\rangle
\ ;$

$\bullet ~$ The Wick's theorem gives a sum of terms, each of which is a
product of a total of $(n+km)/2$ Feynman propagators,
$i\Delta_F(x_j-y_k)$,  whose arguments give the `pairing' of space-time
points from the set: $x_1,\dots,x_n, k-\mbox{copies of} \ y_1, \dots,
k-\mbox{copies of} \ y_m$. The normal ordered form of interaction
Hamiltonian ensures that the paired points are necessarily {\em
distinct}. The $x_1,\dots,x_n$ points are assumed to be distinct. A
`pairing' is also called a `Wick contraction'. The sum of terms is
generated from any particular one by permutations. If $n + km$ is odd,
the contribution vanishes;

$\bullet ~$ There would also be combinatorial factor counting distinct
ways of getting the same set of pairing (or Feynman propagators). The
denominator has similar contributions except the $x_1,\dots,x_n$ points.

There is a convenient diagrammatic representation to keep track of the
generated terms as well as associating specific factors and integrals
with them. The diagrams are the {\em Feynman Diagrams} while the
prescription to construct associated integral is given by the {\em
Feynman rules}.

\underline{Procedure to generate Feynman diagrams for Green's function:} 

$\bullet~$ For each distinct $x_1, \dots, x_n$, draw a {\em vertex} with
a single edge.  Call these vertices as {\em external vertices};

$\bullet~$ For normal ordered monomial of order $k$, draw a vertex $y_j$
with $k$ edges sticking out. Call these as the {\em internal vertices};

$\bullet~$ Wick contract or pair, by joining two edges from two distinct
vertices, internal or external. If the pair has one or both vertices to
be external vertices, the line is called an {\em external line} while a
pairing with both vertices being internal, is called an {\em internal
line}. This constructs a Feynman diagram with vertices and
internal/external lines.

\underline{Procedure to associate integral with a Feynman diagram:}

$\bullet~$ With each contraction, associate a Feynman propagator,
$i\Delta_F(z'-z'')$.

$\bullet~$ With each internal vertex, associate a factor of $(-i)\times
$coefficient of the monomial of the fields in $H_I(\Phi(y))$ and an
integral over its space-time location, $\int d^4y$. 

$\bullet~$ Associate a numerical factor as a ratio. In the numerator,
count the number of {\em distinct} ways of generating the same diagram
(or the same contractions). In the denominator, put $m!$ if $m$ is the
total number of internal vertices.

For the above example of $G_2(x_1,x_2)$ to second order for $H_I =
g:\Phi(y)^3:$, here is a summary: 

\begin{center}
	\begin{tabular}{|c|c|c|}
\hline 
$o(g)$ & 
\parbox[c]{3.0cm}{
	\begin{tikzpicture}
		\begin{feynman}
                        \vertex (v);
			\vertex [above =0.4of v] (a) ;
			\vertex [below left=0.4of v] (b) ;
			\vertex [below right=0.4of v] (c) ; 
			\vertex [left =of v] (d) ; 
			\vertex [left =0.95of v] (d') ; 
			\vertex [right =of v] (e) ; 
			\vertex [right =0.95of v] (e') ; 

			\diagram* { (d') -- [scalar,half
				left,looseness=1.5 ] (a),
				(e') -- [scalar,half left,looseness=1.5] (c) , 
				(v) -- (a) ,
				(v) -- (b) ,
				(v) -- (c) ,
				(d) -- (d'),
				(e) -- (e'),
                        };
                \end{feynman}
        \end{tikzpicture}
}
& \parbox[c]{0.5\textwidth}{Has one loose end and vev is zero.} \\
\hline 
$o(g^2)$ & & \\
\hline 
(a) & 
\parbox[c]{4.8cm}{
	\begin{tikzpicture}
		\begin{feynman}
			\vertex (v) ;
			\vertex [left =1.5of v] (a) ;
			\vertex [right =0.25of a] (a') ;
			\vertex [left =0.1of a] (a0) {\(x_1\)};
			\vertex [right =1.5of v] (b) ;
			\vertex [left =0.25of b] (b') ;
			\vertex [right =0.1of b] (b0) {\(x_2\)};
			\vertex [above =1.0of v] (c) ;
			\vertex [above =0.25of c] (c0) {\(y_1\)};
			\vertex [below =1.0of v] (d) ;
			\vertex [below =0.25of d] (d0) {\(y_2\)};

			\vertex [below left=0.25of c] (c1) ; 
			\vertex [below right=0.25of c] (c2) ; 
			\vertex [above =0.25of c] (c3) ; 

			\vertex [above left=0.25of d] (d1) ; 
			\vertex [above right=0.25of d] (d2) ; 
			\vertex [below =0.25of d] (d3) ; 

			\diagram* { (a) -- (a'),
				(b') -- (b), 
				(c) -- (c3), (c) -- (c1), (c) -- (c2),
				(d) -- (d3), (d) -- (d1), (d) -- (d2),
				(b') -- [scalar,half
				left,looseness=0.75,edge label=\tiny{3}] (d3),
				(a') -- [scalar,half
					left,looseness=0.75,edge
				label=\tiny{3}] (c3),
				(c1) -- [scalar,half
					right,looseness=0.5,edge
				label=\tiny{1}] (d1),
				(c2) -- [scalar,half
			left,looseness=0.5,edge label=\tiny{2}] (d2),
                        };
                \end{feynman}
        \end{tikzpicture}
}
& \parbox[c]{0.5\textwidth}{ 
	$(-ig)^2 \left(\frac{3.3.2.1}{2!}\right)\int d^4y_1d^4y_2\
(i\Delta_F(x_1-y_1)) \times (i\Delta_F(x_2-y_2))(i\Delta_F(y_1-y_2))^2$
} \\
\hline 
(b) & 
\parbox[c]{3.0cm}{
	\begin{tikzpicture}
		\begin{feynman}
			\vertex (v) ;
			\vertex [left =1.5of v] (a) ;
			\vertex [right =0.25of a] (a') ;
			\vertex [right =1.5of v] (b) ;
			\vertex [left =0.25of b] (b') ;
			\vertex [above =1.0of v] (c) ;
			\vertex [above =0.8of c] (c0) ;
			\vertex [below =1.0of v] (d) ;

			\vertex [below left=0.25of c] (c1) ; 
			\vertex [below right=0.25of c] (c2) ; 
			\vertex [above =0.25of c] (c3) ; 

			\vertex [above left=0.25of d] (d1) ; 
			\vertex [above right=0.25of d] (d2) ; 
			\vertex [below =0.25of d] (d3) ; 

			\diagram* { (a) -- (a'),
				(b') -- (b), 
				(c) -- (c3), (c) -- (c1), (c) -- (c2),
				(d) -- (d3), (d) -- (d1), (d) -- (d2),
				(b') -- [scalar,half
				right,looseness=0.75] (c3),
				(a') -- [scalar,half
				right,looseness=0.75] (d3),
				(c1) -- [scalar,half right,looseness=0.5] (d1),
				(c2) -- [scalar,half left,looseness=0.5] (d2),
				(c3) -- [scalar,opacity=0.1] (c0),
                        };
                \end{feynman}
        \end{tikzpicture}
}
& \parbox[c]{0.5\textwidth}{
	$(-ig)^2
\left(\frac{3.3.2.1}{2!}\right)\int d^4y_1d^4y_2\ (i\Delta_F(x_1-y_2))
\times (i\Delta_F(x_2-y_1))(i\Delta_F(y_1-y_2))^2$ 
}\\
\hline 
(c) & %
\parbox[c]{3.0cm}{
	\begin{tikzpicture}
		\begin{feynman}
			\vertex (a) {\(x_1\)};
			\vertex [right = 0.25of a] (a');
			\vertex [right =2.0of a] (b){\(x_2\)} ;
			\vertex [left =0.25of b] (b') ;

			\vertex [below =1.5of a] (c) ;
			\vertex [above right=0.6of c] (c1) ;
			\vertex [right=0.25of c] (c2) ;
			\vertex [below right=0.6of c] (c3) ;
			\vertex [left =0.1of c] (c0) {\(y_1\)} ;
			\vertex [below =0.8of c3] (c00) ;

			\vertex [below =1.5of b] (d) ;
			\vertex [above left =0.6of d] (d1) ; 
			\vertex [left =0.25of d] (d2) ; 
			\vertex [below left =0.6of d] (d3) ; 
			\vertex [right =0.1of d] (d0) {\(y_2\)} ;

			\vertex [above =0.5of a] (a'') ;
			\vertex [above =0.5of b] (b'') ;
			\vertex [below =0.5of c] (c'') ;
			\vertex [below =0.5of d] (d'') ;

			\diagram* { (a) -- (a'),
				(a') -- [scalar] (b'),
				(b') -- (b),

				(c) -- (c1), (c) -- (c2), (c) -- (c3),
				(d) -- (d1), (d) -- (d2), (d) -- (d3),

				(c1) -- [scalar,half left,
					looseness=0.5,edge label=\tiny{3}] (d1),
				(c2) -- [scalar,edge label=\tiny{2}] (d2),
				(c3) -- [scalar,half right,
					looseness=0.5,edge label=\tiny{1}] (d3),

				(a'') -- [opacity=0.1] (b''),
				(c'') -- [opacity=0.1] (d''),
				(c3) -- [opacity=0.1] (c00),
                        };
                \end{feynman}
        \end{tikzpicture}
}
& \parbox[c]{0.5\textwidth}{
	$(-ig)^2
\left(\frac{1.3.2.1)}{2!}\right)\int d^4y_1d^4y_2\ (i\Delta_F(x_1-x_2))
(i\Delta_F(y_1-y_2))^3$ 
} \\ 
\hline 
\end{tabular}
\end{center}
The diagrams (a) and (b) are examples of {\em connected diagrams} while
(c) is a example of a {\em topologically disconnected diagram}. Here
(dis)connection refers to (dis)connection with {\em external vertices}.
A diagram with no external vertices is called a {\em vacuum bubble}. The
diagram in (c) has a disconnected piece which is a vacuum bubble. The
denominator diagrams are all vacuum bubbles.

It is actually possible to separate the vacuum bubble components of the
diagrams in the numerator and cancel them against those in the
denominator. The proof goes as follows.

In an $n-$point function, at the $k^{th}$ order, the numerator has the
form,
\[
	Nr ~ = ~ \sum_{k=0}^{\infty}\frac{(-i)^k}{k!}\int d^4y_1\dots
	d^4y_k\langle 0|T\{\Phi(x_1) \dots \Phi(x_n) :H_I(y_1): \dots:
	H_I(y_k):\}|0\rangle
\]

Group the contractions into two groups: (i) those that have $l$ of the
internal vertices with a contraction with {\em at least one} of the
external vertices and (ii) the remaining $(k - l)$ internal vertices
which have no segment connecting an external vertex. The above vev then
splits as,
\begin{eqnarray*}
\langle 0|T\{\Phi(x_1) \dots \Phi(x_n) :H_I(y_1): \dots:
H_I(y_k):\}|0\rangle ~ = ~ \mbox{\hspace{6.0cm}} & & \\
\langle 0|T\{\Phi(x_1) \dots \Phi(x_n) :H_I(y_1): \dots:
H_I(y_l):\}|0\rangle \times \mbox{\hspace{4.0cm}} & & \\ 
\langle 0|T\{:H_I(y_1): \dots: H_I(y_{k-l}):\}|0\rangle
\mbox{\hspace{2.0cm}}& & 
\end{eqnarray*}
And this split can happen in $^kC_l$ ways. Hence,
\begin{eqnarray*}
	Nr & = & \sum_{k=0}^{\infty} \frac{(-i)^k}{k!} \sum_{l=0}^k
	\frac{k!}{l!(k-l)!}\int d^4y_1\dots d^4y_l\\
	& & \mbox{\hspace{2.5cm}} \langle 0|T\{\Phi(x_1) \dots \Phi(x_n)
	:H_I(y_1): \dots: H_I(y_l):\}|0\rangle \times \\
	& & \mbox{\hspace{4.0cm}} \int d^4y_{l+1}\dots d^4y_{k} \langle
	0|T\{:H_I(y_{l+1}): \dots: H_I(y_{k}):\}|0\rangle
\end{eqnarray*}
Putting $(-i)^k = (-i)^l(-i)^{(k-l)}$ and  interchanging the order of
summation we can write,
\begin{eqnarray*}
Nr & = & \left[\sum_{l=0}^{\infty} \frac{(-i)^l}{l!} \int d^4y_1\dots
d^4y_l \langle 0|T\{\Phi(x_1) \dots \Phi(x_n) :H_I(y_1): \dots:
H_I(y_l):\}|0\rangle \right] \\
& & \mbox{\hspace{1.0cm}} \times \left[ \sum_{k=l}^{\infty}
	\frac{(-i)^{(k-l)}}{(k-l)!}
\int d^4y_{l+1}\dots d^4y_{k} \langle 0|T\{:H_I(y_{l+1}): \dots:
H_I(y_{k}):\}|0\rangle\right]
\end{eqnarray*}
$k \to k+l$ in the second square bracket, shows it as $\langle 0|T\{exp
-{i\int_{-\infty}^{\infty}dt H(t)}\}|0\rangle$ which is just the
denominator! 

The first factor is called the {\em connected Green's function} and
denoted as $G^c_n$. It consists of only the diagrams connected to
external vertices. The diagrams may be topologically disconnected. {\em
	From now on we focus on the connected Green's functions and drop
the denominator.}

Having gotten Feynman diagrams and Feynman integrals, we further
simplify the expression for the $S-$matrix elements.

To go from Green's functions to $S-$matrix elements (a) we operate on
the Green's function by the equation of motion differential operator for
each external vertex; (b) insert the wave function factors for each
external vertex; (c) divide by $\sqrt{Z}$ for each external vertex and
(d) integrate over the external vertices.

We have only connected diagrams. Among these are topologically
disconnected diagrams too. There is a special subclass of topologically
disconnected diagrams in which two external vertices are Wick
contracted. These represent an incoming particle which goes out {\em
without any scattering}. We can separate such un-scattered processes
from the $S-$matrix (which we will do little later) and focus on
processes wherein every incoming particle necessarily scatters i.e. {\em
every external vertex is necessarily connected to an internal vertex.}

Consider an external vertex $x$ connected with an internal vertex $y$.
Associated with the external vertex is a wavefunction factor
$\Case{e^{ik\cdot x}\ (or\ e^{-ik'x'})} {\sqrt{2\omega_{\vec{k}}
(2\pi)^3}}$, integration over $x$ (or $x'$), Feynman propagator
$\Delta_F(x-y)$ and $(\Box_x - m^2)$ acting on the propagator. Since the
propagator is a Green's function, $(\Box_x - m^2)\Delta_F(x-y) = -
\delta^4(x-y)$. Integration over $x$ then removes the delta function and
$e^{ik\cdot x} \to e^{ik\cdot y}$. For the external vertex of an
out-going particle, $e^{-ik'\cdot x'} \to e^{-ik'\cdot y}$. Thus the
integration over external vertices is trivially carried out and the
propagators involving an external vertex is removed - an ``external leg
is said to be amputated''.

Now use the Fourier representation of the remaining propagators,
$\Delta_F(y_i - y_j) = \int \Case{d^4l}{(2\pi)^4}
\Case{e^{il\cdot(y_i-y_j)}}{l^2 + m^2 -i\epsilon}$. Each internal vertex
$y_i$ is connected to possibly several other internal vertices as well
as possibly several external vertices. For internal vertex, the Fourier
transform supplies a factor of $e^{il\cdot y_i}$ and each external
vertex provides a factor of $e^{k\cdot y_i}$. All of these combine and
the $\int d^4y_i$ then gives a $(2\pi)^4\delta^4(\sum k -\sum k' + \sum
l)$. Thus, the integration over internal vertices result in a momentum
conserving delta function.  The space-time integrations are done but we
are left with integration over the Fourier momenta which are associated
with internal lines. The delta functions trivialize many momentum
integrations leaving an overall delta function involving only the
external momenta and enforcing the momentum conservation. We also left
with some unconstrained {\em loop momenta}. 

To be explicit, let $n_E, n_I$ denote the number of external and
internal lines and let $n_v$ denote the number internal vertices. We
have then $n_E + n_I$ number of momenta and $n_v$ number of conservation
equations. Thus $n_E + n_I - n_v + 1$ is the number of undetermined
momenta or the loop momenta. The $+1$ in the counting signifies the loss
of one conservation equation due to the left over delta function
enforcing conservation of external momenta.

There are no dependence on space-time points, no space-time integrations
left and the scattering matrix element is given by integration over a
bunch of loop momenta and a product of momentum space propagators
forming the integrand and of course the various factors of $i, \pi$ and
numerical combinatorial factors. 
\[
\boxed{	S_{fi} \sim \delta^4(\sum k - \sum {k'}) \prod_j\int d^4l_j
\prod_{n_I}\left( \frac{1}{p^2_i(k, l) + m^2 - i\epsilon}\right) }
\]
We will elaborate the numerical factors in explicit examples. The
structure of the $S-$matrix elements (non-trivial scattering) should be
clear from the above discussion.

To separate out the trivial scattering it is customary to introduce the
the so-called $T-$matrix as,
\[
\boxed{	S := \mathbb{1} + iT ~ ~ , ~ ~\langle k'_{j'}|T|k_j\rangle :=
(2\pi)^4 \delta^4(\Sigma_{j'} k'_{j'} - \Sigma_j k_j)\ \mathcal{M}(k_j
\to k'_{j'}) }
\]
Here the $k, k'$ denote the on-shell momenta of the incoming and
outgoing particles. The delta function enforcing momentum conservation is
a consequence of translation invariance and $\mathcal(M)$ is called the
{\em invariant matrix element}. This is what is computed in practice.
The $S_{fi}$ given above is really the $iT_{fi}$.

In practice, computation of the $T-$matrix elements is only part of what
is needed to compare with experiments which measure {\em
cross-sections}. To compute cross-sections, recall that the $S-$matrix
elements are the transition probability amplitudes whose non-trivial
contributions are identified as $\langle f|i\rangle = (2\pi)^4
\delta^4(k'_{total} - k_{total})(i\mathcal{M}_{i\to f})$. The
cross-section is the ratio of the number of outgoing particles per
second to the incident flux. The numerator is proportional to the
transition probability rate while the denominator is determined by the
initial state.

We have given the $S-$matrix elements in the plane wave basis which is
strictly incorrect. One should use wave packets for representing the
asymptotic states. Alternatively, a commonly used practice is to put the
system in a finite space-time box and take the limit of infinite box at
the end. This is simpler to implement in practice and suffices for most
purposes. We will use this \cite{Srednicky} and refer to
\cite{PeskinSchroder} for the wave packet treatment.

\subsection{Differential Cross-section for $2 \to n$ process}
Imagine the scattering experiment to be enclosed in a large spatial box
of volume $V = L^3$, with periodic boundary conditions imposed on the
mode functions. Let the duration of the experiment be a large time
interval $T$. The probability of transition from an initial state
$|i\rangle$ to a final state $|f\rangle$ is given by,
\[
Prob_{i\to f} = \frac{|\langle f|i\rangle|^2}{\langle i|i\rangle \langle
f|f\rangle} = \frac{\left[(2\pi)^4\delta^4(\Sigma k'_{j'} - \Sigma
k_j)\right]\left[(2\pi)^4\delta^4(0)\right]|i\mathcal{M}_{i\to
f}|^2}{\langle i|i\rangle \langle f|f\rangle}  
\]
The square of the momentum conservation delta function is written with
one factor as $\delta(0)$ which is to be understood as:
$\boxed{(2\pi)^4\delta^4(0) = \int d^4x e^{i0\cdot x} = \int d^4x :=
VT}$.  The initial and final states are normalized using $\boxed{\langle
k|k\rangle := \lim_{k'\to k}\delta^3(\vec{k}' - \vec{k}) = \delta^3(0)
:= \Case{V}{(2\pi)^3}}$. Typically, the initial state consists of {\em
two particles}. So let us specialize to this case of $2 \to n$
processes.  Then, $\langle i|i\rangle\langle f|f\rangle =
[\Case{V}{(2\pi)^3}]^{(n+2)}$. Therefore,
\[
	\frac{Prob_{i\to f}}{T}  ~ = ~
	\frac{\left[(2\pi)^4\delta^4(\Sigma k'_{j'} - \Sigma
	k_j)\right]|i\mathcal{M}_{i\to f}|^2 V}{[V/(2\pi)^3]^{n+2}}
\]

Real detectors have finite aperture and hence the detected particle's
momentum is anywhere within a small window around the central value. The
corresponding probabilities are thus to be added. An estimate for such a
window follows from the box normalization we have taken. The momenta are
given by $\vec{k}_j = \Case{2\pi}{L}\vec{n}_j$. Summing over the momenta
within a window is same as summing over the integers within a window and
for large volume, $\Sigma_{\vec{n}_j} ~\approx~ \Case{V}{(2\pi)^3}\int
d^3k$, for each final state particle. This leads to the total
probability per unit time for the transition,
\begin{eqnarray*}
\frac{Prob_{2\to n}}{T}d\Gamma_n & = &
\frac{\left[(2\pi)^4\delta^4(\Sigma k'_{j'} - \Sigma
k_j)\right]|i\mathcal{M}_{2\to n}|^2 V}{[V/(2\pi)^3]^{n+2}} \prod_{j =
1}^n \left[\frac{d^3k_j}{(2\pi)^3}V\right]  \\
& = & \left\{\frac{\left[(2\pi)^4\delta^4(\Sigma k'_{j'} - k_1 -
k_2)\right]\left|i\mathcal{M}_{2\to n}\sqrt{\prod_{j=1}^{n+2}\left[
2\omega_{\vec{k}_j}(2\pi)^3\right]} \right|^2}{V(2\omega_{\vec{k}_1})
(2\omega_{\vec{k}_2})}\right\} \prod_{j = 1}^n \left[\frac{d^3k_j}
{2\omega_{\vec{k}}(2\pi)^3}\right]
\end{eqnarray*}
In going from the first to the second line, we have divided and
multiplied by the product of $(2\omega_{\vec{k}})$ for all the $(n+2)$
particles. The last product is the $d\Gamma_n$ which is the {\em Lorentz
invariant phase space volume}.  The square root factor will get absorbed
in the invariant matrix element and will get rid of similar factors
coming from the wave functions of the incoming and outgoing particles. 

We also need the incident flux. For the initial state of two articles,
consider the {\em laboratory frame} where the particle `2' is at rest.
The particle `1' has a speed $|\vec{k}_1|/E_1$ and gives the number
density of one particle per unit volume, $1/V$. Hence, the incident
number flux is $|\vec{k}_1|/(E_1 V)$. Dividing by the flux, the {\em
exclusive, differential cross-section for $2 \to n$ process} is given
by,
\begin{center} \fbox{ \begin{minipage}{0.8\textwidth} \begin{eqnarray*}
	d\sigma_{lab} & = & \frac{(2\pi)^4\delta^4(k_1 + k_2 -
	\Sigma_jk_j)|i\tilde{\mathcal{M}}_{2\to
n}|^2}{4m_2|(\vec{k}_1)_{lab}|} d\Gamma_n  ~ \mbox{\hspace{1.0cm}
where,} \\
\tilde{\mathcal{M}}_{2\to n} & := & \mathcal{M}_{2\to n}
\sqrt{\prod_{j=1}^{n+2}\left[ 2\omega_{\vec{k}_j}(2\pi)^3\right]} ~
\mbox{\hspace{3.5cm} and,}\\
d\Gamma_n & := & \prod_{j = 1}^n \frac{d^3k_j} {2\omega_{\vec{k}}
(2\pi)^3}
\end{eqnarray*}
\end{minipage}
}
\end{center}
We have used $\omega_{\vec{k}_2} = m_2$ and $\omega_{\vec{k}_1} = E_1$.
Notice that all factors of the volume and the duration $T$ have
disappeared. The explicit square root factors will also cancel out. 

Many basic calculations involve scattering processes with two particles
going into two particles. We have already used the initial state of two
particles to obtain the incident flux. We will now also specialize to
$n=2$ for out going particles and write the phase space integrals more
explicitly.
\subsubsection{The Special case of $2 \to 2$ processes}
Let the initial momenta be denoted by $k_1, k_2$ and the final momenta
be denoted by $k_1', k_2'$. Lorentz invariance implies that the
scattering amplitude will have Lorentz indices (tensorial or spinorial)
carried by the wavefunctions and the invariant amplitude will be a
function of Lorentz invariants. We have three independent momenta thanks
to the overall conservation due to translation invariance and we can
form 6 Lorentz scalars of the form $p_i\cdot p_j$. Of these three are
masses and the remaining three are conveniently defined as ``center of
mass energy'' and two types of ``squared momentum transferred''. These
are known as the {\em Mandelstam variables} and are defined as ($k_1 +
k_2 = k_1' + k_2'$),
\begin{eqnarray}
	s & := & - (k_1+k_2)^2 = - (k_1'+ k_2')^2 ~ ~
\mbox{\hspace{1.0cm} (Centre of Mass Energy)} \\
	t & := & - (k'_1-k_1)^2 = - (k_2'- k_2)^2 ~ ~
\mbox{\hspace{1.0cm} (Squared Momentum transfer)} \\
	u & := & - (k'_2-k_1)^2 = - (k_2- k'_1)^2 ~ ~
\mbox{\hspace{1.0cm} (Squared Momentum transfer)} 
\end{eqnarray}
The definitions imply the {\em Mandelstam identity:} $\boxed{s + u + t =
m^2_{k_1} + m^2_{k_2} + m^2_{k'_1} + m^2_{k'_2} } $.

\underline{Laboratory and Centre of Mass Frames:} 

The lab frame, is defined by regarding particle 2, say, at rest while
particle 1 is incident on it.  Thus, 
\begin{equation*}
\boxed{ k^{lab}_1 = (\sqrt{\vec{k}_1^2 + m_1^2}, \ \vec{k}_1^{lab})\ , \
k^{lab}_2 = (m_2, \ \vec{0})} 
\end{equation*}
The lab frame is convenient for expressing the incident flux. 

The center of mass frame is defined by $\vec{k}_{total} := \vec{k}_1 +
\vec{k}_2 = 0 = \vec{k}'_1 + \vec{k}'_2$. Thus, 
\begin{equation*}
\boxed{k^{cm}_1 = (\sqrt{\vec{k}^2_{cm} + m_1^2}, \ \vec{k}_{cm})\ , \
k^{cm}_2 = (\sqrt{\vec{k}^2_{cm} + m_2^2}, \ -\vec{k}_{cm})} 
\end{equation*}
This frame is more convenient for defining scattering angle. The common
momentum direction singles out say, the z-axis. Orienting the frame
accordingly, we take
\begin{eqnarray}
	k_1 = (E_1, k_{cm}\hat{z}) & , & k_2 = (E_2, -k_{cm}\hat{z}) ~ ~
	, \nonumber \\
	k'_1 = (E'_1, \vec{k}') & , & k'_2 = (E'_2, -\vec{k}') ~ ~ , ~ ~
	\hat{k'}\cdot\hat{z} =: cos(\Theta_{cm}). 
\end{eqnarray}

The Mandelstam invariant $s$ can be used to relate the lab frame
momentum $|k_1|^{lab}$ and the center of mass momentum $|k|_{cm}$. The
invariant $s := - (k_1 + k_2)^2 = - (-m_1^2 - m^2_2 - 2E_1E_2 +
2\vec{k}_1\cdot\vec{k}_2)$ has two equivalent expressions:
\begin{eqnarray*}
	s|_{lab} & = & m^2_1 + m^2_2 + 2m_2\sqrt{m^2_1 + \vec{k}_1^2} \\
	s|_{cm} & = & m_1^2 + m_2^2 + 2\sqrt{m_1^2 + \vec{k}^2_{cm}}
	\sqrt{m_2^2 + \vec{k}_{cm}^2} + 2\vec{k}_{cm}^2 \ .
\end{eqnarray*}

These can be solved for the momenta and give manifestly Lorentz
invariant expressions,
\begin{eqnarray}
	|k_1|_{lab} & = & \frac{1}{2m_2}\sqrt{s^2 - 2s(m_1^2 + m_2^2) +
	(m_1^2 - m_2^2)^2} \\
	|k|_{cm} & = & \frac{1}{2\sqrt{s}}\sqrt{s^2 - 2s(m_1^2 + m_2^2) +
	(m_1^2 - m_2^2)^2} 
\end{eqnarray}

\underline{Phase space volume in center of mass frame:} We have the
Lorentz invariant definition,
\[
d\Gamma_2 = (2\pi)^4\delta^4(k'_1+k'_2 - k_1-k_2)
\Case{d^3k'_1}{2\omega_{k'_1}(2\pi)^3}
\Case{d^3k'_2}{2\omega_{k'_2}(2\pi)^3} .
\]
In the CM frame we simplify it using: $\delta^4 \to \delta(E'_1 + E'_2 -
\sqrt{s})\delta^3(k'_1+k'_2)$ since $\vec{k}_1 + \vec{k}_2 = 0, s =
(E_1+E_2)^2$. We can remove the momentum delta function by integrating
over $\vec{k}'_2$. Denoting $d^3k'_1 := d^3k'_{cm} = dk'_{cm}(k')^2_{cm}
d\Theta_{cm}sin^2(\Theta_{cm})d\phi$, we write,
\begin{eqnarray*}
\int d^3k'_2d\Gamma_2 & = & \frac{1}{(2\pi)^2}\delta(E'_1+E'_2 -
\sqrt{s}) \frac{1}{4E'_1E'_2} dk'_{cm}(k')^2_{cm} d\Theta_{cm}
sin^2(\Theta_{cm}) d\phi 
\end{eqnarray*}
Since $E'_1+E'_2 = \sqrt{(m')^2_1 + (k')^2_{cm}} + \sqrt{(m')^2_2 +
(k')^2_{cm}}$ is a function of $k'_{cm}$, we simplify the delta function
using $\delta(f(x)) = \sum_i\Case{\delta(x - x_i)}{f'(x_i)}, f(x_i) =
0$.  We have
\begin{eqnarray*}
f'(k'_{cm}) & = & \frac{k'_{cm}}{E'_1} + \frac{k'_{cm}}{E'_2} ~ = ~
k'_{cm}\frac{E'_1 + E'_2}{E'_1E'_2} = \frac{k'_{cm}\sqrt{s}}{E'_1E'_2},
\\
f(k'_{cm}) = 0 & \Rightarrow & k'_{cm} = \frac{\sqrt{s^2 - 2s( (m')^2_1
+ (m')^2_2) + ( (m')_1^2 - (m')^2_2)^2}}{2\sqrt{s}}
\end{eqnarray*}	
Doing the $dk'_{cm}$ integration using the delta function gives,
\begin{equation}
	\int d(k')^2_{cm}\delta(E'_1 + E'_2 - \sqrt{s}) =
	\frac{(k')^2_{cm}E'_1E'_2}{k'_{cm}\sqrt{s}} ~ \mbox{which
	gives}~ d\Gamma_2 =
	\frac{1}{16\pi^2}d^2\Omega_{cm}\frac{k'_{cm}}{\sqrt{s}} \ .
\end{equation}

Thus, for the special case of a process with 2 particles going to 2
particles, one gets the {\em differential cross-section} as,
\begin{eqnarray}\label{DiffCrossSection}
\frac{d\sigma}{d\Omega_{cm}} & = &
\frac{1}{64\pi^2}\frac{1}{s}\frac{k'_{cm}}{k_{cm}}|\mathcal{M}|^2
\mbox{\hspace{0.5cm}where we have used\hspace{0.3cm}} 2m_2|k_1|^{lab} =
2\sqrt{s}|k|_{cm}\ , ~ ~ \mbox{and}~ ~\\
\frac{k'_{cm}}{k_{cm}} & = & \sqrt{\frac{s^2 - 2s( (m')^2_1 + (m')^2_2)
+ ((m')_1^2 - (m')^2_2)^2}{s^2 - 2s( m^2_1 + m^2_2) + (m_1^2 -
m^2_2)^2}} \ .
\end{eqnarray}
The invariant amplitude has dependence on the scattering angle
$\Theta_{cm}$ and the total cross-section is obtained by integrating
over the center of mass solid angle.

\newpage 
\section{Diagrammatic recipe for S-matrix elements}\label{Diagrammatics}

We now specify explicit interacting fields by giving an interaction
Lagrangian and state the Feynman rules to complete the diagrammatic
recipe for the $T-matrix$ elements. We will state the rules for the
$\Phi^4$ theory, the Yukawa theory and the Quantum electrodynamics
(QED). In the next section we will compute specific processes.  There is
a good deal of conventions of normalization etc and they have to be kept
track of carefully. The free (quadratic) part of the Lagrangian density
sets the normalization of the propagators while the interaction terms
(beyond quadratic in fields) contribute to the numerical factors. Here
are the terms in the {\em Lagrangian densities}.
\begin{eqnarray}
\mbox{Free Scalar~} & : & -\frac{1}{2} \partial^{\mu}{\Phi}
\partial_{\mu}{\Phi} -\frac{1}{2}m^2\Phi^2(x)  \ ; \\
\mbox{Free Spinor~} & : & -i\bar{\Psi}\dsl{\partial}\Psi + m
\bar{\Psi}\Psi \ ; \\
\mbox{Massless Vector~} & : & -\frac{1}{4}F^{\mu\nu}F_{\mu\nu} ~ ~ , ~ ~
F_{\mu\nu} := \partial_{\mu}A_{\nu} - \partial_{\nu}A_{\mu}  \ ; \\
\mbox{Scalar self coupling~} & : & - \frac{g}{3!}\Phi^3(x) -
\frac{\lambda}{4!}\Phi(x)^4  \ ; \\
\mbox{Yukawa coupling~} & : & - g\Phi(x)\bar{\Psi}(x)\Psi(x) \ ; \\
\mbox{QED coupling~} & : & -ie A_{\mu}(x) \bar{\Psi}(x) \gamma^{\mu}
\Psi(x) \ . \\
\mbox{In the interaction~} & & \mbox{~ {\em Hamiltonian}, all coupling
terms will change signs.}
\end{eqnarray}

The propagators are the Feynman Green's functions for the free equations of
motion. Consider the massless vector field as this has a new feature. 

The free Lagrangian (Maxwell) can be expressed as $\Case{1}{2}A_{\mu}(
\eta^{\mu\nu}\Box - \partial^{\mu}\partial^{\nu})A_{\nu} + $ divergence
terms. The equation of motion is $(\eta^{\mu\nu}\Box -
\partial^{\mu}\partial^{\nu})A_{\nu}(x) = 0$. Equivalently, in Fourier
space equation takes the form $-(\eta^{\mu\nu}k^2 -
k^{\mu}k^{\nu})\tilde{A}_{\nu}(k) = 0$. However, the differential
operator is not invertible and hence does not admit a Green's function!
The non-invertibility follows because every $A_{\nu}$ of the form
$\partial_{\nu}f(x)$, solves the equation. That is, $\partial_{\nu}f$ is
a non-trivial eigenvector  of the differential operator, with zero
eigenvalue. For perturbation theory though we need a propagator for
which extra terms are added to the action to break its gauge invariance
-~invariance under $\delta A_{\mu}(x) = \partial_{\mu}\Lambda(x)$. This
can be done in several ways and each choice corresponds to a gauge. 

A common and convenient choice is to add $-\Case{1}{2}(\partial_{\mu}
A^{\mu})^2$ term to the action. Up to a divergence term, this is just
$+\Case{1}{2}A_{\mu}\partial^{\mu}\partial^{\nu}A_{\nu}$ and precisely
cancels the term in the equation of motion operator, making it
$\eta^{\mu\nu}\Box$ which {\em is} invertible. The added term is
manifestly Lorentz invariant and the corresponding gauge is called the
{\em Lorentz gauge}. For our purposes, this gauge will suffice.  The
propagator is an inverse of the differential operator since
$(\eta^{\mu\nu}\Box_x)(D_F)_{\nu\lambda}(x-y) = \delta^{\mu}_{~\lambda}
\delta^4(x-y)$.

The QED coupling arises from the minimal substitution rule: $\vec{P} \to
\vec{P} + e\vec{A} \leftrightarrow  -i\partial_{\mu} \to -i\partial_{\mu}
+ eA_{\mu} \leftrightarrow \partial_{\mu} \to \partial_{\mu} +
ieA_{\mu}$. Substitution in the free Dirac action gives a $-ieA_{\mu}$ as
the QED coupling. 

Let us gather the various factors for the scalar field in the $T-$matrix
elements. 
\begin{itemize} 
	\item Each in/out particle gives a wave function factor of
		$[2\omega_{\vec{k}}(2\pi)^3]^{-1/2}e^{\pm i k\cdot x}$,
		an $\int d^4x$ and $(\Box_x - m^2)$ acting on the
		$n-$point function. For a spinor field, the amputation
		operator changes to $(-i\dsl{\partial} + m)$ and we have
		additionally the $u, v, \bar{u}, \bar{v}$ spinors. For a
		vector field, the amputation operator changes and we
		have the polarization vectors in addition. Other factors
		remain the same.
	\item Each order in $H_I$, gives $(-i)\int d^4y$  from the
		$T-$ordered exponential and a $(m!)^{-1}$ for the
		$m^{th}$ order, from the exponential;
	\item Each Wick contraction gives $(i\Delta_F(z-z')) := \int
		\Case{d^4l}{(2\pi)^4} \Case{ie^{il\cdot(z-z')}} {l^2 +
		m^2 - i\epsilon}$ ;
	\item Each amputation, action of $(\Box -m^2)$ gives a
		$(-\delta^4(x-y))$ while each integration over $x$ or
		$y$ gives $(2\pi)^4$ times a momentum conserving
		$\delta^4$. The momentum integrations do not produce any
		factors.
	\item Let $E, I, V$ denote the number of external lines (=
		number of external vertices), number of internal lines
		and number of internal vertices (= order of $H_I$)
		respectively.  Then, the factors of $[2\omega_{\vec{k}}
		(2\pi)^3] ^{-1/2}$, precisely cancel the explicit factor
		we found in the amplitude $\tilde{M}_{2\to n}$. This is
		due to the normalization choices which cancel out in
		$\Case{|\langle f|i\rangle|^2}{\langle i|i\rangle\langle
		f|f\rangle}$ and can {\em now be dropped from both
		$\tilde{M}$ and from the $T-$matrix element.}

		Factors of $i$: $(-i)^V(i)^I = (-1)^Vi^{V+I}$;

		Factors of $2\pi$: $(2\pi)^{-4I + 4E + 4V -4}$. The last
		$-4$ is because it has been taken out in the definition
		of the $\mathcal{M}$ due to the overall momentum
		conservation.

		Vertex factors, including the $i$ in the QED vertex are
		to be taken case-by-case. There is the $(m!)^{-1}$ and a
		combinatorial factor that will come for each diagram.
		These too are taken case-by-case.
\end{itemize}
With these, we now state the Feynman rules for Feynman diagrams.

\begin{center}
	\begin{tabular}{|l|c|c|c|c|}
		\hline
		External Line:& \mbox{\hspace{3.5cm}} &
		\mbox{\hspace{2cm}} & \mbox{\hspace{3.5cm}} &
		\mbox{\hspace{2cm}} \\
		\hline
		\mbox{scalar} &
		\parbox[c]{3.0cm}{
			\begin{tikzpicture}
                            \begin{feynman}
                                \vertex (a) ;
                                \vertex [right=1.5 of a, blob] (b){};
                                \diagram*{
                                     (a) -- [scalar,momentum'=\(p\)] (b),
                                };
                            \end{feynman}
                        \end{tikzpicture}
		} & $1$ &
		\parbox[c]{3.0cm}{
			\begin{tikzpicture}
                            \begin{feynman}
                                \vertex (b) ;
                                \vertex [left=1.5 of b, blob] (a){};
                                \diagram*{
                                     (a) -- [scalar,momentum'=\(p\)] (b),
                                };
                            \end{feynman}
                        \end{tikzpicture}
		} & $1$ \\
		\hline
		\mbox{fermion} & 
		\parbox[c]{3.0cm}{
			\begin{tikzpicture}
                            \begin{feynman}
                                \vertex (a) ;
                                \vertex [right=1.5 of a, blob] (b){};
                                \diagram*{
                                     (a) -- [fermion,momentum'=\(p\)] (b),
                                };
                            \end{feynman}
                        \end{tikzpicture}
		} &
		$u(p,\sigma)$ & 
		\parbox[c]{3.0cm}{
			\begin{tikzpicture}
                            \begin{feynman}
                                \vertex (b) ;
                                \vertex [left=1.5 of b, blob] (a){};
                                \diagram*{
                                     (a) -- [fermion,momentum'=\(p\)] (b),
                                };
                            \end{feynman}
                        \end{tikzpicture}
		} &
		$\bar{u}(p,\sigma)$ \\
		\hline
		\mbox{anti-fermion} & 
		\parbox[c]{3.0cm}{
			\begin{tikzpicture}
                            \begin{feynman}
                                \vertex (a) ;
                                \vertex [right=1.5 of a, blob] (b){};
                                \diagram*{
                                     (a) -- [anti fermion,momentum'=\(p\)] (b),
                                };
                            \end{feynman}
                        \end{tikzpicture}
		} &
		$\bar{v}(p,\sigma)$ & 
		\parbox[c]{3.0cm}{
			\begin{tikzpicture}
                            \begin{feynman}
                                \vertex (b) ;
                                \vertex [left=1.5 of b, blob] (a){};
                                \diagram*{
                                     (a) -- [anti fermion,momentum'=\(p\)] (b),
                                };
                            \end{feynman}
                        \end{tikzpicture}
		} &
		$v(p,\sigma)$\\
		\hline
		\mbox{photon} & 
		\parbox[c]{3.0cm}{
			\begin{tikzpicture}
                            \begin{feynman}
                                \vertex (a) ;
                                \vertex [right=1.5 of a, blob] (b){};
                                \diagram*{
                                     (a) -- [photon, edge
				     label=\(\gamma\), momentum'=\(p\)] (b),
                                };
                            \end{feynman}
                        \end{tikzpicture}
		} &
		$\varepsilon^*_{\mu}(p,\lambda)$ & 
		\parbox[c]{3.0cm}{
			\begin{tikzpicture}
                            \begin{feynman}
                                \vertex (b) ;
                                \vertex [left=1.5 of b, blob] (a){};
                                \diagram*{
                                     (a) -- [photon, edge
				     label=\(\gamma\), momentum'=\(p\)] (b),
                                };
                            \end{feynman}
                        \end{tikzpicture}
		} &
		$\varepsilon_{\mu}(p,\lambda)$  \\
		\hline
	\end{tabular}
\end{center}

\begin{center}
	\begin{tabular}{|l|c|c|}
		\hline
		Internal Line: & \mbox{\hspace{4.5cm}} &
		\mbox{\hspace{4.5cm}}\\
		\hline
		\mbox{scalar} & 
		\parbox[c]{3.0cm}{
			\begin{tikzpicture}
                            \begin{feynman}
                                \vertex (a) ;
                                \vertex [right=2.0 of a] (b);
                                \diagram*{
                                     (a) -- [scalar] (b),
                                };
                            \end{feynman}
                        \end{tikzpicture}
		} &
		$\frac{-i}{k^2 + m^2 - i\epsilon}$ \\
		\hline
	\mbox{fermion} & 
		\parbox[c]{3.0cm}{
			\begin{tikzpicture}
                            \begin{feynman}
                                \vertex (a) ;
                                \vertex [right=2.0 of a] (b);
                                \diagram*{
                                     (a) -- [fermion] (b),
                                };
                            \end{feynman}
                        \end{tikzpicture}
		} &
	$\frac{(-i)(-\dsl{k} +m)}{k^2 + m^2 - i\epsilon}$ \\
		\hline
		\mbox{photon (Lorentz gauge)} & 
		\parbox[c]{3.0cm}{
			\begin{tikzpicture}
                            \begin{feynman}
				    \vertex (a){\(\mu\)} ;
				    \vertex [right=2.0 of a] (b){\(\nu\)};
                                \diagram*{
					(a) -- [photon] (b),
                                };
                            \end{feynman}
                        \end{tikzpicture}
		} &
		$\frac{-i\eta^{\mu\nu}}{k^2 - i\epsilon}$ \\
		\hline
	\end{tabular}
\end{center}

\begin{center}
	\begin{tabular}{|l|c|c|}
		\hline
		Vertex: & \mbox{\hspace{5.8cm}} & \mbox{\hspace{5.8cm}}\\
		\hline
		\mbox{$\Phi^3$} & 
		\parbox[c]{3.0cm}{
			\begin{tikzpicture}
                            \begin{feynman}
				    \vertex (v) ;
				    \vertex [left=of v] (a);
				    \vertex [above right=of v] (b);
				    \vertex [below right=of v] (c);
                                \diagram*{
					(a) -- [scalar] (v),
					(b) -- [scalar] (v),
					(c) -- [scalar] (v),
                                };
                            \end{feynman}
                        \end{tikzpicture}
		} &
		$i\frac{g}{3!}$ \\
		\hline
		\mbox{$\Phi^4$} & 
		\parbox[c]{3.0cm}{
			\begin{tikzpicture}
                            \begin{feynman}
				    \vertex (v) ;
				    \vertex [above left=of v] (a);
				    \vertex [above right=of v] (b);
				    \vertex [below right=of v] (c);
				    \vertex [below left=of v] (d);
                                \diagram*{
					(a) -- [scalar] (v),
					(b) -- [scalar] (v),
					(c) -- [scalar] (v),
					(d) -- [scalar] (v),
                                };
                            \end{feynman}
                        \end{tikzpicture}
		} &
		$i\frac{\lambda}{4!}$ \\
		\hline
		\mbox{Yukawa} & 
		\parbox[c]{3.0cm}{
			\begin{tikzpicture}
                            \begin{feynman}
				    \vertex (v) ;
				    \vertex [left=of v] (a);
				    \vertex [above right=of v] (b);
				    \vertex [below right=of v] (c);
                                \diagram*{
					(v) -- [scalar] (a),
					(v) -- [fermion] (b),
					(c) -- [fermion] (v),
                                };
                            \end{feynman}
                        \end{tikzpicture}
		} &
		$ig$  \\
		\hline
		\mbox{QED} & 
		\parbox[c]{3.0cm}{
			\begin{tikzpicture}
                            \begin{feynman}
				    \vertex (v);
				    \vertex [left=of v] (a){\(\mu\)};
				    \vertex [above right=of v] (b);
				    \vertex [below right=of v] (c);
                                \diagram*{
					(v) -- [photon] (a),
					(v) -- [fermion] (b),
					(c) -- [fermion] (v),
                                };
                            \end{feynman}
                        \end{tikzpicture}
		} &
		$ie\gamma^{\mu}$ \\
		\hline
	\end{tabular}
\end{center}

\newpage
\section{Elementary processes in Yukawa and QED: NR
limit}\label{ElementaryProcesses}

We begin with scattering of a fermion off another fermion, interacting
via the Yukawa coupling and compute the $T-$matrix element to the
leading order. We will take the non-relativistic limit and identify an
{\em equivalent potential}. We will consider anti-fermion scattering and
fermion-anti-fermion scattering as well. We will then compare the qed
coupling and briefly the gravitational coupling. This will lead to
appreciate the dependence the attractive/repulsive nature of the
interaction on spin of the exchanged particle.

Consider a general process depicted below together with its `expansion'.
\begin{equation}
	\parbox[c]{2.85cm}{
		\begin{tikzpicture}
                \begin{feynman}
			\vertex [blob] (v){};
			\vertex [above left=of v] (a) {\(p\)};
			\vertex [above right=of v] (b) {\(p'\)};
			\vertex [below right=of v] (c) {\(k'\)};
			\vertex [below left=of v] (d) {\(k\)};;

			\diagram* { (a) -- [fermion] (v),
                                (v) -- [fermion] (b) , 
				(v) -- [fermion] (c) ,
				(d) -- [fermion] (v),
			};
                \end{feynman}
        \end{tikzpicture}
	} 
	~  = ~ 
	\parbox[c]{2.85cm}{
		\begin{tikzpicture}
                \begin{feynman}
			\vertex (v);
			\vertex [above left=0.5of v] (a) {\(p\)};
			\vertex [above right=0.5of v] (b) {\(p'\)};
			\vertex [below right=0.5of v] (c) {\(k'\)};
			\vertex [below left=0.5of v] (d) {\(k\)};;

			\diagram* { (a) -- [fermion] (b),
				(d) -- [fermion] (c),
			};
                \end{feynman}
        \end{tikzpicture}
		%
	} 
	~  +  ~  
	\parbox[c]{2.85cm}{
		\begin{tikzpicture}
                \begin{feynman}
			\vertex (v1);
			\vertex [below=0.75of v1] (v2);
			\vertex [above left=1.0of v1] (a) {\(p\)};
			\vertex [above right=1.0of v1] (b) {\(p'\)};
			\vertex [below right=1.0of v2] (c) {\(k'\)};
			\vertex [below left=1.0of v2] (d) {\(k\)};;

			\diagram* { 
				(a) -- [fermion] (v1),
				(v1) -- [fermion] (b),
				(v1) -- [scalar,momentum=\(p'-p\)] (v2),
				(v2) -- [fermion] (c),
				(d) -- [fermion] (v2),
			};
                \end{feynman}
        \end{tikzpicture}
	%
	}
	~ ~ ~ ~ + 
	\parbox[c]{2.85cm}{
		\begin{tikzpicture}
                \begin{feynman}
			\vertex (v1);
			\vertex [below=1.0of v1] (v2);
			\vertex [above left=0.8of v1] (a) {\(p\)};
			\vertex [above right=0.8of v1] (b) {\(p'\)};
			\vertex [below right=0.8of v2] (c) {\(k'\)};
			\vertex [below left=0.8of v2] (d) {\(k\)};;

			\diagram* { 
				(a) -- [fermion] (v1),
				(v1) -- [fermion] (c),
				(v1) -- [scalar,momentum'=\(k'-p\)] (v2),
				(v2) -- [fermion] (b),
				(d) -- [fermion] (v2),
			};
                \end{feynman}
        \end{tikzpicture}
	%
	}   ~ ~ +  ~ \dots
\end{equation}
The first term ($o(g^0)$) denotes `no scattering'. There are two
contributions at $o(g^2)$ corresponding to exchange of two out-going
fermions. The overall momentum conservation delta function is the same,
enforcing $p + k = p' + k'$. The Feynman rules give the expression as,
\begin{eqnarray}
i\mathcal{M} & = & (ig)^2\left\{ \bar{u}(p')u(p) \frac{-i}{(p'-p)^2 +
m_{\varphi}^2 - i\epsilon} \bar{u}(k')u(k) \right. \nonumber \\
& & \mbox{\hspace{2.0cm}} \left. -  \bar{u}(k')u(p) \frac{-i}{(k'-p)^2 +
m_{\varphi}^2 - i\epsilon} \bar{u}(p')u(k) \right\} 
\end{eqnarray}
The {\em relative} minus sign between the two terms is due to the
T-ordering definition for the fermions. The overall sign of the amplitude
is determined by the convention adopted for ordering the initial/final
state labels for the fermions. See \cite{PeskinSchroder,Srednicky}.
Notice how the fermion arrows are followed.

\underline{Note:} If we had anti-fermion scattering, then all the
fermion arrows will be reversed and their momenta will be denoted as
minus the previous momenta\footnote{Thus, we may adopt a convention that
	the diagram displays the fermion arrow and the momentum is also
	in the same direction. For a fermion the momentum is $p$ and for
anti-fermion it is $-p$.}. Apart from the reversal of fermion arrows,
the $u(p), \bar{u}(p')$ spinors go to $v(p'), \bar{v}(p)$ spinors. 

If it is a fermion-anti-fermion scattering, then the second exchange
diagram will be absent.

A specific scattering arrangement may permit the initial and the final
state fermions to be {\em distinguishable}. Then only one of the two
scattering diagrams will contribute.

\underline{Note:} The reduction formula for fermions gave a factor of
$2m$ for the wavefunctions since we had normalized the spinors as
$\bar{u}(p,\sigma)u(p,\sigma') = \delta_{\sigma,\sigma'}$. With the
Feynman rules we have adopted in the table the wavefunctions have only
the spinors. {\em This is equivalent to using the normalization:
$\bar{u}(p,\sigma)u(p,\sigma') = 2m\delta_{\sigma,\sigma'} = -
\bar{v}(p,\sigma)v(p,\sigma')$}, $m$ is of course the fermion mass.

With this noted, the amplitude becomes,
\begin{eqnarray}
i\mathcal{M} & = & (ig)^2 (-i) (2m)^2 \left\{
	\frac{\delta_{\sigma_{p},\sigma_{p'}}
	\delta_{\sigma_{k},\sigma_{k'}}} {(p'-p)^2 + m_{\varphi}^2 -
	i\epsilon} 
-~\frac{\delta_{\sigma_{p},\sigma_{k'}} \delta_{\sigma_{p'},\sigma_{k}}}
{(k'-p)^2 + m_{\varphi}^2 - i\epsilon} \right\} 
\end{eqnarray}
Noting that the scalar field momentum is just the momentum transfer in
both the diagrams, we denote it by $q$. And $q^2 = -(q^0)^2 +
|\vec{q}|^2 = |\vec{q}|^2$ in both cases. We thus write the amplitude
more conveniently as,
\begin{eqnarray*}
i\mathcal{M} & = & \frac{+ig^2\ 4m^2}{|\vec{q}|^2 + m_{\varphi}^2 -
i\epsilon} \left\{ \delta_{\sigma_{p},\sigma_{p'}}
\delta_{\sigma_{k},\sigma_{k'}} - \delta_{\sigma_{p},\sigma_{k'}}
\delta_{\sigma_{p'},\sigma_{k}} \right\}. 
\end{eqnarray*}
As noted above, only one of the two terms will contribute if the
fermions are distinguishable.

In the non-relativistic scattering theory, we have $\langle
\vec{k}'|S|\vec{k}\rangle := \langle \vec{k}'|\vec{k}\rangle - 2\pi i
\delta(E_{k'} - E_k)\langle \vec{k}'|T|\vec{k}\rangle$ with, 
\[ 
	\langle \vec{k}'|T|\vec{k}\rangle =\langle
	\vec{k}'|H'|\vec{k}\rangle =\langle
	\vec{k}'|V(\vec{q})|\vec{k}\rangle ~ = ~ V(|\vec{q} = \vec{k}' -
	\vec{k}|)\ .
\]
We can read-off the potential as, 
\begin{equation}
	V(|q|) =  - \frac{4m^2 g^2}{|\vec{q}|^2 +
	m^2_{\varphi}}\left(\delta \delta - \delta \delta\right)  := -
	\frac{g^2_{eff}}{\vec{q}^2 + m_{\varphi}^2}. 
\end{equation}
The inverse Fourier transform gives,
\begin{eqnarray}
	V(\vec{x}) & = & -g^2_{eff}\int \frac{d^3q}{(2\pi)^3}
	\frac{e^{i\vec{q}\cdot \vec{x}}}{\vec{q}^2 + m^2_{\varphi}} ~ =
	~ - \frac{g^2_{eff}}{(2\pi)^2}\int_0^{\infty} dq \frac{q^2}{q^2 +
	m^2_{\varphi}}\int_{-1}^{1}d(cos(\theta))e^{iqrcos(\theta)}
	\nonumber \\
	& = & - \frac{g^2_{eff}}{4\pi^2(ir)}\int_{-\infty}^{\infty}dq
	\frac{qe^{iqr}}{q^2 + m^2_{\varphi}} ~ \mbox{\hspace{2.0cm}
	contour integrate to pick up $q = im_{\varphi}$,} \nonumber \\
	\therefore V(r) & = & - \frac{g^2_{eff}}{4\pi}\
	\frac{e^{-m_{\varphi}r}}{r} ~ \mbox{\hspace{4.0cm} (Yukawa
	Potential)}   
\end{eqnarray}
This is a simple illustration how an effective potential can be inferred
from the underlying relativistically specified interaction. Several
remarks are in order.

\underline{Remark:} The most important point is the sign of the
potential which renders it {\em attractive}. But {\em is the sign
unambiguous?} 

We have already noted that the overall sign of the amplitude is
determined by the convention adopted for ordering of the fermion labels
in the in/out states while the relative sign is due to the Pauli
principle. When the fermions are distinguishable, either of the two
diagrams may be chosen, there is no preference. But the diagrams
contribute opposite signs!

The convention arises as follows \cite{PeskinSchroder}. We have taken
$|p,k\rangle \sim b^{\dagger}_pb^{\dagger}_k|0\rangle$ (sequence of
creation operators follows label order in the in-state) and
$\langle p',k'| \sim \langle 0|b_{k'} b_{p'}$ which is opposite to that
of the in-state, but consistent with the Hermitian conjugation. Hence,
\begin{eqnarray*}
\langle p',k'|P,k\rangle & \sim & \langle 0 |b_{k'} b_{p'} b^{\dagger}_p
b^{\dagger}_k |0\rangle ~ = ~ \langle 0|b_{k'}b^{\dagger}_{k}|0\rangle
\delta^3(p'-p) - \langle 0|b_{k'} b^{\dagger}_{p}b_{p}
b^{\dagger}_{k}|0\rangle \\
& \approx & \delta^3(k'-k)\delta^3(p'-p) - \delta^3(p'-k)\delta^3(k'-p)
\end{eqnarray*}
Actually there are $\bar{\Psi}\Psi$ fields in between, but being a
bilinear it does not change the relative sign. This explains the overall
sign.

In a non-relativistic comparison, $|q| \ll m_{\varphi}$, the small
momentum transfer means large spatial separation and hence fermions are
distinguishable. If so, there is no Pauli principle or anti-commutation
and no relative sign either! So either of the two diagrams will give the
same answer.

{\em Ideally, we should consider one of the fermions to be very massive
to mimic a source of potential (means $\vec{q} \to \vec{0} $ limit) and
match all the factors for a precise comparison.} 

The effective coupling contains the Kronecker deltas enforcing
conservation of the spin projections. Hence, the Yukawa potential
preserves spin projection. For matching with non-relativistic
normalization we have to take $2m_{fermion} \delta_{\sigma_{p},
\sigma_{p'}} \to 1$. This is needed for inferring the {\em strength} of
the Yukawa potential.

\underline{Remark:} Suppose we consider fermion-anti-fermion scattering.
Then we will have one of the Kronecker deltas to have a minus sign.
However, in this case, $b^{\dagger} \to d^{\dagger}$ and doing the
contractions with $\Psi's$ from the Yukawa coupling, we pick up another
minus sign. Hence the overall sign does not change. Hence, the Yukawa
potential between fermion-anti-fermion is also {\em attractive}. The
same holds for anti-fermion scattering.

Thus, the inferred Yukawa potential is always attractive. Its underlying
QFT description is an exchange of a (virtual) massive scalar particle of
mass $m_{\varphi}$. Yukawa proposed this potential as a model for binding
of nucleons and working with the properties of nucleons, estimated the
scalar mass to be about 200 MeV which is close to the mass of the pion.
Of course pions come in three varieties, $\pi^{\pm}, \pi^0$ while we
have taken only a single scalar.

Consider fermion scattering in QED, by replacing the Yukawa coupling by
the QED coupling. The relevant diagrams are:
\begin{eqnarray*}
	\parbox[c]{3.0cm}{
		\begin{tikzpicture}
                \begin{feynman}
			\vertex (v1); 
			\vertex [below=1.0of v1] (v2); 
			\vertex [above left=0.8of v1] (a) {\(p\)};
			\vertex [above right=0.8of v1] (b) {\(p'\)};
			\vertex [below right=0.8of v2] (c) {\(k'\)};
			\vertex [below left=0.8of v2] (d) {\(k\)};;

			\diagram* { 
				(a) -- [fermion] (v1),
				(v1) -- [fermion] (b),
				(v1) -- [photon,momentum=\(p'-p\)] (v2),
				(v2) -- [fermion] (c),
				(d) -- [fermion] (v2),
			};
                \end{feynman}
        \end{tikzpicture}
        %
	}
	& \mbox{\hspace{1.0cm} + \hspace{1.0cm}} &
	\parbox[c]{3.0cm}{
		\begin{tikzpicture}
                \begin{feynman}
			\vertex (v1); 
			\vertex [below=1.0of v1] (v2); 
			\vertex [above left=0.8of v1] (a) {\(p\)};
			\vertex [above right=0.8of v1] (b) {\(p'\)};
			\vertex [below right=0.8of v2] (c) {\(k'\)};
			\vertex [below left=0.8of v2] (d) {\(k\)};;

			\diagram* { 
				(a) -- [fermion] (v1),
				(v1) -- [fermion] (c),
				(v1) -- [photon,momentum'=\(k'-p\)] (v2),
				(v2) -- [fermion] (b),
				(d) -- [fermion] (v2),
			};
                \end{feynman}
        \end{tikzpicture}
        %
	}
\end{eqnarray*}
The corresponding expression is,
\begin{eqnarray}
i\mathcal{M} & = & (ie)^2\left\{ \bar{u}(p')\gamma^{\mu}u(p)
\frac{-i\eta_{\mu\nu}}{(p'-p)^2 - i\epsilon} \bar{u}(k')\gamma^{\nu}u(k)
\right. \nonumber \\
& & \mbox{\hspace{2.0cm}} \left. -  \bar{u}(k')\gamma^{\mu}u(p)
\frac{-i\eta_{\mu\nu}}{(k'-p)^2 - i\epsilon} \bar{u}(p')\gamma^{\nu}u(k)
\right\}  \ ,
\end{eqnarray}
with the same relative minus sign.

As before, consider the case of non-relativistic limit with
distinguishable fermions and consider only the first diagram.  The
denominator simplifies the same way: $(p'-p)^2 - i\epsilon \to
|\vec{q}|^2 - i\epsilon$. In the numerator we have $[
\bar{u}(p',\sigma') \gamma^{\mu} u(p,\sigma)][ \bar{u}(k',\rho')
\gamma_{\mu} u(k,\rho)]$.

\underline{Claim:} In the NR limit, $\bar{u}(p',\sigma') \gamma^{\mu}
u(p,\sigma) \to \bar{u}\gamma^0u$.

\underline{Proof:} The spinor $u$ satisfies, $(\dsl{p} + m)u = 0$.
Consider say the Dirac representation of the gamma matrices. Then in
terms of the two component notation we have,
\begin{eqnarray*}
\left( \begin{array}{cc} -\omega_{\vec{p}} + m &
\vec{p}\cdot\vec{\sigma} \\ -\vec{p}\cdot\vec{\sigma} & \omega_{\vec{p}}
+ m \end{array}\right) \left( \begin{array}{c} u_1 \\ u_2
\end{array}\right) & = & 0  \Rightarrow  
%
- \vec{p}\cdot\vec{\sigma}u_1 = (\omega_{\vec{p}} + m)u_2 ,
\vec{p}\cdot\vec{\sigma}u_2 = (-\omega_{\vec{p}} + m)u_1
\end{eqnarray*}
The NR limit $\omega_{\vec{p}} \approx m + \frac{\vec{p}^2}{2m}$ then
implies $u_2 \sim o(|p|/m) u_1$.  Also, in the two component notation,
we have 
\begin{eqnarray*}
	\bar{u}'\gamma^i u & = & ( (u_1')^{\dagger}, (u_2')^{\dagger} )
	\left( \begin{array}{cc} 1 & 0 \\ 0 & -1 \end{array}\right)
	\left(\begin{array}{cc} 0 & \sigma^i \\ -\sigma^i & 0
	\end{array}\right)
	\left( \begin{array}{c} u_1 \\ u_2 \end{array}\right)  =  
	(u_1')^{\dagger}\sigma^iu_2 + (u_2')^{\dagger}\sigma^i u_1 ~
	\sim ~ o(p/m)	\ ; \\
	\bar{u}'\gamma^0 u & = & ( (u_1')^{\dagger}, (u_2')^{\dagger} )
	\left( \begin{array}{cc} 1 & 0 \\ 0 & -1 \end{array}\right)
	\left(\begin{array}{cc} 1 & 0 \\ 0 & -1 \end{array}\right) 
	\left( \begin{array}{c} u_1 \\ u_2 \end{array}\right)  =  
	(u_1')^{\dagger}u_1 + (u_2')^{\dagger}u_2 \sim o(1) + o(|p|/m) .
\end{eqnarray*}
This proves the claim.

Hence the numerator approximates to $[\bar{u}(\vec{p}',\sigma') \gamma^0
u(\vec{p},\sigma)] [\bar{u}(\vec{k}',\rho') \gamma^0 u(\vec{k},\rho)]$
and we get,
\begin{equation}
	i\mathcal{M} \approx
	\frac{ie^2\eta_{00}[u^{\dagger}(\vec{p}'\sigma')
	u(\vec{p},\sigma)] [u^{\dagger}(\vec{k}',\rho')
u(\vec{k},\rho)]}{\vec{q}^2 - i\epsilon} \ .
\end{equation}

Since the square brackets are positive ($p' \approx p, k' \approx k$ for
small momentum transfer) and the $\eta_{00} < 0$, relative to the Yukawa
matrix element we have an opposite sign and hence, the effective
potential from the QED coupling will be {\em repulsive} for fermion
scattering. The form of the potential itself can be obtained from the
Yukawa one by taking $m_{\varphi} \to 0$ and as expected, we get the
{\em Coulomb potential}, $V(r) = \Case{(e')^2}{4\pi} =:
\Case{\alpha}{4\pi}$. We have absorbed the normalization factors in $e'$.
By redefining the original $e$, we can take $e' \to e$, the measured
electric charge (in the natural units). The constant $\alpha :=
\Case{e^2}{4\pi} \approx \Case{1}{137}$ is called the {\em fine
structure constant}.

If we consider fermion-anti-fermion scattering via the QED coupling, the
$u \to v$ does {\em not} introduce a negative sign since it is
$u^{\dagger}u$ and not $\bar{u}u$. The $b^{\dagger} \to d^{\dagger}$
introduces a minus sign as before and hence the overall sign changes.
Hence, for fermion-anti-fermion, the potential is {\em attractive}. For
anti-fermion scattering of course the potential is repulsive.

\underline{Note:} It seems the sign depends on the $\eta_{00}$ which is
convention dependent. However, the signs in the propagator will also
change accordingly and the sign of the potential will be metric
signature independent.

\underline{Note:} For a tensorial interaction like gravity, helicity 2,
we will have two factors of $\eta_{00}$ and the overall sign will {\em
not} change. The {\em Newtonian potential} will be {\em attractive} and
will be so for fermion-anti-fermion scattering as well. This argument is
to be taken as heuristic since we need to be explicit about the
gravitational coupling to infer the numerator factors of the spinors
which are sensitive to $u \to v$ changes.

These examples indicate that exchange of quanta can be interpreted as
giving an equivalent potential (and hence force) in the non-relativistic
limit (where the concept of potential is meaningful). The qualitative
properties of attractive/repulsive are already encoded in the Feynman
rules. The comparison with effective potential also allows the coupling
parameters to be determined experimentally.

\newpage
\section{Basic QED processes} \label{QEDProcesses}

The quantum electrodynamics has basic scattering of two charged
particles, the scattering of light by a charged particle,
particle-anti-particle annihilation and pair creation. These are
analyzed as: (i) electron-muon scattering, (ii) electron-positron
scattering as well as production of muon-anti-muon, (iii)
electron-photon scattering (Compton scattering) and (iv)
electron-positron annihilation into two photons. We will evaluate these
in the leading $e^2$ approximation. We will also compute the
cross-sections for comparison with experiments. These cross-sections
will be for the {\em un-}polarized particles in the initial state, for
which we will {\em average the cross section over the initial
spins/polarizations}. We will also not detect the final particles
spin/polarizations and hence {\em sum the cross-section over the final
spins/polarizations}. 
 
Here is the list of the processes with their diagrams and the
corresponding invariant amplitudes.

\begin{center}
\begin{tabular}{|c|c|}
\hline
$e^-(p) \mu^-(k) \to e^-(p') \mu^-(k')$ & (for $\mu$ replaced by $e$ it
is {\em Moller Scattering}) \\
\hline
\parbox[c]{4.0cm}{ 
		\begin{tikzpicture}
                \begin{feynman}
			\vertex (v1); 
			\vertex [below=1.0of v1] (v2); 
			\vertex [above left=0.8of v1] (a) {\(e^-(p)\)};
			\vertex [above right=0.8of v1] (b) {\(e^-(p')\)};
			\vertex [below right=0.8of v2] (c) {\(\mu^-(k')\)};
			\vertex [below left=0.8of v2] (d) {\(\mu^-(k)\)};;

			\diagram* { 
				(a) -- [fermion] (v1),
				(v1) -- [fermion] (b),
				(v1) -- [photon,momentum'=\(q\)] (v2),
				(v2) -- [fermion] (c),
				(d) -- [fermion] (v2),
			};
                \end{feynman}
        \end{tikzpicture}
} &
		\parbox[c]{10.0cm}{$ \hfill i\mathcal{M} =
		[\bar{u}(p')(ie\gamma^{\mu})u(p)] \left[\frac{-i
	\eta_{\mu\nu}}{q^2 - i\epsilon}\right]
[\bar{u}(k')(ie\gamma^{\mu})u(k)]\hfill $} \\
\hline
$e^+(p') e^-(p) \to \mu^+(k') \mu^-(k)$ & (for $\mu$ replaced by $e$ it
is {\em Bhabha Scattering})
\\
\hline
\parbox[c]{4.0cm}{ 
	\begin{tikzpicture}
                \begin{feynman}
                        \vertex (v1); 
                        \vertex [right=0.6of v1] (v2); 
                        \vertex [above left=0.8of v1] (a) {\(e^+(p')\)};
                        \vertex [below left=0.8of v1] (d) {\(e^-(p)\)};
                        \vertex [above right=0.8of v2] (b) {\(\mu^+(k')\)};
                        \vertex [below right=0.8of v2] (c) {\(\mu^-(k)\)};;

                        \diagram* {
                                (v1) -- [fermion] (a),
                                (d) -- [fermion] (v1),
                                (v1) -- [photon,momentum=\(q\)] (v2),
                                (v2) -- [fermion] (b),
                                (c) -- [fermion] (v2),
                        };
                \end{feynman}
        \end{tikzpicture}
} &
		\parbox[c]{10.0cm}{$ \hfill i\mathcal{M} =
		[\bar{v}(p')(ie\gamma^{\mu})u(p)] \left[\frac{-i
	\eta_{\mu\nu}}{q^2 - i\epsilon}\right]
[\bar{u}(k')(ie\gamma^{\mu})v(k)] \hfill $} \\
\hline
\end{tabular}
\end{center}

In the next two processes, there are two diagrams contributing, obtained
by exchanging the initial and final state photons in the Compton
scattering and the two final state photons in the annihilation process.
Although the diagrams may not look `crossed', they are!

\begin{center}
\begin{tabular}{|c|c|}
\hline
$e^-(p) \gamma(k) \to e^-(p') \gamma(k') $ & ({\em Compton Scattering})
\\
\hline
\parbox[c]{4.5cm}{ 
	\begin{tikzpicture}
                \begin{feynman}
                        \vertex (v1); 
                        \vertex [right=0.6of v1] (v2); 
                        \vertex [above left=0.8of v1] (a) {\(e^-(p)\)};
                        \vertex [below left=0.8of v1] (d) {\(\gamma(k)\)};
                        \vertex [above right=0.8of v2] (b) {\(e^-(p')\)};
                        \vertex [below right=0.8of v2] (c) {\(\gamma(k')\)};;

                        \diagram* {
                                (a) -- [fermion] (v1),
                                (v1) -- [fermion,momentum=\(p+k\)] (v2),
                                (v2) -- [fermion] (b),
                                (d) -- [photon] (v1),
                                (c) -- [photon] (v2),
                        };
                \end{feynman}
        \end{tikzpicture}
} &
		\parbox[c]{10.0cm}{ \begin{eqnarray*} i\mathcal{M} & = &
				\varepsilon_{\nu}(\vec{k}',\lambda')
				\bar{u}(p')(ie\gamma^{\nu})
				\left[\frac{-i (-(\dsl{p} + \dsl{k}) +
				m)}{(p+k)^2 + m^2_e - i\epsilon}\right]
				\\
				& & \mbox{\hspace{2.0cm}} \times
				(ie\gamma^{\mu})u(p) \varepsilon^*_{\mu}
				(\vec{k},\lambda) 
		\end{eqnarray*} } \\
\hline
\parbox[c]{4.5cm}{ 
	\begin{tikzpicture}
                \begin{feynman}
                        \vertex (v1); 
                        \vertex [right=1.0of v1] (v2); 
                        \vertex [above left=0.8of v1] (a) {\(e^-(p)\)};
                        \vertex [below left=0.8of v1] (d) {\(\gamma(k)\)};
                        \vertex [above right=0.8of v2] (b) {\(e^-(p')\)};
                        \vertex [below right=0.8of v2] (c) {\(\gamma(k')\)};;

                        \diagram* {
                                (a) -- [fermion] (v1),
                                (v1) -- [fermion,momentum=\(p'-k\)] (v2),
                                (v2) -- [fermion] (b),
                                (c) -- [photon] (v1),
                                (d) -- [photon] (v2),
                        };
                \end{feynman}
        \end{tikzpicture}
} &
		\parbox[c]{10.0cm}{ \begin{eqnarray*} i\mathcal{M} & = &
				\varepsilon^*_{\nu}(\vec{k},\lambda)
				\bar{u}(p')(ie\gamma^{\nu})
				\left[\frac{-i (-(\dsl{p}' - \dsl{k}) +
				m)}{(p-k')^2 + m^2_e - i\epsilon}\right]
				\\
				& & \mbox{\hspace{2.0cm}} \times
				(ie\gamma^{\mu})u(p) \varepsilon_{\mu}
				(\vec{k}',\lambda') 
		\end{eqnarray*} } \\
\hline
$e^+(p') e^-(p) \to \gamma(k) \gamma(k') $ & ({\em Annihilation
process}) \\
\hline
\parbox[c]{4.0cm}{ 
	\begin{tikzpicture}
                \begin{feynman}
                        \vertex (v1); 
                        \vertex [below=0.6of v1] (v2); 
                        \vertex [above left=0.8of v1] (a) {\(e^-(p)\)};
                        \vertex [above right=0.8of v1] (b) {\(\gamma(k)\)};
                        \vertex [below left=0.8of v2] (d) {\(e^+(p')\)};
                        \vertex [below right=0.8of v2] (c) {\(\gamma(k')\)};;

                        \diagram* {
                                (a) -- [fermion] (v1),
                                (v1) -- [fermion,momentum=\(p-k\)] (v2),
                                (v2) -- [fermion] (d),
                                (b) -- [photon] (v1),
                                (c) -- [photon] (v2),
                        };
                \end{feynman}
        \end{tikzpicture}
} &
		\parbox[c]{10.0cm}{ \begin{eqnarray*} i\mathcal{M} & = &
				\varepsilon_{\nu}(\vec{k}',\lambda')
				\bar{v}(p')(ie\gamma^{\nu})
				\left[\frac{-i (-(\dsl{p} - \dsl{k}) +
				m)}{(p-k)^2 + m^2_e - i\epsilon}\right]
				\\
				& & \mbox{\hspace{2.0cm}} \times
				(ie\gamma^{\mu})u(p) \varepsilon_{\mu}
				(\vec{k},\lambda) 
		\end{eqnarray*} } \\
\hline
\parbox[c]{4.0cm}{ 
	\begin{tikzpicture}
                \begin{feynman}
                        \vertex (v1); 
                        \vertex [below=1.0of v1] (v2); 
                        \vertex [above left=0.8of v1] (a) {\(e^-(p)\)};
                        \vertex [above right=0.8of v1] (b) {\(\gamma(k)\)};
                        \vertex [below left=0.8of v2] (d) {\(e^+(p')\)};
                        \vertex [below right=0.8of v2] (c) {\(\gamma(k')\)};;

                        \diagram* {
                                (a) -- [fermion] (v1),
                                (v1) -- [fermion,momentum'=\(p-k\)] (v2),
                                (v2) -- [fermion] (d),
                                (c) -- [photon] (v1),
                                (b) -- [photon] (v2),
                        };
                \end{feynman}
        \end{tikzpicture}
} &
		\parbox[c]{10.0cm}{ \begin{eqnarray*} i\mathcal{M} & = &
				\varepsilon_{\nu}(\vec{k},\lambda)
				\bar{v}(p')(ie\gamma^{\nu})
				\left[\frac{-i (-(\dsl{k} - \dsl{p}') +
				m)}{(p-k')^2 + m^2_e - i\epsilon}\right]
				\\
				& & \mbox{\hspace{2.0cm}} \times
				(ie\gamma^{\mu})u(p) \varepsilon_{\mu}
				(\vec{k}',\lambda') 
		\end{eqnarray*} } \\
\hline
\end{tabular}
\end{center}
		
Since we will be computing un-polarized (average over initial
spin/polarizations), inclusive (sum over final spin/polarizations)
cross-sections, we have to evaluate $\Case{1}{2} \Case{1}{2}
\sum_{\sigma,\sigma'} |\mathcal{M}|^2$ for the fermion scattering and
likewise for the other cases. For the Compton scattering and pair
annihilation where two diagrams contribute, we need to add them {\em
before} taking the mod-square.

\subsection{Electron-muon processes:}
Consider the electron-muon scattering. We have,
\begin{eqnarray*}
(i\mathcal{M})(-i\mathcal{M}^*) & = & \frac{e^4}{(q^2 - i\epsilon)(q^2 +
i\epsilon)} \{(\bar{u}(p')\gamma^{\mu}u(p)) (u^{\dagger}(p)
(\gamma^{\nu})^{\dagger} (\gamma^0)^{\dagger} u(p'))\} \\
& & \mbox{\hspace{3.0cm}} \times \{(\bar{u}(k')\gamma_{\mu}u(k))
(u^{\dagger}(k) (\gamma_{\nu})^{\dagger} (\gamma^0)^{\dagger} u(k'))\}
\end{eqnarray*}
The spins, $\sigma,\sigma', \rho, \rho'$ are implicit in the above
expression.  Summing over all of these replaces $u\bar{u}$ using the
completeness relation. Each of the braces can be expressed as traces
over the gamma matrices. With the new normalizations used, the
completeness relations take the form,
\begin{equation} \label{NewCompletenessReln}
\boxed{\sum_{\sigma}u(p,\sigma)\bar{u}(p,\sigma) = -\dsl{p} + m ~ , ~
\sum_{\sigma}v(p,\sigma)\bar{v}(p,\sigma) = -\dsl{p} - m}.
\end{equation}
Thus the electron and muon spin sums become,
\[
	\{\dots\}_{p,p'} = Tr[\gamma^{\mu}(-\dsl{p} +
		m_e)\gamma^{\nu}(-\dsl{p}' + m_e] ~ ~ , ~ ~
		\{\dots\}_{k,k'} = Tr[\gamma_{\mu}(-\dsl{k} +
			m_{\mu})\gamma_{\nu}(-\dsl{k}' + m_{\mu}] \ . 
\] This leads to,
\begin{eqnarray}
	\frac{1}{4}\sum_{\sigma,\sigma',\rho,\rho'}
	|\mathcal{M}|^2_{e\mu\to e\mu}  & = & \frac{1}{4}\frac{e^4}{(q^2
	- i\epsilon)(q^2 + i\epsilon)} Tr[\gamma^{\mu}(-\dsl{p} +
		m_e)\gamma^{\nu}(-\dsl{p}' + m_e] \nonumber \\
& & \mbox{\hspace{3.0cm}}\times Tr[\gamma_{\mu}(-\dsl{k} +
	m_{\mu})\gamma_{\nu}(-\dsl{k}' + m_{\mu}]
\end{eqnarray}

For the $e^+ e^- \to \mu^+ \mu^-$ we will have,
\begin{eqnarray*}
(i\mathcal{M})(-i\mathcal{M}^*) & = & \frac{e^4}{(q^2 - i\epsilon)(q^2 +
i\epsilon)} \{(\bar{v}(p')\gamma^{\mu}u(p)) (u^{\dagger}(p)
(\gamma^{\nu})^{\dagger} (\gamma^0)^{\dagger} v(p'))\} \\
& & \mbox{\hspace{3.0cm}} \times \{(\bar{u}(k')\gamma_{\mu}v(k))
(v^{\dagger}(k) (\gamma_{\nu})^{\dagger} (\gamma^0)^{\dagger} u(k'))\}
\end{eqnarray*}
This is the same as before except the $p', k'$ spinors are $v$ spinors.
Using the corresponding completeness relation we get,
\begin{eqnarray}
	\frac{1}{4}\sum_{\sigma,\sigma',\rho,\rho'}
	|\mathcal{M}|^2_{e^+e^-\to \mu^+\mu^-}  & = &
	\frac{1}{4}\frac{e^4}{(q^2 - i\epsilon)(q^2 + i\epsilon)}
	Tr[\gamma^{\mu}(-\dsl{p} + m_e)\gamma^{\nu}(-\dsl{p}' - m_e]
		\nonumber \\
& & \mbox{\hspace{3.0cm}}\times Tr[\gamma_{\mu}(-\dsl{k} +
	m_{\mu})\gamma_{\nu}(-\dsl{k}' - m_{\mu}]
\end{eqnarray}

The Compton scattering and annihilation processes have two diagrams each
and each of these has a {\em single} Dirac trace. But each diagram also
has photon polarization completeness relations. Incidentally, for Moller
and Bhabha processes, there are `crossed diagrams' too. We will do these
later.

To proceed with the two processes above, we need to traces of 2 and 4
gamma matrices. Here are the relevant formulae.
%
\begin{eqnarray}\label{DiracTraces}
	Tr\mathbb{1} & = & 4 ; \\
	Tr \gamma^{\mu}\gamma^{\nu} & = & -4 \eta^{\mu\nu} ; \\
	Tr \gamma^{\mu}\gamma^{\nu}\gamma^{\alpha}\gamma^{\beta} & = &
4\left(\eta^{\mu\nu}\eta^{\alpha\beta} - \eta^{\mu\alpha}\eta^{\nu\beta}
+ \eta^{\mu\beta}\eta^{\nu\alpha}\right) \\ \nonumber
\end{eqnarray}
%

Using these, we get
\begin{eqnarray*}
	Tr[\gamma^{\mu}(-\dsl{p} + m_e)\gamma^{\nu}(-\dsl{p}' + m_e] & =
		& p_{\alpha}p'_{\beta} Tr[\gamma^{\mu} \gamma^{\alpha}
		\gamma^{\nu} \gamma^{\beta}] +
		m_e^2Tr[\gamma^{\mu}\gamma^{\nu}] \\
		& = & 4(p^{\mu}p^{'\nu} - \eta^{\mu\nu}p\cdot p' +
		p^{\nu}p^{'\mu}) - 4m_e^2\eta^{\mu\nu} ~ ~ ~ ~
		\mbox{and,} \\
		Tr[\gamma_{\mu}(-\dsl{k} +
			m_{\mu})\gamma_{\nu}(-\dsl{k}' + m_{\mu}] & = &
			k_{\alpha}k'_{\beta} Tr[\gamma_{\mu}
			\gamma^{\alpha} \gamma_{\nu} \gamma^{\beta}] +
			m_{\mu}^2Tr[\gamma_{\mu}\gamma_{\nu}] \\
		& = & 4(k_{\mu}k'_{\nu} - \eta_{\mu\nu}k\cdot k' +
		k_{\nu}k'_{\mu}) - 4m_{\mu}^2\eta_{\mu\nu}. 
\end{eqnarray*}
Dotting the traces gives, 
\begin{eqnarray}
Tr[\dots]Tr[\dots] & = & 32[(p\cdot k)(p'\cdot k') + (p\cdot k')(p'\cdot
k) + m_{\mu}^2p\cdot p'+ m^2_ek\cdot k'+ 2 m_e^2m_{\mu}^2]
\mbox{\hspace{1cm}} \nonumber \\
\therefore \frac{1}{4}\sum_{spins} |\mathcal{M}|^2_{e\mu\to e\mu}  & = &
\frac{8e^4}{q^4} [(p\cdot k)(p'\cdot k') + (p\cdot k')(p'\cdot k) +
m^2_{\mu}p\cdot p' + m_{e}^2k\cdot k' + 2 m_e^2m_{\mu}^2] 
\end{eqnarray}
where $q  =  p' -p$ . 

For the $e^+ e^- \to \mu^+ \mu^-$ process, the only difference is
$m^2_e, m^2_{\mu}$ changing signs. Thus we have,
\begin{equation}
\frac{1}{4}\sum_{spins} |\mathcal{M}|^2_{e^+e^-\to\mu^+\mu^-}  =
\frac{8e^4}{q^4} [(p\cdot k)(p'\cdot k') + (p\cdot k')(p'\cdot k) -
m^2_{\mu}p\cdot p' - m_{e}^2k\cdot k' + 2 m_e^2m_{\mu}^2] 
\end{equation}
where  $q = p + p'$.

To relate to cross-section, we can use the center of mass frame. For the
electron-muon scattering, we may choose 
\[
	p^{\mu} = (E_e,p\hat{z})\ ,\ k^{\mu} = (E_{\mu}, -p\hat{z}) ~ ~ , ~
	~ p^{'\mu} = (E'_e, \vec{p}')\ ,\ k^{'\mu} = (E'_{\mu}, -\vec{p}')
	~ , ~ cos(\Theta_{cm}) := \hat{z}\cdot\hat{\vec{p}}' .
\]
Notice that $E_e + E_{\mu} = \sqrt{p^2 + m_e^2} + \sqrt{p^2 + m_{\mu}^2}
= E'_e + E'_{\mu} = \sqrt{(p')^2 + m_e^2} + \sqrt{(p')^2 + m_{\mu}^2}$,
implies that $|p'| = |p|$. The magnitude of the center of mass momentum
and the scattering angle are the only two independent parameters given
the electron and muon masses.

For $e^+e^- \to \mu^+\mu^-$, a good deal of convenience ensues. The
equality of magnitudes of momenta and masses being identical in both
initial and final states, implies that the energies of individual
particles in initial and final state are equal and total energy
conservation implies these energies are equal too. The initial and final
momenta magnitudes are then simply related by the masses. Thus we may
choose,
\[
	p^{\mu} = (E,p\hat{z})\ ,\ (p')^{\mu} = (E, -p\hat{z}) ~ ~ , ~ ~
	k^{\mu} = (E, \vec{k})\ ,\ (k')^{\mu} = (E, -\vec{k}) ~ , ~
	cos(\Theta_{cm}) := \hat{z}\cdot\hat{\vec{k}} .
\]
Again only the center of mass energy and the scattering angle are the
only independent parameters given the masses.

It remains to simplify the expression for the cross-sections. We will
give it for the muon production process and leave the $e\mu$ scattering
process as an exercise. We will also neglect the electron mass compared
to the muon mass.

The dot products of momenta, in the center of mass take the form, $q^2 =
(p+p')^2 = - (2E)^2 = -4(k^2 + m^2_e) \approx -4 k^2$. Next, $p\cdot p'
= -E^2 - p^2 \approx -2E^2$ and 
\[
p\cdot k = p'\cdot k' = -E^2 + Ekcos(\Theta_{cm})~ ~,~ ~ p\cdot k' =
p\cdot k = -E^2 - Ekcos(\Theta_{cm})\ .
\]

Substituting in the invariant amplitude gives ($m_e = 0$ set),
\begin{eqnarray}
\frac{1}{4}\sum_{spins} |\mathcal{M}|^2_{e^+e^-\to\mu^+\mu^-}  & = &
\frac{8e^4}{(-4E^2)^2}\left\{(E^2 + Ekcos(\Theta_{cm}))^2 + (E^2 -
Ekcos(\Theta_{cm}))^2 - m_{\mu}^2(-2E^2)\right\} \nonumber \\
& = & e^4\left\{ 1 + \frac{k^2 cos^2(\Theta_{cm})}{E^2} +
\frac{m^2_{\mu}}{E^2} \right\} \nonumber \\
& = & e^4\left[ \left(1 + \frac{m^2_{\mu}}{E^2}\right) +  \left(1 -
\frac{m^2_{\mu}}{E^2}\right) cos^2(\Theta_{cm}) \right]
\end{eqnarray}

The differential cross-section is given by (\ref{DiffCrossSection}),
\begin{eqnarray}
	\frac{d\sigma}{d\Omega_{cm}} & = &
	\frac{1}{64\pi^2}\frac{1}{s}\frac{k'_{cm}}{k_{cm}}\left(\frac{1}{4}\sum_{spins}
	|\mathcal{M}|^2\right)  \mbox{\hspace{1.0cm} where, $s = 4E^2,
	m_e = 0, m'_1 = m'_2 = m_{\mu}$} ~ ~ \Rightarrow \nonumber \\
	& = & \frac{1}{64\pi^2}\frac{1}{s}\sqrt{\frac{s^2 -
	4sm_{\mu}^2}{s}}\left[ \left(1 + \frac{m^2_{\mu}}{E^2}\right) +
	\left(1 - \frac{m^2_{\mu}}{E^2}\right) cos^2(\Theta_{cm})
\right] \nonumber \\
& = & \left(\frac{e^2}{4\pi}\right)^2 \frac{1}{4\cdot 4E^2}\sqrt{1 -
\frac{m^2_{\mu}}{E^2}}\left[ \left(1 + \frac{m^2_{\mu}}{E^2}\right) +
\left(1 - \frac{m^2_{\mu}}{E^2}\right) cos^2(\Theta_{cm}) \right] 
\end{eqnarray}
Putting $\alpha := \Case{e^2}{4\pi}$, the fine structure constant and
$E_{cm} = 2E$, we write the {\em unpolarized}, $e^+e^-\to\mu^+\mu^-$
cross-section as,
\begin{eqnarray} \label{MuProdCrossSecns}
\frac{d\sigma}{d\Omega_{cm}}(E,\Theta_{cm}) & = &
\frac{\alpha^2}{4E_{cm}^2}\sqrt{1 - \frac{m^2_{\mu}}{E^2}}\left[ \left(1
+ \frac{m^2_{\mu}}{E^2}\right) + \left(1 - \frac{m^2_{\mu}}{E^2}\right)
cos^2(\Theta_{cm}) \right] \\
\sigma_{total} := \int d\Omega \frac{d\sigma}{d\Omega} & = &
\frac{\pi\alpha^2}{3E_{cm}^2}\sqrt{1 - \frac{m^2_{\mu}}{E^2}} \left(1 +
\frac{m^2_{\mu}}{2E^2}\right) 
\end{eqnarray}

\underline{Remarks:} The square root factor shows that $E \ge m_{\mu}$
must hold for the muon production to proceed. In the ultra-relativistic
limit, $E \gg m_{\mu}$, the differential cross-section has the
characteristic $(1 + cos^2(\Theta))$ angular dependence.

Since the cross-section is a measured quantity and the center of mass
energy is under our control, by comparing the pair production
cross-sections for two different final states, eg $\mu, \tau$, and
taking the ratio, we can obtain bounds on the masses of the heavier
leptons.

{\em Exercises: Obtain the differential and the total cross-section for
$e\mu \to e\mu$ process. Also for the Moller and the Bhabha processes.}

\subsection{Compton Scattering:}
Recall the total amplitude from the two diagrams (we take the photon
polarizations to be real),
\begin{eqnarray} 
i\mathcal{M} & = & \varepsilon_{\nu}(\vec{k}',\lambda')
\bar{u}(p')(ie\gamma^{\nu}) \left[\frac{-i (-(\dsl{p} + \dsl{k}) +
m)}{(p+k)^2 + m^2_e - i\epsilon}\right]
(ie\gamma^{\mu})u(p) \varepsilon_{\mu} (\vec{k},\lambda) \nonumber \\
 & & +	\varepsilon_{\nu}(\vec{k},\lambda) \bar{u}(p')(ie\gamma^{\nu})
 \left[\frac{-i (-(\dsl{p}' - \dsl{k}) + m)}{(p'-k)^2 + m^2_e -
 i\epsilon}\right]
(ie\gamma^{\mu})u(p) \varepsilon_{\mu} (\vec{k}',\lambda')  \\
\mathcal{M} & = & e^2\varepsilon_{\mu}(\vec{k},\lambda)
\varepsilon_{\nu} (\vec{k}',\lambda') \bar{u}(p')\left[
	\frac{\gamma^{\nu}(-(\dsl{p} + \dsl{k}) +
	m)\gamma^{\mu}}{(p+k)^2 + m^2_e - i\epsilon} +
	\frac{\gamma^{\mu}(-(\dsl{p}' - \dsl{k}) +
m)\gamma^{\nu}}{(p'-k)^2 + m^2_e - i\epsilon}\right]u(p) \nonumber \\
\end{eqnarray}
In the second (crossed diagram) terms, we have used $p - k' = p' -k$
which is more convenient.

\underline{Simplification:} Since the momenta are on shell, $p^2 = -m^2,
k^2 = 0 = (k')^2$, we get 
\[ (p+k)^2 + m^2 = p^2 + m^2 + k^2 + 2p\cdot k = 2p\cdot k ~ ~ , ~ ~ 
(p'-k)^2 + m^2 = -2p'\cdot k ;
\]
\[
	(-\dsl{p} + m)\gamma^{\mu}u(p) = (2p^{\mu} +
	\gamma^{\mu}(\dsl{p} + m))u(p) = 2p^{\mu}u(p) ~ , 
\] 
\[
	\bar{u}(p')\gamma^{\mu}(-\dsl{p}' + m) = \bar{u}(p')(2(p')^{\mu}
	+ (\dsl{p}' + m))\gamma^{\mu} = 2(p')^{\mu}\bar{u}(p') ~ .
\] 
\begin{eqnarray}
\mathcal{M} & = & e^2\varepsilon_{\mu}(\vec{k},\lambda)
\varepsilon_{\nu} (\vec{k}',\lambda') \bar{u}(p')\left[
\frac{\gamma^{\nu}(2p^{\mu} - \dsl{k}\gamma^{\mu})} {2p\cdot k -
i\epsilon} + \frac{(2(p')^{\mu} + \gamma^{\mu}\dsl{k})\gamma^{\nu})}
{-2p'\cdot k - i\epsilon}\right]u(p) \\
\mathcal{M}^{\dagger} & = & e^2\varepsilon_{\alpha}(\vec{k},\lambda)
\varepsilon_{\beta} (\vec{k}',\lambda') \bar{u}(p)\left[
	\frac{(2p^{\alpha} - \gamma^{\alpha}\dsl{k})\gamma^{\beta}}
	{2p\cdot k + i\epsilon} + \frac{\gamma^{\beta}(2(p')^{\alpha} +
\dsl{k}\gamma^{\alpha})} {-2p'\cdot k + i\epsilon}\right]u(p') 
\end{eqnarray}
In the second equation, we have used 
\[
[\bar{u}(p')\Gamma u(p)]^* =
\bar{u}(p)(\gamma^0)\Gamma^{\dagger}\gamma^0u(p') =
\bar{u}(p)\Gamma_{reversed}u(p') .
\]
Here $\Gamma_{reversed}$ has the order of the gamma matrices reversed
due to the $\dagger$ operation.

Summing and averaging over the spins and polarizations the
$|\mathcal{M}|^2$, we get,
\begin{eqnarray}\label{ComptonAmpSquared}
\frac{1}{4}\sum_{\sigma,\sigma',\lambda,\lambda'}|\mathcal{M}|^2 & = &
\frac{e^4}{4}\left[Tr\left\{ \left(   \frac{\gamma^{\nu}(2p^{\mu} -
			\dsl{k}\gamma^{\mu})} {2p\cdot k - i\epsilon} +
			\frac{(2(p')^{\mu} +
			\gamma^{\mu}\dsl{k})\gamma^{\nu}} {-2p'\cdot k -
i\epsilon}\right)(-\dsl{p} + m) \right.\right. \nonumber \\
& & \mbox{\hspace{1.0cm}}\left. \left(\frac{(2p^{\alpha} -
\gamma^{\alpha}\dsl{k})\gamma^{\beta}} {2p\cdot k + i\epsilon} +
\frac{\gamma^{\beta}(2(p')^{\alpha} + \dsl{k}\gamma^{\alpha})}
{-2p'\cdot k + i\epsilon}\right)(-\dsl{p}' + m)\right\}\times
\nonumber\\
& & \mbox{\hspace{1.0cm}} \left. \left\{
\sum_{\lambda}\varepsilon_{\mu}(\vec{k},\lambda)
\varepsilon_{\alpha}(\vec{k},\lambda) \right\} \left\{
\sum_{\lambda'}\varepsilon_{\nu}(\vec{k}',\lambda')
\varepsilon_{\beta}(\vec{k}',\lambda') \right\}\right]
\end{eqnarray} 
We have used the completeness relation (\ref{NewCompletenessReln}) to
convert the spin sum. We have a trace over strings of $\gamma$ matrices,
but now we also have sums over the photon polarizations.  An explicit
expression can be obtained by introducing an explicit set of 4
orthonormal vectors.

Consider the polarization sum.  Given a $k, k^2 = 0$ we can write
$k^{\mu} = (|\vec{k}|, \vec{k})$.  Introduce another vector $\tilde{k}
:= C(|\vec{k}|, -\vec{k})$ so that $\tilde{k}^2 = 0$. Fix $C$ by
demanding $k\cdot\tilde{k} = -2 \leftrightarrow C = 1/|\vec{k}|^2$. We
have two transverse directions and we take polarizations
$\varepsilon(\vec{k},\lambda)$ along these two directions and mutually
orthonormalized. These are space-like vectors and have no time
component. These 4 vectors, $k, \tilde{k}, \varepsilon(\vec{k},1),
\varepsilon(\vec{k},2)$ are independent and the completeness relation
takes the form,
\[
	-\frac{k^{\mu}\tilde{k}^{\nu} + k^{\nu}\tilde{k}^{\mu}}{2} +
	\varepsilon^{\mu}(\vec{k},1)\varepsilon^{\nu}(\vec{k},1) +
	\varepsilon^{\mu}(\vec{k},2)\varepsilon^{\nu}(\vec{k},2) ~ = ~
	\eta^{\mu\nu} \\
\]
Thus we get the sum over photon polarizations as,
\begin{equation}
	\sum_{\lambda=1,2} \varepsilon_{\mu}(\vec{k},\lambda)
	\varepsilon_{\nu}(\vec{k},\lambda) ~ = ~ \eta_{\mu\nu} +
	\frac{k_{\mu}\tilde{k}_{\nu} + k_{\nu}\tilde{k}_{\mu}}{2} 
\end{equation}

\underline{Claim:} The terms containing $k, \tilde{k}$ in the
polarization sum, do not contribute to the $\sum|\mathcal{M}|^2$.

\underline{Proof:} Observe that we can always express $\mathcal{M} =
\varepsilon_{\mu}(\vec{k},\lambda)M^{\mu}$, by just taking out the
$\varepsilon_{\mu}$. Then $\sum_{\lambda}|\mathcal{M}|^2 =
M^{\mu}(M^*)^{\nu}\{\eta_{\mu\nu} + \Case{1}{2}(k_{\mu}\tilde{k}_{\nu} +
k_{\nu}\tilde{k}_{\mu})\}$. The $k$ terms contracting with $M$ can be
seen by replacing $\varepsilon_{\mu} \to k_{\mu}$. It is more convenient
to write the Fermion propagators using, $\Case{- \dsl{q} + m}{q^2+m^2
-i\epsilon} = \Case{1}{\dsl{q}+m}$ since $(-\dsl{q}+m)(\dsl{q}+m) =
q^2+m^2$. Then,
\begin{eqnarray*}
M|_{\varepsilon_{\mu} = k_{\mu}} & = & e^2\bar{u}(p')
\left[\dsl{\varepsilon}(k') \frac{1}{\dsl{p}+\dsl{k}+m}
\{\dsl{k}+\dsl{p}+m - \dsl{p}-m\} \right. \\ 
& & \mbox{\hspace{2.0cm}}\left.  + \{\dsl{k}-\dsl{p}'-m + \dsl{p}'+m\}
\frac{1}{\dsl{p}'-\dsl{k}+m} \dsl{\varepsilon}(k') \right]u(p) \\
& = & e^2\left[\bar{u}(p')\dsl{\varepsilon}(k')u(p) -\bar{u}(p')
\dsl{\varepsilon}(k') \frac{1}{\dsl{p}+\dsl{k}+m}(\dsl{p}+m)\})u(p)
\right.\\
& & \mbox{\hspace{1.0cm}}\left.  \bar{u}(p')(-\dsl{\varepsilon}(k'))u(p)
+ \bar{u}(p')(\dsl{p}'+m) \frac{1}{\dsl{p}'-\dsl{k}+m}
\dsl{\varepsilon}(k')u(p)\right]  ~ ~ = 0 ~ !
\end{eqnarray*}
We have used $p -k' = p'-k$ in the second term in the first equation. In
the last equation, the first and the third terms cancel while the second
and the fourth terms vanish by equation of motion. This proves the
claim. Hence, {\em effectively, each of the polarization sums give only the
$\eta_{\mu\nu}$}.

\underline{Note:} The {\em total scattering amplitude} vanishes when the
photon transverse polarization is replaced by a longitudinal one, is a
general result known as a `Ward identity'. This is a consequence of
gauge invariance. While we do not discuss the general proof here, it
suffices to note that (a) we used the Dirac equation: $(\dsl{p}+m)u(p) =
0 = \bar{u}(p')(\dsl{p}' + m)$; (b) the QED coupling has the form $\sim
A_{\mu}J^{\mu}$ with $J^{\mu} \sim \bar{\Psi}\gamma^{\mu}\Psi \sim
\bar{u}(p')\gamma^{\mu}u(p)$; (c) $\delta A_{\mu} = \partial_{\mu}
\Lambda$ is a gauge transformation ($\leftrightarrow
\varepsilon_{\mu}(k) \to \varepsilon_{\mu}(k) + k_{\mu}\lambda(k)$) and
gauge invariance of the interaction implies $\partial_{\mu}J^{\mu} = 0.$
Thus, the vanishing of the amplitude is related to gauge invariance.

The averaged $|\mathcal{M}|^2$ in equation (\ref{ComptonAmpSquared})
becomes, 
\begin{eqnarray}\label{ComptonAmpSquared4}
\sum_{\sigma,\sigma',\lambda,\lambda'}|\mathcal{M}|^2 & = &
e^4\left[Tr\left\{ \left(   \frac{\gamma^{\nu}(2p^{\mu} -
			\dsl{k}\gamma^{\mu})} {2p\cdot k - i\epsilon} +
			\frac{(2(p')^{\mu} +
			\gamma^{\mu}\dsl{k})\gamma^{\nu}} {-2p'\cdot k -
i\epsilon}\right)(-\dsl{p} + m) \right.\right. \nonumber \\
& & \mbox{\hspace{1.0cm}}\left. \left. \left(\frac{(2p_{\mu} -
\gamma_{\mu}\dsl{k})\gamma_{\nu}} {2p\cdot k + i\epsilon} +
\frac{\gamma_{\nu}(2(p')_{\mu} + \dsl{k}\gamma_{\mu})} {-2p'\cdot k +
i\epsilon}\right)(-\dsl{p}' + m)\right\}\right] \\
& = & e^4\left[Tr\left\{ \frac{(2p^{\mu}\gamma^{\nu} -
\gamma^{\nu}\dsl{k}\gamma^{\mu})\dsl{p}(2p_{\mu}\gamma_{\nu} -
\gamma_{\mu}\dsl{k}\gamma_{\nu})\dsl{p}'}{(2p\cdot k)^2}
\right\}_{\circled{1}} ~ ~ + \right. \nonumber \\
& & \hspace{0.75cm} Tr\left\{ \frac{(2(p')^{\mu}\gamma^{\nu} +
\gamma^{\mu}\dsl{k}\gamma^{\nu})\dsl{p} (2p'_{\mu}\gamma_{\nu} +
\gamma_{\nu}\dsl{k} \gamma_{\mu})\dsl{p}'}{(-2p'\cdot k)^2}
\right\}_{\circled{2}} ~ ~ + \nonumber \\
& & \hspace{0.75cm} Tr\left\{ \frac{(2p^{\mu}\gamma^{\nu} -
\gamma^{\nu}\dsl{k}\gamma^{\mu})\dsl{p}(2p'_{\mu}\gamma_{\nu} +
\gamma_{\nu}\dsl{k} \gamma_{\mu})\dsl{p}'}{-4 p\cdot k p'\cdot
k}\right\}_{\circled{3}} ~ ~ + \nonumber \\
& & \hspace{0.75cm} \left. Tr\left\{ \frac{(2(p')^{\mu}\gamma^{\nu} +
\gamma^{\mu}\dsl{k} \gamma^{\nu})\dsl{p} (2p_{\mu}\gamma_{\nu} -
\gamma_{\mu}\dsl{k} \gamma_{\nu})\dsl{p}'} {-4 p\cdot k p'\cdot
k}\right\}_{\circled{4}} \right]  \nonumber \\
& & + m^2 e^4\left[ Tr\left\{ \frac{(2p^{\mu}\gamma^{\nu} - \gamma^{\nu}
\dsl{k}\gamma^{\mu})(2p_{\mu}\gamma_{\nu} -
\gamma_{\mu}\dsl{k}\gamma_{\nu})}{(2p\cdot k)^2} \right\}_{\circled{5}}
~ ~ + \right.  \nonumber \\
& & \hspace{0.75cm} Tr\left\{ \frac{(2(p')^{\mu}\gamma^{\nu} +
\gamma^{\mu}\dsl{k}\gamma^{\nu}) (2p'_{\mu}\gamma_{\nu} +
\gamma_{\nu}\dsl{k} \gamma_{\mu})}{(-2p'\cdot k)^2}
\right\}_{\circled{6}} ~ ~ + \nonumber \\
& & \hspace{0.75cm} Tr\left\{ \frac{(2p^{\mu}\gamma^{\nu} -
\gamma^{\nu}\dsl{k}\gamma^{\mu})(2p'_{\mu}\gamma_{\nu} +
\gamma_{\nu}\dsl{k} \gamma_{\mu})}{-4 p\cdot k p'\cdot
k}\right\}_{\circled{7}} ~ ~ + \nonumber \\
& & \left. \hspace{0.75cm} Tr\left\{ \frac{(2(p')^{\mu}\gamma^{\nu} +
\gamma^{\mu}\dsl{k} \gamma^{\nu}) (2p_{\mu}\gamma_{\nu} -
\gamma_{\mu}\dsl{k} \gamma_{\nu})} {-4 p\cdot k p'\cdot
k}\right\}_{\circled{8}} \right]
\end{eqnarray} 
The terms linear on $m$ have odd number of $\gamma$'s and hence vanish
under trace.

It may be checked easily that the $\circled{2}$ is obtained from
$\circled{1}$ by $p \leftrightarrow - p'$ and likewise for $5
\leftrightarrow 6$. Noting the identity
$Tr(\gamma_1\gamma_2\cdots\gamma_n) =
Tr(\gamma_n\gamma_{n-1}\cdots\gamma_2\gamma_1)$, it follows that $3 = 4$
and $7 = 8$.  The first group of 4 terms involves trace of a maximum of
8 gamma matrices while the last 4 terms involve a maximum of 6 gamma's.
These are simplified using various identities among the gamma matrices.

Consider $\circled{1}$. We have the traces,
\[
	Tr\left\{ 4p^2\gamma^{\nu}\dsl{p}\gamma_{\nu}\dsl{p}'
		-2\gamma^{\nu}\dsl{p}\dsl{p}\dsl{k}\gamma_{\nu}\dsl{p}'
		-2\gamma^{\nu}\dsl{k}\dsl{p}\dsl{p}\gamma_{\nu}\dsl{p}'
		+\gamma^{\nu}\dsl{k}\gamma^{\mu}\dsl{p}\gamma_{\mu}\dsl{k}\gamma_{\nu}\dsl{p}'
	\right\}
\]
Use: $\boxed{\gamma^{\nu}\dsl{p}\gamma_{\nu} = +2\dsl{p}, ~
\dsl{p}\dsl{p} = -p^2, ~ \dsl{k}\dsl{p} + \dsl{p}\dsl{k} = -2p\cdot k, ~
Tr\dsl{p}\dsl{p}' = -4p\cdot p'}$ and the cyclic property of the trace
to get the above trace as,
\begin{eqnarray}
	Tr\{\cdots\}_{\circled{1}} & = & (4p^2)(2)(-4p\cdot p') -
	2p^2(2)(-4k\cdot p')\times 2 +
	(2)(2)Tr(\dsl{k}\dsl{p}\dsl{k}\dsl{p}') \nonumber \\
	& = & 32 m^2(p\cdot p' - k\cdot p') + 32(k\cdot p)(k\cdot p')
\end{eqnarray}
We have used $p^2 = -m^2$ and $k^2 = 0$.

Similarly we get the other traces (without the denominator factors) as,
\begin{eqnarray}
Tr\{\cdots\}_{\circled{3}} & = & 4p\cdot p'Tr(\gamma^{\nu} \dsl{p}
\gamma_{\nu} \dsl{p}') - 2 Tr(\dsl{k}\dsl{p}'\dsl{p}
\gamma_{\nu}\dsl{p}'\gamma^{\nu}) + 2 Tr(\gamma^{\nu} \dsl{p}
\gamma_{\nu} \dsl{k}\dsl{p}\dsl{p}') \nonumber \\
& & ~ ~ ~ ~ - Tr(\gamma^{\nu}\dsl{k}\gamma^{\mu}\dsl{p}\gamma_{\nu}
\dsl{k}\gamma_{\mu}\dsl{p}') \\
& = & -32(p\cdot p')^2 + (k\cdot p - k\cdot p') (16m^2 + 32 p\cdot p') +
0 \\
Tr\{\cdots\}_{\circled{5}} & = & 4p^2 Tr(\gamma^{\nu}\gamma_{\nu}) -2
Tr(\dsl{k}\dsl{p}(-4\mathbb{1})) +8 Tr(\dsl{p}\dsl{k}) + 16Tr(\dsl{k}
	\dsl{k}) \nonumber \\
& = & 64 m^2 - 64 k\cdot p \\
Tr\{\cdots\}_{\circled{7}} & = & -64p\cdot p' + 32(k\cdot p - k\cdot p')
- Tr(\gamma^{\nu}\dsl{k}\gamma^{\mu}\gamma_{\nu}\dsl{k}\gamma_{\mu})
\nonumber \\
& = &  -64p\cdot p' + 32(k\cdot p - k\cdot p') + 0
\end{eqnarray}

To simplify the expression further, we need to use: $p\cdot p' = -m^2 +
k\cdot p - k\cdot p'$ which follows from $0 = (k')^2 = (p + k - p')^2$.
For comparison with \cite{PeskinSchroder}, it is useful to note: $k\cdot
p = k'\cdot p'\ ,\ p\cdot k' = p'\cdot k$.


The final expression is,
\begin{equation}\label{ComptonAmpSquared2}
\frac{1}{4}\sum_{spin/Pol}|\mathcal{M}|^2 = 2e^4\left[\left(
\frac{p\cdot k'}{p\cdot k} + \frac{p\cdot k}{p\cdot k'}\right) +
2m^2\left( \frac{1}{p\cdot k'} - \frac{1}{p\cdot k}\right) +
m^4\left(\frac{1}{p\cdot k} - \frac{1}{p\cdot k'}\right)^2\right]		
\end{equation}
{\em Exercise:} Check the algebra!

The Compton cross-section is usually presented in the lab frame with the
electron initially at rest, i.e. $p^{\mu} = (m,\vec{0}), k^{\mu} =
(\omega, \omega\hat{z}), (p')^{\mu} = (E',\vec{p}]), (k')^{\mu} =
(\omega', \omega'sin{\theta}, 0, \omega'cos(\theta))$. The final
electron and photon momenta define a plane which is taken to be the
$z-x$ plane with the final photon making an angle $\theta$ to the
$z-$axis. These choices give: $p\cdot k' = -m\omega', p\cdot k = -
m\omega$ and, 
\begin{equation}\label{ComptonAmpSquared3}
\sum_{spin/Pol}|\mathcal{M}|^2 = 2e^4\left[ \frac{\omega'}{\omega} +
\frac{\omega}{\omega'} + 2m\left(- \frac{1}{\omega'} +
\frac{1}{\omega}\right) + m^2\left(-\frac{1}{\omega} +
\frac{1}{\omega'}\right)^2\right]		
\end{equation}

Next, $(p')^2 = (p + k - k')^2$ implies, 
\[
	-m^2 = -m^2 + 2p\cdot(k-k') + (k-k')^2 \Rightarrow 0 =
	m(\omega-\omega') - \omega\omega'(1-cos(\theta)) ,
\]
leading to the Compton formula for shift in the photon wavelength with
the scattering angle:
\begin{equation}
	\mbox{Compton Formula:~ ~} \boxed{\frac{1}{\omega'} -
	\frac{1}{\omega} = \frac{1-cos\theta}{m}} ~ \leftrightarrow ~
	\omega' = \frac{\omega}{1+\Case{\omega}{m}(1-cos\theta)} \ .    
\end{equation}

To get the differential cross-section, we simplify the phase space
integral as,
\begin{eqnarray*}
\int \frac{d^3k'}{2\omega'(2\pi)^3}
\int\frac{d^3p'}{2E'(2\pi)^3}(2\pi)^4 \delta^4(p;+k'-p-k) =
\mbox{\hspace{3.0cm}} & & \\
\int \frac{d\omega'
(\omega')^2d\Omega}{2\omega'(2\pi)^3}\frac{1}{2E'(2\pi)^3}(2\pi)^4\delta(\omega'+E'-m-\omega)    
\end{eqnarray*}
We have used up a $\delta^3$ to get $\vec{p}' = \vec{k}' - \vec{k} -
\vec{p} = (\omega'sin\theta, 0, \omega'cos\theta - \omega)$ which gives
$(E')^2 - (\omega')^2 + \omega^2 - 2\omega'\omega cos\theta + m^2$.
Hence
\[
	\int d\omega'\frac{\omega'}{4E'(\omega')}\delta(\omega' +
	E'(\omega') - m - \omega) =
	\frac{\omega'_*}{4E'(\omega'_*)}\frac{1}{|1+\Case{dE'}{d\omega'_*}|}
	~ ~,~ ~ \frac{dE'}{d\omega'} = \frac{\omega'-\omega
	cos\theta}{E'} \ .  
\]
Inserting in the phase space integral,
\begin{eqnarray}
\int_{\Gamma_2} & = & \int \frac{dcos\theta}{2\pi}
\frac{\omega'_*}{4E'_*}\frac{1}{|1 + \case{\omega'-\omega
cos\theta}{E'}|_*}  ~ ~ , ~ ~ \omega'_* + E'(\omega'_*) = m + \omega
\nonumber \\
& = & \frac{1}{8\pi}\int dcos\theta \frac{\omega'^2}{m\omega} \ . ~ ~
\mbox{\hspace{2.0cm} We have,} \\
\frac{d\sigma}{dcos\theta} & = &
\frac{1}{4m\omega}\frac{1}{8\pi}\frac{(\omega')^2}{m\omega}[\sum
|\mathcal{M}|^2 ]   \nonumber \\
& = &
\frac{e^4}{16\pi}\frac{1}{m^2}\left(\frac{\omega'}{\omega}\right)^2
\left[\frac{\omega'}{\omega} + \frac{\omega}{\omega'} + \left\{2m
\frac{-1+cos\theta}{m} + m^2 \frac{(1-cos\theta)^2}{m^2}\right\}\right]
\nonumber 
\end{eqnarray}
This give the differential cross-section for the Compton scattering,
known as {\em Klein-Nishina formula},
\begin{equation}
\boxed{\left(\frac{d\sigma}{dcos\theta_{lab}}\right) =
	\frac{\pi\alpha^2}{m^2} \left(\frac{\omega'}{\omega}\right)^2
	\left[\frac{\omega'}{\omega} + \frac{\omega}{\omega'} -
	sin^2\theta_{lab}\right]\ , \ \mbox{with}\
	\frac{\omega'}{\omega} = \frac{1}{1 +
\Case{\omega}{m}(1-cos\theta)}  .} 
\end{equation}

\underline{Remarks:}

(a) There has been no approximation at the $\alpha^2$ level computation. It
can now be used for various limits. At $\theta = 0$, forward scattering,
there is no frequency change and the cross-section equaling
$\Case{\pi\alpha^2}{m}$ is independent of the photon energy. For a {\em
massive charged particle}, one defines its {\em Compton length} to be
$\lambda_c := h/(mc) = 2\pi/m$ in the natural units, $\hbar = 1 = c$.
The forward scattering cross-section is then $\alpha^2(\pi
\cross{\lambda}^2)$, where $\cross{\lambda} := \lambda/(2\pi)$.

(b) There are two scales in the problems: $m, \omega$. This gives two
natural limits: (i) $\omega \to 0 \leftrightarrow \omega/m \ll 1$ and
(ii) $\omega \to \infty \leftrightarrow \omega/m \gg 1$.

For a generic $\theta$, consider $\omega \to 0$. Then $\omega/\omega'
\to 1 + (\omega/m)(1-cos\theta)$ and 
\begin{equation}\label{ThomsonCrossSection}
	\left. \frac{d\sigma}{dcos\theta}\right|_{\omega/m \ll 1} =
	\frac{\pi\alpha^2}{m^2}\left[1 + cos^2\theta - 4sin^2(\theta/2)\
	\frac{\omega}{m}  + \dots\right]  ~ ,  ~
	\sigma_{tot} = \frac{8\pi\alpha^2}{3m^2} + \dots
\end{equation}
The leading term is the {\em Thomson cross-section} for classical
radiation scattering off free electrons.

In the opposite limit, $m/\omega \ll 1$, we have 
\[
\omega'/\omega = \frac{m}{\omega}\frac{1}{1-cos\theta}\left(1 -
\frac{m}{\omega}\frac{1}{1-cos\theta} + \dots\right) ~ := ~
\epsilon(1-\epsilon \dots)~ ~,~ ~ \epsilon(\theta) :=
\frac{m}{\omega}\frac{1}{1-cos\theta} .
\]

In this limit,
\begin{eqnarray}
	\frac{d\sigma}{dcos\theta} & \approx &
	\frac{\pi\alpha^2}{m^2}\epsilon(\theta) ~ = ~
	\frac{\pi\alpha^2}{m^2}\frac{m}{\omega(1-cos\theta)} , \\
	\sigma_{tot}(\delta) := \int_{-1}^{1-\delta}dcos\theta
	\frac{d\sigma}{dcos\theta} & \approx &
	\frac{\pi\alpha^2}{m\omega}\left[ -\ell n(\delta/2) \right] ~
	\to \infty\ , \ \mbox{as $\theta\to 0$}.     
\end{eqnarray}
For the $\epsilon(\theta) \ll 1$ condition to hold so that the expansion
is meaningful, $\theta \gg \sqrt{2m/\omega}$ must hold. At $\theta = 0$
we know the exact answer which is finite.

\subsection{Electron-Positron Annihilation:}
Having evaluated the Compton scattering amplitude, this is actually
simple to evaluate. From the table giving the amplitudes (notice the use
of $p'-k$ and $k-p'$ momenta in the crossed diagrams), we can see that
the annihilation diagrams expressions are obtained by making the
substitutions: $\boxed{p \to p, k \to -k, p' \to -p', k' \to k' \
\mbox{and}\ \bar{u} \to \bar{v} .}$ These substitutions preserve the
momentum conservations appropriate for the processes: $p+k = p'+k' \to
p-k = -p'+k'$. The photon polarizations are insensitive to the `sign' of
momentum and will give $\eta_{\mu\nu}$ as before. One of the spin sums
however changes from $[-\dsl{p}'+m] \to [-(-\dsl{p}'+m)] = [\dsl{p}-m]$
which generates an {\em overall minus sign in the summed, squared
amplitude}. This is actually an example of what is known as the {\em
crossing symmetry} of the $S-$matrix: {\em A scattering amplitude for a
particle with momentum $k$ in the initial state is the same as the
amplitude with the initial particle moved to the final state
anti-particle with momentum $-k$.}

Without any explicit calculation we can write down the summed squared
amplitude for the annihilation process as,
\begin{equation}\label{AnnihilationAmpSqured}
\sum_{spin/Pol}|\mathcal{M}|^2 = (-)2e^4\left[-\left( \frac{p\cdot
k'}{p\cdot k} + \frac{p\cdot k}{p\cdot k'}\right) + 2m^2\left(
\frac{1}{p\cdot k'} + \frac{1}{p\cdot k}\right) +
m^4\left(\frac{1}{p\cdot k} + \frac{1}{p\cdot k'}\right)^2\right]		
\end{equation}

It is more convenient to express the cross-section in the center of mass
frame: 
\begin{eqnarray*}
p^{\mu} = (E,p\hat{z}) & , & (p')^{\mu} = (E, -p\hat{z}) ~ , ~ E^2 = p^2
+ m^2 ~ ~,~ ~ s = -(p+p')^2 = 4E^2 \\
k^{\mu} & = & (E, Esin\Theta, 0, Ecos\Theta) ~ , ~ (k')^{\mu} = (E,
-Esin\Theta, 0, -Esin\Theta)  
\end{eqnarray*} 
This gives,
\begin{eqnarray*} 
p\cdot k & = & -E^2 + pEcos\Theta = -E(E - pcos\Theta) ~ ,\\
p\cdot k' & = & -E^2 - pEcos\Theta = -E(E + pcos\Theta) , \\
\frac{|k'|_{cm}}{|k|_{cm}} & = & \sqrt{\frac{s^2 -0 + 0^2}{s^2 -2s(2m^2)
+ 0}} ~ = ~ \frac{1}{1-m^2/E^2} ~ = ~ \frac{E}{\sqrt{E^2 - m^2}} =
\frac{E}{p} \\
\end{eqnarray*} 
The differential cross-section takes the form,
\begin{eqnarray} 
\frac{d\sigma}{d\Omega_{cm}} & = & \frac{1}{64\pi^2} \frac{1}{4E^2}
\frac{E}{|p|} \left(\sum_{spin/Pol}|\mathcal{M}|^2\right)
\mbox{\hspace{2.0cm}with} \nonumber \\
\sum_{spin/Pol}|\mathcal{M}|^2 & = & (-2e^4)\left[
-\left(\frac{E+pcos\Theta}{E-pcos\Theta} +
\frac{E-pcos\Theta}{E+pcos\Theta}\right) \right. \\
& & \mbox{\hspace{1.5cm}} + \frac{2m^2}{-E}\left( \frac{1}{E+pcos\Theta}
+ \frac{1}{E-pcos\Theta}  \right) \\
& & \left. \mbox{\hspace{3.0cm}} + \frac{m^4}{E^2} \left(
\frac{1}{E+pcos\Theta} + \frac{1}{E-pcos\Theta} \right)^2 \right]
\nonumber \\
& = & 4e^4\left[ \frac{E^2+p^2cos^2\Theta}{E^2-p^2cos^2\Theta} +
\frac{2m^2}{E^2-p^2cos^2\Theta} - \frac{2m^4}{(E^2-p^2cos^2\Theta)^2}
\right]
\end{eqnarray} 
The differential cross-section for unpolarized, $e^+e^-$ annihilation
process is then given by,
\begin{equation}
\boxed{	\frac{d\sigma}{d\Omega_{cm}}  = \frac{\alpha^2}{s}\frac{E}{|p|}
\left[ \frac{E^2+p^2cos^2\Theta+2m^2}{m^2+p^2sin^2\Theta} -
\frac{2m^4}{(m^2+p^2sin^2\Theta)^2} \right]  ~ ~ , ~ ~ s = 4E^2\ .}
\end{equation}

In the high energy limit, $E \gg m$, this reduces to,
\begin{equation}
	\frac{d\sigma}{d\Omega_{cm}} ~ \to ~ \frac{\alpha^2}{s}\left(
	\frac{1+cos^2\Theta}{sin^2\Theta} \right)
\end{equation}

To obtain the total cross-section, we need to integrate over the final
state. Since the two photons a identical, $\Theta$ needs to be
integrated between $0, \pi/2$. The Bose factor of 2 (which we did not
include in the amplitude) cancels the factor of 2 from the angular
integration. We get,
\begin{equation}
	\sigma_{tot} ~ = ~ \frac{\alpha^2}{s}(2\pi)\int_{0}^{1}dx
	\frac{1+x^2}{1-x^2}  
\end{equation}
The integral is clearly divergent from the `forward' direction ($\Theta
= 0$). This is artificial since we set $m = 0$ in the integrand. Split
the integration as $\int_0^{1-\Delta} + \int_{1-\Delta}^1$. The first
term is a finite number and this contribution falls off as $s^{-1}$. For
the second term, we go back to the exact differential cross-section,
approximate the integrand for $1-\Delta \le cos(\Theta) \le 1$. Then,
the factors of $E$ all cancel and second term $\sim \alpha^2 (1 -
\Delta)m^{-2}$ where, $\Delta \ll m^2/p^2$ so that the denominator can
be approximated. Thus, the total cross-section does {\em not} diverge
but is bounded by $m^{-2}$. This is an example of bounds satisfied by
total cross-sections at very high energies.


\newpage
\section{Numerical Estimates of Cross-sections and
Applications}\label{Estimates}

We have obtained examples of cross-sections for some of the basic
processes in quantum electrodynamics. We also saw how they relate to
non-relativistic versions of the processes eg potential scattering and
inferred the modification due to relativistic effects as well as
estimate `strengths' of interactions. Before we study the higher order
corrections, it is useful to have a sense of order of magnitudes of
cross-sections and how these numbers are used in applications. We will
focus on the QED processes. 

{\em What is the value of the fine structure constant?}

This is related to the electric charge, the Planck constant due to
quantum framework and the speed of light due to special relativity. We
first obtain the expression in terms of ``engineering units" and then
use the measured values to compute the fine structure constant.

Observe that the Feynman rules are derived from the Lagrangian density
which has properly normalized kinetic (quadratic) terms,
$-\Case{1}{4}F_{\mu\nu}F^{\mu\nu} + i\bar{\Psi} \gamma^{\mu}
\partial_{\mu}\Psi - m\bar{\Psi}\Psi$ and the interaction term
introduces `$e$' as $e\bar{\Psi}\gamma^{\mu}\Psi A_{\mu}$ and thus is
dimensionless. This definition gives the equation of motion,
$\partial_{\mu}F^{\mu\nu} = -e \bar{\Psi}\gamma^{\nu}\Psi
\leftrightarrow \partial_{i}E^i = e\Psi^{\dagger}\Psi$, using the
identifications: $F^{0i} := E^i, \partial_{\mu}F^{\mu\nu} = -J^{\nu},
J^{\mu} = (\rho, J^i)$.  Furthermore, the Hamiltonian and hence the
energy density is $\vec{E}^2/2$.

The MKSA (SI) units has the Gauss law in the form: $\vec{\nabla} \cdot
\vec{E}_{SI} = \rho_{SI}/\epsilon_0$. The field energy density (as
inferred from the Poynting theorem) is $\epsilon_0\vec{E}_{SI}^2/2$.
Comparing the energy densities, we introduce the identification, $E :=
\sqrt{\epsilon_0}E_{SI}$ Then $\rho = \vec{\nabla}\cdot \vec{E} =
\sqrt{\epsilon_0}\vec{\nabla}\cdot \vec{E}_{SI} := \sqrt{\epsilon_0}
\Case{\rho_{SI}}{\epsilon_0}$. This leads to the identification,
$\boxed{e := \Case{e_{SI}}{\sqrt{\epsilon_0}}. }$ Thus, we identify the
variables used in the action with those used in engineering variables by
comparing the equations of motion {\em and} the energy densities.

In the natural units, $\hbar = 1 = c$ used in writing the action, the
fine structure constant was defined as $\alpha := e^2/(4\pi)$ which is
dimensionless as noted above. Substituting $e$ in terms of $e_{SI}$ and
introducing $\hbar, c$, we write $\alpha := \Case{e^2_{SI}}
{4\pi\epsilon_0} \hbar^{a}c^b$ and determine the powers $a,b$ by
requiring $\alpha$ to be dimensionless. For this, we need the dimensions
of $e_{SI}$ and $\epsilon_0$! Now we also use the SI expression of the
Lorentz force, $F = e_{SI}E_{SI}$. Hence, {\em dimensionally}
(Force)$^2$/(energy density) $\sim e^2_{SI}/\epsilon_0 \sim
M^1L^3T^{-2}$. This and $\alpha$ being dimensionless gives $a = b = -1$
and $\boxed{\alpha = \Case{e^2_{SI}}{4\pi\epsilon_0\hbar c} \simeq
1/137\ ,}$ using the values: $(4\pi\epsilon_0)^{-1} = c^210^{-7}, e_{SI}
= 1.6\times 10^{-19}$ Coulomb, $\hbar \simeq 1.05\times 10^{-34}$
Joules.sec and $c \simeq 3\times 10^8$ m/sec. \\

{\em An Aside:} Another set of units are commonly used in gravitational
physics, the so called {\em geometrized units}. These are defined by
setting $G = c = 1$. The choice of $c=1$ gives: 1 sec = 3$\times 10^{8}$
meters. Newton's constant in SI units is, $G \approx 6.67\times 10^{-11}
kg^{-1} meter^3 sec^{-2}$. Hence setting $G = 1$ along with $c=1$, gives
1 kg = $7.4\times 10^{-28}$ meters. 

It is conventional in gravitational physics to use the Gauss law in the
form $\vec{\nabla}\cdot\vec{E} = 4\pi\rho$ and the electric field energy
density as $E^2/(8\pi)$. This gives the identifications: $E_{geom} =
\sqrt{4\pi\epsilon_0} E_{SI}\ ,\ q_{geom} = q_{SI}/\sqrt{4\pi\epsilon_0}
$.  As before, the Lorentz force equation gives the dimensions as:
$[q^2_{SI}{\epsilon_0}] = [4\pi q^2_{geom}] = M^1 L^3 T^{-2} = L^2$.

Setting $q_{geom} = \Case{q_{SI}}{\sqrt{4\pi\epsilon_0}} G^{\alpha}
c^{\beta}$ must have dimensions of $L$, we infer $\alpha = 1/2$ and
$\beta = -2$ and $\boxed{q_{geom} = \Case{q_{SI}}{\sqrt{4\pi\epsilon_0}}
	G^{1/2}c^{-2} = q_{SI}\sqrt{c^2 \times 10^{-7}} G^{1/2}c^{-2}
\approx (8.6\times 10^{-18}) q_{SI}}$ meters. \\

Looking at the expressions for the cross-sections and noting that in the
natural units energy $\sim$ length$^{-1}$, we see that the
cross-sections are of the form $\alpha/E^2$. This gets more and more
accurate at ultra relativistic energies where masses can be neglected.
Since the cross-section has dimensions of area while energies are
typically given in MeV/GeV/TeV etc, it is useful to have a conversion
between energy units and length units. As already notes, $c = 1
\Rightarrow 1$ second $ = 3\times 10^8$ meters. $\hbar = 1 \Rightarrow
1$ Joule $ = 10^{34} sec^{-1} \simeq 3.3\times 10^{25} meter^{-1}$.
Next, $\boxed{100 MeV = 1.6\times 10^{-19+8}J \simeq 5\times 10^{14}
meter^{-1}}$ . (This is also expressed as $200 MeV.fermi = 1$). Clearly,
a cross-section at 100 MeV (electron mass being 0.5 Mev) is
approximately equal to $\boxed{\sigma_{100Mev} \simeq \alpha^2/E^2
\simeq 2\times 10^{-33} m^2.}$ The conventional unit for these
cross-sections is {\em 1 barn := 100 fermi$^2$ = $10^{-28}m^2$.} Typical
numbers encountered in high energy processes (around 1 GeV) dominated by
strong interactions $\sim 10^{-30}m^2$, electromagnetic interactions
$\sim 10^{-36}m^2$ and weak interactions $\sim 10^{-42}m^2$.

So, scattering experiments will provide us with such numbers. How are
they {\em useful}?

In many applications, say passage of a collimated set of particles
through some medium, we have the estimate of the cross-section of
scattering of individual particles comprising the `beam' and the medium.
Intuitively, a cross-section may be viewed as a target disk of area
$\sigma$ which is bombarded by `marbles' which get reflected from the
disk. They are reflected if they hit the disk or pass by if not. Thus,
$\sigma$ gives the likelihood of a scattering interaction. Suppose we
have multiple targets (medium) and multiple beam particles. The
probability of interaction is clearly proportional to the ratio of the
effective area of the target particles and the area exposed to the beam
\begin{figure}[htbp]
	\begin{center}
	\scalebox{0.5}{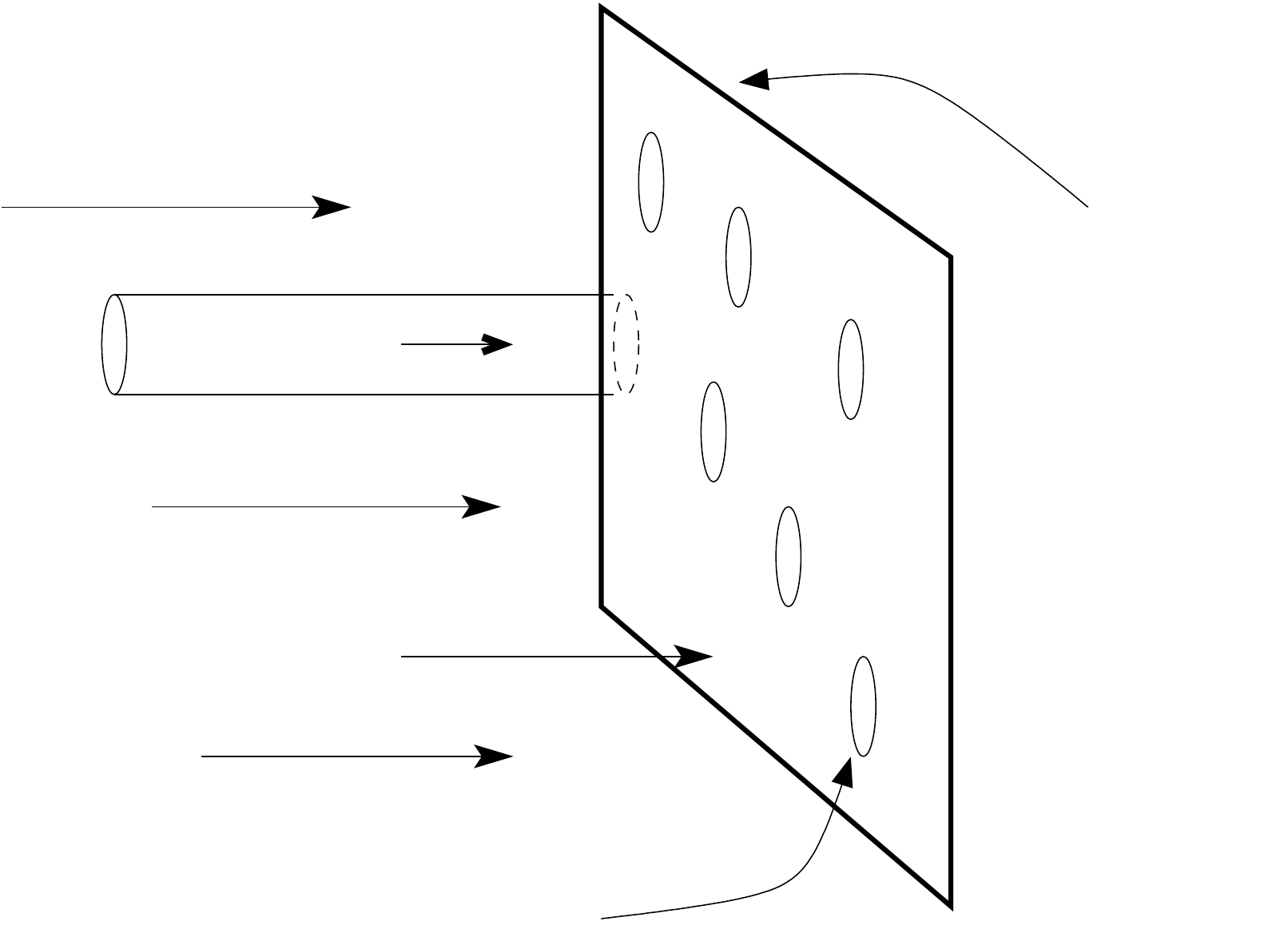}
	\end{center}
\end{figure}
Thus, probability of scattering = (number of target particles exposed o
the beam)$\times \sigma$/Total area presented by the medium). Let $n$ be
the number of target particles exposed per unit area perpendicular to
the beam. Then the number of particles exposed = n$\times$ total area
and the $\boxed{\mbox{probability of scattering\ } = n\cdot\sigma.}$

Let $\rho$ be the number density of the target particles i.e. number per
unit volume. Let $d$ be the thickness of the medium. Then the areal
density $n = \rho\cdot d$. The probability of scattering is $1 n\sigma =
\rho\cdot d\cdot\sigma$ for $d = \hat{d} = (\rho\cdot\sigma)^{-1}$.  Let
$\langle v\rangle$ be the average speed of the beam particles. Then the
average {\em reaction time := } $\tau_r := \hat{d}/\langle v\rangle =
1/(\rho\langle v\rangle\sigma)$. 

Thus, if we have a confined plasma of say electrons and we bombard it
with a beam of positrons and look for production of $\mu^{\pm}$ pairs,
then we should expect to wait for a time $\tau = 1/(\rho\langle
v\rangle\sigma_{e^+e^-\to\mu^+\mu^-})$. This type of use of microscopic
cross-sections is required in discussing bulk processes eg in
nuclear physics or astrophysics	and are useful in estimating
thermalization times.

Sometimes, we aren't interested in tracking individual processes, any
interaction may suffice. In such cases we need to sum over {\em all
possible final states}, $\sigma_{tot} = \sum_f \sigma_{i\to f}$. The
$\sigma_{i\to f} \sim |\mathcal{M}_{i\to f}|^2$. We already have
expressions using phase space integration for fixed number of final
state particles. Now we have to sum over the number of particles and the
labels,
\[
	\sigma_{tot} = \frac{1}{4E_kE_p}v_{lab}\sum_{f,\alpha}\int
	\prod_{i=1}^{N_f} \frac{d^3k'_j}{2\omega_{k'}(2\pi)^3} (2\pi)^4
	\delta^4(\Sigma k' - P_{in})|\mathcal{M}|^2\ .
\]
This can be related to the forward scattering amplitude via the
``Optical Theorem''.

We have $S^{\dagger} S = \mathbb{1} \Rightarrow \sum_f\langle
i|S^{\dagger}|f\rangle\langle|S|i\rangle = 1 = \sum_f|S_{fi}|^2$.
Substituting for $S_{fi}$ we get,
\[
|S_{fi}|^2 = \delta_{fi}\delta_{fi} + (2\pi)^4\delta^4(P_f-P_i)
\delta_{fi} [i (\mathcal{M}_{fi} - \mathcal{M}_{fi}^*)] +
(2\pi)^8[\delta^4(P_f-P_i)]^2|\mathcal{M}_{fi}|^2 \ .
\]
Summing over $f$ and using unitarity gives, 
\begin{equation*}
	(2\pi)^4\sum_f\delta^4(P_f-P_i)\delta_{fi}[i(\mathcal{M}_{fi} -
	\mathcal{M}_{fi}^*)] = -
	(2\pi)^8\sum_f\delta^4(P_f-P_i)\delta^4(P_f-P_i)|\mathcal{M}_{fi}|^2
\end{equation*}
Or, 
\begin{equation} \label{OpticalTheorem}
	\boxed{ 2 Im(\mathcal{M}_{ii}) = (2\pi)^4\sum_f\delta^4(P_f-P-i)
	|\mathcal{M}_{fi}|^2 . }
\end{equation}
Substituting the rhs in the expression for $\sigma_{tot}$ gives,
\[
	\sigma_{tot} = \frac{2Im(\mathcal{M}_{ii})}{4k\sqrt{s}} ~ ~ , ~ ~
		\mbox{where, $k = k_{CM}$ and $s = -(\sum P_i)^2$ \ .}
\]
The $\mathcal{M}_{ii}$ is the forward scattering, $i \to i$, {\em
amplitude} and $\sigma_{tot}$ is the inclusive cross-section for $i \to
\mbox{any} f$ processes. This could also be used to estimate the total
cross-section.

\underline{Remark:} Let us note that we defined the scattering matrix in
the basis of number states or Fock basis. Further, we chose the
idealization of plane waves for both in and out states. This is
convenient but looses the information about {\em spatial} locations of
the scattering particles. This also looses any notion of ``impact
parameter'' - how far the projectile particle is from the target
particle or equivalently the `orbital angular momentum' of the
projectile about the target particle. The information about the
transverse position is of course contained in the parameters of the
outgoing particles. This should be kept in mind while interpreting some
of the divergences of cross-sections.

\newpage
\section{Radiative corrections in QED} \label{RadiativeCorrections}

Consider now corrections due to higher powers of interaction
Hamiltonian, also called {\em radiative corrections}. Having seen the
correspondence between the correction terms and Feynman diagrams, we can
list the corrections directly in terms of the diagrams. A Feynman
diagram is made up of lines (free propagators) and vertices. In QED,
there are two types of propagators and one types of vertex. So consider
corrections to two point functions and the three point function. 
\subsection{The fermion propagator: self-energy}
\begin{eqnarray}
	\vcenter{\hbox{
	\begin{tikzpicture}
		\begin{feynman}
			\vertex (a) ;
			\vertex [right=0.75of a,blob] (b){All} ;
			\vertex [right=1.0of b] (c) ;
			\diagram* {
				(a) -- [fermion] (b) -- [fermion]
				(c),
			};
		\end{feynman}
	\end{tikzpicture}
}}
	& = & 
	\vcenter{\hbox{
	\begin{tikzpicture}
		\begin{feynman}
			\vertex (a) ;
			\vertex [right=of a] (b) ;
			\diagram* {
				a -- [fermion] b,
			};
		\end{feynman}
	\end{tikzpicture}
}}
	~ + ~ 
	\vcenter{\hbox{
	\begin{tikzpicture}
		\begin{feynman}
			\vertex (a) ;
			\vertex [right =0.75of a] (b) ;
			\vertex [right =0.75of b] (c) ;
			\vertex [right =0.75of c] (d) ;
			\diagram* {
				(a) -- [fermion] (b),
				(b) -- [fermion] (c),
				(c) -- [fermion] (d),
				(b) -- [photon,half right,looseness=1.5]
					(c),
					(b) -- [photon,half
					left,looseness=1.5,opacity=0.1]
					(c),
			};
		\end{feynman}
	\end{tikzpicture}
}}
	~ + ~ 
	\vcenter{\hbox{
	\begin{tikzpicture}
		\begin{feynman}
			\vertex (a) ;
			\vertex [right=0.6of a] (b) ;
			\vertex [right=0.6of b] (c) ;
			\vertex [right=0.6of c] (d) ;
			\vertex [right=0.6of d] (e) ;
			\vertex [right=0.6of e] (f) ;
			\diagram* {
				(a) -- [fermion] (b),
				(b) -- [fermion] (c),
				(c) -- [fermion] (d),
				(d) -- [fermion] (e),
				(e) -- [fermion] (f),
				(b) -- [photon,half right,looseness=1.5]
				(d),
				(c) -- [photon,half left,looseness=1.5]
				(e),
			};
		\end{feynman}
	\end{tikzpicture}
}}
	~ + ~ 
\nonumber \\
& & 
	\vcenter{\hbox{
	\begin{tikzpicture}
		\begin{feynman}
			\vertex (a) ;
			\vertex [right=0.75of a] (b) ;
			\vertex [right=0.75of b] (c) ;
			\vertex [right=0.75of c] (d) ;
			\vertex [right=0.75of d] (e) ;
			\vertex [right=0.75of e] (f) ;
			\diagram* {
				(a) -- [fermion] (b),
				(b) -- [fermion] (c),
				(c) -- [fermion] (d),
				(d) -- [fermion] (e),
				(e) -- [fermion] (f),
				(b) -- [photon,half right,looseness=1.5]
				(c),
				(d) -- [photon,half left,looseness=1.5]
				(e),
			};
		\end{feynman}
	\end{tikzpicture}
}}
+ \cdots \\
	& = & 
	\begin{tikzpicture}
		\begin{feynman}
			\vertex (a) ;
			\vertex [right=of a] (b) ;
			\diagram* {
				(a) -- [fermion] (b), 
			};
		\end{feynman}
	\end{tikzpicture}
	~ + ~ 
	\vcenter{\hbox{
	\begin{tikzpicture}
		\begin{feynman}
			\vertex (a) ;
			\vertex [right=0.75of a,blob] (b){1PI} ;
			\vertex [right=1.0of b] (c) ;
			\diagram* {
				(a) -- [fermion] (b) -- [fermion] (c), 
			};
		\end{feynman}
	\end{tikzpicture}
}}
	~ + ~ 
	\vcenter{\hbox{
	\begin{tikzpicture}
		\begin{feynman}
			\vertex (a) ;
			\vertex [right=0.75of a,blob] (b){1PI} ;
			\vertex [right=1.25of b,blob] (c){1PI} ;
			\vertex [right=1.0of c] (d){ } ;
			\diagram* {
				(a) -- [fermion] (b) -- [fermion] (c) --
				[fermion] (d), 
			};
		\end{feynman}
	\end{tikzpicture}
}}
	~ + ~ \cdots
\end{eqnarray}

Here is a graphical expansion of the exact two point function of the
fermions. In the second line, the diagrams have been grouped into a
series of terms involving {\em 1-particle irreducible (1PI)} blobs,
connected by free propagators. The 1PI blob represents the sum of all
diagrams with two external legs and which cannot be disconnected by
removing a single fermion line. The 2nd and the 3rd diagrams in the
first line are 1IP (their extreme propagators being the external lines)
while the last diagram is 1-particle reducible. The second line follows
from first one by inspection. 

Let $iS'_F(p)$ denote the lhs - the full 2-point function or full
propagator - and $iS_F(p)$ denote the free propagator. Let the 1PI blob
be denoted by $i\Sigma(p)$. Then the series is:
\begin{eqnarray*}
	-iS'_F(p) & = & -iS_F(p) + -iS_F\cdot i\Sigma(p)\cdot -iS_F +
	-iS_F\cdot i\Sigma\cdot-iS_F\cdot i\Sigma\cdot -iS_F + \dots \\
	& = & -iS_F + -iS_F\cdot i\Sigma(-iS_F - -iS_F\cdot i\Sigma\cdot
	-iS_F \dots ) \\
	\therefore -iS'_F & = & -iS_F + -iS_F\cdot i\Sigma\cdot -iS'_F
	\\
	\therefore S_F & = & (1 - S_F\Sigma)S'_F ~ \leftrightarrow ~
	\mathbb{1} = S^{-1}_F( - \Sigma)S'_F ~ ~ ~ ~ \mbox{and} ~ ~
	S_F^{-1} = \dsl{p} + m , \\
	\therefore & & \boxed{S_F'(p) = \frac{\mathbb{1}}{\dsl{p} + m -
	\Sigma(p)} ~ , ~ ~ ~ \mbox{$\Sigma(p)$ is called the fermion
self energy.}}
\end{eqnarray*}
The $S'_F(p), \Sigma(p)$ are matrices with the Dirac spinor indices
though we may not always be explicit about it. What can we say about its
singularities?

We can appeal to Lorentz covariance and decompose $\Sigma(p) =
a(p^2)\dsl{p} + b(p^2)\mathbb{1} + c(p^2)\gamma_5$. With parity
conserving interactions such as qed, we can take $c(p^2) = 0$.
Furthermore, {\em to the leading order}, $a(p^2) = 0 = b(p^2)$. Thus the
denominator of the exact propagator is,
\begin{eqnarray*}
[S'_F(p)]^{-1} & = & (1-a(p^2))\dsl{p} + (m-b(p^2))  =  (1-a(p^2))\left[
\dsl{p} + \frac{m-b(p^2)} {1 - a(p^2)} \right] ~ ~ ~ \Rightarrow \\
S'_F(p) & = & \frac{1}{1 - a(p^2)}\left[\frac{-\dsl{p} +
\frac{m-b(p^2)}{1 - a(p^2)}}{p^2 +
\left(\frac{m-b(p^2)}{1-a(p^2)}\right)^2 } \right]  
\end{eqnarray*}
Clearly, the denominator vanishes for $p^2 = - m_{ph}^2, m_{ph} := 
\Case{m-b(-m_{ph}^2)}{1 - a(-m{ph}^2)}$. Since the numerator has the
form $-\dsl{p} + m_{ph}$, we can regard the pole to be defined by
$\dsl{p} = -m_{ph}$. Since $a, b$ are functions of $p^2 =
-\dsl{p}\dsl{p}$, we may regard the self energy as a function of
$\dsl{p}$ and define the pole by the condition: $\boxed{\dsl{p} + m -
\Sigma(\dsl{p})|_{\dsl{p} = -m_{ph}} = 0. } $ Clearly, $m_{ph} \neq m$.

The exact propagator can be expanded about its pole as,
\begin{eqnarray*}
\dsl{p} + m - \Sigma(\dsl{p}) & = & 0 + (\dsl{p} + m_{ph})\left[1 -
\frac{d\Sigma(\dsl{p})}{d\dsl{p}}|_{\dsl{p} = -m_{ph}}\right] + o(
(\dsl{p}+m_{ph})^2 ) \\
\therefore S'_F(p) & \simeq & \frac{1}{\dsl{p} + m_{ph}}\left[\frac{1}{1 -
\frac{d\Sigma(\dsl{p})}{d\dsl{p}}|_{\dsl{p}=-m_{ph}}}\right]  ~ := ~
\frac{Z_2}{\dsl{p}+m_{ph}} 
\end{eqnarray*}

The shift of the position of the pole in the exact propagator from that
of the free propagator, $\boxed{\delta m := m_{ph} - m = -
\Sigma(\dsl{p} = -m_{ph})}$ is important in an $S-$matrix element which
has the spinors on the external lines. These spinors satisfy the
equation of motion with {\em physical mass} (= $m_{ph}$) and {\em not}
the mass parameter in the free propagator inferred from the Lagrangian.
In particular this means that in the LSZ formula for $S-$matrix element,
the amputation of external fermion legs is to be done with
$(-i\dsl{\partial} + m_{ph})$. This in turn means that in the Green's
functions with external fermion legs, the external lines must include
the self-energy corrections i.e. should have $S'_F$ propagator instead
of the free propagator $S_F$ (and of course no more self energy
corrections on the external lines). We will return to this point later
again while discussing renormalized perturbation series.

We also see that the residue at the pole in the exact propagator is not
1 and we identify it with the $Z_2$ which we know from the
Kallen-Lehmann representation to be {\em less than 1}.
\subsection{The photon propagator: photon self-energy}
\begin{eqnarray}
	\vcenter{\hbox{
	\begin{tikzpicture}
		\begin{feynman}
			\vertex (a) ;
			\vertex [right=0.75of a,blob] (b){All} ;
			\vertex [right=1.0of b] (c) ;
			\diagram* {
				(a) -- [photon] (b) -- [photon]
				(c),
			};
		\end{feynman}
	\end{tikzpicture}
}}
	& = & 
	\vcenter{\hbox{
	\begin{tikzpicture}
		\begin{feynman}
			\vertex (a) ;
			\vertex [right=of a] (b) ;
			\diagram* {
				(a) -- [photon] (b),
			};
		\end{feynman}
	\end{tikzpicture}
}}
	~ + ~ 
	\vcenter{\hbox{
	\begin{tikzpicture}
		\begin{feynman}
			\vertex (a) ;
			\vertex [right =0.75of a] (b) ;
			\vertex [right =0.75of b] (c) ;
			\vertex [right =0.75of c] (d) ;
			\diagram* {
				(a) -- [photon] (b),
				(b) -- [fermion,half right,looseness=1.5] (c),
				(c) -- [photon] (d),
				(c) -- [fermion,half right,looseness=1.5]
					(b),
			};
		\end{feynman}
	\end{tikzpicture}
}}
	~ + ~ 
	\vcenter{\hbox{
	\begin{tikzpicture}
		\begin{feynman}
			\vertex (v) ;
			\vertex [left=0.4of v] (b);
			\vertex [left=0.4of b] (a);
			\vertex [right=0.4of v] (c);
			\vertex [right=0.4of c] (d);
			\vertex [above=0.4of v] (e) ;
			\vertex [below=0.4of v] (f) ;
			\diagram* {
				(a) -- [photon] (b),
				(c) -- [photon] (d),
				(e) -- [photon] (f),
				(b) -- [fermion,half left,looseness=1.5]
				(c),
				(c) -- [fermion,half left,looseness=1.5]
				(b),
			};
		\end{feynman}
	\end{tikzpicture}
}}
	~ + ~ 
	\vcenter{\hbox{
	\begin{tikzpicture}
		\begin{feynman}
			\vertex (a) ;
			\vertex [right=0.5of a] (b) ;
			\vertex [right=0.5of b] (c) ;
			\vertex [right=0.5of c] (d) ;
			\vertex [right=0.5of d] (e) ;
			\vertex [right=0.5of e] (f) ;
			\diagram* {
				(a) -- [photon] (b),
				(c) -- [photon] (d),
				(e) -- [photon] (f),
				(b) -- [fermion,half right,looseness=1.5]
				(c),
				(c) -- [fermion,half right,looseness=1.5]
				(b),
				(d) -- [fermion,half right,looseness=1.5]
				(e),
				(e) -- [fermion,half right,looseness=1.5]
				(d),
			};
		\end{feynman}
	\end{tikzpicture}
}}
	\\
	& = & 
	\vcenter{\hbox{
	\begin{tikzpicture}
		\begin{feynman}
			\vertex (a) ;
			\vertex [right=of a] (b) ;
			\diagram* {
				(a) -- [photon] (b), 
			};
		\end{feynman}
	\end{tikzpicture}
}}
	~ + ~ 
	\vcenter{\hbox{
	\begin{tikzpicture}
		\begin{feynman}
			\vertex (a) ;
			\vertex [right=0.75of a,blob] (b){1PI} ;
			\vertex [right=1.0of b] (c) ;
			\diagram* {
				(a) -- [photon] (b) -- [photon] (c), 
			};
		\end{feynman}
	\end{tikzpicture}
}}
	~ + ~ 
	\vcenter{\hbox{
	\begin{tikzpicture}
		\begin{feynman}
			\vertex (a) ;
			\vertex [right=0.75of a,blob] (b){1PI} ;
			\vertex [right=1.25of b,blob] (c){1PI} ;
			\vertex [right=1.0of c] (d){ } ;
			\diagram* {
				(a) -- [photon] (b) -- [photon] (c) --
				[photon] (d), 
			};
		\end{feynman}
	\end{tikzpicture}
}}
	~ + ~ \cdots
\end{eqnarray}

Here is a graphical representation of the exact photon propagator. As
before, we group together the 1-PI diagrams connected by free
propagators. Let the exact propagator be denoted by $-iD'_{\mu\nu}(q)$
and the free propagator by $-iD_{\mu\nu}(q)$. Let the 1PI blob be
denoted by $i\Pi_{\mu\nu}(q)$. In the Lorentz gauge, the free propagator
is: $D_{\mu\nu}(q) = \eta_{\mu\nu}/(q^2 - i\epsilon)$. While the series
can be formally summed as before, it is more convenient to separate the
tensor indices in the $\Pi_{\mu\nu}(q)$ tensor.

Lorentz covariance implies, $\Pi_{\mu\nu}(q) = \eta_{\mu\nu}A(q^2) +
q_{\mu}q_{\nu}B(q^2)$. The $\Pi_{\mu\nu}$ tensor may be thought of as
$\gamma \to \gamma$ process, although $q^2 \ne 0$. We now appeal to a
{\em Ward identity:} $q^{\mu}\Pi_{\mu\nu}(q) = 0$. This is done
separately below. Presently, it implies $A(q^2) = -q^2B(q^2)$ and we
define: $\boxed{\Pi_{\mu\nu}(q) := [\eta_{\mu\nu}q^2 -
q_{\mu}q_{\nu}]\Pi(q^2)}$ . With this notation, the exact 2-point
function takes the form,
\begin{eqnarray*}
	-iD'_{\mu\nu}(q) & = & -i \frac{\eta_{\mu\nu}}{q^2-i\epsilon} +
	-i
	\frac{\eta_{\mu\alpha}}{q^2-i\epsilon}\cdot[i(q^2\eta^{\alpha\beta}
	- q^{\alpha}q^{\beta})\Pi(q)]\cdot -i
	\frac{\eta_{\beta\nu}}{q^2-i\epsilon} + \\
	& & \mbox{\hspace{3.0cm}}
	-iD_{\mu\alpha}\cdot[i(\dots)^{\alpha\beta}]\cdot
	-iD_{\beta\rho}\cdot[i(\dots)^{\rho\sigma}]\cdot -iD_{\sigma\nu}
	+ \dots
\end{eqnarray*}
Define, 
\[
i(q^2\eta^{\alpha\beta} - q^{\alpha}q^{\beta})(-i\eta_{\beta\nu}q^{-2})
= \left(\delta^{\alpha}_{~\nu} - \frac{q^{\alpha} q_{\nu}}{q^2}\right)
=: \Delta^{\alpha}_{~\nu} ~ \Rightarrow ~ \Delta^{\alpha}_{~\nu}
\Delta^{\nu}_{~\beta} = \Delta^{\alpha}_{~\beta} \ .
\] 
Then the exact propagator series takes the form,
\begin{eqnarray*}
-iD'_{\mu\nu}(q) & = & -i \frac{\eta_{\mu\nu}}{q^2-i\epsilon} + -i
\frac{\eta_{\mu\alpha}}{q^2-i\epsilon} \Delta^{\alpha}_{~\nu}\Pi(q^2) +
-i \frac{\eta_{\mu\alpha}}{q^2-i\epsilon} \Delta^{\alpha}_{~\beta}
\Delta^{\beta}_{~\nu}\Pi(q^2) + \dots \\
& = & -i \frac{\eta_{\mu\nu}}{q^2-i\epsilon} + -i
\frac{\eta_{\mu\alpha}}{q^2-i\epsilon} \Delta^{\alpha}_{~\nu}(\Pi +
\Pi^2 + \Pi^3 + \dots) \\
& = & \frac{-i}{q^2-i\epsilon} \left[\left\{\eta_{\mu\nu} -
\frac{q_{\mu}q_{\nu}}{q^2} + \frac{q_{\mu}q_{\nu}}{q^2}\right\} +
\left(\eta_{\mu\nu} - \frac{q_{\mu}q_{\nu}}{q^2}\right)(\Pi + \Pi^2 +
\Pi^3 \dots) \right] \\
\therefore D'_{\mu\nu}(q^2) & = & \frac{1}{q^2(1 -
\Pi(q^2))}\left(\eta_{\mu\nu} - \frac{q_{\mu}q_{\nu}}{q^2}\right) +
\frac{1}{q^2}\frac{q_{\mu}q_{\nu}}{q^2} .   
\end{eqnarray*}
In any $S-$matrix calculation, the $D'$ propagator will land on a
fermion line and thanks to the Ward identity, the terms proportional to
$q^{\mu}, q^{\nu}$ will vanish. Hence, for {\em $S-$matrix element
computations}, we can take $\boxed{D'_{\mu\nu}(q) = \Case{\eta_{\mu\nu}}
{(q^2 - i\epsilon)(1 - \Pi(q^2))}}$. The $\Pi(q^2)$ is called the photon
self energy. To the leading order, the self energy vanishes and hence in
perturbation theory, it can never equal 1 and cause another pole at some
other $q^2$. As long as it is regular at $q^2 = 0$, the exact propagator
{\em continues to have a simple pole at} $q^2 = 0$, just like the free
propagator. Hence, the photon wavefunction, $\epsilon_{\mu}(\vec{k})$
continues to be transverse and longitudinal polarizations will decouple. 

Unlike the fermion self energy which shifts the mass, the photon self
energy does {\em not} shift the photon mass from zero thanks to the Ward
identity.  However, like the fermion self energy, the residue at the
pole does shift away from 1, $\boxed{ (Z_3)^{-1} := 1 - \Pi(q^2 = 0).}$ 
\subsubsection{The Ward identity claim: $q^{\mu}\Pi_{\mu\nu}(q) = 0$:}
\label{WardIdentity}
\[
\vcenter{\hbox{
\begin{tikzpicture}
	\begin{feynman}
		\vertex (v);
		\vertex [right=of v] (1);
		\vertex [right=2.0of v] (11){\(\gamma_{1}\)};
		\vertex [below right=of v] (2);
		\vertex [below right=2.0of v] (22){\(\gamma_{2}\)};
		\vertex [below left=of v] (i);
		\vertex [below left=2.0of v] (ii){\(\gamma_{i}\)};
		\vertex [left=of v] (j);
		\vertex [left=2.0of v] (jj){\(\gamma_{j}\)};
		\vertex [above left=of v] (i+1);
		\vertex [above left=2.0of v] (ii+1){\(\gamma_{i+1}\)};
		\vertex [above right=of v] (n);
		\vertex [above right=2.0of v] (nn){\(\gamma_{n}\)};

		\diagram*{
			(1) --[photon] (11),
			(2) --[photon] (22),
			(i) --[photon] (ii),
			(j) --[photon] (jj),
			(i+1) --[photon] (ii+1),
			(n) --[photon] (nn),
		};

		\draw[fermion, domain=0:-45] (v) plot
		({1.5*cos(\x)},{1.5*sin(\x)}) 
		node[label={[label distance=8.0] 15:$p_1$}];

		\draw[fermion, domain=-45:-135] (v) plot
		({1.5*cos(\x)},{1.5*sin(\x)}) 
		node[label={[label distance=5.5] 160:$p_i$}];

		\draw[fermion, domain=-135:-180] (v) plot
		({1.5*cos(\x)},{1.5*sin(\x)}); 

		\draw[fermion, domain=-180:-225] (v) plot
		({1.5*cos(\x)},{1.5*sin(\x)}) 
		node[label={[label distance=4.0] 190:$p_i+q$}];

		\draw[fermion, domain=-225:-315] (v) plot
		({1.5*cos(\x)},{1.5*sin(\x)}) ;

		\draw[fermion, domain=-315:-360] (v) plot
		({1.5*cos(\x)},{1.5*sin(\x)}) 
		node[label={[label distance=2.8] 80:$p_n+q$}];

	\end{feynman}
\end{tikzpicture}
}}
\]
\underline{Proof:} Recall that $\Pi_{\mu\nu}$ is defined without
external propagators i.e. it is an amputated 2-point function. The
momentum $q$ however, need not be on-shell and hence it does {\em not}
correspond to an $S-$matrix element (even if we disregard the
polarization factors). Consider diagrams contributing to the photon self
energy with some fixed number of vertices (or order in the coupling).
The set of these diagrams will be generated by vertex injecting momentum
$q^{\mu}$ at various points on the available fermion lines. The fermion
lines will necessarily be loops as there are no external fermions
contributing to the photon self energy. Consider the subset of diagrams
wherein the $q^{\mu}$ vertex is on some particular loop which
contributes a factor of the form: (the fermion arrow points from $i=1$
to $i=n$ and so do the fermion momenta)
\[
Tr\left[S_F(p_n+q)\gamma^{\mu_n} \dots S_F(p_{i+1}+q) \gamma^{\mu_{i+1}}
\left\{S_F(p_i+q) \dsl{q} S_F(p_i)\right\} \gamma^{\mu_{i}} \dots
S_F(p_1)\gamma^{\mu_1}\right]
\]
The $q^{\mu}$ vertex is inserted between the $i^{th}$ and $(i+1)^{th}$
vertices and adds the propagator $S_F(p_i+q)$. All momenta from
$p_{i+1}$ to $p_n$ are shifted by $q$. We have also contracted with
$q^{\mu}$. The expression between the braces simplifies as
\footnote{This identity holds as long as both the fermion propagators
have the same mass, $m_{ph}$ or $m$. This is relevant while dealing with
external fermion lines.}, 
\[
	\frac{1}{\dsl{p}_{i} +\dsl{q }+ m}\dsl{q}\frac{1}{\dsl{p}_{i} +
	m}  = \frac{1}{\dsl{p}_{i} + m}- \frac{1}{\dsl{p}_i + \dsl{q} +
	m} = S_F(p_i) - S_F(p_i+q) \ .
\]
The trace thus becomes,
\begin{eqnarray*}
Tr\left[S_F(p_n+q)\gamma^{\mu_n} \dots S_F(p_{i+1}+q)
\gamma^{\mu_{i+1}}\left\{S_F(p_i)\right\}\gamma^{\mu_i} \dots
S_F(p_1)\gamma^{\mu_1}\right] \mbox{\hspace{2.0cm}} & & \\
- Tr\left[S_F(p_n+q) \gamma^{\mu_n} \dots S_F(p_{i+1}+q)
\gamma^{\mu_{i+1}} \left\{ S_F(p_i +q)\right\} \gamma^{\mu_{i}} \dots
S_F(p_1) \gamma^{\mu_1}\right]
\end{eqnarray*}
Note that the momenta after $i$ (and $i-1$) have $q$ added to them.
Summing over the diagrams where the $q^{\mu}$ vertex is on this fermion
loop i.e from $i = 1, \dots, n$ , there is a pairwise cancellation. The
first term at $i$ cancels the second term at $i+1$ and so on. We are
left with the second term at $i=1$ and the first term at $i=n$, i.e.
\[
\sum_{i=1}^{n} Tr\left[\dots\right] = Tr\left[S_F(p_n)\gamma^{\mu_n}
\dots S_F(p_1) \gamma^{\mu_{1}}\right] -
Tr\left[S_F(p_{n}+q)\gamma^{\mu_{n}} \dots S_F(p_1+q)
\gamma^{\mu_{1}}\right] 
\]
Because we have a fermion loop, there is an integration over say $p_1$.
{\em Provided the integral is finite, we can shift the integration
variable and absorb away the $q$}. The integrals of the two traces, then
cancel out and the claim is proved. 

\underline{Remark:} We have not yet introduced loop diagrams. The above
fermion loop, before the $q$ insertion, has $n$ vertices and $n$ momenta
entering at those vertices. The conservation of momenta, taking out the
overall conservation, thus leaves one momentum undetermined and this is
to be integrated.

\underline{Remark:} As an extension of the result, consider an open
fermion line with incoming momentum $p_0$, outgoing momentum $p_n$ and
insertions on it. 
\[
	\vcenter{\hbox{
	\begin{tikzpicture}
		\begin{feynman}
			\vertex (0);
			\vertex [left=of 0] (1);
			\vertex [below=of 1] (11){\(\gamma_1\)};
			\vertex [left=of 1] (2);
			\vertex [below=of 2] (22){\(\gamma_2\)};
			\vertex [left=3.0of 1] (i);
			\vertex [below=of i] (ii){\(\gamma_i\)};
			\vertex [left=1.0of i] (j);
			\vertex [above=of j] (jj){\(\gamma_j\)};
			\vertex [left=1.0of j] (i+1);
			\vertex [below=of i+1] (ii+1){\(\gamma_{i+1}\)};
			\vertex [left=3.0of i+1] (nm1);
			\vertex [below=of nm1] (nnm1){\(\gamma_{n-1}\)};
			\vertex [left=2.0of nm1] (n);
			\vertex [below=of n] (nn){\(\gamma_n\)};
			\vertex [left=1.5of n] (n+1);
			\diagram*{
				(0) --[fermion,edge label=\(p_0\)] (1),
				(1) --[fermion,edge label=\(p_1\)] (2),
				(2) --[fermion] (i),
				(i) --[fermion,edge label=\(p_i\)] (j),
				(j) --[fermion,edge label=\(p_i+q\)] (i+1),
				(i+1) --[fermion] (nm1),
				(nm1) --[fermion,edge label=\(p_{n-1}+q\)] (n),
				(n) --[fermion,edge label=\(p_n+q\)] (n+1),
				(1) --[photon] (11),
				(2) --[photon] (22),
				(i) --[photon] (ii),
				(j) --[photon] (jj),
				(i+1) --[photon] (ii+1),
				(nm1) --[photon] (nnm1),
				(n) --[photon] (nn),
			};
		\end{feynman}
	\end{tikzpicture}
	}}
\]
For the insertion between the $i, i+1$ vertices, we will have a string
of factors as,
\[
S_F(p_{n}+q)\gamma^{\mu_n}S_F(p_n+q) \dots \gamma^{\mu_{i+1}}
\left\{S_F(p_{i}+q) \dsl{q} S_F(p_{i})\right\} \gamma^{\mu_{i}} \dots
S_F(p_1)\gamma^{\mu_1}S_F(p_{0})
\]
Replacing $S_F(p_i+q)\dsl{q}S_F(p_i) = S_F(p_i) - S_F(p_i+q)$ and
summing over the insertions from $i=0,\dots,n$, we get
\begin{eqnarray*}
\sum_{i=0}^{n} \left[\dots\right] & = &
S_F(p_n)\gamma^{\mu_n}S_F(p_{n-1}) \dots S_F(p_{1}) \gamma^{\mu_{1}}
S(p_0) \\
& & \mbox{\hspace{2.0cm}} - S_F(p_n+q)\gamma^{\mu_n}S_F(p_{n-1}+q) \dots
S_F(p_{1}+q) \gamma^{\mu_{1}} S(p_0+q)
\end{eqnarray*}
To get an $S-$matrix contribution, we have to multiply by
$\bar{u}(p_n+q)[S_F(p_n+q)]^{-1}]$ on the left and by
$[S_F(p_0)]^{-1}u(p_0)$ on the right and take the limits $p_0^2 \to
-m^2, (p_n+q)^2 \to -m^2$. Now we see that each of the two terms has
only one of the poles which can cancel the inverse propagator.  In the
on-shell limit then, each of the terms vanishes, thereby proving that
{\em for $S-$matrix elements with external fermion lines, sum over
insertion of photon lines followed by contraction with the photon
momentum gives zero.}

\underline{Note:} There is one subtlety here. For external lines, we
should be using the exact propagator as per the LSZ rules. The
$S\dsl{q}S$ identity requires both propagators to have the same mass,
the physical mass. This would require all internal fermion lines to use the
exact propagator (and then do not consider the fermion self energy
correction). For fermion loop, this issue does not arise. Notice however
that $S'_F(p)^{-1} = S_F(p)^{-1} -\Sigma(p)$ and $\sigma(p)$ is always
of order $\alpha$ and higher. Thus, in using free inverse propagator
instead of exact inverse propagator, we are dropping a higher order
contribution which we can do in a perturbation theory.

The conclusion is then that $q_{\mu}\mathcal{M}^{\mu}(q,\dots) = 0$
which is a statement of Ward identity.
\subsection{The Vertex function: Form factors} \label{VertexFn}
The expectation is that the electron-photon coupling will undergo a
change due to higher order corrections. Unlike the 2-point functions
above which were considered for arbitrary momentum, we consider the
3-point function with the fermion momenta ``on-shell'' (and this means
the $q^2$ is necessarily space-like (prove this)) and hence an internal
line. When viewed as part of an $S-$matrix element, the external fermion
legs are amputated with {\em physical mass} inverse propagator and
replaced by the $\bar{u}(p'), u(p)$ spinors respectively. Here is a
graphical view of the possible corrections.

\begin{eqnarray}
	\vcenter{\hbox{
	\begin{tikzpicture}
		\begin{feynman}
			\vertex (a) ;
			\vertex [left=0.15of a] (a'){\(\mu\)} ;
			\vertex [right=0.75of a,blob] (b){All} ;
			\vertex [above right=1.0of b] (c) ;
			\vertex [below right=1.0of b] (d) ;
			\diagram* {
				(a') -- (a),
				(a) -- [photon,momentum=\(q\)] (b), 
				(b) -- [fermion,edge label=\(p'\)] (c),
				(d) -- [fermion,edge label=\(p\)] (b),
			};
		\end{feynman}
	\end{tikzpicture}
}}
	& = & 
	\vcenter{\hbox{
	\begin{tikzpicture}
		\begin{feynman}
			\vertex (a) ;
			\vertex [right=0.75of a] (b) ;
			\vertex [above right=1.0of b] (c) ;
			\vertex [below right=1.0of b] (d) ;
			\diagram* {
				(a) -- [photon] (b), 
				(b) -- [fermion] (c),
				(d) -- [fermion] (b),
			};
		\end{feynman}
	\end{tikzpicture}
}}
	~ + ~ 
	\vcenter{\hbox{
	\begin{tikzpicture}
		\begin{feynman}
			\vertex (a) ;
			\vertex [right =0.75of a] (b) ;
			\vertex [above right =0.75of b] (c1) ;
			\vertex [above right =1.5of b] (c2) ;
			\vertex [below right =1.5of b] (d2) ;
			\vertex [below right =0.75of b] (d1) ;
			\diagram* {
				(a) -- [photon] (b),
				(b) -- [fermion] (c1) -- [fermion] (c2),
				(d2) -- [fermion] (d1) -- [fermion] (b),
				(d1) -- [photon,half
				right,looseness=1.0] (c1),
			};
		\end{feynman}
	\end{tikzpicture}
}}
	~ + ~ 
	\vcenter{\hbox{
	\begin{tikzpicture}
		\begin{feynman}
			\vertex (a) ;
			\vertex [right =0.5of a] (b) ;
			\vertex [above right =0.5of b] (c1) ;
			\vertex [above right =1.0of b] (c2) ;
			\vertex [above right =1.5of b] (c3) ;
			\vertex [below right =1.5of b] (d3) ;
			\vertex [below right =1.0of b] (d2) ;
			\vertex [below right =0.5of b] (d1) ;
			\diagram* {
				(a) -- [photon] (b),
				(b) -- [fermion] (c1) -- [fermion] (c2)
				-- [fermion] (c3),
				(d3) -- [fermion] (d2) -- [fermion] (d1)
				-- [fermion] (b),
				(d1) -- [photon] (c2),
				(d2) -- [photon,half
				right,looseness=1.0] (c1),
			};
		\end{feynman}
	\end{tikzpicture}
}}
	~ + ~ 
	\vcenter{\hbox{
	\begin{tikzpicture}
		\begin{feynman}
			\vertex (a) ;
			\vertex [right =0.5of a] (b) ;
			\vertex [above right =0.5of b] (c1) ;
			\vertex [above right =1.0of b] (c2) ;
			\vertex [above right =1.5of b] (c3) ;
			\vertex [above right =2.0of b] (c4) ;
			\vertex [above right =2.5of b] (c5) ;
			\vertex [below right =1.5of b] (d3) ;
			\vertex [below right =1.0of b] (d2) ;
			\vertex [below right =0.5of b] (d1) ;
			\diagram* {
				(a) -- [photon] (b),
				(b) -- [fermion] (c1) -- [fermion] (c2)
				-- [fermion] (c3) -- [fermion] (c4) --
				[fermion] (c5),
				(d3) -- [fermion] (d2) -- [fermion] (d1)
				-- [fermion] (b),
				(d1) -- [photon] (c2),
				(d2) -- [photon] (c4),
				(c1) -- [photon,half
				left,looseness=1.5] (c3),
			};
		\end{feynman}
	\end{tikzpicture}
}}
	\\
	ie\Gamma^{\mu}(p,p') & = & 
	ie \gamma^{\mu} ~ + ~ 
	\vcenter{\hbox{
	\begin{tikzpicture}
		\begin{feynman}
			\vertex (a) ;
			\vertex [left=0.15of a] (a'){\(\mu\)} ;
			\vertex [right=0.75of a,blob] (b){1PI} ;
			\vertex [above right=1.0of b] (c) ;
			\vertex [below right=1.0of b] (d) ;
			\diagram* {
				(a') -- (a),
				(a) -- [photon,momentum=\(q\)] (b), 
				(b) -- [fermion,edge label=\(p'\)] (c),
				(d) -- [fermion,edge label=\(p\)] (b),
			};
		\end{feynman}
	\end{tikzpicture}
}}
	~ + ~ 
	\vcenter{\hbox{
	\begin{tikzpicture}
		\begin{feynman}
			\vertex (a) ;
			\vertex [right=0.5of a, blob] (b){1PI} ;
			\vertex [right=2.0of a] (c) ;
			\vertex [above right=0.75of c, blob] (c1){1PI} ;
			\vertex [above right=2.0of c] (c2) ;
			\vertex [below right=0.75of c, blob] (d1){1PI} ;
			\vertex [below right=2.0of c] (d2) ;
			\diagram* {
				(a) -- [photon] (b) -- [photon] (c), 
				(c) -- [fermion] (c1) -- [fermion] (c2),
				(d2) -- [fermion] (d1) -- [fermion] (c),
			};
		\end{feynman}
	\end{tikzpicture}
}}
	~ + ~ \cdots
\end{eqnarray}

Notice that some corrections on the fermion lines and the photon line
are just self-energy corrections. The remaining corrections have photon
lines necessarily connecting the two fermion lines and these diagrams
are already 1PI. The last diagram in the second equation is 1PR.

We can simplify the form of the exact vertex function,
$\Gamma^{\mu}(p,p')$ by appealing to Lorentz covariance.
\[
	\Gamma^{\mu}(p, p') = \gamma^{\mu}A(q^2) + (p^{\mu} + p^{'\mu})
	B(q^2)+ (p^{'\mu} - p^{\mu})C(q^2) \ .
\]
With two independent momenta, we have 3 Lorentz scalars, $p^2, (p')^2,
p\cdot p'$. The first two equal $-m^2_{ph}$ while the third is traded
for $q^2$ for convenience.

\underline{Claim:} $~ \bar{u}(p')q_{\mu}\Gamma^{\mu}u(p) = 0$.

\underline{Proof:} ~ This is the Ward identity argument given above.

The above form for $\Gamma^{\mu}$ then implies, 
\begin{eqnarray*}
	0 & = & \bar{u}(p')\dsl{q}u(p)A(q^2) + (-m^2_{ph} +
	m^2_{ph})\bar{u}(p')B(q^2)iu(p) + q^2\bar{u}(p')C(q^2)u(p) \\ &
	= & 0 + 0 + q^2\bar{u}(p')C(q^2)u(p) ~ \Rightarrow ~
	\bar{u}(p')C(q^2)u(p)  = 0 ~ .
\end{eqnarray*}

The $A, B$ terms can be rewritten using ``Gordon identity'',
\[
\boxed{ \bar{u}(p')\gamma^{\mu}u(p) = \frac{1}{2m_{ph}}\bar{u}(p')\left[
(p+p')^{\mu} - 2i\sigma^{\mu\nu}q_{\nu} \right] u(p) }
\]
This can be established as follows ($m_{ph} \to m$ for convenience):

We have, $u(p) = - \Case{1}{m}\dsl{p}u(p)\ , \ \bar{u}(p') = -
\Case{1}{m} \bar{u}(p')\dsl{p}'$. Therefore 

\begin{eqnarray*}
	\bar{u}(p')\gamma^{\mu}u(p) & = & -\frac{1}{2m}(
	\gamma^{\mu}\dsl{p} + \dsl{p}'\gamma^{\mu} ) ~ ~ \mbox{But,} ~ ~
	\gamma^{\mu}\dsl{p}  =  p_{\nu}\gamma^{\mu}\gamma^{\nu} =
	p_{\nu}\left(-\eta^{\mu\nu} + \frac{1}{2} [\gamma^{\mu},
	\gamma^{\nu}]\right) \\
	\therefore \gamma^{\mu}\dsl{p} & = & -p^{\mu} -2i
	\Sigma^{\mu\nu}p_{\nu} ~ ~ , ~ ~ \dsl{p}'\gamma^{\mu} = -
	(p')^{\mu} - 2i \Sigma^{\nu\mu}p'_{\nu} ~ ~ , ~ ~
	\Sigma^{\mu\nu} := \frac{i}{4}[\gamma^{\mu}, \gamma^{\nu}]\ ; \\
	\therefore \bar{u}(p')\gamma^{\mu}u(p) & = &
	-\frac{1}{2m}\bar{u}(p')\left[-p^{\mu} - (p')^{\mu} -
	4i\Sigma^{\mu\nu}(p_{\nu} - p'_{\nu})\right]u(p)  \\
	& = & \frac{1}{2m}\bar{u}(p')[(p+p')^{\mu}
	-2i\sigma^{\mu\nu}q_{\nu}] u(p) ~ ~ ~ , ~ ~ ~ \sigma^{\mu\nu} :=
	2\Sigma^{\mu\nu} ~ , ~ q := p'-p \ .
\end{eqnarray*}

Eliminating the $(p+p')^{\mu}$ from the $B(q^2)$ term, and identifying
$A + 2m_{ph}B =: F_1(q^2)$ and $4\Sigma^{\mu\nu}B(q^2) =:
\Case{\sigma^{\mu\nu}}{2m_{ph}}F^2(q^2)$, we take the general form of
$\Gamma^{\mu}(p,p')$ as,
\[
\boxed{ \Gamma^{\mu}(p,p') = \gamma^{\mu}F_1(q^2) + i
\frac{\sigma^{\mu\nu}q_{\nu}}{2m_{ph}}F_2(q^2) ~ , ~ q := p' - p\ . } 
\]
The $F_i(q^2)$ are called {\em form factors.} To the lowest order,
$F_1(q^2) = 1, F_2(q^2) = 0$ . These get corrected at higher orders.
\subsection{Electric Charge and Anomalous Magnetic Moment}
\label{FormFactorIdentification}
The form factor decomposition helps us identify the {\em electric
charge} and the {\em anomalous magnetic moment} of electron (and other
charged fermions). This is seen as follows.

Consider the scattering of an electron off a heavy charged particle. We
imagine the scattering to be effected by exchanging a single photon
between the heavy charge and the electron. Schematically the amplitude
can be written as, 
\[
	i\mathcal{M} =
	ie^2\ [\bar{u}(p')\Gamma^{\mu}(p',p)u(p)]\ \frac{1}{q^2}\ [\bar{u}(k')\gamma_{\mu}u(k)] 
\]
The first factor is from electron, second is the single photon
propagator and the third factor is from the heavy charge. The $\Gamma$
vertex has only 1PI diagrams and excludes the photon self energy on the
propagator. This represents the modification of the electron response
due to the 1PI corrections, the photon self energy would indicate
modification in the electromagnetic field felt by the electron and hence
is also referred to as {\em vacuum polarization}. This is ignored in
this calculation. 

In the limit of infinitely heavy fermion, the second
and third factors are replaced by an external classical potential
$A_{\mu}(q)$ and the amplitude is defined as,
\[
	i\mathcal{M}\ (2\pi)\ \delta( (p'-p)^0 ) = -ie
	\bar{u}(p')\Gamma^{\mu}u(p)\tilde{A}^{cl}_{\mu}(p'-p)
\]
We consider two cases - Coulomb potential and magnetic field, and
consider non-relativistic limit of the scattering.

\underline{Coulomb potential:} $A^{cl}_{\mu}(x) = (\phi(x), \vec{0})$.
Coulomb field is static and we also assume it to be slowly varying
spatially. This means that we take $\tilde{A}^{cl}_{\mu}(q) =
2\pi\delta(q^0)(\tilde{\phi}(q), \vec{0})$ with the
$\tilde{\phi}(q)$ having support near $q \simeq 0$. We can thus take the
limit $q \to 0$ of the first factor. This in turn means that only the
$F_1$ term contributes. In the non-relativistic limit,
$\bar{u}(p')\gamma^{\mu}u(p) \to \bar{u}(p')\gamma^0u(p) =
2mu^{\dagger}(p)u(p)$ and the amplitude takes the form, $i\mathcal{M}
\to -ieF_1(0)\tilde{\phi}(q)(2mu^{\dagger}(p)u(p))$. Comparing with the
Born approximation for scattering off a potential leads to
the identification, $V(x) = eF_1(0)\phi(\vec{x})$. Thus,
$\boxed{\mbox{electric charge} = eF(0). }$ Since $F_1(0) = 1$ at the
leading order, the radiative corrections to $F_1(q^2)$ should vanish as
$q^2 \to 0$.

\underline{Magnetic field:} $A_{\mu}^{cl} = (0, \vec{A}^{cl}_i(x))$.
Again this is taken to be time independent and spatially slowly varying.
Thus we consider the limit $q \to 0$. Earlier, we dropped the $F_2$
term, because $F_1$ term was non-zero. Now however $\bar{u}(p')
\gamma^iu(p) \simeq o(p/m)$ in the non-relativistic limit which is
comparable to the $F_2$ term. 

Recall from the NR limit of the Dirac spinors, $u^T := (u_1^T, u_2^T),
\ (\dsl{p} + m)u(p) = 0 \Rightarrow u_2(p) \simeq - \Case{\vec{p} \cdot
\vec{\sigma}}{2m}u_1$ and we have the normalization $u^{\dagger}u(p) =
2m$. We also had, 
\[
	\bar{u}(p')\gamma^i u(p)  \simeq u_1^{\dagger}(p')\sigma^iu_2(p)
	+ u_2^{\dagger}(p')\sigma^iu_1(p) \simeq -\frac{1}{2m}
	u_1^{\dagger}(p')\{ \sigma^i\vec{p}\cdot\vec{\sigma} +
	\vec{p}'\cdot\vec{\sigma}\sigma^i \}u_1(p) .
\]
Using $\sigma^i\sigma^j = \delta^{ij} + i\epsilon^{ij}_{~k}\sigma^k$, we
get,
\[
\bar{u}(p')\vec{\gamma}\cdot\vec{A}^{cl}u(p) = -\frac{1}{2m}
u_1^{\dagger}(p')\left( \vec{p}\cdot \vec{A}^{cl} + \vec{p}'\cdot\vec{A}^{cl}
-iA_i^{cl}\epsilon^{ij}_{~k}q_j\sigma^k \right)u_1(p)
\]

Noting that $q^0 = 0$, the $F_2$ term has $\sigma^{ij}q_j$. Going over
to the two component spinors, we get 
\begin{eqnarray*}
	\frac{i}{2m} \bar{u}(p')\sigma^{ij}u(p)q_j & \approx &
	\frac{i}{2m}
		\epsilon^{ij}_{~k}[u_1^{\dagger}(p')\sigma^ku_1(p)]
\end{eqnarray*}
The contribution of the $u_2$ spinors is negligible in the NR limit.

Combining the terms we get in the NR limit,
\[
\bar{u}(p')\Gamma^iu(p)\tilde{A}_i^{cl}(q) \approx
u_1^{\dagger}(p')\left[\frac{(\vec{p}+\vec{p}')\cdot\vec{A}^{cl}}{2m}F_1(0) -
\frac{i}{2m} \epsilon^{ij}_{~k}q_jA^{cl}_i\sigma^k(F_1(0) +
F_2(0))\right]u_1(p)
\]

In the non-relativistic Hamiltonian, $(p - eA)^2/(2m)$ we have the
$e(p\cdot  + A\cdot p)$ terms which are recovered. The remaining terms
however are new. Focusing only on those, we write the amplitude as,
\[
	i\mathcal{M} = -ieu_1^{\dagger}(p')\left[-\frac{1}{2m} (F_1(0) +
	F_2(0))\sigma^kB_k\right]u_1(p) ~ ~ , ~ ~ B_k :=
	-i\epsilon_k^{~ij}q_i\tilde{A}_j^{cl}(q) \ .
\]
This amplitude is interpreted as the Born approximation to a potential
scattering with potential, $V(x) = -\langle\vec{\mu}\rangle\cdot
\vec{B}(\vec{x})$, with the effective magnetic moment,
\[
\boxed{\langle\vec{\mu}\rangle = \frac{e}{m_{ph}}[F_1(0) +
F_2(0)]u_1^{\dagger}(p) \frac{\vec{\sigma}}{2} u_1(p) := g
\frac{e}{2m_{ph}}\vec{S} ~ , ~ g := 2[F_1(0) + F_2(0)] = 2[1 +
F_2(0)].}
\]

$g$ is called the {\em Lande's g-factor}, its value being $2$ is
prediction of Dirac equation while $F_2(0) \neq 0$ is a prediction of
QED. It is called the {\em anomalous magnetic moment}.

Without actually calculating the higher order corrections, we noted what
they mean for 2-point functions (self-energies). We used Lorentz
covariance to get a form factor decomposition and interpreted them by
appealing to the NR limit. One of the immediate prediction was the
anomalous magnetic moment. 

We also note that to the leading order (see diagrams), we have $Z^{-1}_2
= 1, Z^{-1}_3 = 1, F_1(q^2) = 1, F_2(q^2) = 0$. This is the reason that
at the tree level calculations that we did, did not have any factors of
$Z$'s. This will change when we do explicit evaluations of the leading
corrections in the next section.

\newpage
\section{Radiative corrections at 1-loop: Divergences}\label{1LoopDivergences}

We now calculate the leading radiative corrections to the self energies
and the QED vertex function i.e. the 1-loop contributions to
$\Sigma(\dsl{p}), \Pi(q^2), F_1(q^2), F_2(q^2)$. Here are the Feynman
diagrams and the corresponding invariant amplitudes.
\begin{center}
\begin{tabular}{|c|c|}
\hline
& Fermion self energy \\
\hline
\parbox[c]{5.0cm}{ 
	\begin{tikzpicture}
                \begin{feynman}
                        \vertex (v1); 
                        \vertex [right=1.5of v1] (v2); 
                        \vertex [right=1.5of v2] (v3); 
			\vertex [right=1.5of v3] (v4); 

                        \diagram* {
				(v1) -- [fermion,edge label=\(p\)] (v2),
				(v2) -- [fermion,edge label=\(k\)] (v3),
                                (v3) -- [fermion,edge label=\(p\)] (v4),
				(v2) -- [photon,half left,
				looseness=1.5, momentum=\(p-k\)] (v3),
                        };
                \end{feynman}
        \end{tikzpicture}
} &
\parbox[c]{11.0cm}{ \[ i\Sigma(p) = (ie)^2\int \frac{d^4k}{(2\pi)^4}
\gamma^{\mu} \frac{-i (-\dsl{k}+m)}{k^2+m^2-i\epsilon} \gamma^{\nu}
\frac{-i\eta_{\mu\nu}}{(p-k)^2+\mu^2-i\epsilon} \] }\\
\hline
& Photon self energy (vacuum polarization) \\
\hline
\parbox[c]{5.0cm}{ 
	\begin{tikzpicture}
                \begin{feynman}
                        \vertex (v1); 
                        \vertex [right=1.5of v1] (v2); 
                        \vertex [right=1.5of v2] (v3); 
			\vertex [right=1.5of v3] (v4); 

                        \diagram* {
				(v1) -- [photon,momentum=\(q\)] (v2),
				(v2) -- [fermion,half
				left,looseness=1.5, edge label=\(k+q\)] (v3),
                                (v3) -- [photon,momentum=\(q\)] (v4),
				(v3) -- [fermion,half left,
				looseness=1.5, edge label=\(k\)] (v2),
                        };
                \end{feynman}
        \end{tikzpicture}
} &
\parbox[c]{11.0cm}{ \[ 
		i\Pi^{\mu\nu}(q) = (ie)^2 (-1)\int \frac{d^4k}{(2\pi)^4}
		Tr\left[\gamma^{\mu}\frac{-i}{\dsl{k}+m}\gamma^{\nu}\frac{-i}{\dsl{k}+\dsl{q}+m}\right]
	\] } \\
\hline
& QED Vertex function, $\delta\Gamma^{\mu} := \Gamma^{\mu} -
\gamma^{\mu}$  \\
\hline
\parbox[c]{5.0cm}{ 
	\begin{tikzpicture}
                \begin{feynman}
                        \vertex (v1) {\(\mu\)};
                        \vertex [right=1.5of v1] (v2); 
                        \vertex [above right=1.5of v2] (v3); 
			\vertex [below right=1.5of v2] (v4); 
			\vertex [above right=1.0of v3] (v5); 
			\vertex [below right=1.0of v4] (v6); 

                        \diagram* {
				(v1) -- [photon,momentum'=\(q\)] (v2),
				(v2) -- [fermion, edge label=\(p'-k\)] (v3),
                                (v4) -- [photon,momentum'=\(k\)] (v3),
				(v4) -- [fermion, edge label=\(p-k\)] (v2),
				(v3) -- [fermion,edge label=\(p'\)] (v5),
				(v6) -- [fermion,edge label=\(p\)] (v4),
                        };
                \end{feynman}
        \end{tikzpicture}
} &
\parbox[c]{11.0cm}{ \begin{eqnarray*}
		\delta\Gamma^{\mu}(p,p') & = & (ie)^2\int
		\frac{d^4k}{(2\pi)^4}
		\frac{-i\eta_{\alpha\beta}}{(k)^2+\mu^2-i\epsilon}\times\\
		& & \left[\gamma^{\alpha}
		\frac{-i(-(\dsl{p}'-\dsl{k})+m)}
	{(p'-k)^2+m^2-i\epsilon} \gamma^{\mu}
\frac{-i(-(\dsl{p}-\dsl{k})+m)} {(p-k)^2+m^2-i\epsilon}
\gamma^{\beta}\right]    
	\end{eqnarray*} } \\
\hline
\end{tabular}
\end{center}
In the photon self energy expression, the $(-1)$ is due to the fermion
loop. The photon propagator is added a small mass $\mu^2$ in
anticipation. In the vertex correction, the expression is understood to
be sandwiched between $\bar{u}(p')$ and $u(p)$.

There are some new points to be noted.

$\bullet$ There is a momentum integration over $k$, unrelated to the
external momenta. This $k$ can be space-like, time-like or light-like.
In any frame, we may consider the region where $k^{\mu} \to \infty$,
then: (i) the electron self-energy $\sim \int d^4k k^{\alpha}/k^4 \sim
\int dk$ which is naively {\em linearly divergent}; (ii) the photon
self-energy $\sim \int d^4k/k^2 \sim \int kdk$ which is naively
{\em quadratically divergent} and (iii) the vertex correction $\sim d^4k
k^2/k^6 \sim \int dk/k $ which is naively {\em logarithmically
divergent}.

$\bullet$ There are also other regions of integration space which give
divergent contributions, but these will be visible after the integrand
is put in a convenient form.

$\bullet$ The $i\epsilon$ in the denominator is quite significant now.
It implies that the 4-dimensional integration must be defined with the
$k^0$ integration being done first and the spatial integrations done
subsequently.

$\bullet$ The naively divergent integrals need to be {\em regularised}
i.e. a prescription for the integration must be supplied which will
manifest the divergence in an explicit form. The divergent contribution
so obtained, must be {\em subtracted} to obtain finite answers. This is
the process of {\em renormalization}.

$\bullet$ The integrand has a numerator which has tensor/spinor indices
while the denominator is a product of scalar factors. One combines the
denominator factors into a single factor form using the so called {\em
Feynman parameters}. This is followed by the momentum integration with a
regularization and a finite part is identified. The integration over the
Feynman parameters is done last to get the answer. We explain these
steps now.

\underline{The Feynman/Schwinger trick:}
\[
	\mbox{Claim:\hspace{0.5cm}} \frac{1}{A_1\dots A_n} = (n-1)!
	\int_0^{\infty}dx_1\dots dx_n \frac{\delta(\Sigma_i x_i -
	1)}{(\Sigma_i x_iA_i)^n}  . 
\]
Proof: Observe that $A^{-1} = \int_0^{\infty}d\alpha e^{-\alpha A}~, ~
Re(A) > 0$. Therefore,
\begin{eqnarray*}
	\frac{1}{A_1\dots A_n} & = & \int_0^{\infty}d\alpha_1\dots
	d\alpha_n e^{-\Sigma_i \alpha_iA_i}\cdot 1 ~ ~ , ~ ~ 1 =
	\int_0^{\infty}dt \delta(t - \Sigma_i\alpha_i) ~ ~ \mbox{and} ~
	~ \alpha_i \to t x_i ,\\
	& = & \int_0^{\infty}dt \int_0^{\infty} dx_1\dots dx_n \ t^n
	e^{-t\Sigma_i x_iA_i} \ \frac{1}{t} \ \delta(\Sigma_i x_i - 1)
	\\
	& = & \int_0^{1} dx_1\dots dx_n \delta(\Sigma_i x_i -
	1)\int_0^{\infty}dt \ t^{n-1}e^{-t(\Sigma_i x_iA_i)} \\
	& = &  (n-1)!  \int_0^{\infty}dx_1\dots dx_n
	\frac{\delta(\Sigma_i x_i - 1)}{(\Sigma_i x_iA_i)^n} ~ ~
	\because ~ \int_0^{\infty}dt \ t^{n-1}e^{-t \Lambda} =
	\Lambda^{-n}\Gamma(n) \ .
\end{eqnarray*}
\underline{Note:} Since each of the denominator factors are of the form
$A_i = (p_i-k)^2 + m_i^2 - i\epsilon$, we see that 
\begin{eqnarray*}
\Sigma_i x_i A_i & = & k^2 -2k\cdot(\Sigma_i x_ip_i) + \Sigma_i
x_i(p_i^2 + m_i^2) - i\epsilon \\
& = & (k - \Sigma_i x_ip_i)^2 + M^2 - i\epsilon ~ ~ , ~ M^2(x_i,
p_i,m_i) := \Sigma_i x_i(p_i^2+m_i^2) - (\Sigma_i x_ip_i)^2 \ .
\end{eqnarray*}
The obvious step is to shift the integration variable, $k \to k +
\Sigma_i x_i p_i$ which simplifies the denominator - it has the same
form of a single propagator with the same $-i\epsilon$.

%
Let us begin with the fermion self energy.

\subsection{Isolation of Divergence: Fermion Self Energy}
\[
i\Sigma(p) = (ie)^2\int \frac{d^4k}{(2\pi)^4} \gamma^{\mu} \frac{-i
(-\dsl{k}+m)}{k^2+m^2-i\epsilon} \gamma^{\nu} \frac{-i\eta_{\mu\nu}}
{(p-k)^2+\mu^2-i\epsilon} 
\] 
The Feynman trick gives,
\[
\frac{1}{(k^2+m^2-i\epsilon)( (p-k)^2+\mu^2-i\epsilon)}  = \int_0^1 dx
\frac{1}{[(k-xp)^2+M^2-i\epsilon]^2} ~ , 
\]
with, $ M^2 := x(1-x)p^2 + (1-x^2)m^2 +x\mu^2 $. Shifting $k \to k +
xp$, and using $\gamma^{\mu}\gamma_{\mu} = -4 \mathbb{1},
\gamma^{\mu}\dsl{k}\gamma_{\mu} = 2\dsl{k}$, gives
\[
	i\Sigma = e^2\int_0^1\int \frac{d^4k}{(2\pi)^4} \frac{-4m
	-2(\dsl{k}+x\dsl{)p}}{[k^2 + M^2(x,p) -i\epsilon]^2} 
\]
Recalling that the $k^0$ integration is to be done first and that the
poles are at $k^0 = \pm \sqrt{\vec{k}^2+M^2} \mp i\epsilon$, we rotate
the contour anti-clockwise without crossing any singularity. This is
equivalent to putting $k^0 = i \underline{k}^0$ and $k^2 \to
(\underline{k}^0)^2+\vec{k}^2 = $Euclidean vector norm. This sends $\int
d^4k \to i\int d^4 \underline{k}$. For Euclidean integrals $\int d^4
\underline{k} \Case{\underline{k}^{\alpha}\gamma_{\alpha}}{(
\underline{k}^2 + M^2)^2} = 0 $, since the integrand is odd under $k \to
-k$. We are left with,
\[
	i\Sigma = -ie^2\int_0^1dx \int \frac{d^4
	\underline{k}}{(2\pi)^4} \frac{4m +
	2x\dsl{p}}{(\underline{k}^2+M^2)^2} ~ ~ , ~ ~ M^2 := x(1-x)p^2 +
	(1-x^2)m^2 +x\mu^2
\]
The $k-$integral is logarithmically divergent and needs to be regulated.
There are several ways of doing this. One is the {\em Pauli-Villars}
regularization which subtracts from the photon propagator, another
identical piece with arbitrarily large mass squared, $\Lambda^2$, i.e.
\[
	\frac{1}{(p-k)^2+\mu^2} \to \frac{1}{(p-k)^2+\mu^2} -
	\frac{1}{(p-k)^2+\Lambda^2} . 
\]
For large $\underline{k}$, both terms go as $\underline{k}^{-2}$ and
cancel each other leaving a finite answer. Thus $i\Sigma \to
i\Sigma_{reg} := i\Sigma(\Lambda=\infty) - i\Sigma(\Lambda)$. The same
combining of denominators will produce identical terms with $M^2(\mu^2)
\to M^2(\Lambda^2)$. The momentum integration becomes,
\[
	\int \frac{d^4\underline{k}}{(2\pi)^4}\left[\frac{1}
	{(\underline{k}^2+M_{\mu}^2)^2} - \frac{1}
{(\underline{k}^2+M_{\Lambda}^2)^2}\right] = \int
\frac{d\Omega_3}{(2\pi)^4}\int_0^{\infty}dk |k|^3\left[\dots\right] 
\]
The angular integration gives: $\Omega_{n-1} = \Case{2\pi^{n/2}}
{\Gamma(n/2)} = 2\pi^2$ for $n = 4$ and the integral becomes,
\begin{eqnarray*}
\int \frac{d^4 \underline{k}}{(2\pi)^2}\left[ \dots - \dots\right] & = &
\frac{1}{8\pi^2} \int_0^{\infty} \frac{dy}{2}
\left[\frac{y}{(y+M_{\mu}^2)^2} -\frac{y}{(y+M_{\Lambda}^2)^2}\right] ~
~ , ~ ~ k^2 =: y~ \mbox{substituted.} \\
& = & \frac{1}{16\pi^2} \int_0^{\infty} dy \left[\frac{1}{y+M_{\mu}^2} -
\frac{1}{y+M^2_{\Lambda}} - \frac{M^2_{\mu}}{(y+M_{\mu}^2)^2} +
\frac{M^2_{\Lambda}}{(y+M^2_{\Lambda})^2}\right] \\
& = & \frac{1}{16\pi^2}\left[ ln\left( \frac{y+M_{\mu}^2}
{y+M^2_{\Lambda}} \right)_0^{\infty} + \left( \frac{M_{\mu}^2}
{y+M^2_{\mu}} - \frac{M_{\Lambda}^2}{y+M^2_{\Lambda}}
\right)_{0}^{\infty} \right] \\
\therefore \int \frac{d^4 \underline{k}}{(2\pi)^2}\left[ \dots -
\dots\right] & = &
\frac{1}{16\pi^2}ln\left(\frac{M_{\Lambda}^2}{M_{\mu}^2}\right) \ . 
\end{eqnarray*}
In the limit $\Lambda \to \infty, M^2_{\Lambda} \to x\Lambda^2$ and we
get,
\begin{equation}\label{FermionSelfEnergyCorrection}
\boxed{ i\Sigma(p) = -i \frac{e^2}{8\pi^2}\int_0^1dx \ (2m+x\dsl{p})
ln\left[ \frac{x\Lambda^2}{x(1-x)p^2 + (1-x)m^2 + x\mu^2}\right] }  
\end{equation}

Recall that the mass shift as well as the $Z_2$ are obtained from the
self energy as, $\delta m := m_{ph} - m = -\Sigma(\dsl{p} = -m_{ph}),
Z_2^{-1} = 1 - \Case{\Sigma(\dsl{p})}{d\dsl{p}}|_{\dsl{p} = -m_{ph}}$.
Since the self energy already has an explicit factor of $e^2$, in the
integrand, we can use $\dsl{p} = - m_{ph} \approx -m$. The derivative of
$\Sigma$ is obtained as,
\begin{eqnarray*}
	\left. \frac{d\Sigma}{d\dsl{p}}\right|_{\dsl{p} = -m_{ph}} & = &
		-\frac{\alpha}{2\pi}\int_0^1dx\ \left[ x
		ln\left(\frac{x\Lambda^2}{-x(1-x)m^2 + (1-x)m^2 +
	x\mu^2}\right) + \right. \\
	& & \left. (2m - x\cdot m) \frac{-1}{-x(1-x)m^2 + (1-x)m^2 +
	x\mu^2} (-2x(1-x)(-m)) \right]
\end{eqnarray*}
Simplification leads to,
\begin{equation}
\boxed{
\left. \frac{d\Sigma}{d\dsl{p}}\right|_{\dsl{p} = -m_{ph}} =
	\frac{\alpha}{2\pi}\int_0^1dx\ \left[ -x
		ln\left(\frac{x\Lambda^2}{(1-x)^2m^2+x\mu^2}\right) + 
2(2 - x)\frac{x(1-x)m^2}{(1-x)^2m^2+x\mu^2} \right] 
}
\end{equation}
\begin{equation}
	\boxed{
	\delta m = -\Sigma(\dsl{p} = -m_{ph}) = \frac{\alpha}{2\pi}
	\int_0^1dx\
	(2-x)ln\left(\frac{x\Lambda^2}{(1-x)^2m^2+x\mu^2}\right) 
}
\end{equation}
The integrands are well behaved at both end points so the integrals are
finite and the leading contribution is given by the $\Lambda^2$
dependent part alone and for dimensional reasons we divide by the
fermion mass. Thus, the {\em divergent contributions} as $\Lambda \to
\infty$ take the form,
\begin{eqnarray*}
	\delta m & \rightsquigarrow & \frac{\alpha}{2\pi}\int_0^1dx
	(2-x)ln(\Lambda^2/m^2) = \frac{3\alpha}{4\pi}ln(\Lambda^2/m^2)
	\\
	Z_2^{-1} & \rightsquigarrow & 1 -
	\frac{\alpha}{2\pi}\int_0^1(-xln(\Lambda^2/m^2)) = 1 +
	\frac{\alpha}{4\pi}ln(\Lambda^2/m^2) \Rightarrow Z_2 \approx 1 -
	\frac{\alpha}{4\pi}ln(\Lambda^2/m^2)  
\end{eqnarray*}
The logarithmic divergence is thus manifested in terms of
$ln(\Lambda^2/m^2)$. There are of course finite pieces, but until we
take care of the divergent parts the finite parts are irrelevant. 
\subsection{Isolation of Divergence: Photon Self-Energy}
\[
	i\Pi^{\mu\nu}(q) = -e^2 \int \frac{d^4k}{(2\pi)^4} \frac{Tr[
	\gamma^{\mu}(-\dsl{k}+m)\gamma^{\nu}(-\dsl{k}
	-\dsl{q}+m)]}{(k^2+m^2-i\epsilon)( (k+q)^2+m^2-i\epsilon)} 
\]

Using the trace formulae for the Dirac matrices (\ref{DiracTraces}), we
get
\begin{eqnarray*}
\mbox{Numerator:} & = &
Tr[\gamma^{\mu}\dsl{k}\gamma^{\nu}(\dsl{k}+\dsl{q}) +
m^2\gamma^{\mu}\gamma^{\nu}] \\
& = & 4[ k^{\mu}(k+q)^{\nu} - \eta^{\mu\nu}k\cdot(k+q) +
(k+q)^{\mu}k^{\nu} -m^2\eta^{\mu\nu}] \\
& = & 4[ 2k^{\mu}k^{\nu} - \eta^{\mu\nu}(k^2+m^2) + (k^{\mu}k^{\nu} +
k^{\nu}q^{\mu} - \eta^{\mu\nu}k\cdot q)] \\
\mbox{Denominator:} & = & \frac{1}{k^2+m^2-i\epsilon}\frac{1}{(k+q)^2 +
m^2 -i \epsilon)}  \\
& = & \int_0^1 dx\ \frac{1}{[(k+xq)^2 + x(1-x)q^2 +m^2]^2}  ~ ~ ~ ~
\mbox{shift~} k \to k -xq, 
\end{eqnarray*}
The shift generates additional terms linear in $k$ in the numerator.
Under integration, these terms drop out and we are left with,
\begin{eqnarray*}
	i\Pi^{\mu\nu}(q) & = & (-4e^2)\int \frac{d^4k}{(2\pi)^4}
	\frac{\left\{2k^{\mu}k^{\nu} + 2x^2q^{\mu}q^{\nu} -
	\eta^{\mu\nu}(k^2+x^2q^2+m^2) -x(2q^{\mu}q^{\nu} -
	\eta^{\mu\nu}q^2) \right\}}{(k^2 + M^2 -i\epsilon)^2}
\end{eqnarray*}
where, $\boxed{M^2 = x(1-x)q^2 + m^2}$. Doing Wick rotation as before,
\begin{eqnarray*}
	i\Pi^{\mu\nu}(q) & = & (-4e^2)i\int \frac{d^4k}{(2\pi)^4}
	\frac{\left\{2\underline{k}^{\mu}\underline{k}^{\nu} - 2x(1-x)
	q^{\mu}q^{\nu} + x(1-x)\eta^{\mu\nu}q^2 -
	\eta^{\mu\nu}(\underline{k}^2+m^2) \right\}}{(\underline{k}^2 +
	M^2 -i\epsilon)^2}
\end{eqnarray*}

The tensorial integral,
\[
	\int \frac{d^4 \underline{k}}{(2\pi)^4}
	\frac{\underline{k}^{\mu}\underline{k}^{\nu}}{(\underline{k}^2 +
	M^2)^2}  = A \delta^{\mu\nu} ~ ~ , ~ ~ \mbox{with}~ ~ A =
	\frac{1}{4}\int \frac{d^4 \underline{k}}{(2\pi)^4}
	\frac{\underline{k}^2}{(\underline{k}^2 + M^2)^2}
\]

There are several problems to be faced now. First, the integral is
quadratically divergent and shift in the momentum variable is
unjustified.  Second, we have now $\delta^{\mu\nu}$ and $\eta^{\mu\nu}$
in the numerator, confusing Lorentz covariance. Third, naively we see a
quadratically divergent piece with coefficient $\eta^{\mu\nu}$ but only
logarithmically divergent one with coefficient $q^{\mu}q^{\nu}$ casting
doubts on possibility of gauge invariance. We could try the
Pauli-Villars regulation, but we will introduce a different one, the
{\em dimensional regularization}. The Pauli-Villars may be seen in
\cite{PeskinSchroder}. The dimensional regularization has the main
virtue of keeping the Ward identity satisfied, even for non-abelian
gauge theories.

The idea is to think of the perturbation theory to be a member of a
class of similar theories  formulated in general $n$ dimensions. One
chooses a value of $n$ where the integrals are well defined ($n < 4$)
and analytically continues $n \to 4$. This isolates the original
divergence in a particular form (poles) providing the needed
regularization. Since the mass/length dimensions also depend on
space-time dimensions we need the coupling constants to be
dimensionfull. To maintain space-time covariance, we need to regard the
external momenta and the $\gamma$ matrices etc also to be
$n-$dimensional. We set $\epsilon := 4 -n$ and consider the limit
$\epsilon \to 0$. Since the integrals are over Euclidean momenta due to
the Wick rotation, {\em the external momenta and the metric tensor need to
continue back to Minkowski signature at the end of the calculations}.

In this regularization scheme, the following rules are adopted.
\begin{eqnarray}
	\int \frac{d^nk}{(2\pi)^n}f(|k|^2) & = & \int
	\frac{d\Omega_{n-1}}{(2\pi)^n}\int dk|k|^{n-1}f(|k|^2) ~ ~ , ~ ~
	\boxed{\int d\Omega_{n-1} = \frac{2\pi^{n/2}}{\Gamma(n/2)}}
	\label{DimensionalAngularIntgral}\\
	\therefore \int \frac{d^nk}{(2\pi)^n}f(|k|^2) & = &
	\left[\frac{2\pi^{n/2}}{\Gamma(n/2)}\right]\frac{1}{(2\pi)^n}\int
	dk k^{n-1}f(k^2) \\
	\int_0^{\infty}dk \frac{k^{n-1}}{(k^2 + M^2)^{\alpha}} & = &
	\frac{1}{2}\int_0^{\infty}dx \frac{x^{\Case{n}{2}
	-1}}{(x+M^2)^{\alpha}}  ~ ~ , ~ ~ \mbox{put}~ y =
	\frac{M^2}{x+M^2} \nonumber \\
		& = & \frac{M^{n-2\alpha}}{2}\int_0^1dy
		y^{\alpha-1-\Case{n}{2}}(1-y)^{\Case{n}{2}-1}  =
		\frac{M^{n-2\alpha}}{2}\beta(\alpha-n/2, n/2) \nonumber
\end{eqnarray}
\begin{equation} \label{DimensionalRadialIntgral}
\boxed{\therefore \int_0^{\infty}dk \frac{k^{n-1}}{(k^2 + M^2)^{\alpha}}
= \frac{M^{n-2\alpha}}{2}\frac{\Gamma(\alpha - n/2) \Gamma(n/2)}
{\Gamma(\alpha)} }
\end{equation}
and
\begin{equation} \label{DimensionalMomIntgral}
\boxed{ \int\frac{d^nk}{(2\pi)^n}\frac{1}{(k^2+M^2)^{\alpha}} ~ = ~
\frac{1}{(4\pi)^{n/2}}\frac{\Gamma(\alpha-n/2)}{\Gamma(\alpha)}
(M^2)^{n/2 - \alpha} }
\end{equation}
The $\Gamma$ function has isolated poles at $\alpha - n/2 = 0, -1,
-2,\dots$ i.e.  at $n = 2(\alpha + m), m = 0,1,\dots .$ The Gamma
function has the expansion $\Gamma(\epsilon/2) = 2/\epsilon -\gamma +
o(\epsilon)$ where, $\gamma = 0.57721 \dots$ is the {\em
Euler-Mascheroni Constant}.

Since finite parts can be affected we need to keep $\epsilon$ in various
place too eg $\{\gamma^{\mu}, \gamma^{\nu}\} =
-2\delta^{\mu\nu}\mathbb{1}, Tr(\mathbb{1}) = n = 4-\epsilon,
\gamma^{\mu}\gamma^{\nu}\gamma_{\mu} = (2-\epsilon)\gamma^{\nu}$. These
terms will give contributions from the $1/\epsilon$ terms. The coupling
constant dimensions work as follows. $[\Psi] = (n-1)/2, [A_{\mu}] =
(n-2)/2 \Rightarrow [e] = n - (n-1) - (n/2 -1) = \epsilon/2$. Therefore
we write $e \to e\mu_0^{\epsilon/2}$ where $\mu_0$ is an arbitrary mass
scale which should disappear from physical quantities. The dimensions
are in mass units. The tensorial integral will now follow from
$k^{\mu}k^{\nu} \to \Case{1}{n}\delta^{\mu\nu}k^2$. We use these in
evaluating the $\Pi^{\mu\nu}(q)$.

\begin{eqnarray*}
\Pi^{\mu\nu}(\underline{q}) & = & (-ne^2\mu_0^{\epsilon})\int_0^1dx
\left[ \left\{ -2x(1-x)\underline{q}^{\mu}\underline{q}^{\nu} +
\delta^{\mu\nu}( x(1-x)\underline{q}^2 -m^2)\right\}\times \right. \\
& & \left. \int \frac{d^n \underline{k}}{(2\pi)^n} \frac{1}
{(\underline{k}^2 + M^2)^2} + \delta^{\mu\nu}(2/n-1) \int \frac{d^n
\underline{k}}{(2\pi)^n}\frac{\underline{k}^2 + M^2 -
M^2}{(\underline{k}^2 + M^2)^2}\right] ~ , \mbox{where,} \\
& & \mbox{\hspace{4.0cm}} \boxed{M^2 = x(1-x)q^2 + m^2}\\
& = & (-ne^2\mu_0^{\epsilon}\int_0^1 dx\left[
		\delta^{\mu\nu}\left(\frac{2}{n}-1\right)\underbrace{\int
		\frac{d^n
	\underline{k}}{(2\pi)^n}\frac{1}{\underline{k}^2 +
M^2}}_{\circled{1}} + \underbrace{\int \frac{d^n
	\underline{k}}{(2\pi)^n}\frac{\underline{k}^2}{(\underline{k}^2
	+ M^2)^2}}_{\circled{2}}\times \right. \\
	& & \left. \mbox{\hspace{1.0cm}} \left\{
	-2x(1-x)\underline{q}^{\mu}\underline{q}^{\nu} +
	\delta^{\mu\nu}\left(-m^2 - x(1-x)\underline{q}^2
	-\left(\frac{2}{n}-1\right)M^2\right)\right\} \right] \\
\circled{1} & = & \frac{1}{(4\pi)^{n/2}} \frac{\Gamma(1-n/2)}{\Gamma(1)}
(M^2)^{n/2-1} ~ = ~ \frac{M^2}{(4\pi)^2} \left(\frac{M^2}{4\pi}\right)
^{-\epsilon/2} \frac{\Gamma(\epsilon/2)}{(\epsilon/2-1)} \\
\circled{2} & = & \frac{1}{(4\pi)^{n/2}} \frac{\Gamma(2-n/2)}{\Gamma(1)}
(M^2)^{n/2-2} ~ = ~ \frac{1}{(4\pi)^2} \left(\frac{M^2}{4\pi}\right)
^{-\epsilon/2} \Gamma(\epsilon/2) \\
\therefore \circled{1} & = &
\left(\frac{M^2}{\epsilon/2-1}\right)\circled{2}  ~ ~ ~ , ~ ~ ~
\frac{\epsilon}{2} -1 = 1-\frac{n}{2} = \frac{n}{2}\left(\frac{2}{n} -
1\right) \\
\Pi^{\mu\nu}(q) & = & (-ne^2\mu_0^{\epsilon})\int_0^1dx \left[
\delta^{\mu\nu} \left( \frac{2}{n}-1 \right)
\frac{M^2}{\frac{n}{2}(\frac{2}{n}-1)} \right. \\
& & \mbox{\hspace{0.5cm}} \left. + \delta^{\mu\nu}\left(-m^2 -
\left(\frac{2}{n}-1\right)M^2 + x(1-x)q^2\right)
-2x(1-x)q^{\mu}q^{\nu}\right]\times\circled{2} \\
But, [\dots] & = & \delta^{\mu\nu}\left\{\cancel{\frac{2}{n}M^2}
\underbrace{-m^2 -\left(\cancel{\frac{2}{n}M^2}-M^21\right) +
x(1-x)q^2}_{2x(1-x)q^2}\right\} -2x(1-x)q^{\mu}q^{\nu} \\
&  = & 2x(1-x)(q^2\delta^{\mu\nu} - q^{\mu}q^{\nu}) ~ ~ \Rightarrow ~ ~
\boxed{\Pi^{\mu\nu}(\underline{q}) = (\underline{q}^2\delta^{\mu\nu} -
\underline{q}^{\mu}\underline{q}^{\nu})\Pi(\underline{q}^2) ~
\mbox{with,}  } \\
\Pi(q^2) & := & -ne^2\mu_0^{\epsilon}\int_0^1
dx\frac{1}{(4\pi)^2}\left(\frac{M^2}{4\pi}\right)^{-\epsilon/2}\Gamma(\epsilon/2)
(2x)(1-x) \\
& = & -\frac{2(4-\epsilon)e^2}{(4\pi)^2}\int_0^1dx\
x(1-x)\left\{\frac{2}{\epsilon}-\gamma\right\}
\left\{ln\left(\frac{4\pi\mu_0^2}{M^2}\right) \frac{\epsilon}{2}+1\right\}
\end{eqnarray*}
Using $a^{\epsilon/2} = 1 + \Case{\epsilon}{2}ln(a) + o(\epsilon^2)$, we
write, 
\begin{equation}
\boxed{ \Pi(\underline{q}^2) = -\frac{2	\alpha}{\pi}\int_0^1dx\
x(1-x)\left(\frac{2}{\epsilon}-\gamma\right)\left(1 +
\frac{\epsilon}{2}ln \left(\frac{4\pi\mu_0^2}{M^2}\right)
\left(1-\frac{\epsilon}{4}\right)\right) }
\end{equation}

Notice that for $\underline{q}^2$, $M^2 = m^2$ and 
\begin{equation}
\boxed{\Pi(0) = -\frac{2\alpha}{\pi}\frac{2}{\epsilon}\int_0^1dx(x-x^2)
+ \mbox{constant}\ = -\frac{2\alpha}{3\pi}\frac{1}{\epsilon},} ~
\Rightarrow ~ \boxed{Z_3 = 1-\frac{2\alpha}{3\pi\epsilon}\ }
\end{equation}
We see the divergence in $\Pi(0), Z_3$ and there is of course no mass
shift. We return to the interpretation of $\Pi(q^2)$ later.

\underline{Remark:} The $\epsilon^{-1}$ pole in dimensional
regularization corresponds to logarithmic divergence in Pauli-Villars.
this is indicated by comparing the regulated integral
$\int[(k^2+M^2)^{-2} - (k^2\Lambda^2)^{-2}]$ in Pauli-Villars, which
goes over to $\Case{1}{4\pi^2}ln(x\Lambda^2/M^2)$. In the dimensional
regularization, our integral $\circled{2}$ goes as
$\Case{1}{4\pi^2}\Case{2}{\epsilon}$. Hence, $\epsilon^{-1}
\leftrightarrow ln(\Lambda/M)$. Thanks to the dimensional regularization
maintaining the Ward identity, the $q^2\delta^{\mu\nu} - q^{\mu}q^{\nu}$
got pulled out and the superficially quadratically divergent integral go
converted to a logarithmically divergent one. 
\subsection{Isolation of Divergence: Vertex function}
We compute $\delta\Gamma^{\mu} = \Gamma^{\mu} - \gamma^{\mu}$, with $p^2
= (p')^2 = -m^2$ and sandwiching by $\bar{u}(p'), u(p)$ is implicit.
\begin{equation*}
\delta\Gamma^{\mu}(p,p')  =  -ie^2\int \frac{d^4k}{(2\pi)^4}
\frac{\eta_{\alpha\beta}}{k^2+\mu^2-i\epsilon} \left[\gamma^{\alpha}
	\frac{(-(\dsl{p}'-\dsl{k})+m)} {(p'-k)^2+m^2-i\epsilon}
	\gamma^{\mu} \frac{(-(\dsl{p}-\dsl{k})+m)}
{(p-k)^2+m^2-i\epsilon} \gamma^{\beta}\right]
\end{equation*}
We have three denominators and the Feynman trick leads to,
\begin{eqnarray*}
	\frac{1}{k^2+\mu^2-i\epsilon} \frac{1}{(p-k)^2+m^2-i\epsilon}
	\frac{1}{(p'-k)^2+m^2-i\epsilon} \hspace{3.0cm} & & \\
	= \int_0^1dx\int_0^1dy\int_0^1dz \frac{2}{\left[(k-xp-yp')^2
	+M^2-i\epsilon\right]^3} & & \mbox{where,} \\
M^2 := x(p^2+m^2)+y((p')^2+m^2) +z\mu^2 - (xp+yp')^2 & & \\
\therefore M^2 = 0+0++z\mu^2 + m^2(x+y)^2 + q^2xy \hspace{2.3cm} & & 
\end{eqnarray*}
Since $q^2 > 0$ (space-like) for this diagram, $\boxed{M^2 = (x+y)^2m^2
+ xyq^2+z\mu^2}$ is manifestly positive. As usual shifting the momentum
$k \to k+xp+xp'$, doing the Wick rotation and dropping the terms linear
in $k$ in the numerator, we get,
\begin{equation*}
	\delta\Gamma^{\mu}(p,p') =
	e^2\int_0^1dx\int_0^1dy\int_0^1dz\delta(1-x-y-z)\int
	\frac{d^4\underline{k}}{(2\pi)^4}
	\frac{2\times\mbox{Nr}}{[\underline{k}^2+M^2]^3}
\end{equation*}
\begin{eqnarray*}
Nr & = & \gamma^{\alpha} \left\{(\dsl{p}'-\dsl{k}-x\dsl{p}-y\dsl{p}')
\gamma^{\mu}
(\dsl{p}'-\dsl{k}-x\dsl{p}-y\dsl{p}')\right\}\gamma_{\alpha} +
m^2\gamma^{\alpha}\gamma^{\mu}\gamma_{\alpha} \\
& & -m\gamma^{\alpha}\left\{\gamma^{\mu} (\dsl{p}-x\dsl{p}-y\dsl{p}')+
(\dsl{p}'-x\dsl{p}-y\dsl{p}') \gamma^{\mu}\right\}\gamma_{\alpha} + \
\mbox{linear in $k$} 
\end{eqnarray*}
Now we need to use the identities,
\begin{center}
\fbox{
\begin{minipage}{\textwidth}
\begin{eqnarray}
	\gamma^{\alpha}\gamma^{\mu}\gamma_{\alpha} & = &
	-2\eta^{\alpha\mu}\gamma_{\alpha} - \gamma^{\mu}(-4) =
	2\gamma^{\mu} \\
	\gamma^{\alpha}\gamma^{\rho}\gamma^{\sigma}\gamma_{\alpha} & = &
	4\eta^{\rho\sigma} \\
	\gamma^{\alpha}\gamma^{\rho}\gamma^{\mu}\gamma^{\sigma}\gamma_{\alpha}
	& = & 2\gamma^{\sigma}\gamma^{\mu}\gamma^{\rho} \\
	\therefore
	\gamma^{\alpha}\dsl{k}\gamma^{\mu}\dsl{k}\gamma^{\alpha} & = & 2
	\dsl{k}\gamma^{\mu}\dsl{k} = -4k^{\mu}\dsl{k} + 2\gamma^{\mu}k^2
	~ ~ \because ~ ~ \dsl{k}\dsl{k} = -k^2 \\
	\gamma^{\alpha}\{\dots\}\gamma^{\mu}\{\dots\}\gamma_{\alpha} & =
	& 2\{(1-x)\dsl{p}-y\dsl{p}'\}\gamma^{\mu}\{(1-y)\dsl{p}' -
	x\dsl{p}\} \\
	\gamma^{\alpha}\{\gamma^{\mu}(\dots) + (\dots)\gamma^{\mu}\}
	\gamma_{\alpha} & = & 4\left\{(1-x)p^{\mu}-y(p')^{\mu} +
	(1-y)(p')^{\mu}-xp^{\mu}\right\} \\ \nonumber
\end{eqnarray}
\end{minipage}
}
\end{center}

Using these, the numerator takes the form,
\begin{eqnarray*}
	\bar{u}(p')(Nr)u(p) & = & \bar{u}(p')\left([-4k^{\mu}\dsl{k} +
	2k^2\gamma^{\mu}]_{\circled{1}} + 2[\left\{(1-x)
\dsl{p}-y\dsl{p}' \right\} \gamma^{\mu} \left\{(1-y)\dsl{p}'-x\dsl{p}
\right\}]_{\circled{2}} \right. \\ & & \hspace{2.0cm}\left.
+2[m^2\gamma^{\mu}]_{\circled{3}} - 4m[(1-2x)p^{\mu} +
(1-2y)(p')^{\mu}]_{\circled{4}}\right)u(p)
\end{eqnarray*}
Observe that $\dsl{p}u(p) = -mu(p), \bar{u}{p'}\dsl{p}' =
-m\bar{u}{p'}$. In $\circled{1}, \int d^4k k^{\mu}\dsl{k} \propto
\gamma^{\mu}$ (after Pauli-Villars). We simplify $\circled{2}$ as,
\begin{eqnarray*}
	\frac{1}{2}\bar{u}(p')\circled{2}u(p) & = &
	\bar{u}(p')[(1-x)(1-y)\dsl{p} \gamma^{\mu}\dsl{p}'
	+xym^2\gamma^{\mu}+x(1-x)m\dsl{p}\gamma^{\mu} +
y(1-y)m\gamma^{\mu}\dsl{p}']u(p)\\
	\dsl{p}\gamma^{\mu} & \to & -2p^{\mu} - (-m)\gamma^{\mu} ~ , ~
	\gamma^{\mu}\dsl{p}' \to -2(p')^{\mu} + m\gamma^{\mu}  ~ ~
	\mbox{between the spinors}.
\end{eqnarray*}
Thus we see that, between the spinors, all the terms in the numerator
can be put in the form of $(p+p')^{\mu}[\dots] + \gamma^{\mu}[\dots]$
and therefore the integral can be arranged as contributing to the form
factors $F_1, F_2$. The intermediate steps, including the Pauli-Villars
regularization, are left as an exercise. The result is (from
$\delta\Gamma^{\mu}$):
\begin{eqnarray}
	F_1(q^2) - 1 & = & \frac{\alpha}{2\pi}\int_0^1dx\ dy\ dz
	\delta(1-x-y-z)\left[ln\left(\frac{z\Lambda^2}{M^2}\right) +
	\right. \nonumber \\
	& & \hspace{2.0cm}\left. \frac{1}{M^2}\left( (1-x)(1-y)q^2 +
	(1-4z+z^2)m^2\right)\right] \\
	F_2(q^2) & = & \frac{\alpha}{2\pi}\int_0^1dx\ dy\
	dz\delta(1-x-y-z)\left[\frac{2m^2z(1-z)}{M^2}\right], ~ ~ ~
	\mbox{with} \\
	& & \hspace{2.0cm} \boxed{M^2 := q^2xy + m^2(1-z)^2 +z\mu^2} \nonumber
\end{eqnarray}

\underline{Note:} 

(i) In the $\delta F_1$, there is a UV divergence ($\Lambda \to \infty$) coming
from the large $k$ region, even for $q^2 = 0$. This violates the
$\delta F_1(0) = 0$ condition and thus changes the electric charge. The
condition can be naturally restored by defining $\boxed{\delta
F_{1,ren}(q^2):= \delta F_1(q^2) - \delta F_1(0). }$ This is a {\em
renormalization prescription}.

(ii) The $\delta F_1$ also has an ``infrared divergence'' ($\mu^2 \to
0$) at $q^2 = 0$ coming from the $M^{-2}(0) = m^{-2}(1-z)^{-2}$ from the
$z\to 1$ region.

(iii) The $F_2(q^2)$ has no divergences. Its value at $q^2=0$ gives the
anomalous magnetic moment and can be evaluated explicitly as,
\begin{eqnarray}\label{AnomalousMagMoment}
F_2(0) &  = &  \frac{\alpha}{2\pi}\int_0^1dx\ dy\
dz\delta(1-x-y-z)\frac{2z}{1-z} ~ = ~
\frac{\alpha}{\pi}\int_0^1dz\int_0^{1-z}\frac{z}{1-z}  ~ = ~
\frac{\alpha}{2\pi} \nonumber \\
& \Rightarrow & \boxed{\frac{g-2}{2} = \frac{\alpha}{2\pi}}  ~ ~
\hspace{2.5cm} \mbox{One of the triumphs of QED!}
\end{eqnarray}

The {\em renormalized} $F_1(q^2)$ is obtained as,
\begin{eqnarray}\label{Renormalized-F1}
F_{1,ren}(q^2) & = & 1 + \frac{\alpha}{2\pi}\int_0^1dx\ dy\
dz\delta(1-x-y-z) \left[ ln\left( \frac{m^2(1-z)^2} {m^2(1-z)^2+q^2xy}
\right) \right.  \nonumber \\
& & \left. \hspace{1.5cm} +\frac{m^2(1-4z+z^2) +
q^2(1-x)(1-y)}{m^2(1-z)^2+q^2xy+\mu^2z} -
\frac{m^2(1-4z+z^2)}{m^2(1-z)^2+\mu^2z}\right] 
\end{eqnarray}

We have to face the divergences now. All three radiative corrections
have UV divergence while the vertex function has IR divergence as well.
For dealing with the IR divergence, we need to look at another process
called ``Bremsstrahlung'' (breaking radiation in German).
\subsection{Bremsstrahlung Cross-section to $o(\alpha)$}
\label{Bremsstrahlung}
We know from the classical electrodynamics that accelerated charge
radiates. A quantum mechanical view of such a radiation is depicted in
the diagrams below.
\[ 
	\parbox[c]{3.0cm}{
		\begin{tikzpicture}[scale=0.5]
		\begin{feynman}
			\vertex (a);
			\vertex [right=of a] (c) ;
			\vertex at ($(a)!0.5!(c)$) (b) ;
			\vertex [right=of c] (d) ;
			\vertex [below=0.75of c] (e) ;
			\vertex [left=of e] (f);
			\vertex [right=of e] (g);
			\vertex [above right=0.75of b] (h) ;
			
			\diagram* {
				(a) -- [fermion] (b) -- [fermion] (c) --
				[fermion] (d) ,
				(f) -- [fermion] (e) -- [fermion] (g) ,
				(c) -- [photon] (e) ,
				(b) -- [photon,momentum=\(k\)] (h) ,
			};
		\end{feynman}
	\end{tikzpicture}
}
	\hspace{0.5cm} + \hspace{0.5cm}
	\parbox[c]{3.0cm}{
	\begin{tikzpicture}
		\begin{feynman}
			\vertex (a);
			\vertex [right=of a] (b) ;
			\vertex [right=of b] (d) ;
			\vertex at ($(b)!0.5!(d)$) (c) ;
			\vertex [below=0.75of b] (e) ;
			\vertex [left=of e] (f);
			\vertex [right=of e] (g);
			\vertex [above right=0.75of c] (h) ;
			
			\diagram* {
				(a) -- [fermion] (b) -- [fermion] (c) --
				[fermion] (d) ,
				(f) -- [fermion] (e) -- [fermion] (g) ,
				(b) -- [photon] (e) ,
				(c) -- [photon,momentum=\(k\)] (h) ,
			};
		\end{feynman}
	\end{tikzpicture}
}
	\hspace{0.5cm} + \hspace{0.5cm} \mbox{emission of any number of
	photons}
\]

The diagrams depict a scattering process which causes the top electron
to `accelerate' and `emit' a (on-shell) photon. This is the QFT view of
`radiation from an accelerated charge'. The bottom charged line
represents some heavy fermion/boson/source, while the connecting wavy
line denotes mediation of the interaction. The first diagram suggests
photon emission {\em before} the kick while the second diagram suggests
photon emission {\em after} the kick. As per the rules of the
relativistic QFT, both contributions are to be added and mod-squared to
get to the cross-section. To the leading order - single photon emission
- the cross-section is order $\alpha$. This cross-section also has a an
IR divergence as $k \to 0$. An on-shell photon with arbitrarily small
momentum is called a `soft photon'.
%
%
To isolate the divergence, focus on the following diagrams:
\[
	\parbox[c]{3.0cm}{
	\begin{tikzpicture}
		\begin{feynman}
			\vertex (a);
			\vertex [right=0.75 of a, blob] (b) {};
			\vertex [above right=of b] (c) ;
			\vertex [below right=of b] (d) ;
			\vertex at ($(b)!0.5!(d)$) (d') ;
			\vertex [above right=0.75 of d'] (e) ;
			
			\diagram* { (a) -- [scalar] (b) ,
				(b) -- [fermion,edge label=\(p'\)]
				(c) , (b) -- [anti fermion, edge
				label'=\(p-k\)] (d') -- [anti
				fermion,edge label'=\(p\)] (d) , (d') --
				[boson,edge label'=\(k\)] (e), };
		\end{feynman} 
	\end{tikzpicture}
}
		\hspace{1.0cm} + \hspace{1.0cm}
		\parbox[c]{3.0cm}{
	\begin{tikzpicture}
		\begin{feynman}
			\vertex (a);
			\vertex [right=0.75 of a, blob] (b) {};
			\vertex [above right=of b] (c) ;
			\vertex at ($(b)!0.5!(c)$) (c') ;
			\vertex [below right=of b] (d) ;
			\vertex [right=0.75 of c'] (e) ;
			
			\diagram* { (a) -- [scalar] (b) ,
				(b) -- [fermion,edge label=\(p'+k\)]
				(c'),
				(c') -- [fermion, edge label=\(p'\)] (c), 
				(b) -- [anti fermion, edge label'=\(p\)]
				(d), 
				(c') -- [boson,edge label'=\(k\)] (e), 
			};
		\end{feynman} 
	\end{tikzpicture}
}
\]

We are computing the amplitude for the process $e \to e + \gamma$. Let
$\mathcal{M}_0(k',l')$ denote the part of the amplitude of electron
interacting with the external source, it is depicted as the blob. Here
$k', l'$ denote the out-going and in-coming momenta respectively. The
full amplitude is then given by (polarizations
$\varepsilon(\vec{k},\lambda)$ taken to be real),
\begin{eqnarray*}
i\mathcal{M}(p',p,k) & = & (ie) \bar{u}(p')\left[ \mathcal{M}_0(p',p-k)
\frac{-i(-(\dsl{p}-\dsl{k}) + m)}{(p-k)^2+m^2-\epsilon} \gamma^{\mu}
\varepsilon_{\mu} \right. \\
& & \hspace{3.0cm} \left. \varepsilon_{\mu}\gamma^{\mu}
\frac{-i(-(\dsl{p}'+\dsl{k}) + m)}{(p'+k)^2+m^2 -i\epsilon}
\mathcal{M}_0(p'+k,p)\right]u(p) 
\end{eqnarray*}

To compare with the classical radiation formula, consider the limit
wherein the external source interaction approximates as,
$\mathcal{M}_0(p',p-k) \approx \mathcal{M}_0(p'+k,p) \approx
\mathcal{M}_0(p',p)$. Next, since the photon is soft, we neglect the
$\dsl{k}$ in the fermion numerators. The denominators simplify as,
$(p-k)^2+m^2 = -m^2 -2p	\cdot k + 0 = m^2 = -2p\cdot k$ and likewise,
$(p'+k)^2+m^2 = +2p'\cdot k$. Furthermore, $(-\dsl{p} +
m)\gamma^{\mu}u(p) \to 2p^{\mu}u(p)$, and
$\bar{u}(p')\gamma^{\mu}(-\dsl{p}'+m) \to \bar{u}(p')(p')^{\mu}$. Thus
the amplitude takes the form,
\[
	i\mathcal{M}(p',p,k) = e \underbrace{\left[\bar{u}(p')
	\mathcal{M}_0(p',p) u(p)\right]}_{\mbox{elastic scattering
	amplitude}} \times \underbrace{\left(-\frac{\varepsilon\cdot
			p}{p\cdot k} + \frac{\varepsilon\cdot
			p'}{p'\cdot k} \right)}_{\mbox{extra $k$
			dependent factor}}
\]
The total, unpolarized cross-section is obtained as,
\begin{equation}
d\sigma(p\to p'+k) = d\sigma(p\to p')\int \frac{d^3k}{2|k|(2\pi)^3}
\sum_{\lambda=1,2} e^2 \left| \frac{\varepsilon\cdot p'}{p'\cdot k} -
\frac{\varepsilon\cdot p}{p'\cdot k} \right|^2
\end{equation}
The integration is over soft photons i.e. $|\vec{k}| < k_{max}$. Notice
that The integral is dimensionless.

The integrand is the probability density of radiating a photon of
momentum $k$ within $d^3k$ and any transverse polarization, accompanying
the electron scattering with $p\to p'$ (``acceleration kick'', $p'\neq
p+k$) while {\em the integral is the total probability of emission of a
soft photon with energy up to} $k_{max}$. 

Multiplying the integrand by the photon energy, $|\vec{k}|$ gives the
{\em expectation value of the energy carried away by soft photons up to
energy} $k_{max}$,
\[
	\mathcal{E}_{soft}(k_{max}) := \langle E\rangle=
	\int\frac{d^3k}{(2\pi)^3} \sum_{\lambda} \frac{e^2}{2}
	\left|\vec{\varepsilon}\cdot \left(\frac{\vec{p}'}{p'\cdot k} -
	\frac{\vec{p}}{p\cdot k}\right)\right|^2
\]

The polarization sum in the cross-section can be done as follows. Noting
that $k\cdot(\Case{p'}{p'\cdot k} - \Case{p}{p\cdot k}) = 0$, we can
drop the $k\tilde{k}$ terms in the completeness relation for the
polarizations, effectively taking $\Sigma_{\lambda} \varepsilon_{\mu}
\varepsilon_{\nu} = \eta_{\mu\nu}$. Hence,
\[
	\sum_{\lambda}\left[\left(\frac{p'}{p'\cdot k} - \frac{p}{p\cdot
	k}\right)^{\mu}\varepsilon_{\mu}\right]
	\left[\left(\frac{p'}{p'\cdot k} - \frac{p}{p\cdot
	k}\right)^{\nu}\varepsilon_{\nu}\right] =
	\left(\frac{p'}{p'\cdot k} - \frac{p}{p\cdot
	k}\right)\cdot\left(\frac{p'}{p'\cdot k} - \frac{p}{p\cdot
	k}\right) ,
\]
and the expectation value of energy carried by soft photons is given by,
\begin{equation} \label{ProbRadn}
	\boxed{ \mathcal{E}_{soft}(k_{max}) =
	\int\frac{d^3k}{(2\pi)^3}\sum_{\lambda}\frac{e^2}{2}
\left[-\frac{m^2}{(p'\cdot k)^2} - \frac{m^2}{(p\cdot k)^2} -
2\frac{p\cdot p'}{(p\cdot k)(p'\cdot k)}\right] }
\end{equation}
Note that the integral represents average radiated energy inferred from
the quantum mechanical probability distribution for emission of a
\underline{single} soft photon with $|\vec{k}| < k_{max}$. 

\underline{An Aside:} Incidentally, integrand in the r.h.s. is also the
{\em classical } formula for energy carried away by the $k^{th}$ Fourier
mode of the electromagnetic field. To see this, consider a classical
trajectory $z^{\mu}(\tau) = p^{\mu}/m$ for $\tau < 0$ and equals
$(p')^{\mu}/m$ for $\tau > 0$. The corresponding current is,
\begin{eqnarray*}
	J^{\mu}(x) & = &
	\int_0^{\infty}d\tau\frac{(p')^{\mu}}{m}\delta^4(x - z(\tau)) +
	\int_{-\infty}^0d\tau\frac{p^{\mu}}{m}\delta^4(x - z(\tau)) ~ ~
	\Rightarrow \\
	A^{\mu}(\vec{k}) & = & -
	\frac{e}{|\vec{k}|}\left(\frac{(p')^{\mu}}{p'\cdot k} -
	\frac{p^{\mu}}{p\cdot k} \right) ~ ~ ~ ~  \mbox{(using the
	Retarded Green function.)}
\end{eqnarray*}
The energy of the above radiation field is the precisely as given by the
integrand in the r.h.s. of eq.(\ref{ProbRadn}) \cite{PeskinSchroder}.
The integral now represents {\em the total energy carried by Fourier modes
with momenta} $|\vec{k}| < k_{max}$. 

{\em The integral, $\mathcal{E}_{soft}(k_{max})$,  thus has two
different interpretations.} 

Returning to the evaluation of the integral in eq.(\ref{ProbRadn}), let
us choose a frame in which  the initial and final energies of the
electron are the same: $E' = E \Rightarrow (p')^0 = p^0$ and we write $p
= E(1, \vec{v}),\ p' = E(1,\vec{v}'),\ k = (k := |k|, \vec{k})$. This is
equivalent to $p^2 = -m^2$ and $m^2/E^2 = 1-\vec{v}^2$.  With this
choice, $p\cdot k = Ek(-1+\hat{k}\cdot \vec{v}),\ p'\cdot k =
Ek(-1+\hat{k} \cdot \vec{v}')$ and $p\cdot p' =
E^2(-1+\vec{v}\cdot\vec{v}')$. The $k^2$ from the $d^3k$ cancels and the
integrated emitted energy takes the form,
\begin{eqnarray*}
\mathcal{E}_{soft}(k_{max}) & = & -\frac{e^2}{2}\int dk\int
\frac{d\Omega}{8\pi^3}\left[
	\frac{m_0^2}{E^2}\left(\frac{1}{(1-\hat{k}\cdot \vec{v})^2}    +
	\frac{1}{(1-\hat{k}\cdot \vec{v}')^2} \right) -
	\frac{2(1-\vec{v}\cdot \vec{v}')}{(1-\hat{k}\cdot
\vec{v})(1-\hat{k}\cdot \vec{v}')} \right]  \\
I(\vec{v}, \vec{v}') & := & \int\frac{d\Omega_{\hat{k}}}{4\pi}\left[
	\frac{2(1-\vec{v}\cdot \vec{v}')}{(1-\hat{k}\cdot
	\vec{v})(1-\hat{k}\cdot \vec{v}')} -
	\frac{m^2}{E^2}\left(\frac{1}{(1-\hat{k}\cdot \vec{v})^2}    +
\frac{1}{(1-\hat{k}\cdot \vec{v}')^2} \right) \right] ~ ~ \Rightarrow \\
\mathcal{E}_{soft}(k_{max}) & = & \frac{e^2}{4\pi^2}\int dk I(\vec{v},
\vec{v}') ~ ~ \approx ~ ~ \frac{\alpha}{\pi}k_{max} I(\vec{v}, \vec{v}')
\end{eqnarray*}
The integral is actually divergent at the upper end. It has been cut off
at $k_{max}$ which is provided by the {\em inverse of the duration over
which the electron receives the kick}. This is given by $k_{max} \sim
|\vec{p} - \vec{p}'| := |\vec{q}|$. This is a physical cutoff.

The angular integration receives the maximum contribution from when $
\hat{k}$ is parallel to $\vec{v}$ or $\vec{v}'$ depending upon if it is
the initial or the final state bremsstrahlung. Given the directions $v
\hat{v},\ \hat{v}'$, the $\hat{k}$ varies between these two directions
(smaller angle). Thus, to pick up contribution when $\hat{k}$ is
parallel to $\vec{v}$, we define $cos\theta$ by $\hat{k}\cdot\vec{v} =
vcos\theta$ which varies between $|\vec{v}|cos\theta =
\vec{v}\cdot\vec{v}'$ and $cos\theta = 1$. Similarly for contribution
from nearly parallel to $\vec{v}'$, we define $cos\theta$ by
$\hat{k}\cdot\vec{v}' = |\vec{v}'|cos\theta$ which varies from
$|\vec{v}'|cos\theta = \vec{v}\cdot\vec{v}'$ to $\cos\theta = 1$.

Additionally, in the extreme relativistic limit where we can neglect the
$m^2/E^2$ terms, we can approximate
$(1-\hat{k}\cdot\vec{v})(1-\hat{k}\cdot\vec{v}') \approx
(1-|\vec{v}|cos\theta)(1-\vec{v}\cdot\vec{v}')\ or\ \approx
(1-|\vec{v}'|cos\theta)(1-\vec{v}\cdot\vec{v}')$.  The angular integral
then approximates as,
\begin{eqnarray*}
	I(\vec{v}, \vec{v}') & \approx & \int^1_{|\vec{v}|cos\theta =
	\vec{v}\cdot\vec{v}'} dcos\theta \frac{1-\vec{v}\cdot
	\vec{v}'}{(1-vcos\theta)(1-\vec{v}\cdot\vec{v}')}  +
	\int^1_{|\vec{v}'|cos\theta = \vec{v}\cdot\vec{v}'} dcos\theta
	\frac{1-\vec{v}\cdot
	\vec{v}'}{(1-v'cos\theta)(1-\vec{v}\cdot\vec{v}')} \\
	& \simeq &
	ln\left(\frac{1-\vec{v}\cdot\vec{v}'}{1-|\vec{v}|}\right) +
	ln\left(\frac{1-\vec{v}\cdot\vec{v}'}{1-|\vec{v}'|}\right)
	\approx ln\left( \frac{(p\cdot p')^2}{E^2(E-p)(E-p')} \right) ~
	~ \\
	& \simeq  & 2 ln \frac{q^2}{m^2}   ~ ~ \mbox{where} ~ ~ q :=
	p'-p 
\end{eqnarray*}
We have used: $(1-\vec{v}\cdot\vec{v}') = -p\cdot p'/E^2\ ,
(1-|\vec{v}|) = (E-|\vec{p}|)/E, \ (E-|\vec{p}|)(E - |\vec{p}'|) \approx
(E -|\vec{p}|)^2$ and $E(E-|\vec{p}|) \approx E^2 - \vec{p}^2 = m^2$ in
getting to the last equation. 

Thus the radiated energy in the soft modes is given by,
\begin{equation}
	\mathcal{E}_{soft}(k_{max}) ~ = ~
	\frac{\alpha}{\pi}\int_0^{k_{max}} dk I(\vec{v}, \vec{v}') ~
	\xrightarrow[E \gg m]{} ~ \frac{2\alpha}{\pi}\int_0^{k_{max}}dk\
	ln(q^2/m^2) \ . 
\end{equation}

Interpreting the above energy classically, what would be the total
number of photons emitted, $N_{\gamma}$? Notice that $I(\vec{v},
\vec{v}')dk$ is the contribution to the radiated energy from the Fourier
modes with energy, $|\vec{k}|$. The equivalent number of photons would
be $\Case{I(\vec{v}, \vec{v}')dk}{|\vec{k}|}$ and the total number
$N_{\gamma}$ is obtained by integrating over the energies: 
\[
	N_{\gamma} =
	\frac{\alpha}{\pi}\int_0^{k_{max}}\frac{dk}{k}I(\vec{v},
	\vec{v}') \approx
	\frac{2\alpha}{\pi}\int_0^{k_{max}}\frac{dk}{k}\ ln(q^2/m^2) 
\]
This is clearly divergent from the lower limit.

But this is the same expression for the total quantum mechanical
probability for emission of a soft photon up to energy $k_{max}$ and thus
is divergent. 

{\em This is the IR divergence of the QED cross-section for
the bremsstrahlung process.}

If the photon is given a small mass $\mu$, then the $\int dk/k$ will
give $ln((|q|\simeq k_{max})/\mu)$ while $I(\vec{v}, \vec{v}')$ gives
the $ln(q^2/m^2)$.  Hence,
\begin{equation}
	\boxed{ d\sigma(p\to p'+k) = d\sigma(p\to p')\cdot
	\frac{\alpha}{\pi}\underbrace{\left[ln(q^2/\mu^2)\cdot
ln(q^2/m^2)\right]}_{\mbox{Sudakov double log}} }
\end{equation}

\newpage
\section{Treatment of Divergences:} \label{TreatingDivergences}

At the first non-trivial attempt at computing radiative corrections, we
encounter divergences of the UV type (from large loop momentum) and of
IR type (from small loop momentum in massless propagators). How do we
understand the physical origin of these? How do we adjust the
computational procedure so as to make {\em unambiguous predictions} to
be confronted with observations? Let us recapitulate what we have got.

$\bullet$ Quite generally, using only Poincare covariance and assumption
about the possible spectrum of a theory with mass gap, the
Kallen-Lehmann representation gave us,
\begin{eqnarray*}
	\Delta'_F(p) & = & Z\Delta_0(p, m_{ph}) +
	\int_{m^2_{th}}^{\infty} d\sigma^2 \rho(\sigma^2) \Delta_0(p,
	\sigma)  ~ ~ , ~ ~ 0 < Z < 1 \\
	& & 	\Delta_0(p,\sigma) ~ := ~ \frac{-i}{p^2 +
	\sigma^2-i\epsilon} \hspace{1.5cm} \mbox{(Free propagator of
	mass $\sigma$)} \\
	S'_F(p) & = & Z_2 S_0(p,m_{ph}) + \int_{m^2_{th}}^{\infty}
	d\sigma^2 \frac{-i\{\rho_1(\sigma^2)\dsl{p} +
	\rho_2(\sigma^2)\}} {p^2+\sigma^2-i\epsilon} \\
	& & 	S_0(p, \sigma) ~ := ~ -i\frac{(-\dsl{p}+
	\sigma)}{p^2+\sigma^2-i\epsilon} \hspace{1.0cm} \mbox{(Free
	propagator of mass $\sigma$)} \\
	(D'_F)_{\mu\nu}(q) & = & \left(\eta_{\mu\nu} -
	\frac{q_{\mu}q_{\nu}}{q^2}\right)\left\{
	\frac{Z_3}{q^2-i\epsilon}  + \int_{0}^{\infty} d\sigma^2
\frac{\Pi(\sigma^2)} {q^2+\sigma^2-i\epsilon}\right\} 
\end{eqnarray*}
The $Z$'s are the field (or wavefunction) renormalization constants.
They are determined as the residues at the (isolated) pole at the {\em
physical mass}, $m_{ph}$. This is without any perturbation series.

In perturbation series though we obtained (for fermion and photon),
\[
	S'_F(p) = \frac{1}{\dsl{p}+m-\Sigma(\dsl{p})} ~ ~ ~ \& ~ ~ ~
	D'_F(q) = \left(\eta_{\mu\nu} -
	\frac{q_{\mu}q_{\nu}}{q^2}\right)\left[\frac{1}{q^2}\
	\frac{1}{1-\Pi(q^2)}\right]    
\]
This is consistent with the general expectation that the physical masses
are determined by the poles in the exact propagator while the
corresponding residues determine the $Z$'s. Thus,
\[
	\delta m := m_{ph} - m = -\Sigma(\dsl{p}=-m_{ph}) ~ , ~ Z_2^{-1}
	= 1 - \frac{d\Sigma(\dsl{p})}{d\dsl{p}}|_{\dsl{p} = -m_{ph}} ~ ,
	~ Z_3^{-1} = 1 - \Pi(0). 
\]
and of course the photon physical mass is zero.

$\bullet$ The vertex function for on shell fermion masses takes the
form, 
\[
	\gamma^{\mu}(p,p') := \gamma^{\mu}F_1(q^2) +
	i\sigma^{\mu\nu}q_{\nu}F_2(q^2) ~ , ~ \sigma_{\mu\nu} :=
	\frac{i}{2}[\gamma{\mu},\gamma_{n}] ~ , ~ \dsl{p} = -m_{ph} =
	\dsl{p}'\ . 
\]
This decomposition is understood to be sandwiched between $\bar{u}(p')$
and $u(p)$. Since $m_{ph} = m + \delta m = m + o(\alpha)$, for the first
order corrections, we can take $m_{ph} = m$. Here are the expressions we
obtained:
\begin{center}
	\fbox{
		\begin{minipage}{\textwidth}
\begin{eqnarray}
	(I) \hspace{1.3cm} \Sigma(p) & = &
	\frac{\alpha}{2\pi}\int_0^1dx\ (2m+x\dsl{p}) ln
	\frac{x\Lambda^2}{x(1-x)p^2+(1-x)m^2+x\mu^2-i\epsilon}\\
\delta m & = & \frac{\alpha}{2\pi}\int_0^1dx\ (2-x)ln
\frac{x\Lambda^2}{(1-x)^2m^2 + x\mu^{2}} , \\
Z_2 & = & 1 + \frac{\alpha}{4\pi}ln(\Lambda^2/m^2) ~ ~ \\
(II) \hspace{1.1cm} \Pi(q^2) & = & -\frac{2\alpha}{\pi}\int_0^1dx\
x(1-x)\left\{\frac{2}{\epsilon} - \gamma -
ln\left(\frac{m^2+q^2x(1-x)}{\mu_0^2}\right)\right\} \\
& = & -\frac{2\alpha}{3\pi}\frac{1}{\epsilon} + o(\epsilon^0) ~
\Rightarrow ~ Z_3 = 1-\frac{2\alpha}{3\pi}\frac{1}{\epsilon} \\  
(III) \hspace{0.9cm} F_1(q^2) & = & 1 + \frac{\alpha}{2\pi}\int dx\ dy\
dz\ \delta(1-x-y-z)\left\{ln \left(\frac{z\Lambda^2}{M^2}\right) \right.
\nonumber  \\
& & \left. \hspace{2.0cm} \frac{-(1-x)(1-y)q^2 +
(1-4z+z^2)m^2}{M^2}\right\} \\
F_2(q^2) & = & \frac{\alpha}{2\pi}\int dx\ dy\ dz\delta(1-x-y-z)\left\{
\frac{2m^2z(1-z)}{M^2} \right\} ~ , \\
& & \hspace{3cm} M^2 := q^2xy + m^2(1-z)^2 + \mu^2z \\
\therefore F_1(0) & = & 1 + \frac{\alpha}{2\pi}\int dx\ dy\ dz
\delta(1-x-y-z) \left\{ln\left( \frac{z\Lambda^2}{(1-z)^2m^2} \right)
\right. \nonumber \\ & & \left. \hspace{4.0cm} +
\frac{1-4z+z^2}{(1-z)^2}\right\} \\
F_2(0) & = & \frac{\alpha}{2\pi}\int dx\ dy\ dz\delta(1-x-y-z)
\left(\frac{z}{1-z}\right) = \frac{\alpha}{2\pi}  ~ =: ~ \frac{g-2}{2}
\\
(IV) \hspace{0.5cm} d\sigma_{p\to p'+k} & = & d\sigma_{p\to
p'}\int\frac{d^3k}{2k(2\pi)^2} \sum_{\lambda} e^2\left( \frac{
\vec{\varepsilon}\cdot \vec{p}'}{\vec{p}'\cdot \vec{k}} - \frac{
\vec{\varepsilon}\cdot \vec{p}}{\vec{p}\cdot \vec{k}}\right)^2 \\
& = & d\sigma_{p\to p'}\left[\frac{e^2}{4\pi^2}\int\frac{dk}{k}I(\vec{v},
\vec{v}')\right] ~ ~ \mbox{where,} \\
I(\vec{v}, \vec{v}') & = & \int\frac{d\Omega}{4\pi}\left\{
\frac{2(1-\vec{v}\cdot\vec{v}')}{(1-\hat{k}\cdot\vec{v})
(1-\hat{k}\cdot\vec{v}')} - \frac{m^2/E^2}{(1-\hat{k}\cdot\vec{v})^2} -
\frac{m^2/E^2}{(1-\hat{k}\cdot\vec{v}')^2}\right\} \hspace{0.8cm} \\
& & \mbox{where} ~ p^{\mu} = (E, E\vec{v})\ , \ (p')^{\mu} = (E,
E\vec{v}')\ , \ k^{\mu}(|\vec{k}| , \vec{k}) ~ \mbox{is used.} \\
& & \mbox{For $|\vec{v}| \approx |\vec{v}'| \approx 1\ $},
I(\vec{v},\vec{v}') \approx 2ln(\vec{q}^2/m^2), \ q := p' - p ~
\Rightarrow \nonumber \\
d\sigma_{p\to p'+k} & = & d\sigma_{p\to
p'}\left[\frac{\alpha}{\pi}ln(q^2/\mu^2)\ ln(q^2/m^2)\right]
\end{eqnarray}
\end{minipage}
}
\end{center}

\newpage
\subsection{Treatment of the IR divergences} \label{IRDiv}
Let us consider the IR problem first. Since $F_1$ has both the UV and
the IR divergence, we will separate these by using the UV renormalized
$F_{1,ren}$ which was defined through, $\delta F_{1,ren}(q^2) := \delta
F_1(q^2) - \delta F_1(0)$. This guarantees that the renormalized $F_1$
satisfies the condition $F_1(0) = 1$. Using the expressions above, we
see that,
\begin{eqnarray*}
	\delta F_{1,ren}(q^2) & = & \frac{\alpha}{2\pi}\int dx\ dy\ dz\
	\delta(1-x-y-z) \left\{\frac{-(1-x)(1-y)q^2 +
	(1-4z+z^2)m^2}{M^2(q^2)} \right. \\
	& & \left. \hspace{4.0cm} - \frac{(1-4z+z^2)m^2}{(1-z)^2m^2+z
	\mu^2}\right\}
\end{eqnarray*}
The IR divergence comes from $z\to 1 \leftrightarrow x\sim y\sim 0
\leftrightarrow M(q^2,x,y)$ would vanish but for the photon mass
$\mu^2$. To isolate the divergence, it suffice to take $x=y=0,\ z=1$ in
the numerator. Doing the $x$ integration using the delta function gives
$x = 1-y-z > 0 \Rightarrow y < 1-z$ and leads to,
\[
\delta F_{1,ren} \approx \frac{\alpha}{2\pi}\int_0^1dz \int_0^{1-z}dy
\left\{ \frac{-2m^2-q^2}{(1-z)^2m^2+y(1-y-z)q^2+\mu^2}
\frac{2m^2}{(1-z)^2m^2+\mu^2} \right\}
\]
Substituting $y := (1-z)u, v:= 1-z$, the leading contribution in the
limit $\mu \to 0$ is expressed as \cite{PeskinSchroder},
\begin{eqnarray}
& & \boxed{F_{1,ren}(q^2) \approx 1 -\frac{\alpha}{2\pi}f_{IR}(q^2)\
ln\left( \frac{m^2 \ or\ q^2}{\mu^2} \right) } ~ ~ ~ \mbox{where,}\\
& & \boxed{f_{IR}(q^2) \ = \ \int_0^1du
\left[\frac{m^2+q^2/2}{m^2+u(1-u)q^2}\right] - 1 ~ }
\end{eqnarray}
Since $F_1$ is the coefficient of the $\gamma^{\mu}$ term, we can
replace $e \to eF_{1,ren}$ in the electron scattering off a {\em
classical potential}. The cross-section is then given by,
\[
\left(\frac{d\sigma}{d\sigma}\right) \simeq
\left(\frac{d\sigma}{d\sigma}\right)_{tree}\left[1-\frac{\alpha}{\pi}
f_{IR}(q^2) ln\left( \frac{m^2 \ or\ q^2}{\mu^2} \right)\right] .
\]
Notice that $e\to eF_1 \Rightarrow e^2\to e^2F_1^2 \Rightarrow
\alpha/(2\pi) \to \alpha/(\pi)$.

Not only is this divergent as $\mu \to 0$, for non-zero $\mu$ it is
actually {\em negative} implying negative cross-section! In the limit of
$q^2 \to \infty$ ($q^2$ is space-like and hence positive), $f_{IR}(q^2)
\to ln(q^2/m^2) \Rightarrow F_{1,ren}(q^2) \simeq 1 -
\Case{\alpha}{2\pi}ln(q^2/m^2)\,ln(q^2/\mu^2)$ and hence,
\begin{equation}\label{CrossIR}
	\boxed{ d\sigma(p\to p') \simeq d\sigma_{tree}(p\to
	p')\left[1-\frac{\alpha}{\pi}ln(q^2/m^2)\ ln(q^2/\mu^2)\right] ~
~ , ~ ~ q^2 \to \infty\ , \ \mu^2\to 0. }
\end{equation}

The Bremsstrahlung cross-section on the other hand is,
\begin{equation}\label{BremCross}
\boxed{ d\sigma(p\to p'+k) = d\sigma(p\to
p')\left[\frac{\alpha}{\pi}ln\frac{q^2}{m^2}\,
ln\frac{q^2}{\mu^2}\right] = d\sigma(p\to
p')_{tree}\left[\frac{\alpha}{\pi}ln\frac{q^2}{m^2}\,
ln\frac{q^2}{\mu^2}\right]. }
\end{equation}
Both the cross-sections above are IR divergent and both suffer from
ambiguity from contamination of soft photons.

We already noted that detectors are unable to distinguish a charged
particle accompanied by soft photons below detector sensitivity. What is
the appropriate theoretically computed quantity which reflects this
limitation of the detection process?

When an experimenter reports a detection of a scattered electron, he/she
is actually giving an estimate of the probability that {\em an electron,
$e(p')$ is detected \underline{and} a photon is not detected.} This
probability is the probability for a process with no emitted photon plus
the probability that there are accompanying soft photons with energies
below the detection threshold, $\varepsilon_{th}$. In equation,
\begin{equation} \label{MeasuredCross}
	\boxed{ (d\sigma)_{measured} = d\sigma\ (\ p\to p'\ ) + d\sigma\
	(\ p\to p'+k\ ;\  |\vec{k}| < \varepsilon_{th}\ )\  . }
\end{equation}
But this is precisely the sum of the two cross-sections given in
eqns.(\ref{CrossIR},\ref{BremCross}). We see that the leading
contribution as $q^2 \to \infty, \mu^2\to 0$ is exactly canceled out in
the observed cross-section!

\underline{Note:} For a general $q^2$, the sum of the two cross-sections
is given by \cite{PeskinSchroder},
\[
(d\sigma)_{measured} = d\sigma_{tree}(\ p\to p'\ )\left\{ 1 -
\frac{\alpha}{\pi} f_{IR}(q^2) ln\left( \frac{q^2\ or\ m^2\
}{\mu^{2}}\right) + \frac{\alpha}{2\pi}\ I(\vec{v}, \vec{v}')\ ln(
\varepsilon_{th}/\mu^2 ) \right\}
\]
It turns out that without the limit $q^2\to \infty$, it still holds that
$I(\vec{v}, \vec{v}') \to 2 f_{IR}$ \cite{PeskinSchroder}. For general
$q^2$, the coefficient of $f_{IR}$ is $ln(q^2)$ or $ln(m^2)$. For large
$q^2$ we can of course drop  $m^2$. Furthermore, experimentally it is
easier to track the behavior of the cross-section as a function of
$q^2$, so taking $q^2 \gg m^2$, we write the unambiguous and measurable
prediction as,
\begin{equation}\label{FinalMeasured}
\boxed{ \left(\frac{d\sigma}{d\Omega}\right)_{measured} =
\left(\frac{d\sigma}{d\Omega}\right)_{tree}(p\to p')\left[1 -
\frac{\alpha}{\pi}ln\frac{q^2}{m^2}\ ln\frac{q^2}{\varepsilon_{th}^2} +
o(\alpha^2)\right]_{q^2\gg m^2} }
\end{equation}

Appreciate that we had parametrised the IR divergence in terms of the
photon mass $\mu^2$; we used the renormalized $F_{1,ren}(q^2)$ to
separate the IR from the UV and finally, identified the quantity which
is actually reported by experiments which is the sum of the two
cross-sections.

This has been a demonstration at the 1-loop. It is non-trivial result of
a great deal of work, that the basic mechanism of cancellation works to
all orders in $\alpha$. There are other types of divergences analogous
to the IR divergences, the so called ``mass singularities'' which do not
cancel but can be factorised in a convenient form and then eliminated
from the observed cross-sections. See the book by G. Sterman
\cite{Sterman}.

UV divergences is a different ball game and needs a different procedure.
\subsection{Treatment of the UV divergences} \label{UVDiv}
All the three radiative corrections in, $\Sigma, \Pi$ and $F_1$ have the
UV divergences parametrised in terms of the Pauli-Villars cut-off
$\Lambda$ or the dimensional regularization $\epsilon^{-1}$ pole. Note
that the bremsstrahlung process does not have a UV divergence as there
is physical cut-off for the energy of the ``soft photons''. We need to
pay attention to the $Z$ factors now.

Recall that the $S-$matrix element definition, via the LSZ reduction
procedure, had a factor of $\Case{1}{\sqrt{Z}}$ for each external line.
There was also an amputation of the external line, effected by the
equation of motion operator eg $\Box-m^2$ acting on the Feynman
propagator $\Delta_F$ for the external line. The $\sqrt{Z}$ entered from
the asymptotic condition relating the interacting field to the in/out
fields. The in/out fields satisfy their respective field equations with
{\em physical masses} and also have the normalization factors with
$\omega_k$ which also contain the physical masses. In the momentum
space, these factors associated with external lines are of the form
$\Case{1}{\sqrt{Z}}(p^2+m^2)\Delta'_{F}(p)$ where the $\Delta_F$
contains the self-energy giving it the form $\Delta'_F(p) =
(p^2+m^2-\Pi(p^2))^{-1}$ (for scalars). 

In perturbative computation of the self-energy however, the $m$ is the
mass parameter in the $\cal{L}$ which is not the physical mass. In fact
from the Kallen-Lehmann representation, we know that the physical mass
is determined from $p^2+m^2-\Pi(p^2)|_{p^2=-m_{ph}^2} = 0
\leftrightarrow -m^2_{ph}+m^2 = \Pi(-m^2_{ph})$. Expanding $\Pi(p^2)$
about $-m^2_{ph}$ gives us,
\begin{eqnarray*}
	\Pi(p^2) & = & \pi(-m^2_{ph}) + (p^2+m^2_{ph})\Pi'(-m^2_{ph}) +
	\dots  ~ ~ \Rightarrow \\
	(\Delta'_{F})^{-1}(p) & = & p^2+m^2-\Pi(-m^2_{ph}) -
	(p^2+M^2_{ph})\Pi'(-m^2_{ph}) + \dots \\
	& = & (p^2 + m^2_{ph})\left\{1 - \Pi'(-m^2_{ph})+\dots\right\} ~
	= ~ (p^2 + m^2_{ph})\left\{Z_{\Phi}^{-1}+\dots\right\} \\
	\therefore \frac{1}{\sqrt{Z_{\Phi}}}(p^2+m^2_{ph})\Delta'_F(p) &
	= & \sqrt{Z_{\Phi}}\left\{1 + o(\ (p^2+m^2_{ph})^2\ ) \right\} 
\end{eqnarray*}
Since we evaluate the $S-$matrix elements on shell, all the higher order
terms vanish. The net result is that: 

\noindent {\em In an $S-$matrix element, for each external line
introduce a factor of $\sqrt{Z_{field}}\times\ $ appropriately
normalized wavefunction and now do \underline{not} include self-energy
corrections on the external lines.}

Consider the electron scattering off external potential. The
corresponding invariant amplitude is given by $e(\sqrt{Z_2})^2
\bar{u}(p') \Gamma^{\mu}(p',p) u(p)$ with $\Gamma^{\mu} = \gamma^{\mu}
F_1 + (\dots)F_2$. The $F_2$ has no divergences while $Z_2, F_1$ both
are divergent. Thus we write, $Z_2 = 1 + \delta Z_2, F_1 = 1 + \delta
F_1$ with the $\delta$'s representing the $o(\alpha)$ divergent
corrections.  Thus we write,
\begin{eqnarray*}
\mathcal{M}_{e\to e}(p',p) & = & e(1+\delta Z_2)\left\{
\bar{u}(p')\gamma^{\mu}u(p)\ (1+\delta F_1(q^2)) +
\bar{u}(p')\frac{i\sigma^{\mu\nu}q_{\nu}}{2m_{ph}}u(p) \ \delta
F_2(q^2)\right\} \\
& = & e\left[\bar{u}(p')\gamma^{\mu}u(p)\right]\left\{1+\delta
F_1+\delta Z_2\right\} + \bar{u}(p')
\frac{i\sigma^{\mu\nu}q_{\nu}}{2m_{ph}}u(p) \ \delta F_2(q^2) ~ ~
\mbox{But,}\\
\delta Z_2 & = & \frac{\alpha}{2\pi}\int_0^1dz \left[-zln
\frac{z\Lambda^2}{M^2} + 2(2-z)\frac{z(1-z)m^2}{M^2}\right] ~ , ~ M^2 =
(1-z)^2m^2 + z\mu^2\\
\delta F_1(0) & = & \frac{\alpha}{2\pi}\int_0^1dz(1-z)\left[ln
\frac{z\Lambda^2}{M^2} + (1+z^2-4z)\frac{m^2}{M^2}\right] \\
\therefore \delta F_1(0)+\delta Z_2 & = & \frac{\alpha}{2\pi}\int_0^1dz
\left[(1-2z)ln \frac{z\Lambda^2}{M^2} + \frac{m^2}{M^2}\times \right. \\
& & \left. \hspace{4.0cm} \left\{(4-2z)(1-z)z +
(1-z)(1+z^2-4z)\right\}\right]
\end{eqnarray*}
Integrate the first  term by parts,
\begin{eqnarray*}
\int_0^1dz(1-2z)ln \frac{z\Lambda^2}{M^2} & = & (z-2z^2)\left.
	ln\frac{z\Lambda^2}{M^2}\right|_0^1 -
	\int_0^1dz(z-z^2)\left\{\frac{1}{z} - \frac{2(1-z)(-1)m^2 +
	\mu^2}{M^2}\right\} \\
& = & -\int_0^1dz \frac{1-z}{M^2}\{ M^2 + 2z(1-z)m^2 - z\mu^2\}\\
& = & -\int_0^1dz\frac{1-z}{M^2}\{m^2(1-z)(1+z)\} = -\int_0^1dz
\frac{m^2(1-z)^2(1+z)}{m^2\left((1-z)^2 +z\frac{\mu^{2}}{m^2}\right) }\\
%
%
\therefore \delta F_1(0) + \delta Z_2 & = & \frac{\alpha}{2\pi}
\int_0^1dz\left[-\frac{m^2(1-z)(1-z^2)}{M^2} + \frac{m^2}{M^2}(1-z)\{ 4z
- 2z^2 + 1+z^2-4z\}\right] \\
& = & \frac{\alpha}{2\pi}\int_0^1dz\frac{m^2}{M^2}(1-z)(1-z^2)\{-1 + 1\}
= 0 ~ \mbox{(!)}
\end{eqnarray*}
Thus not only has the $ln\Lambda^2$ divergence canceled, we have got
$\delta Z_2 = - \delta F_1(0)$ and thus the entire correction takes the
form,
\[
\mathcal{M}_{e\to e}(p',p) = e\bar{u}(p')\gamma^{\mu}u(p)\left\{1 +
\underbrace{\delta F_1(q^2) - \delta F_1(0)}_{\delta F_{1,ren}}\right\}
+i\bar{u}(p')\frac{i\sigma^{\mu\nu}q_{\nu}}{2m_{ph}}u(p)\delta F_2(q^2) 
\]
We recover the {\em ad hoc} prescription of using $\delta F_{1,ren}$
introduced earlier while discussing the IR divergences. Note that the
subtracted term was crucial for the IR divergence cancellation. It was
of course  physically expected since $\delta F_1(q^2=0) = 0$ should
actually hold to all orders. Note that the UV divergence in the vertex
function, $\delta F_1(q^2)$, is canceled against the divergence in
$\delta Z_2$ coming from electron self-energy.

Now we are left with the divergence in $Z_3$ from the $\Pi(q^2=0)$ for
the photon self-energy. We still have the electron mass-shift $\delta m$
which is divergent.
\subsubsection{1-Loop Renormalization: Charge Screening and Lamb shift}
Consider charged particle scattering by exchanging a photon, necessarily
off-shell (space-like in fact). Replace the free propagator
$(D_F)_{\mu\nu}(q^2)$ by the exact propagator which includes the photon
self-energy, $(D'_F)_{\mu\nu}(q^2)$. The exact propagator has additional
terms $q_{\mu}q_{\nu}$. But thanks to the Ward identity, these do not
contribute to $S-$matrix elements and effectively, $(D'_F)_{\mu\nu}(q^2)
= -i\eta_{\mu\nu}[q^2(1-\Pi(q^2))]^{-1}$. This has a pole at $q^2=0$
with residue $Z_3 = [1-\Pi(0)]^{-1}$. The replacement thus gives a
factor of $Z_3$. The photon internal line connects two vertices
contributing $e^2$. Thus, if we {\em identify} $\boxed{e^2Z_3 =:
e^2_{ph}}$\ , then all divergences coming from the photon self-energy
can be neatly absorbed in the {\em physical, measured charge} which is
finite.  For the mass parameter too, we identified $m^2_{ph} = m^2 +
\delta m$, which absorbs the divergence in $\delta m$ into the
Lagrangian parameter $m$. The identification $e_{ph} := e\sqrt{Z_3}$ is
called {\em charge renormalization} while $m^2_{ph} := m^2 + \delta
m$ is called {\em mass renormalization}. For external photon line, we
will just have $\sqrt{Z_3}$ factor and no self-energy corrections and of
course no mass shift. 

Thus the exact photon propagator may be replaced by the free propagator
by simultaneously replacing $\alpha$ by $\alpha_{eff}$ defined below.
\begin{eqnarray*}
	\alpha (D'_F)_{\mu\nu}(q^2) & = & (D_F)_{\mu\nu}(q^2)
	\frac{\alpha}{1-\Pi(q^2)} = \frac{\eta_{\mu\nu}} {q^2-i\epsilon}
	\left[ \frac{\alpha_{ph}(1-\Pi(0))}{1-\Pi(q^2)} \right] :=
	\alpha_{eff}(q^2)\left[ \frac{\eta_{\mu\nu}} {q^2-i\epsilon}
	\right] \\
	& & \mbox{with,} \hspace{1.0cm} \boxed{\alpha_{eff}(q^2) =
	\frac{\alpha_{ph}}{1-(\Pi(q^2) - \Pi(0))}}
\end{eqnarray*}

To appreciate the implication of the above procedure, recall the
discussion of the effective potential inferred from the tree level
scattering for both the Yukawa and the QED coupling. Let us use the same
formula (for QED), but include the self-energy correction for the photon
propagator. The inferred potential is,
\begin{eqnarray}
	V(\vec{x}) & = & \int\frac{d^3q}{(2\pi)^3}e^{i\vec{q}\cdot
	\vec{x}} \frac{-e^2_{ph}}{\vec{q}^2(1-\Pi(q^2)+\Pi(0))} \\
	\Pi(q^2) & = & -\frac{2\alpha}{\pi}\int_0^1dx\ x(1-x)\left\{
	\frac{2}{\epsilon} -\gamma -
ln\left(\frac{m^2+q^2x(1-x)}{\mu_0^2}\right)\right\} \\
\Pi(q^2)-\Pi(0) & = & -\frac{2\alpha}{\pi}\int_0^1dx\ x(1-x)\left\{
ln\left(\frac{m^2}{m^2+q^2x(1-x)}\right)\right\} \\
& \xrightarrow[]{\vec{q}^2 \ll m^2} & \frac{2\alpha}{\pi}\int_0^1dx\
x^2(1-x)^2\frac{q^2}{m^2} = \frac{\alpha}{15\pi}\frac{q^2}{m^2} ~ ~
\mbox{(NR limit)}\\
\therefore V(\vec{x}) & \approx & \int\frac{d^3q}{(2\pi)^3}
e^{i\vec{q}\cdot \vec{x}} \frac{-e^2_{ph}} {\vec{q}^2 (1-
\frac{\alpha}{15\pi} \frac{q^2}{m^2})} \\
& \approx & -\frac{\alpha_{ph}}{r} \underbrace{-
\frac{4\alpha^2_{ph}}{15m^2}\delta^3(\vec{x})}_{\mbox{perturbation}}
\end{eqnarray}
The perturbed Coulomb potential induces a shift in the energy levels of
the Coulomb potential, say in Hydrogen atom. The first order
perturbation theory gives the shift as,
\begin{equation}\label{LambShift}
\boxed{ \Delta E = -\frac{4\alpha^2_{ph}} {15m^2}
\frac{\alpha^3_{ph}m^3} {8\pi} = -\frac{\alpha^5_{ph}m}{30\pi} \simeq
-1.123\times 10^{-7} \mathrm{eV} (\mbox{Lamb Shift}) }
\end{equation}
This is the contribution of the photon self-energy to the famous Lamb
shift. The photon self-energy $ \leftrightarrow \alpha_{eff}$ is thus an
observable and observed effect.  More exact calculation may be seen in
\cite{PeskinSchroder}.

We can also consider the ultra-relativistic regime $q^2 \gg m^2$.  Now
\begin{eqnarray} \label{AlphaRunning}
	\Pi(q^2)-\Pi(0) & = & -\frac{2\alpha}{\pi}\int_0^1dx\
	x(1-x)\left\{ln(\frac{m^2}{q^2}) -
	ln\left(x(1-x)+\frac{m^2}{q^2}\right)\right\} \nonumber \\
	& \approx & -\frac{2\alpha}{\pi}\left[\frac{1}{6}ln(m^2/q^2) -
	\int_0^1dx\ x(1-x)ln\{x(1-x) + o(m^2/q^2)\}\right] \nonumber \\
	& \simeq & \frac{\alpha}{3\pi}\left[ln(q^2/m^2) - 5/3 +
	o(m^2/q^2)\right] \nonumber \\
	& & \boxed{ \therefore \alpha_{eff} ~ = ~
	\frac{\alpha_{ph}}{1-\frac{\alpha}{3\pi}ln\left(\frac{q^2}{m^2
	e^{5/3}}\right)} \xrightarrow{q^2\gg m^2}
-\frac{3\pi}{ln(q^2e^{-5/3}/m^2)} }
\end{eqnarray}

Note that the denominator is less than 1 and hence effective coupling is
{\em larger} than the physical coupling. Thus, the effective coupling
gets stronger at shorter distances. This is interpreted as saying that
the photon self-energy correction polarizes the space between say the
nucleus and the electron, shielding the nuclear charge. As the shielding
cloud is penetrated, higher nuclear charge is seen and hence
$\alpha_{eff}$ is larger. For this reason, the photon self-energy
$\Pi(q^2)$ is also called the {\em vacuum polarization}.

To summarize:
\begin{enumerate}
	\item The UV divergences in the self energies and the vertex
		correction are absorbed away by introducing the physical
		mass for electron, physical charge for the electron and
		electron wavefunction renormalization constant to cancel
		the divergence in $F_1(q^2)$;
	\item The IR divergences in the amplitudes, imply IR divergence
		in the cross-section. For the $e-e$ scattering, the IR
		divergence is canceled against the bremsstrahlung
		process, once the measured quantity is correctly
		identified taking into account the finite detector
		resolution;
	\item The hiding of divergences in $\alpha_{ph}$ and $m^2_{ph}$
		imply that $\alpha_{eff}, \delta m^2$ acquire $q^2$
		dependence - they ``run'' with $q^2$.
\end{enumerate}

This seems satisfactory within the context of the $o(\alpha)$
corrections and the magic of the Ward identity. But it raises more
questions.

$\bullet$ Are the divergences generic? What are the mathematical and
physical reasons for their existence?

$\bullet$ Can we sometimes/always take care of the divergences and make
unambiguous predictions?

$\bullet$ What kind of predictions can be made?

$\bullet$ What are the effective/running parameters?

$\bullet$ Are the divergences an artifact of perturbation theory? Etc,
etc \dots .

Addressing these questions is the genesis of the renormalization theory. 

As a matter of strategy, we will stay within perturbative framework and
try to make sense of the divergences. After all, despite divergences,
we could obtain prediction from QED at 1-loop which have been well
tested!

We begin with a comment that the UV divergences persist at higher loops
as well and have to be faced. Can we always absorb them away in
physical masses, couplings and the field renormalizations in $S-$matrix
elements/cross-sections automatically and make finite, unambiguous
predictions? Indeed it is so for QED and for the class of
(super-)renormalizable theories. The proof needs a somewhat modified
procedure, introducing additional diagrams, representing the so called
`counter terms' and adjusting its coefficients/Feynman rules to
systematically absorb the divergences. We discuss this procedure in the
context of a scalar field theory, specifically the $\Phi^4$ theory in 4
dimensions and $\Phi^3$ theory in 6 dimensions.
\subsection{The Method of Counter terms} \label{CounterTerms}
Consider the $\Phi^4$ theory given by,
\[
	\cal{L} = -\frac{1}{2}\partial_{\mu}\Phi_0\partial^{\mu}\Phi_0
	-\frac{1}{2}m_0^2\Phi_0^2 -\frac{\lambda_0}{4!}\Phi_0^4
\]
The corresponding Feynman rules would be,
\[ 
	%
	\begin{tikzpicture}
		\begin{feynman}
			\vertex (v);
			\vertex [right=of v] (a);
			\diagram* {
				(v) -- [momentum=\(p\)] (a),
			};
		\end{feynman}
	\end{tikzpicture}
	%
	\hspace{1.0cm} \frac{-i}{p^2+m_0^2-i\epsilon} \hspace{1.0cm} ,
	\hspace{1.0cm}
	\vcenter{\hbox{
	\begin{tikzpicture}
		\begin{feynman}
			\vertex (v);
			\vertex [right=of v] (a);	
			\vertex [below=of v] (b);	
			\vertex [right=of b] (c);	
			\diagram*{
				(v) -- (c) ,
				(b) -- (a) ,
			};
		\end{feynman}
	\end{tikzpicture}
	}}	
	\hspace{1.0cm} -i\frac{\lambda_0}{4!}
\]
As seen in QED, we will have divergences in the radiative corrections,
\[ 
\vcenter{\hbox{
\begin{tikzpicture}
	\begin{feynman}
		\vertex (v);
		\vertex [right =of v,blob] (a){All} ;
		\vertex [right =of a] (b) ;
		\diagram*{
			(v) -- [fermion,edge label=\(p\)] (a) -- (b) ,
		};
	\end{feynman}
\end{tikzpicture}
}}	
~ ~ = ~ ~ 
\vcenter{\hbox{
\begin{tikzpicture}
	\begin{feynman}
		\vertex (v);
		\vertex [right =of v] (a) ;
		\diagram*{
			(v) -- (a),
		};
	\end{feynman}
\end{tikzpicture}
}}
~ ~ + ~ ~ 
\vcenter{\hbox{
\begin{tikzpicture}
	\begin{feynman}
		\vertex (v);
		\vertex [right =of v] (a) ;
		\vertex [right =of a] (b) ;
		\vertex [right =of b] (c) ;
		\diagram*{
			(v) -- (a) -- (b) -- (c),
			(a) -- [half right,looseness=1.5] (b),
			(a) -- [half left,looseness=1.5] (b),
		};
	\end{feynman}
\end{tikzpicture}
}}
\]
The suffix $0$ quantities are called ``bare quantities''. Define:
$\Phi_0 := \sqrt{Z}\Phi$, arbitrary scaling of the field. This expresses
the Lagrangian density as,
\begin{eqnarray*}
	\cal{L} & = & -\frac{Z}{2}\partial_{\mu}\Phi\partial^{\mu}\Phi
	- \frac{Z}{2}m_0^2\Phi^2 - \frac{\lambda_0Z^2}{4!}\Phi^4 \\
	& = & -\frac{1}{2}\partial_{\mu}\Phi\partial^{\mu}\Phi
	- \frac{1}{2}m^2\Phi^2 - \frac{\lambda}{4!}\Phi^4 \\
	&  & -\frac{Z-1}{2}\partial_{\mu}\Phi\partial^{\mu}\Phi
	- \frac{Zm_0^2-m^2}{2}\Phi^2 - \frac{\lambda_0Z^2-\lambda}{4!}\Phi^4 
\end{eqnarray*}
The last line constitute ``counter terms'', the field $\Phi$, the mass $m$
and the coupling $\lambda$ are called {\em renormalized} quantities. The
corresponding Feynman rules would be:
\[ 
	\begin{tikzpicture}
		\begin{feynman}
			\vertex (v);
			\vertex [right=of v] (a);
			\diagram* {
				(v) -- [momentum=\(p\)] (a),
			};
		\end{feynman}
	\end{tikzpicture}
	%
	\hspace{1.0cm} \frac{-i}{p^2+m^2-i\epsilon} \hspace{1.0cm} ,
	\hspace{1.0cm}
	\vcenter{\hbox{
	\begin{tikzpicture}
		\begin{feynman}
			\vertex (v);
			\vertex [right=0.9of v] (a);	
			\vertex [below=0.9of v] (b);	
			\vertex [right=0.9of b] (c);	
			\diagram*{
				(v) -- (c) ,
				(b) -- (a) ,
			};
		\end{feynman}
	\end{tikzpicture}
	}}	
	\hspace{1.0cm} -i\frac{\lambda}{4!}
\]
\[ 
	\vcenter{\hbox{
	\begin{tikzpicture}
		\begin{feynman}
			\vertex (v);
			\vertex [right =0.75of v, crossed dot] (a){ };
			\vertex [right =0.85of a] (b) ;
			\diagram*{
				(v) -- [fermion, edge label=\(p\)] (a) -- (b),
			};
		\end{feynman}
	\end{tikzpicture}
	}}
	\hspace{1.0cm} -i \{ \underbrace{(Z-1)}_{\delta_Z}\
	p^2- \underbrace{Zm_0^2 -m^2}_{\delta_m} \} \hspace{0.5cm} ,
	\hspace{0.5cm}
	\vcenter{\hbox{
	\begin{tikzpicture}
		\begin{feynman}
			\vertex [crossed dot] (v){ };
			\vertex [above left=0.75of v] (a);	
			\vertex [above right=0.75of v] (b);	
			\vertex [below right=0.75of v] (c);	
			\vertex [below left=0.75of v] (d);	
			\diagram*{
				(v) -- (a) ,
				(v) -- (b) ,
				(v) -- (c) ,
				(v) -- (d) ,
			};
		\end{feynman}
	\end{tikzpicture}
	}}	
	\hspace{1.0cm} -\frac{i}{4!} \underbrace{Z^2\lambda_0 -
	\lambda}_{\delta_{\lambda}} 
\]
We have used the renormalized quantities and have {\em additional}
vertices with {\em adjustable} coefficients $\delta_Z, \delta_m,
\delta_{\lambda}$. The new vertices are generated by the counter terms.

Note that we have only rewritten the bare Lagrangian density split into
two sets of terms. Since we have not stipulated any conditions, the
split is completely arbitrary. Two sets of conditions, called {\em
renormalization conditions}, are now provided:
(i) The exact propagator is given by $\Case{-i}{p^2+m^2-i\epsilon} \ + \
$terms regular at $ p^2 = -m^2 $. That is, the renormalized mass should
be identified with the physical mass {\em and} the residue should be
$1$;

(ii) The exact 4-point, 1PI function should equal $-i\lambda$ at $s =
4m^2, t = u = 0$. $s,t,u$ are the Mandelstam variables.

The definition of $\lambda$ could be changed, but essentially it is the
value of the exact 4-point scattering amplitude which is measurable.

Diagrams generated by these vertices will again have divergences which
must be regulated. The counter terms coefficients, $\delta_Z, \delta_m,
\delta_{\lambda}$ are to be so chosen as to ensure that the
renormalization conditions are satisfied, order-by-order. Since the
conditions are finite and cut-off independent (regularization
independent), all divergences are spent in defining the counter term
coefficients. By construction, we have generated finite quantities to
all orders! 

\underline{Note:} There are different ways of absorbing the divergences
in the counter term coefficients, especially when massless particles are
involved. The different methods of absorbing the UV divergences is
generically referred to as a {\em subtraction scheme}. This will be
illustrated below.

The perturbation series generated using the bare form of the Lagrangian
is called ``bare perturbation series'' while that generated using the
renormalized quantities together with the renormalization conditions, is
called ``renormalized perturbation series''.

Does such a simple splitting procedure always generate UV finite
(cut-off independent) quantities in terms of {\em finitely many}
renormalization conditions? The answer is YES for a class of theories
called ``renormalizable theories''. There are non-renormalizable
theories (= Lagrangians) for which this procedure fails.

\newpage 
\section{Renormalized perturbation series} \label{RenPertSeries}

Let a theory be specified by a (bare) Lagrangian density, $\cal{L}$ as,
\[
	\cal{L}(\Phi_0,m_0,g_{0,k}) =
	-\frac{1}{2}\partial_{\mu}\Phi_0\partial^{\mu}\Phi_0 -
	\frac{1}{2}m_0^2\Phi_0^2 - \sum_{k\ge
	3}\frac{g_{0,k}}{k!}\Phi_0^k  \ .
\]
Introduce renormalized quantities, $\Phi, m^2, g_k$ and {\em scaling
parameters}, $Z_{\phi}, Z_m, Z_{g_k}$  through the definitions: $\Phi_0
=: \sqrt{Z_{\phi}}\Phi,\ m_0^2Z_{\phi} = Z_m m^2,\ g_{0,k}Z_{\phi}^{k/2}
=: Z_{g_k} g_k$. Writing the $Z$'s in $1 + \delta$ form, we recast the
Lagrangian density as,
\begin{eqnarray*}
	\cal{L}(\Phi,m,g_k;\delta_{\Phi},\delta_m,\delta_k) & = &
	-\frac{1}{2}\partial_{\mu}\Phi\partial^{\mu}\Phi -
	\frac{1}{2}m^2\Phi^2 - \sum_{k\ge 3}\frac{g_k}{k!}\Phi^k \\
	& & -\frac{1}{2}\delta_{\Phi} \partial_{\mu}\Phi\partial^{\mu}\Phi -
	\frac{1}{2}\delta_m m^2\Phi^2 - \sum_{k\ge 3}\delta_k
	\frac{g_k}{k!}\Phi^k ~ ~ ~ \longleftarrow ~ \mbox{(Counter
	terms)}
\end{eqnarray*}
We {\em choose} the renormalization conditions to be:

(i) Exact propagator = $\frac{-i}{p^2+m^2-i\epsilon}$, which is
equivalent to the two conditions: (a) $\Pi(p^2= -m^2) = 0,$ and (b)
$\Pi'(p^2= -m^2) = 0$ ;

(ii) $g_k = $ k-point amputated function, also called {\em k-point vertex
function}, at some chosen values of its momentum arguments.

\underline{Note:} Due to the conditions (i), this scheme is called {\em
on-shell renormalization}.

Calculations of the vertex functions will be functions of the external
momenta, $m, g_k, \delta_{\Phi}, \delta_m, \delta_k$ {\em and} the
regularization parameter - $\Lambda$ for a momentum cut-off, $\epsilon =
4-n$ for the dimensional regularization. The renormalization conditions
will serve to define the counter term coefficients in terms of the
regularization parameter and eliminate them from the vertex functions.
We will be left with vertex function having dependence on the momenta,
$m$ and the couplings $g_k$. This is what we seek to understand.  
\subsection{Necessary Conditions for UV divergence: Power Counting}
\label{PowerCounting}
We begin by finding {\em necessary conditions} for occurrence of UV
divergence. These are obtained by estimating the Feynman integrals in
the region where {\em all} loop momenta become large. Feynman integrands
are rational functions of momenta and in the regime of large momenta,
simply give a power of the large momenta.

Consider an arbitrary, connected and topologically connected diagram
made up of $E$-external lines, $I-$internal lines, $n_k-$ vertices of
$k^{th}$ order and $L-$loops. The loops arise because we have $I$ number
of momenta with $V := \Sigma_{k\ge 3}n_k$ number of vertices enforcing
momentum conservation (and $-1$ since an overall momentum conservation)
condition which leaves some momenta undetermined.  All internal momenta
are linear functions of some external momenta, $p_i$ and some loop
momenta, $k_l$. There are several integration regions in the $d\times L$
dimensional space ($d$ is the space-time dimension). These regions
correspond to various subsets of internal lines vanishing as
$q(p_i,k_l)^{-2}$. It suffices to consider region wherein {\em all} loop
momenta diverge which means that all internal momenta also vanish. If we
take the loop momenta to diverge as $k_l = \Lambda \hat{k}_l, \Lambda
\to \infty$, then we get the {\em superficial degree of divergence},
$\boxed{D = d L - 2 I.}$

For vertex functions, there are no external lines and $\boxed{2I =
\Sigma_{k\ge 3}kn_k -E.}$ The Green's functions will have $+E$. As noted
before, the number of loop momenta is given by $\boxed{L = I -V + 1.}$
Using these the superficial degree of divergence is given by, 
\[
	\boxed{ D = d - \frac{d-2}{2}E + \sum_{k\ge
	3}n_k\left\{k\left(\frac{d-2}{2}\right) -d \right\} }
\]
A necessary condition for divergence is that $D \ge 0$.

For $D < 0$, there is no UV divergence, $D = 0$ is (possibly)
logarithmically divergent, $D=1$ is (possibly) linearly divergent, $D =
2$ is (possibly) quadratically divergent. We have taken $k \ge 3$ to
have some interaction, we take at least {\em one} $n_k \neq 0$ for a
non-trivial diagram and of course $d \ge 2, E \ge 1$.

As an example, consider $d = 4$. Then $D = 4 - E + \Sigma_{k\ge
3}n_k(k-4) \ge 0$ for divergence. That is $4 \ge E + n_3 - \Sigma_{k\ge
5}n_k(k-4)$. If we have $n_{k\ge 5} = 0$, i.e. only $\Phi^3, \Phi^4$
terms, then the condition for divergence is independent of the number of
$\Phi^4$ vertices while with increasing $\Phi^3$ vertices, the $E$ must
decrease correspondingly.

$\bullet$ For $n_3 = 0$, we must have $E \le 4$ i.e. the $2-$point and
the $4-$point vertex functions are superficially divergent,
quadratically and logarithmically respectively. For pure $\Phi^4$
theory, the vertex functions with odd number of external lines vanish
identically. Thus, in 4 dimensional $\Phi^4$ theory, the self-energy and
the $4-$point vertex function are divergent to {\em all orders}. Since
we do have $\delta_{\Phi}, \delta_m, \delta_4$ coefficients, we can
satisfy the renormalization conditions to all orders. This theory is
{\em renormalizable}. 

$\bullet$ For pure $\Phi^3$ theory, $n_4 \ge 0$, $D = 4 -E -n_3$. For
$n_3 = 1$, we can have only tree level, $3-$point function. This gives
$D = 0$. But of course there is no loop integration and hence no
divergence (hence superficial degree of divergence is not a sufficient
condition). Let us also require $L \ge 1$. 

For $n_3 = 2$, we can have $2-$point function diagrams 
$
\vcenter{\hbox{
\begin{tikzpicture}
	\begin{feynman}
		\vertex (v);
		\vertex [right=0.5of v] (a) ;
		\vertex [right=0.75of a] (b) ;
		\vertex [right=0.5of b] (c) ;
		\vertex [right=0.25of c] (z){\(,\)} ;
		\vertex [right=1.0of c] (e) ;
		\vertex [above=0.4of e] (d) ;
		\vertex [below=0.4of e] (f) ;
		\vertex [right=0.5of e] (g) ; 
		\diagram*{
			(v) --(a), (b) --(c),
			(a) --[half right,looseness=1.5] (b),
			(b) --[half right,looseness=1.5] (a),
			(d) -- (e) -- (f),
			(e) -- (g),
		};
		\draw (g) arc [start angle=180, end angle=-180,
		radius=0.35cm];
	\end{feynman}
\end{tikzpicture}
}}
$
with $E=2,L=1, D=0$ which 
are logarithmically divergent and also %
$
\vcenter{\hbox{
\begin{tikzpicture}
	\begin{feynman}
		\vertex (v1);
		\vertex [right=0.5of v1] (v2);
		\vertex [above left=0.5of v1] (a);
		\vertex [below left=0.5of v1] (b);
		\vertex [above right=0.5of v2] (c);
		\vertex [below right=0.5of v2] (d);

		\diagram*{
			(a) --(v1)--(v2)--(c),
			(b) --(v1), (v2)--(d),
		};
	\end{feynman}
\end{tikzpicture}
}}
$
with $E=4,D=-1,L=0$ which are (trivially) ``convergent''. 
For $n_3 = 3, E=2$ we can have the diagram
$
\vcenter{\hbox{
\begin{tikzpicture}
	\begin{feynman}
		\vertex (v1);
		\vertex [right=0.5of v1] (v2);
		\vertex [above left=0.5of v1] (a);
		\vertex [above =0.3of a] (a');
		\vertex [below left=0.5of v1] (b);
		\vertex [below =0.3of b] (b');

		\diagram*{
			(v1) -- (v2), (a) -- (a'), (b) -- (b'),
		};
		\draw (v1) arc [start angle=0, end angle=360,
		radius=0.35cm];
		\draw (v2) arc [start angle=180, end angle=-180,
		radius=0.35cm];
	\end{feynman}
\end{tikzpicture}
}}
$
which is divergent even though $D < 0$. But this is
because of the 1PR nature of the diagram. To exclude this triviality, we
stipulate that contributing diagrams should all be 1PI. Then, for all
$n_3 \ge 4$, all $E-$point vertex functions are convergent. This theory
thus has divergences (eg 2-point function), but only up to a {\em finite
order}. Beyond that, there are no divergences. Such a theory is called
{\em super-renormalizable}.

$\bullet$ For a pure $\Phi^k, k \ge 5,\ D = 4 -E +(k-4)n_k$. With
increasing $n_k$, the number of loops also increase and $D$ keeps
increasing.  Equivalently, more and more $E-$point functions turn
divergent and we will need infinitely many counter terms to absorb the
divergences, in the {\em same} $E-$point function. Such a theory is {\em
non-renormalizable}.

$\bullet$ For sake of variety, take $d = 3$, so that $D = 3-E/2 +
n_k(k/2 -3)$. For $k = 6$, $D$ is independent of $n_k$. The $2,4,6$
point vertex functions need counter terms to all orders and $\Phi^6$
theory is renormalizable.

$\bullet$ Take $d = 6, \Rightarrow D = 6 - 2E +n_k(2k-6) = 2[(3-E) +
n_k(k-3)]$. For $k = 3$, $D$ is independent of $n3$. The $1,2,3$ point
functions are divergent to all orders and the theory is renormalizable.

\underline{Remarks:} 

$\bullet$ The formula for the superficial degree of
divergence can be generalised to include fermions and photons. The
fermion internal line contributes $-1$ (instead of $-2$) to the power
counting. The fermion number conservation ust also be paid attention to
by restricting the types of vertices. If there are derivative couplings,
the numerator contributes positive powers to $D$.

$\bullet$ As noted above, $D\ge 0$ indicates the possibility of a UV
divergence. The actual diagram may be less divergent due to symmetries
eg $\Pi(q^2)$ in QED is only log divergent though $D=2$ suggests
quadratic divergence. It is also possible the coefficient of the
indicated divergence is actually zero! An example of this is the photon
$3-$point vertex function in QED, at 1-loop. 
\[
\vcenter{\hbox{
\begin{tikzpicture}
	\begin{feynman}
		\vertex (v);
		\vertex [left =0.25of v] (v'){\(\mu\)};
		\vertex [right=1.25of v] (v0) ;
		\vertex [left=0.5of v0] (a) ;
		\vertex [above right=0.5of v0] (b) ;
		\vertex [above right=1.25of v0] (b') ;
		\vertex [above right=1.50of v0] (b''){\(\lambda\)} ;
		\vertex [below right=0.5of v0] (c) ;
		\vertex [below right=1.25of v0] (c') ;
		\vertex [below right=1.50of v0] (c''){\(\nu\)} ;
		\diagram*{
			(v) --[photon] (a),
			(b) --[photon] (b'),
			(c) --[photon] (c'),
			(a) --[fermion,half left,looseness=0.9] (b),
			(b) --[fermion,half left,looseness=0.9] (c),
			(c) --[fermion,half left,looseness=0.9] (a),
		};
	\end{feynman}
\end{tikzpicture}
}}
\]
Here $D = 4 -3 = 1$. However due to {\em Furry's theorem}, in QED,
any photon amplitude with  odd number of external lines is zero. Hence
the diagram is identically zero. However, what is true is that {\em if
	$D < 0$ for a diagram and each of its sub-diagram, then the
diagram has no UV divergence} (Dyson-Weinberg theorem).

The superficial degree of divergence, with the caveats mentioned above,
suggest a classification of theories as:

(i) \underline{Super-renormalizable:} Only a finitely many diagrams have
$D \ge 0$. The $D$ has a dependence on the number of vertices such that
it decreases with increase in $n_k$;

(ii) \underline{Renormalizable:} Only a finite number of $E-$point
vertex functions have $D \ge 0$. For these functions though, $D$ is
non-negative at all orders;

(iii) \underline{Non-renormalizable:} All vertex functions are
superficially divergent at sufficiently high orders. The $D$ increases
with increase in $n_k$.

A given theory, may be in any one of the classes in different
dimensions.

Dimensional analysis gives another convenient criterion for
renormalizability. This goes as follows.

\underline{Dimensional Analysis:}

In $d$ dimensions $[\Phi] = \Case{d-2}{2}, \ [\Phi^k] = k\Case{d-2}{2}\
\therefore \ [g_k] = d - k\Case{d-2}{2} = d(1-\Case{k}{2}) + k$. An
$E-$point vertex function can come from a term in the Lagrangian as
$g_E\Phi^{E}$ with $[g_E] = d(1-\Case{E}{2}) + E$. If a diagram
contributing to such as vertex function has a momentum cut-off $\Lambda$
and the number of vertices of order $k$ is $n_k$, then the divergent
part is proportional to $g_k^{n_k} \Lambda^D$. For $k = E$ and $n_E =
1$, there is no loop integration and dimension comes only from the
coupling $g_E$. Hence $[g_E] = [g_k^{n_k}\Lambda^D]$, i.e. 
\begin{eqnarray*}
d(1-E/2) + E & = & {n_k}\left\{d - k\frac{d-2}{2})\right\} + D ~ ~ ~
\Rightarrow \\
	%
	%
	or, ~ D & = & d - \frac{d-2}{2}E - \{\underbrace{d -
	\frac{d-2}{2}k}_{[g_k]} \}n_k
\end{eqnarray*}
Thus, $D$ is independent of $n_k$ if $[g_k] = 0$ and we have finitely
many vertex functions potentially divergent. If $[g_k] > 0$, then only
finitely many diagram can be divergent. If $[g_k] < 0$, then every
vertex function has divergence at some order. Thus, the dependence on
the space-time dimension, $d$ can be hidden in the dimensions of the
couplings and 
\begin{center}
	\fbox{ 
		\begin{tabular}{lll}
		$g_k\Phi^k$ & is renormalizable & if $[g_k] = 0]$ ; \\
		$g_k\Phi^k$ & is super-renormalizable & if $[g_k] > 0]$
		; \\
		$g_k\Phi^k$ & is non-renormalizable & if $[g_k] < 0]$ .
		\\
		\end{tabular} }
\end{center}

Having identified simple criteria for presence of UV divergences, we see
now how the counter terms are used for renormalization. 
\subsection{An Example: The $(\Phi^3)_6$ theory}\label{Phi36Example}
As specific example, we will consider the $(\Phi^3)_6$. Here are the
Feynman rules for the renormalized perturbation series.
\begin{center}
\begin{tabular}{ccccc}
	\hspace{1.75cm} $ \vcenter{\hbox{ \begin{tikzpicture}
	\begin{feynman} \vertex (v); \vertex [right =0.8of v] (a) ;
	\diagram*{ (v) -- [edge label=\(p\)] (a), }; \end{feynman}
	\end{tikzpicture} }} $ \hspace{1.75cm}
	& \hspace{1.75cm}  $\frac{-i}{p^2+m^2-i\epsilon}$
	\hspace{1.75cm} & , & 
	\hspace{0.75cm} $ \vcenter{\hbox{ \begin{tikzpicture}
				\begin{feynman} \vertex (v); \vertex
					[above =0.6of v] (a) ; \vertex
					[below left =0.6of v] (b) ;
					\vertex [below right =0.6of v]
					(c) ; \diagram*{ (v) -- (a), (v)
						-- (b), (v) -- (c), };
		\end{feynman} \end{tikzpicture} }} $ \hspace{0.75cm} 
	& \hspace{0.0cm} $+ig$ ,  \\
	$ \vcenter{\hbox{ \begin{tikzpicture} \begin{feynman} \vertex
			(v); \vertex [right =0.7of v, crossed dot] (a){
	} ; \vertex [right =0.8of a] (b); \diagram*{ (v) -- [edge
	label=\(p\)] (a) -- (b), }; \end{feynman} \end{tikzpicture} }} $
	& $-i\left( \delta_{\Phi}\ p^2 + \delta_m\ m^2\right)$  & , & 
	$ \vcenter{\hbox{ \begin{tikzpicture} \begin{feynman} \vertex
					[crossed dot] (v){ }; \vertex
					[above =0.6of v] (a) ; \vertex
					[below left =0.6of v] (b) ;
					\vertex [below right =0.6of v]
					(c) ; \diagram*{ (v) -- (a), (v)
						-- (b), (v) -- (c), };
		\end{feynman} \end{tikzpicture} }} $
	& $+ig\delta_3$ .
\end{tabular}
\end{center}

$D = 6-2E \ge 0 \Rightarrow E = 2$ (quadratically divergent) and $E = 3$
(logarithmically divergent) are the only vertex functions with UV
divergence. Consider the 1-loop divergences first.

\[ 
	\vcenter{\hbox{ \begin{tikzpicture} \begin{feynman} \vertex (v);
					\vertex [right =of v] (a);
					\vertex [right =of a] (b);
					\vertex [right =of b] (c);
					\diagram*{ (v) -- [fermion,edge
						label=\(p\)] (a), (b) --
						[fermion,edge
	label=\(p\)] (c), (a) -- [fermion,half left,looseness=1.5, edge
	label=\(p+k\)] (b), (b) -- [fermion,half left,looseness=1.5,
	edge label=\(k\)] (a), }; \end{feynman} \end{tikzpicture} }}
	\hspace{0.5cm} + \hspace{0.5cm}
	\vcenter{\hbox{ \begin{tikzpicture} \begin{feynman} \vertex (v);
	\vertex [right =0.9of v, crossed dot] (a) { }; \vertex [right
	=of a] (b) ; \diagram*{ (v) -- [fermion] (a) -- [fermion] (b),
	}; \end{feynman} \end{tikzpicture} }}
	\hspace{0.5cm} + \hspace{0.5cm} \cdots
\]
\begin{eqnarray*}
	i\Pi(p^2) & = &
	\left[\left(\frac{1}{2}\right)(ig)^2\int\frac{d^dk} {(2\pi)^d}
	\left(\frac{-i}{(p+k)^2+m^2-i\epsilon}
\frac{-i}{k^2+m^2-i\epsilon}\right) \right]_{\circled{1}} \\
	& & \hspace{5.0cm} + \left[1\cdot(-i)(\delta_{\Phi}p^2 +
	\delta_m m^2) + \dots \right]_{\circled{2}} \\
	& & \\
	& & \mbox{The}~ \frac{1}{2} ~ \mbox{comes from:}~\hspace{0.75cm}  
	\vcenter{\hbox{ \begin{tikzpicture} \begin{feynman} \vertex (a)
					; \vertex [right =0.75of a]
					(b1'); \vertex [right =0.25of
					b1'] (b1); \vertex [above right
					=0.25of b1] (b2); \vertex [below
					right =0.25of b1] (b3); \vertex
					[right =1.5of b1] (c1); \vertex
					[above left =0.25of c1] (c2);
					\vertex [below left =0.25of c1]
					(c3); \vertex [right =0.25of c1]
					(c1'); \vertex [right =0.75of
					c1'] (d); \diagram*{ (a) --
						[scalar,edge
						label={\tiny{6}}] (b1'),
						(b1') -- (b1), (b1) --
						(b2), (b1) -- (b3), (c1)
						-- (c2), (c1) -- (c3),
						(c1) -- (c1'), (c1')
						--[scalar,edge
						label={\tiny{3}}] (d),
						(b2) -- [scalar,half
	left,looseness=1.2,edge label={\tiny{2}}] (c2), (b3) --
	[scalar,half right,looseness=1.2,edge label={\tiny{1}}] (c3), };
	\end{feynman} \end{tikzpicture} }} 
	~\hspace{0.75cm} \frac{6\times 2\times 3}{2!(3!)^2} =
	\frac{1}{2} ~ \mbox{and} \\
	& & \mbox{The}~ 1 ~ \mbox{comes from:} ~ 
	\hspace{1.25cm} \vcenter{\hbox{ \begin{tikzpicture}
				\begin{feynman} \vertex (v); \vertex
					[right =0.25of v] (a); \vertex
					[right =0.75of a] (b); \vertex
					[right =0.25of b, crossed dot]
					(b'){ }; \vertex [right =0.75of
					b'] (c); \vertex [right =0.75of
				c] (d); \vertex [right =0.25of d] (d');
			\diagram*{ (v) -- (a) -- [scalar,edge
		label={\tiny{2}}] (b), (b) -- (b') -- (c), (c) --
[scalar,edge label={\tiny{1}}] (d), (d) -- (d'), }; \end{feynman}
\end{tikzpicture} }} \hspace{0.75cm}
	~ \frac{2\times 1}{1!2!} = 1 \\
	& & \delta_{\Phi}, \delta_m ~ \mbox{to be determined from} ~
	\Pi(-m^2) = 0 = \pi'(-m^2)\ .
\end{eqnarray*}
The first term, after using the Feynman parameters and Wick rotation and
equations (\ref{DimensionalAngularIntgral},
\ref{DimensionalRadialIntgral}, \ref{DimensionalMomIntgral}) gives,
\begin{eqnarray*}
\circled{1} & = & \frac{g^2}{2} I(p^2) ~ , ~ I(p^2) := i\int_0^1 dx\int
\frac{d^d \underline{k}}{(2\pi)^d}\frac{1}{(\underline{k}^2 + M^2)^2}~ ,
~ \boxed{M^2 := x(1-x)\underline{p}^2 + m^2} \\
I(p^2) & = & \frac{\Gamma(-1+\epsilon/2)}{(4\pi)^3} \int_0^1dx\ M^2
(4\pi/M^2)^{\epsilon/2} ~ ~ , ~ ~ \epsilon := 6-d,~ \mbox{put}~ g\to
g(\bar{\mu})^{\epsilon/2}\, \alpha := \frac{g^2}{(4\pi)^3} \\
\circled{2} & = & -i (\delta_{\Phi} p^2 + \delta_m m^2)  \\
\therefore \Pi(p^2) & = &
-\frac{\alpha}{2}\left[(1+\frac{2}{\epsilon})(\frac{p^2}{6}+m^2) +
\int_0^1dx\ M^2 ln
\left(\frac{4\pi\bar{\mu}^2}{e^{\gamma}M^2}\right)\right]
-\delta_{\Phi} p^2 - \delta_m m^2 + o(\alpha^2).
\end{eqnarray*}
Putting $\boxed{\mu := \sqrt{4\pi}e^{-\gamma}\bar{\mu}}$, the logarithm
becomes $ln(\mu^2/M^2)$. We continue the Euclidean momenta back to
Minkowski momenta, $\underline{p}^2 \to p^2$. Club the first group of
terms with the counter terms. This gives the self energy as,
\cite{Srednicky} 
\begin{eqnarray} \label{ScalarSelfEnergy}
	\Pi(p^2) & = & \frac{\alpha}{2}\int_0^1dx\ M^2 ln(M^2/\mu^2)
	\nonumber \\
	& & - \left\{ \frac{\alpha}{6}\left(\frac{1}{\epsilon} +
	\frac{1}{2}\right) + \delta_{\Phi}\right\} p^2 
	- \left\{ \alpha\left(\frac{1}{\epsilon} + \frac{1}{2}\right) +
	\delta_{m}\right\} m^2  + o(\alpha^2)
\end{eqnarray}
Now, in the first term, use $ln(M^2/\mu^2) = ln(M^2/m^2) +
ln(m^2/\mu^2)$ and absorb the contribution of the second term into the
counter term coefficients along with the UV divergences (the
$\epsilon^{-1}$ pole) by {\em choosing},
\begin{equation}
\delta_{\Phi} := - \frac{\alpha}{6}\left(\frac{1}{\epsilon} +
\frac{1}{2} + C_{\Phi}\right) + o(\alpha^2) ~ ~ , ~ ~ 
\delta_m := - \alpha\left(\frac{1}{\epsilon} + \frac{1}{2} + C_m \right)
+ o(\alpha^2) \ .
\end{equation}
Note that we have only absorbed the pole together with some, $\mu-${\em
dependent finite parts} (the $ln(m^2/\mu^2)$ terms) and subsumed these
in the undetermined, finite constants $C_{\Phi}, C_m$. These constants
are determined by the renormalization conditions. The self-energy then
takes the form,
\begin{equation}
	\Pi(p^2) = \frac{\alpha}{2}\int_0^1dx\ M^2\ln(M^2/m^2) +
	\frac{\alpha}{6}C_{\Phi}p^2 + \alpha C_m m^2 + o(\alpha^2).
\end{equation} 
The $C$'s are now determined by imposing $\Pi(-m^2) = 0 = \Pi'(-m^2)$.
This leads to,
\begin{eqnarray}
	C_m - \frac{C_{\Phi}}{6} & = & \int_0^1dx\
	\frac{M^2_0}{m^2}ln(M^2_0/m^2) ~ ~ ~ , ~ ~ ~ \frac{M^2_0}{m^2} =
	1-x+x^2 \ , \\
	-\frac{C_{\Phi}}{6} & = & \frac{\alpha}{2}\int_0^1dx\
	x(1-x)\left\{ ln(M^2_0/m^2) + 1 \right\} \ . 
\end{eqnarray}
Notice that the $C$'s so determined, {\em are independent of the
arbitrary parameter} $\mu$ that entered in the dimensional
regularization scheme. We have expressed the 2-point vertex function
explicitly free of UV divergence {\em and} the arbitrary $\mu$
parameter. In $\Pi(p^2)$ we had two terms with the $\epsilon^{-1}$ pole,
one with coefficient $p^2$ and one with coefficient $m^2$. The two
counter terms $\delta_{\Phi}, \delta_m$ sufficed to absorbed these. 

Consider now the 3-point vertex function, $\Gamma(p_1, p_2, p_3)$:
\begin{eqnarray*}
	i\Gamma_3 & : & 
\vcenter{\hbox{
\begin{tikzpicture}
	\begin{feynman}
		\vertex [blob] (v){ };
		\vertex [above=0.85of v] (a);	
		\vertex [below left=0.85of v] (b);	
		\vertex [below right=0.85of v] (c);	
		\diagram*{
			(a) --[fermion,edge label=\(p_1\)] (v),
			(b) --[fermion,edge label=\(p_2\)] (v),
			(c) --[fermion,edge label'=\(p_3\)] (v),
		};
	\end{feynman}
\end{tikzpicture}
}}	 
~ := ~ 
\vcenter{\hbox{
\begin{tikzpicture}
	\begin{feynman}
		\vertex (v);
		\vertex [above=0.85of v] (a);	
		\vertex [below left=0.85of v] (b);	
		\vertex [below right=0.85of v] (c);	
		\diagram*{
			(a) --[fermion,edge label=\(p_1\)] (v),
			(b) --[fermion,edge label=\(p_2\)] (v),
			(c) --[fermion,edge label'=\(p_3\)] (v),
		};
	\end{feynman}
\end{tikzpicture}
}}
~ ~ + ~ ~ 
\vcenter{\hbox{
\begin{tikzpicture}
	\begin{feynman}
		\vertex (v);
		\vertex [above=0.35of v] (a);
		\vertex [above=0.5of a] (a');
		\vertex [below left=0.35of v] (b);
		\vertex [below left=0.85of v] (b');
		\vertex [below right=0.35of v] (c);
		\vertex [below right=0.85of v] (c');
		\diagram*{
			(a') --[fermion,edge label=\(p_1\)] (a),
			(b') --[fermion,edge label=\(p_2\)] (b),
			(c') --[fermion,edge label'=\(p_3\)] (c),
			(a) --[fermion,half left,looseness=0.9,edge
			label=k] (c),
			(c) --[fermion,half left,looseness=0.9] (b),
			(b) --[fermion,half left,looseness=0.9] (a),
		};
	\end{feynman}
\end{tikzpicture}
}}
~ ~ + ~ ~ 
\vcenter{\hbox{
\begin{tikzpicture}
	\begin{feynman}
		\vertex [crossed dot] (v){ };
		\vertex [above=0.85of v] (a);	
		\vertex [below left=0.85of v] (b);	
		\vertex [below right=0.85of v] (c);	
		\diagram*{
			(a) --[fermion,edge label=\(p_1\)] (v),
			(b) --[fermion,edge label=\(p_2\)] (v),
			(c) --[fermion,edge label'=\(p_3\)] (v),
		};
	\end{feynman}
\end{tikzpicture}
}}
~ ~ + ~ ~  \cdots 
\end{eqnarray*}
At 1-loop we have,
\begin{eqnarray*}
	i\delta\Gamma_3 & : & 
\vcenter{\hbox{
\begin{tikzpicture}
	\begin{feynman}
		\vertex [crossed dot] (v){ };
		\vertex [above=0.85of v] (a);	
		\vertex [below left=0.85of v] (b);	
		\vertex [below right=0.85of v] (c);	
		\diagram*{
			(a) --[fermion,edge label=\(p_1\)] (v),
			(b) --[fermion,edge label=\(p_2\)] (v),
			(c) --[fermion,edge label'=\(p_3\)] (v),
		};
	\end{feynman}
\end{tikzpicture}
}}
~ ~ + ~ ~
\vcenter{\hbox{
\begin{tikzpicture}
	\begin{feynman}
		\vertex (v);
		\vertex [above=0.75of v] (a);
		\vertex [above=0.5of a] (a');
		\vertex [below left=0.75of v] (b);
		\vertex [below left=1.25of v] (b');
		\vertex [below right=0.75of v] (c);
		\vertex [below right=1.25of v] (c');
		\diagram*{
			(a') --[fermion,edge label=\(p_1\)] (a),
			(b') --[fermion,edge label=\(p_2\)] (b),
			(c') --[fermion,edge label'=\(p_3\)] (c),
			(a) --[fermion,half left,looseness=0.8,edge
			label=k] (c),
			(c) --[fermion,half left,looseness=0.8,edge
			label=\(p_2+k\)] (b),
			(b) --[fermion,half left,looseness=0.8,edge
			label=\(k-p_1\)] (a),
		};
	\end{feynman}
\end{tikzpicture}
}} \\
& = & [ig\delta_3]_{\circled{1}} \hspace{1.0cm} + \hspace{1.0cm}
(ig)^3(-i)^3 \times\\
	& & \hspace{1.5cm} \int \frac{d^dk}{(2\pi)^d} \left[
	\frac{1}{k^2+m^2-i\epsilon} \frac{1}{(p_2+k)^2+m^2-i\epsilon}
\frac{1}{(k-p_1)^2+m^2-i\epsilon}\right]_{\circled{2}}
\end{eqnarray*}
In the second term, Feynman parameterization, shifting momentum and Wick
rotation gives,
\[
	\left[\dots\right] = 2\int dx\int dy\int
	dz\frac{\delta(1-x-y-z)}{(k^2 + M^2)^3} ~ ~ , ~ ~ \boxed{M^2 :=
	zxp_1^2 + zyp_2^2+xyp_3^2 + m^2} \ .
\]
As before replacing $g\to g\bar{\mu}^{\epsilon/2}, \epsilon = 6-d$ and
using the (\ref{DimensionalMomIntgral}) gives,
\[
	\int\frac{d^dk}{(2\pi)^d}\frac{1}{(\underline{k}^2+M^2)^3} =
	\frac{\Gamma(3-d/2)}{2(4\pi)^{d/2}}M^{-(3-d/2)} \ .
\]
Hence, putting $\alpha := g^2/(4\pi)^3, \mu^2 := 4\pi
e^{-\gamma}\bar{\mu}^2 $
\begin{eqnarray}
\frac{\delta \Gamma(p_1,p_2,p_3)}{g} & = & \delta_g + \frac{\alpha}{2}
\left[\frac{2}{\epsilon} + 2\int dx\ dy\ dz\delta(1-x-y-z)
\left(\frac{4\pi\bar{\mu}^2}{M^2}\right)^{\epsilon/2}\right] ; \\
& = & \left\{\frac{\alpha}{\epsilon} + \delta_g\right\} \nonumber \\
& & \hspace{1.0cm} - \alpha\int dx dy dz\delta(1-x-y-z)ln (M^2/m^2) +
o(\alpha^2) \nonumber\\
\mbox{Choosing} & & \boxed{\delta_g := -\frac{\alpha}{\epsilon} - \alpha
C_g + o(\alpha^2) } ~ ~ \Rightarrow \\
\Gamma(p_1,p_2,p_3) & = & g\left[1 - \alpha \int dx\ dy\
dz\delta(1-x-y-z)ln(M^2/m^2) - \alpha C_g + o(\alpha^2) \right] ~ ~ ~ ~ 
\end{eqnarray}
As before, we have obtained a completely finite and $\mu-$independent
vertex function with one undetermined constant$C_g$. This is to be fixed
by a suitable renormalization condition.  What condition do we choose?

In QED we had a natural choice $\Gamma^{\mu}(q^2\to 0) = e_{ph}$.  Here
however, there is no natural choice.  {\em Any} cross-section will
involve both $g$ and $C_g$ and the value of a cross-section gives one
condition.  A {\em convenient} choice is $\Gamma_3(0,0,0) = g
\leftrightarrow C_g = 0$.

Thus, at the 1-loop level, we see how the counter terms absorb the
divergence and how the renormalization conditions determine the constants
$C$'s. Note that we could have added any finite constant to the pole in
defining the counter terms. The renormalization conditions then would
give different expressions for the constants.
\subsubsection{At 2-Loops} \label{2Loops}
A natural question is: How does this work at higher orders? In our
$(\Phi^3)_6$ theory, only the 2-point and 3-point vertex functions are
divergent. At 2-loops the contributing diagrams are:
%
\begin{center}
\begin{tabular}{|l|l|l|}
	\hline
	& & \\
	&
	$
	\vcenter{\hbox{
	\begin{tikzpicture}
		\begin{feynman}
			\vertex (v);
			\vertex [right=0.75of v] (a);
			\vertex [right=0.5of a] (c0);
			\vertex [above left=0.5of c0] (c1);
			\vertex [right=0.5of c0] (d);
			\vertex [right=0.75of d] (e);

			\diagram*{
				(v) --[fermion,edge label=\(p\)] (a),
				(d) --[fermion,edge label=\(p\)] (e),
			};
			\draw (a) arc [start angle=180, end angle=0,
			radius=0.5cm];
			\draw (d) arc [start angle=0, end angle=-180,
			radius=0.5cm];
			\draw (c1) arc [start angle=225, end angle=-45,
			radius=0.5cm];
		\end{feynman}
	\end{tikzpicture}
	}}
	$
	~ + ~ 
	$
	\vcenter{\hbox{
	\begin{tikzpicture}
		\begin{feynman}
			\vertex (v);
			\vertex [right=0.75of v] (a);
			\vertex [right=0.5of a] (c0);
			\vertex [above=0.35of c0,crossed dot] (c1){ };
			\vertex [above =0.5of c1]
			(c1'){\( (\delta_{\phi})_{1}\)};
			\vertex [right=0.5of c0] (d);
			\vertex [right=0.75of d] (e);
	
			\diagram*{
				(v) -- (a),
				(d) -- (e),
			};
			\draw (a) arc [start angle=180, end angle=0,
			radius=0.5cm];
			\draw (d) arc [start angle=0, end angle=-180,
			radius=0.5cm];
		\end{feynman}
	\end{tikzpicture}
	}}
	$
	~ + ~ 
	$
	\vcenter{\hbox{
	\begin{tikzpicture}
		\begin{feynman}
			\vertex (v);
			\vertex [right=0.75of v] (a);
			\vertex [right=0.5of a] (c0);
			\vertex [above=0.35of c0,crossed dot] (c1){ };
			\vertex [above =0.5of c1]
			(c1'){\( (\delta_{m})_{1}\)};
			\vertex [right=0.5of c0] (d);
			\vertex [right=0.75of d] (e);
	
			\diagram*{
				(v) -- (a),
				(d) -- (e),
			};
			\draw (a) arc [start angle=180, end angle=0,
			radius=0.5cm];
			\draw (d) arc [start angle=0, end angle=-180,
			radius=0.5cm];
		\end{feynman}
	\end{tikzpicture}
	}}
	$
	~ + ~ $\cdots$
	& I \\ 
	& & \\ \cline{2-2}
	& & \\
	$2-$point & 
	$
	\vcenter{\hbox{
	\begin{tikzpicture}
		\begin{feynman}
			\vertex (v);
			\vertex [right=0.75of v] (a);
			\vertex [right=0.5of a] (c0);
			\vertex [above =0.5of c0] (c1);
			\vertex [below =0.5of c0] (c2);
			\vertex [right=0.5of c0] (d);
			\vertex [right=0.75of d] (e);

			\diagram*{
				(v) -- (a),
				(d) -- (e),
				(c1) -- (c2),
			};
			\draw (a) arc [start angle=180, end angle=0,
			radius=0.5cm];
			\draw (d) arc [start angle=0, end angle=-180,
			radius=0.5cm];
		\end{feynman}
	\end{tikzpicture}
	}}
	$
	~ + ~ 
	$
	\vcenter{\hbox{
	\begin{tikzpicture}
		\begin{feynman}
			\vertex (v);
			\vertex [right=0.75of v,crossed dot] (a) { };
			\vertex [below=0.75of a] (a') {\( (\delta_3)_1\) };
			\vertex [right=0.5of a] (c0);
			\vertex [right=0.5of c0] (d);
			\vertex [right=0.75of d] (e);
	
			\diagram*{
				(v) -- (a),
				(d) -- (e),
			};
			\draw (a) arc [start angle=180, end angle=0,
			radius=0.5cm];
			\draw (d) arc [start angle=0, end angle=-180,
			radius=0.5cm];
		\end{feynman}
	\end{tikzpicture}
	}}
	$
	~ + ~ 
	$
	\vcenter{\hbox{
	\begin{tikzpicture}
		\begin{feynman}
			\vertex (v);
			\vertex [right=0.75of v] (a);
			\vertex [right=0.5of a] (c0);
			\vertex [right=0.35of c0, crossed dot] (d){ };
			\vertex [right=0.75of d] (e);
			\vertex [below =0.75of d]
			(d'){\( (\delta_3)_1\)};
	
			\diagram*{
				(v) -- (a),
				(d) -- (e),
			};
			\draw (a) arc [start angle=180, end angle=0,
			radius=0.5cm];
			\draw (d) arc [start angle=0, end angle=-180,
			radius=0.5cm];
		\end{feynman}
	\end{tikzpicture}
	}}
	$
	~ + ~ $\cdots$
	& II \\ 
	& & \\ \cline{2-2}
	& & \\
	& 
	$
	\vcenter{\hbox{
	\begin{tikzpicture}
		\begin{feynman}
			\vertex (v);
			\vertex [right=1.0of v, crossed dot] (a) { };	
			\vertex [right=1.25of a] (b) { };	
			\vertex [below=0.5of a] (c) {\(
			(\delta_{\phi})_2\) };	
			\diagram*{
				(v) -- (a) -- (b),
			};
		\end{feynman}
	\end{tikzpicture}
	}}
	$
	~ + ~ 
	$
	\vcenter{\hbox{
	\begin{tikzpicture}
		\begin{feynman}
			\vertex (v);
			\vertex [right=1.0of v, crossed dot] (a) { };	
			\vertex [right=1.25of a] (b) { };	
			\vertex [below=0.5of a] (c) {\(
			(\delta_{m})_2\) };	
			\diagram*{
				(v) -- (a) -- (b),
			};
		\end{feynman}
	\end{tikzpicture}
	}}
	$
	~ + ~ $\cdots$
	& III \\
	& & \\ 
	\hline
	& & \\
	$3-$point & 
	$
	\vcenter{\hbox{
	\begin{tikzpicture}
		\begin{feynman}
			\vertex (v);
			\vertex[right=0.75of v] (a);
			\vertex[above right=0.5of a] (b);
			\vertex[above right=0.5of b] (c);
			\vertex[above right=0.5of c] (d);
			\vertex[below right=0.5of a] (b');
			\vertex[below right=0.5of b'] (c');
			\vertex[below right=0.5of c'] (d');
	
			\diagram*{
				(v) -- (a),
				(a)--(b)--(c)--(d),
				(a)--(b')--(c')--(d'),
				(b)--(b'), (c)--(c'),
			};
		\end{feynman}
	\end{tikzpicture}
	}}
	$
	~ + ~ 
	$
	\vcenter{\hbox{
	\begin{tikzpicture}
		\begin{feynman}
			\vertex (v);
			\vertex[right=0.75of v] (a);
			\vertex[above right=0.5of a] (b);
			\vertex[above right=0.5of b] (c);
			\vertex[above right=0.5of c] (d);
			\vertex[below right=0.5of a] (b');
			\vertex[below right=0.5of b'] (c');
			\vertex[below right=0.5of c'] (d');
	
			\diagram*{
				(v) -- (a),
				(a)--(b)--(c)--(d),
				(a)--(b')--(c')--(d'),
				(b)--(c'), (c)--(b'),
			};
		\end{feynman}
	\end{tikzpicture}
	}}
	$
	~ + ~ 
	$
	\vcenter{\hbox{
	\begin{tikzpicture}
		\begin{feynman}
			\vertex (v);
			\vertex[right=0.75of v] (a);	
			\vertex[above right=0.8of a] (b);	
			\vertex[above right=0.5of b] (b');	
			\vertex[below right=0.8of a] (c);	
			\vertex[below right=0.5of c] (c');	
			\vertex[below=0.4of b,crossed dot] (d){ };
			\vertex[right=0.85of d] (d'){\(
			(\delta_{\phi,m})_1\) };

			\diagram*{
				(v)--(a)--(b)--(b'),
				(a)--(c)--(c'),
				(b)--(d)--(c),

			};
		\end{feynman}
	\end{tikzpicture}
	}}
	$
	~ + ~ 
	$
	\vcenter{\hbox{
	\begin{tikzpicture}
		\begin{feynman}
			\vertex (v);
			\vertex[right=0.75of v,crossed dot] (a){ };
			\vertex[above right=0.8of a] (b);
			\vertex[above right=0.5of b] (c);
			\vertex[below right=0.8of a] (b');
			\vertex[below right=0.5of b'] (c');
	
			\diagram*{
				(v) -- (a) -- (b) -- (c),
				(a) -- (b') -- (c'),
				(b) -- (b'),
			};
		\end{feynman}
	\end{tikzpicture}
	}}
	$
	~ + ~ $\cdots$
	& IV \\
	& & \\
	\hline
\end{tabular}
\end{center}
The counter terms have the same form as the original bare terms. When
they are clubbed together into the `interaction' Lagrangian or
Hamiltonian, one would expect that from $H_{int}^k/k!$ terms, we would
get the counter vertices also appearing $k$ times. For example, at
$o(g^2)$, we would have 
$
\vcenter{\hbox{
\begin{tikzpicture}
	\begin{feynman}
		\vertex (v1);
		\vertex [right=1.5of v1] (v2);
		\vertex [left=0.5of v1] (a);
		\vertex [above right=0.5of v1] (b);
		\vertex [below right=0.5of v1] (c);
		\vertex [right=0.5of v2] (d);
		\vertex [above left=0.5of v2] (e);
		\vertex [below left=0.5of v2] (f);

		\diagram*{
			(v1) --(a), (v1) --(b), (v1)--(c),
			(v2) --(d), (v2) --(e), (v2)--(f),
		};
	\end{feynman}
\end{tikzpicture}
}}
$
and 
$
\vcenter{\hbox{
\begin{tikzpicture}
	\begin{feynman}
		\vertex [crossed dot] (v){ };
		\vertex [left=0.5of v] (a) ;
		\vertex [right=0.5of v] (b) ;
		\vertex [right=1.5of v,crossed dot] (v'){ };
		\vertex [left=0.5of v'] (a') ;
		\vertex [right=0.5of v'] (b') ;

		\diagram*{
			(a) --(v) -- (b),
			(a')--(v')--(b'),
		};
	\end{feynman}
\end{tikzpicture}
}}
$
~ , ~ 
$
\vcenter{\hbox{
\begin{tikzpicture}
	\begin{feynman}
		\vertex [blob] (v1){ };
		\vertex [right=of v1,blob] (v2){ };
		\vertex [left=0.75of v1] (a);
		\vertex [above right=0.75of v1] (b);
		\vertex [below right=0.75of v1] (c);
		\vertex [right=0.75of v2] (d);
		\vertex [above left=0.75of v2] (e);
		\vertex [below left=0.75of v2] (f);

		\diagram*{
			(v1) --(a), (v1) --(b), (v1)--(c),
			(v2) --(d), (v2) --(e), (v2)--(f),
		};
	\end{feynman}
\end{tikzpicture}
}}
$
vertices. In vertex functions, the 1PR diagrams from the
self-energy counter terms are omitted. Secondly, the coefficients of the
counter terms are explicitly instructed to be adjusted to enforce the
renormalization conditions at any given order in $g$. Hece, the counter
vertices are {\em not} on par with the elementary vertices. The order of
perturbation is determined by the number of elementary vertices and not
by counter vertices.

In the first of the group of diagrams, we have a subdiagram that is
divergent (only $|k_1| \to \infty, \ k_2$ remains finite). This
divergence is absorbed by the 1-loop counter vertices - the seond and
the third diagrams of group I. To absorb the divergence when both $k_1,
k_2$ loop momenta become large, we need the {\em new 2-loop counter
term} as shown in III.

The group II exhibit the so called ``overlapping divergences'': the
vertical line's momentum diverges when either of the loop momenta
diverge. There are two subdiagrams which are divergent and to absorb
these, we need  the $\delta_3$ 1-loop counter term. The group III
counter terms are needed to absorb the divergence coming from both
momenta becoming large.

Ar higher loops, the procedure is thus recursive. The counter vertices
themselves have an expansion in $g^2$ with coefficients absorbing
divergences from the subdiagrams (lower loop orders). The
renormalization conditions determine the highest loop order
coefficients. 

This explains how the divergences arise and how they are absorbed
systematically via the counter terms and the renormalization conditions.

We have taken for illustration the on-shell renormalization condition.
This method fails when the physical masses vanish and the procedure
needs to be adopted suitably. This impacts what exactly is subtracted -
(divergent part) or (divergent part + a finite piece)? This leads to
different {\em schemes of renormalization}. Let us see how the probem
arises and how different renormalization conditions can be chosen. We
continue with the $(\Phi^3)_6$ theory.
\subsection{Renormalization with massless particles}
We have the self energy in equation (\ref{ScalarSelfEnergy}). Taking
derivative w.r.t. $p^2$ gives,
\[
\Pi'(p^2) = -\left[\delta_{\Phi} + \frac{\alpha}{6}
\left(\frac{1}{\epsilon} + \frac{1}{2}\right)\right] +
\frac{\alpha}{2}\int_0^1dx\ x(1-x)\{ln(M^2/\mu^2)+1\} + o(\alpha^2) \ .
\]
As $m \to 0, M^2 \to x(1-x)p^2$. Then $\Pi(0) = 0$ identically while
$\Pi'(0)$ is ill-defined (depends on how $p^2, m^2 \to 0$ is taken).
Thus the previous renormalization conditions {\em cannot} be used and
the counter term coefficients would remain undetermined! We noted while
discussing the Kallen-Lehmann representation, that the physical spectrum
was assumed to have a mass gap i.e.  single particle pole being
separated from the multi-particle branch point. Whenever this is
violated, the above {\em on-shell renormalization procedure fails}.

All is not lost though. We can try different renormalization conditions.

There are two commonly used {\em subtraction schemes} within dimensional
regularization - the so called {\em Minimal Subtraction scheme (MS)} and
the {\em Modified Subtraction Scheme ($\overline{MS}$)}. They are defined as:

$\boxed{MS \leftrightarrow \mbox{define the counter term constants by
absorbing only the $\epsilon$ pole(s)}; }$ while the $\overline{MS}$
scheme uses $\mu^2 := 4\pi e^{-\gamma}\bar{\mu}^2$ in the MS scheme.

With these definitions, we give the finite self-energy in the three
schemes:
\begin{eqnarray}
	\Pi_{on-shell}(p^2) & = & -\frac{\alpha}{12}(p^2 + m^2) +
	\frac{\alpha}{2}\int_0^1dx\ M^2ln(M^2/M_0^2) + o(\alpha^2)  ,
	\nonumber \\
	\mbox{where,}& & M^2 = x(1-x)p^2+ m^2 ~ \mbox{and} ~ M^2_0 :=
	M^2(p^2=-m^2) \\
	\Pi_{MS}(p^2) & = & -\frac{\alpha}{12}(p^2+m^2) +
	\frac{\alpha}{2}\int_0^1dx M^2ln(M^2/\bar{\mu}^2) + o(\alpha^2)
	\\
	\Pi_{\overline{MS}}(p^2) & = & -\frac{\alpha}{12}(p^2+m^2) +
	\frac{\alpha}{2}\int_0^1dx M^2ln(M^2/\mu^2) + o(\alpha^2) 
\end{eqnarray}

The $\Pi_{on-shell}$ is ill-defined as $m\to 0$ as noted above. For
non-zero $m$ though, it is unambiguous, free of the arbitrary scale
$\mu$ and $m$ is also the physical mass.

By contrast, $\Pi_{\overline{MS}}$ is well defined as $m\to 0$, but it
depends on the mathematical artifact parameter $\mu$! Now the $m$
parameter {\em cannot} be the pole of the propagator and hence is {\em
not} the physical mass. The physical mass is determined from $[p^2+m^2 -
\Pi(p^2)]_{p^2=-m^2_{ph}} = 0 \leftrightarrow m^2_{ph} = m^2 -
\Pi_{\overline{MS}}(-m^2_{ph})$. Since $\Pi$ is order $\alpha$, we can
write, to order $\alpha$, $\boxed{m^2_{ph} = m^2 -
\Pi_{\overline{MS}}(-m^2)}$.  Substitution gives,
\begin{eqnarray}
	m^2_{ph} & = & m^2 + \frac{\alpha}{12}m^2(1-6) +
	\frac{\alpha}{2}\int_0^1dx\ m^2(1-x+x^2)ln( (1-x+x^2)m^2/\mu^2)
	\nonumber \\
	or & & \boxed{m^2_{ph} = m^2\left[1 + \frac{5}{12}\alpha\left\{
	ln\frac{\mu^2}{m^2} + C'\right\} + o(\alpha^2)\right] ~ ~ , ~ ~
C' = \frac{34-3\pi\sqrt{3}}{15} \approx 1.18} \hspace{1.0cm}
\end{eqnarray}
Observe that the physical mass explicitly depends on $\mu$ which it
should not. Hence $\boxed{\Case{d}{d\mu^2} m^2_{ph}(m^2,\alpha,\mu) =
0}$ must hold. Clearly this would be possible if $\alpha$ and/or $m$
also have $\mu$ dependence. The independence of the physical mass then
relates the $\mu$ dependences of $m, \alpha$.

The residue, $R$, at the physical pole is as before, $R^{-1} =
\Case{\Delta'_F}{dp^2}|_{p^2 = -m^2_{ph}} = 1 -
\Pi'_{\overline{MS}}(-m^2_{ph}) = 1 - \Pi'_{\overline{MS}}(-m^2) +
o(\alpha^2)$. This evaluates to,
\begin{equation}
	\boxed{R^{-1} = 1 + \frac{\alpha}{12}\left\{ln(\mu^2/m^2) +
	C''\right\} ~ , ~ C'' = \frac{17-3\pi\sqrt{3}}{\sqrt{3}} \approx
0.23\ . }
\end{equation}
The residue gives the wave function renormalization constant $Z_{\Phi} =
R$.

Finally, in the vertex function, we choose $\delta_g = -\alpha/\epsilon$
in the $\overline{MS}$ scheme. This leads to,
\begin{eqnarray}
\Gamma_{3,\overline{MS}}(p_1, p_2, p_3) & = & g\left[1 - \alpha \int dx\
dy\ dz\delta(-x-y-z)ln(M^2/\mu^2) + o(\alpha^2)\right] , ~ \mbox{where,}
\nonumber \\
& & \hspace{1.0cm} M^2 = xyp_1^2 + yzp_2^2 + zxp_3^2 + m^2 . 
\end{eqnarray}

To see a possible $\mu-$dependence of the coupling $\alpha$, we need to
identify a {\em physical quantity} which has explicit $\mu$ dependence.
As discussed by Srednicky \cite{Srednicky}, we consider the  two
particle going to two particle process which is related to the $4-$point
vertex function which is UV finite. In the limit $m\to 0$, this process
too suffers from IR divergence which is handled as in QED - carefully
considering what is observed and including the soft particles
contribution. From \cite{Srednicky}, we borrow the formulae below.

The amplitude for the process, $p_1+ p_2 \to p_3+p_4$ is depicted below
to order $\alpha$. 
\begin{eqnarray*}
\vcenter{\hbox{
\begin{tikzpicture}
	\begin{feynman}
		\vertex (a);
		\vertex [above left=0.75of a](a');
		\vertex [right=of a] (b);
		\vertex [above right=0.75of b](b');
		\vertex [below=of b] (c);
		\vertex [below right=0.75of c](c');
		\vertex [below=of a] (d);
		\vertex [below left=0.75of d](d');

		\diagram*{
			(a) -- (b) -- (c) -- (d) -- (a),
			(a')--[fermion,edge label=\(p_1\)] (a),
			(b)--[fermion,edge label=\(p'_1\)] (b'),
			(c)--[fermion,edge label=\(p'_2\)] (c'),
			(d')--[fermion,edge label=\(p_2\)] (d),
		};
	\end{feynman}
\end{tikzpicture}
}}
\hspace{1.0cm} & , & \hspace{1.0cm}
\vcenter{\hbox{
\begin{tikzpicture}
	\begin{feynman}
		\vertex (a);
		\vertex [above left=0.75of a](a');
		\vertex [above =0.35of a'](a1);
		\vertex [left =0.35of a'](a2);

		\vertex [right=of a] (b);
		\vertex [above right=0.75of b](b');
		\vertex [above =0.35of b'](b1);
		\vertex [right=0.35of b'](b2);

		\vertex [below=of b] (c);
		\vertex [below right=0.75of c](c');
		\vertex [right=0.35of c'](c1);
		\vertex [below=0.35of c'](c2);

		\vertex [below=of a] (d);
		\vertex [below left=0.75of d](d');
		\vertex [left=0.35of d'](d1);
		\vertex [below=0.35of d'](d2);

		\diagram*{ (a) -- (b) -- (c) -- (d) -- (a),
			(a')--[fermion,edge label=\(p_1\)] (a),
			(b)--[fermion,edge label=\(p'_1\)] (b'),
			(c)--[fermion,edge label=\(p'_2\)] (c'),
			(d')--[fermion,edge label=\(p_2\)] (d), (a1)
			--[scalar] (a') -- [scalar](a2), (b1) --[scalar]
			(b') -- [scalar](b2), (c1) --[scalar] (c') --
			[scalar](c2), (d1) --[scalar] (d') --
			[scalar](d2), }; \end{feynman} \end{tikzpicture}
	}} \\
(a) \hspace{2.5cm} & & \hspace{2.5cm} (b) 
\end{eqnarray*}
The contribution of $(a)$ in the high energy limit, is given by,
\begin{eqnarray*}
\mathcal{M} & = & \mathcal{M}_0\left[ 1 - \frac{11}{12}\alpha\left(
ln(s/m^2) + o(m^0)\right) + o(\alpha^2)\right] ~ ~ \mbox{with,} \\
\mathcal{M}_0 & = & -g^2\left(\frac{1}{s} + \frac{1}{t} +
\frac{1}{u}\right) ~ ~ , ~ s := -(p_1+p_2)^2,\ t = -(p'_1-p_1)^2,\ u :=
-(p'_2-p_1)^2 .
\end{eqnarray*}
The $o(m^0)$ term is free of $lm(m\to 0)$ singularity. The amplitude
$\mathcal{M}$ is divergent as $m\to 0$ (see equation (26.1) of Srednicky
\cite{Srednicky}.)

In the limit $m\to 0$, the above process is experimentally
indistinguishable from the one in which there are on-shell, {\em
collinear} particles associated with any/all external lines\footnote{For
	mass less particles, $(\sum_i p_i)^2 = \sum_{i\neq j}p_i\cdot
p_j = \sum_{ij} |p_i| |p_j| (-1+cos(\theta_{ij}))$ which vanishes if all
particles are collinear, $\theta_{ij} = 0$.}. This is shown in $(b)$
above, with the dotted lines denoting the collinear particles. We must
integrate the squared amplitude over momenta collinear within
detector's angular resolution. 

For observable cross-section with collinear emission included (from all
4 external lines), we have
\begin{eqnarray*}
|\mathcal{M}|^2_{obs} & = & |\mathcal{M}|^2\left[1 +
\frac{4\alpha}{12}\left( ln\left(\frac{\Delta^2 \vec{k}^2}{m^2}\right) +
C\right) + \dots \right]  ~ , ~ C := 4 - 3\sqrt{3}\pi ~ , ~ \vec{k}^2 =
s/4 , \\
& = & |\mathcal{M}_0|^2\left[1 - \frac{11}{6}\alpha\ ln(s/m^2)
\right]\times\left[1+\frac{\alpha}{3} \left( ln(\Delta^2 s/(4m^2)) +
C\right) + \dots \right] \\
& = & |\mathcal{M}_0|^2\left[1 - \alpha\left\{ \frac{3}{2}ln(s/m^2) +
\frac{1}{3}ln(1/\Delta^2) + o(m^0)\right\} + o(\alpha^2)\right].
\mbox{{\bf 26.15} of \cite{Srednicky} }
\end{eqnarray*}
The $\Delta$ is related to the detector angular resolution. In going to
the second equation, we have included collinear emission from all 4
lines. There are no UV divergences here. The above formulae implicitly
assume on-shell renormalization scheme where we cannot take the $m\to 0$
limit.

The same computation of the amplitude can be done in the $\overline{MS}$
scheme. This changes the amplitude $\mathcal{M}$ by changing the
$ln(s/m^2) \to ln(s/\mu^2)$. The correction due to inclusion of
collinear emission continues to have $ln(\Delta^2s/(4m^2))$.
Additionally, the residue at the physical pole is not 1 but $R$ and
hence the amplitude is multiplied by $R^2$. Including these changes, the
observed squared amplitude is given by,
\begin{eqnarray}
|\mathcal{M}|^2_{obs} & = & |\mathcal{M}_0|^2\left[1-\frac{11}{6}\alpha
ln(s/\mu^2)\right]\left[1+\frac{\alpha}{3}ln(\Delta^2s/m^2)\right]
\left[1-\frac{\alpha}{3}ln(\mu^2/m^2)\right] \nonumber \\
& = & |\mathcal{M}_0|^2\left[1 -\alpha\left\{\frac{11}{6}ln(s/\mu^2) -
\frac{1}{3}ln(\Delta^2s/m^2) + \frac{1}{3}ln(\mu^2/m^2)\right\} +
\dots\right] \nonumber \\
\therefore |\mathcal{M}|^2_{obs} & = & |\mathcal{M}_0|^2\left[1
-\alpha\left\{\frac{3}{2}ln(s/\mu^2) + \frac{1}{3}ln(1/\Delta^2) +
o(m^0)\right\} + o(\alpha^2)\right] 
\end{eqnarray}
There is no dependence on $m$ now and the limit $m\to 0$ can be taken.
However, there is explicit $\mu$ dependence and the observed
cross-section {\em cannot} depend on the arbitrary $\mu$. Hence
$\boxed{\Case{d}{d\mu}|\mathcal{M}|^2_{obs} = 0 }\ $ must hold. Writing
\[
ln|\mathcal{M}|^2_{obs} := ln|\mathcal{M}_0|^2 + [\alpha
\Case{3}{2}ln(\mu^2) + \alpha C_2] := C_1 + 2ln\alpha + 3\alpha ln(\mu)
+ 3\alpha C_2
\]
where $C_1, C_2$ are functions of $s,t,u$ but independent of $\alpha$
and $\mu$, the condition of $\mu-$independence gives,
\begin{eqnarray}
0 & = & \left(\frac{2}{\alpha} + 3(C_2+ln(\mu))\right)
\frac{d\alpha}{dln(\mu)} + 3\alpha ~ ~ \Rightarrow \\
\boxed{\beta(\alpha) := \frac{d\alpha}{dln(\mu)}} & = &
-\frac{3}{2}\alpha^2 + o(\alpha^3) . ~ ~ \mbox{We also recall,}\\
0 & = & \frac{dm^2_{ph}}{dln(\mu)} ~ ~ ~ \Rightarrow \\
\boxed{\gamma_m(\alpha) := \frac{dln(m)}{dln(\mu)}} & = & -\frac{\alpha}{12} +
o(\alpha^2)
\end{eqnarray}
The $\gamma_m$ is called the {\em anomalous mass dimension} of the mass
parameter and $\beta(\alpha)$ is the famous ``beta'' function of the
theory. We have just obtained these functions at 1-loop, for the
$(\Phi^3)_6$ theory and these govern how the renormalized parameter must
change with the ``renormalization scale'', $\mu$, in order that the
physical quantities are independent of the renormalization scale. If we
identify $\mu = s$, then we have the running of the renormalized
parameters with the center-of-mass energy scale.

The differential equations can  be trivially solved giving the running
as, $\mu = \hat{\mu}e^t$,
\begin{equation}
\boxed{ \alpha(t) = \frac{\hat{\alpha}}{1+\frac{3}{2}
\hat{\alpha}ln(\mu/\hat{\mu})} ~ ~ ~ , ~ ~ ~ m(t) =
\hat{m}\left(1+\frac{3}{2}\hat{\alpha}\ t\right)^{-5/18} }
\end{equation}

As $\mu \sim \sqrt{s} \gg \hat{\mu}$, the coupling {\em decreases}. Such
a theory is said to be {\em asymptotically free}.

\underline{To summarize:} We have seen how the introduction of counter
terms makes it possible to absorb the UV divergences into unobservable
coefficients such as $\delta_{\Phi}, \delta_m, \delta_g$. This process
of absorption or \`subtraction' has an inherent ambiguity: exactly what
is subtracted.  This is implicitly determined by imposition of a set of
renormalization conditions (eg the on-shell renormalization) or
explicitly (eg the $MS/\overline{MS}$ scheme). When the pole (in $p^2$)
in the self-energy is not a simple pole, which happens with mass less
particle ($m_{ph} = 0$), alternative subtraction schemes are needed.
Such schemes, typically imply renormalization scale dependence in the
renormalized parameters such as masses and couplings. This is governed by
the beta function(s) and the anomalous dimension(s). For the sub-class
of asymptotically free theories, this helps in improving perturbative
predictions at high energies. 

A more general discussion of the renormalization group and its
application is given in subsection (\ref{RGEqn}). 

\newpage
\section{Path Integrals in Quantum Mechanics}\label{QMPathIntegral}

We consider an alternative strategy to view quantum dynamics.

Recall that in a quantum framework, we have a (projective) Hilbert space
(or more generally a density operator on a Hilbert space) that encodes
the kinematics while a family of unitary operators encodes the dynamics.
Observable quantities are provided by operators whose expectation values
and uncertainties in a given state (or density operator), provide
numbers to be matched against experiments. To track time evolution of
observable quantities, the central quantity of interest is the {\em
transition probability amplitude, for a state $|\Psi\rangle$ at time $t$
to make a transition at $t' >t$ to another state $|\Psi'\rangle$}. The
state $\Psi\rangle$ evolves to $exp{-i(t'-t)H}|\Psi\rangle$ by
Schrodinger equation (H is taken time independent for simplicity). Its
inner product with $|\Psi'\rangle$ gives the probability amplitude.
This is denoted as: 
\[
	\mbox{Transition probability amplitude} :=\langle \Psi'|\left(
	e^{-i(t'-t)H}|\Psi\rangle\right) = \langle
	\Psi'|e^{-i(t'-t)H}|\Psi\rangle
\]
Assuming the quantum system to be describing a particle with a
configuration space $\{\vec{q}\}$, we can use the position
representation completeness relation and express the amplitude as,
\begin{eqnarray*}
	\langle \Psi'|e^{-i(t'-t)H}|\Psi\rangle & = & \int
	d\vec{q}'\int d\vec{q}\ \langle\Psi'| \vec{q}'\rangle
	\langle\vec{q}'| e^{-i(t'-t)H} |\vec{q} \rangle \langle
	\vec{q}|\Psi \rangle \\
	& := & \int d\vec{q}'\int d\vec{q}\
	\Psi^*(\vec{q}')K(\vec{q}',t';\vec{q},t)\Psi(\vec{q}) 
\end{eqnarray*}
The idea is to focus on the kernel $K$ and get a convenient
representation for it.

\underline{Note:} We have already presumed the usual framework of
quantum theory. It is possible to develop the ideas ab initio taking the
Kernel as a central quantity with a proposed form and arrive at the
usual quantum framework. The book of Feynman and Hibbs follows this
line. We will first sketch the heuristic approach and then relate it to
the usual quantum framework.
\subsection{The Ab Initio Path Integral} \label{AbInitio}
Consider a classical system with a configuration space $Q$ which, for
convenience of notation we take to be $\mathbb{R}$. Let $q_1, q_2$ be
two points of it. The particle is supposed to make a transition from
$q_1$ at $t_1$ to $q_2$ at $t_2$. Such a transition takes place
classically as well and we use it in the Lagrangian framework and deduce
that the transition takes place along a curve $q(t)$ which extremises
the action. In a departure from the classical dynamics, it is proposed
that there is certain {\em probability for the transition} ($q_1, t_1)
\to (q_2, t_2)$. This is to be computed by squaring the total
probability amplitude obtained by summing over the probability
amplitudes for every path connecting the $(q_1,t_1)$ and $(q_2,t_2)$.
The amplitude for each path is given as a certain phase. Thus,
heuristically, $K(q_2,t_2;q_1,t_1) \sim \sum_{paths}e^{i \varphi[q(t)]}$

{\em Questions:} What do we choose for the phase? How do we restrict the
paths? How is the `sum' to be performed?  

$\bullet$ We expect to recover classical dynamics in the limit $\hbar
\to 0$. So we expect the phase must be such as the amplitude is
dominated by a single path corresponding to the classical transition.
the obvious choice is $\varphi[q(t)] \sim S[q(t)]/\hbar$!

$\bullet$ First guess about the paths would be smooth paths, but this
turns out to be not enough. Since the action involves derivatives of
$q(t)$, we expect at least piecewise differentiability. But even this
may not suffice - we can define derivatives as differences. It seems,
continuity suffices. Without worrying about this too much, let us
proceed by {\em discretizing } the time interval. This will also lead us
to answer the third question.

Let $T := t_2-t_1 := N\epsilon$ where $N$is a large number eventually to
be taken to infinity. For definiteness, let the action be 
\[
S = \int_0^Tdt \left[\frac{m}{2}\dot{q}^2 - V(q)\right] \to
\sum_{k=0}^{N-1}\left\{\frac{m}{2}\frac{(q_{k+1}-q_k))^2}{\epsilon} -
\epsilon V\left(\frac{q_{k+1}+q_k}{2}\right)\right\} ~ , ~ q_k := q(t_k)
\]
For uniformity of notation, denote $q_0 =: q_{initia}, q_N =:
q_{final}$. We have used the natural discretization of the action. The
space of paths is now described by the $N-1$ $q_k$'s varying
independently over $\mathbb{R}$. The integration measure may be taken as
the product measure $\prod_{k=1}^{N-1}\Case{dq_k}{C(\epsilon)}$, where
the constant $C(\epsilon)$ is left free and will be chosen shortly.
Since $N$ is arbitrary, we define a {\em path integral} as:
\begin{equation}
	\boxed{ K(q_{in}, q_{fi};T) := \lim_{\epsilon\to
0}\frac{1}{C(\epsilon)} \prod_{k=0}^{N-1} \int_{-\infty}^{\infty}
\frac{dq_k}{C(\epsilon)}exp\frac{i}{\hbar}\left\{
\frac{m}{2}\frac{(q_{k+1}-q_k))^2}{\epsilon} - \epsilon
V\left(\frac{q_{k+1}+q_k}{2}\right) \right\} }
\end{equation}

We can rewrite it by separating the $k=(N-1)^{th}$ integral as,
\begin{eqnarray*}
	K(q_{in},q_f; T)  = \hspace{12cm} & & \\
\int_{-\infty}^{\infty} \frac{dq'}{C(\epsilon)} \left[
exp\frac{i}{\hbar}\left\{ \frac{m}{2} \frac{(q_f-q')^2} {\epsilon} -
\epsilon V \left(\frac{q_{f}+q'}{2} \right) \right\} \cdot
K(q_{in},q';T-\epsilon) \right] & & 
\end{eqnarray*}

As $\epsilon\to 0$, the rapid oscillations of the $\Case{1}{\epsilon}$
term imply that dominant contribution to the integral comes from $q'$
very close to $q_f$. Therefore Taylor expanding the potential and the
$K(q_{in}, q';T-\epsilon)$ about $q_f$, we get
\begin{eqnarray*}
K(q_{in},q_f;T) & \approx &
\int_{-\infty}^{\infty}\frac{dq'}{C(\epsilon)}e^{\frac{i}{\hbar}
\frac{m}{2\epsilon}(q_f-q')^2}\cdot \left\{1 + \frac{-i}{\hbar}\epsilon
	V(q_f) + \left(\frac{-i}{\hbar}\right)^2 \frac{\epsilon^2}{2}
\frac{V^2_f}{2} \dots \right\} \\
& & \left. \times \left\{1 + (q'-q_f)\partial_{q'} +
	\frac{(q'-q_f)^2}{2!} \partial^2_{q'} + \dots \right\}K(q_{in},
	q';T-\epsilon)\right|_{q'=q_f}
\end{eqnarray*}
Notice that the first braces are independent of $q'$ ($V(q)$ is assumed
to be reasonably well behaved). The second braces is a power series in
$(q'-q_f)$. 
\begin{eqnarray*}
K(q_{in},q_f;T) & \approx & e^{\frac{-i}{\hbar}\epsilon V(q_f)}\
\int_{-\infty}^{\infty}\frac{dx}{C(\epsilon)}e^{\frac{i}{\hbar}
\frac{m}{2\epsilon}x^2}\cdot \left\{1 + x\partial_{q_f} + \frac{x^2}{2!}
\partial^2_{q_f} + \dots \right\}K(q_{in}, q_f;T-\epsilon)
\end{eqnarray*}
In effect, we have a series of Gaussian integrals of the form:
$\int_{-\infty}^{\infty}dx x^m e^{\Case{i}{\hbar} \Case{mx^2}{2\epsilon}
}$. The Gaussian integrals are all well known,
\[
	\int_{-\infty}^{\infty}dx e^{-ax^2} = \sqrt{\frac{\pi}{a}} ~ , ~
	\int_{-\infty}^{\infty}dx x^{2k}e^{-ax^2} =
	\frac{1}{a^{k+1/2}}\Gamma(k+1/2) ~ , ~ \int_{-\infty}^{\infty}dx
	x^{2k+1}e^{-ax^2} = 0\ .
\]
For us, $a = \Case{im}{2\hbar\epsilon}$. For this to be defined, we have
to define the usual Gaussian integral for $a>0$ and continue
analytically in the complex $a-$plane with real part of $a$ being
positive. Hence, we must put $m\to m+i\varepsilon$ to provide the
convergence factor. Thus,
\begin{eqnarray*}
	K(q_{in}, q_f;T) \approx \sqrt{\frac{\pi}{-im/(2\hbar\epsilon)}}
	\frac{1}{C(\epsilon)}\times\left\{1
	-i\frac{\epsilon}{\hbar}V(q_f) +
\frac{i\epsilon\hbar}{2m}\partial^2_{q_f} +
\dots\right\}K(q_{in},q_f;T-\epsilon) \ . 
\end{eqnarray*}
The terms in the braces are regular as $\epsilon\to 0$ and hence the
right hand side can be arranged to have a regular limit as $\epsilon\to
0$ {\em by choosing} $\boxed{C(\epsilon) =
\sqrt{\Case{2\pi\epsilon\hbar}{-im}} .} $ Additionally, we also get,
\begin{eqnarray}
	K(q_{in},q_f;T) - K(q_{in},q_f;T-\epsilon) & = &
	\left(\frac{i\epsilon\hbar}{2m}\partial^2_{q_{f}} -
	\frac{i\epsilon}{\hbar}V(q_f)\right)K(q_{in}, q_f; T) \\
	\mbox{or} \hspace{2.2cm} i\hbar\partial_TK(q_{in},q_f;T) & = &
	\left( - \frac{\hbar^2}{2m}\partial^2_{q_f} + V\right)
	K(q_{in},q_f;T)
\end{eqnarray}
Thus, the $K$ defined above, {\em satisfies the time dependent
Schrodinger equation!}

What about the initial condition? To see the limit $T\to 0$, take $N=1
(T=\epsilon)$. Then there is no $\int dq$. Only the $k=0$ term in the
exponent survives and we get,
\[
	K(q_{in}, q_f; \epsilon\to 0) = \lim_{\epsilon\to
	0}\frac{1}{C(\epsilon)}e^{\frac{i}{\hbar}\frac{m}{2}(q_f-q_{in})^2
	- \epsilon V( (q_f+q_{in})/2 )} ~ ~ , ~ ~ C(\epsilon) =
	\sqrt{\frac{2\pi\epsilon\hbar}{-im}}.
\]
The right hand side is just the representation of $\delta(q_f-q_{in})$.
The usual quantum framework definition of the transition amplitude
satisfies the time dependent Schrodinger equation with the same initial
condition! So we have a strong hint that the $K(q_{in},q_f;T)$ defined
may indeed be identified with $\langle q_f|e^{-iTH}|q_{in}\rangle $ We
can see this directly as well.
\subsection{Derivation From Transition Amplitude} \label{QMDerivation}
Divide the interval $[0,T]$ in to $N$ intervals of size $\epsilon$ each,
$N\epsilon = T$. Since the Hamiltonian is assumed to be time
independent, we can write, 
\[
e^{-\frac{i}{\hbar}TH} = e^{-\frac{i}{\hbar}\sum_{k=0}^{N-1} H
(t_{k+1}-t_k)} = \prod_{k=0}^{N-1}e^{-\frac{i}{\hbar}(t_{k+1}-t_k)H}
\approx \prod_{k=0}^{N-1}\left(1-\frac{i\epsilon}{\hbar}H +
\dots\right)_k \ .
\]

Insert the completeness relation $\mathbb{1} = \int d\vec{q}_k
|\vec{q}_k\rangle\langle \vec{q}_k|$ between each factor so that
\begin{eqnarray*}
	\langle \vec{q}_f|e^{-\frac{iT}{\hbar}H}|\vec{q}_{in}\rangle & =
	& \int d\vec{q}_1\dots d\vec{q}_{N-1} \prod_{k=0}^{N-1} \langle
	\vec{q}_{k+1}| e^{-\frac{i\epsilon H}{\hbar}}|\vec{q}_k\rangle ~
	~ ~ ~ \vec{q}_0 =: \vec{q}_{in}\ , \ \vec{q}_N =: \vec{q}_f \\
	e^{-\frac{i\epsilon H}{\hbar}} & \approx & 1
	-\frac{i\epsilon}{\hbar}H + \dots \hspace{3.0cm} \mbox{for small
	$\epsilon$.}
\end{eqnarray*}
To simplify further, we consider types of Hamiltonian operators (we
suppress the vector arrows).

\underline{(i)} $H(\hat{q}, \hat{p}) = f(\hat{p}) + g(\hat{q})$. Then
\begin{eqnarray*}
	\langle q_{k+1}|f(\hat{p})|q_k\rangle & = & \int dp_k\langle
	q_{k+1}|f( \hat{p})|p_k\rangle\langle p_k|q_k\rangle \\
	& = & \int dp_k\ f(p_k)\langle q_{k+1}|p_k\rangle\langle p_k|
	q_k\rangle \hspace{1.0cm}\mbox{Use: }\langle p_k|q_k\rangle =
	\frac{e^{-\frac{i}{\hbar}\vec{q}_k\cdot
	\vec{p}_k}}{\sqrt{2\pi\hbar}} \\
	\therefore \langle q_{k+1}|f(\hat{p})|q_k\rangle & = & \int dp_k
	f(\hat{p}) \frac{e^{\frac{i}{\hbar} \vec{p}_k
	\cdot(\vec{q}_{k+1} - \vec{q}_k)}}{2\pi\hbar} \hspace{2.0cm}
	\mbox{Likewise,} \\
	\langle q_{k+1}|g(\hat{q})|q_k\rangle & = &
	q(\vec{q}_k)\vec{\delta}(\vec{q}_{k+1} - \vec{q}_k) =
	g(\vec{q}_k)\frac{1}{2\pi\hbar}\int dp_ke^{\frac{i}{\hbar}
	\vec{p}_k \cdot(\vec{q}_{k+1} - \vec{q}_k)} \\
	\therefore \langle q_{k+1}|\hat{H}(\hat{q},\hat{p})|q_k\rangle &
	= & \int\frac{d\vec{p}_k}{2\pi\hbar} H_{cl}
	\left(\frac{\vec{q}_{k+1} + \vec{q}_k}{2}, \vec{p}_k\right)
	^{\frac{i}{\hbar} \vec{p}_k \cdot(\vec{q}_{k+1} - \vec{q}_k)}
\end{eqnarray*}
We have used $g(q_k) \to g(q_{k+1}+q_k)/2)$ when multiplied by the
$\delta(q_{k+1} - q_k)$.

\underline{(ii)} $\hat{H}$ contains the $\hat{q}, \hat{p}$ in various
ordering such as, $\hat{q}\hat{p}, \hat{p}\hat{q}, \hat{q}^2\hat{p},
\hat{q} \hat{p} \hat{q}, \hat{p}\hat{q}^2, \dots $etc. As an example,
consider a term $\hat{q}^a\hat{p}_b$ in the quantum Hamiltonian. Then,
\begin{eqnarray*}
	\langle q_{k+1}|\hat{q}^a\hat{p}_b|q_k\rangle & = & \int
	d\vec{p}_k \langle q_{k+1}|\hat{q}^a|p_k\rangle\langle
	p_k|\hat{p}_b|q_k\rangle = q^a_{k+1}(p_k)_b \int
	\frac{d\vec{p}_k}{2\pi\hbar} e^{\frac{i}{\hbar} \vec{p}_k \cdot
	(\vec{q}_{k+1} - \vec{q}_k)} \\
	\langle q_{k+1}|\hat{p}_b\hat{q}^a|q_k\rangle & = & \int
	d\vec{p}_k \langle q_{k+1}|\hat{p}_b|p_k\rangle\langle
	p_k|\hat{q}^a|q_k\rangle = q^a_{k}(p_k)_b \int
	\frac{d\vec{p}_k}{2\pi\hbar} e^{\frac{i}{\hbar} \vec{p}_k \cdot
	(\vec{q}_{k+1} - \vec{q}_k)} \\
	\therefore \langle q_{k+1}|\frac{\hat{q}^a\hat{p}_b + \hat{p}_b
	\hat{q}^a}{2} |q_k\rangle & = & \left(\frac{q^a_{k+1} +
	q^a_k}{2}\right) (p_k)_b \int \frac{d\vec{p}_k}{2\pi\hbar}
	e^{\frac{i}{\hbar} \vec{p}_k \cdot (\vec{q}_{k+1} - \vec{q}_k)}
\end{eqnarray*}
Thus for the natural ordering, the classical Hamiltonian has its
$\vec{q}$ dependence in the averaged form. Here is another example
involving $\hat{q}^2, \hat{p}^2$, taken in a particular order, called
{\em Weyl order}: $Weyl(\hat{q}^2, \hat{p}^2) := \Case{1}{4}(
\hat{q}^2\hat{p}^2 + 2\hat{q} \hat{p}^2\hat{q} + \hat{p}^2 \hat{q}^2)$.
It is easily seen that,
\begin{eqnarray*}
\langle q_{k+1}|Weyl(\hat{q}^2,\hat{p}^2)|q_k\rangle & = &
\frac{1}{4}\left\{q^2_{k+1}\langle q_{k+1}|p^2|q_k\rangle + 2 q_{k+1}q_k
\langle q_{k+1}|p^2|q_k\rangle + q_k^2 \langle
q_{k+1}|p^2|q_k\rangle\right\} \\
& = & \left(\frac{\vec{q}_{k+1} + \vec{q}_k}{2}\right)^2\langle
q_{k+1}|\hat{p}^2|q_k\rangle 
\end{eqnarray*}
More generally, for monomials in $\hat{q}, \hat{p}$, the Weyl order
$Weyl(\hat{p}^m, \hat{q}^n)$ equals the fully symmetrized and averaged
product. It is a result that
\[
	\langle q_{k+1}|Weyl(\hat{q}^n, \hat{p}^m)|q_k\rangle  =
	\left(\frac{\vec{q}_{k+1} + \vec{q}_k}{2}\right)^n\langle
	q_{k+1}|\hat{p}^m|q_k\rangle
\]

Hence, modulo Weyl ordering, we do get,
\[
\boxed{\langle q_{k+1}|e^{-\frac{i\epsilon}{\hbar}H}|q_k\rangle =  \int
\frac{dp_k}{2\pi\hbar} e^{-\frac{i}{\hbar} \epsilon H\left(p_k,
\frac{q_{k+1}+q_k}{2}\right)} e^{\frac{i}{\hbar} \vec{p}_k \cdot
(\vec{q}_{k+1} - \vec{q}_k) } }
\]
\begin{eqnarray}
	\langle q_f| e^{-\frac{iTH}{\hbar}}|q_{in}\rangle & = & \int
	d\vec{q}_1\dots d\vec{q}_{N-1}\ \int \frac{d\vec{p}_1}
	{2\pi\hbar} \dots \frac{d\vec{p}_N}{2\pi\hbar}\ exp
	\frac{i}{\hbar}\Bigg\{ \nonumber \\
		& & \hspace{2.0cm} \left.\epsilon \sum_{k=0}^{N-1}
		\vec{p}_{k+1}\cdot\frac{(\vec{q}_{k+1}-\vec{q}_k)}{\epsilon}
	- \epsilon H(p_k, \frac{q_{k+1}+q_k}{2})\right\}
\end{eqnarray}
Notice that we have $N, p_i$ integrations, one for each interval and
$N-1, q_i$ integrations, two less than the number of points since $q_0,
q_N$ are fixed. The exponent is clearly a discretized form of $\int_0^T
dt\ \vec{p}(t)\cdot\dot{\vec{q}}(t) - H(\vec{p}(t), \vec{q}(t)) =
\int_0^T dt\ L(\vec{q}(t), \dot{\vec{q}}(t))$. The $d\vec{q}
\frac{d\vec{p}} {(2\pi\hbar)}$ measure suggests that we have paths in
the ``phase space''. What type of paths?

The mis-match in the number of integration variables makes it harder to
view the multiple integrals as a measure on the space of paths in phase
space. Introduce an arbitrary momentum $\vec{p}_0$ so that $(\vec{q}_0,
\vec{p}_0)$ denotes a point in the phase space just as $(\vec{q}_k,
\vec{p}_k), k = 1,\dots,N$ do. The $\int d\vec{p}_N$ shows that the
momentum at the final point is integrated over. Thus, the phase space
path may be specified by an initial point $(\vec{q}_0, \vec{p}_0)$. The
arbitrary momentum does not enter anywhere and just serves to anchor the
initial point. The final point however is the set of all points
$(\vec{q}_N, \vec{p}_N)$ with $\vec{P}_N$ integrated over. Thus, the
space of paths is ($\Gamma$ is a phase space of dimension $2n$): 
\[
	\boxed{ \mbox{Space of paths}~ =: ~ \cal{P}_{\Gamma} ~ :=  ~
	\Big\{\ \gamma(t) \in \Gamma\ /\ \gamma(0) = (\vec{q}_{in},
\vec{p}_0)\ ,\  \gamma(T) = (\vec{q}_f, \vec{p}),\ \vec{p}\in
\mathbb{R}^n.\ \Big\} }
\]
There are arbitrary constants here, the $\vec{p}_0$ is an arbitrary
constant vector. Since the $\vec{p}_k, \vec{q}_k$ are all independent,
the paths are continuous but non-differentiable everywhere. Thus we
denote:
\begin{eqnarray} \label{PhaseSpacePathIntgl}
\langle q_f|e^{-\frac{i T \hat{H}}{\hbar}}| q_{in}\rangle & := &
\lim_{\epsilon\to 0}\int d\vec{q}_1\dots d\vec{q}_{N-1}\int
\frac{d\vec{p}_1}{2\pi\hbar}\dots \frac{d\vec{p}_N}{2\pi\hbar} \nonumber
\\
& & exp \left[ \frac{i}{\hbar}\ \epsilon\ \sum_{k=0}^{N-1} \left\{
\vec{p}_{k+1}\cdot\frac{(\vec{q}_{k+1}-\vec{q}_k)}{\epsilon} -
H\left(\vec{p}_k, \frac{\vec{q}_{k+1}+\vec{q}_k}{2}\right)\right\}
\right] \\
& = & \int_{\cal{P}_{\Gamma}}\cal{D}q(t)\cal{D}p(t)\
exp\left[\frac{i}{\hbar}\int_0^Tdt\left( \vec{p}(t)\cdot\dot{\vec{q}}(t)
- H(\vec{p}(t),\vec{q}(t)) \right)\right] , \\
\int_{\cal{P}_{\Gamma}}\cal{D}q(t)\cal{D}p(t) & := &
\lim_{N\to\infty} \prod_{i=1}^{N-1}\int d\vec{q}_k\ \prod_{j=1}^{N}\int
\frac{d\vec{p}_j}{2\pi\hbar} 
\end{eqnarray}

For the special case of $H(\vec{p},\vec{q}) = \Case{\vec{p}^2}{2m} +
V(\vec{q})$, the momentum dependence is quadratic and the momentum
integrals can be done trivially. For instance,
\begin{eqnarray*}
\int \frac{d\vec{p}_{k+1}}{(2\pi\hbar)^d} e^{\frac{i}{\hbar} \left\{
\vec{p}_{k+1}\cdot(\vec{q}_{k+1}-\vec{q}_k) -\epsilon\left(
\frac{\vec{p}^2}{2m} +
V(\frac{\vec{q}_{k+1}+\vec{q}_k}{2})\right)\right\}} & = &
\frac{1}{(2\pi\hbar)^d}
e^{\frac{i}{\hbar}\frac{m}{2\epsilon}(\vec{q}_{k+1}-\vec{q}_k)^2}\times
\\
& & \int d^dp_ke^{-\frac{i\epsilon}{2m\hbar} \left(\vec{p}-
\frac{m(\vec{q}_{k+1}-\vec{q}_k)} {\epsilon}\right)^2}\
e^{-\frac{i}{\hbar}\epsilon V} \\
& = & \Big(\frac{m}{2\pi\hbar(i\epsilon)}\Big)^{d/2}
e^{\frac{i}{\hbar}\epsilon \left\{
\frac{m}{2}\frac{(\vec{q}_{k+1}-\vec{q}_k)^2}{Wp^2} -
V(\frac{\vec{q}_{k+1}+\vec{q}_k}{2})\right\}}
\end{eqnarray*}
The square root prefactor is just $C(\epsilon)^{-1}$. Doing all the
momentum integrals gives,
\begin{eqnarray}
	\langle q_f|e^{-\frac{i}{\hbar}T\hat{H}}|q_{in}\rangle & = &
	\lim_{\epsilon\to 0}
	\int\frac{d\vec{q}_1}{(2\pi\hbar(i\epsilon/m))^{d/2}}\dots
	\int\frac{d\vec{q}_{N-1}}{(2\pi\hbar(i\epsilon/m))^{d/2}}
	\frac{e^{\frac{i\epsilon}{\hbar}\left\{\frac{m}{2}
	\frac{(\vec{q}_{k+1}-\vec{q}_k)^2}{\epsilon^2}
	-V\right\}}}{(2\pi\hbar(i\epsilon/m))^{d/2}}
\end{eqnarray}
Each $d\vec{q}$ integral gets a factor of $C(\epsilon)^{-d}$ and extra
such factor is left over from the extra momentum integration. This was
the factor introduced and determined by a regular behavior as
$\epsilon\to 0$.

Thus, the upshot is that we can either take the heuristically motivated
definition of $K$ and show that it matched with quantum mechanical
definition of transition amplitude {\em or} directly deduce it from the
quantum mechanical definition. The $K$ of the transition amplitude is
the central quantity in the path integral approach.

\underline{A note:} We defined $K(q_{in}, q_f;T)$ and identified it with
$\langle q_f|exp{-(i/\hbar)TH}|q_{in}\rangle$ and called it the
probability amplitude for a particle at $q_{in}$ at $t_0$ to transit to
$q_f$ at time $t_0+T$. This is sometimes also denoted as $\langle
q_f,T|q_{in},0\rangle$ or $\langle q'',t''|q',t'\rangle$ with $t''-t'=
T$. This notation can be confusing since $|q',t'\rangle$ does {\em not}
denote a solution of the Schrodinger equation. Rather, it denotes the
{\em instantaneous eigenvector of the Heisenberg picture operator}
$\hat{Q}(t) := e^{itH/\hbar}Q_{Sch}(0)e^{-itH/\hbar}$. Thus,
\[
	Q(t)|q,t\rangle = q|q,t\rangle \leftrightarrow
	Q_{Sch}\left(e^{-itH/\hbar}|q,t\rangle\right) =
	q\left(e^{-itH/\hbar}|q,t\rangle\right) \Rightarrow
	e^{-itH/\hbar}|q,t\rangle = |q\rangle
\]
Or, $|q,t\rangle = e^{itH/\hbar}|q\rangle$. If $|q,t\rangle$ were the
time evolution of $|q\rangle$, we should have $e^{itH/\hbar}|q\rangle$.
Thus we have the consistent notation:
\[
\boxed{\langle q_f|e^{-\frac{iTH}{\hbar}}|q_{in}\rangle = \langle
q_f|e^{-\frac{i(t''-t')H}{\hbar}}|q_{in}\rangle = \langle
q_f|e^{-\frac{it''H}{\hbar}}\cdot e^{\frac{it'H}{\hbar}}|q_{in}\rangle
= \langle q_f, t''|q_{in},t'\rangle\ . }
\]

Consider now $\langle q_f,T|Q_H(t)|q_{in}, -T\rangle$, for $t\in
[-T,T]$.
\begin{eqnarray*}
	\langle q_f,T|Q_H(t)|q_{in},-T\rangle & = & \langle q_f|
	e^{-iTH/\hbar} e^{itH/\hbar} Q_S e^{-itH/\hbar} e^{-iTH/\hbar}
	|q_{in}\rangle \\
	& = & \int dq'dq''\langle q_f|e^{i(t-T)H/\hbar}|q'\rangle\langle
	q'| Q_S |q''\rangle\langle q''|e^{-i(t+T)H/\hbar}|q_{in}\rangle
	\\
	& = & \int dq' dq''\langle q_f,T|q',t\rangle\langle
	q'|Q_S|q''\rangle  \langle q'',t|q_{in},-T\rangle \\
	& = & \int dq'\langle q_f,T|q',t\rangle q'(t)\langle
	q',t|q_{in},-T\rangle
\end{eqnarray*}
In the last equation, we have used $\langle q'|Q_S|q''\rangle =
q'\delta(q'-q'')$ - the property of the Schrodinger picture operator,
carried out the $dq''$ integration and as a {\em reminder} inserted the
argument $t$ in $q'(t)$. Using the path integral notation, we write the
last equation as,
\begin{equation}
	\boxed{
	\langle q_f,T|Q_H(t)|q_{in},-T\rangle = \int\cal{D}q\
	q(t)e^{\frac{i}{\hbar}S[-T,T]} = \int_{\mathbb{R}} dq
	\int\cal{D}q e^{\frac{i}{\hbar}S[t,T]} q(t) \int\cal{D}q
	e^{\frac{i}{\hbar}S[-T,t]}
}
\end{equation}

Next, consider the quantity,
\begin{eqnarray*}
	\int\cal{D}q\ q(t_1)q(t_2)e^{\frac{i}{\hbar}S[-T,T]} & := & \int
	dq_1\int dq_2 \left[\int\cal{D}q
	e^{\frac{i}{\hbar}S[t_1,T]}\right] q_1(t_1) \times \\
	& & \left[\int\cal{D}q e^{\frac{i}{\hbar}S[t_2,t_1]}\right]
	q_2(t_2) \left[\int\cal{D}q e^{\frac{i}{\hbar}S[-T,t_2]}\right]
	~ ~ -T \le t_2 \le t_1 \le T ;
\end{eqnarray*}
For $t_2 \ge t_1$ the factors will switch accordingly. We summarize the
formulae as,
%
\begin{eqnarray}
	\langle q_f,t''|q_{in},t'\rangle & = &
	\int\cal{D}qe^{\frac{i}{\hbar}S[t',t'']} ~ =: ~ \langle
	q_f|e^{-\frac{i}{\hbar}(t''-t')\hat{H}}|q_{in}\rangle \\
	\langle q_f,T|Q_H(t)|q_{in},-T\rangle & = & \int\cal{D}q\ q(t)
	e^{\frac{i}{\hbar}S[-T,T]} \nonumber \\
	& := & \int dq_t\langle q_f, T|q,t\rangle q_t\langle
	q,t|q_{in},-T\rangle \\
	\langle q_f,T|T\{ Q_H(t_1) Q_H(t_2) \}|q_{in}, -T\rangle & = &
	\int \cal{D}qq(t_1)q(t_2) e^{\frac{i}{\hbar}S[-T,T]} ~ := ~ \int
	dq_1\int dq_2 \nonumber \\
	& & \left\{ 
		\begin{array}{c}
		\langle q_f,T|q_1,t_1\rangle q_1(t_1)\langle
		q_1,t_1|q_2,t_2\rangle q_2(t_2)\langle
		q_2,t_2|q_{in},-T\rangle\ , \nonumber \\
		t_1 \le t_2 \\
		\langle q_f,T|q_2,t_2\rangle q_2(t_2)\langle
		q_2,t_2|q_1,t_1\rangle q_1(t_1)\langle
		q_1,t_1|q_{in},-T\rangle\ , \nonumber \\
		t_2 \le t_1 
		\end{array} \right. \nonumber \\
\end{eqnarray}
\subsection{Functional Derivative} \label{FnalDiff}
It will be convenient to have the notation of {\em functional
derivative}. Recall from the variational principle of mechanics, the
action functional, $S[q(t)] := \int_{t_1}^{t_2}dt\ L(q(t), \dot{q}(t))$.
For any given function $q:[t_1,t_2]\to q(t) \in \mathbb{R}$, the right
hand side computes a number and that number is the action. It is a {\em
function on the space of paths/curves/functions on } $[t_1,t_2]$, and in
short is regarded as a {\em functional of any given} $q(t)$. This is
{\em not} a functional in the sense of being in the dual of vector space
- it is not linear in $q(t)$'s. Under a variation of a curve, $q(t) \to
q(t)+\delta q(t)$, we compute $\delta S := S[q+\delta q] - S[q] = \int
dt\ \delta q\left\{\Case{\partial L}{\partial q} - \Case{d}{dt}
\Case{\partial L}{\partial \dot{q}}\right\} + $ end-point contributions.
Comparing this with $df(x_1,\dots,x_n) = \Sigma_{i=1}^n dx^i
\partial_if$, we can identify and denote, {\em the functional
derivative} of the action with respect to the curve as: $\boxed{
	\Case{\delta S}{\delta q(t)} := \Case{\partial L}{\partial q} -
\Case{d}{dt} \Case{\partial L}{\partial \dot{q}} } $. Note that $S$ has
no explicit $t$ dependence while its functional derivative does.
Thus the idea of a functional derivative is to consider the first order
variation of a functional and read off the coefficient of the $\delta $.

We can also notice that any function $f(t)$ we can write, $\delta f(t) =
\int dt'\ \delta(t-t')\delta f(t') \leftrightarrow \Case{\delta
f(t)}{\delta f(t')} = \delta(t-t')$. Thus we define the derivative with
respect to a function as:

\begin{center}
	\fbox{
$ \begin{array}{clclr}
	(i) & \frac{\delta}{\delta f(t)}\left(\alpha F[f] + \beta
	G[f]\right) & = & \alpha\frac{\delta F}{\delta f(t)} + \beta
	\frac{\delta G}{\delta f(t)} & \mbox{(linearity)} \\
	(ii) & \frac{\delta}{\delta f(t)} F[f].G[f] & = & \frac{\delta
	F}{\delta f(t)}.G[f] + F[f].\frac{\delta G}{\delta f(t)} &
	\mbox{(Leibnitz rule)} \\
	(iii) & \frac{\delta}{\delta f(t)} F( G[f] ) & = &
	\frac{\partial F}{\partial G}\frac{\delta G}{\delta f(t)} &
	\mbox{ (Chain rule)} \\
	(iv) & \frac{\delta f(t')}{\delta f(t)} & := & \delta(t'-t) & 
\end{array} $
}
\end{center}

For an arbitrary ``source function'', $J(t)$ define the functional,
\begin{equation}
	\langle q'',t''|q',t'\rangle[J] :=
	\int\cal{D}qe^{\frac{i}{\hbar} \int_{t'}^{t''}dt \left\{
	L(q,\dot{q}) + J(t)q(t)\right\}} \ .
\end{equation}
Then it follows that,
\begin{eqnarray*}
\frac{\delta}{\delta J(t)} \langle q'',t''|q',t'\rangle[J] & = &
\int\cal{D}qe^{\frac{i}{\hbar} \int_{t'}^{t''}dt \left\{ L(q,\dot{q}) +
J(t)q(t)\right\}}\cdot \frac{i}{\hbar}q(t) \\
\mbox{Or,} ~ ~ \left(-i\hbar\frac{\delta}{\delta J(t)}\right) \langle
q'',t''|q',t'\rangle[J] & = & \int\cal{D}q\ q(t)e^{\frac{i}{\hbar}
\int_{t'}^{t''}dt\left\{ L(q,\dot{q}) + J(t)q(t)\right\}} \\
& = &\langle q'',t''|Q(t)|q',t'\rangle_J  \ . \\
\left(-i\hbar\frac{\delta}{\delta J(t_1)}\right)
\left(-i\hbar\frac{\delta}{\delta J(t_2)}\right) \langle
q'',t''|q',t'\rangle[J] & = & \int\cal{D}q\ q(t_1)q(t_2)
e^{\frac{i}{\hbar} \int_{t'}^{t''}dt\left\{ L(q,\dot{q}) +
J(t)q(t)\right\}} \\
& = &\langle q'',t''|T\left\{Q(t_1)Q(t_2)\right\}|q',t'\rangle_J  \ . \\
\end{eqnarray*}
The generalization is obvious. Evaluating the functional derivatives at
$J(t) = 0$ gives us,
\begin{equation} \label{CorrlnFn1}
	\boxed{ \langle q'',t''|T\left\{Q(t_1)\dots
	Q(t_n)\right\}|q',t'\rangle = (-i\hbar)^n\frac{\delta^n}{\delta
J(t_1)\dots \delta J(t_n)} \langle q'',t''|q',t'\rangle\Big|_{J = 0} }
\end{equation}
Thus, the ``correlation functions'' (left hand side) are given by the
functional derivatives of the transition amplitude in presence of a
source function, evaluated at vanishing source function. 

To relate it to ground state/vacuum expectation values of time ordered
products of Heisenberg picture operators, we take a closer look at the
transition amplitude with non-zero source function.

Choose the source function to have a compact support, $J(t) = 0$ for $t
< t', t > t''$. Choose $T'<t'$ and $T''>t''$. Then,
\[
	\langle Q'',T''|Q',T'\rangle_J = \int dq''\ dq'\langle
	Q'',T''|q'',t''\rangle_{J=0}\langle
	q'',t''|q',t'\rangle_J\langle q',t'|Q',T'\rangle_{J=0} \ .
\]
The $J=0$ amplitudes can be written in the energy representation (energy
eigenvalues assumed to be discrete for convenience),
\begin{eqnarray*}
\langle q',t'|Q',T'\rangle & = & \langle
q'|e^{-i(t'-T')H/\hbar}|Q'\rangle = \sum_n\langle
q'|\varphi_n\rangle\langle\varphi_n|Q'\rangle e^{-i(t'-T')E_n/\hbar} \\
& = & \sum_n\varphi_n(q')\varphi^*_n(Q') e^{-i(t'-T')E_n/\hbar} \\
\therefore e^{-iT'E_0/\hbar}\langle q',t'|Q',T'\rangle & = & \sum_n
\varphi_n(q') \varphi^*_n(Q') e^{-it'E_n/\hbar} e^{iT'(E_n-E_0)/\hbar}
\\
\therefore \lim_{T'\to i\infty} e^{-iT'E_0/\hbar} \langle
q',t'|Q',T'\rangle & = & \varphi_0(q') \varphi^*_0(Q') e^{-it'E_0/\hbar}
~ ~ \because E_n\neq E_0~ ~ \mbox{terms drop out, likewise} \\
\therefore \lim_{T''\to -i\infty} e^{iT''E_0/\hbar} \langle
Q'',T''|q'',t''\rangle & = & \varphi^*_0(q'') \varphi_0(Q'')
e^{it''E_0/\hbar}
\end{eqnarray*}
We have assumed that the lowest energy eigenvalue is non-degenerate,
otherwise the $\sum_n$ will reduce to the sum over the degenerate
states.  Thus we get,
\begin{equation}
\boxed{ \lim_{\stackrel{T'\to i\infty}{T''\to -i\infty}} \frac{\langle
Q'',T''|Q',T'\rangle_J}{e^{-iE_0(T''-T')} \varphi_0(Q'')\varphi_0^*(Q')}
= \int dq'\ dq'' \varphi_0^*(q'',t'')\varphi_0(q',t')\langle
q'',t''|q',t'\rangle_J } \label{VacToVac}
\end{equation}
\subsection{Ground State-to-ground state Amplitude: $Z[J]$} \label{Zj}
The right hand side of eq. (\ref{VacToVac}) is the ground
state-to-ground state transition amplitude, the one that we were looking
for. The left hand side tells us how to compute it from $\langle
Q'',T''|Q',T'\rangle$ which is similar to $\langle
q'',t''|q',t'\rangle_J$. The factor in the denominator is independent of
$J$ and will not matter. We introduce the definition,
\begin{eqnarray}
	Z[J] & := & \int dq'\ dq''\ \varphi_0^*(q'',t'')\langle
	q''.t''|q',t'\rangle_J \varphi_0(q',t') \label{Z-Defn1}\\
	\frac{\delta^n Z[J]}{\delta J(t_1)\dots \delta
	J(t_n)}\Big|_{J=0} & = & \left(\frac{i}{\hbar}\right)^n\int dq'\
	dq'' \varphi_0^*(q'',t'')\times \nonumber \\
	& & \hspace{2.0cm} \langle q'',t''|T\left\{Q(t_1)\dots
	Q(t_n)\right\}|q',t'\rangle\ \varphi_0(q',t') \ .
\end{eqnarray} 
The previous result tells us that $Z[J]$ may be computed as,
\[
Z[J] \sim \lim_{\stackrel{T'\to i\infty}{T''\to -i\infty}} \langle
Q'',T''|Q',T'\rangle_J = \lim_{\stackrel{T'\to i\infty}{T''\to
-i\infty}} \int\cal{D}q exp\left[\frac{i}{\hbar}\int_{T'}^{T''}
dt\left\{L(q,\dot{q}) + J(t)q(t)\right\}\right]
\]
We have dropped the unimportant denominator on the left hand side.

We have continued $T'\to i\infty, T''\to - i\infty$, but not the
intermediate times, $t_i$ appearing as arguments of the Heisenberg
operators. We will do so now and get to the {\em Euclidean formulation}.

We begin with,
\begin{eqnarray*}
	\langle Q'',T''|T\left\{Q(t_1)\dots
	Q(t_n)\right\}|Q',T'\rangle_{J=0} & \sim & \lim_{\stackrel{T'\to
	i\infty}{T''\to -i\infty}} \int dq_1 \dots dq_n\langle
	Q'',T''|q_1,t_1\rangle q_1\langle q_1,t_1|q_2,t_2\rangle \\
	& & \hspace{3.0cm} \dots q_n \langle q_n,t_n|Q',T'\rangle ~ , ~
	\mbox{where,} 
\end{eqnarray*}
\begin{eqnarray*}
\langle q_i,t_i|q_{i+1},t_{i+1}\rangle & \sim &
\int\prod_{j=1}^{N-1}\frac{dq_j}{\sqrt{2\pi i \epsilon}} exp\left[
\frac{i}{\hbar}\epsilon\sum_j L\left(\frac{q_i+q_{i+1}}{2},
\frac{q_{i+1}-q_i}{\epsilon}\right)\right] ~ , ~ N\epsilon = t_{i+1} -
t_i \\
\langle q_i,-i\tau_i|q_{i+1},-i\tau_{i+1}\rangle & \rightsquigarrow  &
\int\prod_{j=1}^{N-1}\frac{dq_j}{\sqrt{2\pi \epsilon'}} exp\left[
\frac{\epsilon'}{\hbar}\sum_j L\left(\frac{q_i+q_{i+1}}{2},
\frac{q_{i+1}-q_i}{-i\epsilon'}\right)\right] ~ , ~ N\epsilon' :=
\tau_{i+1}-\tau_{i}
\end{eqnarray*} 
The analytic continuation of the $\langle T\{\dots\}\rangle$ is defined
through the $\langle ..|..\rangle$. Thus,
\begin{eqnarray*}
	\langle Q'',T''|T\left\{Q(t_1)\dots
	Q(t_n)\right\}|Q',T'\rangle_{J=0}\Big|_{t_i = -i\tau_i}  & \sim
	& \lim_{\stackrel{\tau_{in} \to -\infty}{\tau_f\to \infty}} \int
	\cal{D}q\ q(\tau_1)\dots q(\tau_N) \times \\
	& & \hspace{2.0cm} exp\left[\int_{\tau_{in}}^{\tau_f}d\tau
	L(q,-\frac{dq}{d\tau})\right]
\end{eqnarray*}

This suggests going over to a {\em Euclidean formulation},
\begin{equation} \label{ZE-Defn}
	Z_E[J] := \int\cal{D}q\
	exp\left[\int_{-\infty}^{\infty}d\tau\left\{
	L\left(q,i\frac{dq}{d\tau}\right) + J(\tau)q(\tau)\right\}
\right]
\end{equation}
and the paths are between some $Q' = \lim_{\tau\to - \infty}q(\tau)$ and
$Q'' = \lim_{\tau\to\infty}q(\tau)$. The Euclidean $Z_E$ and the
Minkowskian $Z$ are related through,
\[
	\boxed{ \frac{1}{Z[J]}\frac{\delta^n Z[J]}{\delta
	J(t_1)\dots\delta J(t_n)}\Big|_{J=0} = i^n
\frac{1}{Z_{E}[J]}\frac{\delta^n Z_{E}[J]}{\delta J(\tau_1)\dots\delta
J(\tau_n)}\Big|_{J=0}  }
\]
This will be used in the field theory Green's functions.
\subsection{Explicit evaluation of a path integral} \label{ExplicitEval}
We would like to see the various expressions above explicitly for a
1-dimensional harmonic oscillator with a source function. We have $L =
\Case{1}{2}(\dot{q}^2 - \omega^2q^2) + J(t)q(t)$ and we want to compute
$\langle q'',t''|q',t'\rangle_J$. Discretizing the corresponding action,
the definition gives,
\begin{eqnarray*}
\langle q'',t''|q',t'\rangle_J & = & \lim_{\epsilon\to
0}\frac{1}{C(\epsilon)} \int_{-\infty}^{\infty} \prod_{k=1}^{N-1}
\frac{dq_k}{C(\epsilon)} exp\left[ \frac{i}{\hbar} \left\{
		\sum_{k=0}^{N-1}\frac{1}{2}
		\frac{(q_{k+1}-q_k)^2}{\epsilon} -
\frac{\epsilon\omega^2}{2} \left(\frac{q_{k}+q_{k+1}}{2}\right)^2
\right. \right. \\
& & \hspace{2.5cm}\left.\left.  + \epsilon J_k\left(
\frac{q_k+q_{k+1}}{2} \right) \right\} \right] ~ , ~ q_0 := q(t'), q_N
:= q(t''), J_k ;= J(t_k)
\end{eqnarray*}
All are Gaussian integrals with coupled variables. It is more convenient
to discretize a {\em different} action.

The equations of motion are: $\ddot{q} + \omega^2q^2 = J(t)\ ,\ q(t') =
q', q(t'') = q''$. Let $q_{cl}(t)$ be a classical solution of this
equation satisfying the end point conditions. Let us assume that there
is just {\em one} such solution. Introduce $\eta(t)$ via the definition:
$q(t) := q_{cl}(t) + \eta(t)$. Then $\eta(t') = 0 = \eta(t'')$. The
action becomes,
\begin{eqnarray*}
S(t',t'') & = & \int_{t'}^{t''}dt\left\{\frac{1}{2}(\dot{q}_{cl} +
\dot{\eta})^2 - \frac{\omega^2}{2}(q_{cl}+\eta)^2 +
J(t)(q_{cl}+\eta)\right\} \\
& = & \int_{t'}^{t''}dt\left[\left\{\frac{1}{2}(\dot{q}_{cl}^2-
	\omega^2q_{cl}^2) + J(t)q_{cl}\right\} + \left\{\frac{1}{2}
	(\dot{\eta}^2- \omega^2\eta^2)\right\} + \left\{\dot{\eta}
		\dot{q}_{cl} - \omega^2\eta q_{cl} + J
\eta\right\}\right]
\end{eqnarray*}
Up to a total derivative, the last term is $-\eta(\ddot{q}_{cl} +
\omega^2q_{cl} -J) = 0$. The total derivative term gives
$\eta\dot{q}_{cl}|_{t'}^{t''} = 0$ thanks to the end point condition
satisfied by $\eta(t)$. Thus, $\boxed{S(t',t'') = S_{cl}(t',t'') +
\int_{t'}^{t''}\Case{1}{2}(\dot{\eta}^2-\omega^2\eta^2), }$ where
$S_{cl}$ is the action evaluated at the presumed solution $q_{cl}$. The
terms linear in $\eta$ have vanished thanks to $q_{cl}$ being a
solution. We now ``quantize'' the $\eta$ variable, i.e. we {\em define}
\begin{eqnarray}
	\langle q'',t''|q',t'\rangle_J & := &
	\left[e^{\frac{i}{\hbar}S^J_{cl}(t',t'')}\right]
	\left[\lim_{\epsilon\to 0}\frac{1}{C(\epsilon)}
		\int_{-\infty}^{\infty} \prod_{k=1}^{N-1}
		\frac{d\eta_k}{C(\epsilon)} exp\left\{ \frac{i}{\hbar}
	\left( \sum_{k=0}^{N-1}\frac{1}{2}
	\frac{(\eta_{k+1}-\eta_k)^2}{\epsilon} \right.\right.\right.
	\nonumber \\
	& & \left.\left. \left.\hspace{5.7cm} - \frac{\epsilon\
	\omega^2}{2}((\eta_{k}+\eta_{k+1})/2)^2 \right) \right\}\right] 
\end{eqnarray}
The paths $\eta(t)$ begin and end at $\eta = 0$. The second factor is
independent of both $J$ and $t',t''$. It depends on $T = t''-t'$. Our
task is to evaluate the first factor and carry out the coupled Gaussian
integral.

The exponent in the second factor is of the form,
\begin{eqnarray*}
\frac{i}{2\hbar}\Big(\dots\Big) & = & \sum_{k=0}^{N-1}
\frac{1}{\epsilon} (\eta_{k+1}-\eta_k)^2 - \epsilon\
\omega^2(\eta_{k+1}+\eta_k)/2)^2 ~ := ~ \sum_{k,l=0}^{N-1}\eta_k
A_{kl}\eta_l ~ ~ , ~ ~ \mbox{where,}\\
A_{lk} & := & \underbrace{2(1/\epsilon-\epsilon\
\omega/4)}_{a}\delta_{lk} - \underbrace{(1/\epsilon + \epsilon\
\omega^2/4)^2}_{b}\left\{ \delta_{l,k+1} + \delta_{l,k-1} \right\}~ , ~
l,k = 1, \dots N-1 . \\
& = & \left( \begin{array}{ccccc} a & -b & 0 & \dots & 0 \\ -b & a & -b
& \dots & 0 \\ 0 & -b & a & -b & \dots 0 \\ . & . & . & . & . \\ . & . &
. & . & . \\ 0 & 0 & \dots & -b & a \end{array} \right) \hspace{1.0cm}
\mbox{A tri-diagonal, symmetric matrix.} \\
\therefore \langle q'',t''|q',t'\rangle_J & := &
\left[e^{\frac{i}{\hbar}S^J_{cl}(t',t'')}\right] \left[\lim_{\epsilon\to
	0}\frac{1}{C(\epsilon)} \int_{-\infty}^{\infty}
	\frac{d^{N-1}\eta}{C(\epsilon)^{N-1}} exp\left\{
\frac{i}{2\hbar} \sum_{i,j=1}^{N-1}\eta_iA_{ij}\eta_j \right\}\right]
\end{eqnarray*}
For a single variable we have $\int_{-\infty}^{\infty}dx e^{-ax^2} =
\sqrt{\pi/a} \Rightarrow \int_{-\infty}^{\infty}dx e^{iax^2} =
\sqrt{\pi/(-ia)}$. Its multi-dimensional generalization is
\[
	\int_{-\infty}^{\infty}d^nx e^{i\sum_{ij} x_iA_{ij}x_j} =
	\frac{(i\pi)^{n/2}}{\sqrt{det\ A}}
\]
Using this, our transition amplitude takes the form,
\begin{eqnarray*}
\langle q'',t''|q',t'\rangle_J & = &
\left[e^{\frac{i}{\hbar}S^J_{cl}(t',t'')}\right] \left[\lim_{\epsilon\to
0}\frac{1}{C(\epsilon)} \frac{1}{(\epsilon)^{N-1}}
\frac{(i\pi\hbar)^{(N-1)/2}}{\sqrt{det\ A}}\right]
\end{eqnarray*}
Using $C(\epsilon) = \sqrt{2i\pi\epsilon\hbar}$ since we have taken unit
mass, $m = 1$, we have,
\begin{equation}
\boxed{ \langle q'',t''|q',t'\rangle_J  =
\left[e^{\frac{i}{\hbar}S^J_{cl}(t',t'')}\right] \left[\lim_{\epsilon\to
0} \frac{1}{\sqrt{2\pi i\hbar}(\epsilon)^{N/2}} \frac{1}{\sqrt{det\
A(\epsilon)}}\right] }
\end{equation}

Now we need to compute the determinant of the tridiagonal, symmetric
matrix. This is usually solved by using a recursion relation. Denote
$D_n := det\ A_{n\times n}$ where $A$ has the form given above. Clearly,
$D_1 = a$ and $D_2 = a^2-b^2$. By checking for $4\times 4, 5\times 5$
matrices, it is easy to see that $D_n$ satisfies the recursion relation,
\[
\boxed{D_n = a D_{n-1} - b^2D_{n-2} ~ , ~ D_0 := 1, D_1 = a ~ ~ ; ~ ~ a
= 2\left(\frac{1}{\epsilon}-\frac{\epsilon\ \omega^2}{4}\right) ~ , ~ b
= \frac{1}{\epsilon} + \frac{\epsilon\ \omega^2}{4}\ .  }
\]
We can pull out a factor of $\epsilon^{-1}$ from $A$ and since our $A$
is of order $(N-1)$, we pull out a factor of $(\epsilon)^{-(N-1)/2}$.
This replaces the second $[\dots]$ by $\lim_{\epsilon\to 0}(2\pi\hbar
\sqrt{\epsilon det A'})^{-1}$. The $A'$ matrix has elements $a' := 2 -
\epsilon^2\omega^2/2\ , \ b' := 1+\epsilon^2\omega^2/4$.

\underline{Note:} For a given $N, \epsilon = T/N$, hence (suppressing
the primes)  $a, b$ have an $N$ dependence. The matrix itself is also
$(N-1)\times(N-1)$ and our notation $D_n$ as the determinant of
$A_{n\times n}$ is valid for $n \le N-1$. For a given $N$, the $D_{n\ge
N}$ is {\em not} defined. Hence, the recursion relation is a {\em
difference equation with constant coefficients} \cite{Elaydi}. It is
important to keep the distinction between fixed $N$ and a variable $n$.
The $a,b$ are functions of $N$ but independent of $n$.

Such difference equations are solved by the ansatz, $D_n = \lambda^n$.
Shifting $n\to n+2$, we write the difference equation as,
\[
	\forall\ 0 \le n \le N-3\ : ~ D_{n+2} - aD_{n+1} + b^2D_n = 0 ~
	~ , ~ ~ D_0 = 1, D_1 = a .
\]
Substitution gives the characteristic equation $\lambda^2 -a\lambda +
b^2 = 0$. Its solutions, for $\epsilon \ll 1$,  are
\begin{eqnarray*}
\lambda_{\pm} & \approx & (1-\frac{\epsilon^2\omega^2}{4}) \pm \sqrt{
(1-\frac{\epsilon^2\omega^2}{2}) - (1+\frac{\epsilon^2\omega^2}{2})} ~ ~
\approx ~ 1 - \frac{\epsilon^2\omega^2}{4} \pm i\epsilon\ \omega \\
\therefore \lambda_{\pm} & \simeq & ~ 1 \pm i\epsilon \omega \ . ~
\Rightarrow  ~ D_n = \alpha \lambda_+^n + \beta \lambda_-^n ~ ~ ~ ~
\forall ~ n \ \in\ [0, N-1].\\
& & \mbox{Initial conditions give,} ~ ~ \alpha =
\frac{1}{2}\left(1-\frac{i}{\epsilon\omega}\right)~ , ~ \beta =
\frac{1}{2}\left(1+\frac{i}{\epsilon\omega}\right)\ .
\end{eqnarray*}

The desired determinant is then given by,
\begin{eqnarray*}
D_{N-1} & = & \frac{1}{2}\left(1-\frac{i}{\epsilon\omega}\right)
(1+i\epsilon\omega)^{N-1} +
\frac{1}{2}\left(1+\frac{i}{\epsilon\omega}\right)
(1-i\epsilon\omega)^{N-1} ~ , ~ \epsilon = T/N\ \\
& \xrightarrow[N\to\infty]{} & \frac{1}{2}i \left(1-\frac{iN}{\omega T}\right)
(1+i\omega T/N)^{N} +
\frac{1}{2}\left(1+\frac{iN}{\omega T}\right)
(1-i\omega T/N)^{N}  \\
& \to & \left(-\frac{iN}{2\omega T}\right)\left(e^{i\omega T} -
e^{-i\omega T}\right) ~ = ~ \frac{N}{\omega T}sin(\omega T) \\
\therefore \epsilon D_N & \to & \frac{sin(\omega T}{\omega}\ .
\end{eqnarray*}
We finally get,
\begin{equation}
	\boxed{
\langle q'',t''|q',t'\rangle_J  = 
\left[e^{\frac{i}{\hbar}S^J_{cl}(t',t'')}\right]\left[\frac{1}{\sqrt{2\pi
i\hbar}}\sqrt{\frac{\omega}{sin(\omega T)}}\right] . } 
\end{equation}

\underline{Note:} All the $J$ dependence is in the first factor only
while the ``quantum correction'' are in the second factor and
independent of $J$.

\underline{Note:} This method of evaluating the amplitude near a
classical solution can also be adopted for more general (non-linear)
equations of motion. The $S_{cl}$ always comes out, the term linear in
$\eta$ always vanishes while the $o(\eta^2)$ terms always gives the
(determinant)$^{-1/2}$. In the general context, it is termed as a
semi-classical approximation which is exact for the oscillator.

The calculation of the first factor requires solving the equation of
motion with the source function and then evaluating the action for
$q_{cl}(t)$. This is little involved as the equation is inhomogeneous
and requires use of Green's function. The Green's function can be
obtained by directly solving the differential equation with $\delta$-
function source and matching the discontinuity due to the delta
function. We just note the final result [Problem 3.11 of Feynman-Hibbs,
Abers-Lee].
\begin{eqnarray}
	S_{cl}^J(t',t'') & = & \frac{\omega}{2sin(\omega
	T)}\left[((q')^2 + (q'')^2)cos(\omega T) - 2q' q''\right]
	\nonumber \\
	& & + \frac{q''}{sin(\omega T)}\int_{t'}^{t''}dt
	J(t)sin(\omega(t-t')) + \frac{q'}{sin(\omega T)}\int_{t'
	}^{t''}dt J(t)sin(\omega(t''-t)) \nonumber \\
	& & -\frac{1}{\omega sin(\omega
	T)}\int_{t'}^{t''}d\sigma\int_{t'}^{\sigma} d\tau
	J(\sigma)J(\tau) sin(\omega(t''-\sigma). sin(\omega(\tau-t'))
\end{eqnarray}
This exercise was done to illustrate the schematics of evaluating the
path integral directly using the definition.

{\em Note:} If we have a system of two degrees of freedom, the $q(t)$
and the $J(t)$ which are coupled by a linear coupling, $\int_{t'}^{t''}
dt J(t)q(t)$, then the $S_{cl}^J(t',t'')$ can be viewed as an effective
action contribution after integrating out the $q(t)$ degree of freedom.
This has the form $\sim \int dt J(t)\alpha(t) - \int dt J(t)\int^t
d\sigma \beta(t,\sigma)J(\sigma)$. This last term is a {\em non-local
term}.
\subsection{Alternative Expression for $Z[J]$} \label{AlternativeZj}
We now consider another method which is closer to what is done is field
theory. We begin by modifying the Hamiltonian operator as:
$\boxed{\hat{H} \to (1-i\epsilon)\hat{H}}~$\cite{Srednicky}. Then,
\begin{eqnarray*}
|q',t'\rangle & = & e^{\frac{i}{\hbar}t'\hat{H}}|q'\rangle =
\sum_n\varphi_n^*(q')e^{\frac{i}{\hbar}t' E_n}|n\rangle  \to
\sum_n\varphi_n^*(q')e^{\frac{i}{\hbar}t' (1-i\epsilon)E_n}|n\rangle 
\end{eqnarray*}
Assuming $E_0 = 0$ for convenience and taking the limits, we get
\begin{equation}
\boxed{ |q',t'\rangle ~  \xrightarrow[t'\to-\infty]{} ~
\varphi^*_0(q')|0\rangle  \hspace{1.0cm} \mbox{and} \hspace{1.0cm}
\langle q'',t''| ~ \xrightarrow[t''\to \infty]{} ~ \langle
0|\varphi_0(q'') \ . }
\end{equation}
Thus, $|q',t'\rangle, \langle q'',t''|$ both go to the (presumed
non-degenerate) ground state as $t'\to -\infty$ and $t''\to \infty$,
{\em provided} we make the substitution: $\hat{H} \to
(1-i\epsilon)\hat{H}$. We will work with the limits. 

Consider, for our 1-dimensional oscillator, $H = p^2/2 + \omega^2q^2/2$,
\[
	\langle 0|0\rangle_J = \int\cal{D}p\cal{D}q exp\left[
	\frac{i}{\hbar} \int_{-\infty}^{\infty}dt \left\{ p\dot{q} -
(1-i\epsilon)H + J(t)q\right\}\right]
\]
We do momentum integration by using $(1-i\epsilon)p^2/2 \approx
\Case{p^2}{2(1+i\epsilon)}$. This gives the term
$\Case{1}{2}(1+i\epsilon)\dot{q}^2$ and we get,
\begin{equation}
\boxed{
\langle 0|0\rangle_J = \int\cal{D}q exp\left[ \frac{i}{\hbar}
\int_{-\infty}^{\infty}dt \left\{ \frac{1}{2}(1+i\epsilon)\dot{q}^2 -
\frac{1}{2}\omega^2(1-i\epsilon)q^2 + J(t)q \right\}\right] . 
}
\end{equation}

Define the Fourier transform,
\[
\tilde{q}(\nu) := \int_{-\infty}^{\infty}dt e^{i\nu t}q(t)
\leftrightarrow q(t) = \frac{1}{2\pi}\int_{-\infty}^{\infty}d\nu
e^{-i\nu t}\tilde{q}(\nu)
\]
and similarly for $J(t)$. Substitution give,
\begin{eqnarray*}
L & = & \frac{1}{2}\int_{-\infty}^{\infty} \frac{d\nu}{2\pi}
\int_{-\infty}^{\infty} \frac{d\nu'}{2\pi} e^{-i(\nu+\nu')t} \Big[
	\big\{-(1+i\epsilon)\nu\nu' - (1-i\epsilon)\omega^2\big\}
	\tilde{q}(\nu) \tilde{q}(\nu') \\
& & \hspace{5.5cm} + \frac{\tilde{J}(\nu)\tilde{q}(\nu') +
\tilde{J}(\nu')\tilde{q}(\nu)}{2} \Big]
\end{eqnarray*}
The expression in the braces simplifies to $\{\dots\} = \nu^2 - \omega^2
+ i\epsilon(\nu)~ , ~ \epsilon(\nu) := \epsilon(\nu^2+\omega^2)$. Define
$\tilde{x}(\nu) := \tilde{q}(\nu) + \Case{\tilde{J}(\nu)}
{\nu^2-\omega^2 + i\epsilon(\nu)}$. The the action becomes,
\[
S = \frac{1}{2}\int_{-\infty}^{\infty}\frac{d\nu}{2\pi}\left[
\tilde{x}(\nu)\big(\nu^2-\omega^2+i\epsilon(\nu)\big)\tilde{x}(-\nu) -
\frac{\tilde{J}(\nu)\tilde{J}(-\nu)}{\nu^2-\omega^2+i\epsilon(\nu)}
\right]
\]
The shift, $\tilde{q}(\nu) \to \tilde{x}(\nu)$ is a {\em constant
shift}, the shift is independent of $\tilde{q}$. Hence the Jacobian of
the transformation will be 1 and $\cal{D}\tilde{q}(\nu) =
\cal{D}\tilde{x}(\nu)$. Taking Fourier transform, the shift takes the
form $q(t) \to x(t) + f(t)$ with $f(t)$ independent of $q(t)$. This is
also a constant shift and in the time domain too, and we expect
$\cal{D}q(t) = \cal{D}x(t)$.
\begin{eqnarray}
	\therefore\langle0|0\rangle_J & = & exp\left[\frac{i}{2\hbar}
	\int_{-\infty}^{\infty}\frac{d\nu}{2\pi} \left(
\frac{\tilde{J}(\nu)\tilde{J}(-\nu)} {
-\nu^2+\omega^2-i\epsilon(\nu)}\right)\right]\times \\
& & \hspace{1.0cm} \left[\int\cal{D}\tilde{x}(\nu)
exp\frac{i}{2\hbar}\int_{-\infty}^{\infty}\frac{d\nu}{2\pi}
\tilde{x}(\nu)\big(\nu^2-\omega^2+i\epsilon(\nu)\big)\tilde{x}(-\nu)
\right]
\end{eqnarray}
The second factor is a path integral independent of $J$ and hence equals
$\langle 0|0\rangle_{J=0}$. However, without any source interaction, the
ground state remains a ground state and hence $\boxed{\langle 0|0
	\rangle_{J=0} = 1!}$. This gives finally,
\begin{equation}
	\boxed{
	\langle 0|0 \rangle_{J}= exp\frac{i}{2\hbar}
	\int_{-\infty}^{\infty}\frac{d\nu}{2\pi} \left(
	\frac{\tilde{J}(\nu)\tilde{J}(-\nu)} {
-\nu^2+\omega^2-i\epsilon(\nu)}\right)
}
\end{equation}
We never needed to evaluate the path integral!

The above expression can also be expressed in the time domain as,
\begin{eqnarray} \label{Z-Defn2}
	\langle 0|0\rangle_J & = & exp
	\frac{i}{2\hbar}\int_{-\infty}^{\infty} dt' dt''
	J(t'')G(t''-t')J(t') \hspace{1.5cm} \mbox{with,} \\
	G(t''-t') & = & \int_{-\infty}^{\infty}\frac{d\nu}{2\pi}
	\frac{e^{-i\nu(t''-t')}}{-\nu^2+\omega^2-i\epsilon(\nu)} ~ = ~
	\frac{i}{2\omega}e^{-i\omega|t''-t'|} \hspace{0.0cm}\mbox{(by
	contour integration)} ~ ~ ~ ~ ~ 
\end{eqnarray}
\underline{Note:} The $\nu$ dependence in the $\epsilon(\nu)$ does not
affect the contour integration.

We have discussed two different methods of computing the transition
amplitude in the limit of infinite time separation. Notice that the
$\langle 0|0\rangle_J$ above is the same as the $Z[J]$ defined earlier
in eq. (\ref{Z-Defn1}) since $\int dq' |q',t'\rangle \varphi_0(q',t') =
|0,t'\rangle$ and likewise for $\langle 0,t''|$.

The above frequency domain form is very convenient to obtain the correlation
functions as seen below.

Our general formula (\ref{CorrlnFn1}) tells us that $\langle
0|T\{Q(t_1)\dots Q(t_n)\}|0\rangle$ is given by the functional
derivatives of $Z[J]$ evaluated at $J=0$. Choosing the above form of
$Z[J]$ we see that,
\begin{eqnarray}
\langle 0|T\{Q(t_1)Q(t_2)\}|0\rangle & = & (-i\hbar)^2
\frac{\delta^2\langle 0|0\rangle_J}{\delta J(t_1)\delta J(t_2)}\bigg|_{J
= 0} \\
& = & (-i\hbar)^2\frac{\delta}{\delta J(t_1)}\left\{ \frac{i}{2\hbar}
2\int_{-\infty}^{\infty}dt' G(t_2-t_1)J(t')\times \right. \nonumber \\
& & \hspace{4.0cm}\left.  exp\Big(\frac{i}{2\hbar}\int\int J G
J\Big)\right\}\bigg|_{J=0}\nonumber \\
& = & (-i\hbar)^2(i/\hbar) G(t_2-t_1) = -i\hbar G(t_2-t_1).
\end{eqnarray}
The derivative of the $e^{\int JGJ}$ does not contribute in the limit of
$J=0$. Taking one more derivative to get the three point function, we
have
\begin{eqnarray*}
\frac{\delta}{\delta J_1}\frac{\delta}{\delta J_2}\left\{
\frac{i}{\hbar}\int dt G(t_3-t)J(t) \cdot \langle 0|0\rangle_J\right\}
~ ~ = \hspace{7cm} & & \\
\frac{\delta}{\delta J_1}\left\{ \frac{i}{\hbar}G(t_3-t_2)\cdot\langle
0|0\rangle_J + \frac{i}{\hbar}\int dt\ G(t_3-t)J(t)\cdot
\frac{i}{\hbar}\int dt'\ G(t_2-t')J(t')\cdot\langle0|0\rangle_J\right\}
& & \\
~ = ~ 0 + 0 ~ = 0 ~ \hspace{1.0cm} \because ~ \mbox{there is always a
factor of $J$ which kills the term at $J=0$.} & & 
\end{eqnarray*}

For 4 derivatives, we will have,
\begin{eqnarray*}
\frac{\delta}{\delta J_0}\bigg\{ \frac{i}{\hbar}G(t_3-t_2) \cdot
	\frac{i}{\hbar}\int dt G(t_1-t)J(t) \cdot\langle 0|0\rangle_J +
	\hspace{6.0cm} & & \\
\frac{i}{\hbar}G(t_3-t_1) \cdot \frac{i}{\hbar}\int dt G(t_2-t)J(t)
\cdot\langle 0|0\rangle_J + \hspace{6.0cm} & & \\
\frac{i}{\hbar}\int dt G(t_3-t)J(t) \cdot \frac{i}{\hbar} G(t_2-t_1)
\cdot\langle 0|0\rangle_J + \hspace{6.0cm} & & \\
\Big(\frac{i}{\hbar}\Big)^3 \int dt G(t_3-t)J(t)\cdot \int dt'
G(t_2-t')J(t') \cdot \int dt'' G(t_1-t'')J(t'')\cdot\langle
0|0\rangle_J\bigg\} & & 
\end{eqnarray*}
Carrying out the $J_0$ differentiation and putting $J=0$, only the first
3 terms contribute since they have a single factor of $J(t)$. For a
4-point function, this is multiplied by $(-i\hbar)^4$ and we get,
\begin{equation}
	\boxed{ \langle 0|T\{Q(t_0)Q(t_1)Q(t_2)Q(t_3)\}|0\rangle =
	(-i\hbar)^2\Big\{ G_{01}G_{23} + G_{02}G_{31} + G_{03}G_{12}
\Big\} ~, ~ G_{ij} \leftrightarrow G(t_i-t_i)\ . }
\end{equation}
We recognize this as the same pattern seen in the vacuum expectation
value calculation using the Wick's theorem. Indeed, the quantum field
theory Wick's theorem shows up here just as a result of the functional
differentiation. 

We now have all the definitions, notations and pattern of computation
needed in the field theory generalization.

\newpage 
\section{Path Integrals in Quantum Field Theory}\label{QFTPathIntegral}

We consider the path integral formulation of a field theory,
specifically a scalar field theory. While discussing the field as a
dynamical system, we had already noted that the notation
$\phi(t,\vec{x})$ refers to infinitely many degrees of freedom labeled
by $\vec{x} \in \mathbb{R}^3$ (for us). Thus we could view
$\phi(t,\vec{x})$ as $\phi_{\vec{x}}(t)$ and draw analogy with $q_i(t)$.
Each of these degrees of freedom can be quantized a la path integral
exactly as the single degree of freedom discussed before. In place of
$\int\cal{Dq}(t) \xrightarrow[T=\epsilon N]{}
\int\prod_{k=0}^{N-1}\Case{dq_k}{C(\epsilon)}$, we now have
$\int\cal{D}\phi(x) \xrightarrow[T=\epsilon N]{} \int\prod_{\vec{x}\in
\Sigma} \prod_{k=0}^{N-1}d\phi_{\vec{x},k}$. We have dropped the
$C(\epsilon)$ which will be subsumed in the overall normalization
constant. The action for the field theory is as the usual one. Thus, the
vacuum-to-vacuum transition amplitude in presence of a source is denoted
as,
\begin{equation}\label{Z-Field}
	\langle 0|0\rangle_J := \int \cal{D}\phi exp\frac{i}{\hbar}\int
	d^4x\Big\{\mathcal{L}(\phi,\partial_{\mu}\phi)+
	J(x)\phi(x)\Big\} ~ =: ~ Z[J]
\end{equation}
The `paths' implicit in the measure $\cal{D}\phi$ are of course paths in
the configuration space of the field i.e. the space of all
$\phi(\vec{x})$ at any fixed $t$. A path itself connects
$\phi_1(\vec{x},t_1)$ to $\phi_2(\vec{x},t_2),\ \forall\ \vec{x}$. The
paths are {\em not} paths in the space-time.

For the Lagrangian density $\mathcal{L}$, we write $\mathcal{L} =
\mathcal{L}_0 + \mathcal{L}_1$, with $\boxed{\mathcal{L}_0 :=
-\Case{1}{2}\partial^{\mu}\phi\partial_{\mu}\phi -
\Case{1}{2}m^2\phi^2.}$ The $m^2$ has $-i\epsilon$ implicit, in
anticipation. The $\mathcal{L}_1$ will typically involve interaction
terms, polynomials in $\phi$ such as $g_3\phi^2(x) + g_4\phi^4(x) \dots
$ etc. Consider 
\[
	Z_0[J] := \int \cal{D\phi}\ exp\Big\{\frac{i}{\hbar}\int d^4x (
	\mathcal{L}_0 + J(x)\phi(x) )\Big\}\ .
\]
We go to the momentum space as in the case of the single particle.

Let $\tilde{\phi}(k) := \int d^4x e^{-ik\cdot x}\phi(x) \leftrightarrow
\phi(x) = \int \Case{d^4k}{(2\pi)^4} e^{ik\cdot x}\tilde{\phi}(k)$, with
$k\cdot x := -k^0x^0 + \vec{k}\cdot\vec{x}$ and $k^0$ is the integration
variable in $d^4k$ and not $k_0$. Substituting in the action and doing
the $d^4x$ integration gives,
\[
	S_0 = \frac{1}{2}\int\frac{d^4k}{(2\pi)^4}\left[
	-\tilde{\phi}(k)(k^2+m^2)\tilde{\phi}(-k) +
\tilde{J}(k)\tilde{\phi}(-k) + \tilde{J}(-k)\tilde{\phi}(k)\right]
\]
Note that $S_0$ is {\em not} $\int \cal{L}_0$, but includes the source.

Define $\tilde{\phi}(k) := \tilde{\chi}(k) +
\Case{\tilde{J}(k)}{k^2+m^2}$. Then $\cal{D\phi} = \cal{D\chi}$ since we
have a ``constant shift transformation''. This also holds in the
$\phi(x)$ space. Then, as in the case of the single oscillator, we get
\[
S_0 = \frac{1}{2}\int\frac{d^4k}{(2\pi)^4}\left[
\frac{\tilde{J}(k)\tilde{J}(-k)}{k^2+m^2} -
\tilde{\chi}(k)(k^2+m^2)\tilde{\chi}(-k)\right]\ .
\]
The path integral $\int\cal{D\chi}$, just gives a factor independent of
$\tilde{J}$, namely $\langle 0|0\rangle_{J=0} = 1$ and we write,
$\boxed{\hbar=1\ \mbox{from now on}}$,
\begin{eqnarray*}
\langle 0|0\rangle_J & = & Z_0[J] = exp \left[i\int\frac{d^4k}{(2\pi)^4}
\left\{\frac{1}{2} \frac{\tilde{J}(k) \tilde{J}(-k)} {k^2+m^2-i\epsilon}
\right\}\right] \\
& = & exp \left\{i\int d^4x\int d^4x'\
\frac{1}{2}J(x)\Delta(x-x')J(x')\right\} ~ ~ \mbox{where,} \\
\Delta(x-x') & = & \int\frac{d^4k}{(2\pi)^4}\
\frac{e^{ik\cdot(x-x')}}{k^2+m^2-i\epsilon} ~ ~,~ ~ \mbox{(Feynman
propagator)}
\end{eqnarray*}
Note that the Feynman propagator arose because of the $-i\epsilon$
included in the mass term in the $\cal{L}$. 

As discussed for the oscillator, we get
\begin{eqnarray*}
	\langle 0|T\left\{\phi(x_1)\phi(x_2)\right\}|0\rangle & = &
	(-i)\frac{\delta}{\delta J(x_1)}\cdot (-i)\frac{\delta}{\delta
	J(x_2)}\cdot Z_0[J]\Big|_{J=0} \\
	& = & (-i)\frac{\delta}{\delta J(x_1)}\left\{ Z_0[J]\cdot\int
	\Delta(x'-x_2)J(x')d^4x'\right\} (-i)(i)\Big|_{J=0} ~
	\Rightarrow \\
	\langle 0|T\left\{\phi(x_1)\phi(x_2)\right\}|0\rangle & = & -i
	\Delta(x_1-x_2) \ . 
\end{eqnarray*}
Exactly as for the Oscillator, we get 
\begin{eqnarray*}
	\langle 0| T\left\{\phi(x_1)\phi(x_2)\phi(x_3)\phi(x_4)0\right\}
	|0 \rangle = \hspace{9.0cm} & & \\
	(-i)^2\Big[ \Delta(x_1-x_2)\delta(x_3-x_4) +
	\Delta(x_1-x_3)\delta(x_2-x_4) + \Delta(x_1-x_4)\delta(x_2-x_3)
\Big]& &  
\end{eqnarray*}
which easily generalizes to,
\begin{eqnarray*}
\langle 0| T\left\{\phi(x_1)\dots\phi(x_{2n})\right\} |0 \rangle =
\hspace{9.0cm} & & \\
(-i)^n\Big[
\sum_{pairings}\Delta(x_1-x_2)\delta(x_3-x_4)\dots\Delta(x_{2n-1}-x_{2n})
+ \mbox{permutations}\Big]
\end{eqnarray*}
This is what we had obtained as the {\em Wick's theorem}.

\underline{Note:} Wick's theorem expressed the time ordered product of
quantum fields in terms of the normal ordered product plus contractions.
The normal ordering was for all Poincare generators to ensure invariance
of the vacuum. The theorem was also proved for free fields for which we
do have Fourier decomposition.

Here too the ``Wick's theorem'' is again seen for free fields. Here it
is no more than the chain rule of differentiation.

Consider now interacting fields, by which we mean $\mathcal{L} =
\mathcal{L}_0 + \mathcal{L}_1$. Recall that $\langle 0|Q(t)|0\rangle =
\int \cal{Dq}\ q(t)e^{iS}$ and we can get the $q(t)$ in the integrand by
$\Case{\delta}{\delta J(t)} S_J|_{J=0}$ (remember $S = S_J|_{J=0}$).
Thus, insertion of $q(t)$'s or $\phi(x)$'s in field theory, can be
effected by $\Case{\delta}{\delta J(x)}S_J|_{J=0}$. If we expand
$e^{i\int\mathcal{L}_1(\phi)} = \Sigma_{n=0}^{\infty}
[i\int\mathcal{L}_1]^n/n!$, then we have integrals of polynomials of the
fields which can be expressed as $\Case{\delta}{\delta J}S_J$. The
$\delta_J$ can be taken outside of the path integration. Thus we write,
\begin{eqnarray}
	Z[J] & := & \langle 0|0\rangle_J^{int}  :=  \int\cal{D\phi}\
	exp\ i\int d^4x \Big(\mathcal{L}_0(\phi) + \mathcal{L}_1(\phi) +
	J(x)\phi(x)\Big) \nonumber \\
	& \propto & e^{i\mathcal{L}(-i\delta_{J(x)})} \cdot \int
	\cal{D\phi}e^{i\int d^4x \{ \mathcal{L}_0 + J(x)\phi(x)\} }
	\hspace{2.0cm} \mbox{or,} \nonumber \\
	Z[J]	& \propto & \left[exp\ i\int d^4x\mathcal{L}_{1}\left(
	-i\frac{\delta}{\delta J(x)}\right)\right]\cdot Z_0[J] ~ , ~
	\mbox{with} \\
	Z_0[J] & := & exp\Big(\frac{i}{2}\int d^4x\int d^4y
	J(x)\Delta(x-y)J(y)\Big)
\end{eqnarray}

In the oscillator case, for $Z_0[J]$ we had the $exp\Case{i}{2}\int\int
J\Delta J$ term and a Gaussian path integral which turned out to be
equal to $Z_0[J=0]$. By taking the asymptotic time limits, we could
assert this to be equal to 1 since it was vacuum-to-vacuum amplitude
without any source. In presence of interactions, even with $J=0$, the
vacuum may not remain unaffected and we cannot justify $\langle
0|0\rangle^{int}_{J=0} = 1$. Instead, the proportionality constant is
determined by {\em demanding} $Z[J=0] = 1$.

The prescription to compute $Z[J]$ via the
$e^{i\mathcal{L}_1(-i\delta_J)}Z_0[J]$, is in effect the perturbative
prescription. Let us see how this works in an example
$\mathcal{L}_1(\phi) = g\phi^3(x)/3!$. Then $Z[J] \propto
e^{\Case{g}{6}\delta^3_J}\cdot Z_0[J]$. Expand the exponential,
\begin{eqnarray*}
	Z[J] & = & \mathcal{N}\sum_{v=0}^{\infty}\frac{1}{v!}
	\left[\frac{ig}{3!}\int d^4x\left(-i\frac{\delta}{\delta
	J(x)}\right)^3\right]^v \times  \\
	& &
	\hspace{3.0cm}\sum_{p=0}^{\infty}\frac{1}{p!}\left[\frac{i}{2}
	\int d^4y\int d^4z J(y)\Delta(y-z)J(z)\right]^p \ .
\end{eqnarray*} 
This too is a power series in $J$ as is $Z_0[J]$, but with different
coefficients. Consider a particular term in this double sum with a fixed
$v$ and $p$.

$\bullet$ Then we have $3v$ derivatives acting on $2p\ J$'s, leaving
$E = (2p -3v) J$'s. Clearly, $E\ge 0$ must hold and there are several
such term;

$\bullet$ The overall {\em numerical factor} associated with such a group of
terms is:
\[
	\left(\frac{ig}{3!}\right)^v
	\frac{(-i)^{3v}}{v!} \frac{(i/2)^p}{p!} =
	\frac{i^{v-3v+p}}{(3!)^v v! p!}g^v =
	\left(\frac{g}{3!}\right)^v\frac{1}{v! p!}(i)^{E+v-p}
\]

$\bullet$ The {\em combinatorial factor} resulting from the number of ways he
derivative acts is: $3v$ derivatives on $2p J$'s give
$\Case{(2p)!}{(2p-3v)!} $ (This is because the first derivative can act
on $2p$ J's, second on $(2p-1)$, \dots $(3v)^{th}$ on $(2p-3v+1)$ J's.)

We can generate and keep track of the various derivatives by denoting 
\begin{center}
	\fbox{
$
\begin{array}{lcccl}
	\frac{ig}{3!}\frac{\delta^3}{\delta J^3(x)} & \hspace{0.75cm}
	\leftrightarrow \hspace{0.75cm} & \hspace{0.75cm} 
	\frac{ig}{3!} & \hspace{0.75cm}
	\vcenter{\hbox{
	\begin{tikzpicture}
		\begin{feynman}
			\vertex (v);
			\vertex [left =0.5of v] (a);
			\vertex [above right =0.5of v] (b);
			\vertex [below right =0.5of v] (c);
			\diagram*{
				(v) -- (a), (v) -- (b), (v) -- (c),
			};
		\end{feynman}
	\end{tikzpicture}
	}} 
	\hspace{0.75cm} & \hspace{0.75cm} \mbox{Has $\int d^4x$} \\
	\frac{i}{2}J(x)\Delta(x-y)J(y) & \hspace{0.75cm} \leftrightarrow
	\hspace{0.75cm} &
	\hspace{0.75cm} \frac{i}{2} & \hspace{0.75cm}  
	\vcenter{\hbox{
	\begin{tikzpicture}
		\begin{feynman}
			\vertex [dot] (v) { };
			\vertex [right =0.75of v,dot] (a){ };
			\diagram*{
				(v) -- (a),
			};
		\end{feynman}
	\end{tikzpicture}
	}} 
	& \hspace{0.75cm} \mbox{Has $\int d^4y\int d^4z$} \ ,
\end{array}
$
}
\end{center}
and the operation of evaluating the derivative by joining the free ends
from the `vertices' to the free ends of the `propagator lines' - exactly
as we represented the Wick contractions. 

The number of terms with a give $v,p$ is the number of ways of joining
the free ends and generating {\em a diagram}. Some of the diagrams may
have the identical factors associated with them For example, the $3
\delta_J$'s at a given vertex can be permuted in joining up with the
propagator lines. This clearly gives a factor of $3!$. Likewise ends of
propagator lines gives a factor of $2!$. The diagram will look the same
if the $v$ vertices are themselves permuted (the $\int d^4x$ are dummy
variables) and this gives $v!$. Similarly the propagator lines give
$p!$. Thus, all the numerical factors, except $i, g$ cancel out.

Compared to the previous counting based on Wick's theorem, we have a
double expansion and lines (pairs of $J$'s) to be contracted with the
edges of the vertices. Secondly, in the Wick's theorem, thanks to the
normal ordering of the interaction terms, we don't have self
contractions at the vertices ($\Delta(x,x)$ propagators). In the present
approach, there is no ``normal ordering'' and we do get such term. Eg.
\begin{center} $\begin{array}[c]{cl}
		\vcenter{\hbox{
		\begin{tikzpicture}
			\begin{feynman}
				\vertex (v); 
				\vertex [right=0.15of v] (v') {$x$};
				\vertex [left =0.5of v] (a);
				\vertex [above right =0.5of v] (b);
				\vertex [below right =0.5of v] (c);
				\vertex [right =2.5of v] (d);
				\vertex [above =0.75of d, dot] (d1){ };
				\vertex [above =0.25of d1] (d'){$y$};
				\vertex [below =0.75of d, dot] (d2){ };
				\vertex [below =0.25of d2] (d''){$z$};
				\diagram*{
					(v) --(a),
					(v) --(b),
					(v) --(c),
					(b) --[scalar,
					       half left,looseness=0.5,
						edge label={$\delta(x-y)$}
					] (d1),
					(c) --[scalar,
						half right,looseness=0.5,
						edge label'={$\delta(x-z)$}
					] (d2),
					(d1) --[scalar,
						half left,looseness=0.1,
						edge label={$\Delta(x-z)$}
					] (d2),
				};
			\end{feynman}
		\end{tikzpicture}
		}} 
		& \mbox{Such diagrams are called
		`tadpoles'. } ~ ~
		\left(
		\vcenter{\hbox{
		\begin{tikzpicture}
			\begin{feynman}
				\vertex (v);
				\vertex [right=0.5of v] (a){\circled{ ~
				~ ~ }};
				\diagram*{
					(v) -- (a),
				};
			\end{feynman}
		\end{tikzpicture}
		}}
	\right)
	\end{array}$ \end{center}
Thus we have more diagrams and different ways of working out the total
combinatorial factor. The above bulleted points have gotten rid of all
explicit factors, but could involve an over counting. This is
compensated by {\em dividing} by a factor called the {\em symmetry
factor}. This is a geometrical problem of identifying the groups of
diagrams. Here are a couple of examples \cite{Srednicky}.
\begin{center}
	\begin{tabular}[c]{ccl}
		$
		\vcenter{\hbox{
		\begin{tikzpicture}
			\begin{feynman}
				\vertex (v);
				\vertex[left=0.5of v] (a){ \circled{~ ~ ~}
				};
				\vertex[right=0.5of v] (b){ \circled{~ ~ ~}
				};
				\diagram*{
					(v) -- (a),
					(v) -- (b),
				};
			\end{feynman}
		\end{tikzpicture}
		}} 
		$
		& : & \parbox{0.6\textwidth}{
			We can exchange the two loops in 2 ways.
			We can exchange the ends of each loop in $2\times
			2$ ways. Therefore the total symmetry factor is
			$2\cdot 2\cdot 2  = 8$.
		} \\
		& & \\
		$
		\vcenter{\hbox{
		\begin{tikzpicture}
			\begin{feynman}
				\vertex (v);
				\vertex [right =of v] (a);
				\diagram*{
					(v) -- (a),
					(v) --[half right,looseness=1.5]
					(a),
					(v) --[half left,looseness=1.5]
					(a),
				};
			\end{feynman}
		\end{tikzpicture}
		}}
		$
		& : & \parbox{0.6\textwidth}{
			The 3 propagators can be permuted in $3\cdot 2$
		ways. The 2 vertices can be exchanged in 2 ways. So the
		symmetry factor is: $3!2! = 12$.
	}
	\end{tabular}
\end{center}
A general statement and a derivation may be seen in the appendix of
Sterman's book \cite{Sterman}.

In the double expansion, we have terms with no $J$'s, a single $J$,
$2J$'s \dots etc. When $J=0$ is put, only the terms with no $J$ survive.
These are termed as the ``vacuum bubbles'' and are the contributions to 
the $Z[J=0]$. 

A general contribution to a {\em given number} of left over $J$'s would
consist of  product of topologically connected diagrams. Let
$C_{\gamma}$ denote the contribution of a particular topologically
connected diagram, $\gamma$. In a given diagram, $\Gamma$, let $\gamma$
occur $n_{\gamma}$ times. The symmetry factor resulting from
permutations within each $\gamma$, are included in the contribution
$C_{\gamma}$. But now we can also have permutations across different,
topologically connected diagrams. Such permutations can leave the
product diagram invariant if the different copies of a given $\Gamma$
are permuted as a whole. Hence, we have an additional symmetry factor of
$n_{\gamma}!$ by which we have to divide. A product diagram with
different $\gamma$'s thus has a symmetry factor of $\prod_{\gamma}
n_{\gamma}!$.

Returning to the full $Z[J]$ which we can now represent as a
contribution from a sum of diagrams $\Gamma$, each of which can have
several topologically connected $\gamma$'s with $n_{\gamma}$ copies.
Hence,
\begin{eqnarray*}
	Z[J] & \propto & \sum_{\{n_{\gamma}\}} C_{\Gamma} ~ \sim ~
	\sum_{\{n_{\gamma}\}}\prod_{\gamma}\frac{(C_{\gamma})^{n_{\gamma}}}{n_{\gamma}!}
	~ := ~ \mathcal{N} \prod_{\gamma}\sum_{n_{\gamma} = 0}^{\infty}
	\frac{(C_{\gamma})^{n_{\gamma}}}{n_{\gamma}!}\\
	& = &\mathcal{N}\prod_I exp [C_{\gamma}] = \mathcal{N} exp
	\Big[\sum_{\gamma} C_{\gamma}\Big] \ .
\end{eqnarray*}
Thus, $Z[J]$ is proportional to the exponential of the sum of
contributions of topologically connected diagrams. In this sum, are also
the contributions topologically connected vacuum diagrams i.e. $Z[J=0]$.
Imposing the normalization condition, $Z[J=0] = 1$, gives
$\boxed{\mathcal{N} \times exp (\mbox{topologically connected vacuum
bubbles}) = 1.}$. Hence, the normalization is trivially incorporated by
simply dropping the contributions of the vacuum diagrams and we are left
with,
\begin{equation} \label{WDefn}
	\boxed{ Z[J] = exp\Big[ i W[J] \Big] ~ , ~ iW[J] = \sum
	\Big(\mbox{topologically connected diagrams.}\Big)}
\end{equation}

The Green's functions that we get from $Z[J]$, now get expressed as,
\[
	\langle 0|\phi(x)|0\rangle = -i\frac{\delta Z[J]}{\delta
	J(x)}\bigg|_{J=0} = \frac{\delta W[J]}{\delta J(x)}\bigg|_{J=0}
	~ ~ , ~ ~ \because ~  \frac{1}{W[J=0]} = 1.
\]
This gives the contribution of diagrams with a single $J(x)$ and gets
removed when $J=0$ is put.

Similarly other Green's functions can be expressed in terms of the
derivatives of the $W[J]$.

As an illustration of the organization of the topologically connected
diagrams in $W[J]$, consider the two point function.
\begin{eqnarray*}
	G(x_1,x_2) & := & (-i)^2\frac{\delta^2}{\delta J_1\delta
	J_2}Z[J]\Big|_{J=0} = -i\delta_{J_1}\left(
	e^{iW}\delta_{J_2}W\right) \\
	& = & e^{iW}\left(\delta_{J_1}W.\delta_{J_2}W -
	i\delta^2_{J_1J_2}W\right)_{J=0}  \\
	\therefore G(x_1,x_2) & = & G_c(x_1)G_c(x_2) - G_c(x_1,x_2) ~ ~
	, ~ ~ \mbox{where,} \\
	& & G_c(x) := \delta_{J(x)}W[J]|_{J=0}~, ~ G_c(x,y) :=
	i\delta^2_{J(x) J(y)}W[J]|_{J=0} \\
	\vcenter{\hbox{
	\begin{tikzpicture}
		\begin{feynman}
			\vertex (v);
			\vertex [left=0.15of v](v'){\(x_1\)};
			\vertex [right =0.35of v,blob] (a){G} ;
			\vertex [right =0.65of a] (b);
			\vertex [right =0.15of b] (b'){\(x_2\)};
			\diagram*{
				(v')--(v) -- (a) -- (b)--(b'),
			};
		\end{feynman}
	\end{tikzpicture}
	}} 
	& = & 
	\underbrace{
	\vcenter{\hbox{
	\begin{tikzpicture}
		\begin{feynman}
			\vertex (v);
			\vertex [left=0.15of v] (v'){\(x_1\)};
			\vertex [right =0.5of v,blob] (a){\(G_c\)};
			\vertex [right =1.5of a,blob] (b){\(G_c\)};
			\vertex [right=0.85of b] (c);
			\vertex [right=0.15of c] (c'){\(x_2\)};
			\diagram*{
				(v) -- (a),
				(b) -- (c),
			};
		\end{feynman}
	\end{tikzpicture}
	}}
}_{\mbox{Top. disconnected}}
	\hspace{0.5cm} - \hspace{0.5cm} 
	\underbrace{
	\vcenter{\hbox{
	\begin{tikzpicture}
		\begin{feynman}
			\vertex (v);
			\vertex [left=0.15of v](v'){\(x_1\)};
			\vertex [right =0.5of v,blob] (a){\(G_c\)};
			\vertex [right =0.85of a] (b);
			\vertex [right =0.15of b] (b'){\(x_2\)};
			\diagram*{
				(v) -- (a) -- (b),
			};
		\end{feynman}
	\end{tikzpicture}
	}} 
	}_{\mbox{Top. connected}}
\end{eqnarray*}

The $G_c$ denote, contributions from connected (to external points) and
topologically connected diagrams.  This continues to the other $n-$point
functions, as proved by the argument leading to eqn. (\ref{WDefn}). To
conclude,
\begin{center}
	\fbox{ \begin{tabular}{lcccc}
	$G(x_1,\dots,x_n)$ & : & $\Sigma$ of all connected diagrams & :
	& $ (-i)^n\frac{\delta^n}{\delta J(x_1)\dots\delta J(x_n)}
	Z[J]\bigg|_{J=0}$ \\
	& & & & \\
	$G_c(x_1,\dots,x_n)$ & : & $\Sigma$ of topologically connected
	diagrams & : & $ (-i)^{n-1}\frac{\delta^n}{\delta
	J(x_1)\dots\delta J(x_n)} W[J]\bigg|_{J=0}$ \\
	& & & & \\
	$n = 0$ & : & $\Sigma$ of {\em all} vacuum diagrams & : &
	$Z[J=0] = 1$\\
	$n=0$ & : & $\Sigma$ of all topologically connected & : &
	$W[J=0] = 0$ \\
	&  & vacuum diagrams &  & 
\end{tabular}
} \end{center}
The computation of the diagrams uses the same Feynman rules that we
discussed earlier. The $Z$ and $W$ provide a convenient way of dealing
all the diagrams together and serve as {\em generating functions} for
the Greens functions.
\subsection{The 1-point function: Renormalization $\leftrightarrow$
normal ordering} \label{1PtFn}

The 1-point function of a scalar field has a distinction. Since this
represents the $\langle 0|\Phi(x)|0\rangle$, the Poincare invariance
allows this to be {\em a non-zero constant, say $v$}. No other Lorentz
covariant field can have its vacuum expectation value to be non-zero
since there are no Lorentz invariant spinors or vectors\footnote{A
	second rank, symmetric tensor field {\em can have} $\langle
	0|\hat{h}_{\mu\nu}(x) |0\rangle = \Lambda \eta_{\mu\nu}$. But
this typically arises only when gravity is included.}. However, if we
want the quanta exchanged during interactions to have the appropriate
quantum numbers, the vacuum should {\em not have any quantum numbers}.
As noted in the discussion of uniqueness of vacuum, a non-unique vacuum
would have to carry non-trivial representation labels of the symmetry
group which will add to the labels of the exchanged quanta.
Consequently, we must have a unique vacuum and then, by the mode
expansion of the quantum field, its expectation value must vanish. 
Does our path integral definition of $n-$point function satisfy this
property?


Consider computation of the 1-point function in the $(\phi)^3_n$ theory.
Notice that topologically disconnected diagrams contributing to the
$1-$point function can only be the vacuum bubbles giving a
multiplicative contribution of $1$. So we can limit to only
topologically connected diagrams. Furthermore, if we have $1PR$
diagrams, their contribution is again of a product form:
$(1PI)\times\Delta(x-y)\times(1PI)\times\Delta(z-w)\times(1PI)\cdots$.
Hence it suffices to consider only the $1PI$ diagrams. Thus we have,
\[
	\vcenter{\hbox{
	\begin{tikzpicture}
		\begin{feynman}
			\vertex [dot] (v){ };
			\vertex [right=0.85of v,blob] (a){\(G_c(x)\)};
			\diagram*{
				(v) -- (a),
			};
		\end{feynman}
	\end{tikzpicture}
	}} 
	 ~ = ~ 
	 \vcenter{\hbox{
	 \begin{tikzpicture}
	 	\begin{feynman}
			\vertex [dot] (v){ };
			\vertex [right=0.75of v] (a);
			\vertex [right=1.25of a] (b);
	 		\diagram*{
				(v) --(a),
				(a) -- [half left,looseness=1.5] (b),
				(a) -- [half right,looseness=1.5] (b),
	 		};
	 	\end{feynman}
	 \end{tikzpicture}
	 }} 
	 ~ + ~ 
	 \vcenter{\hbox{
	 \begin{tikzpicture}
	 	\begin{feynman}
			\vertex [dot] (v){ };
	 		\vertex [right=0.75of v] (a);
	 		\vertex [right=1.30of a] (b);
	 		\vertex [above right=0.85of a] (c);
	 		\vertex [below right=0.85of a] (d);
	 		\diagram*{
				(v) --(a),
				(a) -- [half left,looseness=1.5] (b),
				(a) -- [half right,looseness=1.5] (b),
				(c) -- (d),
	 		};
	 	\end{feynman}
	 \end{tikzpicture}
	 }} 
	 ~ + ~ \cdots
\]

At 1-loop (and higher), the diagrams are non-zero and in fact divergent.
We invoke the counter term method to absorb away these divergences.
Thus, we add a term in $\mathcal{L}_1$, of the form $Y \phi(x) = -i
Y\Case{\delta}{\delta J(x)}$. This is represented by the diagram
$\vcenter{\hbox{
\begin{tikzpicture}
	\begin{feynman}
		\vertex [crossed dot](v){ };
		\vertex [right=0.5of v] (a);	
		\diagram*{
			(v) -- (a),
		};
	\end{feynman}
\end{tikzpicture}
}} .$
 With its inclusion, the contributing diagrams are:
\begin{eqnarray*}
	\vcenter{\hbox{
	\begin{tikzpicture}
		\begin{feynman}
			\vertex [dot] (v){ };
			\vertex [right=0.85of v,blob] (a){\(G_c(x)\)};
			\diagram*{
				(v) -- (a),
			};
		\end{feynman}
	\end{tikzpicture}
	}} 
	 & = & 
	 \vcenter{\hbox{
	 \begin{tikzpicture}
	 	\begin{feynman}
			\vertex [dot] (v){ };
			\vertex [left=0.25of v] (v'){\(x\)};
			\vertex [right=0.75of v] (a);
			\vertex [right=0.1of a] (a'){\(y\)};
			\vertex [right=1.25of a] (b);
	 		\diagram*{
				(v) --(a),
				(a) -- [half left,looseness=1.5] (b),
				(a) -- [half right,looseness=1.5] (b),
	 		};
	 	\end{feynman}
	 \end{tikzpicture}
	 }} 
	 ~ + ~ 
	 \vcenter{\hbox{
	 \begin{tikzpicture}
	 	\begin{feynman}
			\vertex[dot] (v){ };
			\vertex[right=0.5of v,crossed dot] (a) { };
			\vertex[right=0.4of a] (a') {\(Y\) };
	 		\diagram*{
				(v) -- (a),
	 		};
	 	\end{feynman}
	 \end{tikzpicture}
	 }}
	 ~ + ~ 
	 \vcenter{\hbox{
	 \begin{tikzpicture}
	 	\begin{feynman}
			\vertex [dot] (v){ };
	 		\vertex [right=0.75of v] (a);
	 		\vertex [right=1.30of a] (b);
	 		\vertex [above right=0.85of a] (c);
	 		\vertex [below right=0.85of a] (d);
	 		\diagram*{
				(v) --(a),
				(a) -- [half left,looseness=1.5] (b),
				(a) -- [half right,looseness=1.5] (b),
				(c) -- (d),
	 		};
	 	\end{feynman}
	 \end{tikzpicture}
	 }} 
	 ~ + ~ 
	 \vcenter{\hbox{
	 \begin{tikzpicture}
	 	\begin{feynman}
			\vertex[dot] (v){ };
			\vertex[right=0.5of v,crossed dot] (a) { };
			\vertex[right=0.4of a] (a') {\(Y\) };
	 		\diagram*{
				(v) -- (a),
	 		};
	 
	 		\diagram*{
	 		};
	 	\end{feynman}
	 \end{tikzpicture}
	 }}
	 ~ + ~ \cdots \\
	 & = & \hspace{1.0cm} (a) \hspace{2.5cm} (b)
	 \hspace{5.0cm} \mbox{(2-loop)}
 \end{eqnarray*}
These give, to 1-loop, 
\[
	\langle 0|\phi(x)|0\rangle = \left[\underbrace{iY}_{(b)} +
	\frac{ig}{2}\cdot \underbrace{-i\Delta(0)}_{(a)}\right]\cdot
	\int d^ny ( i \Delta(x-y)) + o(g^3)
\]
For the left hand side to be zero to $o(g)$, we must set $\boxed{ Y =
i\Case{g}{2}\Delta(0) + o(g^3) } $ with,
\begin{eqnarray*}
	\Delta(0) & = & \int
	\frac{d^nk}{(2\pi)^n}\frac{1}{k^2+m^2-i\epsilon} ~ ~
	\mbox{(divergent for $n \ge 2$.)} \\
	& = & i\int \frac{d^n\underline{k}}{(2\pi)^n}
	\frac{1}{\underline{k}^2+m^2-i\epsilon} ~ = ~
	i\frac{1}{(4\pi)^{n/2}} \frac{\Gamma(1-n/2)}{\Gamma(1)}
	(m^2)^{n/2 - 1} ~ ~ \mbox{for} ~ ~ n = 4-\epsilon ~ \mbox{(say)}
	\\
	& = &
	i\frac{1}{16\pi^2}(-1)\frac{2}{\epsilon}(m^2)^{1-\epsilon/2} \
	.\\
	\therefore Y & = & \frac{g}{16\pi^2}\frac{m^2}{\epsilon}
	\left(1- \frac{\epsilon}{2}ln(m^2) + \cdots \right) ~ \approx ~
	\frac{g}{16\pi^2}\frac{m^2}{\epsilon} + o(\epsilon^0) + \cdots \
	.
\end{eqnarray*}
Thus, we fix the counter term $Y$ by demanding that the 1-point function
be zero, as the renormalization condition and this can be continued at
higher orders. The upshot is that we can ensure $\langle
0|\phi(x)|0\rangle = 0$ by means of a counter
term. Hence, {\em the sum of all connected diagrams with a single
external line vanishes.}

Clearly, a tadpole can attach to any other diagram only in a 1PR manner
(eg replacing the source end of the tadpole by the other diagram). This
is a multiplicative contribution and immediately renders all such
diagrams to be zero.  Hence {\em all diagrams with a tadpole attached,
when summed to a given order, vanish.} We may thus simply remove the
tadpole diagrams and obtain results equivalent to those obtained by
normal ordering prescription employed before.

\underline{Note:} Had the $1-$point function been {\em finite}, we would
still need to use the counter term method to set it to zero to ensure
uniqueness of the vacuum. In the dimensional regularization, massless 
tadpole diagrams are zero.

\underline{Qn:} Normal ordering eliminates {\em all} self contractions -
$1-$point function in $\Phi^3$  as well as $2-$point function in
$\Phi^4$ and likewise if higher order vertices are included. For
instance, the 1-loop diagram contributing to the $2-$point function in
$\Phi^4$. Now there will be a 1-loop counter term for the 2-point
function (absent in the normal ordered version). {\em Does it affect the
finite part?} 

\underline{Ans:} The self contractions at any vertex, in any diagram
produces a 1-loop contribution as a multiplicative factor, $\Delta_F(0)$
and this multiplicative factor is independent of any momenta
entering/exiting the vertex. Consider such a contribution to the
self-energy, $\Pi(p^2)$, in the $\Phi^4$ theory. Unless a counter vertex
is introduced, it is not possible to satisfy the renormalization
condition, $\Pi(-m^2) = 0$ at the order $\lambda^1$ (here we consider
on-shell renormalization for simplicity, so that $m^2 = m^2_{ph}$). So
introduce an order $\lambda$ counter vertex. Then at 1-loop, the self
energy equals $\Delta_F(0) + \delta_1$. Both are {\em independent} of
$p^2$ and the renormalization condition implies that two must add to
zero. The self energy thus receives no contribution at 1-loop, exactly
as if normal ordering has been invoked. In any other diagram (1PI), if a
self contraction appears, then the same counter vertex, $\delta_1$ will
again automatically cancel the contribution, effectively enforcing
normal ordering. {\em Again this is independent of UV divergence and is
a consequence of the requirement of the renormalization condition on
self-energy}.
\subsection{Path Integrals and Statistical Mechanics} \label{PartitionFn}
We discuss briefly an interesting connection of the path integral
representation of the transition amplitude and the (grand) partition
function of statistical mechanics \cite{GibbonsHawking}.

The basic observation is that the transition amplitude, schematically
$\langle Q_2,t_2|Q_1,t_1\rangle = \int \cal{D}q\ exp\{\Case{i}{\hbar}
\int_{t_1}^{t_2}dt L(q(t),\dot{q}(t))\}$, can be taken as a {\em
postulate} rather than a derivation from matrix element of the quantum
unitary evolution operator. We can also separately postulate that the
transition amplitude {\em is} the matrix element $\langle
Q_2|e^{-\Case{i}{\hbar}(t_2-t_1)H}|Q_1\rangle$. Mathematically, set
$t_2-t_1 =: -i\beta$, set $Q_2 = Q_1 =: Q$ and integrate over $Q$. On
the left hand side, we get $Tr e^{-\beta H}$ which is the usual
canonical partition function. On the right hand side, the path integral
turns to a path integral over {\em all closed paths} of an integrand
which is exponential of a {\em Euclidean continuation} ($t \to -i\tau$)
of the action. We thus obtain a path integral representation of the
statistical mechanical canonical partition function, in terms of
Euclidean continuation of a classical action. This is used in the
context of the black hole entropy which may be seen in
\cite{GibbonsHawking}.

\newpage
\section{Path Integrals as Generating Functionals}
\label{GeneratingFunctionals}

We have noted the two basic quantities, $Z[J]$ and $W[J] = -ilnZ[J]$.
Their functional derivatives with respect to $J(x)$ give the
contributions to the $n-$point ($n \ge 1$) functions: $(-i)^n\delta^n
Z[J]|_{J=0}$ contains the contributions of all the connected diagrams
while $(-i)^{:n-1}\delta^n W[J]$ contains the contributions of all the
connected and topologically connected diagrams. The normalization
condition: $Z[J=0] = 1 \leftrightarrow W[J=0] = 0$ omits the
contributions of the vacuum bubbles. We had encountered the 1PI diagrams
while discussing the self-energies. Their contributions too can be
obtained by functional differentiation of another quantity which we
obtain below. Note that the $n-$point functions being obtained by
functional derivatives also means that $Z[J],\ W[J]$ can be viewed as
{\em generating functional for Green's functions}.
\subsection{The Generating Functionals: $Z[J],\ W[J],\ \Gamma[\Phi]$}
\label{ZWGamma}
Consider a connected and topologically connected diagram contributing to
$\delta^nW[J]|_{J=0}$. It is said to be {\em 1-particle reducible}
(1PR), if it can disconnected by cutting {\em 1 internal line}. A
diagram which is not 1PR is {\em 1-particle irreducible} (1PI). Any
given diagram can be viewed as a collection of 1PI sub-diagrams connected
by one internal line. As noted in the discussion of self energy
diagrams, all 1PI diagrams with two external lines when strung together
by an internal line (free propagator) lead to the exact propagator.
Thus, the {\em sum of all connected and topologically connected
diagrams} contributing to an $n-$point function can be viewed as new
connected and topologically connected diagrams whose ``vertices'',
$\gamma_k$ are all 1PI diagrams with $k > 2$ edges, connected by lines
representing the {\em exact propagator}. These new diagrams all have a
``tree topology'' - there are no loops in such a reorganization of
diagrams since any loop would make a 1PI sub-diagram and will already be
included in one of the $\gamma_k$'s. These vertices can be of any order,
unlike elementary vertices dictated by the Lagrangian defining a theory.
The set of these new diagrams can be generated from another generating
functionals,
\begin{eqnarray*}
Z_{\gamma}[J] & := & \int \cal{D\varphi}\ exp\left[
i\left(\gamma[\varphi] + i\int d^4x J(x)\varphi(x)\right)\right] ~ := ~
e^{iW_{\gamma}[J]} \ , \\
\gamma[\varphi] & = & \sum_{n\ge 2}\int dx_1\cdots dx_n \frac{
\gamma_n(x_1,\cdots,x_n)}{n!} \varphi(x_1)\cdots\varphi(x_n) \ .
\end{eqnarray*}

{\em Note:} The coefficients of the $\varphi^n$ terms are not simple
numbers and this action is {\em non-local}. The coefficient of the $n=2$
term, $\gamma_2(x_1,x_2)$ is the Fourier transform of the inverse of the
exact propagator. Note that this establishes that $\boxed{W_S[J] =
W_{\gamma}[J]\Big|_{tree}.}$

However, such a $Z_{\gamma}[J], W_{\gamma}[J]$ will generate diagrams
with loops as well and we are interested in generating only the tree
diagrams. To pick out these alone, introduce a fictitious dimensionless
parameter, $\lambda$ and define
\[
Z_{\gamma,\lambda}[J] := \int \cal{D\varphi}\ exp\left[
\frac{i}{\lambda}\left(\gamma[\varphi] + i\int d^4x
J(x)\varphi(x)\right)\right] ~ := ~ exp{\left[i
W_{\gamma,\lambda}[J]\right]} \ .
\]
Let us count the powers of $\lambda$ in a any connected and
topologically connected diagram.  Noting  that the $\gamma_2$ term in
$\gamma[\varphi]$ is the (Fourier transform of the) inverse of the exact
propagator, the scaling by $\lambda^{-1}$ means that that each
propagator gives a factor of $\lambda$ while each vertex and every J,
gives a factor of $\lambda^{-1}$. Thus, for any diagram contributing to
$W_{\gamma,\lambda}[J]$ we get the factor $(\lambda)^{P-V-E}$, where $P,
V, E$ are the number of propagator lines, number of vertices and the
number ``external'' lines (lines connected to a source) respectively.
The number of ``internal'' lines is $P-E$. In the diagram below, we have
$P = 6, V= 2, E=4$ giving $\lambda^{0}$. 
\[
\vcenter{\hbox{
\begin{tikzpicture}
	\begin{feynman}
		\vertex [blob](v1){ };
		\vertex [left=0.75of v1](v1'){\(\lambda^{-1}\) };
		\vertex [right=2.5of v1, blob](v2){ };
		\vertex [right=0.75of v2](v2'){\(\lambda^{-1}\) };
		\vertex [above left=1.5of v1, dot](a){ };
		\vertex [below left=1.5of v1, dot](b){ };
		\vertex [above right=1.5of v2, dot](c){ };
		\vertex [below right=1.5of v2, dot](d){ };
		\vertex [above left=0.35of a] (a'){\(\lambda^{-1}\)};
		\vertex [below left=0.45of b] (b'){\(\lambda^{-1}\)};
		\vertex [above right=0.45of c] (c'){\(\lambda^{-1}\)};
		\vertex [below right=0.45of d] (d'){\(\lambda^{-1}\)};

		\diagram*{ (a) --[edge label=\(\lambda\)] (v1), (b) --
			[edge label'=\(\lambda\)](v1), (v2) --[edge
			label=\(\lambda\)] (c), (v2) --[edge
			label'=\(\lambda\)](d), (v1) --[half
			right,looseness=0.8,edge label=\(\lambda\)] (v2),
			(v2) --[half right,looseness=0.8,edge
			label=\(\lambda\)] (v1), }; \end{feynman}
\end{tikzpicture} }}
\]
For topologically connected diagrams we also have $L = I-V+1 = P -E -V +
1$. Hence the factor of $\lambda$ is $(\lambda)^{L-1}$. We can thus write,
$W_{\gamma,\lambda}[J] = \sum_{L=0}^{\infty} \lambda^{L-1} W_{\gamma,L}[J]$
and organize the diagrams by the number of loops. In the formal limit,
$\lambda \to 0$, $W_{\gamma,\lambda\to 0}[J] \to \Case{1}{\lambda}
W_{\gamma,L=0}$ which is the contribution of the tree diagrams (with
exact propagators and exact vertices alone).

On the other hand, we may evaluate $Z_{\gamma,\lambda}[J]$ in the limit
$\lambda \to 0$ by stationary phase approximation. Clearly, the path
integral is dominated by the fields, $\varphi(x)$ which satisfy
$\Case{\delta\gamma}{\delta\varphi} + J = 0$. Denoting $
\varphi_{cl}(x)$ as its solution, we get,
\[
	Z_{\gamma,\lambda \to 0}[J] \simeq exp \frac{i}{\lambda}\left\{
	\gamma[\varphi_{cl}] + \int d^4x J(x)\varphi_{cl}(x)\right\} =
	e^{iW_{\gamma,\lambda\to 0}[J]} \simeq
	exp{\frac{i}{\lambda}W_{\gamma,L=0}[J]}. 
\]
\[
	\therefore \gamma[\varphi_{cl}] + \int d^4x J(x)\varphi_{cl}(x)
	= W_{\gamma}[J]\Big|_{tree} ~ = ~ W_S[J] .
\]
The last equation shows that $\gamma[\varphi]$ {\em is just the Legendre
transform of} $W[J]$ and by construction, it is the generating function
of connected, topologically connected, 1PI diagrams with external legs
amputated! The last property follows because the vertex functions
$\gamma_n(x_1,\cdots,x_n)$ explicitly do not have external propagator
lines included in them. 

{\em Note:} We have followed a route of organizing the diagrams as a
tree of 1PI diagrams connected by exact propagators and inferred the
corresponding generating function $\gamma[\varphi]$ as a Legendre
transform of the $W[J]$. The usual approach is to begin with a Legendre
transform definition and arrive at its interpretation. This conventional
approach goes as follows.

Define, $\boxed{\Phi(x)[J] := \Case{\delta W[J]}{\delta J(x)} = \langle
0|\hat{\Phi}(x)|0\rangle_J , }$ $J$ is {\em not set to zero} and hence
this is not the 1-point function. We have the path integral
representation as,
\[
	\Phi(x)[J] = \frac{-i\delta ln(Z[J])}{\delta J(x)} =
	\frac{-i}{Z[J]}\frac{\delta Z[J]}{\delta J(x)} = \frac{\int
	\cal{D\varphi}\varphi(x)e^{iS[J]}}{\int \cal{D\varphi} e^{S[J]}}
	\hspace{0.5cm},\hspace{0.5cm} \boxed{v := \Phi(x)[J=0]} \ .
\].

Define, $\bar{\Phi}(x) := \Phi(x) - v$. Then $\langle
0|\hat{\bar{\Phi}}(x)|0\rangle = 0$. This generalizes to,

\underline{Claim:} 
\[
	\frac{\delta^n}{\delta J(x_1)\dots \delta J(x_n)}
	W[J]\bigg|_{J=0} = i^{n-1}\langle 0|T\left\{\bar{\Phi}(x_1)\dots
	\bar{\Phi}(x_n)\right\} |0\rangle ~ , ~ \forall ~ n \ge 2 .
\]
The proof is by induction starting at $n=2$. For $n=2$ we
have,
\begin{eqnarray*}
\frac{\delta^2}{\delta J_1\delta J_2} W[J]\bigg|_{J=0} & = &
\left[\frac{\delta}{\delta J_2}\left(-\frac{i}{Z[J]}\frac{\delta
Z[J]}{\delta J_1}\right)\right]_{J=0} = \left[ \frac{i}{Z^2}\frac{\delta
	Z[J]}{\delta J_2}\frac{\delta Z[J]}{\delta J_1} -
i\frac{\delta^2 Z[J]}{\delta J_2 \delta J_1}\right]_{J=0} \\
& = & \frac{i}{1}(iv)(iv) - i(i^2)\langle 0|\Phi(x_1)\Phi(x_2)|0 \rangle
\\
& = & i\left\{\langle 0|\Phi(x_1)\Phi(x_2)|0\rangle - v^2\right\}
=\langle 0|\bar{\Phi}(x_1)\bar{\Phi}(x_2)|0\rangle .
\end{eqnarray*}
Which verifies the claim for $n=2$. The pattern repeats and the proof
follows. 

As a corollary, we can write
\begin{eqnarray}
	W[J] & = & \sum_{n\ge 2}\frac{1}{n!}\int dx_1\dots dx_n
	J(x_1)\dots J(x_n)\frac{\delta^n W}{\delta J_1\dots
	J_n}\bigg|_{J=0} \\
	& = & \sum_{n\ge 2}\frac{(i)^{n-1}}{n!}\int dx_1\dots dx_n
	\langle 0|T\left\{\bar{\Phi}(x_1)\dots \bar{\Phi}(x_n)\right\}
	|0\rangle J(x_1)\dots J(x_n)
\end{eqnarray}

Using $\Phi(x)[J] = \delta_{J(x)}W[J]$, {\em define the Legendre
transform of} $W[J]$ ,
\begin{equation} \label{LegendreTransform}
	\boxed{ \Gamma[\Phi] := W[J] - \int d^4x J(x) \Phi(x) }
	\hspace{0.5cm},\hspace{0.5cm}\mbox{(compare with: $-H(p) =
	L(\dot{q}) - p\dot{q}$~ )}
\end{equation}
Notice that this is exactly the same as the $\gamma[\varphi]$ defined
above.

The $\Gamma[\Phi]$ is defined by expressing the right hand side as a
function of $\Phi$, in particular $J = J[\Phi]$ is understood which is
obtained by inverting the relation $\Phi[J] = \delta_JW[J]$. It follows
as usual for a Legendre transform, 
\begin{eqnarray*}
	\frac{\delta\Gamma[\Phi]}{\delta \Phi(x)} & = & \cancel{\int d^4y
	\frac{\delta W}{\delta J(y)}\frac{\delta J(y)}{\delta\Phi(x)}} -
	\cancel{\int d^4y \frac{\delta J(y)}{\delta\Phi(x)}\Phi(y)} -
	\int d^4y J(y)\frac{\delta\Phi(y)}{\delta\Phi(x)} = -J(x)
\end{eqnarray*}
In the last equality we have used $\Case{\delta \Phi(y)}{\delta \Phi(x)}
= \delta^4(x-y)$. Since $\Phi(x)[J=0] = v$, we also have
$\boxed{\Case{\delta \Gamma[\Phi]}{\delta \Phi(x)}\Big|_{\Phi(x) = v} =
0. }$ Thus, $v$ is that value which extremises $\Gamma[\Phi(x)]$ and
hints at $\Gamma[\Phi]$ being some sort of action whose extremization
leads to a solution. This is supported further as follows.

We have $Z[J] = e^{iW[J]} = \int \cal{D\varphi}e^{iS[J,\varphi]}$. Let
$\varphi_{cl}(x)$ be a classical solution i.e. $\delta S[J,\varphi_{cl}]
= 0$. Expand the action around such a solution by setting $\varphi(x) =
\varphi_{cl}(x) + \eta(x)$. We get (schematically to avoid clutter),
\begin{eqnarray*}
S[\varphi] & = & S[\varphi_{cl}+\eta] = S[\varphi_{cl}] +
\underbrace{\frac{\delta S}{\delta\varphi_{cl}}}_{0}\cdot\eta +
\frac{1}{2}\frac{\delta^2 }{\delta\varphi^2}\eta^2 + \dots \\ 
& = & S[\varphi_{cl}] + \frac{1}{2}\int d^4x\ d^4y \frac{\delta^2
S[J,\varphi]}{\delta\varphi_{cl}(x)\delta\varphi_{cl}(y)}
\eta(x)\eta(y)\\
\therefore e^{iW[J]} & = & e^{iS[J,\varphi_{cl}]} \int\cal{D\eta} \
exp\left[\frac{i}{2}\int d^4x\int d^4y\left\{\eta(x)
\frac{\delta^2S}{\delta\varphi_{cl}(x)\delta\varphi_{cl}(y)}\eta(y)\right\}
+ \dots \right]
\end{eqnarray*}
The path integral does have an implicit dependence on $J(x)$ through the
$\varphi_{cl}(x)$, unless the action itself is quadratic as was the case
of the oscillator.

Momentarily let us {\em just ignore} the path integral altogether. Then,
$W[J] \approx W_0[J] := S[J,\varphi_{cl}] = \int d^4x [
\mathcal{L}(\varphi_{cl}) + J(x)\varphi_{cl}(x)]$ where the classical
solution is to first obtained by solving $\delta S[J,\Phi] = 0$ and then
substituted back in the action.

Applying our definition, $\Phi(x) = \delta_{J(x)}W[J] \approx
\delta_{J(x)}W_0[J] = \delta_{J(x)}S_{cl}[J]\ , \ S_{cl} :=
S[J,\varphi_{cl}(x)[J]]$. Straight forward evaluation gives,
\[
	\Phi(x) = \int d^4y\underbrace{\frac{\delta S_{cl}}{\delta
	\varphi_{cl}(y)}}_{=0} \frac{\delta \varphi_{cl}(y)}{\delta
	J(x)} + \varphi_{cl}(x)
\]
the last term coming from $\delta_{J} \int J\varphi$. Hence, in this
approximation, $\Phi(x) = \varphi_{cl}(x)$. 

Proceeding with the Legendre transform, we get 
\begin{eqnarray}
	\Gamma_0[\Phi] & = & W_0[J] - \int d^4x J(x)\Phi(x) \nonumber \\
	& = & \underbrace{S[\varphi_{cl}]}_{\mbox{no explicit $J$}} +
		\cancel{\int d^4x J(x)\varphi_{cl}(x)} - \cancel{\int
		d^4x J(x)\Phi(x)} =
		\underbrace{S[\varphi_{cl}]}_{\mbox{no explicit $J$}}.
\end{eqnarray}
Thus, within the approximation, $\Gamma[\Phi]$ is just the classical
action ($\varphi_{cl}(x) \to \Phi(x)$).

But this also shows that when the path integral included, $W[J]$ will be
very different and so also the corresponding $\Gamma[\Phi]$. Thus,
$\Phi(x)$ is identified as the ``quantum corrected solution'' while the
$\Gamma[\Phi]$ is called a ``quantum action'' or an ``effective action''
incorporating quantum corrections.

Consider Taylor expanding the effective action about $\Phi[x] = v$.
$\Gamma[v] = W[J=0] - \int 0\cdot\phi = 0$ since $W[0] = 0$ by our
normalization, $Z[0] = 1$. We have also seen that $\delta_{v}\Gamma =
0$. The remaining terms are a power series in $(\Phi(x) - v) =:
\bar{\Phi}(x)$ with the coefficients evaluated at $v$.  This is the same
as regarding $\Gamma$ as a function of $\bar{\Phi}(x)$ and expanding it
about $\bar{\Phi}(x) = 0$ i.e. 
\[
\Gamma[\bar{\Phi}] = \sum_{n\ge 2}\int dx_1\cdots dx_n \frac{
\Gamma(x_1,\cdots,x_n)}{n!} \bar{\Phi}(x_1)\cdots\bar{\Phi}(x_n) \ .
\]
What do these coefficients represent? We already have the answer. We
just need to recognize that $\Gamma[\bar{\Phi}] = \gamma[\varphi]$! The
coefficients are the contribution of all connected, topologically
connected, 1PI diagrams without external legs. An explicit
demonstration of amputation of the external legs may be seen in
\cite{AbersLee}. The generating functions are summarized in the table
below.

\begin{center}
\fbox{
	\begin{minipage}{0.99\textwidth}
		\begin{eqnarray}
			& & \hspace{2.0cm}\mbox{In summary} \nonumber \\
			& & \nonumber \\
			Z[J] & := & \int\cal{D\varphi}e^{i S[J,\varphi]}
			~ ~ , ~ ~ Z[J=0] = 1 ~ , ~
			\mbox{(normalization)}\\
			G(x_1,\dots,x_n) & := &
			\left(-i\frac{\delta}{\delta J(x)}\right)^n
			Z[J]\Big|_{J=0} ~ = ~\langle
			0|T\{\Phi(x_1)\dots\Phi(x_n)\}|0\rangle \\
			W[J] & := & -i\ ln(Z[J]) ~ ~ , ~ ~ W[J=0] = 0 \\
			G_c(x_1,\dots,x_n) & := &
			(-i)^{n-1}\frac{\delta^n W[J]}{\delta
			J(x_1)\dots\delta J(x_n)}\Big|_{J=0} ~ = ~
			\langle
			0|T\{\Phi(x_1)\dots\Phi(x_n)\}|0\rangle_c \\
			\Phi(x)[J] & := & \frac{\delta W[J]}{\delta
			J(x)} ~ ~ , ~ ~ v := \Phi(x)[J=0] ~ , ~
			\bar{\Phi}(x) := \Phi(x) - v\\
			\Gamma[\Phi] & := & W[J] - \int d^4x
			J(x)\varphi(x) \\
			J(x) & = &
			-\frac{\delta\Gamma[\Phi]}{\delta\Phi(x)} ~ ~ ,
			~ ~ \frac{\delta
			\Gamma[\Phi]}{\delta\Phi(x)}\Big|_{\Phi(x) = v}
			= 0 \\
			Z[J] & = & \sum_{n\ge 1}\frac{i^{n-1}}{n!}\int
			d^4x_1\dots d^4x_n G(x_1,\dots,x_n)J(x_1)\dots
			J(x_n) \\
			W[J] & = & \sum_{n\ge 1}\frac{i^{n-1}}{n!}\int
			d^4x_1\dots d^4x_n G_c(x_1,\dots,x_n)J(x_1)\dots
			J(x_n) \\
			\Gamma[\bar{\Phi}] & = & \sum_{n\ge
			2}\frac{1}{n!}\int d^4x_1\dots d^4x_n
			\Gamma_n(x_1,\dots,x_n)\bar{\Phi}(x_1)\dots
			\bar{\Phi}(x_n) \\
			G(x_1,\dots,x_n) & \leftrightarrow & \mbox{all
			connected diagrams}\\
			G_c(x_1,\dots,x_n) & \leftrightarrow & \mbox{all
			connected and topologically diagrams}\\
			\Gamma(x_1,\dots,x_n) & \leftrightarrow &
			\mbox{all connected, topologically connected, 1PI
			diagrams} \nonumber \\
			& & \mbox{with external legs amputated}\\ \nonumber 
		\end{eqnarray}
	\end{minipage} } 
\end{center}

\subsection{The Renormalization Group Equation}\label{RGEqn}
Recall that while discussing the renormalized perturbation series, we
introduced $\phi_0, m_0, g_0$ as the `bare' quantities with
$\mathcal{L}(\phi_0, m_0, g_0)$ generating the diagrams containing the
UV divergences. We then introduced the renormalized variables $\phi, m,
g_k$ defined through the scaling:  $\phi_0 := \sqrt{Z_{\phi}}\phi,\
m^2_0Z_{\phi} := m^2Z_m,\ g_{0k}Z_{\phi}^{k/2} := Z_{g_k}g_k$ which were
finite by definition. We also introduced counter terms to take care of
the UV divergences. With the $Z, G, \Gamma$ we have the formal
representation of the totality of all diagrams. Going over to the
momentum space labels we have (we take the 1-point function to be zero
and a single coupling constant for convenience), 
\begin{eqnarray}
\Gamma_{0,n}(p_1,\dots,p_n) := \frac{\delta \Gamma_0[\phi_0]}
{\delta\phi_0(p_1) \dots \delta\phi_0(p_n)} \Big|_{\phi_0 = 0} ~ & , & ~
\Gamma_{n}(p_1,\dots,p_n) := \frac{\delta \Gamma_0[\sqrt{Z_{\phi}}
\phi]} {\delta\phi(p_1)\dots\delta\phi(p_n)}\Big|_{\phi = 0} \nonumber
\\
\Rightarrow ~ ~ ~ ~  \Gamma_{0n}(p_i,m_0,g_0) & ~ = ~ & Z_{\phi}^{-n/2}
\Gamma_n(p_i, m, g)
\end{eqnarray}
The renormalized parameters are defined by a set of renormalization
conditions, which introduce a scale, say, $\mu$ (eg a $MS$ or
$\overline{MS}$ scheme). The bare quantities know nothing about this
scale and therefore $\boxed{ \mu\frac{d \Gamma_0(p_i,m_0,g_0)}{d\mu} = 0
. }$. Substitution gives,
\begin{eqnarray*}
0 & = & \left(\mu\frac{d}{d\mu} Z_{\phi}^{-n/2}\right)\Gamma_n +
Z_{\phi}^{-n/2}\mu \frac{d}{d\mu} \Gamma_n  ~ ~ \mbox{and,} \\
\mu\frac{d}{d\mu} \Gamma_n & = & \left[ \mu\frac{\partial}{\partial\mu}
+ \mu\frac{d m}{d\mu}\frac{\partial}{\partial m} + \mu\frac{d g}{d\mu}
\frac{\partial}{\partial g}\right] \Gamma_n(p_i, m, g)
\end{eqnarray*} 
Thus we have the renormalization group (RG) equation for the
renormalized vertex function as,
\begin{equation}\label{RGEqnGamma}
\boxed{ 
\mu\frac{\partial}{\partial\mu} \Gamma_n(p_i,m(\mu),g(\mu))\Big|_{m,g} +
\beta(g)\frac{\partial}{\partial g} \Gamma_n\Big|_{\mu,m} - \gamma_m(g)
m\frac{\partial}{\partial m}\Gamma_n\Big|_{\mu,g} -
n\gamma_{\phi}\Gamma_n = 0, }
\end{equation}
where,
\begin{equation} \label{BetaGammaFns}
	\boxed{ \beta(g,m) := \mu\frac{d g}{d\mu} ~ ~ , ~ ~ \gamma(g,m)
	:= -\mu\frac{d ln(m)}{d\mu} ~ ~ , ~ ~ \gamma_{\phi}(g,m) :=
\frac{1}{2}\mu\frac{d ln(Z_{\phi})}{d\mu} .  }
\end{equation}
The first is the {\em beta function}, the second is called the {\em mass
anomalous dimension} while the last is called the {\em anomalous
dimension}. These are computed typically in a perturbation theory as a
power series in the renormalized coupling (see below). Different schemes
give different simplifications and there is some {\em scheme dependence}
in the coefficients of these functions. 

As mentioned above, we describe a method of computation of the
renormalization constants and their dependence on the couplings in the
MS scheme \cite{Gross}.

We have the defining equations (\ref{RGEqnGamma},\ref{BetaGammaFns}).
These were obtained by noting that the bare vertex functions as a
function of the bare parameters and the regulator are independent of
$\mu$, {\em for fixed $g_0, m_0$ and $\epsilon$.} When the renormalized
parameters are {\em not} defined in terms of physically measured
quantities, how do we compute the $\mu$ dependence of these parameters?
For this, we go back to the basic definitions and recall the relation
between the bare and the renormalized parameters, in particular, $g_0 :=
\mu^{\epsilon}Z_g g$. Note that we have absorbed the $Z_{\phi}$ factor
into $Z_g$ and have also introduced the $\mu$ dependence by taking $g$
to be dimensionless. The $\mu\partial_{\mu}$ is now evaluated with the
bare parameters and the regulator $\epsilon$ held fixed i.e.  
\[
	\beta(g) = \mu\partial_{\mu}
	g(g_0\mu^{-\epsilon},\epsilon)\Big|_{g_0, \epsilon} =
	\mu\partial_{\mu}\left(g_0\mu^{-\epsilon}Z_g^{-1}\right) =
	-\epsilon g - g\mu\partial_{\mu} ln(Z_g) \ . 
\]
Noting that $Z_g$ is determined as a function of $g$, $Z_g(g,\epsilon) =
1 + \sum_{k\ge 1}\epsilon^{-k} Z_g^k$, we can write the defining
equation for the $\beta$ function in the form,
\[
	\beta(g,\epsilon) = -\epsilon g - g\beta(g,\epsilon)\partial_g
	ln(Z_g(g,\epsilon)) \leftrightarrow
	\beta(g,\epsilon)\partial_g(gZ_g(g,\epsilon)) + \epsilon
	gZ(g,\epsilon) = 0 .
\]
The $\beta(g, \epsilon)$ must have a smooth limit as $\epsilon \to 0$
and we may consider $\beta \sim \beta_0 + \beta_1\epsilon +
\beta_2\epsilon^2+\cdots$. The $\partial_g(gZ_g(g,\epsilon)) \sim 1 +
\epsilon^{-1} + \cdots$ while $\epsilon g Z_g(g,\epsilon) \sim \epsilon
+ \epsilon^0 + \epsilon^{-1} + \cdots$. Clearly, the coefficients of the
positive powers of $\epsilon$ greater than 1 must vanish. Hence we take,
$\beta(g, \epsilon) := \beta_0(g) +\beta_1 \epsilon $. It follows that
$\beta_1 = -g$ and the equation reduces to,
\begin{eqnarray}\label{BetaFnCaln}
	0 & = & (\beta_0 - \epsilon g)(Z_g + g\partial_g Z_g) \\
	0 & = & \beta_0\left[1 + \sum_{k\ge
	1}\frac{1}{\epsilon^k}\partial_g(gZ_g^k)\right] -
	\left[g^2\partial_gZ_g^1 + \sum_{k\ge 1}\frac{g^2\partial_g
	Z_g^{k+1}}{\epsilon^k}\right] \\
	\therefore & & \boxed{\beta_0(g) = g^2\partial_gZ_g^1 ~ ~ , ~ ~
	\beta_0\partial_g(gZ_g^k) = g^2\partial_g Z_g^{k+1}. }
\end{eqnarray}
The$\beta_0(g)$ is the usual beta function and is obtained from the
renormalization constant $Z_g$. The $Z_g^{k >2}$ coefficients are also
determined recursively.

\underline{Note:} The above expressions are in the context of the
$\Phi^4$ coupling. For Yukawa or the Yang-Mills cubic coupling, the
coupling  has $g_{Yukawa} = \mu^{\epsilon/2}g_{dimensionless}$ and the
above equations will change accordingly.

As a practical application of the renormalization group equation,
consider a theory which is massless and remains so perturbatively. Then
the anomalous mass dimension term is absent. Putting $t :=
lm(\mu/\mu_0)$ for some arbitrary scale $\mu_0$ at which the theory is
defined (renormalization conditions are imposed), we write
$\mu\Case{\partial}{\partial\mu} = \Case{\partial}{\partial t}$ and the
equation takes the form,
\[
	\left[ \frac{\partial}{\partial t} +
	\beta(g)\frac{\partial}{\partial g} - n\gamma(g)\right]
	\Gamma_n(g,t,p_i)  =  0 ~ ~ \mbox{with} ~ ~ \beta(g) =
	\frac{dg(t)}{dt} 
\]
This immediately gives, $\frac{d\Gamma_n(g(t),t,p_i)}{d t}  =  0 .$. Its
solution is obtained as,
\begin{equation}
	\boxed{ \Gamma_n(t,g_0,p_i) = \Gamma_{n,0}\left( g(t,g_0),p
	\right)\ exp\left\{ n\int_0^t dt'\ \gamma(g(t',g_0))\right\} } ~
	~ \mbox{where,} 
\end{equation}
$g(t,g_0)$ is the solution: $\boxed{\frac{d g(t,g_0)}{d t} =
\beta(g(t,g_0)) ~ , ~ g(0,g_0) = g_0. }$

Although typically the beta function is a power series in the coupling.
Once $g(t)$ is known, the $t$ dependence of {\em all} vertex functions
is determined. A good deal of qualitative behavior can be gleaned from
studying the beta function - especially near its zeros.  We know it has
a zero at $g=0$ (perturbative calculation), but at may have other zeros.
The derivative of $\beta(g)$ near its fixed points, controls how $g(t)$
evolves with $t$.

Let us assume it to be some given function of the form as shown in the
figure. 

%
\begin{figure}[htbp]
	\begin{center}
		\scalebox{0.5}{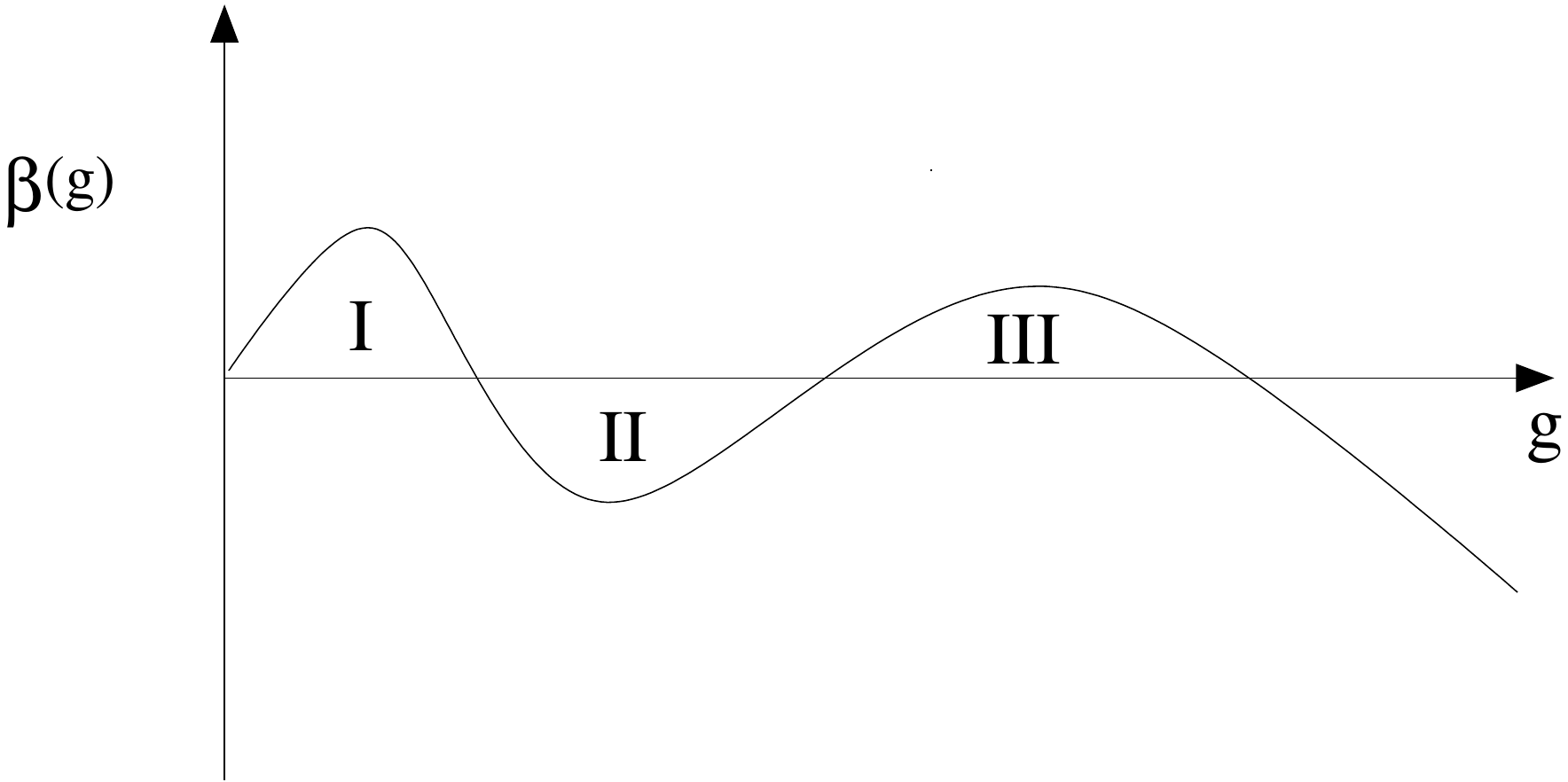}
	\end{center}
\end{figure}

It has multiple zeros, say $g_0=0<g_1<g_2<g_3\dots$.  Let $I,
II, III, \dots$ denote the intervals on the $g-$axis, bounded by the
zeros. Clearly, given a $g_0$ in one of these intervals, it will remain
so for all $t$. If $\beta(g)> 0$ in an interval, then $g(t)$ will evolve
to its upper bound and the opposite if $\beta(g)< 0$. A coupling will
always flow to a UV-stable (we are considering increasing $t$) fixed
point. In the figure, $g_1, g_3$ are the stable fixed points while $g_0,
g_2$ are unstable ones. 

Let $g_0$ be in region I so that as $t\to \infty, g(t,g_0) \to g_1$. The
asymptotic behavior of $\Gamma_n$ is controlled by the integral of the
anomalous dimension. Let us write,
\[
\int_0^{t}dt' \gamma(g(t',g_0)) = \int_0^{t}dt' \left[\gamma(g(t',g_0))
- \gamma(g_1)\right] + \int_0^{t}dt' \gamma(g_1)) .
\]
In the first term, we may take the upper limit $t\to\infty$ since the
integrand provides exponential suppression. This integral is then some
constant, $C$. In the second term, $\gamma(g_1)$ is independent of $t'$
and thus evaluates to $\gamma(g_1)\cdot t$. Hence,
\[
	\Gamma_n(t,g_0,p_i) \xrightarrow[t\to\infty]{ }
	\Big[\Gamma_{n,o}(g_1,p_i)\cdot C\Big]\cdot exp\{ \gamma(g_1)
	t\} ~ ~ , ~ ~ t = ln(\mu/\mu_0) \ .
\]
$\mu$ is some large scale eg some of the invariants $p_i\cdot p_j$.
Thus, we get $\Gamma_n(ln(\mu/\mu_0),g_0, p_i) \xrightarrow
[\mu\to\infty]{ } (\mu/\mu_0)^{\gamma(g_1)} ~ \forall ~ n$.  The
exponent is the same for all $n-$point vertex functions and is governed
by a stable fixed point in the vicinity of $g_0$. All these vertex
functions have the same asymptotic behavior. For further application, I
refer you to \cite{Coleman}.  


\subsection{The Background Field Method}
As an illustration of the utility of the formal advantage of the
functional methods, we discuss the so called {\em background field
method}, extremely useful for non-abelian gauge theories \cite{Abbot}. 
For illustration, we consider the simplest case of a scalar field $\varphi$.

We have the basic definitions 
\[
	e^{iW[J]} = Z[J] := \int \cal{D}\varphi e^{iS[\varphi] + \int
	J(x)\varphi(x)} ~ , ~ \bar{\varphi}(x) := \frac{W[J]}{J(x)} ~ ,
	~ \Gamma[\bar{\varphi}] := W[J(\bar{\varphi})] - \int
	J(x)\bar{\varphi}(x) \ .
\]

Introduce a new, arbitrary {\em background field}, $\chi(x)$ and define,
\begin{eqnarray*}
	e^{i\tilde{W}[J,\chi]} = \tilde{Z}[J,\chi] & := & \int
	\cal{D}\varphi e^{iS[\varphi + \chi] + \int J(x)\varphi(x)} ~ ,
	\\
	\tilde{\varphi}(x) := \frac{\tilde{W}[J,\chi]}{J(x)} & , &
	\tilde{\Gamma}[\tilde{\varphi},\chi] := W[J(\varphi,\chi)] -
	\int J(x)\tilde{\varphi}(x) \ .
\end{eqnarray*}
In the new generating functionals defined, shift the integration
variable as $\varphi \to \varphi - \chi$. This is a simple translation
and gives the Jacobian to be 1. It is immediate that $\tilde{Z}[J,\chi]
= Z[J] e^{-i\int J(x) \chi(x)}$ and $\tilde{W}[J,\chi] = W[J] - \int
J(x)\chi(x)$. Clearly, $\tilde{\varphi}(x) = \bar{\varphi}(x)- \chi(x)$
and 
\[
	\tilde{\Gamma}[\tilde{\varphi}, \chi] = W[J] - \int J(x)\chi(x)
	- \int J(x)(\bar{\varphi}(x) - \chi(x)) = \Gamma[\bar{\varphi}]
	~ = ~ \Gamma[\tilde{\varphi} + \chi] \ .
\]
If we had arranged that $\bar{\varphi} = 0$, then $\Gamma[\chi] =
\tilde{\Gamma}[0,\chi]$. The background field $\chi$ being arbitrary, we
get an alternate method of computing the effective action $\Gamma[\chi]$
using the shifted fields.

The shifted action has the form:
\[
	S[\varphi+\chi] = S[\chi] + \int \cal{L}_{1}(\chi)\varphi(x) +
	\int \cal{L}_2(\chi)\varphi^2(x) + \cdots .
\]
Since there is no functional integration over $\chi$, the first term is
just the classical action. Since $\chi$ is arbitrary, the second term
does {\em not} vanish (it vanishes if $\chi$ is an exact solution of the
classical equations of motion). The third term gives the propagator for
the $\varphi$ field and the higher order terms give interactions among
the $\varphi, \chi$ fields.

Now, the shifted effective action, $\tilde{\Gamma}$ is a generator of
1PI diagrams in presence of $\tilde{\varphi}$ and $\chi$. That is, its
derivatives with respect to $\tilde{\varphi}$ will give the 1PI diagrams
in presence of $\chi$. The $\varphi$ propagator, which comes from the
terms in the action which are quadratic in $\varphi$,  will in general
be $\chi-$dependent and the vertices will have factors of $\chi(x)$.
When $\tilde{\varphi} = 0$, only the diagrams with no external
$\tilde{\varphi}$ lines i.e. only the {\em vacuum diagrams} will
contribute.  Thus the {\em desired effective action can be computed
using only vacuum diagrams, albeit with vertex factors having the $\chi$
field}.

There are two ways to approach the computation. If we treat the
background field exactly, then in obtaining the Feynman rules the
propagator of the $\varphi$ field will have the $\chi$ field as well
(the terms quadratic in $\varphi$ in $S[\varphi+\chi]$). Except when
$\chi$ is space-time independent, this propagator is complicated. And
for a general background field, this method is not useful. An
alternative is to treat $\chi$ also perturbatively i.e. the $\varphi$
propagator remains exactly same as before and the additional
$\chi-$dependent terms are treated as additional couplings (the
$\varphi$ factors are replaced as $\Case{\delta}{\delta J}$). The
Feynman rules have the same $\varphi-$propagator, there is no propagator
for the $\chi$ field and the vertices have extra edges denoting the
background field.  Since there is no $\chi$ propagator, these vertices
generate {\em only} external $\chi$ lines.

As an explicit example, consider the $(\varphi^3)_6$ theory. The shift
by a background field gives, 
\[
S[\varphi+\chi] = S[\chi] + S[\varphi] + \int d^4x\left[
-\partial_{\mu}\varphi \partial^{\mu}\chi -m^2\varphi(x)\chi(x)
-\frac{g}{2!}\left(\varphi(x)\chi^2(x) +
\varphi^2(x)\chi(x)\right)\right]
\]
The $S[\chi]$ comes out of the path integral, the $S[\varphi]$ gives the
usual $\varphi-$propagator and the $\varphi^3$ vertex while the last
four terms give the additional interaction vertices. 

Since in the 1PI diagrams we are interested in, internal lines are only
$\varphi-$lines and the external lines are only $\chi$ lines, the first
three of the four additional vertices are irrelevant and only the
$\varphi^3$ and $\varphi^2\chi$ vertices remain. It is a simple exercise
in power counting to determine the superficially divergent 1PI diagrams
with no internal $\chi-$lines and no external $\varphi-$lines. Consider
a 1PI diagram with $n_{0,3}$ number of $\varphi^3$ vertices and
$n_{1,2}$ number of $\chi\varphi^2$ vertices. Then the internal
$\varphi-$lines is given by $2I_{\varphi} = 3n_{0,3} + 2n_{1,2}$ and the
external $\chi-$lines if given by $E_{\chi} = n_{1,2}$. It is easy to
see that the superficial degree of divergence, $D$, is then given by $D
= 6 - 2E_{\chi}$.  Notice that $n_{0,3}$ drops out of the degree of
divergence. The renormalizability of the theory is thus manifest and
only the $2-$point and the $3-$point vertex functions are divergent.

How does renormalization work in background field method? Re-scale the
fields as $\varphi \to \sqrt{Z_{\varphi}}\varphi\ , \chi \to
\sqrt{Z_{\chi}}\chi$.  Then the $\varphi-$propagator will get a factor
of $Z_{\varphi}^{-1}$.  The $n_{0,3}$ vertices give a factor of
$(Z_{\varphi}^{3/2})^{ n_{0,3}}$ and the $n_{1,2}$ vertices  give a
factor of $Z_{\varphi}^{n_{1,2}} \cdot (Z_{\chi}^{1/2})^{n_{1,2}}$.  The
$I_{\varphi}$ lines give a factor of $Z_{\varphi}^{-I_{\varphi}}$.
Clearly, the factors of $Z_{\varphi}$ cancel out since $-I_{\varphi} +
\Case{3}{2}n_{0,3} + \Case{2}{2}n_{1,2} = 0$. We may thus choose not to
renormalise the $\varphi$ fields. The left over factor is
$Z_{\chi}^{n_{1,2}/2} = Z_{\chi}^{E_{\chi}/2}$ as expected.

Notice that the diagrams that need to be computed for any vertex
function, are exactly the same as without background field except for
the vertex factors from the vertices connecting the external lines. The
renormalization procedure(s) then proceed as usual eg as discussed in
the subsection \ref{Phi36Example}.

For scalar fields used in the illustration, there is no particular
advantage. But with gauge theories the method allows enormous
simplification \cite{Abbot}. This is beyond the scope of this course
though.

Returning to our generating functionals $Z, W, \Gamma$; we have seen how
the perturbative diagrams for the Green's functions and the $S-$matrix
elements (through the vertex functions) can be subsumed by formal
manipulations with the basic path integral. The utility of the formalism
goes beyond perturbation theory and computations of cross-sections.
When it comes to gauge theories, especially the non-abelian ones
with/without spontaneous symmetry breaking, the generating functional
provide a convenient tool to establish renormalizability and unitarity.
The proof of the Ward identities - constraints or relations enforced by
gauge invariance on different $n-$point functions - is most economical
in this framework. The existence of {\em anomalies} - violation of
classical symmetries at the quantum level and their treatment is also
much more transparent in these functional methods.

The basic path integral is defined without any pre-supposition of
perturbation theory. If the basic definition can be implemented  eg via a
lattice discretization, we can potentially have a non-perturbative
handle through the computation of the vertex functions.


\newpage
\section{Closing Remarks}\label{Closing}

The lectures were organized around four strands. The main references
used for each strand are also cited. 
\begin{enumerate}
	\item Poincare symmetry realization and its consequences
		\cite{Weinberg, Wetterich}:
	
	This involved the particle representations which identified the
	attributes of the permissible quanta; 

	Manifest covariance needs the field representation and the
	irreducibility condition imposed the {\em linear} field equations; The
	unitarity analysis revealed anti-particles;

	Action formulation showed the free classical fields as a
	dynamical system of infinitely many independent oscillators;
	Introduction of interaction with non-dynamical source brought in
	the various Green's function, the Feynman propagator conflicting
	with a causally consistent classical interpretation;

	Quantization of free fields required their covariance to be
	formulated differently and led to the CPT theorem; requirement
	of causality lead to the spin-statistics theorem; 

	States of Free fields naturally contain particles as the relativistic
	wave packets and pave the way for recovering the usual
	non-relativistic limit. 

	\item Interacting quantum fields \cite{ReedSimon, Newton,
		BjorkenDrellII}:

	This was discussed within the context of scattering experiments;
	corresponding scattering theory postulates were discussed
	leading to the Kallen-Lehmann spectral representation; 
	vanishing of the 1-point function is required for the 
	stability of (unique) vacuum; 

	LSZ reduction was discussed to obtain the S-matrix elements to
	vacuum expectation values of time ordered fields; covariant
	perturbation theory was premised on the interacting fields have
	the same canonical commutation relation as the in/out fields
	which are deemed {\em physical - masses and residues at poles};
	S-matrix elements are given entirely in terms of free quantum
	fields with arbitrarily specified but Lorentz invariant
	interactions and a diagrammatic recipe follows; 

	This characterizes interacting quantum fields as facilitating
	discrete transactions of energy/momentum/spin/charge etc via
	{\em virtual} quanta.  The off-shell quanta do not satisfy the
	equations of motion and hence represent a model of quantum
	fluctuations.

	\item Application to the Yukawa interaction and QED up to 1-loop
		\cite{PeskinSchroder, Srednicky}: 

	The early successes at tree level were discussed followed by the
	Radiative corrections; 

	The IR divergences highlighted the care needed in applying the
	formalism to what is actually measured;

	The UV divergences highlighted the unacceptable, unbounded
	dominance of the quantum fluctuations and questioned the purely
	theoretical parameters in the Lagrangian (`bare'); Some
	renormalization procedure needs to be adopted to identify the
	physically measured parameters; 

	The generic problem is to be handled recursively using counter
	terms which were illustrated at the 2-loop level; The
	renormalization process also led to the renormalization group
	equation.

	\item The path integral formulation \cite{AbersLee, Srednicky,
		Gross}: 

	The classical action has a direct and central role in computing
	quantum transition amplitudes; The path integral provides
	generating functionals for various n-point functions and is a
	powerful tool for both formal studies as well as for richer
	non-abelian gauge theories with/without spontaneous symmetry
	breaking (SSB);

	When extended to the more general theories - non-abelian gauge
	theories with/out SSB, it follows that one needs to identify the
	`correct fields' and their interactions {\em before computing
	the various n-point functions}; The qualitative physics guesses
	are invoked to propose the choice of {\em the unique vacuum}.
	In the gauge theory context, for instance, gauge invariance
	under large gauge transformations (non-trivial at infinity)
	reveal the $\theta-$vacua and some choice is made by nature. 

%

	%
\end{enumerate}

\newpage

\end{document}